\documentclass[12pt,a4paper,titlepage,pdftex]{book}
\title{\Huge \scshape TESIS }
\author{\scshape David Yllanes Mosquera}
\date{ }

\usepackage{tesis}

\newcommand{\bx}{{\boldsymbol x}}
\newcommand{\by}{{\boldsymbol y}}
\newcommand{\tw}{t_{\mathrm w}}
\newcommand{\Var}{\mathrm{Var}}
\newcommand{\Cov}{\mathrm{Cov}}

\newindex[myThePage]{default}{idx}{ind}{Alphabetic index}
\makenomenclature

\renewcommand{\sectionmark}[1]{\markright{\thesection\ --- #1}}
\fancypagestyle{plain}{%
\fancyhf{} 
\fancyfoot[C]{\oldstylenums\thepage} 

}

\makeatletter
\preto\@tabular{\fontfamily{pplx}\selectfont}
\makeatother

\titleformat{\chapter}[display]
{\bfseries\LARGE} {\filleft\MakeUppercase{\chaptertitlename}
\Huge\Roman{chapter}} {2ex} {\titlerule
\vspace{1.5ex}%
\filright}
[\vspace{1.5ex}%
]

\titleformat{\section}[block]
{\Large\normalfont}
{\bfseries\thesection}{.5em}{\titlerule\\[.8ex]\bfseries}

\begin{document}



\thispagestyle{empty}
\vspace*{2cm}

\begin{center}
{\Huge \bfseries
Rugged Free-Energy Landscapes\\ in Disordered Spin Systems}



\vspace*{1.3cm}
{\large 
Memoria de tesis doctoral presentada por\\
\textsc{David Yllanes Mosquera}
\vspace*{1cm}

Directores\\
\textsc{Luis Antonio Fernández Pérez}\\
\textsc{Víctor Martín Mayor}
\vspace*{2.5cm}}

\includegraphics[scale=0.5]{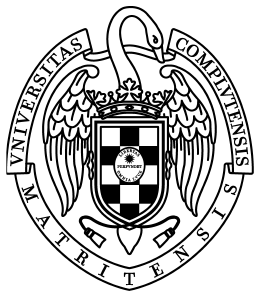}
\vspace*{1cm}

Universidad Complutense de Madrid\\
Facultad de Ciencias Físicas\\
Departamento de Física Teórica I\\
MMXI

\end{center}


\newpage
\thispagestyle{empty}
$ $
\newpage

\thispagestyle{empty}
\vspace*{7cm}
\begin{flushright}
\itshape A mi hermano
\end{flushright}

\chapter*{Preface}
This dissertation reports the research I have carried out 
as a PhD student in the Statistical Field 
Theory Group of the UCM,
where I have been privileged to work and learn under
the guidance of Luis Antonio Fernández and Víctor 
Martín. They not only form a truly remarkable scientific partnership, 
but have also been the best advisors I could ever have hoped to have.

My work during this time constitutes an attempt to make
some headway in the field of complex condensed matter systems, \index{complex systems}
concentrating on disordered spin models and
taking a Monte Carlo approach. In particular, 
this dissertation deals with two archetypical systems:
the diluted antiferromagnet in a field and the Edwards-Anderson \index{spin glass}
spin glass. The former is studied with Tethered Monte Carlo,  \index{DAFF}
a formalism developed during this thesis.                     \index{tethered formalism}

As to the Edwards-Anderson spin glass, I have been fortunate
to have access to \textsc{Janus}, a special-purpose \index{Janus@\textsc{Janus}}
machine that outperforms conventional computing architectures
by several orders of magnitude in the Monte Carlo simulation
of spin systems. This would be akin to being one of a few 
particle physicists with access to a newer, vastly more powerful
collider (\emph{si parva licet componere magnis}) and has allowed
our group to tackle head-on a much studied model and still see 
some new physics. From the point of view of a PhD student, 
it has been a unique learning opportunity. \textsc{Janus} is the fruit of a collaboration
of physicists and engineers from five universities in Spain
and Italy. The project is directed by Alfonso Tarancón
and has as scientific coordinators Víctor Martín Mayor, Giorgio
Parisi and Juan Jesús Ruiz Lorenzo.  My own participation has been, of course,
only as a very junior member of a large collaboration, so I only include
in this dissertation those physical studies where I carried out 
a major fraction of the work.

\section*{Agradecimientos}
Esta tesis sólo ha sido posible gracias al constante
e inestimable apoyo de muchas personas. En primer lugar, 
Luis Antonio  Fernández y Víctor Martín merecen una nueva mención
por su dedicación y accesibilidad, que ha ido mucho más allá de 
la habitual relación entre directores y doctorando. Mi grupo
inmediato de trabajo lo completa Beatriz Seoane, brillante estudiante
a quien espera una muy prometedora carrera científica.

Durante estos años he sido miembro del Departamento 
de Física Teórica~I de la Universidad Complutense de Madrid, 
donde recibí una acogida ejemplar. Siento un agradecimiento
especial hacia nuestro director, Antonio Muñoz, en cuyo despacho 
fui un intruso durante tres años, hasta que llegó el prometido traslado al
módulo oeste de la facultad. No puedo dejar de recordar a Chon, una 
auténtica institución de la facultad que ha dejado huella en 
generaciones de físicos teóricos.

Agradezco también a todos los miembros de la Janus Collaboration
todo lo que he aprendido de ellos. Dentro de ella, quiero hacer
mención especial a los estudiantes que me han precedido, Antonio
Gordillo y Sergio Pérez; así como a los que me siguen, Raquel Álvarez,
José Miguel Gil y Jorge Monforte. también a Juan Jesús Ruiz,
quien se tomó la molestia de preparar un cursillo intensivo
sobre vidrios de espín para los miembros noveles. Estoy seguro
de que no hablo solamente por mí cuando digo que en ese par de días
nos aclaró muchas cosas con las que llevábamos peleando largo
tiempo. Valoro, asimismo, enormemente haber tenido la oportunidad
de aprender a través de la interacción con 
científicos de primer nivel como Enzo Marinari, 
Denis Navarro, Giorgio Parisi y Lele Tripiccione.

Por supuesto, Janus no habría sido posible sin la visión
y carácter emprendedor de su director, Alfonso Tarancón,
quien, no contento con ello, dirige también el Instituto
de Biocomputación y Física de Sistemas Complejos (BIFI)
en la Universidad de Zaragoza. A esta institución, de la que 
soy miembro, debo agradecer los enormes recursos computacionales
que ha puesto a mi disposición durante esta tesis
(que, aparte de Janus, se cuentan
por varios millones de horas de CPU). Quiero recordar 
especialmente a los administradores de \emph{Piregrid}:  \index{Piregrid@\emph{Piregrid}}
Jaime Ibar, Patricia Santos y Rubén Vallés.

Agradezco mucho a Leticia Cugliandolo y a Claudio
Chamon, así como a todos los miembros de sus grupos, 
su hospitalidad y enseñanzas  en sendas estancias
en el LPTHE de la Université Pierre et Marie Curie (París) y
en el Condensed Matter Theory Group de la Boston 
University.

También quiero agradecer a José María Martín y a Luis Garay, con quienes
di mis primeros pasos hacia una carrera investigadora durante la 
licenciatura. Asimismo, agradezco a Antonio Dobado y a Felipe Llanes
haberme introducido en el mundo de la docencia, con mi participación
como ayudante en sus asignaturas de Electrodinámica Clásica
y Mecánica Cuántica.

Mis amigos en Madrid y La Coruña han sido un pilar imprescindible
durante todo este tiempo. Muchas gracias a todos, en especial
a Antonio Arévalo, Eva Béjar, Pedro Feijoo, Rosa Gantes, Carlos Lezcano, Jesús
Pérez y Antón Sanjurjo.

Dejo para el final al grupo más importante: mi familia.
He sido afortunado en muchas cosas en mi vida, pero en nada
tanto como en el cariño y apoyo de la familia que me ha tocado
tener. Todos ellos tendrán siempre mi más sentido agradecimiento, en especial
mis padres, que tantos sacrificios han hecho por mí. A mi hermano, y amigo,
Daniel va dedicada esta tesis.

Durante este trabajo he estado financiado primero por 
una beca del BIFI y luego por una beca FPU del Ministerio 
de Educación. También he recibido apoyo de los proyectos
\textsc{fis}2006-08533 y \textsc{fis}2009-12648
del MICINN y de los Grupos UCM -- Banco de Santander.
Agradezco finalmente a la Red Española de Supercomputación
el haberme concedido alrededor de cuatro millones de horas  
de cálculo en el ordenador \emph{Mare Nostrum}. \index{Mare Nostrum@\emph{Mare Nostrum}}

\begin{flushright}
\textsc{David Yllanes Mosquera}\\
\itshape
Universidad Complutense, Madrid, junio de 2011
\end{flushright}


\clearpage
\tableofcontents
\phantomsection

\clearpage
\listoffigures
\addcontentsline{toc}{chapter}{\protect\numberline{}Figures}
\phantomsection

\clearpage
\listoftables
\addcontentsline{toc}{chapter}{\protect\numberline{}Tables}

\phantomsection
\renewcommand{\nompreamble}{\markboth{Notation}{Notation}}
\printnomenclature[2.6cm]
\addcontentsline{toc}{chapter}{\protect\numberline{}Notation}


\part{Introduction}\label{part:intro}
\chapter{General introduction}\label{chap:intro} \index{complexity}\index{glasses}\index{protein folding}
Modern physics is steadily broadening its scope and     \index{complex systems|(}
tackling increasingly complex systems, whose rich collective
behaviour is not easily explained from the 
often simple nature of  their constituent parts.
Thus, a lot of attention is being
focused on understanding, from a fundamental point of view, 
an extremely diverse class of problems, 
ranging from vortex glasses in high-temperature        \index{superconductors}
superconductors to biological macromolecules.  \index{biomolecules}
The featured physical phenomena 
can be as exotic as the colossal magnetoresistance of some
manganites~\cite{dagotto:01,coey:09,levy:02}, \index{colossal magnetoresistance}
or as familiar as the formation of glass~\cite{angell:95,debenedetti:97,debenedetti:01}. 
 The latter
constitutes a particularly conspicuous example of an everyday material
whose microscopic description remains, in the words
of P. W. Anderson~\cite{anderson:95}, `probably
the deepest and most interesting unsolved problem
in solid state theory'.
On a different vein, the study complex physical systems
has deep relations to the field of computational complexity \index{computational complexity}
and NP-incompleteness~\cite{mezard:02,zecchina:06}. 

The enormous variety of problems, often straddling the boundaries
between physics, chemistry and biology, seems to suggest that the label
of `complex system' bears little meaning, since it seems that each class of systems
must surely be studied separately. Actually, behind the diversity
 we can find key unifying features, which has motivated attempts to find 
some solid common ground for a joint treatment of complexity.

\begin{figure}[p]
\centering
\includegraphics[width=.7\linewidth]{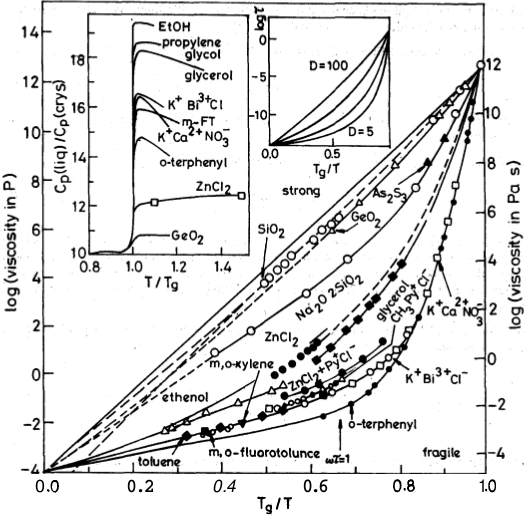}
\caption[Angell plot]{Angell plot,
taken from~\cite{angell:95}, showing
the viscosity of many glass-forming liquids.
The horizontal axis is rescaled in terms
of a glass temperature $T_\text{g}$, defined as 
that where the system's viscosity reaches the 
value of $10^{13}$~P. In this representation, 
the deviation from the ideal Arrhenius law (leftmost
straight line) seems completely encoded by 
the derivative at $T_\text{g}$. Notice 
that the values of the viscosity span $15$ orders
of magnitude.
\label{fig:INTRO-angell}
\index{glasses}
\index{viscosity|indemph}
\index{Angell plot|indemph}}
\vspace*{2cm}

\includegraphics[width=.7\linewidth]{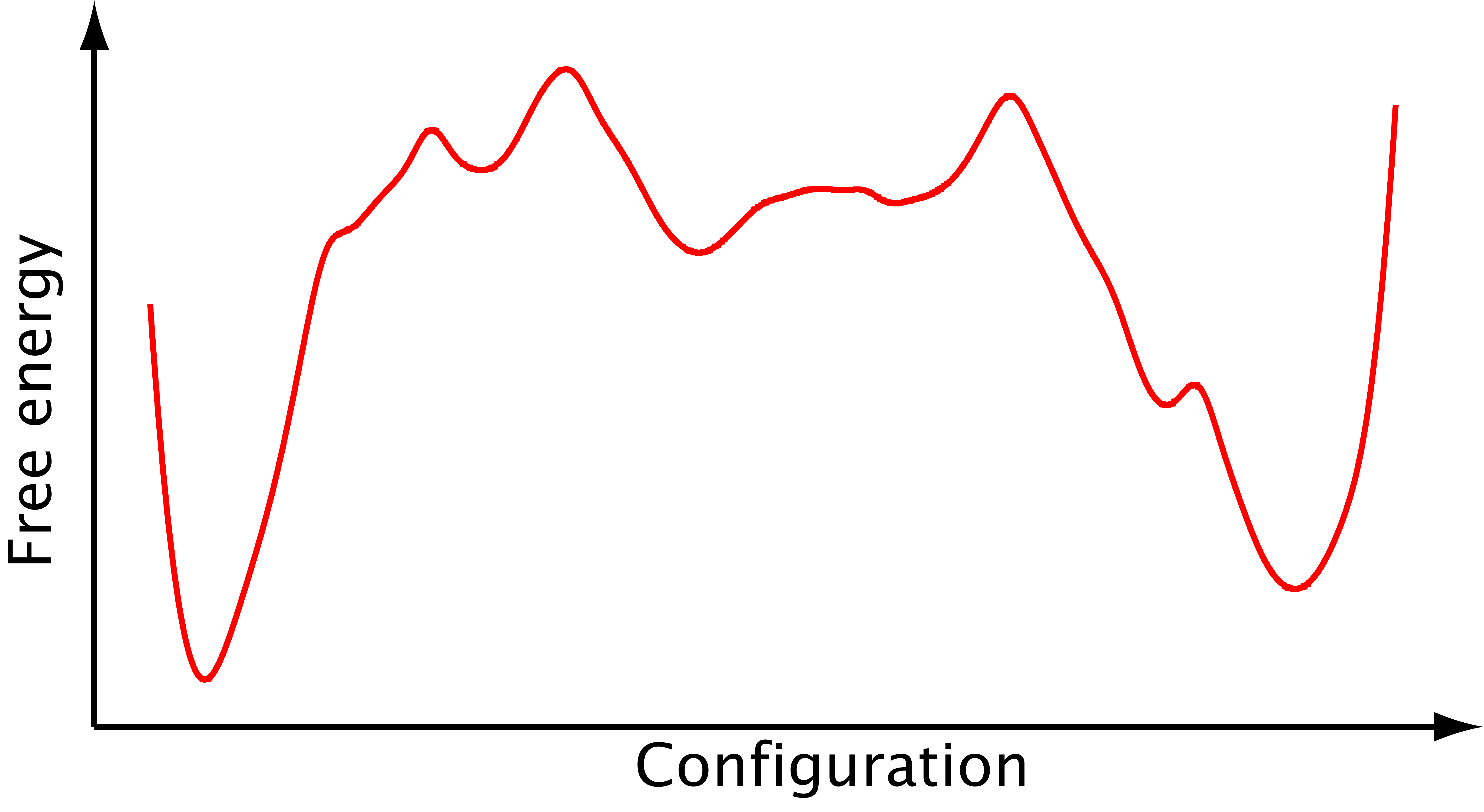}
\caption[Rugged free-energy landscape]{Idealised picture of a complex system. The free-energy
profile contains many local minima, which define metastable states
with exponentially long escape times.
\label{fig:INTRO-landscape}
\index{free energy!landscape|indemph}}
\end{figure}

In this sense, the best hope of the fundamental physicist \index{universality}
is the notion of universality~\cite{cardy:96,amit:05,zinn-justin:05}. In general, a strict microscopic
description of a complex system is a daunting task: one has to account
for many degrees of freedom, whose interactions
follow complicated laws. Fortunately, one can often identify
a few crucial scaling variables, whose evolution encodes the behaviour  \index{scaling}
of the whole system. More than that, by expressing all the more complicated
quantities in terms of these, very different systems can be shown 
to have the same qualitative behaviour. Hence, all these systems can be understood through the 
study of their simplest representative.

Perhaps the most striking example of this universal behaviour is 
the celebrated Angell plot, of which we show an example
in Figure~\ref{fig:INTRO-angell}.
In it, glass-forming liquids with completely different compositions 
and qualitatively different temperature dependencies of the viscous flow
are classified according to their `fragility'. This
is  defined as the (logarithmic) derivative of the \index{viscosity}
viscosity at the glass transition temperature and, as it turns out, 
it characterises the material's deviation from the Arrhenius law, 
along 15 orders of magnitude. Scant few physical quantities
have ever been measured along so wide a range but, beyond that,
\index{fluctuation-dissipation}
the figure encloses very deep physics.

For instance,
fluctuation-dissipation relations~\cite{kubo:57} allow
us to translate viscosity into time. Therefore, the plot
is showing us a situation where relaxation times cross
over from a microscopic to a macroscopic range. Notice 
that the scaling temperature in this plot is chosen
as the point where the viscosity reaches $10^{13}$~P
(equivalent to relaxation times of one hour). More than that, and
one has to accept working out of equilibrium, which, 
in some fields, is a tough pill to swallow.

In the context of magnetic systems ---where, unlike with glasses,
one is sure of being below a phase transition--- the off-equilibrium
regime is a completely natural experimental environment and has
been for some time. A classic application is the study of coarsening, \index{coarsening}
a kind of dynamics characterised by the growth of compact coherent
domains.  In this case, an especially powerful version of 
universality operates, aptly called superuniversality~\cite{fisher:86}. \index{superuniversality}
According to it, all the spatial and temporal scales during the dynamics \index{coherence length}
are encoded in the growth of a coherence length, which indicates the size
of the coherent domains (cf. Chapter~\ref{chap:sg}).

In general, we can say that the most
common feature of complex systems is an incredibly slow dynamical  \index{aging}
evolution, or aging~\cite{struick:78,bouchaud:98}. The study of
non-equilibrium relaxation is, thus, very important and often 
the only accessible experimental regime.

In order to explain  this sluggishness, the most 
often invoked defining characteristic of complex 
systems is the picture of a `rugged  \index{free energy!landscape}
free-energy landscape' \cite{frauenfelder:97,janke:08}.
The configuration space is pictured as having many
valleys, defining metastable states  where a configuration
is much more favourable than neighbouring ones (Figure~\ref{fig:INTRO-landscape}).\index{metastability}
The system in its evolution, then, must 
jump from one local minimum to another through 
rare-event states, causing the slow dynamics.

The causes of this ruggedness are diverse. For some materials,
it may be due to the presence of impurities or other defects, 
which hinder the physical evolution. In others, the sluggish
behaviour has been modelled as a hierarchically constrained dynamics,
consisting in the sequential relaxation of different degrees of freedom, 
from the fastest to the slowest~\cite{palmer:84}.

Sometimes one of the valleys dominates and the free-energy profile
is funnel shaped. This is the case, for instance, of protein 
folding, where the native configuration defines an absolute minimum.
For other systems, on the other hand, there can be many 
equally favourable configurations, so one must take several 
metastable states into account even when defining the equilibrium.
The difference between both cases is not idle: proteins quite 
obviously are able to fold into their equilibrium configuration
very quickly (in human terms), while glassy systems with metastable
behaviour are perennially out of equilibrium.

Since in this latter case the equilibrium phase is experimentally unreachable,
determining what (if any) bearing it 
has on the dynamical evolution (what we shall call
the `statics-dynamics relation') is a non-trivial problem. \index{statics-dynamics equivalence}

This discussion notwithstanding, we must caution the reader
that Figure~\ref{fig:INTRO-landscape} is, at best, a 
metaphor. In order to turn it into real physics one must
first, at the very least, identify one (or more) appropriate \index{reaction coordinate}
reaction coordinates capable of actually labelling the different 
minima. This requires a great deal of insight into the system's 
physics and still leaves unresolved the non-trivial step
of actually computing the free energy.                       \index{free energy}

The quantitative investigation of these two issues (statics-dynamics
and the  free-energy landscape)  will constitute
the main themes of this dissertation. We shall work in
the context of disordered magnetic systems,               \index{disorder}
long considered prime examples of complexity.\footnote{We shall give 
a detailed introduction to these systems
in their respective chapters, for now we limit
the discussion to a few general comments.} 
One may think that the introduction of disorder
cannot be responsible for very exciting changes 
in a physical system. This is true in some cases
(after all, even the most perfect experimental
sample has some impurities, yet we can still talk 
of crystals or ferromagnets), but not in general.  
For some strongly disordered systems, we shall see,  \index{ferromagnets}
the impurities have a dramatic effect both
in a technical sense (being a relevant perturbation
in a renormalisation-group  \index{renormalisation group}
setting) and in a very physical sense. Consider, for instance,         
the example of Anderson localisation~\cite{anderson:58}, capable  \index{Anderson localisation}
of turning metallic systems into insulators.

\begin{figure}[t]
\centering
\includegraphics[width=0.3\linewidth]{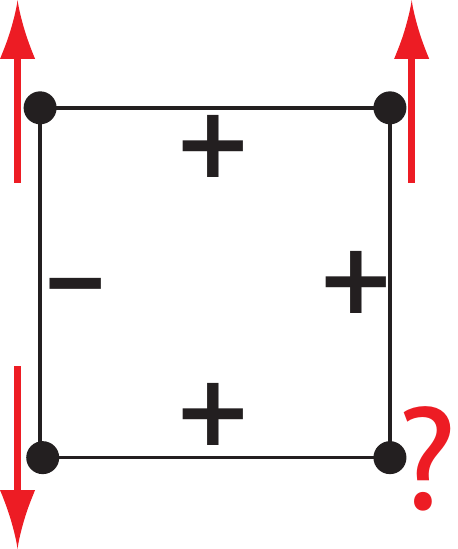}
\caption[Frustrated plaquette]{%
A small portion of a spin-glass lattice. The 
spins on the nodes can only take the `up' or `down' orientation.
They are joined along nearest-neighbour links
by a mixture of ferromagnetic ($+$) and 
antiferromagnetic ($-$) couplings. The former 
favour aligned pairs of spins and the latter
favour antiparallel ones. With the arrangement
shown in the figure there is no way of orienting the 
spins so that all of the bonds are satisfied.
\index{frustration|indemph}
\label{fig:INTRO-frustrado}}
\end{figure}

One of the simplest mechanisms responsible for the complexity
of disordered systems is that of frustration~\cite{toulousse:77}, as very clearly           
illustrated in the case of spin glasses (see, e.g.,~\cite{binder:86,mezard:87}
and cf. Chapter~\ref{chap:sg}).\index{spin glass}
These are magnetic alloys in which the interactions between
the spins are in conflict. The disorder manifests as a mixture
of ferromagnetic and antiferromagnetic couplings. Thus, 
even for the lowest-energy configuration some of the bonds
are necessarily frustrated, as we
see in Figure~\ref{fig:INTRO-frustrado}. This
makes it exceedingly difficult for the physical 
system to find the equilibrium state (and no less for 
the physicist performing a computation). Furthermore, many
configurations have a similar degree of frustration and are
therefore equally favourable, giving rise to a free-energy 
landscape with many relevant metastable states.

Yet not only the equilibrium is complicated but also the dynamical
behaviour. In the case of spin glasses, to the aging
behaviour \index{aging} \index{memory} \index{rejuvenation}
we have to add phenomena such as rejuvenation or memory effects~\cite{jonason:98}.

The choice of disordered magnetic systems as paradigmatic
models for complexity is mainly due to their 
permitting more precise experimental studies
than other classes of complex systems~\cite{mydosh:93,belanger:98}.
There are both technical and physical reasons for this, as we shall
see later. 

On the theoretical front, on the other hand, these systems are at least easy
to model with deceptively simple lattice systems.
The solution of these models is another matter entirely.
Indeed, disordered systems have often defied traditional, 
and powerful, analytical tools such a perturbation           \index{perturbation theory}
theory~\cite{dedominicis:06}. In the case of spin glasses, 
even the solution of such a gross simplification  \index{mean field}
as the mean-field approximation has been 
a veritable \emph{tour de force}~\cite{parisi:79b,parisi:80}.
Other systems, such as the random field Ising model, are  \index{RFIM}
seemingly more amenable to a perturbative treatment~\cite{dedominicis:06,nattermann:97}
but remain very poorly understood, the analytical approach failing to 
analyse the critical behaviour convincingly.

In the last decades, a third avenue has opened for basic research: the 
computational approach, of which the most salient example is Monte Carlo \index{Monte Carlo method}
simulation~\cite{landau:05,rubinstein:07}.
Unfortunately, in the case of disordered systems traditional 
Monte Carlo methods suffer from the same problems as experiments do, to 
an even greater degree. A simulation of a rugged free-energy landscape \index{free energy!landscape}
for a finite lattice will get trapped in the local minima, just as     \index{critical slowing down!exponential}
an experiment, with an escape time that grows as the exponential 
of the free-energy barrier, which in turn goes as a power of the lattice size.
As a consequence, for many models of interest only very small
systems can be thermalised, making extrapolation to the thermodynamical
limit very complicated.  \index{thermalisation} \index{thermodynamical limit}
Many simulation algorithms have been proposed to address this problem.
However, as a general rule efficient innovations require some previous
knowledge, or at least a somewhat detailed working hypothesis, of 
the underlying physics. Therefore, the investigation of new Monte Carlo
dynamics must not be considered in isolation, but as 
an enterprise that should be undertaken jointly with a thorough study
of challenging physical problems. 

An alternative is simply to eschew equilibrium and attempt a Monte Carlo
reproduction of an experimental dynamics. This is simple enough in principle, 
the most straightforward Monte Carlo dynamics being good mock-ups 
of the physical evolution, and has the advantage of considering the system
in more controlled conditions that are possible in a laboratory. Unfortunately, 
current computers are several orders of magnitude too slow to reach 
the experimentally relevant time scales.

In short, the study of complex systems faces significant obstacles,
both experimental and theoretical. From a fundamental point of view,
far from discouraging further effort, these problems are at heart the reasons
why these systems are so interesting, constituting a constant reminder 
that the traditional tools of statistical mechanics must be continuously
complemented
and expanded. This in addition to the fact that by `complex systems' we encompass
such everyday materials as glasses, as well as systems with great technological
or medical relevance (colossal magnetoresistance oxides, proteins, etc.). \index{colossal magnetoresistance} \index{protein folding}

\section{Scope of this thesis}
This thesis is an attempt to provide a new outlook on complex systems, 
as well as some physical answers for certain models, taking a 
computational approach. We have focused on disordered systems,  \index{disorder}
addressing two traditional `hard' paradigmatic models 
in three spatial dimensions: the Edwards-Anderson spin glass  \index{Edwards-Anderson model} 
and the diluted antiferromagnet in a field (the physical realisation  \index{DAFF}
of the random-field Ising model). These systems have been              \index{RFIM}
studied by means of large-scale Monte Carlo simulations, exploiting a
variety of platforms (computing clusters, supercomputing facilities, grid \index{grid computing}
computing resources and even a custom-built special-purpose supercomputer). 
In accordance to the above discussion, the physical study has been taken hand in hand 
with the development of new Monte Carlo methods.                     \index{Monte Carlo method}

Indeed, at the foundation of the work reported herein is the development of
Tethered Monte Carlo, a general strategy for the study of rugged         \index{tethered formalism}
free-energy landscapes. This formalism provides a general method to 
guide the exploration of configuration space by constraining (tethering)
one or more parameters. In particular, one selects a reaction coordinate  \index{reaction coordinate}
(typically, but not necessarily, order parameters) capable of labelling 
the different local free-energy minima. A statistical ensemble is then constructed,
permitting efficient Monte Carlo simulation where these coordinates are fixed, 
avoiding the need to tunnel between competing metastable states. 
From these tethered simulations the Helmholtz potential associated \index{effective potential}
to the reaction coordinates is reconstructed, yielding all the information 
about the system.

This philosophy is applied first to ferromagnetic models (hardly complex \index{ferromagnets}
systems, but extraordinary benchmarks nonetheless) and then to the diluted 
antiferromagnet in a field. There it is showed that the tethered approach, far
from being a mere optimised Monte Carlo algorithm, is capable of providing
valuable information that would be hidden from a traditional study, thus
permitting a more complete picture of the physics. One of the more conspicuous
examples of this is that pictures such as Figure~\ref{fig:INTRO-landscape}, 
long treated as mere conceptual aides, have been turned into real   computations
of free-energy profiles. \index{free energy!landscape}

The next part of this dissertation is concerned with the Edwards-Anderson \index{Edwards-Anderson model}
spin glass. For this system, our physical understanding is not yet at 
a level that would permit a full tethered treatment. This notwithstanding, many
pointers are taken from the tethered philosophy, particularly in regards to 
the analysis of physical results. For the Monte Carlo simulation, the strategy  \index{spin glass}
has been mainly one of sheer brute force. Our work on spin glasses  \index{Monte Carlo method}
has been conducted within the Janus Collaboration, a joint effort
of researchers from five universities in Spain and Italy. This project
has as its main goal the construction and exploitation of \textsc{Janus}, \index{Janus@\textsc{Janus}}
a special-purpose computer optimised for spin-glass simulations, where
it outperforms conventional computers by several orders of magnitude.

Aside from having been carried out  with slightly different methods, our 
work on spin glasses is complementary to the rest of this thesis
in a major physical way: it makes a strong emphasis on off-equilibrium dynamics.  
Indeed, as we advanced in the previous discussion, experiments on spin glasses \index{statics-dynamics equivalence}
(and many other complex systems) are always performed in an off-equilibrium 
regime. Then, a very valid question arises: how relevant is it to know
the unreachable equilibrium phase?  Our outlook, as we said before, has been 
that equilibrium structures, though inaccessible, do condition 
the off-equilibrium evolution. This idea, long accepted as a working hypothesis,
is turned into a quantitative statement with the finite-time scaling 
paradigm. Time is treated as a state variable, much as pressure or temperature
in a traditional thermodynamical setting. A time-length
dictionary links off-equilibrium results 
in the thermodynamical limit for finite times with
equilibrium results for finite lattices.

The next section summarises the organisation of the rest of this dissertation. It should
be noted that each section contains a more detailed topical introduction, expanding
on the issues touched in this General Introduction.                   \index{complex systems|)}

\section{Organisation of this dissertation}
As discussed above, the work reported herein is concerned both with 
the study of paradigmatic complex systems in statistical mechanics
(the DAFF and the spin glass) and with the development of new \index{spin glass}
Monte Carlo and analysis methods. The rest of this dissertation is, therefore,  \index{DAFF}
organised thematically in the following way:
\begin{itemize}
\item Part~\ref{part:intro}, including this General Introduction, has the 
purpose of motivating our study and presenting our outlook 
on complex systems and how they should be treated.
We start by, very briefly, recalling some statistical mechanical concepts
relevant to the study of complex systems (Chapter~\ref{chap:disorder}).
This is followed by Chapter~\ref{chap:tmc}, already concerned with our 
Monte Carlo approach. In it we expand on the practical problems posed \index{Monte Carlo method}
by the numerical simulation of rugged free-energy landscapes and  \index{free energy!landscape}
motivate the tethered formalism as a way of removing or alleviating them.
This last chapter contains some material from our paper~\cite{martin-mayor:11}. \index{tethered formalism}
\item Part~\ref{part:tmc} deals with our work on new Monte Carlo
methods, motivated by the above considerations.
We start by detailing the construction of the tethered formalism in
Chapter~\ref{chap:tmc-formalismo}.
We then present a first demonstration of the method
in a straightforward application:  the $D=2$ Ising model (Chapter~\ref{chap:Ising}). \index{Ising model}
This is, of course, a well understood system, so our aim
is not so much presenting new physics as demonstrating our methods and how a tethered
study can provide a complementary picture to canonical methods.
Finally, Chapter~\ref{chap:cluster} proves that the tethered formalism is compatible \index{cluster methods}
with advanced Monte Carlo algorithms, in this case cluster methods.
We first introduced the Tethered Monte Carlo method in~\cite{fernandez:09}
and we later presented it in a more general context in~\cite{martin-mayor:11}.
Chapters~\ref{chap:Ising} and~\ref{chap:cluster}  contain material
from~\cite{fernandez:09} and~\cite{martin-mayor:09}, respectively.
\item Part~\ref{part:daff} discusses the first class of complex systems \index{DAFF}
we shall study: the diluted antiferromagnet in a field.
In Chapter~\ref{chap:daff-canonical}
we present the situation that existed prior to our work and demonstrate how this 
system is particularly ill-suited to a study with canonical methods. We then
tackle the problem with the tethered formalism in Chapter~\ref{chap:daff-tethered}, 
applying all the techniques introduced in Part~\ref{part:tmc}.
This chapter is a much expanded version of~\cite{fernandez:11b}, including
some material from~\cite{martin-mayor:11}, as well as some unpublished
results.
\item Part~\ref{part:sg} deals with spin glasses. In Chapter~\ref{chap:sg}
we introduce these systems, stressing their peculiarities
from an experimental point of view (with phenomena such as  \index{aging} \index{memory}\index{rejuvenation}
aging, rejuvenation, memory, etc.) and their resulting theoretical
importance as paradigmatic complex systems. Chapter~\ref{chap:sg-2} deals
in depth with one of the main themes of this thesis: the relationship
between equilibrium and non-equilibrium and introduces the
finite-time scaling framework. Finally, Chapter~\ref{chap:sg-3} studies
in detail the structure of the spin-glass phase in three spatial
dimensions. Our work on spin glasses, carried out within the \textsc{Janus}
collaboration, was published in~\cite{janus:08b,janus:09b,janus:10,janus:10b}.
Chapters~\ref{chap:sg-2} and~\ref{chap:sg-3} are a heavily reworked and reorganised \index{spin glass}
version of the results in those papers, including some unpublished material. \index{Janus@\textsc{Janus}}
\item Finally, Chapter~\ref{chap:conclusions} contains our conclusions.
\item We include several appendices. Appendix~\ref{chap:thermalisation} gives some notes
on how to assess thermalisation in Monte Carlo simulations. It starts reporting  \index{thermalisation}
some standard definitions but then describes some 
new techniques for complex systems (first introduced in~\cite{janus:10}).
Appendix~\ref{chap:correlated} presents 
some techniques for analysing the strongly correlated data produced in the Monte Carlo
simulation of disordered systems. The explained methods are illustrated with especially tough
examples and case studies taken from our work.  Appendix~\ref{chap:recipes} contains some practical
notes for an efficient numerical implementation 
of Tethered Monte Carlo. \index{tethered formalism}
Finally, Appendix~\ref{chap:janus} introduces the \textsc{Janus} special-purpose \index{Janus@\textsc{Janus}}
computer used in our spin-glass simulations and Appendix~\ref{chap:runs-sg}
gathers all the technical information on these runs (parameters, thermalisation, etc.).
\end{itemize}

\chapter{Statistical mechanics of disordered systems: basic definitions}\label{chap:disorder}
In this chapter we briefly recall some general definitions that will we employed
throughout this dissertation,  with the main purpose of fixing the notation
and introducing some notions particular to the study of disordered systems.
For general references on statistical mechanics or the theory
of critical phenomena see, e.g.,~\cite{landau:80,huang:87,amit:05,lebellac:91,zinn-justin:05,cardy:96}.
For disordered systems, see~\cite{mezard:87,young:98,dedominicis:06,dotsenko:01}.

\section{Statistical mechanics and critical phenomena}
We consider a system whose configuration can be specified 
by $N$ degrees of freedom $\{s_\bx\}$.
\nomenclature{$\{s_\bx\}$}{Spin configuration}
\nomenclature[N]{$N$}{Number of degrees of freedom (i.e., lattice nodes)}
In the canonical ensemble, the behaviour of the system is encoded 
in the partition function
\begin{equation}\label{eq:INTRO-Z}
Z = \sum_{\{s_\bx\}} \ee^{-\beta E(\{s_\bx\})}.
\nomenclature[E]{$E,e$}{Total energy of the system}
\nomenclature[Z]{$Z$}{Partition function}
\index{partition function}
\index{energy}
\end{equation}
In this equation, the sum is extended to all possible configurations, their 
relative weights depending on their energy $E$. The parameter $\beta=1/k_\text{B} T$
is the inverse temperature. We use units where the Boltzmann constant \index{Boltzmann constant}
is $k_\text{B}=1$, so $\beta = 1/T$.
\nomenclature[T]{$T$}{Temperature}
\nomenclature[beta]{$\beta$}{Inverse temperature, $\beta=1/T$}

Throughout this dissertation we study Ising spins, $s_\bx =\pm1$,
 on a square lattice
of $N=L^D$ nodes, where $D$ is the spatial dimension. Therefore, in the following
we often use the language and notation of magnetic systems, even if much of what we
say can be applied to more general systems.
\nomenclature[D]{$D$}{Spatial dimension of the system}
\nomenclature[L]{$L$}{Linear size of the square lattice}
\nomenclature[s]{$s_\bx$}{Ising spin, $s_\bx=\pm1$}

We define an \emph{observable} $O(\{s_\bx\})$ 
\nomenclature[Observable]{Observable}{A function of the spin configuration, $O(\{s_\bx\})$}
as any function of the spin configuration. In the context of Monte Carlo
simulations, where we estimate $Z$ by means of a random walk in 
configuration space, we use the word \emph{measurement} for a single evaluation 
\nomenclature[Measurement]{Measurement}{A single evaluation of an observable during the simulation}
of the observable during the simulation.

The total energy of the system can often be written in the following way
\begin{equation}\label{eq:INTRO-H}
E(\{s_\bx\}) = U(\{s_\bx\}) - h X(\{s_\bx\}),
\nomenclature[U]{$U,u$}{Interaction energy of the spins}
\nomenclature[h]{$h$}{Applied field}
\index{field, external}
\end{equation}
where $U$ is the interaction energy of the spins and $h$ is an 
external field, coupled to some reaction coordinate $X$. 
In our case, $U$ will be of the form of a two-spin interaction
\begin{equation}
U(\{s_\bx\}) = -\sum_{\bx,\by} J_{\bx \by} s_\bx s_\by,
\nomenclature[J]{$J_{\bx\by}$}{Coupling between sites $\bx$ and $\by$}
\end{equation}
where the $J_{\bx \by}$ are the couplings. For instance, for 
the  ferromagnetic Ising model, \index{Ising model}
$J_{\bx\by}=1$ if $\bx$ and $\by$ are nearest neighbours on the lattice
and zero otherwise. We represent this sort of nearest-neighbours interaction
as
\begin{equation} 
U(\{s_\bx\}) = - \sum_{\braket{\bx,\by}} s_\bx s_\by.
\nomenclature{$\braket{\cdot,\cdot}$}{Summation restricted to first neighbours}
\end{equation}

In this context, the most straightforward reaction coordinate is the 
magnetisation $M$,
\begin{equation}
M(\{s_\bx\}) = \sum_\bx s_\bx,
\nomenclature[M]{$M,m$}{Magnetisation}
\index{magnetisation}
\end{equation}
although we shall consider other kinds.

Therefore, the canonical ensemble describes the system at 
fixed temperature and applied field. The basic 
thermodynamic quantity is the free-energy density\footnote{%
This is often defined with a different normalisation, 
$\mathcal F_N = -\frac{k_\text{B}T}{N} \log Z$, but we shall find 
our definition more convenient later on.}
\begin{equation}
F_N(\beta, h) = \frac{1}{N} \log Z(\beta,h).
\nomenclature[F]{$F_N$}{Gibbs free-energy density}
\index{free energy}
\end{equation}

The average value of an observable in the canonical ensemble
is denoted by
\begin{equation}
\braket{O} = \frac1Z \sum_{\{s_\bx\}} O(\{s_\bx\})\ \ee^{-\beta E(\{s_\bx\})}.
\nomenclature[ 1]{$\langle{O}\rangle$}{Canonical expectation value of $O$}
\index{canonical ensemble}
\end{equation}

For quantities such as $M$ or $E$ we distinguish the extensive 
version from the density by the use of uppercase and lowercase
symbols, respectively,
\begin{align}
M &= N m,& E &= N e.
\end{align}
Notice that we do not use this convention for the free energy, $F_N$, which
is also a density, but only for observables.

\subsection{Legendre transformation}\label{sec:INTRO-legendre}
Throughout Part~\ref{part:tmc}, we shall find it interesting
to consider an alternative ensemble where it is the reaction 
coordinate $m$, and not the field $h$, which is kept fixed.
In classical thermodynamics, this is accomplished through the
Legendre transformation, \index{Legendre transformation}
defining a new basic potential. 
\begin{equation}\label{eq:INTRO-legendre-inf}
\varOmega_N(\beta,m) = \beta m h - F_N(\beta,h).
\nomenclature[Omega]{$\varOmega_N$}{Helmholtz free energy or efective potential}
\index{effective potential}
\end{equation}
We say $\varOmega_N$ is the Helmholtz free energy ($F_N$ is the Gibbs 
free energy). Since these names are often exchanged in the literature, 
we avoid confusion by reserving the name `free energy' for $F_N$,
while we shall call $\varOmega_N$ `effective potential', borrowing
the name from the context of quantum field theory.  
If $F_N$ is a convex function 
of $h$, this operation allows us to define $\varOmega_N$
as a convex function of $m$. However, for  disordered
systems with rugged free-energy landscapes $\varOmega_N$ is never
convex for finite $N$. Therefore, we shall consider the following
alternative representation of the transformation,
\begin{equation}\label{eq:INTRO-legendre}
Z_N(\beta,h)= \ee^{N F_N(\beta,h)}=\int \mathrm{d} m \ \mathrm{e}^{N[\beta hm -\varOmega_N(\beta,m)]}\,,
\end{equation}

The conjugate nature of $m$ and $h$ can be summarised by 
the following formulae
\begin{align}\label{eq:INTRO-conjugate}
\braket{m} &=\left. \frac1\beta \frac{\partial F_N}{\partial h}\right|_\beta, &
\braket{h}_m &=\left. \frac1\beta \frac{\partial \varOmega_N}{\partial m}\right|_\beta,
\end{align}
where by $\braket{\cdots}_m$ we denote the expectation value in the 
fixed-$m$ ensemble defined by $\varOmega_N$.

\subsection{Critical phenomena and exponents}\label{sec:INTRO-critical-exponents}
We shall often consider the behaviour of physical systems
in the neighbourhood of a second-order phase transition, 
where the system approaches continuously a state at which 
the scale of correlations becomes unbounded. In particular, 
if we write the correlation between sites $\bx$ and $\by$
as 
\begin{equation}
\braket{s_\bx s_\by} \xrightarrow{\ \ |\bx-\by|\to\infty\ \ }
\exp(- |\bx-\by| /\xi),
\nomenclature[Tc]{$T_\text{c}$}{Critical temperature}
\nomenclature[xi]{$\xi$}{Correlation length}
\index{correlation function (equilibrium)!spatial}
\index{correlation length}
\end{equation}
then, in the thermodynamic limit, the correlation length $\xi$ diverges as we approach
the critical temperature $T_\text{c}$. We characterise the divergence
by a critical exponent $\nu$
\begin{equation}\label{eq:INTRO-nu}
\xi \sim |T-T_\text{c}|^{-\nu}\, .
\nomenclature[nu]{$\nu$}{Critical exponent of the correlation length}
\index{critical exponent}
\index{critical exponent|nu@$\nu$}
\end{equation}

This behaviour is not exclusive of the correlation length, many
other quantities either diverge or vanish as we approach $T_\text{c}$.
Therefore, one defines additional critical exponents: 
\begin{itemize}
\item Response of the system to an infinitesimal field $h$ (for instance, 
the magnetic susceptibility),
\begin{equation}
\chi \sim |T-T_\text{c}|^{-\gamma}.
\index{critical exponent|gamma@$\gamma$}
\nomenclature[gamma]{$\gamma$}{Critical exponent of the system's response}
\index{response function}
\index{susceptibility}
\end{equation}
\item Specific heat
\begin{equation}
C\sim |T-T_\text{c}|^{-\alpha}.
\index{critical exponent|alpha@$\alpha$}
\index{specific heat}
\nomenclature[alpha]{$\alpha$}{Critical exponent of the specific heat}
\end{equation}
\item Order parameter (for instance, the magnetisation)
\begin{equation}\label{eq:INTRO-beta}
m \sim (T_\text{c}-T)^{\beta}.
\index{critical exponent|beta@$\beta$}
\index{order parameter}
\index{magnetisation}
\nomenclature[beta]{$\beta$}{Critical exponent of the order parameter}
\end{equation}
Unfortunately, this exponent uses the same symbol
as the inverse temperature in what is a completely 
universal usage, but the context should always make
clear which is the referred quantity. 
\item Precisely at $T=T_\text{c}$, the decay of the correlation
function is characterised by the anomalous dimension \index{anomalous dimension}
\begin{equation}\label{eq:INTRO-eta}
\braket{s_\bx s_\by} \sim |\bx-\by|^{-(D-2+\eta)}.
\index{critical exponent|eta@$\eta$}
\index{correlation function (equilibrium)!spatial}
\nomenclature[eta]{$\eta$}{Anomalous dimension}
\end{equation}
\item Finally, again at $T=T_\text{c}$, the order parameter 
has a critical dependence on the applied field $h$
\begin{equation}\label{eq:INTRO-delta}
m \sim h^{1/\delta}.
\index{critical exponent|delta@$\delta$}
\nomenclature[delta]{$\delta$}{Critical exponent of the equation of state at $T=T_\text{c}$}
\end{equation}
\end{itemize}
We have thus defined six different critical exponents. However, 
very general considerations let us establish the so-called
scaling relations \index{scaling relations} \index{hyperscaling}
\index{critical exponent}
\begin{subequations}\label{eq:INTRO-scaling}
\begin{align}
2\beta + \gamma &= 2-\alpha,\\
2\beta\delta - \gamma &= 2-\alpha,\label{eq:INTRO-scaling2}\\
\gamma &= \nu (2-\eta),\\
\nu D &= 2-\alpha. \label{eq:INTRO-hyperscaling}
\end{align}
\end{subequations}
The last of these, involving the dimension $D$, is called a
hyperscaling relation. Using~\eqref{eq:INTRO-scaling}, we 
see that for a general system, only two critical exponents 
are independent.

In mean-field theory \index{mean field}
the exponents are $\beta=\nu=1/2$, $\gamma=1$, $\alpha=\eta=0$, $\delta=3$.
Above some (model-dependent) upper critical dimension $D_\text{u}$ \index{critical dimension}
all systems are described by these mean-field exponents.
Notice that they do not depend on $D$, in contrast
with the hyperscaling law.
Usually, $D_\text{u}=4$, but we shall see that this is not the 
case for disordered systems.
Finally, below 
the lower critical dimension $D_\text{l}$ there is no transition.

\subsection{Finite-size scaling}\index{finite-size scaling|(}\label{sec:INTRO-FSS}
One of the most useful tools for the study of critical phenomena is 
the scaling hypothesis. According to this, in the thermodynamic 
limit  the correlation length $\xi_\infty$ is the only characteristic length of the system in
the neighbourhood of $T_\text{c}$  (in this regime, the correlation length
is large in units of the lattice spacing and the system `forgets' about
the lattice).
\index{scaling}
\index{correlation length}

In a finite lattice, the corresponding finite-size scaling (FSS) ansatz states
\nomenclature[FSS]{FSS}{Finite-size scaling}
that the finite-size behaviour is determined by 
the ratio $L/\xi_\infty$. If the ratio is large, the finite-size 
effects are not important and the finite system is not 
essentially different from the thermodynamic limit. If the ratio
is small, however, we say we are in the FSS regime. There, 
an observable $O$ will behave as
\index{finite-size scaling!ansatz}
\begin{equation}\label{eq:INTRO-FSS1}
\braket{O}^{(L)} \simeq L^{x_O/\nu} f_O(L/\xi_\infty),
\end{equation}
where $x_O$ characterises the critical behaviour of $O$,
\begin{equation}
\braket{O}^{(\infty)} \sim |T-T_\text{c}|^{-x_O}.
\end{equation}
Alternatively, using the definition of the critical exponent $\nu$
we can write
\begin{equation}\label{eq:INTRO-FSS2}
\braket{O}^{(L)} \simeq L^{x_O/\nu} \tilde f_O(L^{1/\nu} t), \qquad t = (T_\text{c}-T)/T_\text{c}.
\end{equation}
The FSS ansatz can be derived using renormalisation-group
techniques (see, for instance,~\cite{amit:05}).

Note that in a finite lattice one cannot really talk of a phase
transition. There are no actual divergences of the physical quantities, only
ever narrowing peaks whose position tends to the real critical point and whose
height grows as $L^{x_O/\nu}$.
In this sense, Eq.~\eqref{eq:INTRO-FSS2} encodes the behaviour in
a  crossover region of width~$\sim~L^{-1/\nu}$ between the two phases.
In the thermodynamical limit, this interval degenerates in a point and the crossover \index{thermodynamical limit}
turns into a proper phase transition.

Finally, let us note that at the critical point $\xi_\infty\to\infty$, so
the scaling hypothesis leads to the conclusion that the system
exhibits scale invariance, \index{scale invariance}
since there is no characteristic length. This is an important
observation for detecting the presence of a second-order transition.
\index{finite-size scaling|)}

\section{Quenched disorder}\label{sec:INTRO-quenched}\index{disorder!quenched|(}
Let us now consider a system with disorder. In addition to 
the spins $s_\bx$, we need to specify additional variables
$\mu_i$ that characterise the randomness. In our model
Hamiltonian~\eqref{eq:INTRO-H}, these can be random couplings
$J_{\bx\by}$, vacancies in the lattice or even a random,
site-dependent field $h_\bx$.

In principle, for each configuration of the disorder variables
\nomenclature[Sample]{Sample}{A particular configuration of the disorder variables}
(disorder realisation or sample) we will have a different 
partition function
\begin{equation}
Z_\mu = \sum_{\{s_\bx\}} \ee^{-\beta \mathcal H_\mu(\{s_\bx\})}.
\end{equation}
If the disorder variables exhibit a dynamical evolution 
in time scales short compared to the observation time
(diffusion of impurities at high temperatures, for instance)
we say the disorder is annealed. \index{disorder!annealed}
In this situation, we can treat the $\mu_i$ as additional
dynamical variables and  average over them, to 
obtain the complete partition function. We denote this 
disorder average by an overline, $\overline{(\cdots)}$, so the free energy
of the system is
\begin{align}
F_N =\frac1N \log \overline{Z_\mu}\,.
\nomenclature{$\overline{(\cdots)}$}{Average over the quenched disorder}
\index{free energy!disordered systems}
\index{partition function!disordered systems}
\end{align}

We are interested in the opposite limit, where the 
impurities show no dynamical evolution in 
experimental time scales. We say the disorder is 
quenched, so the free energy is different for each sample
\begin{align}
F_N(\mu) =\frac1N \log Z_\mu\,.
\end{align}
This does not seem like a useful concept, because it seems to imply
we would need a different model for each particular piece of material.
 What actually
happens is that, for large enough systems, the physical 
properties do not depend on the $\mu_i$ anymore,
\begin{equation}
\lim_{N\to\infty} F_N(\mu) = F_\infty.
\end{equation}
There is a simple argument for this in finite dimension, due to Brout~\cite{brout:59}. We
divide the lattice in many macroscopic systems of size $1\ll R^D\ll N$. 
Then the free-energy density of the whole system will be the average
of those of the $N/R^D$ subsystems, plus a contribution coming
from interactions between them. If we assume that the interactions are 
short-range, this latter contribution is an interface energy, negligible
in the large-$N$ limit. Therefore, computing the free-energy density
of a very large lattice is essentially the same as averaging 
that of many smaller systems and  the central limit theorem
guarantees that \index{central limit theorem}
\begin{equation}
 \overline{F^2_N(\mu)} - \overline{F_N(\mu)} ^2 \sim \frac1N\,.
\end{equation}
We say the free energy self-averages. \index{self-averaging}

So, the concept of quenched disorder is physically sound, but 
it implies a serious difficulty. In order to obtain physically
meaningful results, we have to average the free energy, which
is the same as averaging the logarithm of $Z_\mu$
\begin{equation}
F_N = \overline{F_N(\mu)} = \frac1N \overline{\log Z_\mu}\,.
\end{equation}
The task of computing the average of a logarithm, 
unusual in statistical mechanics, is exceedingly difficult.
There is, however, a way around it: the so-called replica method~\cite{kac:68,edwards:72}. \index{replica trick}
This is based on the elementary relationship
\begin{equation} \label{eq:INTRO-replica-trick}
\log Z = \lim_{n\to0} \frac{Z^n -1}{n}.
\end{equation}
For positive integer $n$, $Z^n_\mu$ can be expressed
in terms of identical replicas of the system (sharing the same   \index{replicas}
configuration of the $\mu_i$)
\begin{equation}
Z^n_\mu = \prod_{a=1}^n Z_\mu^{(a)} = \sum_{\{s_\bx^a\}} \exp\biggl[-\beta \sum_{a=1}^n \mathcal H_\mu^{(a)}(\{s_\bx^a\})\biggr].
\end{equation}
This quantity is easier to average over disorder. The objective then, is 
to obtain a replica partition function 
\begin{equation}
Z_n = \overline{Z^n_\mu},
\end{equation}
that no longer depends on disorder and afterwards take
the limit $n\to 0$ in~\eqref{eq:INTRO-replica-trick}.
This procedure can be mathematically delicate in 
some cases (see, e.g.,~\cite{mezard:87}).

Let us consider now the disorder average of an observable $O(\{s_\bx\})$. In
principle we have to do
\begin{equation}
\overline{\braket{O}} = \overline{\frac{1}{Z_\mu} \sum_{\{s_\bx\}} \ee^{-\beta \mathcal H_\mu(\{s_\bx\})}  O(\{s_\bx\})},
\end{equation}
which has the unpleasant feature of having disorder variables both in the numerator
and denominator. This can be solved multiplying both by $Z_\mu^{n-1}$,
\begin{equation}
\overline{\braket{O}} = \overline{\frac{Z_\mu^{n-1}}{Z_\mu^n} \sum_{\{s_\bx\}} \ee^{-\beta \mathcal H_\mu(\{s_\bx\})}  O(\{s_\bx\})}.
\end{equation}
Now we write the numerator in the replica notation, assigning the original partition
function to replica $1$ (this choice is, of course, arbitrary)
\begin{equation}
\overline{\braket{O}} = \overline{\frac{1}{Z_\mu^n} \sum_{\{s_\bx^a\}} \ee^{-\beta\sum_a \mathcal H^{(a)}_\mu(\{s_\bx^a\})}  O(\{s_\bx^1\})}.
\end{equation}
In the $n \to0$ limit the denominator goes to one and we 
have
\begin{equation}
\overline{\braket{O}} = \lim_{n\to0} \overline{\sum_{\{s_\bx^a\}} \ee^{-\beta \sum_a
\mathcal H^{(a)}_\mu(\{s_\bx^a\})} O(\{s_\bx^1\})}.
\end{equation}
In a successful application of the replica trick, one hopes to integrate 
the dependence on the $\mu_i$ explicitly and define an effective Hamiltonian
$\mathcal H_n$ that no longer depends on the disorder (only on $n$). Then
\begin{equation}\label{eq:INTRO-replica-O}
\overline{\braket{O}} = \lim_{n\to0} \braket{O}_n = \lim_{n\to0}
\sum_{\{s_\bx^a\}} \ee^{-\beta \mathcal H_n(\{s_\bx^a\})} O(\{s_\bx^1\}).
\end{equation}

In this dissertation we do not carry out any such replica computations, but
in Chapter~\ref{chap:sg} 
we give an outline of a particularly famous example: the mean-field
\index{mean field}
theory of spin glasses.\index{spin glass}

\subsection{The relevancy of disorder}
Real systems always have some measure of disorder, either in the form of 
impurities or vacancies in the lattice. However, this disorder
is not always relevant in the sense of changing the universality \index{universality!and disorder}
class of the system.

For simplicity let us consider a ferromagnetic system, where
we introduce some disorder in the couplings. \index{ferromagnets}
We consider a Hamiltonian of the form
\begin{equation}
\mathcal H  = - \sum_{\bx,\by} J_{\bx \by} s_\bx s_\by,
\end{equation}
and let us write the couplings $J_{\bx\by}$ as
a translationally invariant part plus a perturbation \index{pure system}
\begin{equation}\label{eq:INTRO-dis}
J_{\bx \by} = J(|\bx-\by|) + \delta J_{\bx\by}.
\end{equation}
We say that a system described exclusively by 
the translationally invariant part is the `pure' 
system corresponding to our disordered model.

Then, there are two interesting limiting cases. 
In the first, the disorder is strong, so
\begin{equation}
\delta J_{\bx\by} \gg J(|\bx-\by|).
\end{equation}
Therefore, the disorder completely dominates 
the low-temperature properties of the system.
In particular, the low-temperature ferromagnetic
order is destroyed and the system is
described by a `spin glass' phase where 
$\overline{\braket{s_\bx}}=0$ but 
$\overline{\braket{s_\bx}^2}\neq0$. 
The Edwards-Anderson model, which we study
in Part~\ref{part:sg}, is one example.
\index{spin glass}
\index{Edwards-Anderson model}
\index{disorder!strong}
\index{disorder!weak}

On the other hand, the disorder may be weak
\begin{equation}
\delta J_{\bx\by} \ll J(|\bx-\by|).
\end{equation}
In this case, one would not expect the disorder
to effect great changes in the ground-state
properties. The low-temperature phase would 
continue to have a ferromagnetic order, for
instance. However, in the case of a second-order
phase transition for the pure model,
the critical exponents may change. Furthermore, 
if the transition of the pure system is of first order, 
it may become continuous.

There is a useful criterion for determining
whether the weak disorder is going to  be relevant, due \index{Harris criterion}
to Harris~\cite{harris:74}.         
According to it, if the specific-heat exponent of the  \index{critical exponent}
pure system is $\alpha^{(0)}>0$, then the disorder \index{critical exponent!alpha@$\alpha$}
will change the critical behaviour. On the other
hand, if the specific heat of the pure system is 
finite, the disorder will be irrelevant (it will 
not change the critical exponents).

Finally, let us note that in Part~\ref{part:daff} we 
study the random field Ising model, where the disorder
is not of the kind described by~\eqref{eq:INTRO-dis}, but takes
the form of random fields. In that case the disorder
is also very severe, because the randomness couples \index{RFIM}
to the local order parameter.

\subsection{Self-averaging violations}\index{self-averaging|(}
\label{sec:INTRO-self-averaging}
We started our discussion of quenched disorder by 
giving a general argument in favour of the 
self-averaging property. This argument, however, 
breaks down at the critical point, where the          \index{correlation length}
correlation length diverges and our division of the lattice
in smaller subsystems with negligible interaction 
no longer works. Therefore, the issue
of self-averaging becomes non-trivial at
the transition point.

In fact, it has long been known that for
spin glasses there is no self-averaging
in the ordered phase~\cite{binder:86}.
For systems with weak disorder, a
framework analogous to the Harris criterion
\index{Harris criterion}
can be established.

Let us consider some macroscopic quantity $O$ 
(the magnetisation, energy, etc.) 
and  let us consider the probability distribution
of the $\langle O\rangle$ for different samples,
which we characterise by its relative variance,
\begin{equation}\label{eq:DAFF-R-O}
R_O = \frac{\overline{\langle O\rangle^2} - \overline{\langle O\rangle}^2}{
\overline{\langle O\rangle}^2}.
\end{equation}
We say the system is self-averaging if
\begin{equation}
R_O \xrightarrow{\ \ L\to\infty\ \ } 0.
\end{equation}

Aharony and Harris~\cite{aharony:96} reached the following conclusions:
\begin{enumerate}
\item Away from the critical region, we can apply
the Brout argument and 
\begin{equation}
R_O \sim \left({\xi}/{L}\right)^D.
\end{equation}
We say that the system is strongly self-averaging.
\item At the critical point we have to distinguish two
possibilities.
\begin{enumerate}
\item The disorder is irrelevant (in the sense of the Harris
criterion). Then
for the pure system $\alpha^{(0)}<0$ and 
\begin{equation}
R_O \sim L^{\alpha^{(0)}/\nu^{(0)}} =L^{\alpha/\nu}.
\end{equation}
The system is weakly self-averaging.
\item The disorder is relevant. In this case
the system is no longer self-averaging,
\begin{equation}
\lim_{L\to\infty} R_O  \neq 0.
\end{equation}
\end{enumerate}
\end{enumerate}
Soon after the pioneering renormalisation-group work
of Aharony and Harris, several authors studied 
the issue of (lack of) self-averaging in disordered
systems with numerical simulations~\cite{wiseman:95,pazmandi:97,wiseman:98,ballesteros:98b,berche:04}.

This break down of the self-averaging property is 
an additional difficulty for the study 
of the critical behaviour of disordered 
systems. In Chapter~\ref{chap:daff-tethered}, however, 
we shall demonstrate an approach that minimises
its effects, in the context of the random field \index{RFIM}
Ising model (a system where the violation of 
self-averaging is particularly severe~\cite{parisi:02,wu:06,malakis:06,fytas:11}).

\index{self-averaging|)}
\index{disorder!quenched|)}

\chapter[Managing rugged free-energy landscapes: a Tethered Monte Carlo primer]{Managing rugged free-energy landscapes:\\ a Tethered Monte Carlo primer}\label{chap:tmc}
Monte Carlo (MC) simulation (see, e.g., \cite{landau:05,rubinstein:07,sokal:97,newman:96} for \index{Monte Carlo method}
general reference works) constitutes one of the most important 
modern tools of theoretical physics. At a first glance, it seems a very 
inefficient method: its statistical character meaning that the uncertainty
in the result decreases only as $1/\sqrt{\mN}$, where $\mN$ is a measure
of the numerical effort. From a closer inspection, however, comes the realisation
that deterministic numerical methods, typically thought to converge 
with higher powers of $\mN$ or even exponentially, quickly lose their efficiency
when the number of degrees of freedom is increased (think, for instance, of 
the computation of multi-dimensional integrals). 
In contrast, the $1/\sqrt{\mN}$ behaviour of the MC method, a consequence of
the central limit theorem,\index{central limit theorem} is stable. 

In the context of statistical mechanics, we are interested
in extracting system configurations
that follow some complicated probability distribution, with a huge number
of degrees of freedom ---typically $p(\{s_\bx\})\propto \ee^{-\beta E(\{s_\bx\})}$,
for systems in the canonical ensemble~\eqref{eq:INTRO-Z}.
This is accomplished by means of a dynamic Monte Carlo, where the  \index{Monte Carlo method!dynamic}
generation of a new configuration depends on the current one.         \index{Markov chain}
In technical terms, we set up an ergodic Markov chain whose stationary
distribution is the physical distribution describing the equilibrium state
of the system (cf. Section~\ref{sec:THERM-Markov-chain}).  Once the stationary regime is reached, 
we estimate $\langle O\rangle$ as an unweighted average of $O(t) = O(\{s_\bx(t)\})$.
\nomenclature[Ot]{$O(t)$}{Value of observable $O$ along the simulation $O(t)=O(\{s_\bx(t)\})$}
The error then goes as $\sim 1/\sqrt{\mN_\text{steps}}$, but with a potentially 
large prefactor (see Appendix~\ref{chap:thermalisation}).

At a first glance one may think this probabilistic method is a poor alternative
to traditional tools such as perturbation theory. \index{perturbation theory}
Yet, among the most interesting problems in statistical
mechanics and quantum field theory we often find strongly coupled
systems, far from the perturbative regime. 
In these situations, most of our analytical tools break down and MC simulation
emerges as one of a handful of workable methods.

This is not to say that a MC computation does not have its 
difficulties. Chief among these is the issue of thermalisation:     \index{critical slowing down}
before we can even start to worry about the $1/\sqrt{\mN}$ behaviour 
of our numerical precision, the Markov chain has to reach its stationary \index{Markov chain}
distribution.  For the most physically interesting regime, 
the neighbourhood of a phase transition, this turns out to
be difficult, because of critical slowing down~\cite{hohenberg:77,zinn-justin:05}.
This phenomenon consists in the rapid growth of the characteristic \index{thermalisation}
times with the system size. Even for very simple systems, such as 
the Ising model, the thermalisation times of traditional MC methods \index{Ising model}
grow as $L^z$, with $z\approx2$ and $L$ the linear 
size of the system. Only for scant few systems can one  \index{critical exponent!z@$z$}
find optimised dynamics with $z<1$ (cf. our study of cluster methods \index{cluster methods}
in Chapter~\ref{chap:cluster}).                                      

In many other situations the critical slowing down is even worse than 
the $z\approx2$ behaviour. This is the case of the 
rugged free-energy landscapes considered in the General Introduction, \index{free energy!landscape}
where the thermalisation times grow exponentially with the free-energy
\index{critical slowing down!exponential}
barriers. These barriers not only constitute a formidable stumbling block for traditional
MC methods, but are also physically interesting in their own right.
This is because the actual physical evolution of the system is hindered
by these same dynamical bottlenecks.

In the following section we give some precise examples of this 
phenomenon and briefly review some of the methods that have been devised
to address it. In Section~\ref{sec:tmc} we introduce our proposal:
Tethered Monte Carlo, a formalism which we shall develop and employ
throughout Parts~\ref{part:tmc} and~\ref{part:daff} of this thesis.

\section{Free-energy barriers and Monte Carlo simulations}\label{sec:TMC-free-energy}
\index{free energy!barriers|(}
The most straightforward example of a free-energy barrier is encountered
whenever we want to consider first-order phase transitions~\cite{gunton:83,binder:87}.
\index{phase transition!first order}
In these situations two phases (ordered and disordered, or with different kinds
of order)  coexist at the critical point.
In a traditional MC simulation the system must tunnel back and forth between these 
two pure phases by forming a mixed configuration, featuring
an interface of size $L$. This intermediate state has a free-energy cost   \index{interface}
of $\varSigma L^{D-1}$, where $\varSigma$ is the surface tension  \index{surface tension}
and $D$ the spatial dimension of the system. Therefore, the probability \index{metastability}
of crossing the gap between the ordered and disordered states
scales as $\exp[-\varSigma L^{D-1}]$. Equivalently, the simulation
suffers an exponential critical slowing down, where  \index{critical slowing down!exponential}
the characteristic times grow as $\exp[\varSigma L^{D-1}]$.

The situation can be even worse, as demonstrated by crystallisation  \index{crystallisation}
studies.  Here, even for the simplest models there are 
many local free-energy minima. These correspond to crystals with 
different symmetries and varying numbers of defects, or even 
to amorphous solids (glasses). See, e.g.,~\cite{pusey:89} for \index{glasses}
 an experimental example.

Furthermore, the issue of free-energy barriers is not limited
to first-order transitions. A prime example of this is    \index{RFIM} \index{DAFF}
the random field Ising model, which we shall study      
extensively in Part~\ref{part:daff}. Here there still 
exists a free-energy barrier between the ordered 
and the disordered states, but only one of these configurations
defines a stable phase. The difference with the first-order
scenario is that the barriers grow as $L^\theta$, where $\theta< D-1$. \index{critical exponent!theta@$\theta$}
Therefore, we still have a thermally activated critical 
slowing down, with $\log \tau\sim L^\theta$ (we shall see 
that $\theta=1.469(20)$ for $D=3$, so this is very severe).
The difficulty is compounded by the fact that this is a disordered
system, so one must consider many disorder realisations
in order to get a meaningful picture (cf. the discussion
in Chapter~\ref{chap:disorder}).

For the random field Ising model, we at least know
the appropriate order parameter that signals the phase
transition, but this is not always the case. The most 
conspicuous example of this additional complication      \index{spin glass}
is the spin glass.
The problem is patched, but not completely solved, by using
real replicas (clones of the system with the same disorder   \index{replicas!real}
realisation, evolving independently under the same dynamics).
In this case, the actual structure of the ordered
phase is still in dispute but, at least for finite systems,
there are a large number of local minima (see Chapter~\ref{chap:sg}). 
This is a very popular problem, which has prompted the introduction  \index{parallel tempering}
not only of ad-hoc MC methods, such as parallel tempering (see Appendix~\ref{chap:thermalisation}),
 but even of special-purpose computers~\cite{ogielski:85,cruz:01,ballesteros:00}. \index{special-purpose computers}
The latest example of this is the \textsc{Janus} machine (see Appendix~\ref{chap:janus}),  \index{Janus@\textsc{Janus}}
which we have used for our own spin-glass simulations (Part~\ref{part:sg} of this dissertation).
Still, the use of a custom computer may accelerate the simulation by a constant factor, but does 
not change the scaling of the thermalisation times which, even with parallel tempering, 
are believed to suffer an exponential critical slowing down below the critical temperature.

Many optimised schemes and formalisms have been proposed to deal with these
problems. The case of first-order phase transitions, with two clean and easy  \index{Wang-Landau method}
to differentiate phases, is perhaps the best understood. For instance, 
multicanonical~\cite{berg:92} or Wang-Landau~\cite{wang:01} methods      \index{multicanonical methods}
consider a generalised statistical ensemble. The dynamics consists 
in a random walk in energy space, covering the range bounded by the two
competing phases. This strategy is able to overcome the free-energy barriers 
for small systems, but this only delays to larger sizes 
the advent of exponential slowing down~\cite{neuhaus:03}. This is
mainly due to the emergence of geometrical transitions in the energy
gap between the two phases~\cite{biskup:02,binder:03,macdowell:04,
macdowell:06,nussbaumer:06}.

In a more general case, however, the local minima cannot be distinguished 
by their energies and we have to consider alternative reaction coordinates     \index{reaction coordinate}
(typically, but not necessarily, order parameters). Examples abound, perhaps    \index{order parameter}
the best known being the studies of crystallisation in supercooled \index{glasses} \index{crystallisation}
liquids~\cite{tenwolde:95,chopra:06}, where the different phases 
can be labelled with a bond-orientational crystalline order
parameter~\cite{steinhardt:83}. The random-walk strategy can be adapted
to some of these cases, resulting in the so-called umbrella sampling~\cite{torrie:77}. \index{umbrella sampling}
Unfortunately, for sufficiently complex systems considering a single reaction
coordinate is not enough. Tuning the parameters of a Wang-Landau \index{Wang-Landau method}
or umbrella sampling simulation is in these cases rather cumbersome.

A different strategy was first introduced in a microcanonical setting  \index{microcanonical methods}
in~\cite{martin-mayor:07}. In this method, one performs independent
simulations with a fixed energy along the whole gap, which are then 
combined with a fluctuation-dissipation formalism to yield the entropy \index{energy}
of the system. Thus, the need for tunnelling across geometrical transitions \index{entropy}
is eliminated and very large system sizes can be considered.

In~\cite{fernandez:09} we generalised this microcanonical method to consider  \index{reaction coordinate}
any reaction coordinate instead of the energy density. In a similar manner, 
the role played by the entropy in the microcanonical setting is taken  
by the Helmholtz effective potential associated to the chosen reaction coordinate.
Furthermore, the  application of this Tethered Monte Carlo method is not
necessarily more difficult with several reaction coordinates.

In a tethered computation, one simulates a statistical ensemble where the 
reaction coordinate $x$ is constrained (tethered) to a narrow range around 
a fixed parameter $\hat x$. This is accomplished through the introduction
of a bath of Gaussian demons, which absorb the changes   \index{demons}
in the reaction coordinate, so long as these are not too large, 
to keep $\hat x$ constant.  From several such simulations for different values 
of $\hat x$ the Helmholtz potential is readily  \index{effective potential}
reconstructed, yielding all the information about the system.
The tethered formalism is not intended as a mere
thermalisation speed-up, but it also grants us access to precious
information that would remain hidden from a traditional approach.

\index{free energy!barriers|)}
We will explain the construction of
the tethered formalism in Chapter~\ref{chap:tmc-formalismo}.
Before engaging in detailed derivations, however, it is 
useful to understand all the steps of a Tethered Monte 
Carlo simulation. The remainder of this chapter provides such an outline,
actually constituting a self-contained guide to the 
set-up of a TMC computation.

We note that the following was first published as Section~2 of~\cite{martin-mayor:11}, 
which we reproduce here with minor emendations.

\nocite{fernandez:09}
\section{Tethered Monte Carlo, in a nutshell}\label{sec:tmc}\index{tethered formalism|(}
In this section we give a brief overview of the Tethered Monte Carlo (TMC)
\nomenclature[TMC]{TMC}{Tethered Monte Carlo}
method, including a complete recipe for its implementation in a typical
problem.  This is as simple as performing several independent ordinary MC
simulations for different values of some relevant parameter and then averaging
them with an integral over this parameter.  We shall give the complete
derivations and the detailed construction of the tethered ensemble in
Chapter~\ref{chap:tmc-formalismo}.

We are interested in the scenario of a system whose phase
space includes several coexisting states, separated by free-energy 
barriers. The first step in a TMC study is identifying the reaction
coordinate $x$ that labels the different relevant phases. This can be  \index{reaction coordinate}
(but is not limited to) an order parameter. In the remainder of this   \index{order parameter}
section we shall consider a ferromagnetic setting, so the reaction     \index{ferromagnets}
coordinate will be the magnetisation density $m$.                      \index{magnetisation}

The goal of a TMC computation is, then, constructing the Helmholtz potential \index{effective potential}
associated to $m$, $\varOmega_N(\beta,m)$, which will give us all the information about
the system. This involves working in a new statistical ensemble tailored to
the problem at hand, generated from the usual canonical ensemble by Legendre
transformation (cf. Chapter~\ref{chap:disorder}):
\index{Legendre transformation}%
\index{partition function}%
\index{free energy}%
\begin{equation}
Z_N(\beta,h)= \ee^{N F_N(\beta,h)}=\int \mathrm{d} m \ \mathrm{e}^{N[\beta hm -\varOmega_N(\beta,m)]}\,.
\tag{\ref{eq:INTRO-legendre}}
\end{equation}

Since in a lattice system the magnetisation is discrete, we actually couple it
to a Gaussian bath to generate a smooth parameter, called $\hat m$. The \index{demons}
effects of this bath are integrated out in the formalism.

In order to implement this construction as a workable
Monte Carlo method we need to address two different problems:
\begin{itemize}
\item We need to know how to simulate at fixed $\hat m$. 
\item We need to reconstruct $\varOmega_N(\beta,\hat m)$ from simulations at
  fixed $\hat m$ and, afterwards,
  to recover canonical expectation values from~\eqref{eq:INTRO-legendre}\index{canonical ensemble}
  to any desired accuracy.
\end{itemize}
We explain separately how to solve each of the two problems, in the
following two paragraphs.

\subsection{Metropolis simulations in the tethered ensemble}\label{sec:TMC-Metropolis}
Let us denote the reaction coordinate by $m$ (for the sake of concreteness let \index{reaction coordinate}
us think on the magnetisation density for an Ising model). The dynamic degrees
of freedom are $\{s_\bx\}$. Therefore $m$ is an observable  (i.e. a function of the
$\{s_\bx\}$). We wish to simulate at fixed $\hat m$ ($\hat m$ is a parameter
closely related to the average value of $m$).

The canonical weight at inverse temperature $\beta$ and $h=0$ would be \index{canonical ensemble}
$\mathrm{exp}[-\beta U]$ where $U$ is the interaction energy. Instead, the
tethered weight is (see Section~\ref{sec:TMC-tethered-ensemble} for a derivation)
\index{tethered weight}
\begin{equation}\label{eq:weight1}
\omega_N(\beta,\hat m;\{s_\bx\})=\mathrm{e}^{-\beta U + N (m-\hat m)} (\hat
m -m)^{(N-2)/2} \varTheta(\hat m -m)\,.
\end{equation}
The Heaviside step function $\varTheta(\hat m -m)$ imposes the constraint that
$\hat m> m(\{s_\bx\})$.
\nomenclature[Theta]{$\varTheta$}{Heaviside step function}
\nomenclature[omega]{$\omega_N$}{Tethered weight}

The tethered simulations with weight (\ref{eq:weight1}) are exactly like a
standard canonical Monte Carlo in every way (and the balance condition, \index{balance condition}
etc.). For instance, in an Ising model setting, the common Metropolis \index{Ising model}
algorithm~\cite{metropolis:53} is
\begin{enumerate}
\item Select a spin $s_\bx$.
\item The proposed change is flipping the spin, $s_\bx\to -s_\bx$.~\footnote{In an
  atomistic simulation, one would try to displace a particle, or maybe to
  change the volume of the simulation box.}
\item The change is accepted with probability\footnote{In general,
in order to satisfy the balance condition (see Section~\ref{sec:THERM-Markov-chain}) \index{balance condition}
we have to take into account  both 
the weight of the current and proposed configurations
and the probabilities of proposing this particular
change and its reciprocal. However, 
the latter are trivial, because the change is always $s_\bx\to -s_\bx$.
}
\begin{equation}
\mathcal P(s_\bx \to -s_\bx) = \min \{ 1,\omega_\mathrm{new}/\omega_\mathrm{old}\}.
\end{equation}
\item Select a new spin $s{_\bx'}$ and repeat the process. We 
can either pick $s{_\bx'}$ at random or run through the lattice
sequentially. In the work reported in this dissertation we have
always followed the second option, more numerically 
efficient.
\end{enumerate}
Once $N$ spins have been updated (or we have run through the whole lattice,
in the sequential case) we say we have completed one Monte Carlo Sweep (MCS). 
\nomenclature[MCS]{MCS}{Monte Carlo sweep, i.e., update of the whole
lattice} 

We remark that the above outlined algorithm produces a Markov chain entirely
analogous to that of a standard, canonical Metropolis simulation. As such it has
all the requisite properties of  a Monte Carlo simulation (mainly reversibility
and ergodicity). Tethered
mean values can be computed as the time average along the simulation of the
corresponding observables (such as internal energy, magnetisation
density, etc.). Statistical errors and autocorrelation times can be computed
with standard techniques (see Appendices~\ref{chap:thermalisation} and~\ref{chap:correlated}).

The actual magnetisation density is constrained (tethered) in this simulation,
but it has some leeway (the Gaussian bath can absorb small variations in $m$). \index{demons}
In fact, its fluctuations are
crucial to compute an important dynamic function, whose introduction would
seem completely unmotivated from a canonical point of view:
the tethered field $\hat b$
\index{tethered field}
\begin{equation}\label{eq:campo-tethered}
\hat b = -\frac{1}{N} \frac{\partial \log \omega_N(\beta,\hat
  m;\{s_\bx\})}{\partial \hat m}= 1 - \frac{N-2}{2 N [\hat m - m(\{s_\bx\})]}\,.
\end{equation}
One of the main goals of a tethered simulation is the accurate computation of
the expectation value $\langle \hat b\rangle_{\hat m}$.

The case where one wishes to consider two reaction coordinates $m_1$ and $m_2$
is completely analogous:
\begin{equation}
\begin{split}
\omega_N(\beta,\hat m_1,\hat m_2;\{s_\bx\})=&\mathrm{e}^{-\beta U + N (m_1-\hat
  m_1)+ N (m_2-\hat m_2) }\\
 &\times (\hat m_1 -m_1)^{(N-2)/2} \varTheta(\hat m_1-m_1)\\
 & \times  (\hat m_2 -m_2)^{(N-2)/2}\varTheta(\hat m_2 -m_2)\, ,
\end{split}
\end{equation}
where
\begin{align}
\hat b_1&= -\frac{1}{N} \frac{\partial \log \omega_N(\beta,\hat
  m_,\hat m_2;\{s_\bx\})}{\partial \hat m_1}= 1 - \frac{N-2}{2 N [\hat m_1 -
    m_1(\{s_\bx\})]}\,,\\
\hat b_2&= -\frac{1}{N} \frac{\partial \log \omega_N(\beta,\hat
  m_,\hat m_2;\{s_\bx\})}{\partial \hat m_2}= 1 - \frac{N-2}{2 N [\hat m_2 - m_2(\{s_\bx\})]}\,.
\end{align}
\index{tethered field}
\index{tethered weight}

For the Ising model, a Metropolis Tethered Monte Carlo simulation  \index{Ising model}
reconstructs the crucial tethered magnetic field $\hat b$ 
without critical slowing down
(see Chapter~\ref{chap:Ising} for a
benchmarking study). This may be considered surprising for what is 
a local update algorithm, but notice that the constraint on $\hat m$ \index{critical slowing down}
is imposed globally. Non-magnetic observables, such as the energy, \index{critical exponent!z@$z$}
do not enjoy this non-local information and hence show a typical $z\approx2$
critical slowing down (although the correlation times are low enough to
permit equilibration for very large systems, see Chapter~\ref{chap:Ising}).

Let us stress that the above outlined update algorithm is by no means
the only one possible. For instance, \index{Fortuin-Kasteleyn construction}
the Fortuin-Kasteleyn construction~\cite{kasteleyn:69,fortuin:72}
can be performed just as easily in the tethered ensemble, so 
we can consider tethered simulations with cluster
update methods~\cite{swendsen:87,edwards:88,wolff:89}. We demonstrated
this in~\cite{martin-mayor:09} (see also Chapter~\ref{chap:cluster}),
where the tethered version of the Swendsen-Wang   \index{Swendsen-Wang algorithm}
algorithm was shown to have the same critical slowing down
as the canonical one for the $D=3$ Ising model ($z\approx0.47$). \index{cluster methods}
This is an example that the use of the tethered formalism implies
no constraints on the choice of Monte Carlo algorithm, nor does it hinder
it in the case of an optimised method.

\subsection[Reconstructing the effective potential from
tethered simulations]{Reconstructing the 
Helmholtz effective potential from simulations
  at fixed \boldmath $\hat m$}
\begin{figure}
\centering
\includegraphics[height=0.8\linewidth,angle=270]{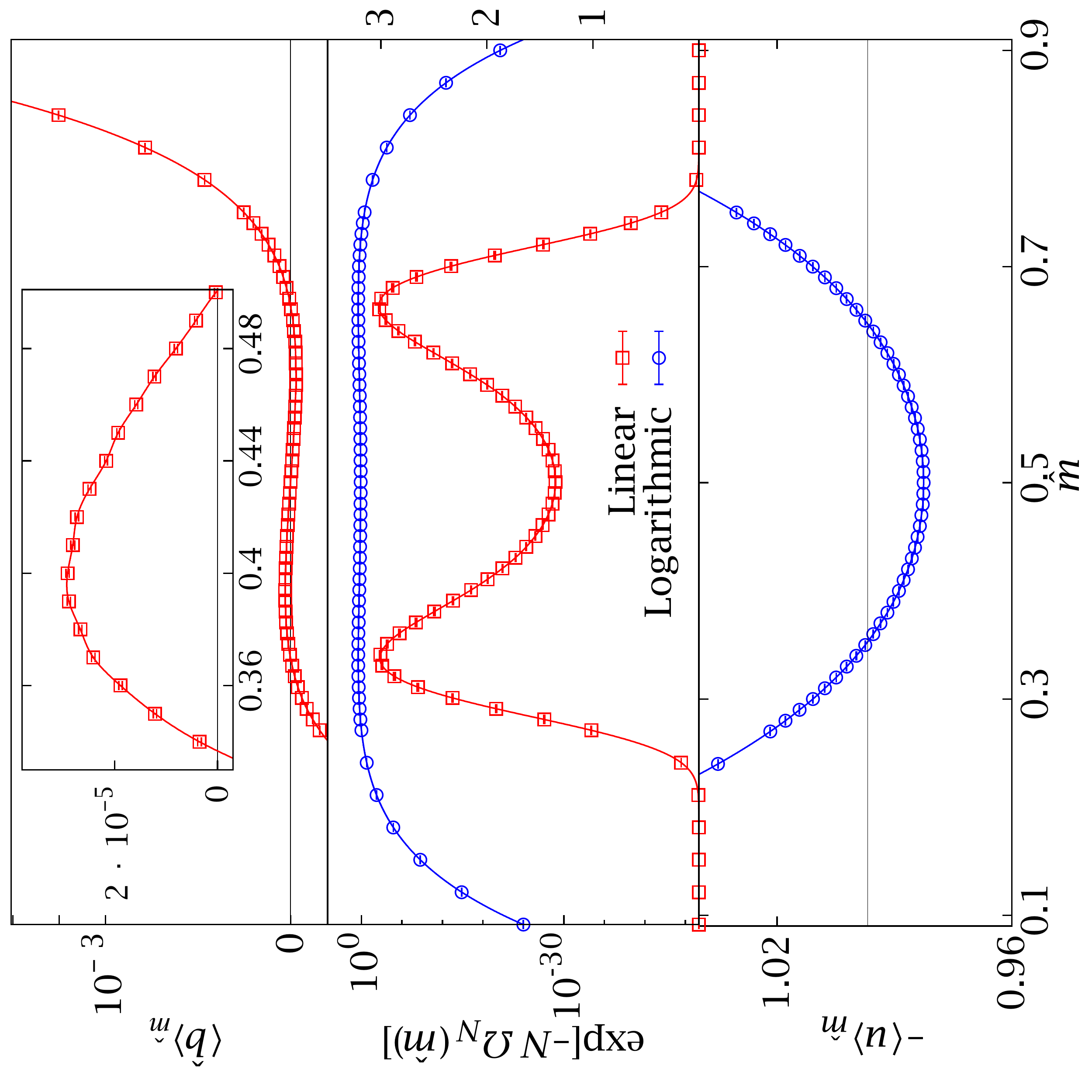}
\caption[Computation of the effective potential
from tethered simulations]{Computation of the Helmholtz potential $\varOmega_N$
and the canonical expectation values from
tethered averages in a $D=3$, $L=64$ ferromagnetic Ising model
at the critical temperature. \emph{Top:}  tethered magnetic field
$\langle \hat b\rangle_{\hat m}$ (we show only the positive tail, in order
to see its structure better at this scale), with an inset zooming in on
the region between its two zeros.
The statistical errors cannot be seen at this scale
(except for the leftmost points in the inset). The integral of this quantity 
is the Helmholtz potential $\varOmega_N(\hat m)$.
\emph{Middle:} $p(\hat m)=\exp[-N\varOmega_N(\hat m)]$
in a linear (right axis) and in a logarithmic scale (left axis).
\emph{Bottom:} tethered expectation values
of the energy density $u$. Their integral over the whole 
$\hat m$  range, weighted with $p(\hat m)$, gives
the canonical expectation value $\langle u \rangle = -0.996\,868(11)$.
(horizontal line). See Chapter~\ref{chap:cluster} for further details
on these simulations.
\label{fig:TMC-tethered}
\index{energy!Ising|indemph}
\index{tethered field!Ising|indemph}
\index{effective potential!Ising|indemph}
}
\end{figure}

\index{effective potential}
The steps in a TMC simulation are, then, (see also Figure~\ref{fig:TMC-tethered})
\begin{enumerate}
\item Identify the range of $\hat m$ that covers 
the relevant region of phase space. 
Select $\mN_{\hat m}$ points $\hat m_i$, evenly spaced 
along this region.
\item For each $\hat m_i$ perform a Monte Carlo simulation where the smooth reaction
  parameter $\hat m$ will be fixed at $\hat m=\hat m_i$.
\item We now have all the relevant physical observables
as discretised functions of $\hat m$. We 
denote these tethered averages at fixed $\hat m$ by
$\langle O\rangle_{\hat m}$. 
\item The average values in the canonical ensemble, denoted by
 $\langle O\rangle$, can be recovered with a simple integration
\begin{equation}
\langle O \rangle = \int_{\hat m_\text{min}}^{\hat m_\text{max}} \dd \hat m\
p(\hat m) \langle O\rangle_{\hat m}.
\end{equation}
In this equation the probability density $p(\hat m)$ is
\begin{equation}
p(\hat m) = \ee^{-N \varOmega_N(\hat m,\beta)},\quad \varOmega_N(\hat m,\beta) = \varOmega_N(\hat m_\mathrm{min})+ 
\int_{\hat m_\text{min}}^{\hat m} \dd \hat m'\
\langle \hat b\rangle_{\hat m'}.
\end{equation}
The tethered field $\langle \hat b\rangle_{\hat m}$ was defined in
Eq.~\eqref{eq:campo-tethered}.  The integration constant $\varOmega_N(\hat
m_\mathrm{min})$ is chosen so that the probability is normalised.
\item If we are interested in canonical averages in the presence
of an external magnetic field $h$, we do not have to run any
new simulations. Indeed, we can reuse the $\langle O\rangle_{\hat m}$
and only recompute $\varOmega_N$ (only the relative weight of the tethered
averages changes).
This is as simple as shifting the tethered magnetic field: \index{field, external}
$\langle \hat b\rangle_{\hat m} \to \langle \hat b\rangle_{\hat m} - \beta h$.
\item In order to improve the precision and avoid systematic errors, 
we can run additional simulations in the region where $p(\hat m)$
is largest.
\end{enumerate}
The whole process is illustrated in Figure~\ref{fig:TMC-tethered}, where we compute
the energy density at the critical temperature in an $L=64$ lattice of 
the $D=3$ Ising model. Notice that the tethered averages $\braket{u}_{\hat m}$ vary in about 
$10\%$ in our $\hat m$ range, but the computation of the effective potential
is so precise that the averaged value for the energy, $\langle u \rangle = -0.996\,868(11)$, 
has a relative error of only $\sim10^{-5}$.

This is the general TMC algorithm for the computation of canonical averages
from the Helmholtz potential. As we shall see in some of the applications,
sometimes the integration over all phase space in step 4 is not needed and one
can use the ensemble equivalence property to recover the $\langle O \rangle$
from the $\langle O\rangle_{\hat m}$ through saddle-point equations, remember \index{saddle point}
Eq.~\eqref{eq:INTRO-legendre}.  In other words, the tethered averages can be
physically meaningful by themselves. For example, the crystallisation
study of~\cite{fernandez:11} is built entirely over the \index{crystallisation}
effective potential, one never uses the $p(\hat m)$.

As will be shown in Chapter~\ref{chap:tmc-formalismo}, the reconstruction of canonical
averages from the combination of tethered averages does not
involve any approximation. We can achieve any desired accuracy, provided
we use a sufficiently dense grid in $\hat m$ (to control systematic errors)
and simulate each point for a sufficiently long time (to reduce
statistical ones). Table~\ref{tab:TMC-energies} and Figure~\ref{fig:TMC-tethered}
show the kind of 
precisions that we can achieve. One could initially think that 
the computation of the exponential in $p(\hat m) = \exp[-N \varOmega_N(\hat m)]$
would produce unstable or imprecise results for large system sizes.
Instead, the combination  of  \index{self-averaging}
self-averaging and no critical slowing down makes the numerical precision \index{critical slowing down}
grow with $N$.

\begin{table}
\small
\centering
\begin{tabular*}{\columnwidth}{@{\extracolsep{\fill}}rrcl}
\toprule
$L$ & $\mN_{\hat m}$ & MCS &\multicolumn{1}{c}{$-\langle u\rangle$}\\
\toprule
$16$ & $91$  & $10^6$ & $1.034\,72(10)$\\ 
$32$ & $91$  & $10^6$ & $1.007\,189(78)$  \\
$64$ & $109$ & $10^6$ & $0.996\,868(11)$\\
$128$ & $50$ & $10^6$ & $0.992\,949\,3(45)$\\
\toprule
\end{tabular*}
\caption[Energy density of the $D=3$ Ising model]{Energy density
of the $D=3$ ferromagnetic Ising
model computed with the Tethered Monte Carlo method, showing 
that the reconstruction of canonical averages can be performed
with great accuracy. The second column shows the  number
of points in the $\hat m$ grid and the third the number
of Monte Carlo sweeps taken on each (we use a cluster
update scheme).  Data from the simulations reported
in Chapter~\ref{chap:cluster}.}
\label{tab:TMC-energies}
\index{tethered formalism!numerical performance}
\index{energy!Ising|indemph}
\end{table}
\index{tethered formalism|)}

\part[The Tethered Monte Carlo formalism,\\ with a new look at ferromagnets]{The 
Tethered Monte Carlo formalism,\\
with a new look at ferromagnets}\label{part:tmc}
\chapter{The tethered formalism}\label{chap:tmc-formalismo}\index{tethered formalism|(}
This chapter presents the tethered formalism in detail, noting its 
relation to the canonical ensemble and introducing some of the
techniques that we will use throughout this dissertation (such as saddle-point \index{saddle point}
equations). We first presented the tethered statistical ensemble in~\cite{fernandez:09}, 
demonstrating its application to the Ising model \index{Ising model}
(cf. Chapter~\ref{chap:Ising}). The exposition in this chapter
also uses the Ising ferromagnet as a model system, since
that will be the first application considered in this thesis (it would be straightforward 
to reproduce the construction for a  different model, as we shall see
in Chapter~\ref{chap:daff-tethered}). However, the treatment of the tethered formalism
is otherwise more general than that of~\cite{fernandez:09}.  For instance, 
we include details 
on how to consider several tethered variables (Section~\ref{sec:TMC-several-tethers}), 
a feature that we will need in our study of the DAFF (Chapter~\ref{chap:daff-tethered}). \index{DAFF}

\section{The tethered ensemble}\label{sec:TMC-tethered-ensemble}\index{Ising model}
As noted above, we consider the $D$-dimensional Ising model,
characterised by the following partition function
\begin{equation}\label{eq:TMC-Z-Ising}
Z(\beta,h) = \ee^{NF_N(\beta,h)}=  \sum_{\{s_\bx\}} \exp\biggl[ \beta  \sum_{\langle\bx,\by\rangle} 
s_\bx s_\by + \beta h \sum_\bx s_\bx\biggr],
\index{partition function!Ising}
\end{equation}
(recall that the angle brackets indicate that the sum is restricted to first
neighbours and that the spins are $s_\bx=\pm1$). As we indicated in Chapter~\ref{chap:disorder},
we shall always consider square lattices of linear size $L$ and 
periodic boundary conditions,
so a system in $D$ spatial dimensions will have $N=L^D$ nodes.
The partition function includes 
\index{boundary conditions}
an applied magnetic field $h$. In this chapter we will work \index{field, external}
at fixed $\beta$. Hence, to lighten the expressions we shall drop the explicit
$\beta$ dependencies. This simplified notation is also employed
throughout our applications.

\newcommand{\sU}{\textsf{\itshape U\/}}
\newcommand{\sM}{\textsf{\itshape M\/}}
\newcommand{\su}{\textsf{\itshape u\/}}
\newcommand{\sm}{\textsf{\itshape m\/}}
\newcommand{\sO}{\textsf{\itshape O\/}}
\newcommand{\sR}{\textsf{\itshape R\/}}
\newcommand{\sr}{\textsf{\itshape r\/}}
\newcommand{\sx}{\textsf{\itshape x\/}}
\newcommand{\sX}{\textsf{\itshape X\/}}
\newcommand{\shatb}{\hat{\textsf{\itshape b\/}\,}\!}
\newcommand{\shath}{\hat{\textsf{\itshape h\/}\,}\!}
\newcommand{\shatm}{\hat{\sm\,}\!}
\newcommand{\shatx}{\hat{\sx\,}\!}
\newcommand{\shatM}{\hat{\sM\,}\!}
\newcommand{\shatX}{\hat{\sX\,}\!}
The spin interaction energy and magnetisation of a given configuration  are
\begin{align}\label{eq:TMC-U-M}
\sU(\{s_\bx\}) &= N \su(\{s_\bx\})= - \sum_{\langle\bx,\by\rangle} s_\bx s_\by,&
\sM(\{s_\bx\}) &= N \sm(\{s_\bx\}) =\sum_\bx s_\bx,
\end{align}
Since this chapter is concerned with the construction of a new statistical 
ensemble, we have to be very precise with our notation.
Therefore, we use sans-serif italics for random variables (i.e., functions
of the spins) and serif italics for real numbers (e.g., expectation values or
arguments of probability density functions). For future chapters, dedicated
to physical results, we return to the usual convention  and will no
longer make this distinction explicit.

For instance, we shall denote the expectation values in the canonical
ensemble, for a given value of the applied magnetic field, by
\begin{align}
U(h) &= N u(h)= \langle\sU\,\rangle(h) , &
M(h) &= N m(h)=\langle\sM\,\rangle(h) .
\end{align}
As noted in Chapter~\ref{chap:disorder}, whenever a symbol has an uppercase and a lowercase version,
they correspond to extensive and intensive quantities, respectively.
These expectation values  are weighted averages over the $2^N$ possible configurations
$\{s_\bx\}$ of the system
\begin{align}\label{eq:TMC-canonical-average}
\langle \sO\, \rangle(h) = \frac{1}{Z(h)} \sum_{\{s_\bx\}} \sO(\{s_\bx\})\  \exp\bigl[-\beta \sU(\{s_\bx\})+\beta h \sM(\{s_\bx\})\bigr].
\nomenclature[ 2]{$\langle{O}\rangle(h)$}{Canonical expectation value of $O$ for the applied field $h$}
\end{align}
Let us for the moment consider the case $h=0$ and use the shorthand
\begin{equation}
\braket{\sO\,} = \langle \sO\,\rangle(h=0).
\end{equation}
Since this is a ferromagnetic system, we may be interested
in considering the average value of $\sO$ conditioned to different 
magnetisation regions. The naive way of doing this  would 
be 
\begin{align}\label{eq:TMC-O-m-fijo}
\langle \sO\, | m \rangle = \frac{\bigl\langle \sO\, \delta\bigl(Nm- \sM(\{s_\bx\})\bigr)\rangle}
{\bigl\langle\delta\bigl(Nm- \sM(\{s_\bx\})\bigr)\bigr\rangle}\, .
\end{align}
The canonical average could then be recovered by a weighted
average of the $\braket{\sO|m}$,
\begin{equation}
\langle \sO\, \rangle = \sum_m \langle \sO\,|m\rangle\ p_1(m),
\end{equation}
where
\begin{equation}\label{eq:TMC-p1}
p_1(m)=\frac{1}{Z(0)}  \sum_{\{s_\bx\}} \delta\bigl(m- \sM(\{s_\bx\})/N\bigr)  \exp\bigl[-\beta \sU(\{s_\bx\})\bigr]\, .
\end{equation}
In the thermodynamical limit, $p_1$ would be a smooth function, \index{thermodynamical limit}
the logarithm of the effective potential associated \index{effective potential}
to the reaction coordinate $m$ (cf. Section~\ref{sec:INTRO-legendre}).
For a finite system, however, there are only $N+1$ possible values
of $m$, so $p_1(m)$ is a comb-like function.

We want to construct a statistical ensemble where a smooth effective potential \index{effective potential}
can be defined in finite lattices. 
The first step is extending the configuration space 
with a bath $\sR$ of $N$  demons and defining the smooth magnetisation $\shatM$, \index{demons}
\begin{equation}\label{eq:TMC-hatM}
\shatM = \sM + \sR,
\index{demons}
\nomenclature[R]{$R$}{Bath of Gaussian demons}
\nomenclature[Mhat]{$\hat M,\hat m$}{Smooth magnetisation}
\end{equation}
where\footnote{Actually, $\sR$ can be defined in different ways, see Section~\ref{sec:TMC-demonios}.} 
\begin{equation}
\sR = \sum_i \eta_i^2/2.\label{eq:TMC-demonios-cuadraticos}
\end{equation}
The demons are statistically independent from the spins 
and Gaussianly distributed \index{Gaussian distribution}
\begin{equation}\label{eq:TMC-p2-r}
p_2(r) = \int  \mathrm{D}\eta \ 
\exp\biggl[- \sum_i \eta^2_i/2\biggr] \delta\biggl(r - \sum_i \eta_i^2/(2N)\biggr),
\qquad  \mathrm{D}\eta = \biggl(\prod_{i=1}^N \frac{\dd\eta_i}{\sqrt{2\uppi}}\biggr).
\end{equation}
Notice that, due to the central limit theorem, \index{central limit theorem}
the above probability distribution approaches a Gaussian of mean $1/2$ and variance $(2N)^{-1}$ 
in the large-$N$ limit.
Now our partition function is
\begin{align}
Z(0)&= \int\mathrm{D}\eta\ 
\sum_{\{s_\bx\}} \exp\biggl[ -\beta \sU(\{s_\bx\}) -\sum_i \eta^2_i/2 \biggr].
\end{align}
The convolution of $p_1(m)$ and $p_2(r)$ then gives the probability 
density function (pdf) for $\shatm$,
\begin{equation}\label{eq:TMC-p-hatm}
p(\hat m) = \int\dd m \int\dd r\ p_1(m) p_2(r)
\delta(\hat m-m-r).
\nomenclature[p]{$p(\hat m)$}{Probability density function of $\hat m$}
\nomenclature[pdf]{pdf}{Probability density function}
\end{equation}
So $p(\hat m)$ is essentially a smooth version of 
$p_1(\hat m-1/2)$. 
Writing $p(\hat m)$ explicitly we have
\begin{align}
p(\hat m) &= \frac{1}{Z(0)} \int\mathrm{D}\eta
\sum_{\{s_\bx\}} \ee^{-\beta \sU(\{s_\bx\})-\sum_i \eta^2_i/2}
\delta\biggl(\hat m- \sm(\{s_\bx\})-\sum_i\eta^2_i/(2N)\biggr)\\
&=\sum_{\{s_\bx\}}\frac{\ee^{-\beta \sU(\{s_\bx\})+\sM(\{s_\bx\}) - N \hat m}}{Z(0)}
\int\mathrm{D}\eta\ \delta\biggl(\hat m- \sm(\{s_\bx\})-\sum_i\eta^2_i/(2N)\biggr)\\
\intertext{%
In the second step we have used the Dirac delta to simplify the integral
over the demons, which has been reduced to the computation
of the area of an $N$-dimensional sphere.  We have, finally
}
p(\hat m)&= \sum_{\{s_\bx\}} \frac{(2\uppi)^{N/2}}{\varGamma(N/2) Z(0)} \omega_N(\hat m;\{s_\bx\})
= C \sum_{\{s_\bx\}} \omega_N(\hat m;\{s_\bx\}),
\end{align}
where 
\begin{equation}\label{eq:TMC-omega}
 \omega_N(\hat m;\{s_\bx\}) =  \ee^{-\beta \sU(\{s_\bx\})+ \sM(\{s_\bx\})-\hat M} [\hat m-\sm(\{s_\bx\})]^{(N-2)/2} \varTheta\bigl(\hat m -\sm(\{s_\bx\})\bigr)\, .
\end{equation}
The Heaviside step function enforces the constraint $\hat m \geq \sm$.
We can now introduce the effective potential $\varOmega_N$
\begin{equation}\label{eq:TMC-Omega}
\ee^{-N\varOmega_N(\hat m)} =  p(\hat m) = C \sum_{\{s_\bx\}} \omega_N(\hat m,\{s_\bx\}).
\index{effective potential}
\end{equation}
We want to construct the tethered statistical ensemble, where $\varOmega_N$ would be the basic physical
quantity, instead of the free energy $F_N$. Comparing~\eqref{eq:TMC-Omega} \index{free energy}
with~\eqref{eq:TMC-Z-Ising} and~\eqref{eq:TMC-canonical-average},
we see that $\omega_N$ is going to take the role of the tethered weight,
just as $\exp[-\beta \sU+\beta h \sM]$ in the canonical case.
We therefore define the tethered expectation value as
\begin{equation}\label{eq:TMC-tethered-average}
\langle \sO \rangle_{\hat m}=  \frac{\sum_{\{s_\bx\}}  \sO(\{s_\bx\}) \ \omega_N(\hat m;\{s_\bx\})}{\sum_{\{s_\bx\}}\omega_N(\hat m;\{s_\bx\})}\,,
\nomenclature[ 4]{$\langle O\rangle_{\hat m}$}{Tethered expectation 
value for smooth magnetisation $\hat m$}
\end{equation}
We can now rewrite the canonical average as 
\begin{equation}
\langle \sO \rangle = \int\dd \hat m\  \langle \sO \rangle_{\hat m} 
\ p(\hat m) = \int \dd \hat m \ \braket{\sO}_{\hat m} \ee^{-N \varOmega_N(\hat m)},
\end{equation}
This has the structure of Eq.~\eqref{eq:TMC-O-m-fijo},
but now $\langle \sO\rangle_{\hat m}$
and $p(\hat m)$ are smooth functions even for finite
lattices. 

Suppose now that we want to reintroduce the applied field $h$. 
It is clear from~\eqref{eq:TMC-O-m-fijo} that $\braket{\sO|m}(h) = \braket{\sO|m}$.
Then, computing $\langle\sO\rangle(h)$ is just a matter of reweighting
the different magnetisation sectors:
\begin{equation}
\langle \sO\rangle(h) = \sum_{m} \braket{\sO| m} p_1(m;h),
\end{equation}
where 
\begin{equation}
p_1(m;h) = \frac{1}{Z(h)} \sum_{\{s_\bx\}} \delta\bigl(m- \sm(\{s_\bx\})\bigr)  \exp\bigl[-\beta \sU(\{s_\bx\})+\beta h \sM(\{s_\bx\})\bigr].
\end{equation}
Analogously, in the tethered notation we would have
\begin{equation}\label{eq:TMC-tethered-to-canonical}
\langle \sO\rangle (h) = \frac{1}{Z(h)} \int \dd m\ \braket{\sO}_{\hat m}\ \ee^{-N\varOmega_N(\hat m)+N\beta h \hat m}\ ,
\end{equation}
where the partition function can be written as
\begin{equation}\label{eq:TMC-Z-Omega}
Z(h) = \ee^{N F_N(h)} = \int \dd\hat m\ \ee^{-N [\varOmega_N(\hat m) -\beta h \hat m]}\ .
\end{equation}
This expression illustrates the fact that the construction of the tethered ensemble 
from the canonical one is a Legendre transformation.\index{Legendre transformation}
We have replaced the free energy, a function of $h$, with the 
effective potential, a function of $\hat m$.

In the canonical ensemble the derivative of $F_N$ with respect to 
$h$ defines the magnetisation, the basic observable. Analogously, 
in the tethered ensemble the $\hat m$-derivative of $\varOmega_N$ 
is going to define the tethered magnetic field ---recall Eq.~\eqref{eq:INTRO-conjugate}.
We differentiate in~\eqref{eq:TMC-Omega} to obtain
\begin{equation}
\frac{\partial \varOmega_N}{\partial \hat m} =  
\frac{\sum_{\{s_\bx\}}  \left(1-\frac{1/2-1/N}{\hat m-\sm(\{s_\bx\})}\right) \ \omega_N(\hat m;\{s_\bx\})}{\sum_{\{s_\bx\}}\omega_N(\hat m;\{s_\bx\})}\,,
\end{equation}
Then, recalling the definition of tethered expectation values
 in~\eqref{eq:TMC-tethered-average}  we define the tethered magnetic 
field\footnote{%
In this thesis, we have changed the sign convention for $\varOmega_N$ 
with respect to that of~\cite{fernandez:09}. 
Therefore, our $\shatb$ would be the $-\shath$ of~\cite{fernandez:09}.}

\begin{equation}\label{eq:TMC-hatb}
\shatb(\{s_\bx\}) = 1  - \frac{1/2-1/N}{\hat m - \sm(\{s_\bx\})}\, ,
\nomenclature[b]{$\hat b$}{Tethered field}
\end{equation}
so that
\begin{equation}
\braket{\shatb\,}_{\hat m} = \frac{\partial \varOmega_N}{\partial \hat m}\, .
\end{equation}
The tethered magnetic field is essentially a measure of the fluctuations
in $m$, which illustrates the dual roles the magnetisation and magnetic
field play in the canonical and tethered formalism.  This
is further demonstrated by the tethered fluctuation-dissipation formula
\begin{equation}\label{eq:TMC-fluctuation-dissipation}
\frac{\partial \langle \sO\,\rangle_{\hat m}}{\partial \hat m}
= \left\langle \frac{\partial \sO}{\partial \hat m}\right\rangle_{\hat m}
+ N [ \langle \sO \shatb\,\rangle_{\hat m} - \langle \sO\, \rangle_{\hat m}
\langle \shatb\,\rangle_{\hat m}].
\index{fluctuation-dissipation}
\end{equation}
This formula is easy to prove differentiating in~\eqref{eq:TMC-tethered-average}.
The values of $\hat m$ where $\langle \hat b\rangle_{\hat m}=0$
will define maxima and minima of the effective potential and, hence,
of the probability $p(\hat m)$. 

Definition~\eqref{eq:TMC-hatb}, together with the condition
that $p(\hat m)= \ee^{- N\varOmega_N(\hat m)}$ be normalised,
allows us to reconstruct the Helmholtz potential from the tethered averages
alone.  Therefore, from the knowledge of the  $\braket{\sO}_{\hat m}$ we 
can obtain all the information about the system (at the working temperature $\beta$),
including the canonical averages for arbitrary applied fields. 
Notice, from~\eqref{eq:TMC-tethered-to-canonical}, that considering
a non-zero $h$ is simply equivalent to shifting the tethered magnetic field,
$\langle \shatb\rangle_{\hat m} \to \langle \shatb \rangle_{\hat m} - \beta h$.

The computation of the $\braket{\sO}_{\hat m}$, as detailed
in the previous chapter, is just a matter of running several
independent Monte Carlo simulations with the weight~\eqref{eq:TMC-omega}
in a sufficiently dense grid of $\hat m$. 

As a final comment, we note that our introduction of demons
is reminiscent of Creutz's microcanonical algorithm~\cite{creutz:83},  \index{microcanonical methods}
but there are several important differences: (\textsc{i}) we include    \index{Creutz method}
one demon for each degree of freedom, (\textsc{ii}) our demons          \index{demons}
are continuous variables and (\textsc{iii}) we explicitly integrate out 
the demons, finding a tractable effective Hamiltonian.
Still, notice that one could also define a version of the tethered 
formalism where the demons are not integrated, but rather treated as 
actual dynamical variables. This has worse numerical performance 
(as we shall see, the non-local nature of the conservation law is \index{critical slowing down}
crucial to break the critical slowing down), but could have advantages \index{demons}
for non-standard computer architectures. \index{special-purpose computers}

\subsection{Several tethered variables}\label{sec:TMC-several-tethers}\index{reaction coordinate!multiple|(}
Throughout this section we have considered an ensemble with only one tethered
quantity. However, as we shall see in Chapter~\ref{chap:daff-tethered}, it is often 
appropriate to consider several reaction coordinates at the same time. 
The construction of the tethered ensemble in such a study presents no 
difficulties. We start by coupling reaction coordinates $\sx_i$, $i=1..n$,  
with $N$ demons each,
\begin{align}
\shatX_1 =N\shatx_1=  \sX_1 + \sR_1\,,\quad \ldots,\quad
\shatX_n =N\shatx_n=  \sX_n + \sR_n\,.
\end{align}
We then follow the same steps of the previous section, with the consequence
that the tethered magnetic field is now a conservative field computed
from an $n$-dimensional potential $\varOmega_N(\hat{\bx})$
\begin{align}
\boldsymbol\nabla \varOmega_N &= \bigl( \partial_{\hat x_1} \varOmega_N,\ldots,
\partial_{ \hat x_n} \varOmega_N\bigr) = \hat{\boldsymbol B}\\
\hat {\boldsymbol B } &= \bigl(\langle \shatb_1\rangle_{\hat{\boldsymbol x}},\ldots,
\langle \shatb_n\rangle_{\hat{\boldsymbol x}}\bigr),
\index{effective potential}
\end{align}
and each of the $\shatb_i$ is of the same form as in the case with only 
one tethered variable. Similarly, the tethered weight of Eq.~\eqref{eq:TMC-omega} is now,
up to irrelevant constant factors,
\begin{equation}
\omega^{(n)}_N (\hat{\boldsymbol x}, \{s_\bx\}) \propto \ee^{-\beta U}
\gamma\bigl(\hat x_1,\sx_1(\{s_\bx\}\bigr) \gamma\bigl(\hat x_2,\sx_2(\{s_\bx\})\bigr)\cdots\gamma\bigl(\hat x_n,\sx_n(\{s_\bx\})\bigr).
\index{tethered weight}
\end{equation}
where
\begin{equation}\label{eq:TMC-gamma}
\gamma(\hat x;\sx) = \ee^{N(\sx-\hat x)} (\hat x-\sx)^{(N-2)/2} \varTheta(\hat x- \sx).
\end{equation}
\index{reaction coordinate!multiple|)}

\subsection{The Gaussian demons}\label{sec:TMC-demonios}\index{demons|(}
In the construction of the tethered ensemble we defined the bath of Gaussian
demons as 
\begin{equation}
\sR = \sum_i \eta_i^2/2,\tag{\ref{eq:TMC-demonios-cuadraticos}}
\end{equation}
adding the $\eta_i$ quadratically to the spins. However, we could 
also have considered linear demons, for instance,
\begin{equation}
\sR^{(\text{L})} = \sum_i \eta_i.
\end{equation}
The whole construction could be followed in the same way, 
but we would have
\begin{align}
\omega_N^{(\text{L})}(\hat m;\{s_\bx\}) &\propto \ee^{-\beta \sU- (\sM-\hat M)^2/(2N)},\\
\shatb^{(\text{L})} &= \hat m-\sm(\{s_\bx\}).
\end{align}
Furthermore, we have assumed there are
as many demons as spins.  While this choice seems natural, it is by no means
a necessity. In fact,  if we were to consider an off-lattice
system, the demons would increase the dispersion
in the already continuous $p_1(x)$. In order 
to control the fluctuations it is useful in such a case
to consider a variable 
number of demons (see~\cite{fernandez:11}). Using the linear
$\sR^{(\text{L})}$ we would have
\begin{align}\label{eq:TMC-sum-demons-lineales}
\sR_\alpha & =\frac1\alpha \sum_{i=1}^{\alpha N} \eta_i,\\
\shatb^{(\text{L})} &= \alpha(\hat m- \sm).
\end{align}

\index{demons|)}

\section{Ensemble equivalence}\label{sec:TMC-ensemble-equivalence}\index{ensemble equivalence|(}
In this chapter we have constructed the tethered ensemble, noting
its relation to the canonical one. In particular, we have showed how
to reconstruct canonical expectation values from the tethered
averages as a function of $\hat m$. However, this
is not always the ultimate goal. 
Throughout this dissertation we shall see several cases
where we obtain physically relevant  results without
considering canonical averages. Good examples \index{critical exponent!theta@$\theta$}
are the computation of the hyperscaling violations \index{hyperscaling}
exponent~$\theta$ for the DAFF in Chapter~\ref{chap:daff-tethered} \index{DAFF}
or the computation of the critical exponents ratio 
$\beta/\nu$ in Chapters~\ref{chap:Ising} and~\ref{chap:cluster}. \index{critical exponent!beta@$\beta$}

Still, most of the time the averages in the canonical ensemble are the ones
with an easiest physical interpretation (fixed temperature, fixed applied
field, etc.). In principle, their computation implies reconstructing the whole
effective potential $\varOmega_N$ and using it to integrate over the whole 
coordinate space, as in~\eqref{eq:TMC-tethered-to-canonical}.
Sometimes, however, 
the connection between the tethered and canonical ensembles is easier to make.
Let us return to our ferromagnetic example, with a single tethered quantity $m$.
Recalling the expression of the canonical partition function in terms of 
$\varOmega_N$ for finite $h$, Eq.~\eqref{eq:TMC-Z-Omega}, we see
that the integral is clearly going to be dominated by a saddle point  \index{saddle point}
such that
\begin{equation}\label{eq:TMC-saddle-point}
\partial_{\hat m} [\varOmega_N(\hat m) - \beta h \hat m] = 0\qquad\Longrightarrow
\qquad \langle \shatb\rangle_{\hat m} = \beta h.
\index{saddle point}
\end{equation}
Clearly, this saddle point, the minimum of the effective potential,
rapidly grows in importance with the system 
size $N$, to the point that we can write
\begin{equation}\label{eq:TMC-ensemble-equivalence}
\lim_{N\to \infty} \langle \sO \rangle(h) =\lim_{N\to \infty} \langle \sO \rangle_{\hat m(h)}.
\end{equation}
That is, in the thermodynamical limit we can identify the canonical average \index{thermodynamical limit}
for a given applied magnetic field $h$ with the tethered average computed 
at the saddle point defined by $h$, which is nothing more than the point
where the tethered magnetic field coincides with the applied magnetic
field (times $\beta$). 

This ensemble equivalence property would be little more than a curiosity if
it were not for the fact that the convergence is actually very fast
(see Chapter~\ref{chap:Ising} for a study). Therefore, in many practical
applications the equivalence~\eqref{eq:TMC-ensemble-equivalence} can 
be made even for finite lattices 
(see Section~\ref{sec:DAFF-disorder-average} below for 
an example).

The saddle-point approach can be applied to systems with spontaneous symmetry
breaking. \index{spontaneous symmetry breaking}
In the typical analysis, one has to perform first a large-$N$
limit and then a small-field
one (this is troublesome for numerical work, where one usually has to 
consider non-analytical observables such as $|m|$). In the tethered ensemble 
we can implement this double limit in an elegant way by considering a restricted
range in the reaction coordinate from the outset (see Chapter~\ref{chap:Ising} for a  \index{reaction coordinate}
straightforward example in the Ising model and Section~\ref{sec:DAFF-disorder-average}
for a more subtle one).

\index{ensemble equivalence|)}
\index{tethered formalism|)}

\chapter{Tethered Monte Carlo study of the ferromagnetic Ising model}\label{chap:Ising}\index{Ising model|(}
This chapter presents a study of the $D=2$ ferromagnetic Ising model
carried out with the tethered formalism.
This is the first
system that we studied with TMC, as a demonstration 
of the method~\cite{fernandez:09}. It is intended as a step-by-step
guide to TMC in a straightforward application as
well as a demonstration of its power.
 We work 
in several regimes, covering most of the techniques that will
be needed in a more sophisticated implementation (saddle-point equations,
grid refinement, etc.). We also illustrate how TMC can provide
complementary information and enable some new analysis techniques
(see, for instance, the computation of $\beta/\nu$ in Section~\ref{sec:ISING-betanu}).

The two-dimensional Ising model was solved by Onsager in 1944~\cite{onsager:44}. 
Since then, many other exact results have been obtained (see~\cite{mccoy:73} for a review), making it
the best understood model with a phase transition. 
Furthermore, the Ising model is the ideal setting for sophisticated 
Monte Carlo methods, particularly cluster algorithms (cf. Chapter~\ref{chap:cluster}). Therefore, we can 
confront our results with accurate numerical computations
 even in those cases where the exact solution is
unknown. 

\section{The model and observables}\label{sec:ISING-model}
We consider the two-dimensional Ising model, introduced
in Chapter~\ref{chap:tmc-formalismo},\footnote{As in Chapter~\ref{chap:tmc-formalismo}, we do not
write explicit $\beta$ dependencies, since we always work at constant temperature.}
\begin{equation}
Z(h) = \ee^{NF_N(h)}=  \sum_{\{s_\bx\}} \exp\biggl[ \beta  \sum_{\langle\bx,\by\rangle} 
s_\bx s_\by + \beta h \sum_\bx s_\bx\biggr],\tag{\ref{eq:TMC-Z-Ising}}
\end{equation}
\index{partition function!Ising}
The infinite-volume system undergoes a second-order phase 
transition at a critical (inverse) temperature $\beta_\text{c}$
\begin{equation}\label{eq:ISING-betac}
\beta_\text{c} = \frac{\log(1+\sqrt{2})}2 = 0.440\,686\,793\,509\,771\ldots \index{critical temperature!Ising}
\end{equation}
The main observables we consider are the energy $U$ and the magnetisation $M$, both
defined on Eq.~\eqref{eq:TMC-U-M}. Other interesting physical quantities
are the specific heat $C$ and the magnetic susceptibility $\chi_2$
\begin{align}
C &= N [\braket{u^2}-\braket{u}^2],\label{eq:ISING-C}\\
\chi_2 &= N [\braket{m^2}-\braket{m}^2].\label{eq:ISING-chi2}
\end{align}
\nomenclature[C]{$C$}{Specific heat}
\nomenclature[chi]{$\chi_2$}{Magnetic susceptibility}
\index{specific heat}
\index{susceptibility!magnetic}
\index{fluctuation-dissipation}
The latter quantity can be generalised to define the higher 
cumulants of the magnetisation,
\begin{equation}\label{eq:ISING-chi-2n}
\chi_{2n}=\frac1{N\beta^{2n}} \left.\frac{\partial^{2n} \log Z}{(\partial h)^{2n}}\right|_{h=0}=
\frac{1}{\beta^{2n}}\left.\frac{\partial^{2n} F_N}{(\partial h)^{2n}}\right|_{h=0} \ ,
\index{cumulants}
\index{free energy}
\end{equation}
These can also be defined as the zero-momentum components of 
the $2n$-point correlation functions. In particular, the
two-point propagator is
\index{propagator|see{correlation function}}
\index{correlation function (equilibrium)!spatial}
\begin{equation}\label{eq:ISING-G2}
G_2(\boldsymbol k) = \frac1N \sum_\bx \braket{s_\bx s_0}\ \ee^{\ii \boldsymbol k \cdot \bx}\ .
\end{equation}
In the limit when $k\to0$, this function can be approximated by its
free-field form~\cite{zinn-justin:05,amit:05}
\begin{align}\label{eq:ISING-free-field}
G_2(\boldsymbol k) &\xrightarrow[\ \ |\boldsymbol k| \to0\ \ ]{} \frac{A}{\mu^2+ \cancel{k}^2},&
\cancel{k}^2&= 4\sum_{i=1}^D \sin^2 k_i/2.
\index{correlation function (equilibrium)!free field}
\end{align}
We can use the correlation function in position space to define 
a correlation length,
\begin{equation}\label{eq:ISING-xi-exp}
\xi_\text{exp} = \lim_{|\bx|\to\infty} \frac{- |\bx|}{\log \tilde G_2(\bx)}\ .
\end{equation}
In a finite lattice this definition is not practical but we can use 
the free-field approximation~\eqref{eq:ISING-free-field} to arrive 
at more easily measured ones. In particular, definition~\eqref{eq:ISING-xi-exp}
would give $\xi_\text{exp}=\mu^{-1}$ so we define
\begin{equation}\label{eq:ISING-xi-generic}
\xi^2 = \frac{G_2(\boldsymbol k_1)- G_2(\boldsymbol k_2)}{\cancel k_2^2 G_2(\boldsymbol k_2)
- \cancel k_1^2 G_2(\boldsymbol k_1)}.
\end{equation} 
We can choose $\boldsymbol k_1 = (0,0,0)$ and 
$\boldsymbol k_2 = \boldsymbol k_\text{min}^{(i)} = (2\uppi/L) \hat{\boldsymbol u}_i$, 
where $\hat{\boldsymbol u}_i$ is one of the unit vectors in reciprocal space.
With this choice, $\boldsymbol k_\text{min}^{(i)}$ is the smallest non-zero
momentum compatible with our periodic boundary conditions.
\nomenclature[kmin]{$k_\text{min}$}{Smallest non-zero momentum}
We thus arrive at the second-moment correlation length $\xi_2$,
\begin{align}\label{eq:ISING-xi-second-moment}
\xi_2 &= \frac{1}{2\sin(\uppi/L)}\left[ \frac{\chi_2}
{G_2(\boldsymbol k_\text{min})}-1\right]^{1/2}\, &
G_2(\boldsymbol k_\text{min}) &= \frac1D \sum_{i=1}^D G_2\bigl( \boldsymbol k^{(i)}_\text{min}\bigr)\, ,
\index{correlation length}
\index{correlation length!second moment}
\nomenclature[xi2]{$\xi_2$}{Second-moment correlation length}
\end{align}
(remember that $G_2(0)=\chi_2$).
This definition of the correlation length
was introduced in~\cite{cooper:82} and has since then proved very useful
in finite-size scaling studies~\cite{caracciolo:95,ballesteros:96,ballesteros:97}.

In the broken symmetry phase (or with non-zero $h$), definition~\eqref{eq:ISING-xi-second-moment}
does not work and we have to use larger momenta (see Section~\ref{sec:ISING-h-neq-zero}).

A final interesting observable is the Binder ratio, related to the fourth cumulant
of the magnetisation,
\begin{equation}\label{eq:ISING-Binder}
B = \frac{\braket{m^4}}{\braket{m^2}^2}\ .
\nomenclature[B]{$B$}{Binder ratio}
\index{Binder ratio}
\end{equation}

For the infinite model, the energy and specific heat 
can be shown to be (see, e.g., \cite{huang:87} for a summary of the 
computation)
\begin{align}
u^\infty(\beta) &= -\coth(2\beta) \left[1+\frac2\uppi \kappa' K_1(\kappa)\right],\label{eq:ISING-E-inf}\\
C^\infty(\beta) &= \frac2\uppi(\beta \coth 2\beta)^2 \left[2K_1(\kappa)-2E_1(\kappa)-(1-\kappa')\left(\frac\uppi2+\kappa' K_1(\kappa)\right)\right]. \label{eq:ISING-C-inf}
\index{Ising model!Onsager solution}
\intertext{%
Here $E_1(x)$ and $K_1(x)$ are the complete elliptic integrals of the first
and second kind (see, e.g., \cite{gradshteyn:00})
and $\kappa=2\sinh(2\beta)/\cosh^2(2\beta)$, $\kappa'=2\tanh^2 (2\beta)-1$.
In addition, C. N. Yang~\cite{yang:52} showed that the spontaneous magnetisation
is}
\label{eq:ISING-m-yang}
m^\infty(\beta)&=\begin{cases} 0, & \beta<\beta_\text{c},\\
\left[ 1-\bigl(\sinh(2\beta)\bigr)^{-4}\right]^{1/8}. & \beta> \beta_\text{c}.
\index{magnetisation!Yang}
\end{cases}
\end{align}
We shall also compare our results with the exact expressions for finite $L$
computed by Ferdinand and Fisher in~\cite{ferdinand:69} (too long to 
reproduce here).

\section[Results at $\beta_{\text{c}}$, $h=0$]{Results at $\boldsymbol \beta_{\text{\bfseries c}}$, $\boldsymbol h \boldsymbol = \boldsymbol 0$}\label{sec:ISING-h-eq-zero}
\begin{figure}
\centering
\includegraphics[height=.7\columnwidth,angle=270]{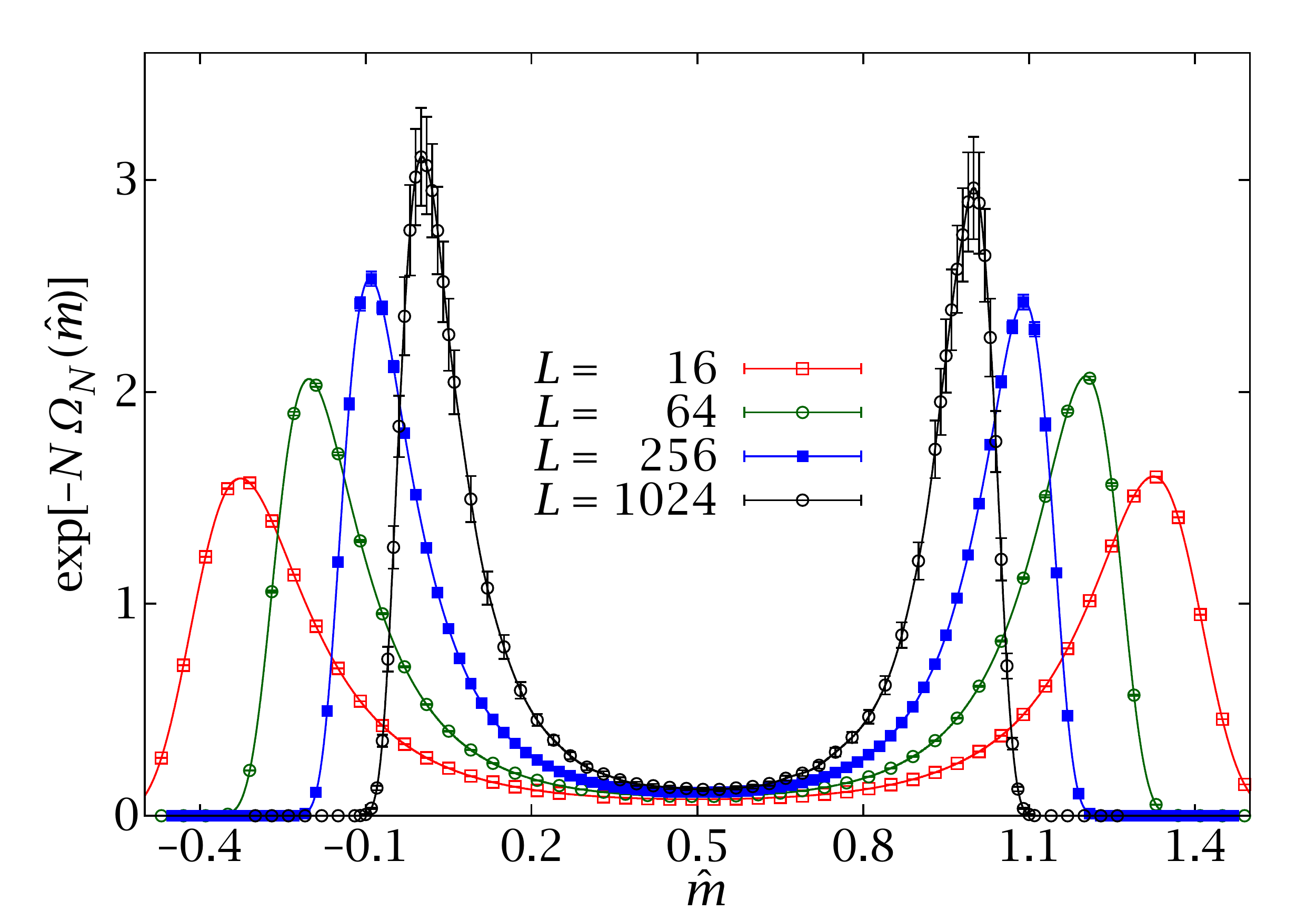}
\caption[Probability density $p(\hat m)$ at $\beta_\text{c}$ for several sizes]{Probability density function $p(\hat m)$ for several lattice sizes
at the critical temperature. The peaks get closer as $L$ grows, eventually 
merging in the thermodynamical limit.\index{p@$p(\hat m)$|indemph}
\label{fig:ISING-p-hatm}
}
\end{figure}
We have used the TMC method, with the simple Metropolis implementation described 
in Chapter~\ref{chap:tmc},
to simulate lattices $L=16,32,\ldots,1024$ at the critical point of the $D=2$
ferromagnetic Ising model. Since the tethered formalism was already constructed
using the Ising model as an example, no further notes on it are needed
here. However, we still have some practical decisions to make, regarding the numerical implementation:
\begin{itemize}
\item We need to decide on a numerical interpolation scheme for the $\braket{O}_{\hat m}$ 
and on a numerical integration method to obtain $\varOmega_N$ from $\braket{\hat b}_{\hat m}$.
\item We need to decide which values of $\hat m$ to simulate.
\end{itemize}
These issues are discussed in detail on  Appendix~\ref{chap:recipes}, which
also contains some practical recipes for the numerical implementation
of the Metropolis algorithm and for the analysis of some delicate physical
observables. Suffice it to say here that the issues of  interpolation and 
integration are not critical (all the tethered averages being very smooth 
functions of $\hat m$).

As to the  $\hat m$ sampling, it is best done starting 
from a uniform grid and, after a first analysis,
perhaps refining the areas that are seen to have
more weight in the $p(\hat m)$.
 In our case, 
for systems with $L\leq256$ we have sampled the $p(\hat m)$  (see Figure~\ref{fig:ISING-p-hatm})
with $51$ uniformly distributed points in the interval $\hat m \in[-0.5,1.5]$ (except
for $L=16$, where the range had to be widened to avoid cutoff errors).
For larger lattices the probability density drops much faster 
at the tails, so we have narrowed the range to $\hat m\in[-0.3,1.3]$ and, since
the peaks are also steeper, we have then added another $26$ points 
around the probability maxima (see Section~\ref{sec:ISING-hatm-grid}).

In all cases we have performed $10^7$ MCS for each value of $\hat m$ and computed
the errors with $100$ jackknife blocks (see Appendix~\ref{chap:correlated}
for our error estimation techniques),\index{jackknife method} after discarding the first \index{error analysis}
fifth of the measurements for thermalisation (although the correlation times are
much smaller, see Section~\ref{sec:ISING-numerical-performance}).

In the following subsection we report the result of computing the canonical
expectation values for several standard observables for zero magnetic field.
We then demonstrate the new options afforded to us by TMC, 
by  carrying out an unconventional, but very precise, computation of the anomalous 
dimension of the system from the position of the peak in $p(\hat m)$.
In Section~\ref{sec:ISING-h-neq-zero} we shall reweight these simulations
to obtain canonical expectation values at non-zero magnetic field.

\subsection{Computation of the canonical expectation values}
\begin{table}[t]
\centering
{\small \begin{tabular*}{\columnwidth}{@{\extracolsep{\fill}}llllll}
\toprule
$L$ & \multicolumn{1}{c}{ $-\braket{u}$} &
\multicolumn{1}{c}{$\chi_2/L^2$} & \multicolumn{1}{c}{$\xi_2/L$} & 
\multicolumn{1}{c}{$C$}& \multicolumn{1}{c}{$B$} \\
\toprule
\fontfamily{pplx}\selectfont
$16$ (TMC)        & 1.453\,08(4) & 0.545\,43(6) & 0.911\,6(2)  & 7.718\,6(14) & 1.165\,62(7)\\
$16$ (CAW)        & 1.452\,9(2)  & 0.545\,1(3)  & 0.910\,4(9)  & 7.718(10)    & 1.165\,9(3) \\
$16$ (E)        & 1.453\,065\ldots&           &              & 7.717\,134\ldots               \\
\midrule
$32$ (TMC)        & 1.433\,69(4) & 0.459\,00(10)& 0.907\,2(4)  & 9.509(3)   & 1.167\,23(14)\\
$32$ (CSW)        & 1.433\,67(12)& 0.459\,1(2)  & 0.907\,8(9)  & 9.493(13)  & 1.167\,1(3)  \\
$32$ (E)        & 1.433\,659\ldots&           &              & 9.509\,379\ldots\\
\midrule
$64$ (TMC)       & 1.423\,97(4) & 0.386\,19(18) & 0.906\,5(9)   & 11.285(6)  & 1.167\,5(3)\\
$64$ (CSW)       & 1.423\,90(6) & 0.386\,0(2)   & 0.905\,6(10)  & 11.293(17) & 1.167\,7(4)\\
$64$ (E)       & 1.423\,938\ldots &           &               & 11.288\,138\ldots\\          
\midrule
$128$ (TMC)      & 1.419\,05(5) & 0.324\,4(3)   & 0.904\,0(18)  & 13.063(10) & 1.168\,4(7)\\
$128$ (CSW)      & 1.419\,06(4) & 0.324\,59(17) & 0.904\,8(10)  & 13.06(2)   & 1.167\,7(4)\\
$128$ (E)      & 1.419\,076\ldots &           &               & 13.060\,079\ldots \\ 
\midrule
$256$ (TMC)      & 1.416\,63(5) & 0.272\,8(6)   & 0.904(4)     & 14.83(2) & 1.168\,7(14) \\
$256$ (CSW)      & 1.416\,64(2) & 0.272\,86(14) & 0.904\,2(9)    & 14.83(2) & 1.168\,2(4)\\
$256$ (E)      & 1.416\,645\ldots&            &              & 14.828\,595\ldots \\
\midrule
$512$ (TMC)      & 1.415\,42(4)  & 0.229\,3(7)  & 0.903(6)     & 16.57(3) & 1.168(2)   \\ 
512 (CSW)      & 1.415\,444(11)& 0.229\,68(13)& 0.905\,9(10) & 16.60(2) & 1.167\,6(4)\\
512 (E)      & 1.415\,429\ldots&            &              & 16.595\,404\ldots&\\
\midrule
1024 (TMC)     & 1.414\,89(4)  & 0.194\,9(15) & 0.919(15)    & 18.28(8) & 1.163(6)   \\
1024 (CSW)     & 1.414\,826(6) & 0.193\,07(12)& 0.904\,6(11) & 18.35(3) & 1.168\,1(4)\\
1024 (E)     & 1.414\,821\ldots&            &              & 18.361\,348\ldots       \\
\bottomrule
\end{tabular*}}
\caption[Results at $\beta=\beta_\text{c}$, $h=0$]{Results at the critical temperature and comparison with a cluster algorithm.
(TMC): Tethered Monte Carlo,
(CSW): Canonical Swendsen-Wang, (E): Exact results at finite $L$ from~\cite{ferdinand:69}.
\label{tab:ISING-results-h-eq-zero}}
\index{specific heat!Ising|indemph}
\index{magnetisation!Ising|indemph}
\index{energy!Ising|indemph}
\index{correlation length!Ising|indemph}
\index{Binder ratio!Ising|indemph}
\index{Swendsen-Wang algorithm|indemph}
\end{table}
Our first physical result is the pdf of $\hat m$, represented in
Figure~\ref{fig:ISING-p-hatm}. This distribution features 
a central minimum in the zero-magnetisation region (recall that $\hat m\simeq m + 1/2$) 
as well as two symmetric peaks that get closer together as $L$ grows. Of course, since
this system experiences a second-order phase transition, the peaks should converge in 
the thermodynamical limit (where the effective potential is a convex function).

In Table~\vref{tab:ISING-results-h-eq-zero} we report our estimates
for the canonical expectation values of several standard observables. We check 
our results against the values obtained from the exact formulas in~\cite{ferdinand:69}
and against a canonical Swendsen-Wang simulation\index{Swendsen-Wang algorithm}.
We use our own implementation, based on the one distributed with~\cite{amit:05}, 
although our results are compatible with those of~\cite{salas:00}. We perform
$10^7$ Swendsen-Wang cluster updates.\index{cluster methods}

From Table~\ref{tab:ISING-results-h-eq-zero} we can confirm that a simple implementation of  TMC is capable
of producing very accurate results. The relative errors for $\chi_2$ and $B$ 
scale as $L$. This can be explained by noticing that both 
are completely determined by $p(\hat m)$ (see the discussion following
Eq.~\eqref{eq:ISING-moments} in Appendix~\ref{chap:recipes}).
In addition, $\hat b$ is self-averaging (cf. Section~\ref{sec:INTRO-self-averaging})  \index{self-averaging} 
and, as we shall see in Section~\ref{sec:ISING-numerical-performance}, virtually
free of critical slowing down\index{critical slowing down} (meaning that for a fixed simulation length
its error scales as $1/\sqrt{N}$). Finally, in the computation
of $p(\hat m)$ we are multiplying $\varOmega_N(\hat m)$ by a factor of $N$, yielding
an overall $\sqrt{N}$ scaling for the errors.

Of course, TMC is not meant to be a competitor to cluster algorithms for the Ising model
without magnetic field. For example, the CPU time to compute 
each of the $77$ simulations at fixed $\hat m$ for $L=1024$ is
similar to what we needed for the whole Swendsen-Wang simulation, which
is also more precise. See, however, Chapter~\ref{chap:cluster}
for a tethered implementation of the Swendsen-Wang algorithm, which
turns out to be as efficient as the canonical Swendsen-Wang for
the Ising model. In any case, our tethered version of the Metropolis
algorithm is much more efficient than canonical Metropolis \index{Metropolis algorithm}
(see Section~\ref{sec:ISING-numerical-performance}).

\subsection{The magnetic critical exponent}\label{sec:ISING-betanu}
\index{finite-size scaling|(}
Let us now see an example of the new kind of analyses
afforded to us by the tethered formalism. 
In particular, we shall compute the critical
parameter $\beta/\nu$ (recall the definitions \index{critical exponent!beta@$\beta$}
of the critical exponents in~\ref{sec:INTRO-critical-exponents})
in a very simple way that, however, would not
be practical in a canonical simulation.

We consider the finite-size scaling formula (recall Section~\ref{sec:INTRO-FSS})
\begin{equation}\label{eq:ISING-FSS}
\braket{O}(h) = L^{-y_O/\nu} \left[ f_O( L^{1/\nu} t, L^{y_h} h)+ \ldots\right],\qquad t =  \frac{\beta_\text{c}-\beta}{\beta_\text{c}}.
\end{equation}
We have included a second variable in the scaling function $f_O$ to 
allow for displacements in the applied field $h$, not
only in temperature. These are regulated by a new critical
exponent $y_h$ (analogous to $y_t=1/\nu$). The dots
represent possible corrections to scaling. 
For the moment, we shall work with no field, $h=0$.
\begin{table}
\small
\centering
\begin{tabular*}{\columnwidth}{@{\extracolsep{\fill}}rll}
\toprule
$L$ & \multicolumn{1}{c}{$-m^{-}_\mathrm{peak}=\bigl|\hat m^{-}_\text{peak} - \tfrac12\bigr|$} &
 \multicolumn{1}{c}{$m_\mathrm{peak}^{+} = \hat m_\text{peak}^+ - \tfrac12$}\\
\toprule
32  & 0.764\,01(10) & 0.764\,31(11) \\
64  & 0.702\,86(18) & 0.703\,0(2)     \\
128 & 0.645\,3(3)   & 0.645\,1(4)   \\
256 & 0.592\,1(7)   & 0.591\,0(7)   \\
512 & 0.541\,9(12)  & 0.542\,7(9)   \\
1024& 0.499(2)      & 0.500(2)      \\
\bottomrule
\end{tabular*}
\caption[Position of the peaks of $p(\hat m)$]{Position of the
positive and negative maxima of the $p(\hat m)$. Since we are
at the critical temperature of a second-order phase transition, 
both columns should extrapolate to zero in the large-$L$ limit
(cf. our study of the ferromagnetic region in Section~\ref{sec:ISING-ordered-phase}). 
\label{tab:ISING-peak}
\index{p@$p(\hat m)$|indemph}}
\end{table}

Applied to the magnetisation, whose associated critical 
exponent is $\beta$, the FSS ansatz implies that
\begin{equation}
\tilde p_1(m,\beta_\mathrm{c};L) = L^{\beta/\nu} \tilde f(L^{\beta/\nu} m),
\end{equation}
where $\tilde p$ is a smooth version of $p_1(m;L)$.  Now, 
the pdf of $\hat m$, $p(\hat m;L)$ ---Eq.~\eqref{eq:TMC-p-hatm}--- 
is precisely a smooth version of $p_1(\hat m-1/2)$. Therefore, we can write
\begin{equation}
p(\hat m,\beta_\mathrm{c};L) = L^{\beta/\nu} f\bigl(L^{\beta/\nu} (\hat m-1/2)\bigr).
\end{equation}
If we concentrate on the peaks of the pdf, we see that  their
height is going to grow as $L^{\beta/\nu}$, while their
position is going to shift as 
\begin{equation}\label{eq:ISING-shift}
|\hat m_\text{peak}^{\pm}-\frac12| \simeq A  L^{-\beta/\nu}.
\end{equation}
Now, computing the maximum of a numerical function is 
usually a delicate operation. However, in our case 
we have measured directly $\hat b$, which is the derivative
of (the logarithm of) $p(\hat m)$. Herein lies our advantage
with respect to a canonical simulation (which cannot access
$\hat b$ directly), because we only 
have to find a zero, a much better conditioned 
operation. In particular, we run through our 
computed tethered averages $\braket{\hat b}_{\hat m_i}$
until we find two consecutive points in the grid
such that $\braket{\hat b}_{\hat m_i} >0$ 
and $\braket{\hat b}_{\hat m_{i+1}} <0$.
Then, we find the single zero of the section of our
cubic spline (see Appendix~\ref{chap:recipes}) that  \index{cubic splines}
joins both points. This operation is performed for 
each jackknife block in order to  \index{jackknife method}
estimate the statistical error (cf. Appendix~\ref{chap:correlated}).
The resulting maxima are collected in Table~\ref{tab:ISING-peak}.

Notice that if we just wanted to obtain the maxima very precisely, 
we would not have performed a simulation of the whole $\hat m$ range.
Instead, we would have done a fast sweep to place the peaks
approximately and then we would have
simulated only a few points in their neighbourhood,
to a very high precision. However, even with our suboptimal  
grid we have determined the position of the maxima with a 
relative ranging from $\sim10^{-4}$ for $L=32$ to $\sim 5\times 10^{-3}$
for $L=1024$.

\begin{table}
\small
\centering
\begin{tabular*}{\columnwidth}{@{\extracolsep{\fill}}rlclc}
\toprule
\multirow{2}{0.5cm}{$L_\mathrm{min}$}&  \multicolumn{2}{c}{$m^{-}_\mathrm{peak}$} &  \multicolumn{2}{c}{$m_\mathrm{peak}^+$} \\
\cmidrule{2-3} \cmidrule{4-5}
& 
\multicolumn{1}{c}{$\beta/\nu$}
 & $\chi^2/\mathrm{d.o.f.}$
 &\multicolumn{1}{c}{$\beta/\nu$} & $\chi^2/\mathrm{d.o.f.}$  \\
\toprule
32  & 0.121\,7(3)   &  23.44/4  &  0.122\,4(3)       &  27.85/4  \\ 
64  & \textbf{0.123\,9}(5)   &  \textbf{2.027/3}  &  \textbf{0.124\,5}(5)       &  \textbf{2.087/3}   \\
128 & 0.125\,0\textbf{(11)}  &  0.7569/2 &  0.124\,6\textbf{(10)}     &  2.053/2   \\
256 & 0.126(4)      &  0.6456/1 &  0.122\,0(23)      &  0.3248/1      \\
\bottomrule
\end{tabular*}
\caption[Computation of $\beta/\nu$]{Fits
 of $\hat m_\text{peak}^{\pm}(L)$ to~\eqref{eq:ISING-shift},
in order to find $\beta/\nu$, for different 
fitting ranges $L\geq L_\text{min}$.
For each fit we give the chi-square estimator
and the degrees of freedom (see Appendix~\ref{chap:correlated}).
Our results converge to the exact
value, $\beta/\nu=0.125$.}\label{tab:ISING-beta-nu}
\index{critical exponent!beta@$\beta$|indemph}
\end{table}

We have fitted the $\hat m_\text{peak}^{\pm}$ to~\eqref{eq:ISING-shift}, 
for different fitting ranges $L\geq L_\text{min}$ (Table~\ref{tab:ISING-beta-nu}).
The fits are good (in the sense of an acceptable $\chi^2$/d.o.f.,
see Appendix~\ref{chap:correlated}) for $L_\text{min}\geq 64$. Still, there
are possible systematic sources of error (corrections to scaling), 
so, following~\cite{ballesteros:96}, we give as our final
estimate the fit for $L_\text{min}=64$, but with the larger 
error of the fit for $L_\text{min}=128$.
For the negative magnetisation peak we find $\beta/\nu = 0.123\,9(11)$
and for the positive peak $\beta/\nu=0.124\,5(10)$. Both
are compatible with the exact value for the $D=2$ Ising model, 
known to be $\beta/\nu=1/8$.

Since  we have a wide range of lattice sizes, we can try 
to characterise the first corrections to scaling. \index{finite-size scaling!corrections}
Following~\cite{salas:00}, we assume that in the $D=2$ Ising model
the dominant corrections to scaling are analytical
\begin{equation}
m_\text{peak}^{\pm} = L^{-\beta/\nu}[ A^\pm + B^\pm L^{-\varDelta}],\qquad \varDelta = 7/4.
\end{equation}
We have fitted our points for all lattices to this expression, fixing 
the exponents to their exact values and varying $A^\pm$ and $B^\pm$.
We have obtained $\chi^2/\text{d.o.f.} = 0.9858/4$ for the negative 
peak and $\chi^2/\text{d.o.f.} = 2.825/4$ for the positive one.

\index{finite-size scaling|)}
\section[Results at $\beta_{\text{c}}$, $h\neq0$]{Results at $\boldsymbol \beta_{\text{\bfseries c}}$, $\boldsymbol h \boldsymbol \neq \boldsymbol 0$}\label{sec:ISING-h-neq-zero}
\begin{figure}
\centering
\includegraphics[height=.7\columnwidth,angle=270]{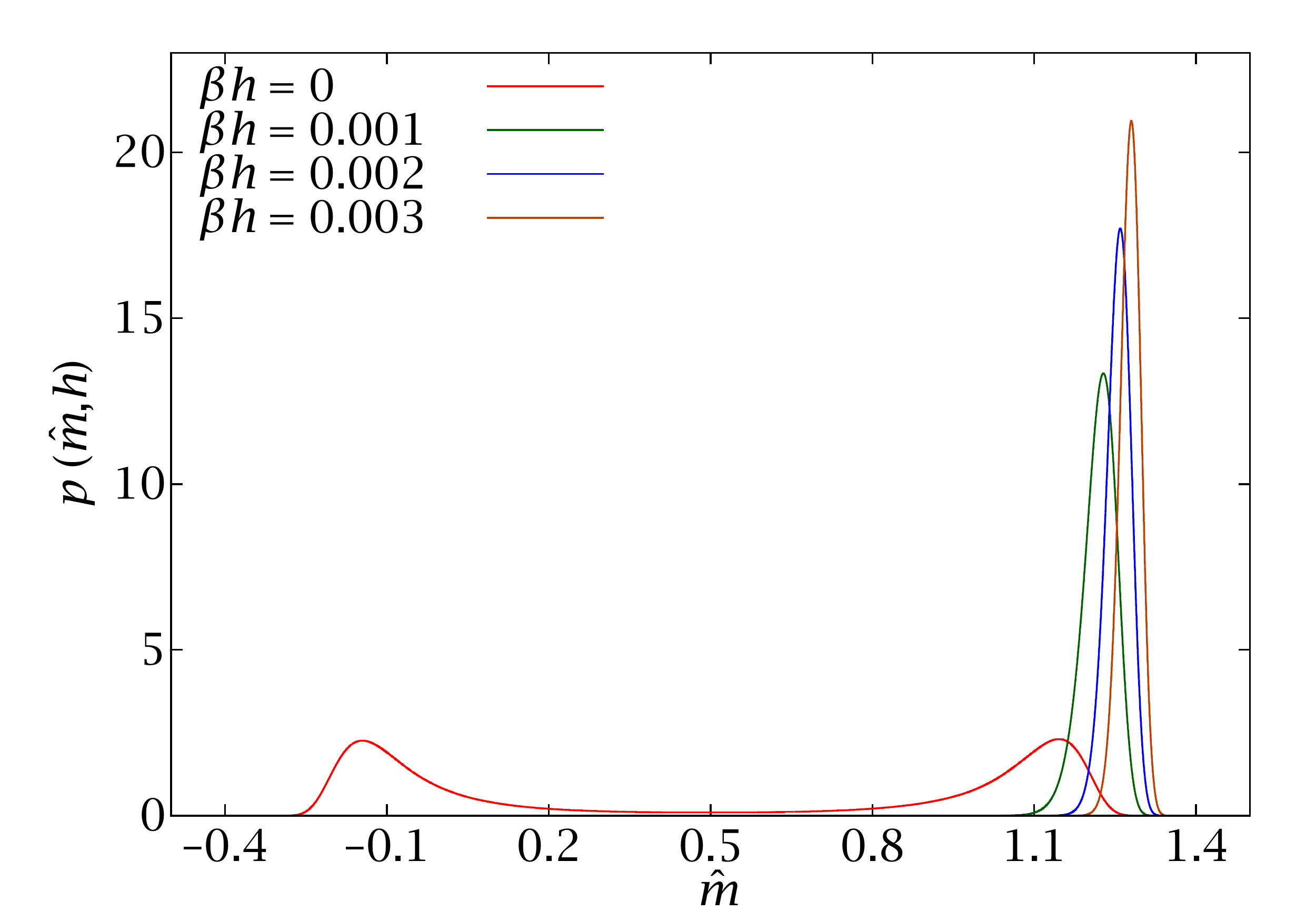}
\caption[Probability density $p(\hat m,h)$ at $\beta_\text{c}$ for several applied fields]%
{Probability density function $p(\hat m,h)$ for $L=128$ at $\beta_\text{c}$,
for several values of the applied magnetic field $h$.
\index{p@$p(\hat m)$|indemph}
\label{fig:ISING-p-hatm-h}
}
\end{figure}
\index{tethered formalism|(}
One of the strengths of TMC is that it can compute the canonical expectation
values for an arbitrary applied field $h$ without any need for new simulations. 
Notice that in the  canonical formalism not only would we have to run a new simulation
for each value of $h$, but we would also lose the possibility of using the 
highly efficient cluster methods. \index{cluster methods}
One simply has to change the weight of 
the tethered expectation values when integrating
over $\hat m$, as indicated by Eq.~\eqref{eq:TMC-tethered-to-canonical}.
Figure~\ref{fig:ISING-p-hatm-h} shows an example of this reweighting. We plot
the pdf $p(\hat m;h)$ for several values of the applied field.
By integrating the same tethered averages computed in Section~\ref{sec:ISING-h-eq-zero} with this new pdf we can obtain the canonical average
at the corresponding applied magnetic field.

Notice that now only a very narrow region has any significant contribution to
the canonical average, which implies a loss of statistical precision.
In general, if one is interested in the  canonical average at a
particular value of $h$, the $\hat m$ grid should be chosen so that the 
resulting peak is appropriately sampled, neglecting the exponentially
suppressed region away from it. This is a simple task, because the position
\index{saddle point}
of the peak is easily determined from a first trial run,
as we saw in the previous section. The only difference is that
now we do not have to solve 
for $\braket{\hat b}_{\hat m}=0$, but for
\begin{equation}\label{eq:ISING-saddle-point}
\braket{\hat b}_{\hat m} = \beta h.
\end{equation} 
Once the peak has been accurately placed, the best strategy is running long
tethered simulations precisely there and at neighbouring points. 

In this section, however, we are not interested in any particular value of
$h$, but rather in studying the behaviour of the 
canonical averages $\langle{O}\rangle(h)$ as smooth functions of the magnetic field.
The simulations we already have will be sufficient, accepting a small 
loss of precision, because the $\hat m$ grid will not be necessarily optimised.  We can
improve our precision somewhat, though,  if we notice that most 
interesting observables are either odd or even functions of the applied magnetic field. 
Therefore, we can (anti)symmetrise the curves:
\begin{align}
\langle O\rangle^\mathrm{odd}(h) &= \frac{\langle{O}\rangle(h) 
- \langle{O}\rangle(-h)}{2}\ ,\label{eq:ISING-odd}\\  
\langle O\rangle^\mathrm{even}(h) &= \frac{\langle{O}\rangle(h) 
+ \langle{O}\rangle(-h)}{2}\ .\label{eq:ISING-even}
\end{align}
For statistically independent data, averaging two equivalent estimates  in this way would
yield an error reduction of $1/\sqrt{2}$. But this is not the case
here: the individual averages for $\pm h$ are very strongly correlated. 
Therefore, the error reduction for even quantities is negligible. 
For odd quantities, on the other hand, since we are computing a difference, 
the fluctuations are greatly suppressed and the resulting error
reduction is very large (around a factor of~$10$, specially
for small values of $h$). We shall use equations~\eqref{eq:ISING-odd} 
and~\eqref{eq:ISING-even}, but
dropping the explicit `odd' or `even' superscripts.
\index{error analysis}
\index{tethered formalism|)}

\subsection{The magnetisation}
We first consider the most straightforward observable, the magnetisation $\langle m\rangle(h)$.
This is  an odd function of $h$, so as discussed above we can greatly increase 
our precision by using formula~\eqref{eq:ISING-odd}. 
\index{magnetisation!Ising}

In this case, unlike previous sections, 
no data for~$\langle m\rangle(h)$ are readily available in a finite lattice, since
canonical cluster methods lose much of their power in the presence of
a magnetic field (in fact, current numerical methods for the 
investigation of the $D=2$ Ising model in a field typically
rely on transfer matrix techniques~\cite{caselle:00,grinza:03,caselle:04}).
\index{transfer matrix}
Therefore, rather than compare our results directly with other computations, 
we shall carry out self-consistency checks.

The first step is recalling Eq.~\eqref{eq:ISING-chi-2n}, 
\index{free energy}
according to which the free energy of the system (an even function of $h$, for obvious
symmetry considerations)
can be written as a Taylor expansion in the following way
\begin{equation}\label{ISING-F_N-Taylor}
F_N(h) - F_N(0) = \sum_{n=1}^\infty \frac{\chi_{2n}}{(2n)!} (\beta h)^{2n}\ .
\end{equation} 
\index{susceptibility!magnetic}
Therefore, the magnetisation $\langle{m}\rangle(h)$ is simply
\begin{equation}\label{eq:ISING-m-h-Taylor}
\langle m\rangle(h)= 
\frac1\beta\frac{\partial F_N}{\partial h} = 
\chi_2 \ \beta h +
\frac{\chi_4}{3!} (\beta h)^3 + \frac{\chi_6}{5!} (\beta h)^5 + \frac{\chi_8}{7!}(\beta h)^7 + \ldots 
\end{equation}
Now, we can compute $\chi_{2n}$ as the cumulants of the magnetisation 
at $h=0$, but this equation provides an alternative way. We generate a reasonable
number of points of the $\langle m\rangle(h)$ curve, which we then parameterise 
with a truncated version of~\eqref{eq:ISING-m-h-Taylor}.  The choice of 
values for $h$ is somewhat delicate: if we use very small 
magnetic fields we will only be able to appreciate the first few coefficients
but if we go too far in $h$ we would need to have sampled the tails of 
the pdf of $\hat{m}$ very precisely. We have found that magnetic fields 
up to $\beta h\sim (\chi_2)^{-1}$ provide a good compromise (notice that 
this value is $L$-dependent). 
\begin{table}[t]
\small
\begin{tabular*}{\columnwidth}{@{\extracolsep{\fill}}rllll}
\toprule
\multicolumn{1}{c}{$L$} & 
\multicolumn{1}{c}{$N^{-1}\chi_2$} &
\multicolumn{1}{c}{$N^{-2}\chi_4$} &
\multicolumn{1}{c}{$N^{-3}\chi_6$} &
\multicolumn{1}{c}{$N^{-4}\chi_8$}\\
\toprule
16 (M)        & 0.545\,43(6)       & $-0.545\,72(13)$     & 2.265\,7(8)        & $-20.059(10)$ \\ 
16 \;(F)      & 0.545\,43(6)       & $-0.545\,7(2)$       & 2.262\,8(19)       & $-19.70(14)$   \\
\midrule
32 (M)        & 0.459\,00(10)      & $-0.386\,1(2)$       & 1.348\,5(10)       & $-10.042(10)$\\
32 \;(F)      & 0.459\,00(10)      & $-0.386\,2(3)$       & 1.348\,4(18)       & $-10.00(2)$  \\
\midrule
64 (M)        & 0.386\,19(18)      & $-0.273\,3(3)$       & 0.803\,1(13)       & $-5.032(11)$ \\ 
64 \;(F)      & 0.386\,19(18)        & $-0.273\,8(5)$       & 0.805(2)           & $-5.02(2)$   \\ 
\midrule
128 (M)       & 0.324\,4(3)        & $-0.192\,8(4)$       & 0.475\,8(17)           & $-2.504(12)$ \\
128 \;(F)     & 0.324\,4(3)        & $-0.194\,3(7)$       & 0.481(3)           & $-2.52(2)$   \\ 
\midrule
256 (M)       & 0.272\,8(6)        & $-0.136\,2(7)  $     & 0.283(2)           & $-1.250(12)$ \\
256 \;(F)     & 0.272\,5(6)        & $-0.135\,8(15) $     & 0.280(6)           &$ -1.20(4)  $ \\
\midrule                                                                                     
512 (M)       & 0.229\,3(7)        & $-0.096\,4(7)  $     & 0.168(2)           &$  -0.625(9)$   \\
512 \;(F)     & 0.229\,3(7)        & $-0.096\,0(12) $     & 0.166(4)           &$ -0.60(2)  $ \\
\midrule
1024 (M)      & 0.194\,9(15)       & $-0.069\,8(13) $     & 0.104(3)           &$  -0.328(12)$  \\
1024 \;(F)    & 0.194(3)           & $-0.063(6)     $     & 0.08(2)            &$ -0.2(3)   $ \\
\bottomrule
\end{tabular*}                       
\caption[Non-linear susceptibilities]{Non-linear susceptibilities from direct measurements at $h=0$ (M) and
from a finite-difference formula for $\langle m\rangle(h)$ as
a function of the magnetic field (F).
\label{tab:ISING-chi_2n}
\index{susceptibility!Ising|indemph}
\index{cumulants|indemph}}
\end{table}                          

As we discuss in Appendix~\ref{chap:correlated}, the correlation 
among the points in the curve has both a beneficial
and a detrimental effect. The former is that, since all
the points  fluctuate more or less coherently, 
the resulting curves are very smooth. The latter is that
these same correlations make it very difficult to 
estimate errors or to check whether a proposed \index{fitting techniques}
fitting function is a good model for the data.
In this section we have taken the approach
of computing an odd interpolating polynomial with a finite-difference formula,
which gives us as many $\chi_{2n}$ as we have points. We estimate
the statistical errors with the jackknife  \index{jackknife method}
method. We still face a systematic error, which we try
to control by varying both the range in $h$ and the number
of points. We have found that the last one or two
coefficients in the fit are typically unstable.
Therefore, if we want to obtain $n$ physically meaningful parameters,
we should compute at least $n+2$ points. 
In our case, we have computed the  non-linear susceptibilities up to $\chi_{8}$,
so, to be safe, we have used $7$ points for each lattice size.
These were equally spaced at intervals of $\Delta(\beta h)=(10\chi_2)^{-1}$,
where $\chi_2$ is the susceptibility computed in the simulation at $h=0$.

In Table~\ref{tab:ISING-chi_2n} we compare the non-linear susceptibilities computed
from measurements of the cumulants at $h=0$ and from the $\langle m\rangle(h)$ 
curve. Both series are compatible, but the former are more precise.

\begin{figure}[p]
\centering
\includegraphics[height=\columnwidth,angle=270]{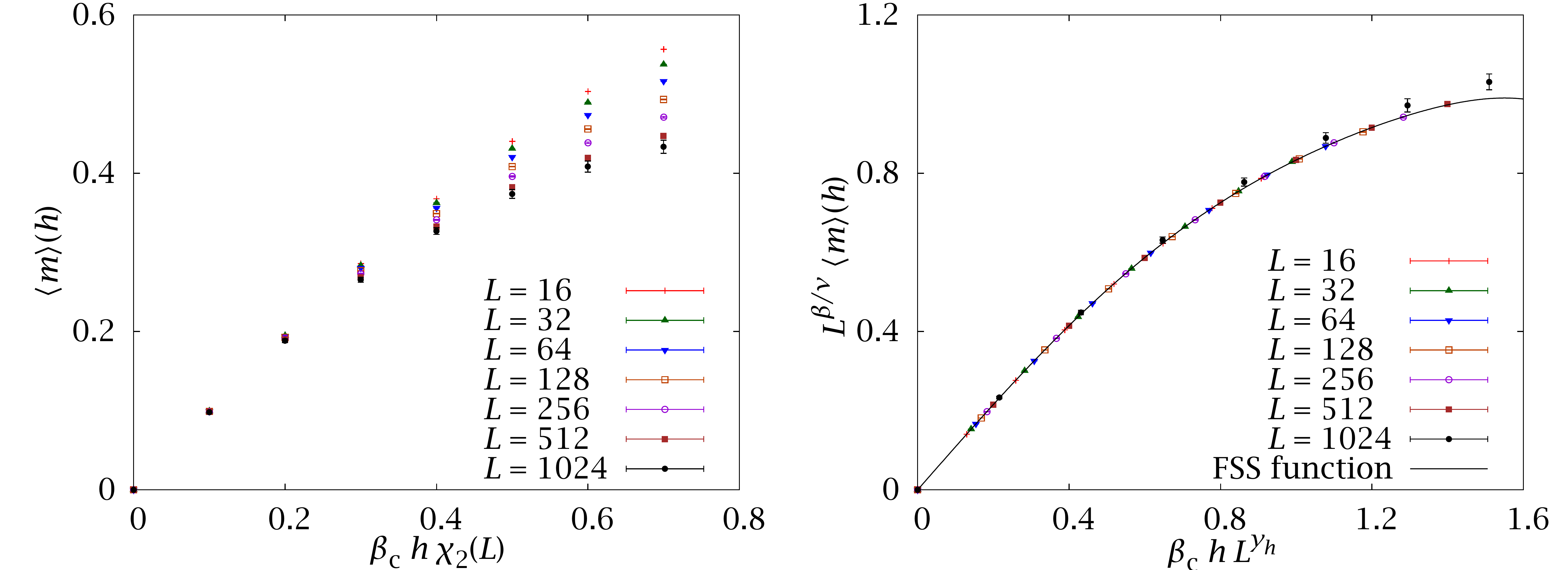}
\caption[Magnetisation as a function of the applied field and FSS plot]{
Magnetisation as function of the applied magnetic 
field. In the left panel we have scaled the graphs by the linear
susceptibility at zero field, $\chi_2(L)$, and appreciate deviations for large
fields. In the right panel we show a scaling plot,
with a much better collapse. The continuous
line is a fit for the scaling function  $f_m$, Eq.~\eqref{eq:ISING-m-FSS},
using a seventh degree polynomial, with $\chi^2_\mathrm{d}/\text{d.o.f}=41.85/31$.
\label{fig:ISING-m-h-FSS}
\index{magnetisation!Ising|indemph}
\index{finite-size scaling|indemph}
\index{scaling plot|indemph}
}
\vspace*{1.5cm}

\centering
\includegraphics[height=.7\columnwidth,angle=270]{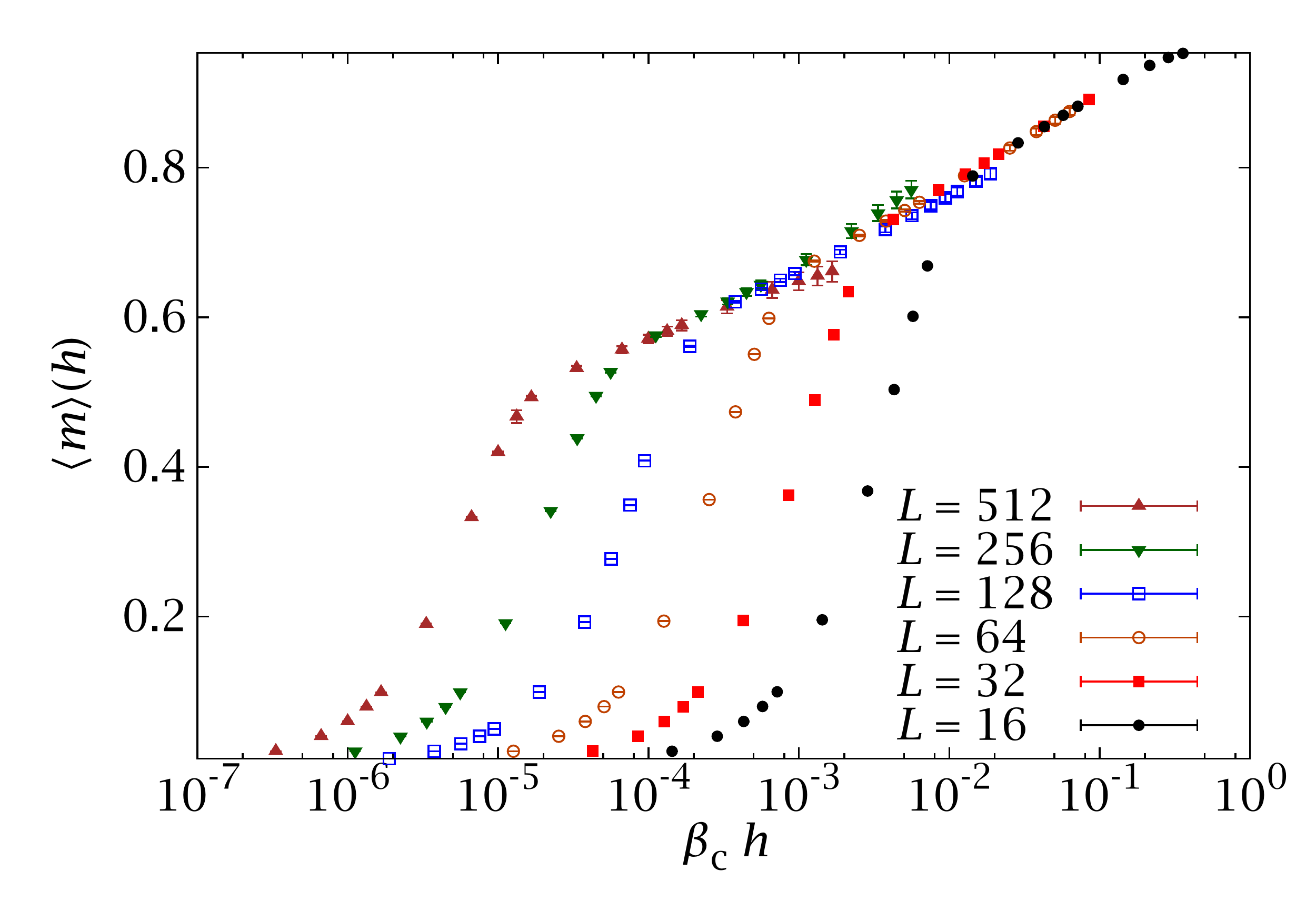}
\caption[Magnetisation as a function of the applied field, logarithmic scale]
{
Magnetisation as a function of the applied field for several lattices
at $\beta_\text{c}$. We use an $L$-independent range 
for $h$ (notice the logarithmic scale of the horizontal axis).
\label{fig:ISING-m-h-grande}
\index{magnetisation!Ising|indemph}
}
\end{figure}

We can also perform a FSS analysis.\index{finite-size scaling}
For small applied fields we can collapse the curves for different sizes 
by plotting $\langle m\rangle(h)/\chi_2$ against $\beta h$. As $h$ grows,
however,  we start to appreciate the deviations from linear behaviour computed
above (Figure~\ref{fig:ISING-m-h-FSS}---left). We can attempt a better collapse of the
curves for different $L$ with the FSS formula (remember Section~\ref{sec:ISING-betanu})
\begin{align}\label{eq:ISING-m-FSS}
\langle m \rangle (h) &\simeq L^{-\beta/\nu} f_m(L^{y_h} h),&
y_h = 15/8.
\end{align}
We have plotted $\langle m\rangle L^{\beta /\nu}$ against $h L^{y_h}$ on 
Figure~\ref{fig:ISING-m-h-FSS}---right. We also include a fit
for the scaling function $f_m$. We use an odd seventh-degree polynomial
and include in the fit the points for $L\geq 64$. The value of the diagonal
$\chi^2_\text{d}$ for this fit is $\chi^2_\text{d}/\text{d.o.f.} = 41.85/31$ (see
Appendix~\ref{chap:correlated} for definitions).
The last point in the curve starts to show a deviation, probably due to corrections
to leading order scaling.  Remember that we spaced our values of $\beta h $ in units 
of $(10\chi_2)^{-1}$, which is not an optimal choice for a FSS study.

We complete our analysis of $\langle m\rangle(h)$ by considering  now $L$-independent 
magnetic fields across several orders of magnitude (Figure~\ref{fig:ISING-m-h-grande}).
We observe two well-differentiated scales: a FSS regime, where the slope 
of the curve is very large, and a saturation regime where the curves merge.
Notice that the only hard limit in the values of $h$ we can consider is given
by the saddle-point equation~\eqref{eq:ISING-saddle-point}. So long as $\beta h$ 
is contained within the bounds set by our measured values of $\braket{\hat b}_{\hat m}$,
the peak in $p(\hat m;h)$ will be contained in our simulated range and we can 
interpolate it with some precision (this is the reason for the rapid error
growth in our largest lattices). Notice that the displacement of the curves in the FSS section
seems linear in $\log L$. This is an artifact of the low value of $\beta/\nu=1/8$ in 
Eq.~\eqref{eq:ISING-m-FSS}.

\begin{figure}
\centering
\includegraphics[height=.7\linewidth,angle=270]{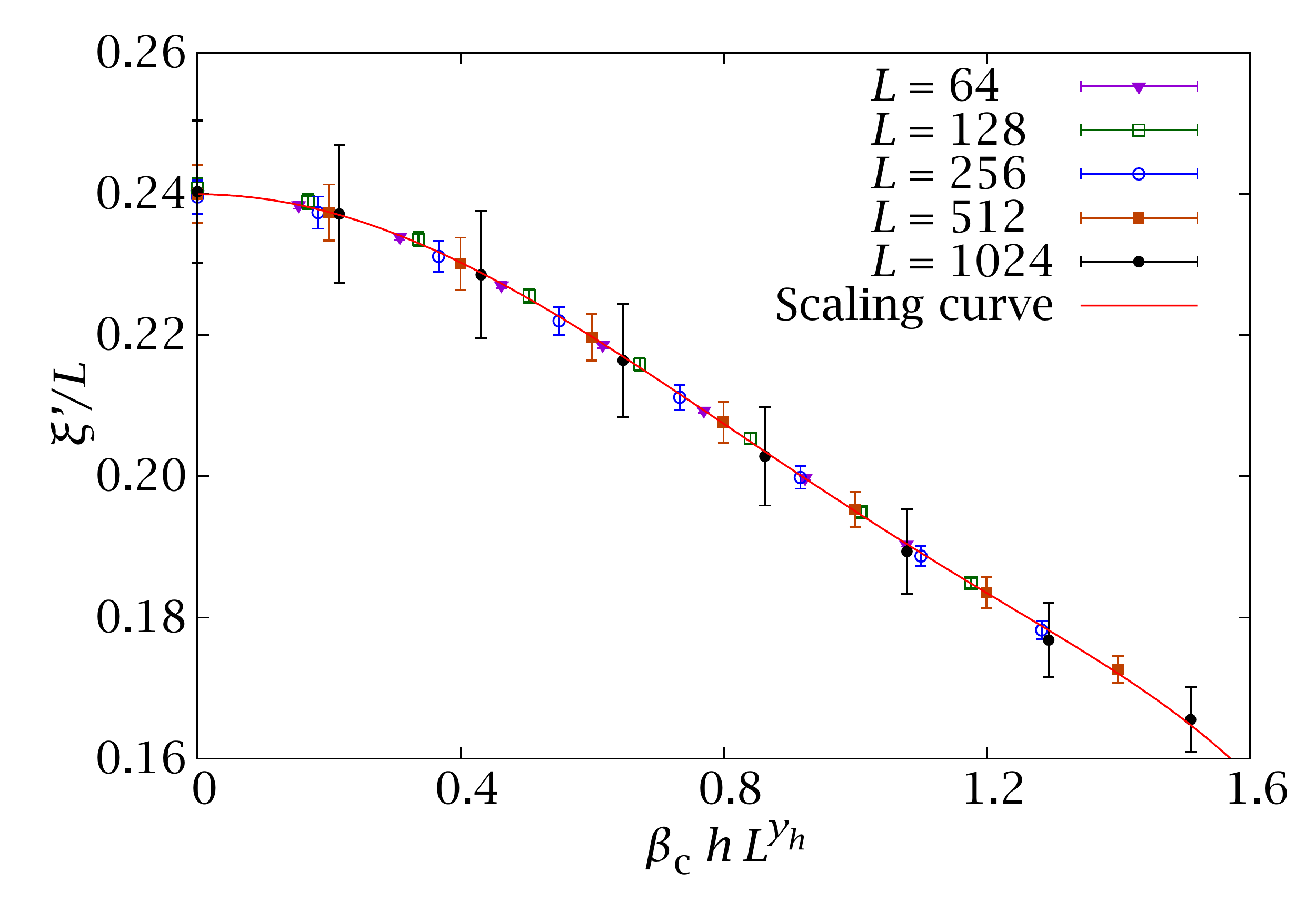}
\caption[Scaling plot of the correlation length in a field]{Correlation 
length $\xi'(h)/L$, Eq.~\eqref{eq:ISING-xi-prime}, against the scaling 
variable $\beta_\text{c} h L^{y_h}$. We also plot a fit to the scaling 
function $f_{\xi'}$ of~\eqref{eq:ISING-xi-h}.
\label{fig:ISING-xi-h}
\index{correlation length|indemph}} 
\end{figure}

\subsection{The correlation length}
We consider now the correlation length of the system for non-zero applied
field. As we discussed in Section~\ref{sec:ISING-model}, the second-moment
\index{correlation length!Ising}
definition~\eqref{eq:ISING-xi-second-moment}  does not work in an applied
field, so we have to go back to the generic formula~\eqref{eq:ISING-xi-generic}
and choose different $\boldsymbol k_1$ and $\boldsymbol k_2$. The simplest 
choice is to use the next smallest momenta: $\boldsymbol k_1 = \boldsymbol k_\text{min}$ and 
$\boldsymbol k_2 = 2\uppi/L (1,\pm 1)$,
\begin{equation}\label{eq:ISING-xi-prime}
\xi'= \frac{1}{2\sin (\uppi/L)} \left[ \frac{G_2(\boldsymbol k_\text{min}) - G_2(\boldsymbol k_2)}
{2G_2(\boldsymbol k_2) - G_2(\boldsymbol k_\text{min})}\right]^{1/2}\, .
\end{equation}
Now, $\xi'(h)$ is an even function of $h$, so we symmetrise it with~\eqref{eq:ISING-even}. 
Again, due to the lack of readily available data in a finite lattice for
this observable, we check our results with a FSS analysis.
\index{finite-size scaling}
To leading order, we have
\begin{equation}\label{eq:ISING-xi-h}
\xi'/L \simeq f_{\xi'}(L^{y_h} h),
\end{equation}
where, as in the case of $m$, $f_{\xi'}$ is expected to be very smooth. 
As we can see in Fig.~\ref{fig:ISING-xi-h}, Eq.~\eqref{eq:ISING-xi-h} is 
perfectly valid in our case, if we discard the data for $L\leq 32$.
At a first glance, it may seem that we
have even overestimated our errors for $L=1024$, but remember
that the points are very strongly correlated. The universal scaling curve
is well represented by a sixth order even polynomial, with 
$\chi^2_\mathrm{d}/\text{d.o.f.}=3.978/36$.
Thus, the universal scaling function $f_{\xi_2}(x)$ 
for $x\lesssim1.5$  is very well approximated by
\begin{equation}
f_{\xi'}(x) = a_0 + a_2 x^2 + a_4x^4+ a_6x^6,
\end{equation}
with
\begin{align}
a_0 &= 0.239\,9(2),& a_2&=-0.063\,9(4), & a_4&=0.023\,5(7), & a_6 &= -0.004\,5(4).
\end{align} 
We have estimated the errors in the fit parameters with the techniques of Appendix~\ref{chap:correlated}.

\section{The ordered phase: spontaneous symmetry breaking}\label{sec:ISING-ordered-phase}
In this section we present new simulations of the $D=2$ Ising model in the ferromagnetic 
phase, using them to illustrate the treatment of spontaneous symmetry breaking in
the tethered formalism. We then examine the approach 
to the thermodynamical limit, studying the equivalence between the canonical
and tethered ensembles.

\subsection{Spontaneous symmetry breaking}\label{sec:ISING-SSB}
\index{spontaneous symmetry breaking|(}
Let us now consider the $\beta>\beta_\mathrm{c}$ regime. In this situation,
the infinite system shows a nonzero expectation value
for the order parameter, $\langle m\rangle\neq 0$,
even in the absence of an external magnetic field, Eq.~\eqref{eq:ISING-m-yang}.
This may seem
incompatible with the partition function~\eqref{eq:TMC-Z-Ising}, where
the configurations $\{s_\bx\}$ and
$\{-s_\bx\}$ occur with equal probability. 
The well-known solution for this apparent paradox 
is spontaneous symmetry breaking, see, e.g., \cite{huang:87,zinn-justin:05},
whose mathematical formulation
involves considering a small magnetic field (which establishes 
a preferred direction) and taking the double limit
\index{thermodynamical limit}
\begin{equation}\label{eq:ISING-symmetry-breaking}
\langle m\rangle^{(\infty)} = \lim_{h\to 0} \lim_{L\to\infty}\langle m\rangle^{(L)}(h).
\end{equation}
The order of the two limits is crucial: were we to reverse it, the magnetisation
would always vanish. We see then that the symmetry of our model 
complicates the definition of a broken symmetry phase for finite lattices
in the canonical ensemble. The traditional workaround consists in considering
not the magnetisation $m$, but its absolute value $|m|$. 

\begin{figure}
\centering
\includegraphics[height=.7\linewidth,angle=270]{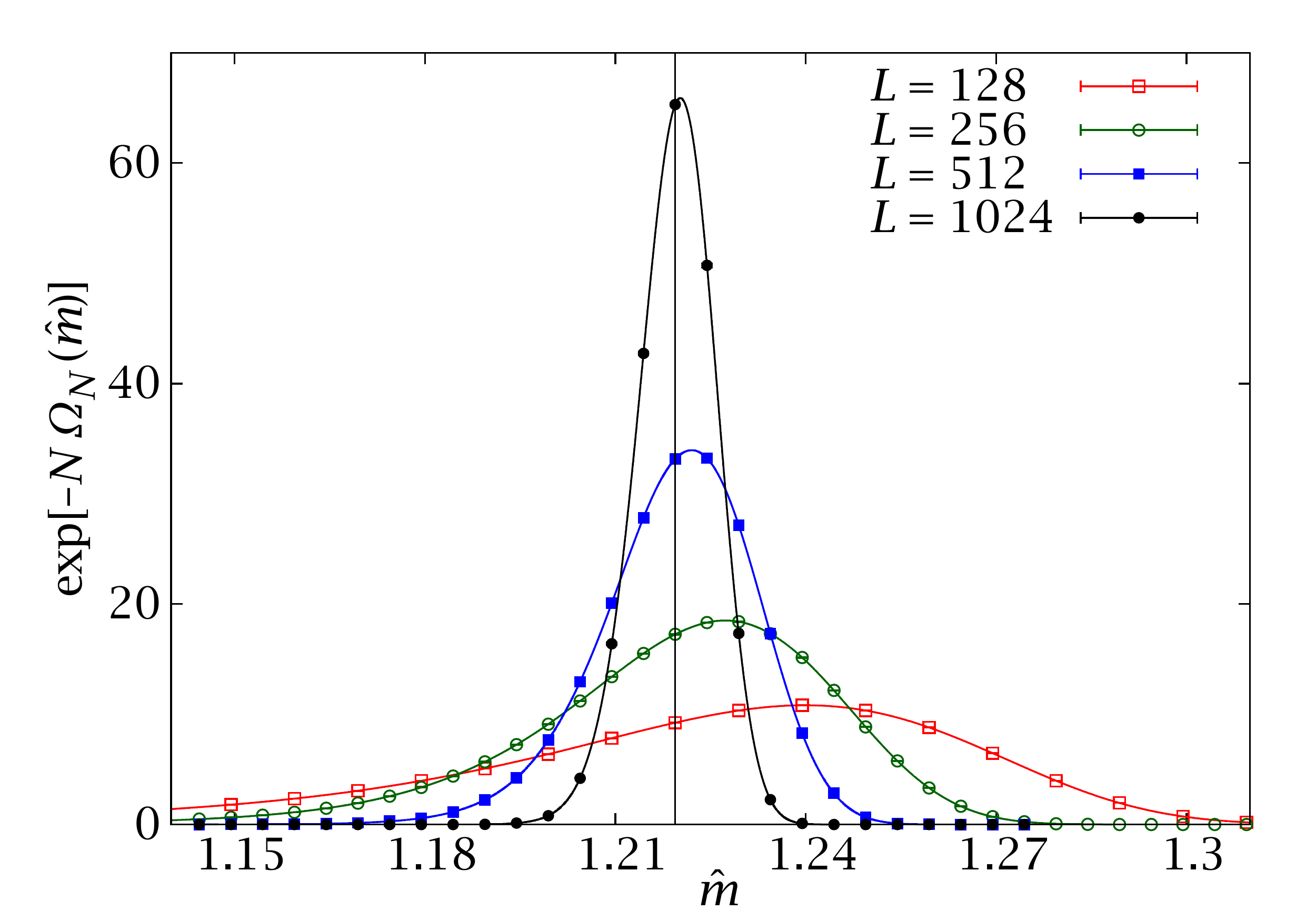}
\caption[Probability density function of $\hat m$ in the ferromagnetic phase]{Computation
at $\beta=0.4473>\beta_{\mathrm{c}}$ (ferromagnetic phase).
The peak of the pdf of $\hat m$ gets narrower and closer
to $m_\textsc{y}+1/2$ as $L$ increases (the vertical line, where
$m_\textsc{y}$ is Yang's magnetisation for the infinite system
at $\beta=0.4473$).
Compare the scale of the $OX$ axis and the height of the peaks
with those of Fig.~\ref{fig:ISING-p-hatm}.}
\label{fig:ISING-p-hatm-yang}
\index{p@$p(\hat m)$|indemph}
\index{magnetisation!Ising|indemph}
\end{figure}
\begin{table}
\centering
\footnotesize
\begin{tabular*}{\columnwidth}{@{\extracolsep{\fill}}rccllll}
\toprule
\multicolumn{1}{c}{$L$} &
 $N_\mathrm{points}$ & $\Delta \hat m$ &
\multicolumn{1}{c}{$-\braket{u}$} & \multicolumn{1}{c}{$C$}
&\multicolumn{1}{c}{$\xi'$}
&\multicolumn{1}{c}{$\braket{m}$}\\
\toprule
128 (T) & 90 & 0.9  &  1.490\,397(18)        & 8.874(4) & 10.394(17) & 0.719\,34(6)\\ 
128 (E) &    &      &  1.490\,409\,763\ldots & 8.877\,363\ldots\\
\midrule
256 (T) & 79 & 0.39  &  1.490\,407(11)        & 8.869(5) & 11.26(4) & 0.719\,41(4) \\
256 (E) &    &      &  1.490\,415\,672\ldots & 8.874\,075\ldots\\
\midrule
512 (T) & 27 & 0.13 &  1.490\,419(5)         & 8.877(5)         &  11.5(3) & 0.719\,45(3)\\
512 (E) &    &      &  1.490\,415\,689\ldots & 8.874\,046\ldots\\
\midrule
1024 (T)& 27 & 0.13 &  1.490\,416(4)         & 8.868(7)         &  11.4(18) & 0.719\,45(2) \\
1024 (E)&    &      &  1.490\,415\,689\ldots & 8.874\,046\ldots\\
\midrule
$\infty$ (E) & &      &  1.490\,415\,689\ldots & 8.874\,046\ldots &          & 0.719\,436\ldots \\
\bottomrule
\end{tabular*}   
\caption[Canonical averages at $\beta=0.4473$]{Canonical averages for several physical quantities
of an Ising lattice at $\beta=0.4473$ computed with 
the tethered method (T). The grid of 
$\hat m$ values is uniform  in the narrow simulated band. Also included
are the exact results for finite lattices from~\cite{ferdinand:69}
and the exact results in the thermodynamical limit from
Eqs.~(\ref{eq:ISING-E-inf}--\ref{eq:ISING-m-yang}).
We appreciate that by simulating
only a very small range $\Delta \hat m$ of values for $\hat m$ we
can obtain very precise values. 
Within our error, we have already reached the thermodynamical limit for $L=512$.}
\label{tab:ISING-results-ferromagnetic}
\index{energy!Ising|indemph}
\index{specific heat!Ising|indemph}
\index{correlation length!Ising|indemph}
\index{magnetisation!Ising|indemph}
\end{table}
The tethered ensemble provides a cleaner concept of broken symmetry phase.
Consider the pdf of $\hat m$, as in Fig.~\ref{fig:ISING-p-hatm}.  In the ferromagnetic
phase the corresponding graph will again have two peaks, but now these will
be much narrower and higher, approaching two Dirac deltas in thermodynamic
limit. Suppose we want to perform the double limit of
equation~\eqref{eq:ISING-symmetry-breaking}. This would involve introducing a small
magnetic field which would shift the origin of $\hat b$~\eqref{eq:TMC-tethered-to-canonical}.
The neighbourhood of one of the peaks would then become exponentially
suppressed and eventually disappear in the thermodynamical limit. Thus, we can mimic
the effect of equation~\eqref{eq:ISING-symmetry-breaking} by considering 
only one of the two peaks from the outset. This would not work
at or below $\beta_{\mathrm{c}}$, as there the peaks extrapolate to $m=0$ (recall
Section~\ref{sec:ISING-betanu}).  This procedure has the 
considerable advantage that it works for any lattice size. In this section we 
have chosen the peak of positive magnetisation. 

We have run simulations for lattice sizes $L=128,256,512,1024$ 
at inverse temperature $\beta=0.4473>\beta_{\mathrm c}$.  We chose
this temperature as we estimated that the correlation length 
would be around $\xi'\approx10$.\footnote{In this section we 
again have to use $\xi'$, Eq.~\eqref{eq:ISING-xi-prime}, instead
of $\xi_2$, Eq.~\eqref{eq:ISING-xi-second-moment}.}

\index{tethered formalism!numerical implementation|(}
Following the previous discussion, we have worked in the
$\hat m > 0.5$ (positive magnetisation) region, where there is only one peak.
An appropriate sampling of $\hat m$ is even more important in this phase (but
easier to optimise) than in the situation described in detail in Section~\ref{sec:ISING-hatm-grid}.
The reason is that the peak is now so narrow that a choice of $\hat m$ spaced
as in the aforementioned section would not only be completely wasteful,
but may also completely fail to sample the peak (remember the loss
of precision for very high magnetic fields in Section~\ref{sec:ISING-h-neq-zero}).

In the case of the Ising model,
we know Yang's exact solution $m^\infty(\beta)$ for the magnetisation of the
infinite system, Eq.~\eqref{eq:ISING-m-yang}.
The positive peak for $p(\hat m)$ will then be very close
to $m^\infty(\beta)+\tfrac12$ and get closer as we increase~$L$. With this
information in hand, we can adequately reconstruct the effective potential
by running simulations in a small neighbourhood of $m^\infty(\beta)+\tfrac12$.
For a different model, where we would lack the knowledge of the peak's position
in the thermodynamical limit, we can just run simulations with a very fine 
grid for some small and essentially costless lattice size and infer from them
an efficient distribution of points for the larger systems. 
\index{tethered formalism!numerical implementation|)}

We have represented $p(\hat m)$ for all the simulated lattices in Fig.~\ref{fig:ISING-p-hatm-yang}, which
plots the whole simulated range of $\hat m$ for $L\geq512$ (for the smaller lattices we have used
a somewhat larger interval). It is interesting to compare the scale on the axes with
that of Fig.~\ref{fig:ISING-p-hatm}. As will be discussed in detail in
Section~\ref{sec:ISING-ensemble-equivalence}, the peak
approaches $m^\infty(\beta)+ \tfrac12$
 (the vertical line) as $L$ increases. Table~\ref{tab:ISING-results-ferromagnetic}
compares the values of the energy and specific heat obtained in our simulations
with the exact values given in~\cite{ferdinand:69}. Notice how very small simulated
ranges of $\hat m$ ($\Delta \hat m = \hat m_\text{max} - \hat m_\text{min}$) yield very accurate results. 
In fact, we can see that for the $L=1024$ lattice we obtain a more precise determination
for the energy with $27$ points than what we obtained at the critical temperature
with $77$ (we still perform $10^7$ Monte Carlo sweeps in each point).
This result is even more impressive if we consider that some of these $27$ points,
being deeply inside the tails of the distribution,
do not have any effect whatsoever 
in the average with our error (of course, we do not know this until we have run the simulation 
and seen the actual width of the peak). 

From Table~\ref{tab:ISING-results-ferromagnetic} we can conclude that the
thermodynamical limit has already been
reached for $L=256$, at least to the level indicated by our errors.
Our whole computation for $L=512$ required about $270$ hours of computer time.
For comparison, a Swendsen-Wang computation with a run time of $30$ hours
\index{Swendsen-Wang algorithm}
for $L=512$ gives $\xi'=11.8(2)$. We see that
the ratio of computation time for both methods has changed significantly 
from the critical point, where the advantage of the cluster algorithm was
much greater.

\index{spontaneous symmetry breaking|)}
\subsection{Ensemble equivalence}\label{sec:ISING-ensemble-equivalence}
\index{ensemble equivalence|(}
Once we wander away from the critical point, the main goal
is finding the value of physical quantities in the thermodynamical limit,
rather than attempting a finite-size scaling.
The ensemble equivalence property discussed in Section~\ref{sec:TMC-ensemble-equivalence}
 suggests a way to reach this limit
without constructing the whole canonical $p(\hat m)$, but 
by concentrating instead on its maximum. From the computational 
point of view, this supposes a dramatic reduction in the needed
effort for a TMC simulation. 

Ensemble equivalence can be expressed in mathematical terms by 
\begin{equation}\label{eq:ISING-ensemble-equivalence}
\lim_{N\to\infty} \braket{O} = \lim_{N\to\infty} \braket{O}_{\hat m_\text{peak}},
\end{equation}
where $\hat m_\text{peak}$ is the saddle point of Section~\ref{sec:TMC-ensemble-equivalence}
(for zero field, in this case).
This equation can be understood
as a more formal way of summarising the behaviour of Fig.~\ref{fig:ISING-p-hatm-yang}. Indeed, 
we saw in the previous section that we could reconstruct the canonical 
averages considering only a very narrow range of $\hat m$; in the thermodynamic
limit a single point would be sufficient. 

For the Ising model we know exactly where this point would be situated,
because, from Yang's spontaneous magnetisation~\eqref{eq:ISING-m-yang}
\begin{equation}
\hat m^\infty(\beta) = \lim_{N \to \infty} \braket{m} + \frac12 =
\left[1- (\sinh 2\beta)^{-4}\right]^{1/8} + \frac12. 
\end{equation}
\index{magnetisation!Yang}
We could then run simulations for several lattice sizes precisely
at $\hat m^\infty$ 
and study the evolution of $\braket{O}_{\hat m^\infty}$ as we increase $L$.
This is not the most practical approach, as for a model other than the $D=2$ Ising
lattice we would not know the position of the peak beforehand. Instead, we will 
follow a more general analysis that would work in more complex situations. 

\index{saddle point|(}
Let us consider the canonical average of some quantity 
and recall that we are using periodic boundary conditions, 
so the approach to the thermodynamical limit is exponential
\begin{equation}\label{eq:ISING-ey}
\braket{O}= \int_{1/2}^\infty \dd\hat m\ p(\hat m,\beta; L) \braket{O}_{\hat m} 
= O^\infty + A_{O} \ee^{-L/\xi_\infty},
\end{equation}
where $A_O$ is a constant amplitude. 
We have just considered the positive magnetisation peak.
\begin{table}[t]
\centering
\small
\begin{tabular*}{\columnwidth}{@{\extracolsep{\fill}}rcll}
\toprule
\multirow{2}{0.5cm}{$L$}   
& \multicolumn{3}{c}{$\beta = 0.4473$ } \\
\cmidrule{2-4} 
 &
 $\hat b^\infty\cdot10^{5}$ & \multicolumn{1}{c}{$\hat m_\mathrm{peak}^+$} &
                \multicolumn{1}{c}{$-u_{\mathrm{peak}}^+$} \\
\toprule
16  &$2984(3)$  & 1.343\,84(4) & 1.577\,07(8)   \\
32  &$924.7(16)$& 1.299\,30(7) & 1.528\,98(9)  \\
64  & $284.9(9)$  & 1.263\,33(7) & 1.505\,48(5)\\
128 & $86.0(6)$ & 1.239\,88(9)& 1.495\,63(4)  \\
256 & $25.5(3)$  & 1.227\,32(8) & 1.491\,99(2)\\
512 & $6.9(2)$  & 1.222\,04(6) & 1.490\,859(16)\\
1024& $2.11(11)$  & 1.220\,24(4) & 1.490\,538(9) \\
\midrule
$\infty$ & 0    & 1.219\,435\ldots & 1.490\,416\ldots\\ 
\bottomrule
\end{tabular*}
\vspace*{\baselineskip}

\begin{tabular*}{\columnwidth}{@{\extracolsep{\fill}}rcll}
\toprule
\multirow{2}{0.5cm}{$L$}   
& \multicolumn{3}{c}{$\beta = 0.6$} \\
\cmidrule{2-4} 
              & $\hat b^\infty\cdot10^5$
              & \multicolumn{1}{c}{$\hat m_\mathrm{peak}^+$} &
                \multicolumn{1}{c}{$-u_{\mathrm{peak}}^+$} \\
\toprule
16   &$ -290.6(14)$ & 1.471\,943(7)        & 1.912\,98(4)           \\ 
32   &$ -52.5(8)  $ & 1.473\,299(5)        & 1.910\,18(2)                \\
64     &$ -11.0(4)  $ & 1.473\,543(2)    & 1.909\,374(9)   \\
128 &$ -2.45(19) $ & 1.473\,594\,0(12)& 1.909\,165(5)    \\
256  &$ -0.67(11) $ & 1.473\,604\,7(2) & 1.909\,107(3)       \\
512  &$ -0.19(5)  $ & 1.473\,607\,7(4) & 1.909\,090\,7(15)     \\
1024  &$ -0.03(2)$   & 1.473\,608\,3(2) & 1.909\,086\,7(4)\\
\midrule
$\infty$ & 0  & 1.473\,608\,7\ldots  & 1.909\,086\,2\ldots\\
\bottomrule
\end{tabular*}
\caption[Tethered averages at the saddle point]{Tethered mean values of several parameters 
at the peak of the probability density function
for $\beta=0.4473$ and $\beta=0.6$, together with the 
value of $\hat b^\infty=\braket{\hat b}_{\hat m^\infty}$
(this observable is zero at the peak and helps characterise
how close we are to it). The exact value for 
an infinite lattice, which coincides with the canonical 
average, is also included for comparison.}
\label{tab:ISING-results-yang}
\index{magnetisation!Ising|indemph}
\index{energy!Ising|indemph}
\index{tethered field!Ising|indemph}
\end{table}

The integral will be dominated by a saddle point at $\hat m_\text{peak}^+$,
with $\hat m_\text{peak}^+ \xrightarrow{L\to\infty} \hat m^\infty$,
so we can approximate the pdf by a Gaussian
\begin{equation}
p(\hat m,\beta;L) \simeq \sqrt{\frac{N}{2\uppi\chi_2}} \exp\left[-\frac{N(\hat m - \hat m_\text{peak})^2}{2\chi_2}\right].
\index{Gaussian distribution}
\end{equation}
Therefore, we expect the tethered average of $O$
at this saddle point to approach the canonical average \eqref{eq:ISING-ey}, with a correction
of order $N^{-1}$,
\begin{equation}
O^\infty= \braket{O} - A_{O} \ee^{-L/\xi_\infty} = \braket{O}_{\hat m_\text{peak}^+} + \mathcal O\left(L^{-D}\right).
\end{equation}
To ease the notation we shall use the definition
\begin{equation}
O^+_{\text{peak}} = \braket{O}_{\hat m_\text{peak}^+}.
\end{equation}
This simple analysis provides a practical way of approaching the thermodynamic
limit without knowing the limiting position of the peak in advance. 

We first run a complete simulation for some small lattice, covering the whole
range of $\hat m$. This provides a first approximation to the position
of the peak. For growing lattices, we just compute two or three points
at both sides of where we think the maximum is going to be. Our objective
is not to reconstruct the whole peak of $p(\hat m)$, just to 
find a good approximation to the point $\hat m_\text{peak}^+$ 
where $\braket{\hat b}_{\hat m}$ vanishes. We use the same
procedure as in Section~\ref{sec:ISING-betanu}, finding the zero of the cubic
spline and interpolating the physical observables. Actually, if the position \index{cubic splines}
of the peak is sufficiently bounded we could just place one point very closely 
at either side and use a linear interpolation.

With this procedure we are able to compute the tethered mean values 
of the relevant physical quantities at the peak with a minimum of
numerical effort. Here we shall apply this method to the energy
and we shall also characterise the approach of the peak to $\hat m^\infty$.
To the latter purpose, we have computed $\hat b^\infty=\braket{\hat b}_{\hat m^\infty}$
for several lattice sizes and studied how fast it approaches zero.
We also give the values for the position of the peak (Table~\ref{tab:ISING-results-yang}).

Following the above analysis we should find that 
\begin{align}
|u^+_{\text{peak}}-u^\infty | &= A_u\cdot L^{-\zeta_u},\nonumber\\
|\hat m^+_{\text{peak},\beta}-\hat m^\infty | &= A_{\hat m}\cdot L^{-\zeta_{\hat m}},\label{eq:ISING-zeta}\\
\hat b^\infty&= A_{\hat b}\cdot L^{-\zeta_{\hat b}}\nonumber,
\end{align}
with $\zeta\approx 2$. We present in Table~\ref{tab:ISING-ensemble-equivalence}
the result of applying the quotients method (see Section~\ref{sec:DAFF-quotients}) 
to these observables.
 As we can see, our results 
are always $\zeta<2$, even though this exponent grows with $L$.

We believe this was caused by the proximity of the critical point, 
so we ran analogous simulations for $\beta=0.6$. We were able to complete
this new computations in very little time, following the above procedure.
For example, for $L=512$ the position of the peak was so tightly bounded
that we just computed one point at either side.

\begin{table}
\centering
\small
\begin{tabular*}{\columnwidth}{@{\extracolsep{\fill}}rllllll}
\toprule
\multirow{2}{0.5cm}{$L$}
& \multicolumn{3}{c}{$\beta = 0.4473$ } &  
 \multicolumn{3}{c}{$\beta = 0.6$} \\
\cmidrule{2-4} \cmidrule{5-7}
& \multicolumn{1}{c}{$\zeta_{\hat b}$}
              & \multicolumn{1}{c}{$\zeta_{\hat m}$}
              & \multicolumn{1}{c}{$\zeta_u$}  
              & \multicolumn{1}{c}{$\zeta_{\hat b}$}
              & \multicolumn{1}{c}{$\zeta_{\hat m}$} &
                \multicolumn{1}{c}{$\zeta_u$} \\
\toprule
16   & 1.690(3) & 0.6394(14)& 1.168(4)  & 2.47(2)  & 2.43(2)  & 1.83(3)\\
32   & 1.699(5)& 0.864(3)  & 1.356(6)  & 2.25(5)  & 2.24(5)  & 1.93(5)\\
64   & 1.729(12)  & 1.102(7)  & 1.531(12) & 2.17(12) & 2.16(13) & 1.87(10)\\
128  & 1.75(2)  & 1.375(16) & 1.73(3)   & 1.9(3)   & 1.89(13) & 1.9(2)  \\
256  & 1.88(5)  & 1.60(4)   & 1.83(6)  & 1.8(4)   & 2.0(5)   & 2.2(5) \\
512  & 1.71(9) & 1.70(8)   & 1.85(12)  & 2.7(12)  & 1.4(10)  & 3.1(12)\\
\bottomrule
\end{tabular*}
\caption[Approach to the thermodynamical limit]{Rate at which several observables approach zero. We consider a functional
form $A\cdot L^{-\zeta}$ and compute the effective exponent $\zeta$
from the ratio of the computed values at consecutive lattice sizes. 
We consider three exponents, $\zeta_{\hat b}$, $\zeta_{\hat m}$ and $\zeta_u$
for the evolution of $\hat b^\infty$, $\hat m_\text{peak}^+$ and $u^+_\text{peak}$, 
respectively, see~\eqref{eq:ISING-zeta}. 
We observe that for $\beta=0.6$ the effective exponent approaches $2$, as
expected from the discussion in the text, while for $\beta=0.4473$ the 
proximity of the critical point complicates the analysis.
}\label{tab:ISING-ensemble-equivalence}
\index{ensemble equivalence|indemph}
\end{table}

Comparing Table~\ref{tab:ISING-results-yang}
with Table~\ref{tab:ISING-results-h-eq-zero} we see that for $\beta=0.6$,
with a computation effort 
almost $40$ times smaller, we have obtained a result
an order of magnitude more precise than what we had at $\beta_{\mathrm{c}}$. 
Recomputing the effective exponents for these new simulations we obtain 
results compatible with $\zeta=2$. Notice that for this temperature
the error in the exponents is much bigger than that for $\beta=0.4473$.
The reason is clear from Table~\ref{tab:ISING-results-yang}. The left-hand
sides of~\eqref{eq:ISING-zeta} are now much closer to zero
than in $\beta=0.4473$, yet their errors are only slightly smaller.
In the computation of the effective exponents only  the relative
errors matter, which explains our bigger uncertainties. Notice, however,
that we have been able to distinguish values for $b^\infty$ of 
order $10^{-6}$ from zero and that we have located the peak
with seven significant figures.

\index{saddle point|)}
\index{ensemble equivalence|)}

\section{Numerical performance analysis}\label{sec:ISING-numerical-performance}
\index{tethered formalism!numerical performance|(}
We consider here the Metropolis update algorithm described in detail
in Section~\ref{sec:TMC-Metropolis}, for a $D=2$ Ising model 
at the critical temperature. Notice that, since this is a local 
update algorithm,
one would expect the autocorrelation times (see Appendix~\ref{chap:thermalisation}
for definitions) to scale as
\begin{equation}\label{eq:ISING-CSD}
\tau \propto L^z
\index{critical slowing down}
\index{critical exponent!z@$z$}
\end{equation}
with $z\approx2$, due to critical slowing down~\cite{hohenberg:77,zinn-justin:05}.

\begin{table}\centering
\small
\begin{tabular*}{\columnwidth}{@{\extracolsep{\fill}}rlc}
\toprule
$L$ & $\ \hat m=0.5$ & $\hat m=1.14$\\
\toprule
32   & 0.491(9)  & 0.564(4)\\
64   & 0.548(11) & 0.614(8)\\
128  & 0.580(14) & 0.654(8)\\
256  & 0.597(13) & 0.641(6)\\
512  & 0.639(4)  & 0.661(4)\\
1024 & 0.632(4)  & 0.665(5)\\
\bottomrule
\end{tabular*}
\caption[Autocorrelation time for magnetic observables]{Autocorrelation
 times for the tethered magnetic field
$\langle \hat b\rangle_{\hat m}$ at the minimum, $\hat m=0.5$,
 and close to one of the maxima
of the $p(\hat m)$ for a critical $D=2$ Ising model, simulated
with a Metropolis algorithm.
There is no critical slowing down for this observable. The times
appear to grow for small sizes, but this is an artificial 
effect of the discrete nature of Monte Carlo time
(notice that $\tau\geq1/2$ by construction and that times 
very close to this limit are difficult to measure).
}\label{tab:ISING-tau-hatb}
\index{autocorrelation time!Ising|indemph}
\end{table}
\begin{figure}
\centering
\includegraphics[height=\linewidth,angle=270]{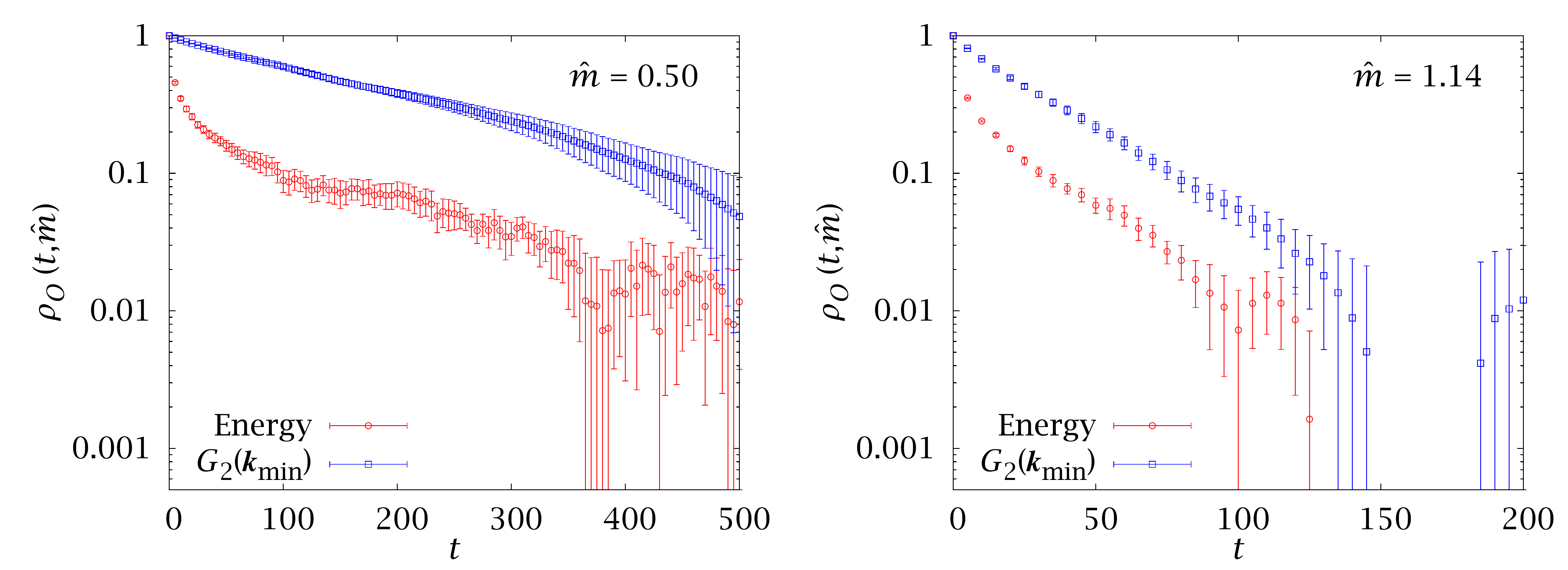}
\caption[Autocorrelation functions for the $D=2$ Ising model]{%
Normalised autocorrelation functions $\rho_O(t,\hat m)$ for the energy
and the propagator $G_2(\bk_\text{min})$ at $T_\text{c}$ in an $L=128$ lattice.
We plot them at 
the central minimum of $p(\hat m)$ (left) and
at one of the peaks (right). In both cases, the correlation 
function for $G_2$ is almost a pure exponential. Notice, however, 
that the correlation functions for both quantities become
parallel, indicating that they share the same exponential 
autocorrelation time.
\index{correlation function (equilibrium)!temporal!Ising|indemph}
\index{energy!Ising|indemph}
\label{fig:ISING-corr}}
\end{figure}

In Table~\ref{tab:ISING-tau-hatb} we show the integrated autocorrelation
times for the tethered magnetic field at the values of 
the smooth magnetisation $\hat m$ corresponding to the minimum
and one of the maxima in $p(\hat m)$ (recall Figure~\ref{fig:ISING-p-hatm}).
As we can see, there is hardly any evolution (in fact, the values are
so small that we cannot even measure them properly, since times smaller
than one are meaningless in a sequential update scheme). The same
situation is reproduced if we consider the autocorrelation times
for other magnetic observables (functions of the magnetisation $m$).
This result is surprising, specially considering that naive 
conserved order parameter dynamics exhibit $z=4-\eta$ critical
slowing down~\cite{hohenberg:77}.

The key is that our method
imposes the constraint globally so, even though the update
algorithm is local, the magnetisation has global information.
In this case, the new dynamical exponent should be
$z_\text{nonlocal} = z-2\approx 0$~\cite{tamayo:89}, as is the case for our method.
Notice that this result implies that Tethered Monte Carlo is able to 
reconstruct the effective potential $\varOmega_N$ without critical 
slowing down for the Ising model. 

Other physical observables, however, do not enjoy the same nonlocal 
information as $\hat b$ and therefore show standard critical slowing down.
This is the case, for instance, of the energy. The quantity with the 
slowest dynamics seems to be to be the two-point propagator $G_2$, 
defined in~\eqref{eq:ISING-G2},
We have plotted in Figure~\ref{fig:ISING-corr} the temporal
autocorrelation for the energy and for $G_2$ in an $L=128$ lattice, at 
the central probability minimum and at the probability peak. 
At both magnetisations, the correlation functions 
for the energy and propagator become parallel for long enough
times. This indicates that they both have the same exponential 
time, which we also expect to be the exponential time of the system
as a whole. In addition, the $\rho_{g_2}(t,\hat m)$ is almost 
a pure exponential, so we can make the approximation
\begin{equation}
\tau_\text{exp}=\tau_{\text{exp},G_2} \simeq \tau_{\text{int},G_2}\, .
\end{equation}

\begin{figure}
\centering
\includegraphics[height=.7\linewidth,angle=270]{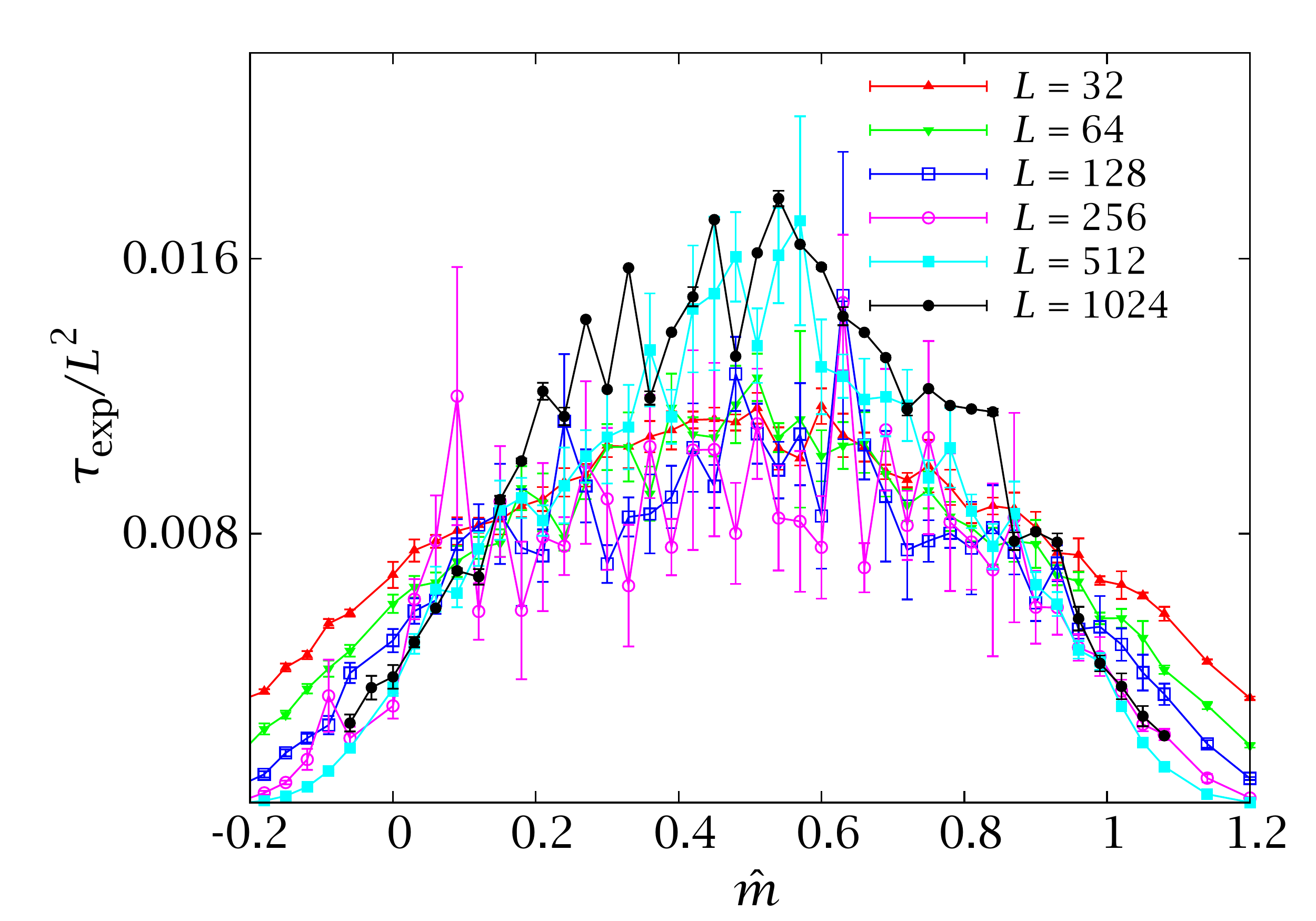}
\caption[Exponential autocorrelation time of TMC for the Ising model]{Exponential
 autocorrelation times for tethered simulations 
of the $D=2$ Ising model at the critical point, as a function of $\hat m$.
The $\tau_\text{exp}$, computed from the correlation function 
of $G_2(\bk_\text{min})$, are represented in units of the system size, showing 
a $z\approx2$ critical slowing down. 
}
\label{fig:ISING-tau_exp}
\index{autocorrelation time!Ising|indemph}
\end{figure}

Figure~\ref{fig:ISING-tau_exp}
shows this exponential time as a function of $\hat m$ for 
several system sizes. As we can see, the central region does exhibit critical
slowing down. Notice, however, that even for an $L=1024$ system 
$\tau_\text{exp}$ is only  $\sim 10^4$, so this simple Metropolis algorithm
can be used safely to study very large systems.
As a comparison, the exponential autocorrelation time
of the canonical version of the Metropolis algorithm
for $L=64$ is already as large as that of the 
tethered version for $L=1024$ (see, e.g.,~\cite{amit:05}).

\index{tethered formalism!numerical performance|)}
\index{Ising model|)}

\chapter{Optimising Tethered Monte Carlo: cluster methods}\label{chap:cluster}\index{cluster methods|(}
In the previous chapter we showed how a straightforward implementation of the tethered formalism,
using as Monte Carlo dynamics the simple Metropolis algorithm, can be very precise and efficient.
Eventually, however, the system size we can consider in the critical regime is capped \index{Metropolis algorithm}
by the appearance of critical slowing down, according to which the thermalisation times \index{critical slowing down}
grow as $\tau\sim L^z$, with $z\approx2$.  \index{critical exponent!z@$z$}

However, this $z\approx 2$ is not a hard limit imposed by the tethered formalism,
but is instead determined by the Metropolis algorithm we have used to explore the tethered
ensemble. If we had run our tethered simulations with some kind of optimised dynamics, the critical
slowing down would have been reduced. In this chapter we consider the most spectacular example 
of optimised update algorithm: cluster methods. These are able to achieve $z<1$, for all intents
and purposes eliminating critical slowing down. Because of this, since their introduction
in the 1980s~\cite{swendsen:87,wolff:89}, they have attracted a considerable interest,
which lasts to the present day~\cite{chayes:98,deng:07c,deng:07d,deng:07,deng:07b,deng:10,garoni:11}. 
Despite this continued research effort, however, efficient implementations of cluster
methods remain restricted to very few situations. In particular, the development
of a cluster update algorithm with conserved order parameter dynamics has long
been considered somewhat of a challenge~\cite{annett:92}.

The work reported in this chapter (whose results were originally published,
 in a reduced form, in~\cite{martin-mayor:09})
shows how, using the tethered formalism, one can actually define a workable 
cluster update with a conserved order parameter. This is demonstrated in the context
of the $D=2,3$ ferromagnetic Ising model, the ideal scenario for canonical
cluster methods. Furthermore, the tethered version of the cluster update is shown to
be  just as
efficient (in the sense of having the same $z$ exponent) as the canonical one.
\newpage

\section{The canonical Swendsen-Wang algorithm}\label{sec:CLUSTER-canonical-SW}\index{Swendsen-Wang algorithm|(}
We summarise here the construction of the Swendsen-Wang algorithm, following~\cite{edwards:88}.
The process is based on the  Fortuin-Kasteleyn transformation~\cite{kasteleyn:69,fortuin:72}, \index{Fortuin-Kasteleyn construction}
which maps Potts models onto bond-percolation \index{Potts model} \index{percolation}
problems.\footnote{The Potts model is a generalisation of the Ising model
where the spins can take $q$ different values, rather than just two.}
 We start by 
rewriting the partition function of the Ising model in the following way
\begin{align}
Z &= \sum_{\{s_\bx\}} \exp\biggl[ \beta \sum_{\braket{\bx,\by}} s_\bx s_\by\biggr]
 = \ee^{\beta} \sum_{\{s_\bx\}} \prod_{\braket{\bx,\by}} \bigl[p \delta_{s_\bx,s_\by} + (1-p)\bigr],
\index{partition function!Ising}
\index{partition function!Fortuin-Kasteleyn-Swendsen-Wang}
\end{align}
where  
\begin{align}
p = 1-\ee^{-2\beta}.
\nomenclature[p1]{$p$}{Probability that a link is occupied (cluster methods)}
\end{align}
We now notice that, trivially, 
\begin{align}
\bigl[p \delta_{s_\bx,s_\by} + (1-p)\bigr] = \sum_{n=0,1} \bigl[p \delta_{s_\bx,s_\by}\delta_{n,1} + (1-p)\delta_{n,0}\bigr].
\end{align}
Finally, we introduce this identity in the partition function by considering one auxiliary
variable $n_{\bx\by}$ for each first-neighbours pair $\braket{\bx,\by}$:
\begin{align}\label{eq:CLUSTER-Z-FKSW}
Z &= \sum_{\{s_\bx\}} \sum_{\{n_{\bx\by}\}} \prod_{\braket{\bx,\by}} \bigl[p \delta_{s_\bx,s_\by}\delta_{n_{\bx\by},1} + (1-p)\delta_{n_{\bx\by},0}\bigr].
\end{align}
This partition function describes a model with $(D+1)N$ dynamic variables: the original $N$ spins
of the Ising model and the $DN$ auxiliary variables $n_{\bx\by}$. We can think of the latter
as bond-occupation variables. We say that a link $\bx\by$ is occupied if $n_{\bx\by}=1$ 
and that it is empty if $n_{\bx\by}=0$. Notice that the spins joined by 
an occupied bond must be aligned, while those joined by an empty bond
can be either aligned or opposed. Therefore, the system has been partitioned into 
several \emph{clusters} (two spins are in the same cluster if they can be 
connected by a path of occupied bonds).

If we sum over the $s_\bx$ for fixed bonds, we see
that, while all spins in the same cluster must be aligned, the orientation of separate
clusters is independent. Consider a bond configuration with $\ell$ occupied bonds, defining
$\mN_\text{C}$ clusters (some of which may be single-spin clusters). We see 
in~\eqref{eq:CLUSTER-Z-FKSW} that each occupied bond contributes 
a factor $p$ to the weight and that each empty one contributes 
a factor $(1-p)$. There are $2^{\mN_\text{C}}$ possible cluster orientations. 
Therefore, the marginal probability of the bonds is just
\begin{equation}
\omega(\{n_{\bx\by}\}) = Z^{-1} p^\ell (1-p)^{DN-\ell} 2^{\mN_\text{C}}.
\end{equation}

More interesting are the conditioned probabilities of the spins given
the bonds and vice versa, which can be read directly from~\eqref{eq:CLUSTER-Z-FKSW},
\begin{itemize}
\item[$(a)$] Given the $\{s_\bx\}$, the bonds are independent from one 
another and $n_{\bx\by}=1$ with probability $p\delta_{s_{\bx},s_\by}$
and $n_{\bx\by}=0$ otherwise.
\item[$(b)$] Given the $\{n_{\bx\by}\}$, all the spins within a cluster
are aligned and two spins in different clusters are independent from each
other. Each cluster has a positive or negative orientation
with $50\%$ probability. 
\end{itemize}

These conditioned probabilities can be used to define a Monte Carlo 
update scheme. Each lattice update consists of two steps:
\begin{enumerate}
\item Given a starting spin configuration $\{s_\bx\}$, the conditioned
probability $(a)$ is used to trace the clusters. This can be implemented
in several ways, with an algorithmic complexity of $\mathcal O(N\log N)$.
See~\cite{landau:05} or the code distributed with~\cite{amit:05} for an example.
\item Once the clusters are traced, they must be flipped to 
update the spin configuration. The conditioned probability $(b)$ 
is trivial, the sign of each cluster is chosen as $\pm1$ with 
$50\%$ probability.
\end{enumerate}
Each of these two steps satisfies the detailed balance condition.  \index{balance condition!detailed}
The combination of the two is irreducible and satisfies balance \index{balance condition} \index{irreducibility}
(see Section~\ref{sec:THERM-Markov-chain} for definitions).

This scheme can be implemented for systems other than the Ising model 
(for instance, in an antiferromagnetic model, the clusters have 
uniform staggered magnetisation), but it is for this system that
it shows its full power. The critical slowing down is practically \index{critical slowing down}
eliminated, with $z<1$ in both two and three dimensions (see below),
meaning that even for very large lattices a few updates are enough
to decorrelate the system completely.

\subsection{Improved estimators}
Performing a single cluster update changes the spin configuration radically, 
almost instantaneously decorrelating the system. Therefore, not 
only is the system thermalised very quickly, but one can also take 
measurements very frequently and obtain independent data (remember 
the discussion of integrated and exponential correlation times 
in Appendix~\ref{chap:thermalisation}).

But the statistical gain of a cluster scheme does not end here. Indeed,
in the Swendsen-Wang scheme, after tracing the clusters, we have to flip them 
randomly, and with equal probability for each orientation, in order
to obtain our new system configuration. However, when taking 
measurements of some physical observables we can consider at once the average
over all the $2^{\mN_\text{C}}$ possible cluster orientations with fixed $\{n_{\bx\by}\}$.
We obtain in this way the so-called improved or cluster estimators~\cite{sweeny:83,wolff:88}.

Let us call $S_i=\pm1$ the spin of the $i$-th cluster and $n_i$ its size. Then, 
the spin estimator for the   magnetisation of the system is \index{cluster estimators}
\begin{equation}
M = \sum_{\bx} s_\bx = \sum_i n_i S_i.
\end{equation}
And the squared magnetisation is
\begin{equation}
M^2 = \sum_{\bx\by} s_\bx s_\by = \sum_{i,j} n_i n_j S_iS_j.
\end{equation}
Let us denote by an overbar the average over the $2^{N_\text{C}}$ 
cluster configurations $\{S_i\}$. Obviously, 
\begin{equation}\label{eq:CLUSTER-improved1}
\overline M= \sum_i n_i \overline S_i = 0.
\end{equation} 
For the squared magnetisation, only the diagonal terms survive
(remember the $S_i$ are independent)
\begin{equation}
\overline{M^2} = \sum_i n_i^2 \overline{S_i^2} = \sum_i n_i^2.
\end{equation}
Obviously, $\braket{M^2} = \braket{\overline{M^2}}$, but the second estimator
has more information and is therefore potentially more precise.
We can do the same for other observables, for instance, 
\begin{subequations}
\begin{align}
\overline{M^4} &= 3\sum_{i,j} n_i^2 n_j^2 - 2 \sum_i n_i^4,\\
\overline{M^6} &= 15 \sum_{i,j,k} n_i^2n_j^2n_k^2 - 30 \sum_{i,j} n_i^2 n_j^4
 + 16 \sum_i n_i^6,\\
\overline{M^8} &= 105 \sum_{i,j,k,l} n_i^2n_j^2n_k^2n_l^2 - 420 \sum_{i,j,k}
n_i^2n_j^2n_k^4 + 448 \sum_{i,j} n_i^2 n_j^6\nonumber\\
&\qquad + 140 \sum_{i,j} n_i^4 n_j^4 - 272 \sum_i n_i^8.
\end{align}
\end{subequations}
Consider now the propagator $G_2(\boldsymbol k)$, Eq.~\eqref{eq:ISING-G2},
with spin estimator for a given configuration\footnote{As explained
in Appendix~\ref{chap:thermalisation}, we use
the Wiener-Khinchin theorem to write the Fourier Transform of  \index{Wiener-Khinchin theorem}
the spatial correlation of $s_\bx$ as the modulus of the  \index{Fourier transform}
Fourier Transform of $s_\bx$.}
\begin{equation}
G_2(\boldsymbol k) =\left|\sum_\bx s_\bx \ee^{\ii \boldsymbol k\cdot \bx}\right|^2
= \sum_{\bx\by} s_\bx s_\by \ee^{\ii\boldsymbol k\cdot\bx} \ee^{-\ii \boldsymbol k\cdot \by}.
\end{equation}
Now, when averaging over the cluster orientations at fixed $\{n_{\bx\by}\}$,
two spins belonging to different clusters are uncorrelated 
and two spins in the same cluster are equal, therefore
\begin{equation}\label{eq:CLUSTER-improved2}
\overline{G_2(\boldsymbol k)}  
= \sum_{i=1}^{\mN_\text{C}} \left| \sum_{\bx\in C_i} \ee^{\ii \boldsymbol k \cdot \bx}\right|^2\, .
\end{equation}
The error reduction due to the adoption of cluster estimators 
is heavily dependent on the cluster distribution (see~\cite{fernandez:09c}),
but it can be very large. See, for instance,
Table~\ref{tab:CLUSTER-cluster-estimators}, where we compute the renormalised
coupling constants of the $D=2$ Ising model with several methods, and
obtain a tenfold error reduction using cluster estimators.

\begin{table}[h]
\small
\begin{tabular*}{\columnwidth}{@{\extracolsep{\fill}}clll}
\toprule
$O$ & TMC & SWS & SWC  \\
\toprule
$-\braket{u}$     & 1.226\,067(7)&  \multicolumn{2}{c}{1.226\,076(8)}   \\     
$\chi_2$  & 203.78(11)   &  204.07(10)    & 203.92(2)      \\        
$\xi_2$  & 11.888(11)   &  11.907(10)    & 11.893\,2(10)  \\
$r_6$    & 3.70(6)      &  3.73(8)       & 3.731(6)       \\   
$r_8$    & 26.2(6)      &  24(3)         & 26.47(18)      \\
$g_4$    & 14.66(5)     &  14.69(9)      & 14.673(8)     \\
$g_6$    & 794(9)       &  806(24)       & 803.3(16)     \\
$g_8/10^{4}$ & 8.25(13)     &  7.5(11)       & 8.34(7)    \\
\bottomrule
\end{tabular*}
\caption[Renormalised coupling constants and cluster estimators]{Several
observables for the $D=2$ Ising model (the $r_i$ and $g_i$ are 
renormalised coupling constants, see~\cite{caselle:01}
for definitions).
We work at $\beta=0.42$, with $L=100$, and take $10^8$ Swendsen-Wang lattice updates.
Compare the results with spin estimators (SWS) and with cluster estimators (SWC).
We also give, for comparison, results with the Metropolis TMC algorithm
($191$ values of $\hat m$, with $10^7$ MCS on each).
\index{renormalised coupling constants|indemph}
\index{susceptibility!Ising|indemph}
\index{energy!Ising|indemph}
\index{correlation length!Ising}
\index{cluster estimators|indemph}
\index{Metropolis algorithm}
\label{tab:CLUSTER-cluster-estimators}}
\end{table}
\section{The tethered Swendsen-Wang algorithm}\index{tethered formalism|(}\label{sec:CLUSTER-tethered-SW}%
\index{Swendsen-Wang algorithm!tethered|(}
The Fortuin-Kasteleyn construction is just as easy to do in the tethered formalism. \index{Fortuin-Kasteleyn construction}
Remember the tethered weight
\begin{equation}
\omega_N(\hat m,\{s_\bx\}) \propto \ee^{-\beta U + M-\hat M} (\hat m-m)^{(N-2)/2} \varTheta(\hat m-m).\tag{\ref{eq:TMC-omega}}
\index{tethered weight}
\end{equation}
Notice that the spin-connectivity term, given by the energy $U$, is exactly the same
as that of the canonical weight $\omega_\text{canonical} = \exp[-\beta U]$,
and can thus be represented as in~\eqref{eq:CLUSTER-Z-FKSW}. Therefore, 
the conditioned probability of the bonds given the spins is exactly the same
in the tethered and the canonical ensembles. Only the conditioned probability 
of the spins given the bond varies,
\begin{itemize}
\item[$(a)$] Given the $\{s_\bx\}$, the bonds are independent from one 
another and $n_{\bx\by}=1$ with probability $p\delta_{s_{\bx},s_\by}$
and $n_{\bx\by}=0$ otherwise.
\item[$(b')$] Given the $\{n_{\bx\by}\}$, all the spins within a cluster
are aligned and two spins in different clusters are independent from each
other. Each of the  $2^{\mN_\text{C}}$ cluster configurations $\{S_i\}$ has
probability
\begin{equation}
p(\{S_i\}) \propto \ee^{M-\hat M} (\hat m-m)^{(N-2)/2} \varTheta(\hat m-m),
\qquad  M =Nm= \sum_{i} n_i S_i.\label{eq:CLUSTER-probabilidad-clusters} 
\index{tethered weight!cluster}
\end{equation}
\end{itemize}
Just as in the canonical case, the tethered Swendsen-Wang lattice 
update has two steps. First we trace the clusters using the conditioned
probability $(a)$. Then we flip them using $(b')$. This second step
was trivial in the canonical ensemble, as all $2^{\mN_\text{C}}$ cluster
configurations were equiprobable. In the tethered ensemble, however, 
we have to pick one using \eqref{eq:CLUSTER-probabilidad-clusters}.
A naive way to make this choice would be to select one configuration
randomly and accept it with a heat-bath probability\index{heat bath algorithm},
\begin{equation}
P = \frac{ p(\{S_i\})}{ \sum_{j=1}^{\mN_\text{C}} p(\{S_j\})}\ .
\index{tethered formalism!numerical implementation|(}
\end{equation}
The problem with this method is obvious: $\mN_\text{C}$ is potentially 
a very large number, so it is impossible to compute all the weights 
and average them. 

Clearly, we need a more manageable scheme. In order to construct it
let us first recall the Wolff or single-cluster algorithm~\cite{wolff:89}.
With this (canonical) method, we randomly pick one spin and immediately trace
and flip the cluster to which it belongs. Since the clusters are \index{Wolff algorithm}
flipped with probability proportional to their size, this method is able
to cause a very large change in the system. 

In the tethered formalism 
we cannot take the same approach, because the weights of the configurations
with $\pm S_i$ are different. We can, however, make use of the notion 
that not all clusters need be flipped. Indeed, we can select 
$\mN'_\text{C}\ll \mN_\text{C}$ clusters and consider a heat bath of 
the $2^{\mN'_\text{C}}$
configurations $\{S'_i\}$ where all the other clusters are fixed. In principle, one
could select one of the $\{S'_i\}$ randomly and accept it with 
probability
\begin{equation}\label{eq:CLUSTER-P-prime}
P' = \frac{ p(\{S'_i\})}{ \sum_{j=1}^{\mN'_\text{C}} p(\{S'_j\})}\ .
\end{equation}
This is already a manageable computation, but there still remains a problem.
Since the weight~\eqref{eq:CLUSTER-probabilidad-clusters} is heavily 
dependent on $M$, only a few of the $2^{\mN'_\text{C}}$ configurations
will have non-negligible weight. Therefore, a method where we choose one randomly and
check whether it is accepted with~\eqref{eq:CLUSTER-P-prime} is going
to have a very low acceptance. It is better to use a modified 
representation of the heat bath algorithm, where we construct a vector
with the cumulative probabilities
\begin{equation}\label{eq:CLUSTER-heat-bath}
A_k = \frac{\sum_{j=1}^k p(\{S'_j\})}{\sum_{j=1}^{\mN'_\text{C}} p(\{S'_j\})},\qquad A_0=0.
\end{equation}
We then extract a uniform random number $0\leq \mathcal R<1$ and choose 
the configuration $\{S'_k\}$ such that 
\begin{equation}
A_{k-1} \leq \mathcal R< A_k.
\end{equation}

We still have to decide how to choose the $\mN'_\text{C}$ clusters
that we will attempt to flip. Two simple choices, both
satisfying the balance condition, present themselves: \index{balance condition}
\begin{itemize}
\item Select the $\mN'_\text{C}$ largest clusters.
\item Select $\mN'_\text{C}$ spins randomly and note their 
clusters (in case of repetition, we keep drawing spins until we 
get $\mN'_\text{C}$ different clusters). 
\end{itemize}
\begin{figure}
\centering
\includegraphics[height=.7\linewidth,angle=270]{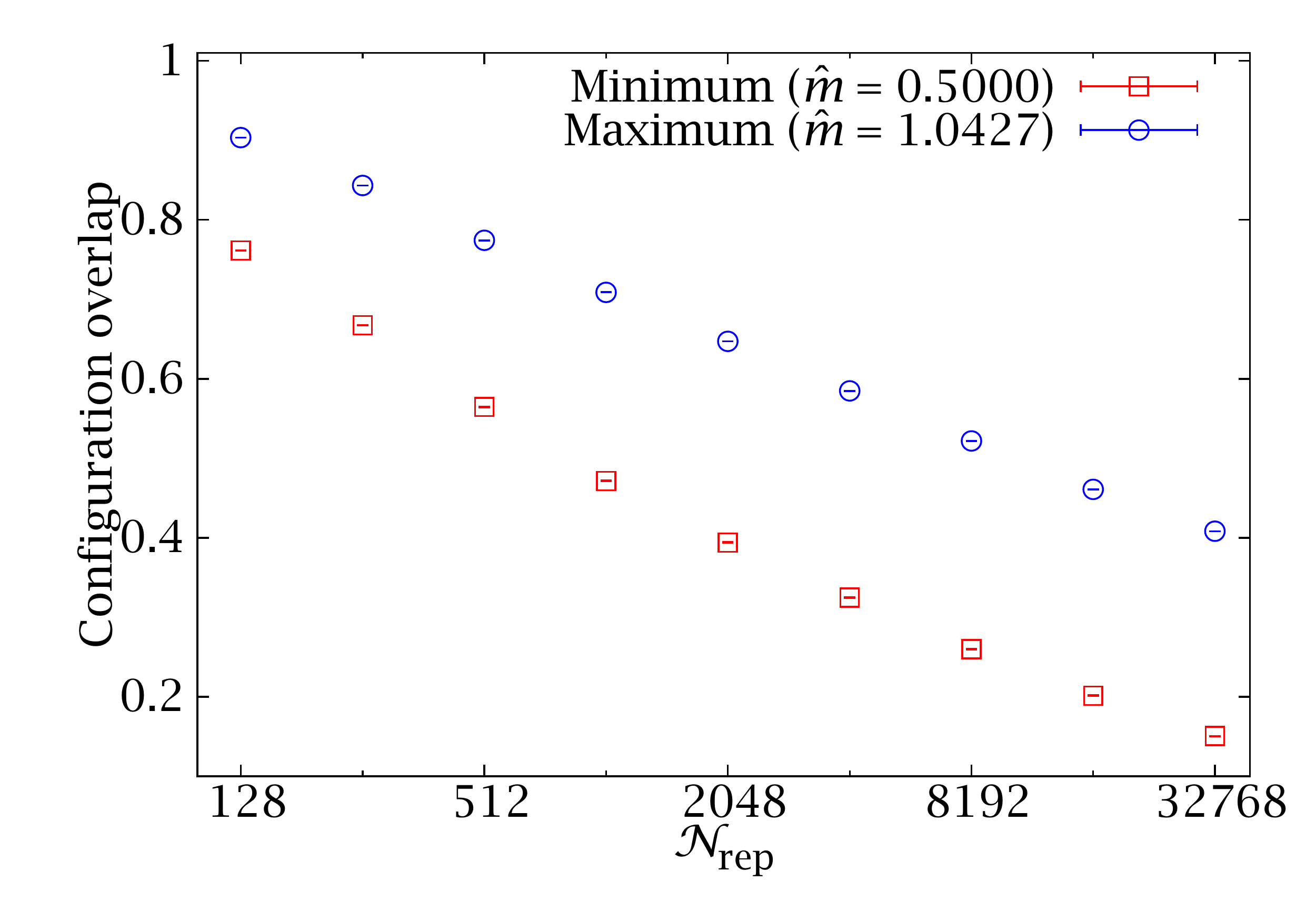}
\caption[Spin overlap against $\mN_\text{rep}$]{Spin overlap, Eq.~\eqref{eq:CLUSTER-overlap}, in a $D=2$
critical $L=512$ Ising model. We plot the results, whose errors
cannot be discerned at this scale, for the central minimum of the 
$p(\hat m)$ and for one of its maxima.
\label{fig:CLUSTER-overlap}}
\end{figure}
It may seem that the first choice is preferable, with more potential 
for changing the system quickly. However, nothing says that we can only
select the $\mN'_\text{C}$ clusters and flip them once, before retracing 
the clusters.  In fact, tracing the clusters is a relatively costly
operation, $\mathcal O(N \log N)$, so the flipping time is negligible.
Therefore, we can adopt the following algorithm as our lattice update
\begin{enumerate}
\item Trace the clusters in the system.
\item Flip the clusters. For $t=1,\ldots,\mN_\text{rep}$ do
\begin{enumerate}
\item Select $\mN'_\text{C}$ clusters with the second of the methods 
described above (we use $\mN'_\text{C}=5$, but this number can be changed).
\item Obtain the $\{S_i^{t}\}$ with the heat bath of Eq.~\eqref{eq:CLUSTER-heat-bath},
fixing the $\mN_\text{C}-\mN_\text{C}'$ remaining clusters in the orientations
they had at $t-1$.
\end{enumerate}
\end{enumerate}
We are therefore using a dynamic Monte Carlo in order to pick \index{Monte Carlo method!dynamic}
the cluster orientations, rather than a static Monte Carlo, as 
is the case for the canonical algorithm.
It may seem counterintuitive that the optimal value of $\mN_\text{rep}$ is rather large.
In order to see it,  let us consider the overlap
\begin{equation}\label{eq:CLUSTER-overlap}
o = \frac{\biggl\langle \sum_\bx \bigl[ {s_\bx^{t=0} s_\bx^{t=\mN_\text{rep}}}
 - \braket{m}^2_{\hat m}\bigr]\biggr\rangle}{N \bigl(1-\braket{m}^2_{\hat m}\bigr)}\, .
\end{equation}
Notice that $o$ would  vanish for completely uncorrelated configurations.
We have plotted this quantity in Figure~\ref{fig:CLUSTER-overlap}. Clearly, 
the configuration can significantly evolve for a fixed distribution
of the bonds. We have empirically found that an $\mN_\text{rep}$ that
equilibrates the cluster tracing and cluster-flipping times
is close to optimal and convenient (one simply scales
$\mN_\text{rep}$ with $N$). For the largest systems that we will consider here, 
with $N=256^3$ spins, this results in $\mN_\text{rep}\approx 5\times10^5$.

In the above discussion we have constructed a Monte Carlo algorithm
capable of quickly and non-locally changing the spin configuration, potentially
accelerating thermalisation. Remember, however, that in a canonical setting
this was not the end of the story: we could obtain significant error
reductions by employing improved estimators. In our case, even if 
we cannot write closed formulas such 
as~(\ref{eq:CLUSTER-improved1}--\ref{eq:CLUSTER-improved2}) we can get 
an even better enhancement by taking measurements of the relevant
observables
at each of the $\mN_\text{rep}$ steps. In Table~\ref{tab:CLUSTER-Nrep}
we can see that for the all-important tethered field $\hat b$
this nets us a factor $29$ in error reduction  
for an $N=256^3$ system, where we use $\mN_\text{rep}= 2^{19}\approx5.2\times10^5$, 
an effect equivalent to considering a simulation $841(=29^2)$ times longer.
\begin{table}[H]
\small
\begin{tabular*}{\textwidth}{@{\extracolsep{\fill}}lrrr}
\toprule
$\mN_\text{rep}$ & $\langle \hat b^\text{MCS}\rangle_{\hat m=0.5}\times10^6$ &  $\langle \hat b^{\mN_\text{rep}} \rangle_{\hat m=0.5}\times10^6$ & Error ratio\\
\toprule
$2^{0}$ &   $-1.9224 \pm 5.2934$  &  $-1.9224 \pm 5.2934$ & 1.00      \\
$2^{1}$ &   $-1.7207 \pm 3.5529$  &  $-2.1219 \pm 3.5332$ & 1.01 \\
$2^{2}$ &   $3.2730 \pm 2.7653$  &  $3.5203 \pm 2.7173$   & 1.02  \\
$2^{3}$ &   $-7.5114 \pm 1.9301$  &  $-8.1937 \pm 1.9104$ & 1.01 \\
$2^{4}$ &   $-1.8460 \pm 1.6477$  &  $-1.8712 \pm 1.5495$ & 1.06 \\
$2^{5}$ &   $-2.2291 \pm 1.3728$  &  $-0.4083 \pm 1.1968$ & 1.15 \\
$2^{6}$ &   $-0.5641 \pm 1.3838$  &  $0.6485 \pm 0.9193$  & 1.50\\
$2^{7}$ &   $-0.4048 \pm 1.3637$  &  $-0.4720 \pm 0.6707$ & 2.03 \\
$2^{8}$ &   $2.2643 \pm 1.3579$  &  $-0.0167 \pm 0.4785$  & 2.84\\
$2^{9}$ &   $0.6270 \pm 1.2368$  &  $0.2684 \pm 0.3510$   & 3.52 \\
$2^{10}$ &   $2.3544 \pm 1.1953$  &  $0.3511 \pm 0.2633$  & 4.54\\
$2^{11}$ &   $-1.0922 \pm 1.3222$  &  $0.2449 \pm 0.1813$ &  7.29 \\
$2^{12}$ &   $0.4522 \pm 1.3126$  &  $0.2188 \pm 0.1359$  &  9.66 \\
$2^{13}$ &   $-0.7033 \pm 1.3030$  &  $-0.0485 \pm 0.1227$& 10.62  \\
$2^{14}$ &   $1.0856 \pm 1.2832$  &  $0.0181 \pm 0.0849$  & 15.11\\
$2^{15}$ &   $1.5875 \pm 1.3750$  &  $0.1504 \pm 0.0864$  & 15.91\\
$2^{16}$ &   $-0.2056 \pm 1.2189$  &  $-0.1113 \pm 0.0671$& 18.17  \\
$2^{17}$ &   $1.9023 \pm 1.4305$  &  $0.0963 \pm 0.0593$  & 24.12\\
$2^{18}$ &   $0.3395 \pm 1.3451$  &  $0.0064 \pm 0.0607$  & 22.16\\
$2^{19}$ &   $-2.5036 \pm 1.3495$  &  $0.0166 \pm 0.0460$ & 29.34  \\
\bottomrule
\end{tabular*}  
\caption[Improved estimators for tethered Swendsen-Wang]{%
Evolution of $\langle \hat b\rangle_{\hat m=0.5}$ and its 
error with increasing $\mN_\text{rep}$ for simulations 
of an $N=256^3$ system at $\beta_\text{c}$ with $\mN_\text{MC}=50\,000$ 
MCS. We compare the result of measuring $\hat b$ only after each MCS 
with  the result of measuring it after each of the 
$\mN_\text{rep}\times \mN_\text{MC}$ cluster flippings. 
The error in the first estimator is quickly saturated but 
the second keeps getting more precise even for very large $\mN_\text{rep}$.
The last column gives the ratio of the errors of both estimates.
Since we are working at the 
probability minimum, we have $\langle \hat b\rangle_{\hat m=0.5}=0$,
so the central
value can be taken as a check that the errors are 
correctly estimated.
\index{Swendsen-Wang algorithm|indemph}
}\label{tab:CLUSTER-Nrep}
\end{table}
\index{tethered formalism!numerical implementation|)}
\index{tethered formalism|)}

\section{Numerical performance analysis}
\index{tethered formalism!numerical performance|(}
\index{critical slowing down}
\begin{table}
\small
\begin{tabular*}{\textwidth}{@{\extracolsep{\fill}}llll}
\toprule
$L$ &  TMC (Cluster) &  TMC (Met. + Cluster) &  Canonical\\
\toprule
 16  & 2.310(14) & 0.775(3)& 3.253(8)  \\   
 32  & 2.758(20) & 1.055(5)& 4.011(11) \\   
 64  & 3.347(22) & 1.417(7)& 4.891(11) \\   
 128 & 4.11(5)   &1.861(12)& 5.510(20) \\
 256 & 4.87(4)   &2.391(16)& 6.928(22) \\
 512 & 5.79(8)   &3.040(24)& 8.107(25) \\
 1024& 6.78(8)   &3.70(4)  &           \\
\midrule                     
$z_E$  & 0.241(7) &        & 0.222(7)\\
$\chi^2$/d.o.f. & 0.36/2   &         \\
\bottomrule
\end{tabular*}  
\caption[Autocorrelation times for the $D=2$ Ising model]{Integrated autocorrelation times for the energy at $\hat m= 0.5$ and $\beta = \beta_\text{c}$
for the $D=2$ Ising model. We compare  the  cluster 
and mixed versions of our TMC algorithm. We also include
the results of~\cite{salas:00} for canonical Swendsen-Wang. 
For the pure cluster algorithm we fit to $\tau_E = AL^{z_E}$,
in the range $L\geq 128 $. Our resulting value 
for the dynamic critical exponent is very similar to
the result of~\cite{salas:00} for the canonical algorithm.
\index{autocorrelation time!Ising|indemph}
\index{critical exponent!z@$z$|indemph}
\index{Swendsen-Wang algorithm|indemph}
\index{Metropolis algorithm}
}\label{tab:CLUSTER-tau-2D}
\end{table}

\begin{table}
\small
\begin{tabular*}{\textwidth}{@{\extracolsep{\fill}}llll}
\toprule
$L$ & TMC (Cluster) & TMC (Met. + Cluster) & Swendsen-Wang \\
\toprule
  16      & 2.135(13)   & 0.782(3)     & 5.459(3)  \\ 
  32      & 2.80(3)     & 1.134(5)     & 7.963(9)  \\    
  48      & 3.467(28)   & 1.427(8)     & 9.831(9)  \\    
  64      & 3.88(3)     & 1.700(10)    & 11.337(12)  \\  
  96      & 4.79(5)     & 2.152(14)    & 13.90(3)    \\        
  128     & 5.46(6)     & 2.566(17)    & 15.90(5)  \\
  192     & 6.54(11)    & 3.32(4)     & 19.10(9)  \\        
  256     & 7.51(13)    & 3.85(5)      & 21.83(10)  \\
\midrule
$z_E$     & 0.472(8)& 0.591(4)  & 0.460(5)  \\
$\chi^2$/d.o.f.     & 5.85/5  & 4.61/5       &   \\           
\bottomrule
\end{tabular*}  
\caption[Autocorrelation times for the $D=3$ Ising model]{Integrated autocorrelation
times for the energy at $\hat m= 0.5$ and $\beta = \beta_\text{c}$
for the $D=3$ Ising model. We compare  the cluster 
and mixed versions of our TMC algorithm 
with the results for the canonical version reported in~\cite{ossola:04}.
Our value of $z_E$ comes from a fit for $L\geq32$.
}\label{tab:CLUSTER-tau-3D}
\end{table}
In order to assess the efficiency of the tethered Swendsen-Wang method, we have computed
the integrated autocorrelation times for several observables. As in the Metropolis
case (Section~\ref{sec:ISING-numerical-performance}), they are largest 
for the $\hat m = 0.5$ ($m\approx 0$) region. However, unlike the local
case, now the energy is the slowest observable (this is a common feature of 
cluster methods). Therefore, in order to evaluate the overall performance
of the method, we concentrate on the energy at $\hat m=0.5$. In the next
section we give a second measurement of the performance, in terms
of the precision of the computed physical observables.

We show $\tau_{\text{int},E}$ at $\hat m=0.5$ 
for the $D=2,3$ Ising 
model in Tables~\ref{tab:CLUSTER-tau-2D} and~\ref{tab:CLUSTER-tau-3D}.
For $D=2$ we have worked at  $\beta=\beta_\text{c}= \log(1+\sqrt2)/2$ \index{critical temperature!Ising}
and for $D=3$ at $\beta =0.221\,654\,59\approx \beta_\text{c}$~\cite{blote:99}. 

We consider both the cluster algorithm described in the previous section
and a mixed scheme, where we take two Metropolis steps  \index{Metropolis algorithm}
between cluster updates (the cluster update takes more time
to perform, so this does not change the running time 
noticeably). We have computed the integrated time with the 
self-consistent window method described in Section~\ref{sec:THERM-recipes}.

We also compare our results with the corresponding autocorrelation 
times for the canonical Swendsen-Wang algorithm, taken from~\cite{salas:00}
for the $D=2$ case\footnote{The quoted 
values are reported by Salas and Sokal only in the preprint
version of~\cite{salas:00}, as a combination
of their simulations and those of~\cite{baillie:91}.}
and from \cite{ossola:04} for $D=3$.  Recall that 
the Ising model at the critical point is the ideal
setting for these methods.

Notice that in both dimensions the integrated times of the 
tethered pure cluster method are smaller than those of its canonical
counterpart, although the dynamic critical exponents are comparable.
The mixed algorithm has even smaller correlation times, but
in $D=2$ it does not follow a power law in $L$ and in $D=3$ the  \index{critical exponent!z@$z$}
resulting $z_E$ is larger than the one for the  pure cluster method.
This probably means that the asymptotic regime has not yet been reached for the
mixed algorithm. Presumably, for large enough systems it will scale
as the pure cluster one. However, since for our range of lattice 
sizes the combined algorithm has smaller $\tau$, 
it is the one we shall use to compute physical results in the next
section.

In short, the critical slowing down of the  Tethered Monte Carlo method,
already absent for magnetic observables with a local update scheme,
can be removed just as completely as in the canonical
case with the application of cluster methods. 
This is a demonstration that adopting the tethered formalism
does not imply abandoning optimised update schemes, nor does
it hinder their performance.

\index{tethered formalism!numerical performance|)}
\index{Swendsen-Wang algorithm!tethered|)}
\index{Swendsen-Wang algorithm|)}
\section{Physical results for the ferromagnetic Ising model}\label{sec:CLUSTER-results}
In this section we compute some physically relevant quantities for the critical 
Ising model in $D=2$ and $D=3$. In the former case, we compare the 
the tethered cluster simulations  with our Metropolis
results of Chapter~\ref{chap:Ising} and in the latter
with the canonical Swendsen-Wang results of~\cite{ossola:04}.

\begin{table}[b]
\centering
{\small \begin{tabular*}{\columnwidth}{@{\extracolsep{\fill}}llllll}
\toprule
$L$ & \multicolumn{1}{c}{ $-\braket{u}$} &
\multicolumn{1}{c}{$\chi_2/L^2$} & \multicolumn{1}{c}{$\xi_2/L$} & 
\multicolumn{1}{c}{$C$}& \multicolumn{1}{c}{$B$}\\
\toprule
512 (MTMC) & 1.415\,42(4)  & 0.229\,3(7)  & 0.903(6)     & 16.57(3)   & 1.168(2)   \\ 
512 (CTMC) & 1.415\,435(7) & 0.229\,71(13)& 0.906\,4(11) & 16.588(10) & 1.167\,5(4)\\
512 (E)   & 1.415\,429\ldots&            &              & 16.595\,404\ldots&\\
\midrule
1024 (MTMC)& 1.414\,89(4)  & 0.194\,9(15) & 0.919(15)    & 18.28(8)   & 1.163(6)   \\
1024 (CTMC)& 1.414\,819(4) & 0.193\,02(15)& 0.905\,3(14) & 18.332(16) & 1.167\,7(6)\\  
1024 (E)  & 1.414\,821\ldots&            &              & 18.361\,348\ldots       \\
\bottomrule
\end{tabular*}}
\caption[Canonical expectation values for the $D=2$ Ising model]{Results at the critical temperature for the $D=2$ Ising model. We compare
the result of taking $10^7$ MCS with a Metropolis implementation of TMC (MTMC) with
that of taking $10^6$ MCS with the tethered Swendsen-Wang scheme (CTMC). 
For the energy and specific heat we also give the exact results 
of~\cite{ferdinand:69}.
\label{tab:CLUSTER-results-2D}}
\index{specific heat!Ising|indemph}
\index{magnetisation!Ising|indemph}
\index{energy!Ising|indemph}
\index{correlation length!Ising|indemph}
\index{Binder ratio!Ising|indemph}
\index{Swendsen-Wang algorithm|indemph}
\index{Metropolis algorithm|indemph}
\end{table}
We give our $D=2$ results in Table~\ref{tab:CLUSTER-results-2D}. We have
used the same grid of $77$ simulation points as in the Metropolis
simulations of Chapter~\ref{chap:Ising}, but now we have 
performed $10^6$ cluster updates instead of $10^7$ Metropolis ones.
As we can see, the errors of the Swendsen-Wang algorithm 
are about $5$ times smaller for $L=512$ and $10$ times smaller for $L=1024$.

For $D=3$ we have simulated an $N=128^3$ lattice
at $\beta_\text{c}$, taking $10^6$ cluster steps on each of the
points  in a grid of $50$ values of $\hat m$.
We compare with the canonical Swendsen-Wang simulation conducted in~\cite{ossola:04},
with a total of $4.8\times10^7$ cluster steps. This results 
in a similar number of total lattice updates (and, therefore, simulation 
time) for both simulations. In accordance
with our autocorrelation time study, the statistical errors in
the canonical averages computed with both methods are similar
(see Table~\vref{tab:CLUSTER-results-3D}.

Let us recall, however, that, while the results for zero
applied field are similarly precise in the canonical and
tethered computations, the latter has more information 
about the system. This is because its results can be
reweighted to yield canonical averages for non-zero
applied field.

As a final test of our method's accuracy and as a demonstration
that it can do more than merely reproduce canonical
averages we shall reproduce the study of Section~\ref{sec:ISING-betanu}.
We shall compute the anomalous dimension of the $D=3$ Ising model \index{finite-size scaling}
from a FSS analysis of the peak position in $p(\hat m)$. \index{critical exponent!eta@$\eta$}

\newpage
Recalling the analysis of Section~\ref{sec:ISING-betanu}
and the scaling relation $2\beta/\nu= D-2+\eta$, Eq.~\eqref{eq:INTRO-scaling}, we have
\begin{equation}\label{eq:CLUSTER-peak}
\index{scaling relations}
\hat m_\text{peak} - \frac12 = A L^{-(D-2+\eta)/2}+\ldots
\end{equation}
Our results for the $D=3$ Ising model can be seen 
in Table~\ref{tab:CLUSTER-peak}. For these simulations 
we have first located the approximate position of the peak
with a short sweep in $\hat m$ and then run long simulations
for only two very close $\hat m$ values, one at either side
of the peak.\footnote{Actually, for each lattice size
we ran many independent runs, whose results were later
averaged. The table gives the total of MCS, adding
those of all the individual runs.} The peak position is then computed from 
the saddle-point equation
\index{saddle point}
\begin{equation}
\braket{\hat b}_{\hat m} = 0,
\end{equation}
with a simple linear interpolation.

We show the result of fitting the data in Table~\ref{tab:CLUSTER-peak}
to~\eqref{eq:CLUSTER-peak} in Table~\ref{tab:CLUSTER-eta}. We obtain good
fits already with $L_\text{min}=48$ and excluding more points results 
in compatible values of $\eta$, with growing errors. Our preferred
final estimate is $\eta=0.036\,0(7)$, combining the central value for
$L_\text{min}=48$ and the more conservative error for $L_\text{min}=64$
to account for systematic effects. This value should
be compared to the best determinations known to us, 
a Monte Carlo computation giving $\eta=0.036\,27(10)$~\cite{hasenbusch:10}
and a high-temperature expansion value of $\eta=0.036\,39(15)$~\cite{campostrini:02}.
Both quoted values, however, were computed with a `perfect'
action, not directly in the Ising model. 

\begin{table}[p]
\small
\begin{tabular*}{\columnwidth}{@{\extracolsep{\fill}}lccccc}
\toprule
Method & MCS & $-\langle u\rangle$ & $C$ & $\chi$ & $\xi_2$ \\
\toprule
Canonical SW & $48\times10^6$ & 0.992\,946\,6(48) & 66.465(54) & 21\,193(13) & 82.20(3)\\
CTMC         & $50\times10^6$ & 0.992\,949\,3(45) & 66.522(40) & 21\,202(13) & 82.20(6)\\
\bottomrule
\end{tabular*}
\caption[Canonical expectation values for the $D=3$ Ising model]{Comparison of canonical Swendsen-Wang 
(data from~\cite{ossola:04}) with TMC for an $N=128^3$ lattice at $\beta_\text{c}$.
We take $10^6$ MC steps at each of the $50$ points 
of our $\hat m$ grid. This results in a 
similar number of MCS for both simulations.
\label{tab:CLUSTER-results-3D}
\index{specific heat!Ising|indemph}
\index{magnetisation!Ising|indemph}
\index{energy!Ising|indemph}
\index{correlation length!Ising|indemph}}
\vspace*{1.5cm}

\small
\begin{tabular*}{\columnwidth}{@{\extracolsep{\fill}}lll}
\toprule
\multicolumn{1}{c}{$L$}& \multicolumn{1}{c}{MCS} &  \multicolumn{1}{c}{$\hat m_\text{peak} -\tfrac12$} \\
\toprule
16  & $1.0\times10^8$ & 0.334\,21(5)  \\ 
32  & $1.0\times10^8$ & 0.233\,77(4)  \\
48  & $1.0\times10^8$ & 0.189\,56(4)  \\
64  & $1.0\times10^8$ & 0.163\,41(4)  \\
96  & $1.0\times10^8$ & 0.132\,40(4)  \\
128 & $1.0\times10^8$ & 0.114\,083(24)\\
192 & $6.0\times10^7$ & 0.092\,46(4)  \\
256 & $8.2\times10^6$ & 0.079\,59(12) \\
\bottomrule
\end{tabular*}
\caption[Peak of $p(\hat m)$ for the $D=3$ Ising model]{
Position of the peak of $p(\hat m;L)$
for the $D=3$ Ising model at $\beta_\text{c}$.
\index{p@$p(\hat m)$|indemph} 
\index{magnetisation!Ising|indemph}
\label{tab:CLUSTER-peak}}
\vspace*{1.5cm}

\small
\begin{tabular*}{\columnwidth}{@{\extracolsep{\fill}}lll}
\toprule
\multicolumn{1}{c}{$L_\text{min}$}& \multicolumn{1}{c}{$\eta$} & \multicolumn{1}{c}{$\chi^2$/d.o.f.}\\       
\toprule
16& 0.033\,92(21)          & 42.6/6\\ 
32& 0.035\,26(30)          & 8.79/5\\
48& \textbf{0.036\,0}(5)   & \textbf{4.24$\boldsymbol/$4}\\
64& 0.036\,8\textbf{(7)}   & 1.02/3\\
96& 0.036\,3(12)           & 0.78/2\\
128& 0.037\,3(19)           & 0.31/1\\
192& \multicolumn{1}{c}{---} & \multicolumn{1}{c}{---} \\
256& \multicolumn{1}{c}{---} & \multicolumn{1}{c}{---} \\
\bottomrule
\end{tabular*}
\caption[Anomalous dimension of the $D=3$ Ising model]{Fits to $\hat m_\text{peak}-\tfrac12 = AL^{-(D-2+\eta)/2}$
of the data in Table~\ref{tab:CLUSTER-peak} for several
ranges $L\geq L_\text{min}$.
\label{tab:CLUSTER-eta}
\index{critical exponent!eta@$\eta$|indemph}}
\end{table}

\index{cluster methods|)}

\part{The Diluted Antiferromagnet in a Field}\label{part:daff}
\chapter{The state of the art: the DAFF with canonical methods}\index{DAFF|(}\label{chap:daff-canonical}%
In this Part we intend to demonstrate the first of our two main 
points: when studying complex systems, one should use a statistical
ensemble tailored to the problem at hand. We do this in the context
of the diluted antiferromagnet in a field (DAFF), a system that
had been extensively studied with canonical methods, but that
remained very poorly understood. In the present Chapter we start
by giving an introduction to the model (and its universality class), 
noting the difficulties that have been encountered in the different 
approaches. We then, in Section~\ref{sec:DAFF-model}, present 
the main observables and summarise the theoretical expectations
for the  critical behaviour. In Section~\ref{sec:DAFF-canonical-simulations}
we demonstrate  the difficulties of a canonical MC approach
to the problem.  The next Chapter, finally, presents 
our study of the DAFF with the tethered formalism. 

\section{Introduction}\label{sec:DAFF-intro}
In Chapter~\ref{chap:disorder} we introduced the concept of quenched disorder
and explained how it can radically affect the collective behaviour
of a physical system, to the point that disordered models are 
prime examples of complex systems. Here we consider a particularly    \index{complex systems}
interesting class: magnetic systems with random fields. 

In particular, we are interested in the random field Ising model (RFIM):  \index{RFIM}
\begin{equation}\label{eq:DAFF-RFIM}                                       \index{disorder!quenched}
\mathcal H = - \sum_{\langle\bx,\by\rangle} s_\bx s_\by - \sum_\bx h_\bx s_\bx
 = U + E_\text{F}.
\end{equation}
The $h_\bx$ are quenched and independent random variables, typically chosen so that 
$\overline{h_\bx} = 0$. Let us in addition assume they are Gaussianly distributed and \index{Gaussian distribution}
\begin{align}
\overline{h_\bx} &= 0, & \overline{h_\bx h_\by} =  \delta_{\bx\by} H.
\end{align}
Recall that the overline denotes the average over the disorder. In the language 
of Chapter~\ref{chap:disorder}, then, a sample is a choice of $\{h_\bx\}$.

Notice that the pure system is in this case the ferromagnetic Ising model \index{Ising model}
studied in Part~\ref{part:tmc}, whose critical behaviour is well understood. \index{pure system}
In particular, the phase diagram for $D>1$ consists in a low-temperature 
phase with ferromagnetic order and a high-temperature paramagnetic phase, 
connected through a second-order phase transition. In principle, the presence 
of a small random field  would disturb the ferromagnetic order and lower
the transition temperature.  For a high enough value of 
the field strength $H$ the critical temperature would go to zero and 
the ferromagnetic phase would be destroyed. 

That this naive qualitative picture of the phase diagram hides a more
complex and interesting situation was first demonstrated 
by Imry and Ma in 1975~\cite{imry:75}. In particular, they showed how, 
for a low spatial dimension, even an infinitesimal $H$ can destroy
the ferromagnetic order completely.
Since this argument is very simple and elegant, it is worth recalling
here.
\begin{figure}
\centering
\includegraphics[width=0.7\linewidth]{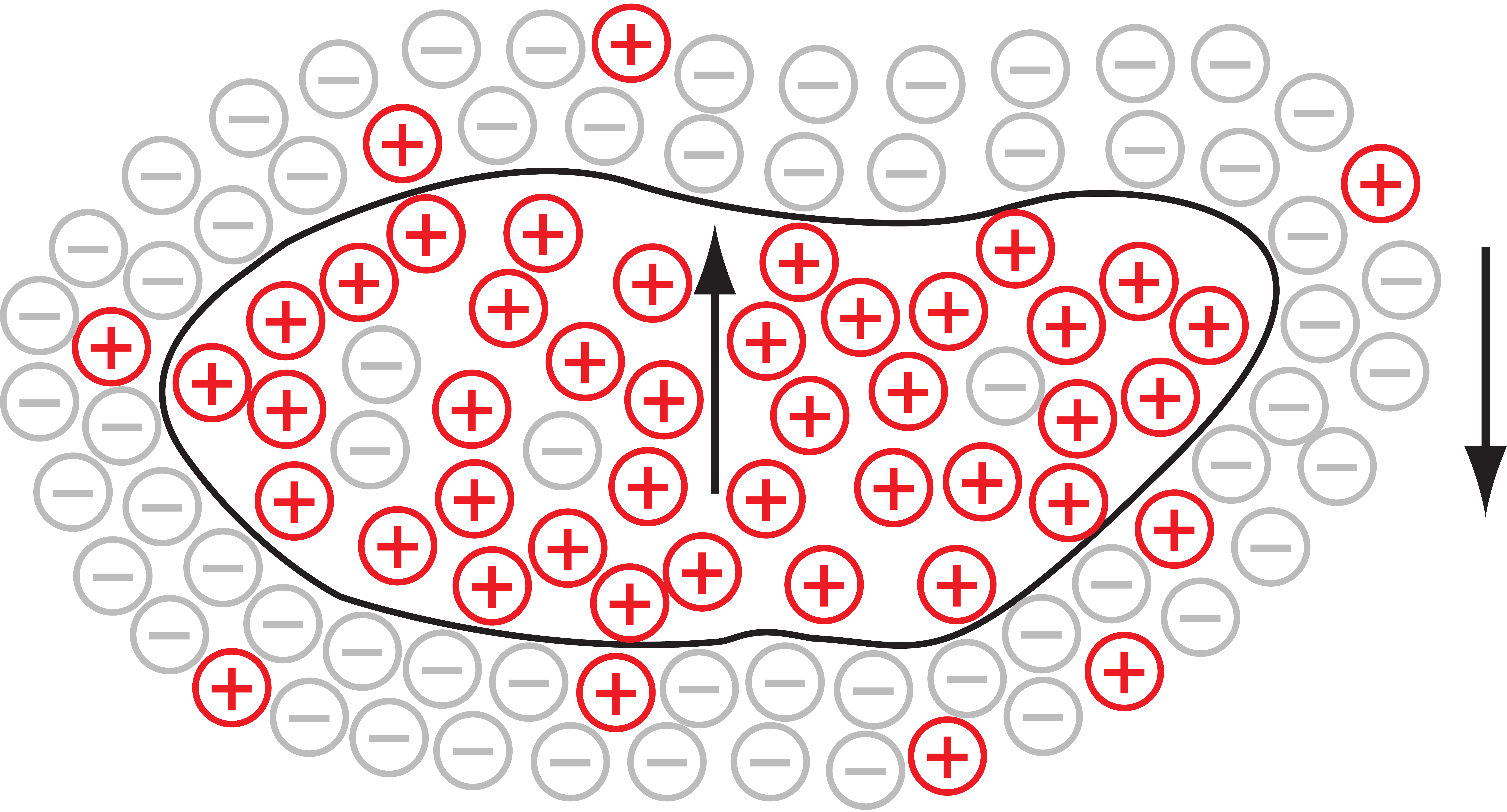}
\caption[Inverted droplet in a ferromagnetic domain]{Inverted
compact domain in a ferromagnetic system.
\label{fig:DAFF-droplet}
\index{Imry-Ma argument|indemph}}
\end{figure}

Let us consider a system with a predominantly
ferromagnetic order, with most of the spins in the $(-)$ orientation. In this
situation, we flip a compact domain of radius $R$ (see Figure~\ref{fig:DAFF-droplet}).
\index{Imry-Ma argument}

The value of the exchange interaction energy in the bulk of this domain does 
not change with this reversal, but all the links across its boundary 
are inverted. Therefore, there is an increase $\Delta U \sim R^{D-1}$.
As to the field-interaction energy
the average value of $h_\bx$ inside the domain is of course
zero. On the other hand, the variance of the sum of $n$ Gaussian
variables, each of variance $H$, is $\sqrt{n}H$. \index{central limit theorem}
That is, we have $\Delta E_\text{F} = 2 E_\text{F} \sim H R^{D/2}$.
Choosing a region with positive $E_\text{F}$, we then have
\begin{equation}
\Delta E = \Delta U + \Delta E_\text{F} = a R^{D-1} - b H R^{D/2},
\end{equation}
where $a,b$ are positive constants. Therefore, if $D-1<D/2$
we see that these perturbations are actually 
energetically favourable, for large enough $R$. Hence, the
ferromagnetic phase is unstable for $D<2$. The marginal case
$D=2$ was considered in~\cite{binder:83} and also found to be unstable.
Therefore, the lower critical dimension for the RFIM 
had to be $D_\text{l} \geq 2$ (other mechanisms than the domain-wall
argument could destroy the long-range order in three or more \index{long-range order}
dimensions). The issue was solved by the rigorous proof that \index{critical dimension}
$D_\text{l}=2$~\cite{imbrie:84,bricmont:87,aizenman:89}.

The disorder in the RFIM was thus showed to be strong enough \index{RFIM}
to modify the phase diagram of the pure model profoundly, 
yet not so radical as to make the phase transition disappear
altogether. The interest in the system was further enhanced 
when it was shown that it could be faithfully realised physically
by a diluted antiferromagnet in a uniform field (DAFF)~\cite{fishman:79,cardy:84}.
This ignited a considerable experimental effort (see~\cite{belanger:98}
for a review).

In short, it was shown more than twenty years ago that the RFIM
experiences a phase transition in three spatial dimensions and that 
the system is also of experimental interest through the DAFF. 
Despite a continuing analytical, experimental and numerical effort, however, 
almost all the details of this transition remain controversial.

In particular, characterising the phase transition
proved to be no easy task from an analytical point
of view (see~\cite{nattermann:97, dedominicis:06} for reviews
of theoretical results). First of all, we have the general observation 
that the upper critical dimension for disordered systems
is $D_\text{u}=6$ (as opposed to $D_\text{u}=4$ for pure models, cf. Chapter~\ref{chap:disorder}
and Section~\ref{sec:DAFF-phase-transition}, below).
Therefore, it is more difficult to predict the behaviour in 
the physically interesting case of $D=3$ from renormalisation-group
expansions in $\epsilon=D_\text{u}-D$.

An example of the problems with perturbation expansions is afforded 
by the issue of dimensional reduction. In 1979, Parisi and Sourlas~\cite{parisi:79}
presented an elegant supersymmetry argument by which the critical behaviour of the RFIM
at dimension $D$  should be equivalent to that of the pure model in $D-2$. But this \index{dimensional reduction}
had to be wrong: we have already seen that there is a phase transition in $D=3$, while
the Ising model has no ordered state in $D=1$. Parisi pointed out a possible
flaw in the supersymmetric argument as soon as 1984~\cite{parisi:84},  \index{perturbation theory}
but the issue has remained an active topic of research (see, e.g.,  \index{renormalisation group}
\cite{brezin:01,tissier:11} and references therein).

The experimental and numerical approaches to the problem are similarly
pla\-gued by severe intrinsic problems. The main one is perhaps
posed by the peculiar critical behaviour of the
DAFF/RFIM~\cite{villain:85,fisher:86b,nattermann:97}. 
In general, most authors work in the framework of an unconventional scaling theory,
which we shall explain in Section~\ref{sec:DAFF-phase-transition},  \index{scaling relations}
where the spatial dimension $D$ is replaced by $D-\theta$
in the traditional hyperscaling relation~\eqref{eq:INTRO-hyperscaling}.
Here $\theta$ is a new critical exponent, believed \index{hyperscaling}
to be $\theta\approx1.5$ from theoretical arguments (cf. Section~\ref{sec:DAFF-phase-transition})  \index{critical exponent!theta@$\theta$}
but inaccessible to a direct computation 
both in experimental studies and in (canonical) Monte Carlo simulations.

The complete uncertainty in the determination of $\theta$ has allowed 
experimental and numerical works to report qualitatively different sets of values
for the other critical exponents.\footnote{Notice that only three exponents are
independent, the other being fixed through scaling and hyperscaling 
relations.} 
One example of this is the experimental claim for a divergence
in the specific heat~\cite{belanger:83,belanger:98}, 
not observed numerically~\cite{hartmann:01}. We must note
that the specific heat exponent is notoriously difficult \index{critical exponent!alpha@$\alpha$}
to estimate numerically~\cite{malakis:06}, especially   \index{specific heat!DAFF}
considering that the expected divergence is very slow. \index{scaling relations}

From the experimental point of view, an additional problem \index{hyperscaling}
is posed by the uncertainty over the analytical form 
of the scattering line shape or structure factor.
Different ansätze lead \index{critical exponent!nu@$\nu$}
to mutually incompatible results for the thermal 
critical exponent, from $\nu=0.87(7)$~\cite{slanic:99}
to $\nu=1.20(5)$~\cite{ye:04}. The first of these 
works also reported a value of $\eta=0.16(6)$ \index{critical exponent!theta@$\theta$}
which, combined with a divergent specific heat ($\alpha\geq0$),
would violate hyperscaling if $\theta\approx1.5$.

On the numerical front, the determination of $\nu$ has
been at least as difficult, with estimates
ranging from $\nu\approx 1$ to $\nu>2$ (see, e.g.,~\cite{nattermann:97,wu:06,vink:10} for 
detailed lists). The most precise values have typically
come from ground state studies. Hartmann and Nowak claimed \index{ground state studies}
$\nu=1.14(10)$ in~\cite{hartmann:99}, although the value
has recently been increased, with a recent
estimate of $1.37(9)$~\cite{middleton:02}
(other recent works~\cite{malakis:06,fytas:11} give similar
results). In general, these latest numerical estimates are
outside the experimental range of values and only
barely compatible with a divergent specific heat. 
Another recent work,~\cite{wu:06}, gives $\nu=1.25(2)$ and $\alpha=-0.05(2)$.
However, these values suffer from similar problems as experimental
estimates (with which they are compatible) in that they do not
satisfy hyperscaling bounds.

An additional source of uncertainty is the extremely low value of
the magnetic critical exponent, believed to be $\beta\sim0.01$.
This exponent is also determined with almost no precision~\cite{nattermann:97,hartmann:99},
the latest estimate of which we have notice being $\beta=0.007(5)$~\cite{fytas:11}.
The fact that $\beta$ is compatible with the first-order result of $\beta=0$
has led some authors to suggest the possibility that the phase  \index{universality}
transition in the DAFF may not be continuous. This claim is typically
based on the finding of metastable signatures~\cite{sourlas:99,maiorano:07,wu:06}
and casts doubts on the supposed universality between the DAFF and the RFIM.

A final issue is the lack of self-averaging in the DAFF, which has been 
extensively studied~\cite{parisi:02,malakis:06,fytas:11} and complicates any numerical 
analysis.

At the root of these problems is the deep physical issue of free-energy barriers, discussed \index{free energy!barriers}
in Chapter~\ref{chap:tmc}. Both experimentally and  in canonical MC simulations, 
the system gets trapped in local minima, with escape  \index{critical slowing down!exponential}
times $\log \tau \sim \xi^\theta$, which makes it exceedingly  
hard to thermalise even relatively small systems and 
ensures that the statistics will be dominated by extremely rare events.

In the remainder of this chapter, after giving some definitions in Section~\ref{sec:DAFF-model},
we shall examine more closely the reasons why canonical MC simulations 
of the DAFF fail.

\newcommand{\Ms}{M_\text{s}}
\newpage
\section{Model and observables}\label{sec:DAFF-model}
We consider the diluted antiferromagnet in a field (DAFF),  defined
by the following Hamiltonian (as always $N=L^D$ Ising spins in a 
cubic lattice)
\begin{align}\label{eq:DAFF}
\mathcal H &= \sum_{\langle \boldsymbol x , \boldsymbol y\rangle} \epsilon_\bx s_\bx
\epsilon_\by s_\by - h \sum_\bx \epsilon_\bx s_\bx - h_\text{s} \sum_\bx \epsilon_\bx s_\bx\pi_\bx,& 
\pi_\bx&=\ee^{\ii \uppi \sum_{\mu=1}^D x_\mu}
\nomenclature[hs]{$h_\text{s}$}{Staggered component of the applied field}
\nomenclature[pi]{$\pi_\bx$}{Parity of site $\bx$}
\nomenclature[epsilon]{$\epsilon_\bx$}{Quenched occupation variables}
\end{align}
In addition to the Ising spins ($s_\bx=\pm 1$), this Hamiltonian includes 
quenched occupation variables $\epsilon_\bx$.  These are  \index{dilution} 
equal to one with probability $p$ and equal to zero with probability $(1-p)$  \index{percolation}
\nomenclature[p2]{$p$}{Probability that a node is occupied (DAFF)}
and characterise the dilution of the system. The value of $p$ is not supposed
to be important to the physics, so long as we keep our distance 
both from the percolation threshold of $p_\text{c}\approx0.31$~\cite{stauffer:84}
and from the pure case.\footnote{%
Some experimental works~\cite{barber:00}
have pointed out possible non-equilibrium effects if the 
concentration is low enough to permit percolation of vacancies
($p<1-p_\text{c} \approx 0.69$), although the issue
is not completely understood~\cite{ye:02,shelton:04}.
}
The $\epsilon_\bx$ are quenched variables, which, recalling 
Section~\ref{sec:INTRO-quenched}, means that we should perform first the thermal
average for each choice of the $\{\epsilon_\bx\}$ and only afterwards compute
the disorder average.  In the canonical ensemble we denote
the thermal average  by $\langle\cdots\rangle$
and the disorder average by $\overline{(\cdots)}$.

The Hamiltonian~\eqref{eq:DAFF} includes a two-component applied field $(h,h_\text{s})$,
coupled to the regular and staggered magnetisations
\begin{align}
M &= Nm = \sum_\bx \epsilon_\bx s_\bx, \label{eq:M}\\
\Ms &= N m_\mathrm{s} = \sum_\bx \epsilon_\bx s_\bx  \pi_\bx,
\label{eq:Ms}.
\nomenclature[Ms]{$M_\text{s},m_\text{s}$}{Staggered magnetisation}
\end{align}
We shall continue to denote by $U$ the spin interaction component
of the energy
\begin{equation}\label{eq:DAFF-U}
U = N u = \sum_{\langle \boldsymbol x , \boldsymbol y\rangle} \epsilon_\bx s_\bx
\epsilon_\by s_\by. 
\end{equation}
Therefore, the total energy $E$ of a given spin configuration can be written as
\begin{equation}\label{eq:DAFF-E}
E = U - h M - h_\text{s} M_\text{s}.
\end{equation}
In general, one is interested in the case $h_\text{s}=0$, but we will find 
this parameter useful later on.

In order to study spatial correlations we need to consider the staggered 
Fourier transform of the spin field
\begin{equation}\label{eq:DAFF-phi-k}
\phi(\bk) = \sum_\bx s_\bx \pi_\bx \ \ee^{-\ii \bk\cdot \bx}.
\end{equation}

\subsection{Phase transition in the DAFF: theoretical expectations}\label{sec:DAFF-phase-transition}
In this section, we give
some heuristic arguments justifying the modified scaling relations expected
for the DAFF, assuming a second-order scenario.\footnote{We follow 
closely the arguments given in~\cite{nattermann:97} and~\cite{fisher:86b}
for the RFIM, but adapt the language to the antiferromagnetic system.}

Let us consider the system at the paramagnetic-antiferromagnetic 
transition point $T_\text{c}(p,h)$ for a given dilution $p$ and field $h$.
Now let us consider the effect of introducing a staggered magnetic field
$h_\text{s}$. Unlike $h$, $h_\text{s}$ is coupled to the order parameter
of the transition, so by definition~\eqref{eq:INTRO-delta} of the critical exponent $\delta$,
\index{critical exponent!delta@$\delta$}
\index{scaling relations}
\begin{equation}
m_\text{s}\sim h_\text{s}^{1/\delta}.
\end{equation}
Now, recalling that $m_\text{s} \sim xi^{-\beta/\nu}$, from~\eqref{eq:INTRO-nu}
and~\eqref{eq:INTRO-beta} we have
\begin{equation}
h_\text{s} \sim \xi^{-\beta \delta/\nu}.
\index{correlation length}
\end{equation}
In addition, if we divide the system in the even and odd sublattices, 
the dilution implies that one is going to have $\sim L^{D/2}$ more spins
than the other. This causes an excess staggered field $\delta h_\text{s} \sim \xi^{-D/2}$.
Now, if we want the system to have a real transition as we approach 
the critical point $h_\text{s}\to0$, we need
\begin{equation}
\frac{\delta h_\text{s}}{h_\text{s}} \xrightarrow{\ \ \xi\to\infty\ \ } 0.
\end{equation}
That is, we need
\begin{equation}\label{eq:DAFF-scaling1}
\frac{D}{2} \geq \frac{\beta \delta}{\nu} = \frac{2+\gamma-\alpha}{2\nu}\ ,
\end{equation}
where we have used the scaling relation~\eqref{eq:INTRO-scaling2}.

If we plug the mean-field exponents \index{mean field} into this 
equation, we obtain $D/2\geq 3$. Therefore, the upper critical \index{critical dimension}
dimension is in this case $D_\text{u}=6$. 

Below $D_\text{u}$ we can in principle apply the hyperscaling \index{hyperscaling}
law~\eqref{eq:INTRO-hyperscaling}. However, if we do this we obtain
\begin{equation}
\left.
\begin{array}{rcl}
\nu D &=&2-\alpha\\
\nu D &\geq& 2+\gamma-\alpha  
\end{array}
\right\}\quad \Longrightarrow\quad \gamma \leq 0.
\index{critical exponent|gamma@$\gamma$}
\end{equation}
Since the susceptibility critical  exponent $\gamma$ must be  \index{susceptibility}
positive, this is not possible.

The answer to this problem was given independently
by Villain~\cite{villain:85} and  by Fisher~\cite{fisher:86b}.
We consider the system at the length scale of the correlation
length. According to Widom scaling (see, e.g.,~\cite{amit:05}),
the free energy of this correlation volume $\xi^{D}$
is 
\begin{equation}\label{eq:DAFF-Fxi}
F(\xi) \sim \xi^{D-(2-\alpha)/\nu}.
\end{equation}
In conventional systems, the scale of variations of $F(\xi)$
is set by the thermal fluctuations, so $F(\xi)$ is of
order one and the usual hyperscaling law $\nu D = 2-\alpha$
follows. In our disordered system, however, the fluctuations due
 to the randomness dominate and
\begin{equation}\label{eq:DAFF-Fxi-random}
F(\xi)\sim \xi^\theta,
\index{critical exponent|theta@$\theta$}
\end{equation}
where $\theta$ is a new, independent, critical exponent.\footnote{%
From the naive Imry-Ma argument of Section~\ref{sec:DAFF-intro} \index{Imry-Ma argument}
we would expect $\theta=D/2$.} 
Comparing~\eqref{eq:DAFF-Fxi} with~\eqref{eq:DAFF-Fxi-random}
we read off the modified hyperscaling relation
\begin{equation}\label{eq:DAFF-modified-hyperscaling}
2-\alpha = \nu (D-\theta).
\nomenclature[theta1]{$\theta$}{Hyperscaling violations exponent}
\end{equation} 
In addition, from~\eqref{eq:DAFF-scaling1} we obtain the inequality
\begin{equation}\label{eq:DAFF-theta-bound}
\theta \geq \gamma/\nu = 2-\eta.
\index{scaling relations}
\index{hyperscaling}
\end{equation}
This last relation has been proven rigorously for a class of models~\cite{schwartz:85}.

Some authors~\cite{aharony:76,schwartz:86} have suggested that the above
inequality is saturated for the DAFF, giving
\begin{equation}\label{eq:DAFF-theta-equality}
\theta = 2-\eta,
\end{equation}
so we would again have only two independent critical exponents.

Several heuristic arguments have been given to motivate
Eq.~\eqref{eq:DAFF-theta-equality} for the RFIM (see, e.g.,~\cite{nattermann:97}),
but we can also examine the issue directly in the DAFF.

Let us consider the system in the low-temperature phase with no field,
so there is a non-zero spontaneous staggered magnetisation $m_\text{s}$
(in one of two orientations). As we said
before, one of the sublattices (let us say the even one), has 
$\sim L^{D/2}$ more spins than the other. When the magnetic
field $h$ is introduced, of the two orientations for the same $|m_\text{s}|$
the one where the even spins are aligned with the field 
will be more favourable.

Now, let us consider a region of radius $R$ where the odd 
sublattice is dominant (always possible to find, if the system
is large enough). Clearly, inverting the magnetisation within 
this region aligns more spins with the field, lowering the total
energy. The energy gain is $\sim - R^{D/2} m_\text{s} h$.

On the other hand, this inversion also has an energy cost, due to the surface
energy. Therefore, the probability of inverting this region
is going to be proportional to the exponential of 
\begin{equation}
F(R) = R^{D/2} m_\text{s} - \varSigma R^{D-1},
\end{equation}
where $\varSigma$ is the surface tension and we have neglected \index{surface tension}
irrelevant constant factors in both terms. By definition
of correlation length, this probability is maximum for $\xi=R$, 
where $F'(R=\xi) = 0$. Therefore, 
\begin{equation}\label{eq:DAFF-Sigma}
F(\xi) \sim \xi^{D/2} m_\text{s} \sim  \varSigma \xi^{D-1}.
\end{equation}
Finally, by definition~\eqref{eq:DAFF-Fxi-random} of $\theta$ and using $m_\text{s} \sim \xi^{-\beta/\nu}$ 
\begin{equation}
\xi^{\theta} \sim \xi^{D/2} m_\text{s} \sim \xi^{D/2-\beta/\nu},
\end{equation}
and we have
\begin{equation}
\theta = D/2 - \beta/\nu,
\end{equation}
which is equivalent to~\eqref{eq:DAFF-theta-equality}.

On the other hand, from $F(\xi)\sim\xi^\theta$ and~\eqref{eq:DAFF-Sigma},
the surface tension is seen to scale as $\varSigma \sim \xi^{-(D-1-\theta)}$.
Since we are assuming a second-order scenario, this has to vanish in the thermodynamical
limit and we have \index{thermodynamical limit}
\begin{equation}
\theta < D-1,
\end{equation}
(this last bound can be obtained without assuming the two-exponent scenario).

Finally, let us close this section by mentioning that the need for a third 
exponent is related to the existence of two correlation functions that scale differently,
\begin{subequations}\label{eq:DAFF-Sk}
\begin{align}
S_\text{c}(\bk) &= \overline{\braket{ \phi(\bk) \phi(-\bk)}} - \overline{\braket{\phi(\bk)}}
\ \overline{\braket{\phi(-\bk)}} \sim \frac{1}{k^{2-\eta}},\\
S_\text{d}(\bk) &=  \overline{\braket{\phi(\bk)}} \ \overline{\braket{\phi(-\bk)}} 
\sim \frac{1}{k^{4-\bar\eta}}.
\index{critical exponent|etabar@$\bar\eta$}
\index{correlation function (equilibrium)!spatial!DAFF}
\end{align}
\end{subequations}
In this equation, $\phi(\bk)$ is the staggered Fourier transform
of the spin field, defined in~\eqref{eq:DAFF-phi-k}, and $\bar \eta$ is
\begin{equation}
\bar \eta = 2+ \eta - \theta.
\end{equation}
Notice that in the two-exponent scenario, $\bar \eta = 2\eta$ and 
the disconnected propagator diverges as the square of the connected one
\begin{equation}
S_\text{d} \sim S_\text{c}^2.
\end{equation}

\section{The DAFF in canonical Monte Carlo simulations}\label{sec:DAFF-canonical-simulations}
In this section we carry out canonical Monte Carlo simulations 
of a DAFF system, to demonstrate the problems inherent in the
traditional approach and motivate our tethered study of Chapter~\ref{chap:daff-tethered}.

\begin{figure}
\centering
\includegraphics[height=.9\columnwidth,angle=270]{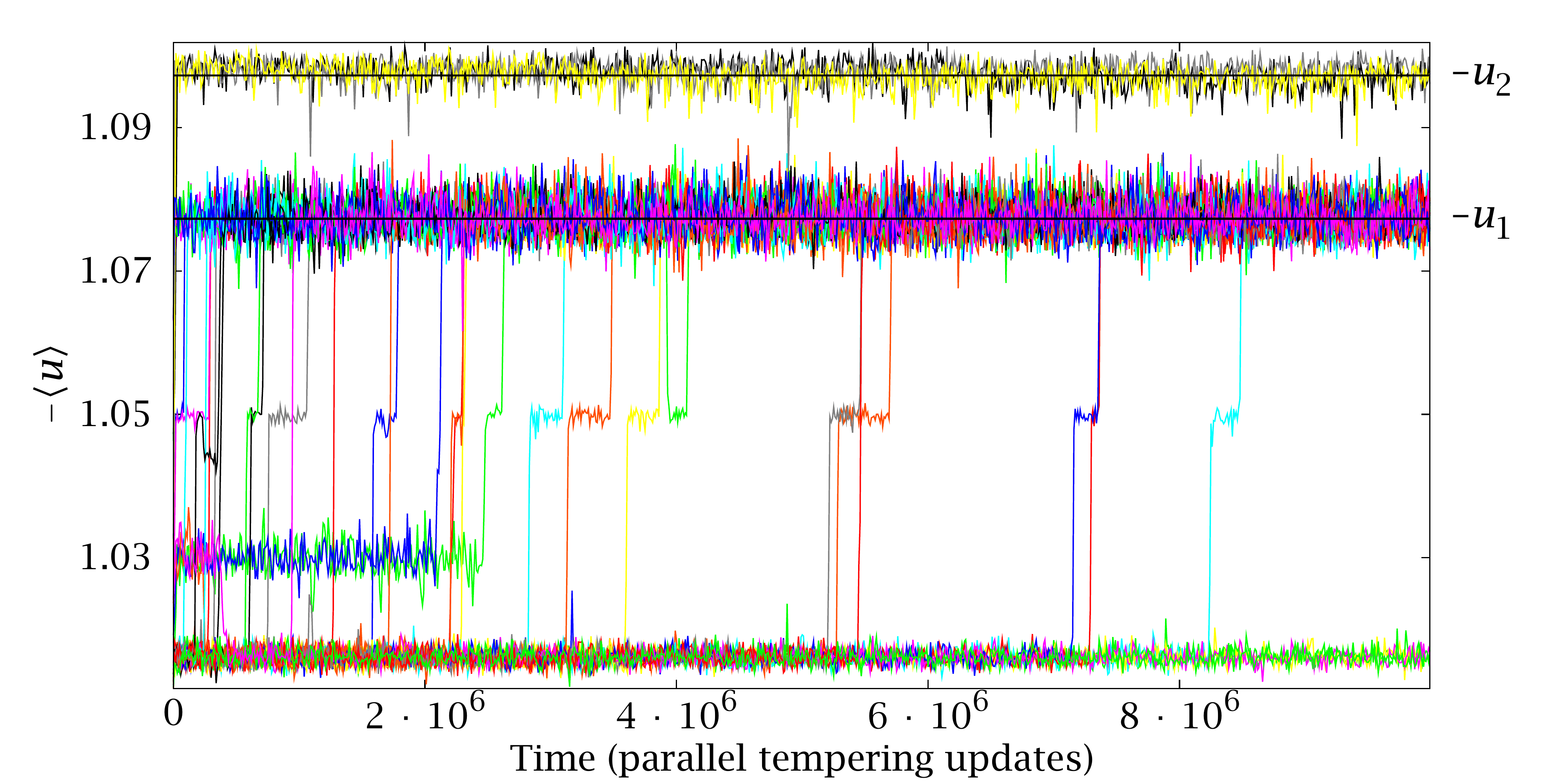}
\caption[Monte Carlo histories for canonical simulations]{
Time evolution of the interaction energy $u(t)$ ---Eq.~\eqref{eq:DAFF-U}---
for the configuration at $T=1.6$, $\beta=2.4$ in $50$ different 
parallel tempering simulations of the same sample. 
We plot $50$ out of a total of $100$ runs. 
After a variable simulation time, most of the simulations fall into one 
of two metastable states, whose energies $u_1$ and $u_2$ we mark
with straight black lines.  In a thermalised situation, we would
see tunnelling between the two states (or all of the runs reaching the same
one, if only one is dominant).
\label{fig:DAFF-historia-canonico}
\index{parallel tempering|indemph}
\index{energy!DAFF|indemph}
}
\end{figure}

In these simulations we have used parallel tempering, \index{parallel tempering}
a common method to thermalise systems with free-energy barriers. \index{free energy!barriers}
The basic idea is running several copies of the same sample of the system
concurrently, each at a different temperature. Every so often, we attempt
to exchange copies at neighbouring temperatures. When the temperature 
is raised, the energy fluctuations increase and copies that would
get trapped in metastable states can escape (see Appendix~\ref{chap:thermalisation}
for a full explanation of this method).

In our case, we have not attempted to conduct a full canonical
study of the phase transition, we have just simulated a few samples.
For each of them we performed many parallel tempering runs, 
using $40$ temperatures evenly spaced in the range $1.6\leq T\leq 2.575$
and choosing the field such that $\beta h = 1.5$ (in this way, we 
intend to cross the phase diagram with a diagonal straight 
line). We considered a system size of $L=24$ and a dilution of $p=0.7$.
These parameters are taken from~\cite{maiorano:07}.\footnote{In fact, 
the authors of~\cite{maiorano:07} go to even lower temperatures.
Here we make things easier for the parallel tempering 
by going no further down than $T=1.6$, which the authors
of~\cite{maiorano:07} found to be the critical temperature
along the line $\beta h=1.5$.}

Figure~\vref{fig:DAFF-historia-canonico} shows the time evolution
of the interaction energy $u(t)$ for the lowest temperature
($T=1.6$, $h=2.4$) for $50$ runs of a single sample (we performed
a total of $100$ such runs, but we only show $50$ to avoid cluttering
the graph). We can see
how, after a variable simulation length, most of the systems
reach one of two metastable states, with \index{metastability}
interaction energies $u^{(1)}=-1.077\,35(8)$ and $u^{(2)}=-1.097\,3(7)$.
The corresponding total energy densities, Eq.~(\ref{eq:DAFF-E}), are
$e^{(1)}=-1.376\,08(2)$ and $e^{(2)}=-1.382\,0(2)$.
The same qualitative picture is obtained for different samples.
\begin{figure}
\centering
\includegraphics[height=.7\columnwidth,angle=270]{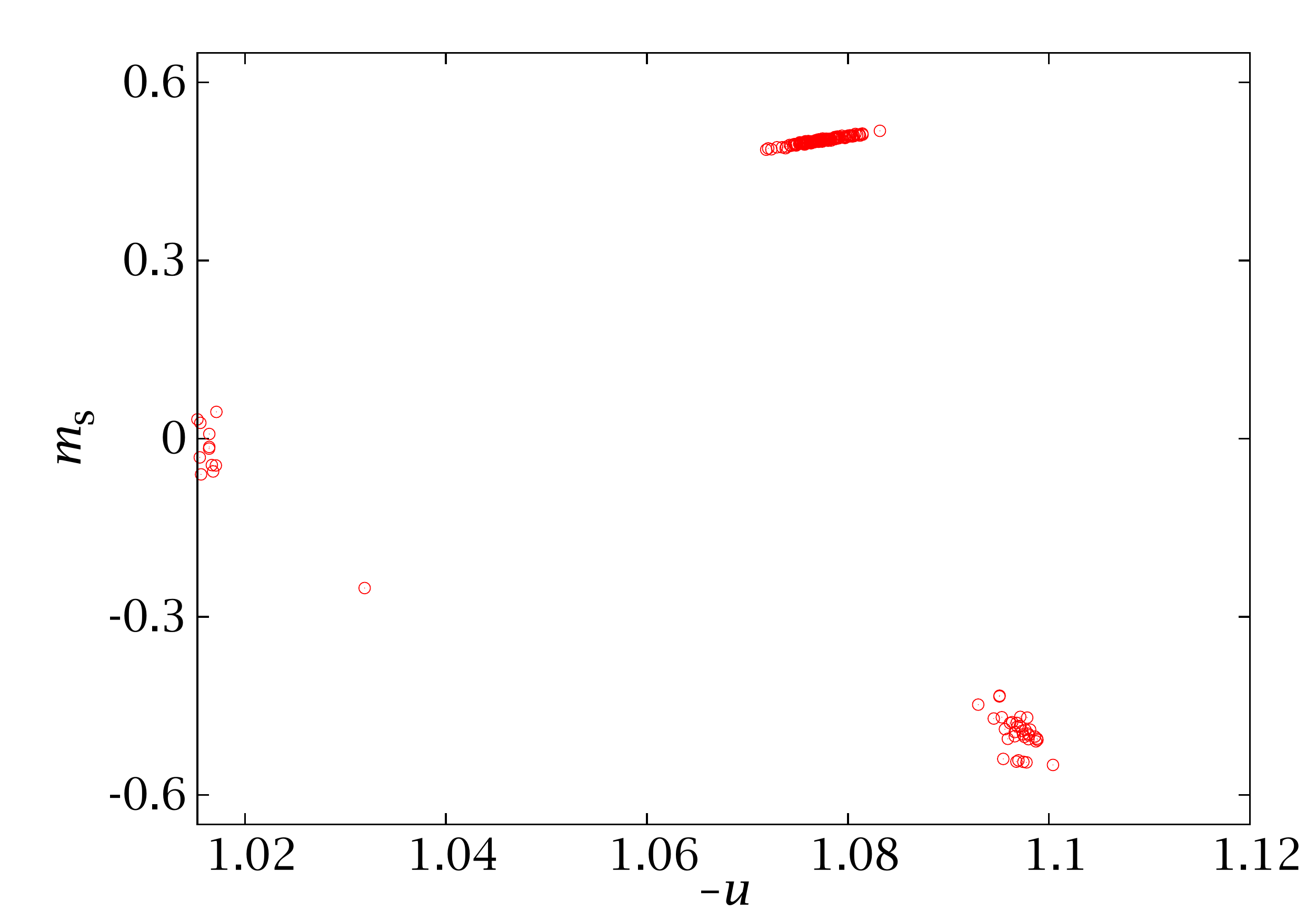}
\caption[Scatter plot of $m_\text{s}$ vs. $u$ in canonical simulations]{Scatter plot of
  the interaction energy $u$ and the staggered
  magnetisation for the $100$ runs of Figure~\ref{fig:DAFF-historia-canonico}. 
  In each case we average the observables over the last $10\,000$
  MCS, corresponding to the last $0.1\%$ of their whole Monte Carlo
  history. We see that the states with energies $u_1\approx-1.077$ and $u_2\approx-1.097$
  correspond to systems in an antiferromagnetic phase (with opposite 
  orientations).}\label{fig:DAFF-scatter-canonico}
\end{figure}

The first conclusion we can draw from this plot is that 
parallel tempering has failed. Assuming $u_1$ and $u_2$ represent  \index{metastability}
intermediate metastable states, not representative of the 
system's equilibrium phase, we have failed to escape from them
and reach the relevant regions of configuration space.
On the other hand, assuming both are important to the equilibrium
phase, then in a thermalised simulation we should see tunnelling
between the two. Notice that obtaining the two estates in separate
runs is not enough, we need to see the tunnelling in order
to know their relative weights. However, we have not seen a single
such jump, once the simulation reaches one of the two states it never
escapes. In total we performed $100$ runs for this sample, each 
with $10^7$ parallel tempering updates, which suggests that the
tunnelling probability is upper bounded by $10^{-9}$.  In other
words, parallel tempering is not able to thermalise the system
in a reasonable amount of time. Notice that some of the simulations
have not even reached one of the two metastable states 

The second  problem with the canonical approach is 
that of interpreting the results. Suppose we had considered
runs many orders of magnitude longer than the ones plotted
in Figure~\ref{fig:DAFF-historia-canonico}. Then, we would eventually
begin to see the quick jumps from one state to the other, separated
by long stays in each of them. This is the sort of metastable behaviour
that one expects in the neighbourhood of a first-order 
phase transition, where two different phases have a similar weight,
but are separated by large tunnelling barriers.

However, this interpretation would be wrong, as evinced by Figure~\ref{fig:DAFF-scatter-canonico}.
In it we represent a scatter plot of the staggered magnetisation against $u$ for the
$100$ runs of Figure~\ref{fig:DAFF-historia-canonico}. It is readily apparent that the
two states correspond to systems with opposite sign
of the order parameter $m_\text{s}^{(1)}=0.502\,3(3)$,
$m_\text{s}^{(2)}=-0.543(3)$. Therefore, the observed 
metastability would not correspond to  jumps between an antiferromagnetic
and a paramagnetic phase (the phase transition we want to study), 
but would rather reflect jumps between two antiferromagnetic states
with different spin orientation.\footnote{One could think 
that these two states should be symmetric and have the same energy, 
but remember that the number of spins in the odd and even sublattices
is not the same for this random system, thus breaking the usual
$\boldsymbol Z_2$ symmetry usually associated with antiferromagnets.}

This latter problem is intrinsic to the canonical description. Even if 
we could devise a more efficient thermalisation algorithm (or use
a much faster computer) our results would still be contaminated 
by this spurious metastability and their statistical 
analysis dominated by extremely rare events.


\chapter{The DAFF in the tethered formalism}\label{chap:daff-tethered}
We have just seen how traditional Monte Carlo methods are not well
suited for the simulation of the DAFF. This was not only 
a case of thermalisation problems, but also revealed intrinsic 
limitations in the canonical approach, where the statistics
is dominated by extremely rare events. In this Chapter
we intend to demonstrate how a tethered approach can solve
many of these difficulties.
\section{The DAFF in the tethered ensemble}\label{sec:DAFF-tethered}\index{tethered formalism!DAFF|(}
Our canonical investigation revealed that in the DAFF 
one has to deal with free-energy barriers \index{free energy!barriers}
separating states with different staggered magnetisations.
Therefore, the appropriate reaction coordinate \index{reaction coordinate}
to tether is $m_\text{s}$. In addition, we are interested in a transition
at non-zero magnetic field, where metastability could also appear.
Therefore, we also tether the regular magnetisation $m$. Notice that by
doing this, we do not need to specify the applied field for the 
tethered simulations. We simply simulate at zero field and then 
reweight the results as we did for the Ising model. \index{Ising model}

Let us consider the tethered description of a single sample.
Recalling Section~\ref{sec:TMC-several-tethers}, we need
a two-variable  effective potential, whose gradient defines the tethered
field, now a two-dimensional vector,
\begin{equation}\label{eq:DAFF-Omega}
\boldsymbol\nabla \varOmega_N(\hat m,\hat m_\text{s}) = \biggl( \frac{\partial \varOmega_N(\hat m,\hat m_\text{s})}
{\partial \hat m},\ \frac{\partial \varOmega_N(\hat m,\hat m_\text{s})}{\partial \hat m_\text{s}}\biggr)
= \bigl( \langle\hat b\rangle_{\hat m,\hat m_\text{s}},\ \langle \hat b_\text{s}\rangle_{\hat m,\hat m_\text{s}}\bigr)
\end{equation}
In this equation
\begin{subequations}
\begin{align}\label{eq:DAFF-tethered-magnetic-field}
\hat b &=  1-\frac{1/2-1/N}{\hat m-m},\\
\hat b_\text{s} &=  1-\frac{1/2-1/N}{\hat m_\text{s}-m_\text{s}}.
\index{tethered field}
\end{align}
\end{subequations}

Then, the  canonical averages in the presence of an external magnetic 
field with a regular component $h$ and a staggered component $h_\text{s}$
are related to the tethered expectation values through
the Legendre transformation \index{Legendre transformation}
\begin{equation}\label{eq:DAFF-tethered-to-canonical}
\langle O\rangle(h,h_\text{s}) = \frac{\int\dd \hat m\int \dd \hat m_\text{s}\ \langle O\rangle_{\hat m,\hat m_\text{s}}\
\ee^{-N [\varOmega_N(\hat m,\hat m_\text{s})-\hat m \beta h-\hat m_\text{s} \beta h_\text{s}]}}
{\int\dd \hat m\int \dd \hat m_\text{s}\ 
\ee^{-N [\varOmega_N(\hat m,\hat m_\text{s})-\hat m \beta h-\hat m_\text{s} \beta h_\text{s}]}},
\end{equation}
although here we are interested in $h_\text{s}=0$, so we shall use
the notation
\begin{equation}
\langle O\rangle(h) = \langle O\rangle (h,h_\text{s}=0).
\end{equation}
And furthermore, we shall use the following shorthand
\begin{equation}\label{eq:DAFF-Omega-h}
\varOmega_N^{(h)}(\hat m,\hat m_\text{s}) = \varOmega_N(\hat m,\hat m_\text{s}) - \beta h \hat m,
\end{equation}
and
\begin{equation}\label{eq:DAFF-B}
\hat{\boldsymbol B} = \boldsymbol \nabla \varOmega^{(h)}_N(\hat m,\hat m_\text{s})
= \bigl( \braket{\hat b}_{\hat m,\hat m_\text{s}}-\beta h,\ \braket{\hat b_\text{s}}_{\hat m,\hat m_\text{s}}\bigr).
\end{equation}
According to the procedure described in Section~\ref{sec:tmc}, 
a tethered study of the DAFF would consist in a set of 
tethered simulations for fixed values 
of $(\hat m,\hat m_\text{s})$ in a two-dimensional grid.
From the value of the tethered field $(\braket{\hat b}_{\hat m,\hat m_\text{s}},
\braket{\hat b_\text{s}}_{\hat m,\hat m_\text{s}})$ 
we would then reconstruct $\varOmega_N(\hat m,\hat m_\text{s})$
and use~\eqref{eq:DAFF-tethered-to-canonical}.
\begin{figure}[p]
\centering
\includegraphics[height=.85\columnwidth,angle=270]{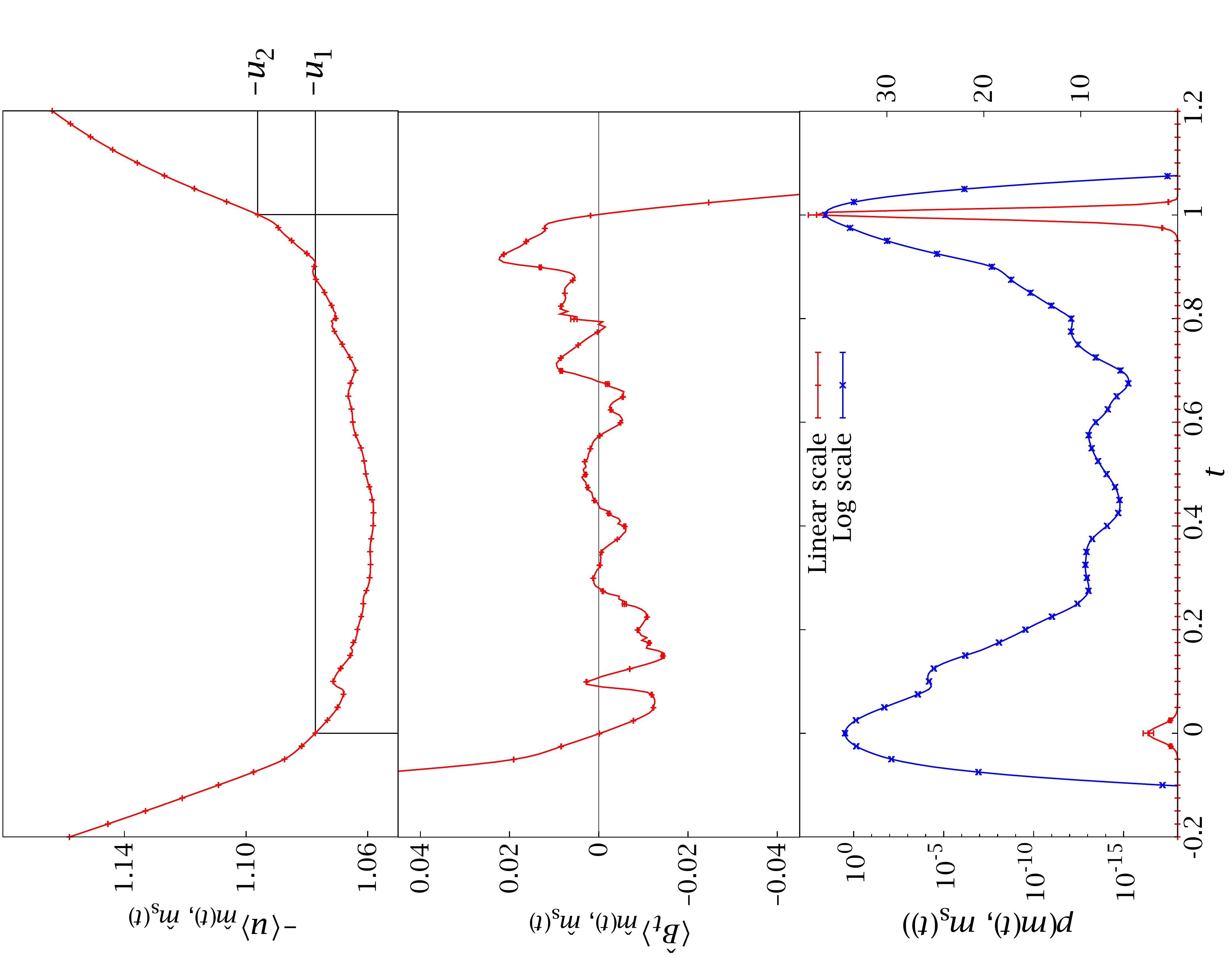}
\caption[Connecting saddle points with the tethered formalism]{%
Result of a tethered investigation of the sample that we studied
with canonical methods in Section~\ref{sec:DAFF-canonical-simulations}.
We join the two metastable states found in the canonical
computation with a straight line and measure
along this line the tethered expectation values
for the energy $u$ and the projection of the tethered field $\hat B_t$
(top and middle, respectively). We simulated $281$ independent 
points along the line, but only plot one in five with error bar to avoid cluttering
the graph.  The tethered field has many zeros, defining saddle points
in~\eqref{eq:DAFF-tethered-to-canonical}. 
Finally, the bottom panel shows the (exponential of the) integral of $\hat B_t$,
defining the relative weight of the points along the path. Two peaks
dominate, where the tethered values of the energy correspond
to the two metastable states $u_1$ and $u_2$ that we found
in Section~\ref{sec:DAFF-canonical-simulations}.
Notice the extremely low probability of the connecting region, which
explains the difficulty to thermalise canonical simulations.
\label{fig:DAFF-segmento}
\index{effective potential!DAFF|indemph}
\index{energy!DAFF|indemph}
\index{free energy!landscape|indemph}
\index{metastability}
}
\end{figure}

This is not, however, the most practical approach. 
For instance, it would involve the non-trivial numerical computation
of a two-dimensional potential from its
gradient, evaluated in a discrete grid. Instead, we note
that, for a given $h$, only a very small region
of the reaction coordinate space $(\hat m,\hat m_\text{s})$ 
will have a relevant weight. In fact, the integral~\eqref{eq:DAFF-tethered-to-canonical}
is going to be dominated by the 
minima of the effective potential (recall our discussion of ensemble equivalence in
Section~\ref{sec:TMC-ensemble-equivalence}). \index{saddle point}
These will be given by the saddle-point equations 
\begin{equation}
\boldsymbol \nabla \varOmega^{(h)}_N(\hat m,\hat m_\text{s}) = \hat{\boldsymbol B} =  0,
\end{equation}
that is 
\begin{equation}\label{eq:DAFF-saddle-point-sample}
\begin{cases}
\displaystyle
\frac{\partial \varOmega_N}{\partial \hat m} &= \langle\hat b\rangle_{\hat m,\hat m_\mathrm{s}} = \beta h,\\ 
\\
\displaystyle
\frac{\partial \varOmega_N}{\partial \hat m_\mathrm{s}} &= \langle\hat b_\mathrm{s}\rangle_{\hat m,\hat m_\mathrm{s}}  = 0. 
\end{cases}
\end{equation}

In principle, some of these equations will correspond to local minima of $\varOmega^{(h)}_N$
and some to local maxima or to saddle points in the strict sense
(a minimum in one direction, but a maximum in the other).
In the case of local minima, the saddle points will correspond
to metastable states, potentially relevant to the description
of the equilibrium phase. Notice that we also obtained several
metastable states in our canonical study of Section~\ref{sec:DAFF-canonical-simulations}, 
but we were unable to know their relative weights. In the tethered
approach, however, this is easily done by considering their
potential difference, i.e., the line integral
of $\hat{\boldsymbol B}$ along a connecting path.

Let us demonstrate this procedure for the sample of Section~\ref{sec:DAFF-canonical-simulations},
where we had identified two metastable states (two local minima of $\varOmega^{(h)}_N$, 
in the tethered nomenclature).
We first have to determine the values
of $(\hat m,\hat m_\text{s})$ that correspond to
these saddle points using~\eqref{eq:DAFF-saddle-point-sample}.
This is easy, because our canonical study has given us 
the values of $m^{(i)}$ and $m_\text{s}^{(i)}$.
We can now use~(\ref{eq:DAFF-tethered-magnetic-field}), 
setting $\hat b_\text{s}=0$ and $\hat b=\beta h = 2.4$.
Therefore, we know that $\hat m_\text{s}^{(i)} \simeq m_\text{s}^{(i)}+1/2$
and $\hat m^{(i)} \simeq m^{(i)} + 1/(2(1-\beta h))$. From
these starting guesses the actual saddle points are readily found.
We can connect them with any path (because $\hat{\boldsymbol B}$ is a conservative
field). Since the system is random, it is difficult to predict which will 
be the optimal connecting curve (in the sense of thermalisation),
so we consider a simple straight line,
\begin{equation}
(\hat m(t), \hat m_\text{s}(t)) = (\hat m^{(1)}, \hat m_\text{s}^{(1)}) (1-t) + (\hat m^{(2)} ,
\hat m_\text{s}^{(2)}).
\end{equation}

The whole computation is depicted in Figure~\ref{fig:DAFF-segmento}.
We performed tethered simulations for $281$ values of 
the parameter $t$ and measured the tethered expectation 
values of the energy $\braket{u}_{\hat m(t),\hat m_\text{s}(t)}$
and the tethered field $\hat{\boldsymbol  B}$. These are
plotted in the top and middle panels of the figure (for the 
tethered field we plot its projection $\hat B_t$ along
the path).

Now, following~\eqref{eq:DAFF-tethered-to-canonical}, the probability
density or relative weight of the points along the path is just
\begin{equation}
p(\hat m(t) ,\hat m_\text{s}(t)) = \ee^{-N \varOmega^{(h)}_N (\hat m(t), \hat m_\text{s}(t))},
\end{equation}
where $\varOmega^{(h)}_N(t)$ is the line integral of $\hat{\boldsymbol B}$, with 
the integration constant chosen so that the whole weight is normalised. 
This probability density is plotted in the bottom panel of Figure~\ref{fig:DAFF-segmento},
in both a linear and a logarithmic scale. We find that one 
of the two metastable has about ten times more weight than the other
(curiously, fewer canonical simulations found the more relevant 
state). In addition, the two resulting peaks in the $p(\hat m(t), \hat m_\text{s}(t))$
are separated by a region with very low probability, explaining
the difficulty of canonical simulations to tunnel between 
the two states. Interestingly enough, even within this 
low-probability sector we can see a rich structure of the $p(\hat m(t),\hat m_\text{s}(t))$
---equivalently, of the effective potential. In other words, we are seeing a  \index{free energy!landscape}
quantitative example of a rugged free-energy landscape.

\subsection{Self-averaging and the disorder average}\label{sec:DAFF-disorder-average}
In order to perform a quantitative analysis of the DAFF, we have to
simulate a large number of samples and perform the disorder average.
The naive way of doing this for a system
with quenched disorder would be to measure $\hat{\boldsymbol B}$
and construct $\varOmega_N$ for each sample,
then use Eq.~(\ref{eq:DAFF-tethered-to-canonical}) 
to compute all the  physically relevant $\langle O \rangle(h)$. 
Only then would we average $\langle O\rangle(h)$ over the disorder.
\index{self-averaging|(}

This approach is, however, paved with pitfalls. First of all, computing 
a two-variable  $\varOmega_N(\hat m,\hat m_\text{s})$ from a two-dimensional 
$(\hat m,\hat m_\text{s})$ grid is not an easy matter. In the previous section
we avoided this problem by computing the saddle points first and then
evaluating $\varOmega_N$ only along a path joining them. But this cannot be done 
efficiently and safely for a large number of samples. Even if it could be done,
the free-energy landscape of each sample is very complicated, with many local 
minima, several of which could be relevant to the problem, so a very high
resolution would be needed on the simulation grid. 

Finally, even reliably and efficiently computing 
the canonical averages $\langle O\rangle(h)$ 
would not be the end of our problems. Indeed, as 
we discussed in Section~\ref{sec:INTRO-self-averaging},
the canonical expectation values suffer from 
severe self-averaging violations.
In this section we address
all these problems and demonstrate the computational strategy
followed in our study of the DAFF.

The first step is ascertaining whether the tethered averages themselves
are self-averaging or not. We already know that, since we are going to 
be working with a large regular external field $h$, the relevant 
region for the regular magnetisation $m$ (and hence $\hat m$)
is going to be exceedingly narrow (cf. Section~\ref{sec:ISING-h-neq-zero}, 
where the fields were much smaller). Therefore, we are going to explore
the whole range of $\hat m_\text{s}$ for a fixed value of $\hat m=0.12$
(this smooth magnetisation is in the range where the saddle points typically
lie for the applied fields we considered in Chapter~\ref{chap:daff-canonical}).

\begin{figure}[p]
\centering
\includegraphics[height=0.7\columnwidth,angle=270]{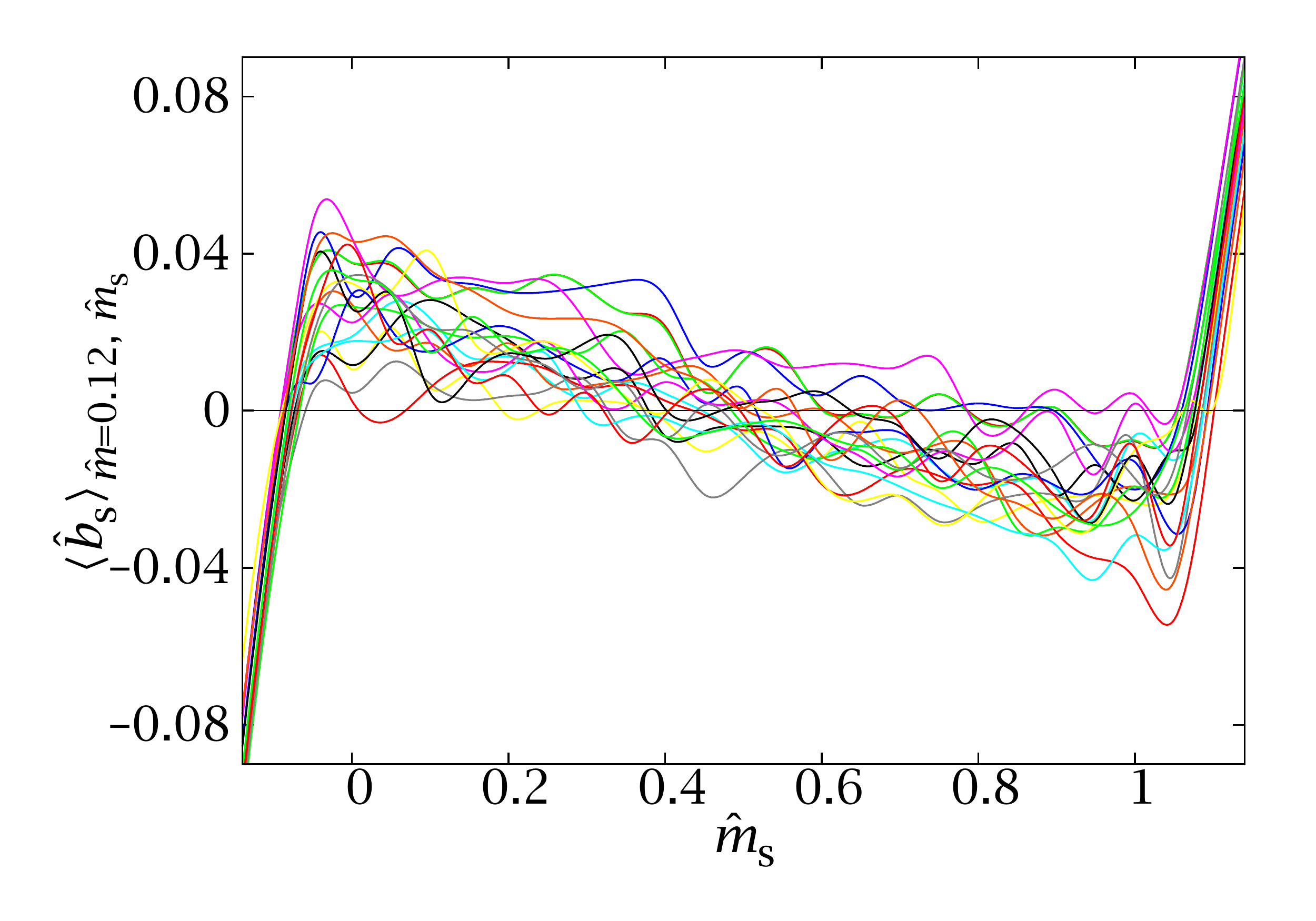}
\vspace*{1cm}
\includegraphics[height=0.7\columnwidth,angle=270]{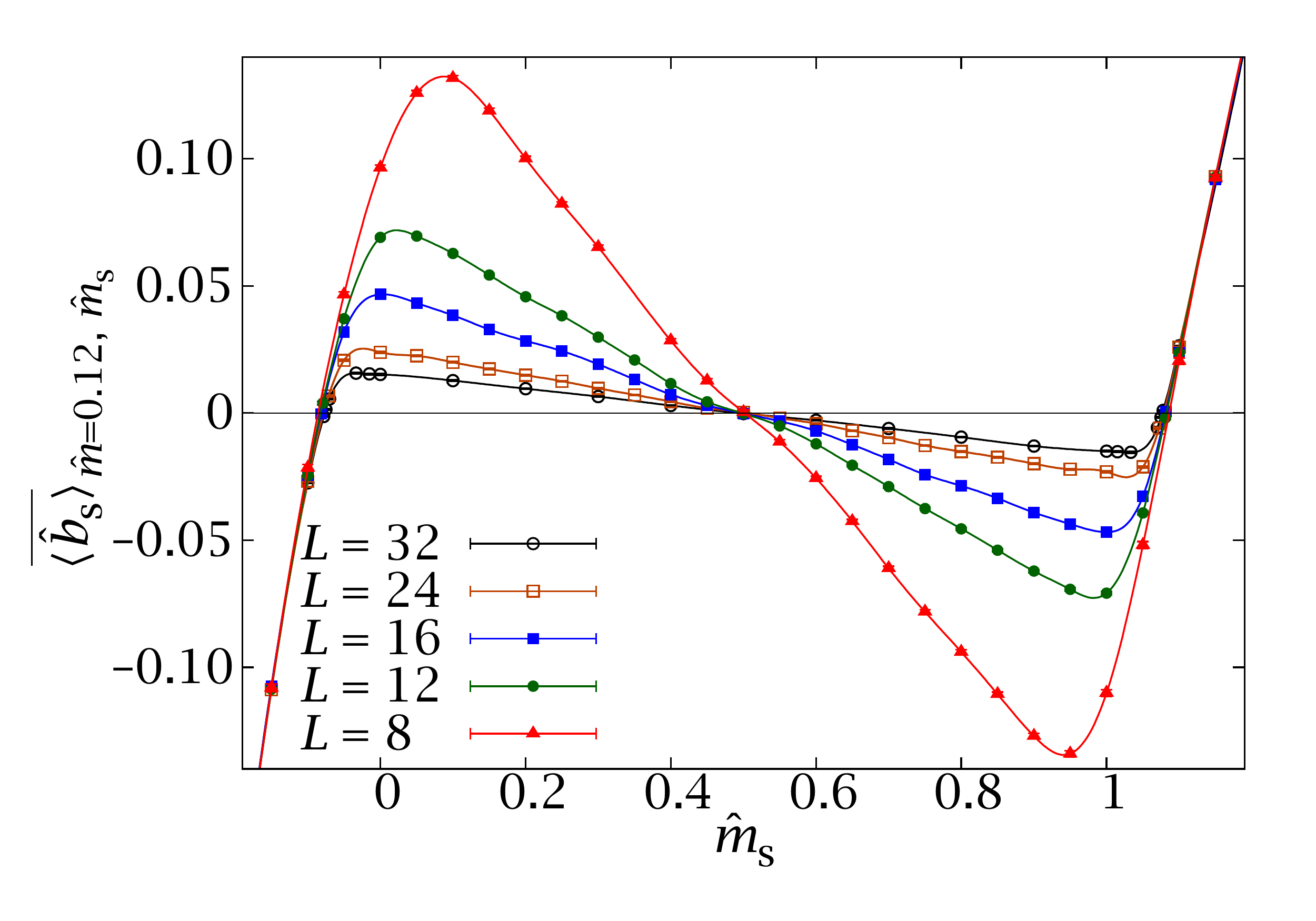}
\caption[Staggered tethered magnetic field]{\emph{Top:} Staggered component of the tethered magnetic field, 
$\langle \hat b\rangle_{\hat m,\hat m_\text{s}}$ for $\hat m=0.12$ 
as a function of $\hat m_\text{s}$. We plot the results for 
several samples of an $L=24$ system at $\beta=0.625$.
The curves are cubic splines interpolated from $33$ simulated points.
We note that the errors are very small (invisible
at this scale), so the  fluctuations of the curves are not artifacts
of the interpolation scheme.
\emph{Bottom:} Disorder-averaged
$\overline{\langle \hat b_\text{s}\rangle}_{\hat m,\hat m_\text{s}}$
for the same smooth magnetisation and temperature as the left 
panel, for all our system sizes. The plot shows the different 
behaviour of the regions inside the gap, where the staggered 
magnetic field goes to zero as $L$ increases,
and outside it, where there is a non-zero enveloping curve.
\label{fig:DAFF-hatbs-samples}
\index{tethered field!DAFF|indemph}}
\index{self-averaging|indemph}
\index{cubic splines}
\end{figure}
We have plotted the staggered tethered magnetic field
 $\langle \hat b_\text{s}\rangle_{\hat m=0.12,\hat m_\text{s}}$
for $20$ samples of an $L=24$ system at $\beta=0.625$
in Figure~\ref{fig:DAFF-hatbs-samples}---top.  The different 
curves have a variable number of zeros, but all of them have at least three:
one in the central region and two roughly symmetrical ones for large 
staggered magnetisation.  The positions
of the two outermost zeros  clearly separate two differently behaved 
regions. Inside the gap the sample-to-sample fluctuations are chaotic,
while outside it the sheaf of curves even seems to have an envelope.
This impression is confirmed in the bottom panel of Figure~\ref{fig:DAFF-hatbs-samples},
where we show the sample-averaged tethered magnetic field for several system
sizes.

\begin{figure}
\centering
\begin{minipage}{.49\linewidth}
\includegraphics[height=1.05\columnwidth,angle=270]{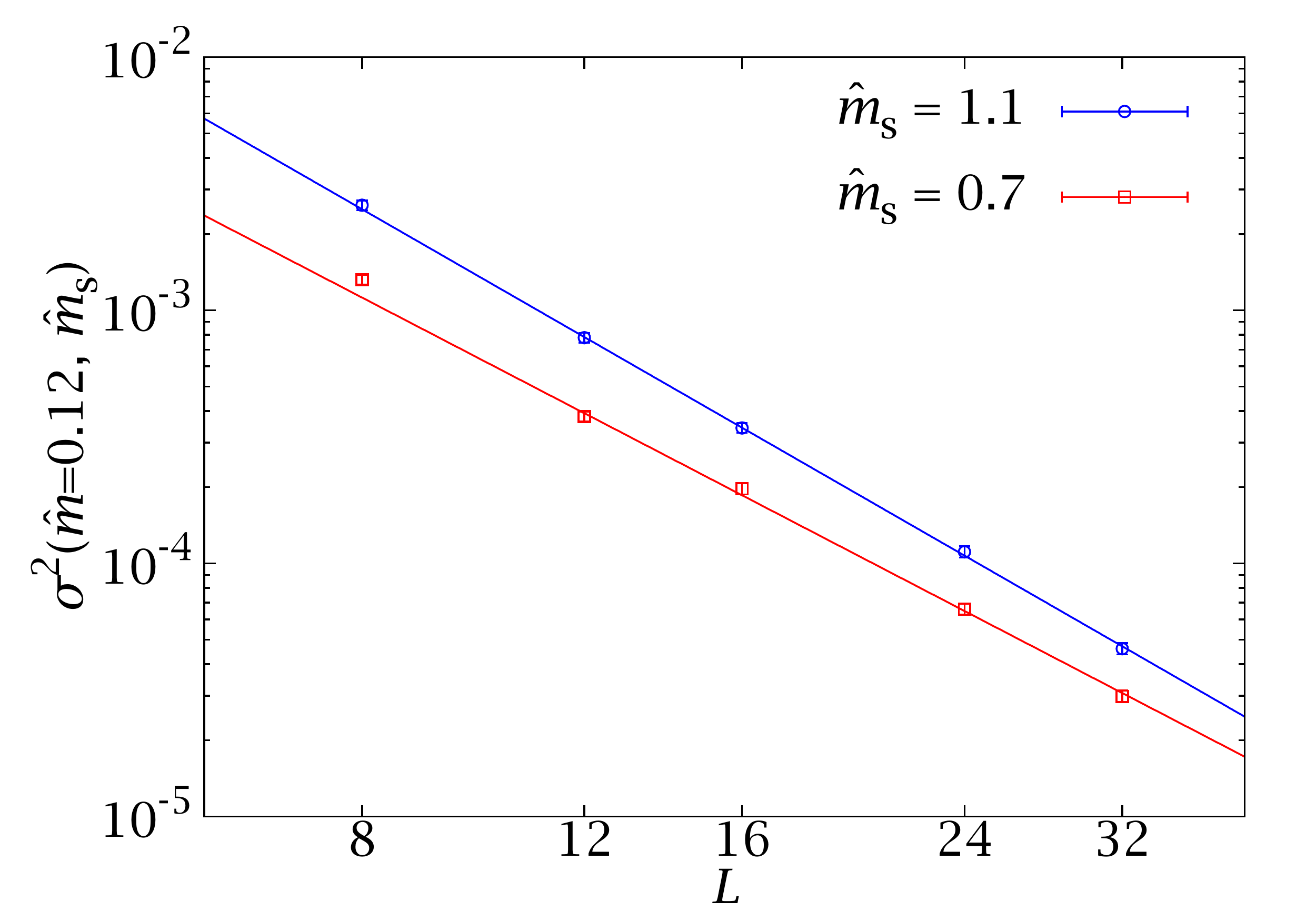}
\end{minipage}
\begin{minipage}{.49\linewidth}
\includegraphics[height=1.05\columnwidth,angle=270]{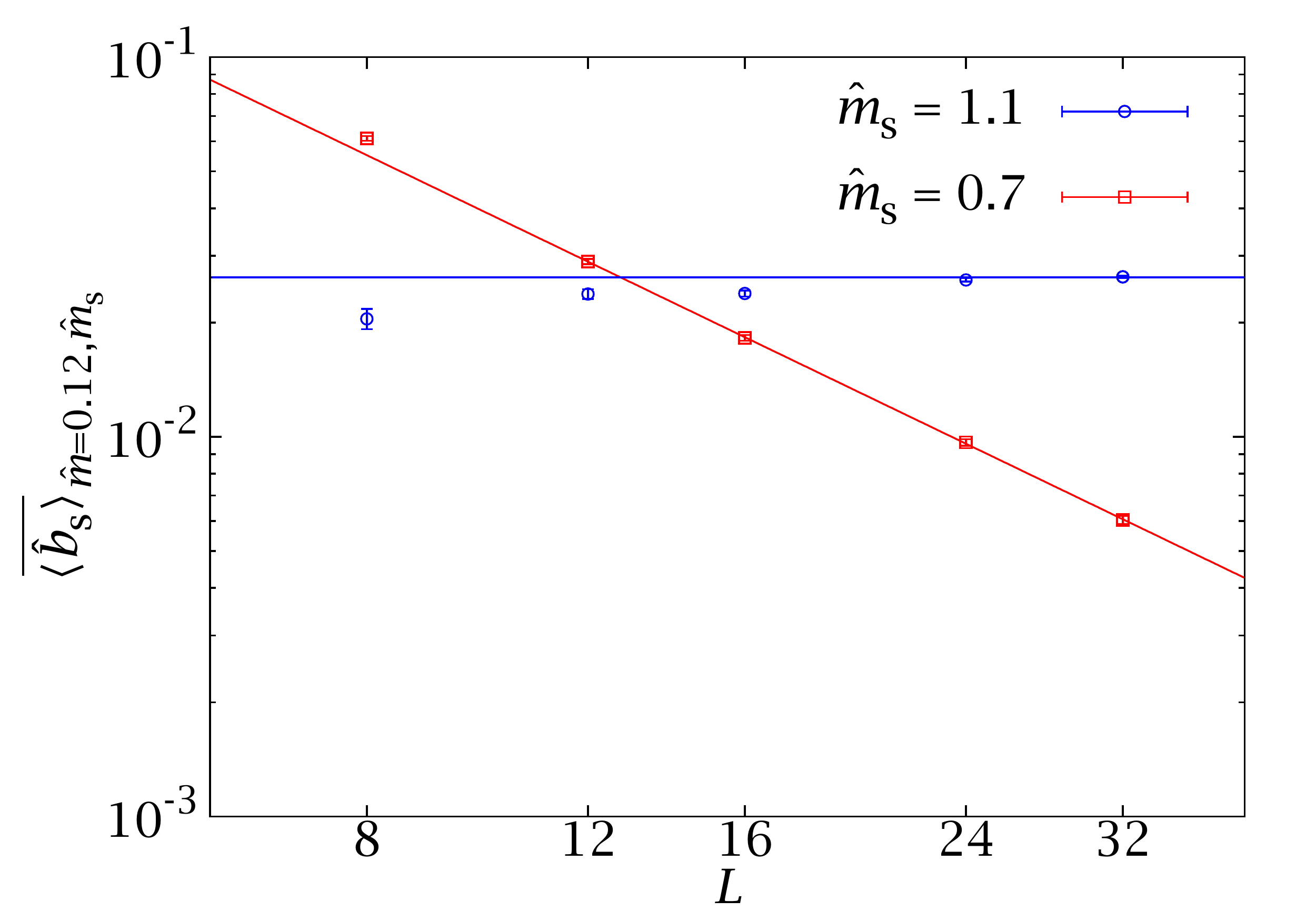}
\end{minipage}
\caption[Self-averaging in the tethered averages]{\emph{Left:} Sample-to-sample fluctuations in the 
staggered component of the tethered magnetic field, Eq.~\eqref{eq:DAFF-sigma},
against system size. We show two sets of points, both at $\beta=0.625$ 
and $\hat m=0.12$, but  for different values of $\hat m_\text{s}$.
The red curve, corresponding to $\hat m_\text{s}=0.7$, is right
inside the gap defined by the outermost zeros in Figure~\ref{fig:DAFF-hatbs-samples},
while the blue curve ($\hat m_\text{s}=1.1$) is outside. Both 
are shown to decay with a power law.
\emph{Right:} Disorder average $\overline{\langle\hat b_\text{s}\rangle}_{\hat m,\hat m_\text{s}}$
for the systems of the left panel. Inside the gap, the field goes to zero with $L$;
outside it has a finite limit.}
\label{fig:DAFF-automediancia} 
\end{figure}
In order to quantify this  observation we can study the fluctuations of the 
disorder-averaged $\overline{\langle \hat b_\text{s}\rangle}_{\hat m,\hat m_\text{s}}$,
\begin{eqnarray}
\sigma^2(\hat m,\hat m_\mathrm{s}) &=&  \overline{\left({\langle\hat b_\mathrm{s}\rangle}_{\hat m,\hat m_\mathrm{s}} - \overline{\langle\hat b_\mathrm{s}\rangle}_{\hat m,\hat m_\mathrm{s}} \right)^2}\ \label{eq:DAFF-sigma}.
\end{eqnarray} 
This quantity is plotted in the left panel of Figure~\ref{fig:DAFF-automediancia}. As we
can see, it goes to zero as a power in $L$, so we also plot fits to
\begin{equation}
\sigma^2(\hat m=0.12,\hat m_\text{s}) = A(\hat m_\text{s}) L^{-c(\hat m_\text{s})}.
\end{equation} 
This would seem like a very good
sign, because it could be indicative of self-averaging behaviour. However,
it is not the whole story. If we recall the bottom panel
of Figure~\ref{fig:DAFF-hatbs-samples}, we see that inside the gap the tethered magnetic
field itself, not only its fluctuations, goes to zero as $L$ increases. In fact, as shown 
in Figure~\ref{fig:DAFF-automediancia}, for
$\hat m_\text{s} = 0.7$,  $\sigma^2\sim L^{-2.5}$, while 
$\overline{\langle \hat b_\text{s}\rangle}_{\hat m,\hat m_\text{s}}\sim L^{-1.6}$.
This means that the relative fluctuations $R_{\hat b_\text{s}}$, Eq.~\eqref{eq:DAFF-R-O},
do not decrease with increasing $L$.
For the point outside the gap, however, the disorder average of the tethered magnetic 
field reaches a plateau. Furthermore, the fluctuations decay
 with $c=2.94(7)$, which
is compatible with the behaviour $R_{\hat b_\text{s}} \sim L^{-D}$ 
of a strongly self-averaging system (in the language of Section~\ref{sec:INTRO-self-averaging}).

\index{saddle point|(}
In physical terms, this analysis means that the saddle point  defined
by a small, but non-zero,
value of the applied staggered magnetic field $h_\text{s}$ would be self-averaging.
We can now recall the well-known recipe for dealing with spontaneous 
symmetry breaking (see Section~\ref{sec:ISING-SSB}):
consider a small applied field and take the thermodynamical
limit \emph{before} making the field go to zero.
Translated to the DAFF, this means that we should solve the saddle-point
equations~\eqref{eq:DAFF-saddle-point-sample}
on average, rather than sample by sample, and then take the $h_\text{s}\to0$
limit on the results,
\begin{equation}\label{eq:DAFF-saddle-point-medio}
\begin{cases}
\displaystyle
\frac{\partial \overline\varOmega_N}{\partial \hat m} &= \overline{\langle\hat b\rangle}_{\hat m,\hat m_\mathrm{s}} = \beta h,\\ 
\\
\displaystyle
\frac{\partial \overline\varOmega_N}{\partial \hat m_\mathrm{s}} &= \overline{\langle\hat b_\mathrm{s}\rangle}_{\hat m,\hat m_\mathrm{s}}  = 0^+. 
\end{cases}
\end{equation}
In other words, we are considering the disorder average of a thermodynamical
potential, $\varOmega_N$, different from the free energy. This approach was
first introduced in~\cite{fernandez:08}, in a microcanonical context
(the averaged potential was in that case the entropy).
\index{disorder!quenched}

In order to understand the limit $h_\text{s}\to0^+$, it is convenient 
to keep in mind the analysis of the $p(\hat m)$ for the Ising model
in Sections~\ref{sec:ISING-h-neq-zero} and~\ref{sec:ISING-SSB}.
\index{Ising model}
A small but positive $h_\text{s}$ exponentially suppresses the negative staggered
magnetisation region and, as its value is decreased, it is
equivalent to considering a `smeared' saddle point
and averaging over the whole sector with $\hat m_\text{s}> 1/2$.
A small but negative $h_\text{s}$ would have the corresponding 
effect in the $\hat m_\text{s}< 1/2$ region. Since most of the interesting 
disorder-averaged observables are symmetric with respect to $\hat m_\text{s}=1/2$,
we can gain statistics by averaging over the whole $\hat m_\text{s}$ range.

\begin{figure}
\centering
\includegraphics[height=.85\columnwidth,angle=270]{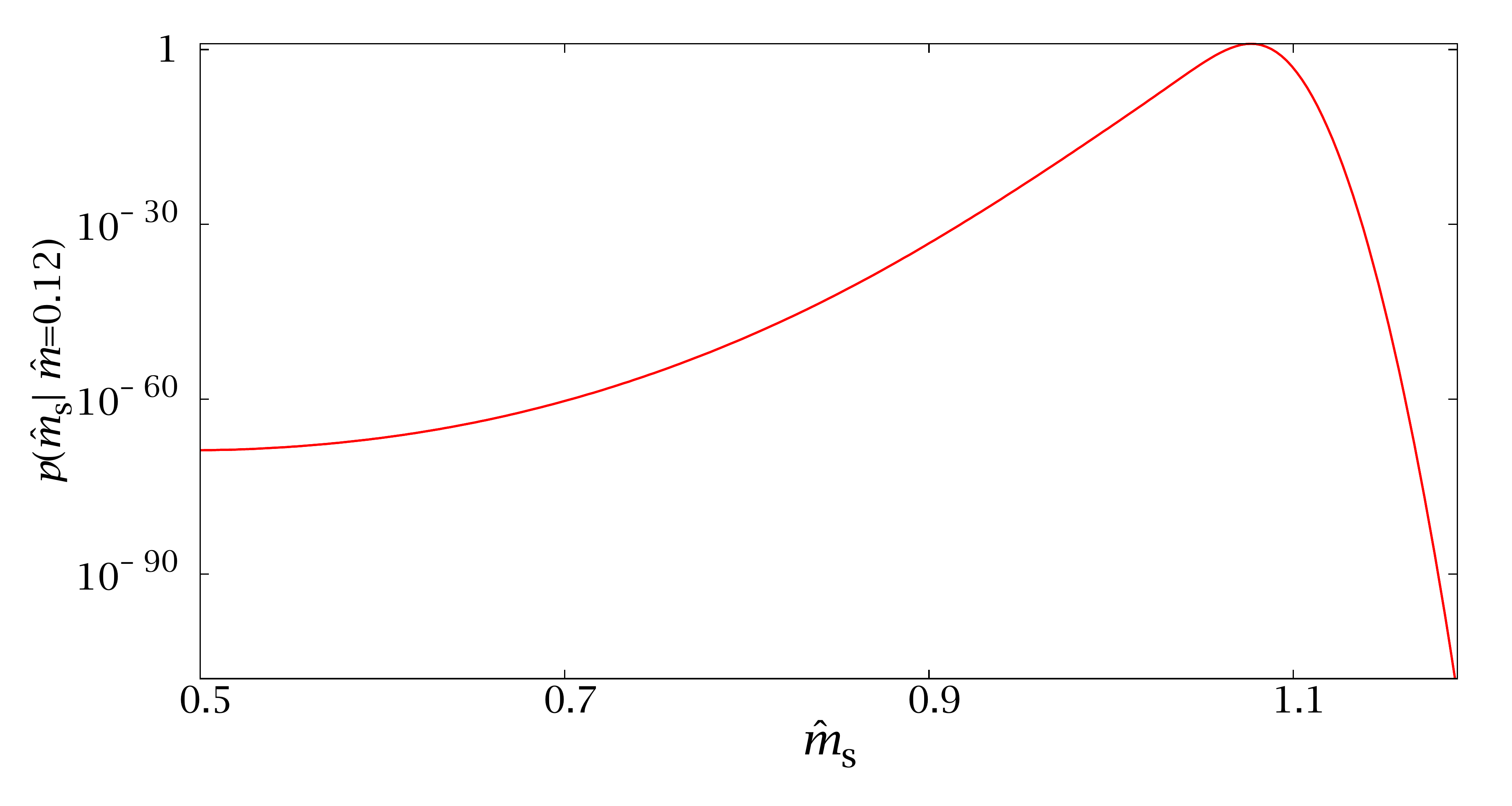}
\caption[Conditioned probability density $p(\hat m_\text{s}|\hat m)$]{Disorder-averaged
pdf of $\hat m_\text{s}$ conditioned 
to $\hat m = 0.12$ for the system of Figure~\ref{fig:DAFF-hatbs-samples}.
Compare the simple structure of this graph with the bottom 
panel of Figure~\ref{fig:DAFF-segmento}.
\label{fig:DAFF-p-hatms}
\index{effective potential!DAFF|indemph}
}
\end{figure}
By doing this for fixed $\hat m=\hat m_0$ we obtain the probability 
distribution of $\hat m_\text{s}$, conditioned to $\hat m =\hat m_0$, which we will
denote by $p(\hat m_\text{s} | \hat m_0)$ (Figure~\ref{fig:DAFF-p-hatms}). This 
probability density function can be used to average over $\hat m_\text{s}$ for 
fixed $\hat m$,
\begin{equation}\label{eq:promedio-hatms}
\overline{\langle O\rangle}_{\hat m} = \int \dd \hat m_\text{s} \ p(\hat m_\text{s} | \hat m)
\overline{\langle O\rangle}_{\hat m,\hat m_\text{s}} =
\int \dd \hat m_\text{s} \ \overline{\langle O\rangle}_{\hat m,\hat m_\text{s}}  \ee^{-N \overline\varOmega_N(\hat m_\text{s} | \hat m)},
\end{equation}
where 
\begin{equation}\label{eq:DAFF-Omega-hatms}
\frac{\partial \overline \varOmega_N(\hat m_\text{s}|\hat m)}{\partial \hat m_\text{s}} = 
\overline{\langle \hat b_\text{s}\rangle}_{\hat m,\hat m_\text{s}}\ .
\end{equation}
The zero in $\overline \varOmega_N(\hat m_\text{s}|\hat m)$ is chosen so that $p(\hat m_\text{s}|\hat m)$
is normalised (so $\overline\varOmega_N(\hat m_\text{s}|\hat m)$
differs from the two-dimensional $\overline\varOmega_N(\hat m,\hat m_\text{s})$
in a constant term). In practice, since
the $p(\hat m_\text{s}|\hat m)$ is not symmetric sample by sample, but 
only for the disorder average, considering the whole $\hat m_\text{s}$ range 
for $\mN_\text{samples}$ disorder realisations is roughly equivalent to considering only 
the positive sector for $2\mN_\text{samples}$.\footnote{Compare
with Section~\ref{sec:ISING-h-neq-zero}, where the average for $\pm h$ 
netted a very small error reduction for even observables, due to the correlation
in the data.} 

With this process we have integrated out the dependence on $\hat m_\text{s}$.
\index{thermodynamical limit}
\index{ensemble equivalence}
We now have a series of smooth functions $\overline{\langle O\rangle}_{\hat m}$,
together with a smooth one-to-one function
$h(\hat m) = \overline{\langle \hat b\rangle}_{\hat m}/\beta$. Recalling our 
analysis of ensemble equivalence in Section~\ref{sec:ISING-SSB},
we see that $\overline{\braket{O}}_{\hat m(h)}$ and $\overline{\langle O\rangle}(h)$ 
are simply two quantities that tend to the same thermodynamical limit.
Furthermore, for finite lattices the tethered definition $\overline{\braket{O}}_{\hat m}$ 
is better behaved statistically and arguably more faithful to the physics 
of an experimental sample. Therefore, we shall make the identification
\begin{equation}
\overline{\langle O\rangle}(h) = \overline{\braket{O}}_{\hat m},
\end{equation}
exact in the thermodynamical limit, even for finite lattices.
\index{saddle point|)}
\index{self-averaging|)}

\index{tethered formalism!DAFF|)}
\section{Our tethered simulations}\label{sec:DAFF-simulations}
\begin{table}
\small
\centering
\begin{tabular*}{\columnwidth}{@{\extracolsep{\fill}}rcccccc}
\toprule
 $L$ & $\mN_\mathrm{samples}$ &$\mN_T$ &  $\mN_{\hat m}$ &
 $\mN_{\hat m_\mathrm{s}} $ & $\mN_\mathrm{MC}^{\mathrm{min}}$ 
& $\mN_\mathrm{MC}^\mathrm{av}$ \\
\toprule
8&  1000 & 20 & 5 &  31 & $7.7\times10^5$ & $7.7\times10^5$  \\
12& 1000 & 20 & 5 &  35 & $7.7\times10^5$ & $7.7\times10^5$  \\
16& 1000 & 20 & 5 &  35 & $7.7\times10^5$ & $7.7\times10^5$  \\
24& 1000 & 40 & 5 &  33 & $7.7\times10^5$ & $9.3\times10^5$  \\
32&  700 & 40 & 4 &  25 & $1.5\times10^6$ & $5.5\times10^6$  \\
\bottomrule
\end{tabular*}
\caption[Parameters of our simulations]{Parameters of our simulations. For each of the $\mN_\mathrm{samples}$
disorder realisations for each $L$,
 we run $\mN_{\hat m} \times \mN_{\hat m_\mathrm{s}}$
tethered simulations with temperature parallel tempering.
 The $\mN_T$ participating temperatures
are evenly spaced in the interval $[1.6,2.575]$. For each size
we report the minimum number of Monte Carlo steps
 (Metropolis sweep + parallel tempering). After the application of
our thermalisation criteria, some of the simulations for $L\geq24$ 
needed to be extended, leading to a higher average number of
\index{parallel tempering}
\index{Metropolis algorithm}
Monte Carlo steps ($\mN_\mathrm{MC}^\mathrm{av}$).}
\label{tab:DAFF-parametros-PT}
\index{parallel tempering}
\end{table}
We can infer several useful conclusions from the analysis of the previous section 
\begin{itemize}
\item The disorder average should be performed on the tethered
observables, before computing the effective potential.
\item It is best to analyse several values of $\hat m$ separately,
since the average over $\hat m_\text{s}$ for each fixed $\hat m$ can be 
unambiguously related to the canonical average $\overline{\langle O\rangle}(h)$ via
$h(\hat m) = \overline{\langle \hat b\rangle}_{\hat m} / \beta$.
In this way, we can study the phase transition that arises by varying 
the applied magnetic field at fixed $\beta$.
\item For fixed $\hat m$ the conditioned probability $p(\hat m_\text{s} | \hat m)$ 
has two narrow, symmetric peaks, separated by a region with extremely 
low probability.
\end{itemize}
Therefore, we have carried out the following steps
\begin{enumerate}
\item Select an appropriate grid of $\hat m$ values. This
should be wide enough to include the critical point for the simulation
temperature, and fine enough to detect the fluctuations of
$\overline{\langle O\rangle}_{\hat m}$. These turn out
to be very smooth functions of $\hat m$, so a few values of this parameter
suffice, as we shall see.
\item For each value of $\hat m$, select an appropriate grid of $\hat m_\text{s}$.
We start with evenly spaced points and after a first analysis add more
values of $\hat m_\text{s}$ in the neighbourhood of the saddle points, as this
is the more delicate and relevant region. In contrast with our study for the Ising 
model (Section~\ref{sec:ISING-hatm-grid}), the steepness of the peaks makes this second
step crucial here, even for small lattices.
\item The simulations for each $(\hat m,\hat m_\text{s})$ are carried out 
with the Metropolis update scheme of section~\ref{sec:TMC-Metropolis}.
\index{Metropolis algorithm}
In addition, we use parallel tempering (cf. Appendix~\ref{chap:thermalisation}). \index{parallel tempering}
This is not needed in order to thermalise the system
for $L<32$, but it is convenient since we also study
the temperature dependence of some observables.
Furthermore, the use of parallel-tempering provides a reliable 
thermalisation check (Section~\ref{sec:THERM-thermalisation-PT}).
Section~\ref{sec:DAFF-numerical-implementation} in Appendix~\ref{chap:recipes}
contains some information about our numerical implementation of TMC for the DAFF.
\end{enumerate}
From Figure~\ref{fig:DAFF-p-hatms} we see that, if we only want to compute
the $\langle O\rangle(h)$, the tethered simulations for $\hat m_\text{s}$ away 
from the peaks are worthless, since their contribution to the 
average over $\hat m_\text{s}$ is exponentially suppressed by a huge
factor. We could have run tethered simulations only in a narrow range of $\hat m_\text{s}$ around the
peaks, as we did in Section~\ref{sec:ISING-SSB}.
As we shall see, however, some physically relevant quantities require that 
the whole range be explored, the most conspicuous being the hyperscaling \index{hyperscaling}
violations exponent $\theta$.\index{critical exponent!theta@$\theta$}

The parameters of our simulations are presented in Table~\ref{tab:DAFF-parametros-PT}. 
The table lists the number $\mN_{\hat m}$ of values in our $\hat m$ grid
and the number of points in the $\hat m_\mathrm{s}$ grid for each, so 
the total number of tethered simulations for each sample is $\mN_{\hat m}\times \mN_{\hat m_\mathrm{s}}$.

The number of Monte Carlo steps in each tethered simulation 
is adapted to the autocorrelation time (see next section), the table
lists the minimum length $\mN_\text{MC}^\text{min}$ 
and the average length $\mN_\text{MC}^\text{av}$ for each lattice size.

\subsection{Thermalisation and metastability}\label{sec:DAFF-thermalisation}
\begin{figure}[p] \centering
\includegraphics[height=0.7\columnwidth,angle=270]{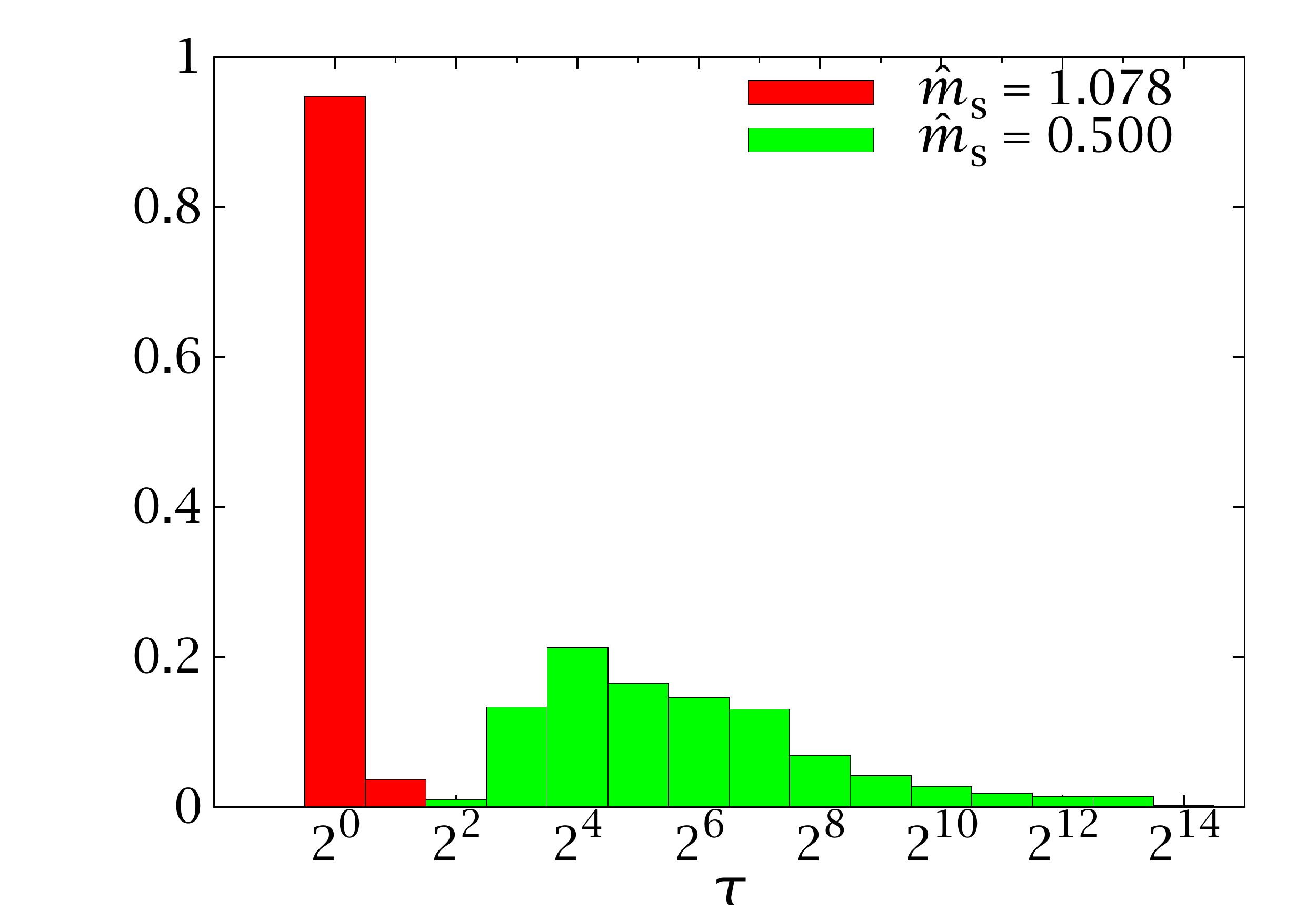}
\caption[Thermalisation times for $\hat m=0.12$ and different $\hat m_\text{s}$]{Histograms of thermalisation times for our $L=32$ simulations,
at two values of $\hat m_\mathrm{s}$ for $\hat m=0.12$. Notice 
the logarithmic scale in the horizontal axis, which is in units 
of $300$ parallel-tempering steps. The saddle point (the peak
in the probability distribution) at this $\hat m=0.12$
corresponds to $\hat m_\mathrm{s}=1.078$. Notice 
that we cannot measure times shorter than our measuring
frequency of $300$ parallel-tempering steps, so the first 
bin should be taken as encompassing all shorter autocorrelation
times. Only the samples with $\tau\gtrsim 2^6$ have to 
be extended from the minimum simulation time.}
\label{fig:DAFF-histograma-hatm-012}
\index{autocorrelation time!DAFF|indemph}

 \centering
\includegraphics[height=0.7\columnwidth,angle=270]{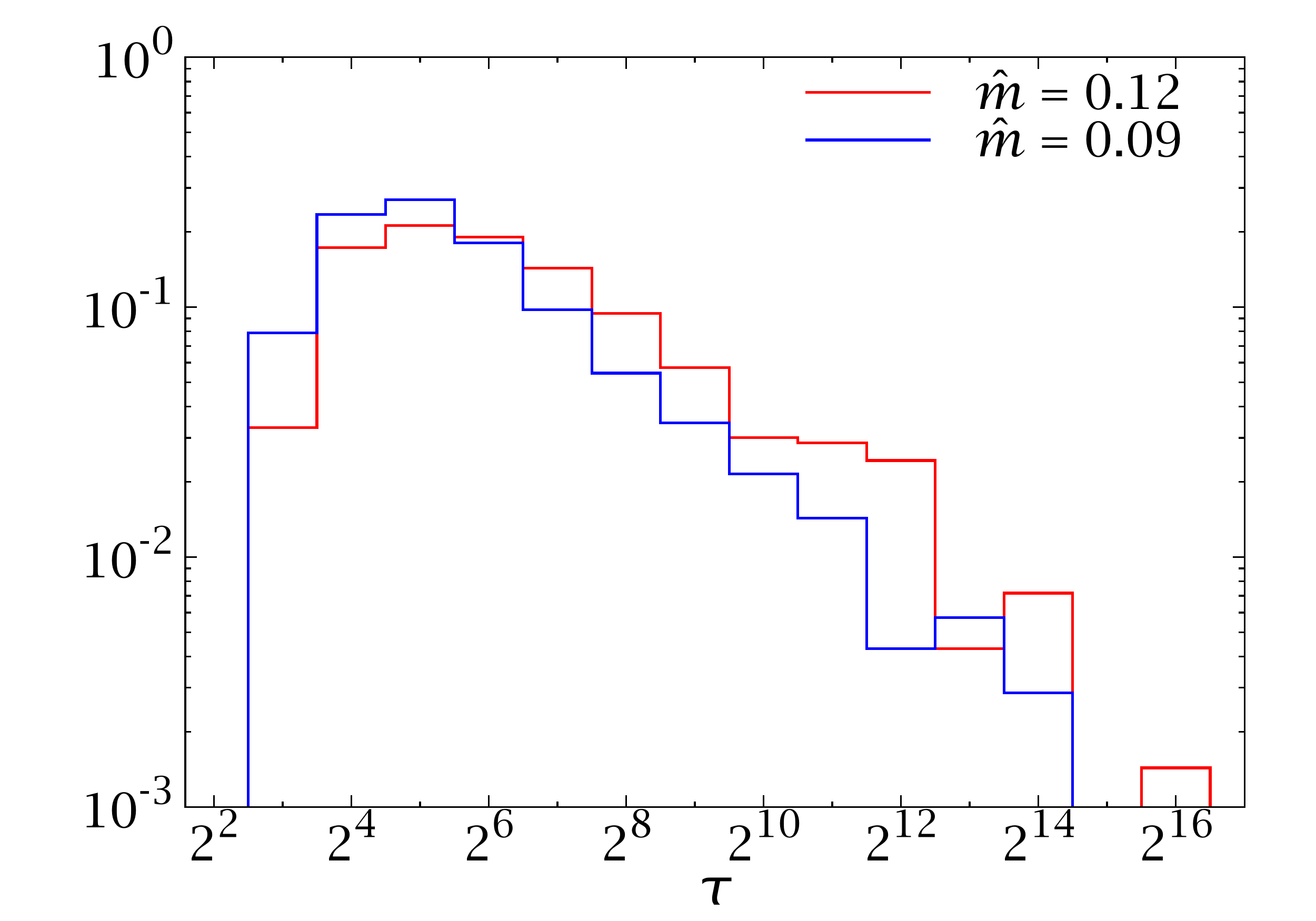}
\caption[Thermalisation times for $\hat m_\text{s}=0.8$ and different $\hat m$.]{Same as Fig.~\ref{fig:DAFF-histograma-hatm-012}, but now we consider
  several values of $\hat m$ for $\hat m_\mathrm{s}=0.8$ (in the hard
  thermalisation region far from the peak). The points for $\hat m=0.12$,
  closer to the critical point, exhibit a heavier long-time  tail (mind
  the vertical logarithmic scale).}
\label{fig:DAFF-histograma-hatms-08}
\end{figure}
We have assessed the thermalisation of our simulations through the 
autocorrelation times of the system, computed from the analysis
of the parallel tempering dynamics (see Section~\ref{sec:THERM-thermalisation-PT}).
We  require a simulation time longer than
$100\tau_\mathrm{int}$.\footnote{For most 
simulations $\tau_\text{int} \simeq \tau_\text{exp}$, so
this value is much larger than what is required 
to achieve thermalisation. This ample choice of minimum
simulation time protects us from the few cases where $\tau_\text{exp}$
is noticeably larger than $\tau_\text{int}$.
Cf. Section~\ref{sec:THERM-DAFF}.}
This process is only followed for $L\geq24$. For smaller
sizes we have simply made the minimum simulation 
time large enough to thermalise all samples.

The distribution of correlation times for our different samples
turns out to be dependent on the value of $(\hat m,\hat m_\mathrm{s})$.
Considering first the variation of the average $\tau_\mathrm{int}$
with $\hat m_\mathrm{s}$ at fixed $\hat m$, we see that the
peak of the $p(\hat m_\text{s}|\hat m)$ and its adjoining region 
are much easier to thermalise~(Figure~\ref{fig:DAFF-histograma-hatm-012}).
This region coincides with the only points that have a non-negligible
probability density~(Figure~\ref{fig:DAFF-p-hatms}), i.e., the only points 
that contribute to the computation of the $\overline{\langle O\rangle}_{\hat m}$.
This fact suggests a possible optimisation, that we will discuss in 
Section~\ref{sec:DAFF-optimization}.
\begin{figure}[p]
 \centering
\includegraphics[height=1.05\columnwidth,angle=270]{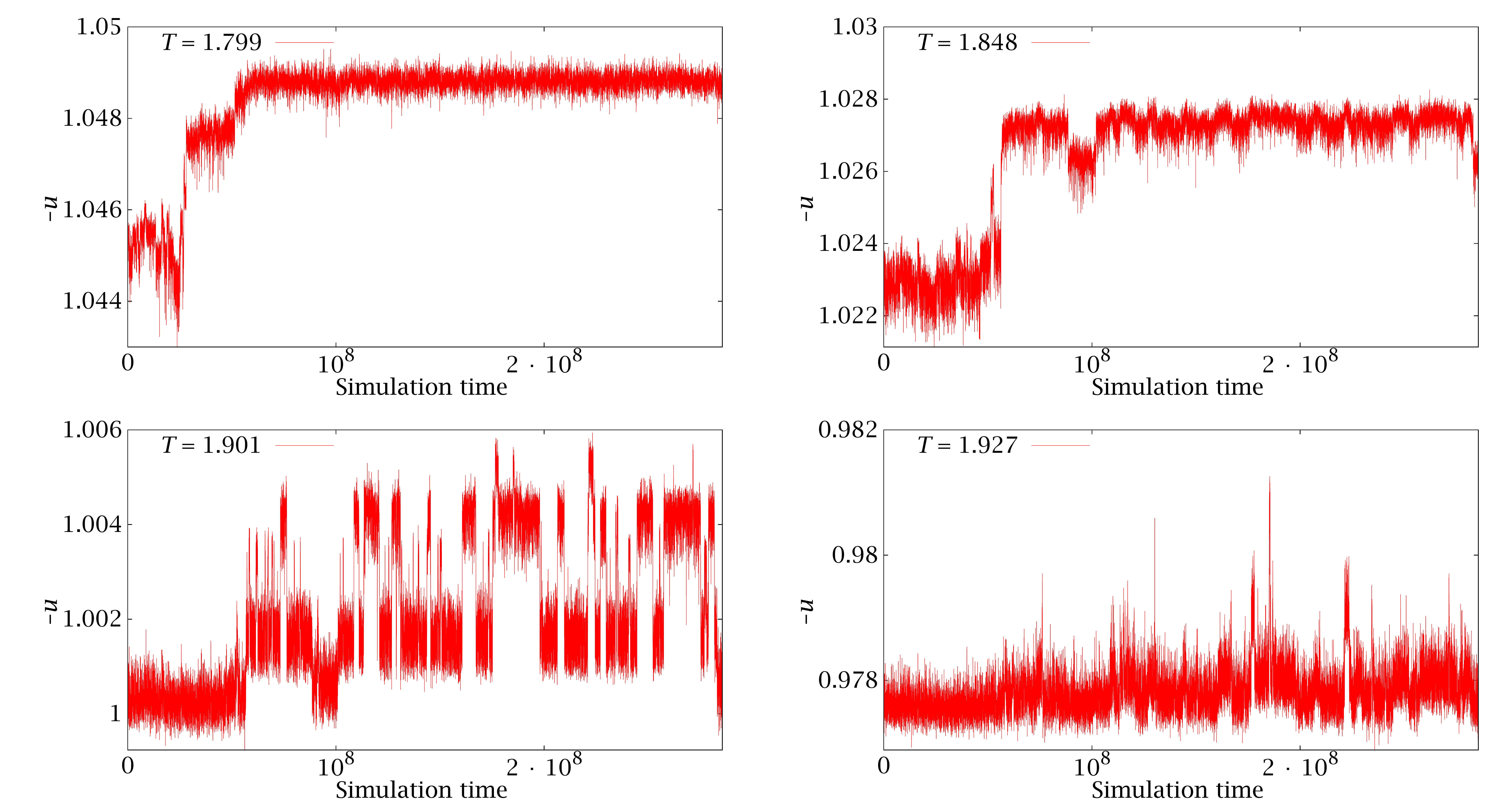}
\caption[Metastability in tethered simulations]{Time evolution (in number of parallel tempering steps) for the energy density $u$ 
for the same sample of an $L=32$ system  at $\hat m=0.12$, $\hat m_\text{s}=0.7$
and different temperatures.
For most temperatures the equilibrium value is quickly reached, but
for a very narrow temperature range there are several competing 
metastable states (bottom-left panel for $T=1.901$).
}
\label{fig:DAFF-metaestabilidad}
\index{metastability|indemph}

\centering
\includegraphics[height=.85\columnwidth,angle=270]{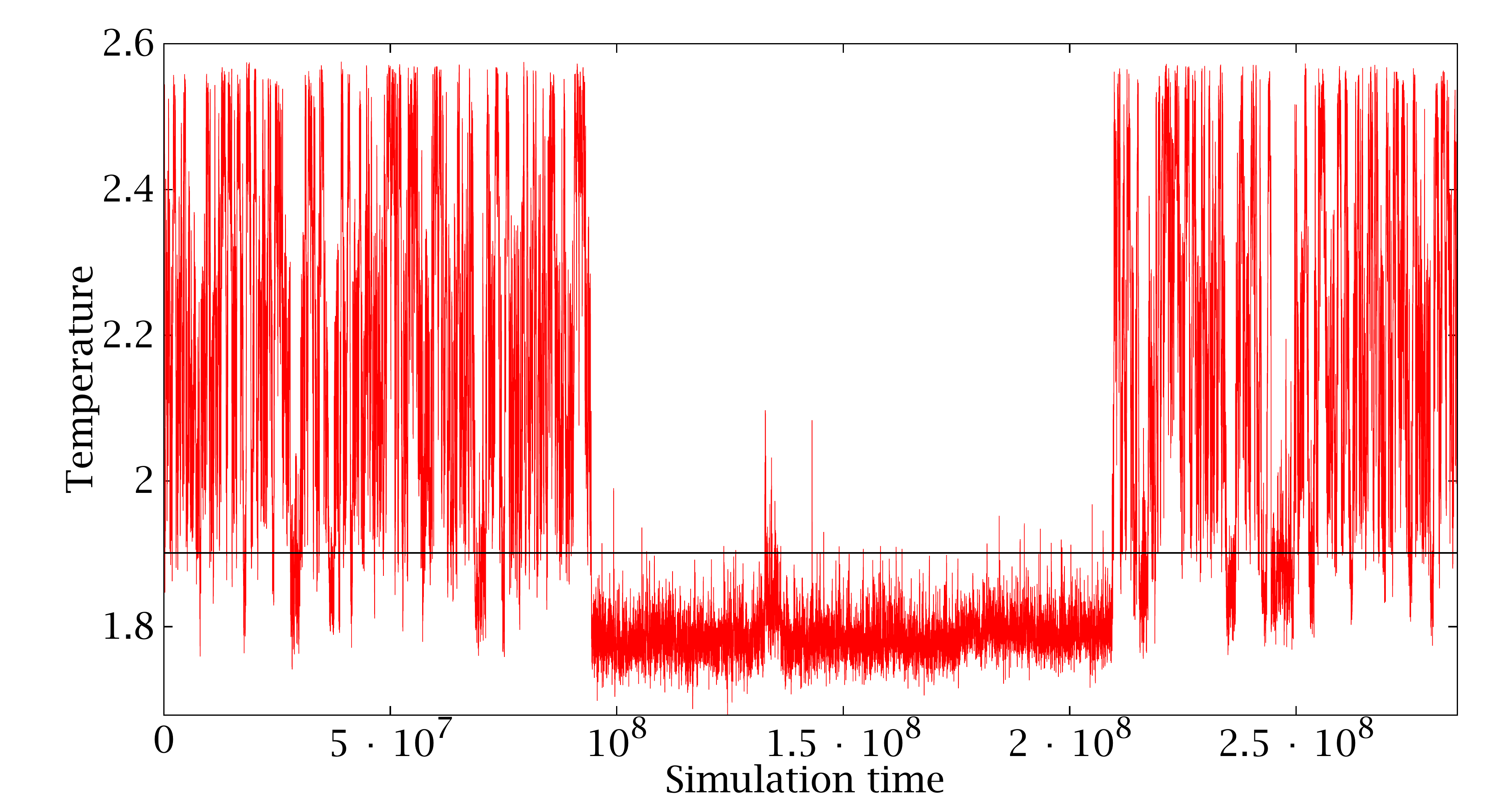}
\caption[Metastability and temperature random walk]{Temperature random walk for one of the 40 replicated systems for the
  parallel tempering simulation of the sample in
  Figure~\ref{fig:DAFF-metaestabilidad}.  The flow is blocked at the same
  temperature that had several metastable states (marked with a horizontal
  line), see bottom-left panel in Fig.~\ref{fig:DAFF-metaestabilidad}.}
\label{fig:DAFF-historia-metaestabilidad}
\index{parallel tempering|indemph}
\end{figure}

A second interesting result comes from studying the evolution of the $\tau_{\mathrm int}$ 
with $\hat m$. Figure~\ref{fig:DAFF-histograma-hatms-08} represents the histogram
of autocorrelation times for $\hat m_\mathrm{s}=0.8$ (in the `hard' region)
for two values of $\hat m$. The distribution for $\hat m=0.12$ has a much 
heavier tail. As we shall see, this is due to the onset of a 
phase transition.

The difficulty in thermalising some samples stems from the coexistence \index{metastability}
of several metastable states, even for fixed $(\hat m,\hat m_\mathrm{s})$.
In Figure~\ref{fig:DAFF-metaestabilidad} we represent the time evolution of the 
energy $u$ for several temperatures of the same sample ($L=32$, $\hat m=0.12$, 
$\hat m_\text{s}=0.8$). As we can see, for a narrow temperature range several
metastable states compete. This has a very damaging effect on the parallel
tempering dynamics, whose flow is hindered whenever a configuration that is metastable
for one temperature is very improbable in the next (see Figure~\ref{fig:DAFF-historia-metaestabilidad}).

\begin{figure}[h]
 \centering
\includegraphics[height=0.7\columnwidth,angle=270]{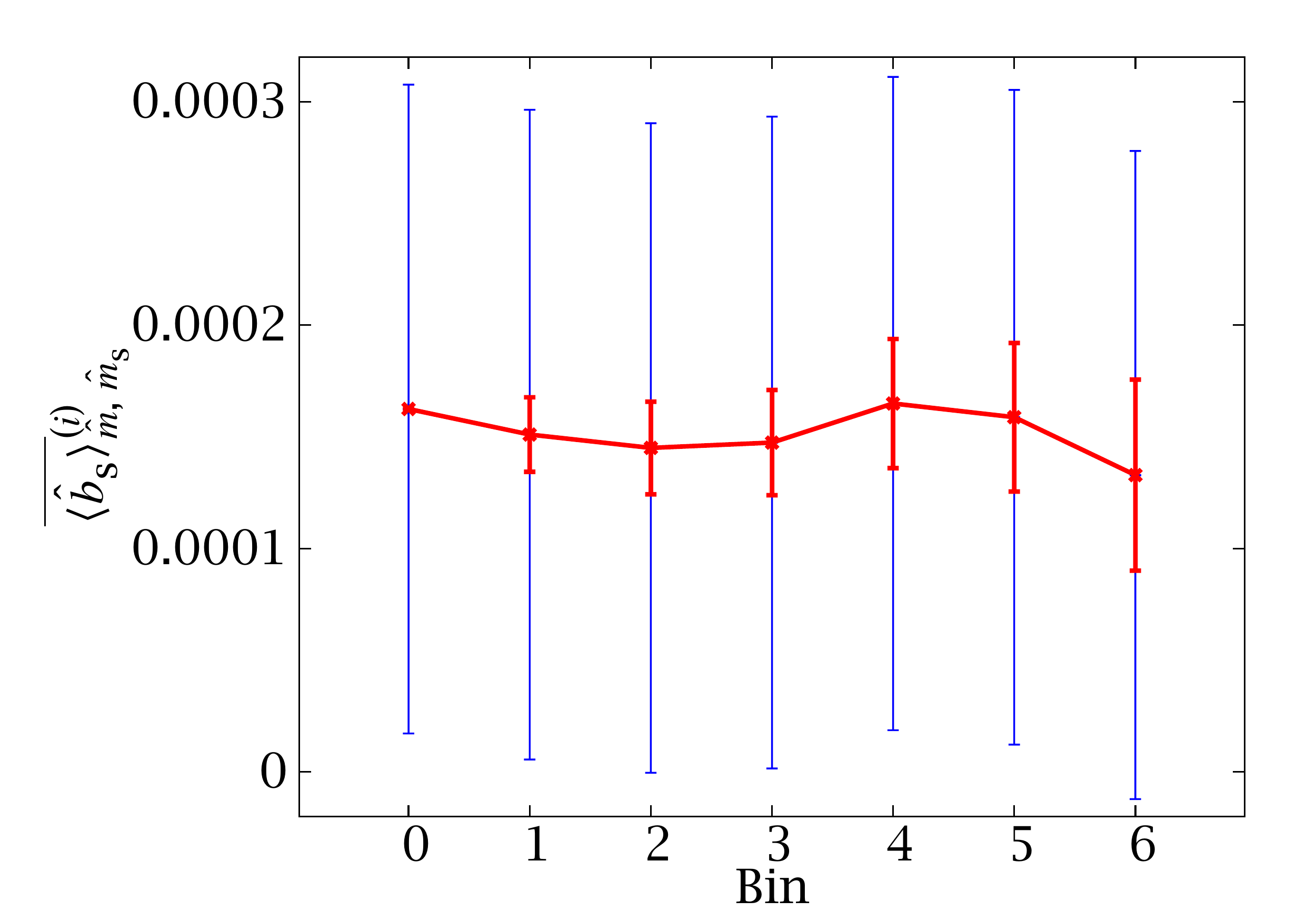}
\caption[$\log_2$-binning of the tethered field]{%
Time evolution of $\overline{\langle\hat b_\mathrm{s}\rangle}_{\hat m,\hat m_\mathrm{s}}$
for $\hat m=0.12$, $\hat m_\mathrm{s}=0.5$, in a logarithmic scale
(the first bin is the average over the second half of the simulation, the second
bin the average over the second quarter and so on, cf. Section~\ref{sec:THERM-disorder}).
This is in the middle of the $(\hat m,\hat m_\mathrm{s})$
region where thermalisation is hardest. The blue bars are the actual errors of our results
and the red bars mark the error in the correlated difference
of each bin with the first. Even for these reduced errors, we have several bins of convergence.}
\label{fig:DAFF-log2}
\index{log2 binning@$\log_2$-binning|indemph}
\end{figure}

For $L=32$, some points\footnote{By `point' we mean
any of the $\mN_\mathrm{samples}\times \mN_{\hat m}\times \mN_{\hat m_\mathrm{s}}$
individual tethered simulations for each $L$.} presented a metastability so severe
that enforcing a simulation time longer than $100\tau_\mathrm{int}$ 
would require a simulation of more than $10^9$ parallel-tempering 
updates (one thousand times longer than 
our minimum simulation of $\sim 10^6$ steps). Thermalising these points 
(which constitute about $0.1\%$ of the total) would have thus required
some $10^6$ extra CPU hours, with a wall-clock of many months. \index{wall-clock}
We considered this to be disproportionate to their physical
relevance (they are all restricted to a region far from the peaks  where the probability density
is $<10^{-40}$, see Figure~\ref{fig:DAFF-p-hatms}).
Therefore, we have stopped these simulations at about $10\tau_\mathrm{int}$. 
This is still a more demanding thermalisation criterion than 
is usual for disordered systems and
does not introduce any measurable bias in the physically relevant 
disorder-averaged observables. This can 
be checked in several ways:
\begin{itemize}
\item First of all, as we have already discussed,
these points are restricted to a region
in $\hat m_\mathrm{s}$ with probability density of at most $10^{-40}$.
Therefore, even if there were a bias it would not have any effect 
in the computation of canonical averages.
\item Even at the most difficult values of $(\hat m,\hat m_\mathrm{s})$ the
  \index{log2 binning@$\log_2$-binning}
  log$_2$-binning plot (the only thermalisation test typically used in
  disordered-systems simulations, see Section~\ref{sec:THERM-disorder})
   presents many logarithmic bins of stability
  (Figure~\ref{fig:DAFF-log2}, blue bars).  Even if we subtract the result of the last
  bin from the others, taking into account statistical
  correlations (see Section~\ref{sec:THERM-disorder} and \cite{fernandez:07}),
  several bins of stability remain
  (Figure~\ref{fig:DAFF-log2}, red bars). This is a very strict test, and one that
  even goes beyond physical relevance (because it reduces the errors
  dramatically from those given in the final results).
\end{itemize}

\section{The effective potential}\label{sec:DAFF-effective-potential}
\index{effective potential!DAFF|(}
\begin{figure}
\centering
\begin{minipage}{.49\columnwidth}
\includegraphics[height=\columnwidth,angle=270]{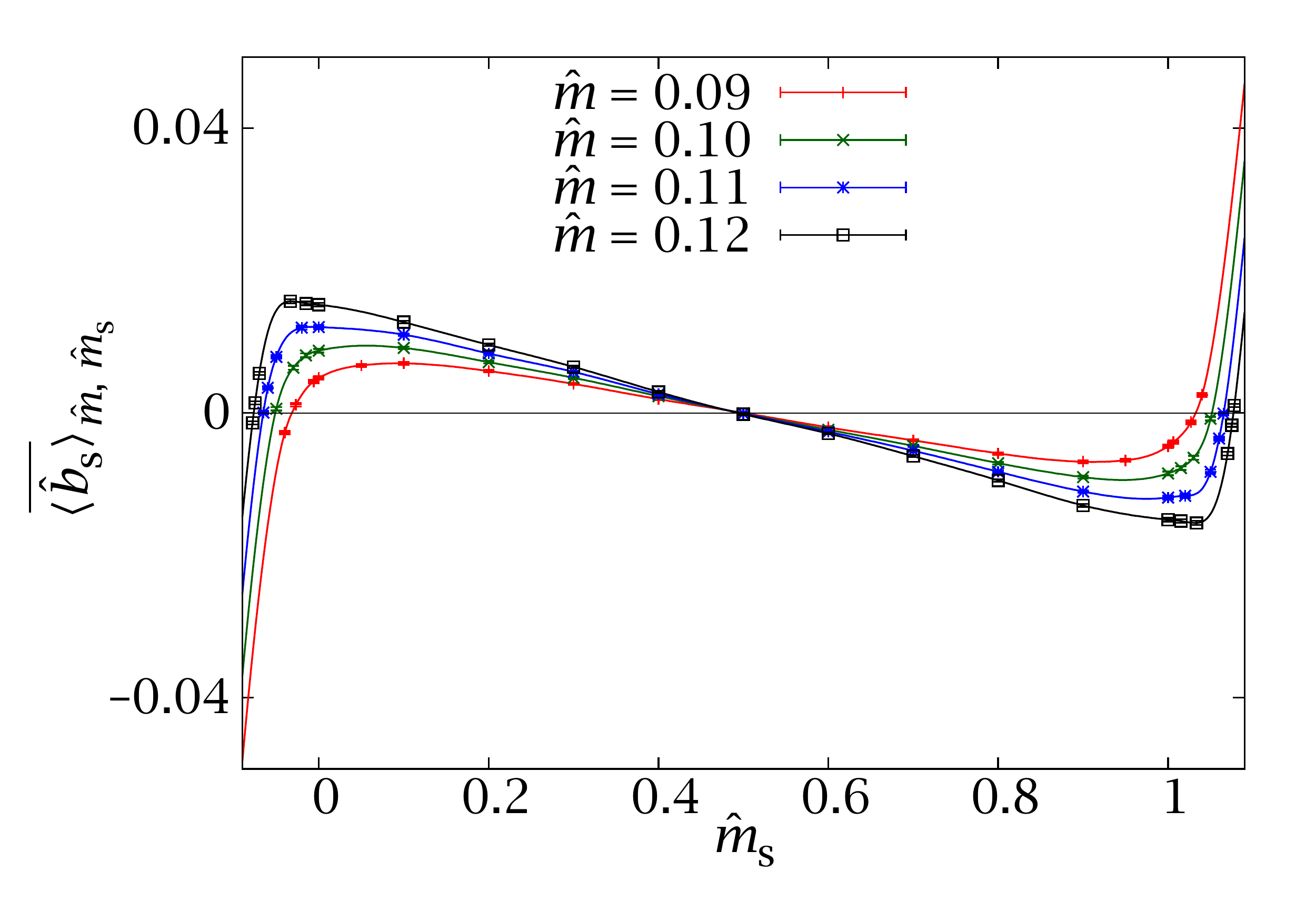}
\end{minipage}
\begin{minipage}{.49\columnwidth}
\includegraphics[height=\columnwidth,angle=270]{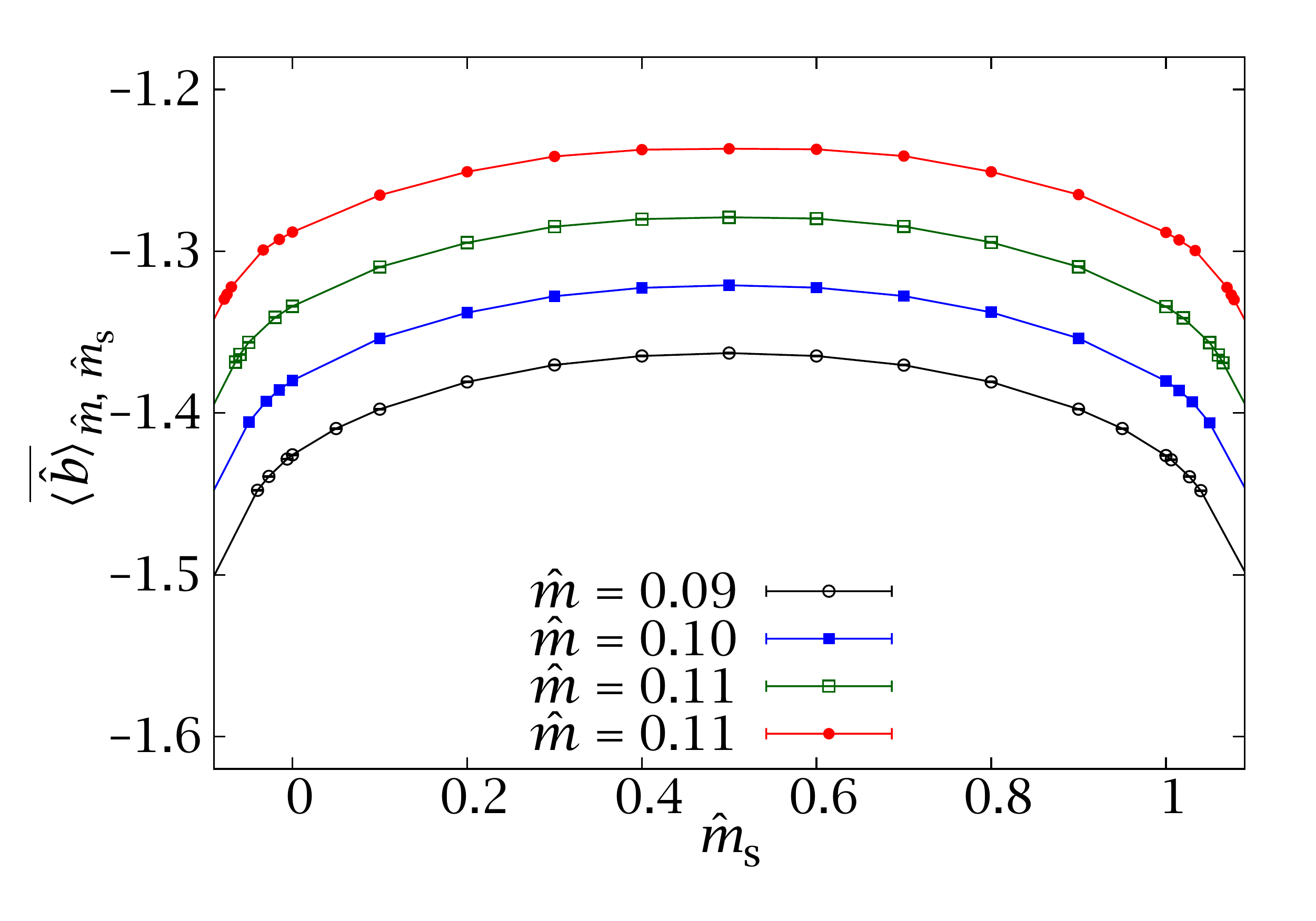}
\end{minipage}
\caption[Tethered field]{\emph{Left:} Staggered  component of the tethered field 
as a function of $\hat m_\text{s}$ for all the values of $\hat m$ in our 
$L=32$ simulations.
\emph{Right:} Same plot for the regular component of the tethered field.
}
\label{fig:DAFF-hatb-hatbs}
\index{tethered field!DAFF|indemph}
\end{figure}

As we discussed in detail in Section~\ref{sec:DAFF-disorder-average}, we can learn
much about the physics of the DAFF by studying the disorder-averaged 
saddle-point equations~\eqref{eq:DAFF-saddle-point-medio},  \index{saddle point}
which amounts to studying the disorder-averaged tethered field. 
To this end, we have plotted both components of the tethered field
in Figure~\ref{fig:DAFF-hatb-hatbs}. Let us first consider the staggered
component.
In the region we are considering, the corresponding equation has three solutions for each
$\hat m$: one in the paramagnetic region at $\hat m_\text{s}=0.5$ and
two symmetric ones
that get closer together as we lower $\hat m$.\footnote{%
The tethered field is actually not exactly symmetric in $\hat m_\text{s}$, 
there are deviations in the tails due to the way we have added the Gaussian demons
when constructing $\hat m_\text{s}$ in~\eqref{eq:DAFF-tethered-magnetic-field}.
However, we can consider the peak positions symmetric with great precision.}
Notice that $\overline{\langle\hat b\rangle}_{\hat m,\hat m_\text{s}}$ 
is negative, therefore when we lower $\hat m$ we are actually
increasing $|m|$.\footnote{%
The minus sign is an awkward remnant from the sign convention
for the tethered field used in~\cite{fernandez:09}, but inverted
in~\cite{fernandez:11b} after, alas, having finished the simulations.
We would obviously have obtained a positive range of $\hat b$ if we 
had simulated the $\hat m$ region for the opposite 
range of $m$.}

\begin{figure}[p]
\centering

\includegraphics[height=.7\columnwidth,angle=270]{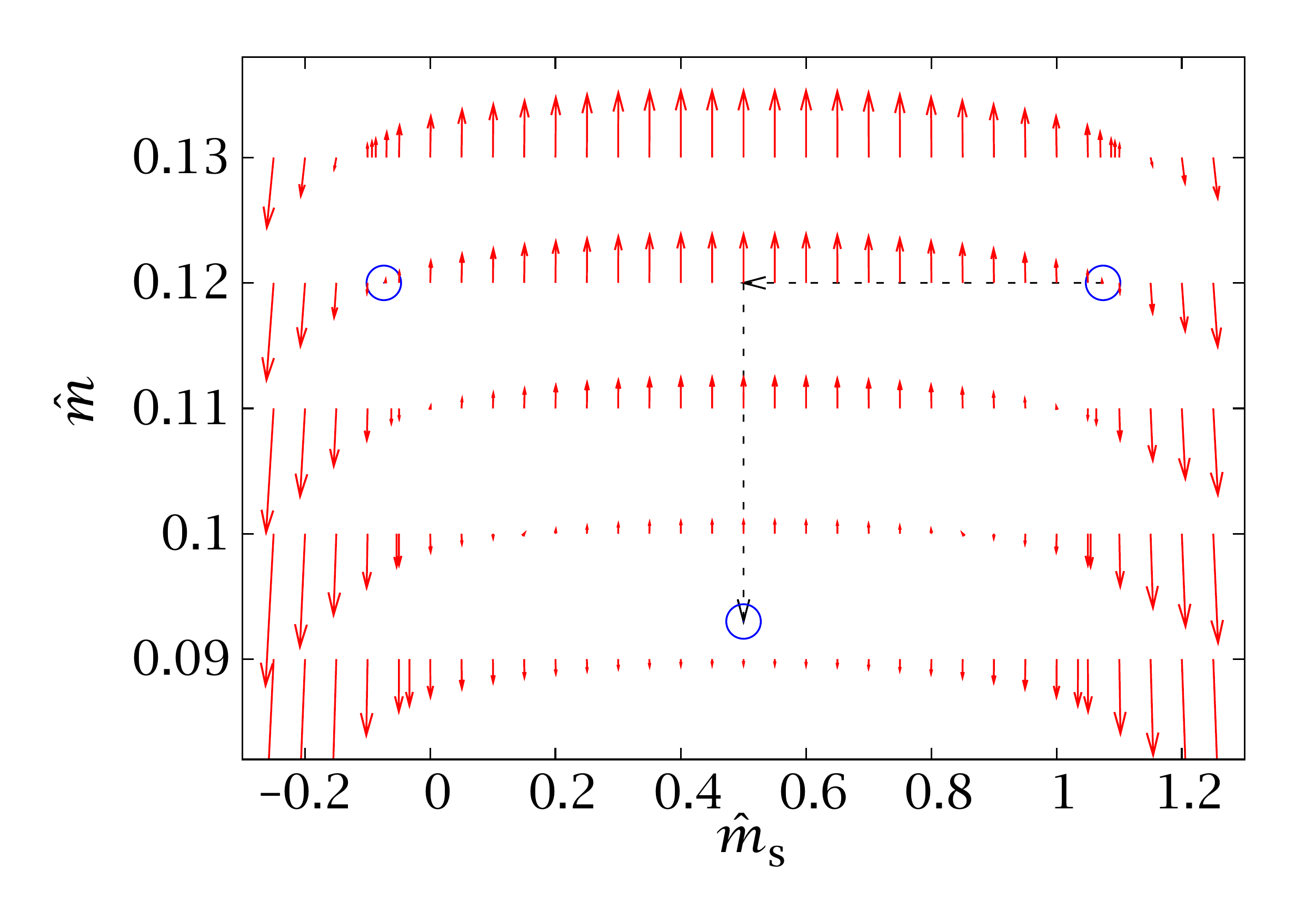}
\caption[Vector plot of the tethered field]{Plot 
of $\hat{\boldsymbol B}=\bigl(\overline{\braket{\hat b}}_{\hat m,\hat m_\text{s}}-\beta h
, \overline{\braket{\hat b_\text{s}}}_{\hat m,\hat m_\text{s}}\bigr)=\boldsymbol\nabla\varOmega^{(h)}$
for $\beta h = -1.33$. There are three stationary points (circles): two symmetric
antiferromagnetic  minima
at $\hat m\approx0.12$  and a paramagnetic saddle point at $\hat m\approx0.093$.
The dashed arrow marks the integration path that we will use in Section~\ref{sec:DAFF-theta}.
\label{fig:DAFF-flujos}}

\centering
\includegraphics[height=.7\columnwidth,angle=270]{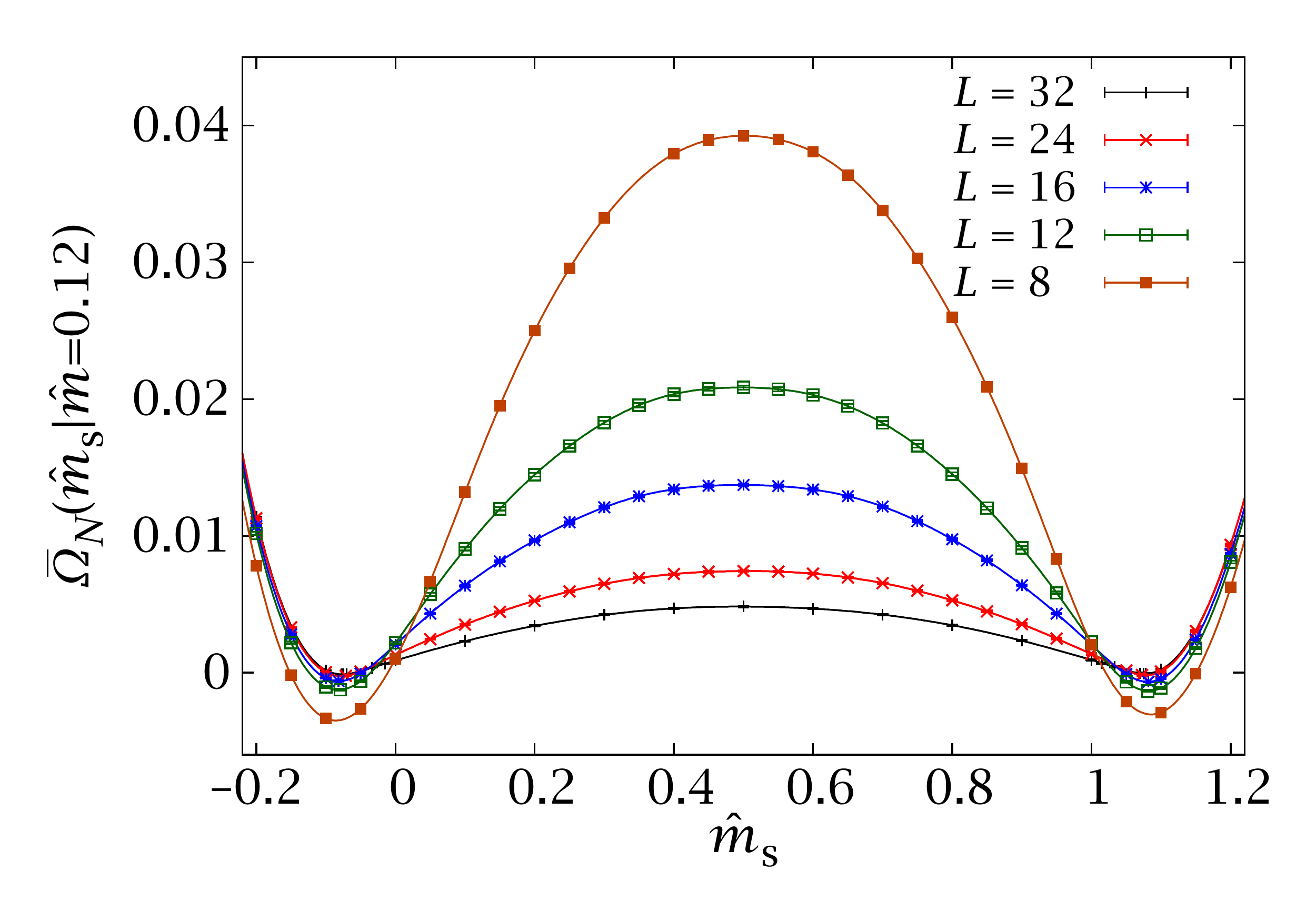}
\caption[Effective potential]{Effective potential $\overline\varOmega_N(\hat m_\text{s} | \hat m=0.12)$, Eq.~(\ref{eq:DAFF-Omega-hatms}).
 The errors can hardly be seen at this scale (they are smaller than the width
 of the lines for the larger sizes).}
\label{fig:DAFF-Omega-L}
\index{effective potential!DAFF|indemph}
\end{figure}

The other equation,
$\overline{\langle \hat b\rangle}_{\hat m,\hat m_\text{s}} = \beta h$, has two 
symmetrical solutions for each $\hat m$ (for values of $h$ in the range).
The resulting structure of the effective potential
is better understood with a two-dimensional plot of
$\hat{\boldsymbol B} = \bigl(\overline{\langle \hat b\rangle}_{\hat m,\hat m_\text{s}}-\beta h,
\overline{\langle \hat b\rangle}_{\hat m,\hat m_\text{s}}\bigr)$,
which is the gradient of $\overline\varOmega_N^{(h)}(\hat m,\hat m_\text{s})=\overline\varOmega_N(\hat m,\hat m_\text{s}) - \beta h \hat m$.
We have plotted this vector in Figure~\ref{fig:DAFF-flujos}, for 
$\beta h=-1.33$.\footnote{This choice seems arbitrary, but we shall
see later that it corresponds to the critical $h$ for this temperature.}
There are three stationary points (circles in the figure):
\begin{itemize}
\item Two in the antiferromagnetic region, symmetric with respect 
to $\hat m_\text{s}$ and at the same $\hat m$. These are minima
of the effective potential and, therefore, maxima of the probability
density $p(\hat m,\hat m_\text{s}; h)$
(cf. Figure~\ref{fig:DAFF-Omega-L}).
\item One in the paramagnetic region, at $\hat m_\text{s} = 0.5$ and
at a different value of $\hat m$ from the other two. This is a saddle 
point, it is a maximum in the $\hat m_\text{s}$ direction 
and a minimum in the $\hat m$ direction (by symmetry considerations, these
are its principal directions).
\end{itemize}

We can use this analysis to make a first qualitative assessment  of
the order of the DAFF's phase transition.
Indeed, one of the most important features of a first-order
transition is the emergence of metastability. The system, close
to the transition point, has two metastable states, one ordered
and the other disordered, characterised by peaks 
in the $p(\hat m,\hat m_\text{s};h)$. As we modify the magnetic 
field, the relative weight of the peaks changes, until,
at the transition point 
$p(\hat m^\text{ordered},\hat m_\text{s}^\text{ordered};h_\text{c})=p(\hat m^\text{disordered},\hat m_\text{s}^\text{disordered};h_\text{c})$.
This is equivalent to the Maxwell construction in a microcanonical setting. \index{Maxwell construction}
However, in our analysis we have found no peak in the paramagnetic region,
which would rule out the first-order scenario.

How, then, can we explain the metastable behaviour observed in previous
work? Actually, our preliminary 
study of the DAFF in the canonical ensemble in Section~\ref{sec:DAFF-canonical-simulations}
had the answer: the observed metastability is caused by the two coexisting 
antiferromagnetic peaks.

Let us consider $\overline{\varOmega}(\hat m_\text{s}|\hat m)$
in Figure~\ref{fig:DAFF-Omega-L}.
In a canonical simulation the system would spend most of its time
in one of the two antiferromagnetic minima, only very rarely tunnelling
from one to the other. Notice that, even though the barrier
$\Delta \overline\varOmega_N$ decreases with $L$, the escape time actually
goes as $\exp[N\Delta \overline\varOmega_N]$, thus causing an exponential \index{critical slowing down!exponential}
critical slowing down (cf. Figure~\ref{fig:DAFF-p-hatms}, showing
$\exp[N \Delta \overline \varOmega_N]\sim 10^{70}$ for an $L=32$ system).
The precise behaviour of these free-energy barriers is of paramount   \index{free energy!barriers}
importance to a characterisation of the physics and we shall dedicate
Section~\ref{sec:DAFF-theta} to its investigation.
\index{effective potential!DAFF|)}

\section{Finite-size scaling study of the phase transition}
The study of the effective potential in the 
previous section found no evidence of first-order behaviour in the DAFF, 
suggesting that its phase transition is continuous.
Of course, promoting this statement to more than a working
hypothesis would be premature, for several 
reasons. Not the least of which is that we have not yet shown that 
there actually is a phase transition in the studied $(\hat m,\hat m_\text{s})$
region. In the remaining sections we shall both find this phase
transition and characterise its properties. We shall first work at
fixed $\beta=0.625$ and find the phase transition that appears 
when varying $h$ and we shall then examine the temperature
behaviour at fixed $h$.
\subsection{Scale invariance}
One of the clearest signals of a continuous phase transition is the presence
\index{correlation length}
\index{scale invariance}
of scale invariance in the system.  Therefore, our first step 
is computing the correlation length.
\begin{figure}[p]
\centering
\includegraphics[height=.7\columnwidth,angle=270]{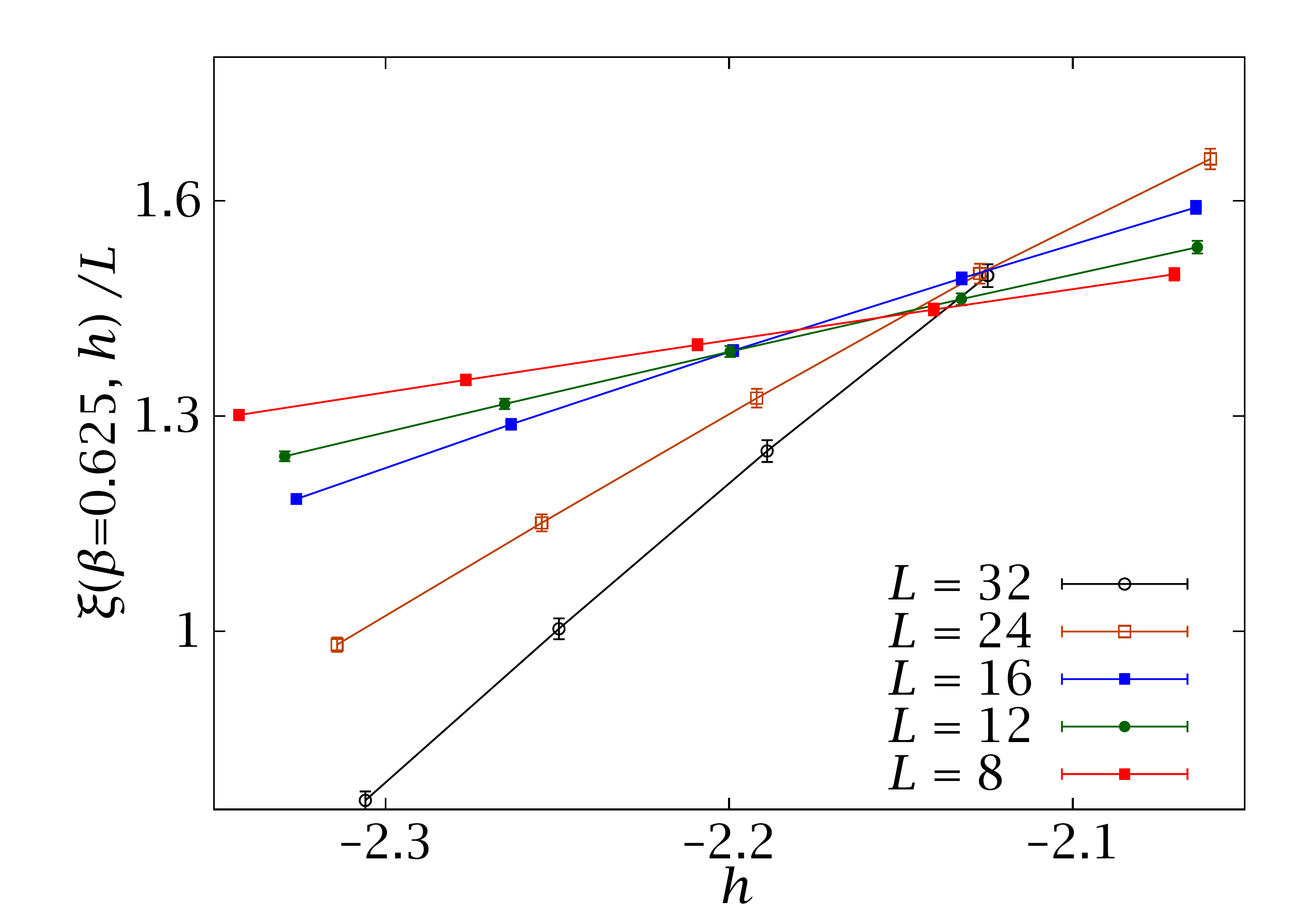}
\caption[Correlation length for $\beta=0.625$]{Correlation length in units of the system size 
as a function of the external magnetic field $h$, for $\beta=0.625$. The clear
crossings at $h\approx -2.13$ signal a second-order
phase transition.
\label{fig:DAFF-xi-h}
\index{correlation length!DAFF|indemph}}
\centering
\vspace*{1.5cm}

\includegraphics[height=.7\columnwidth,angle=270]{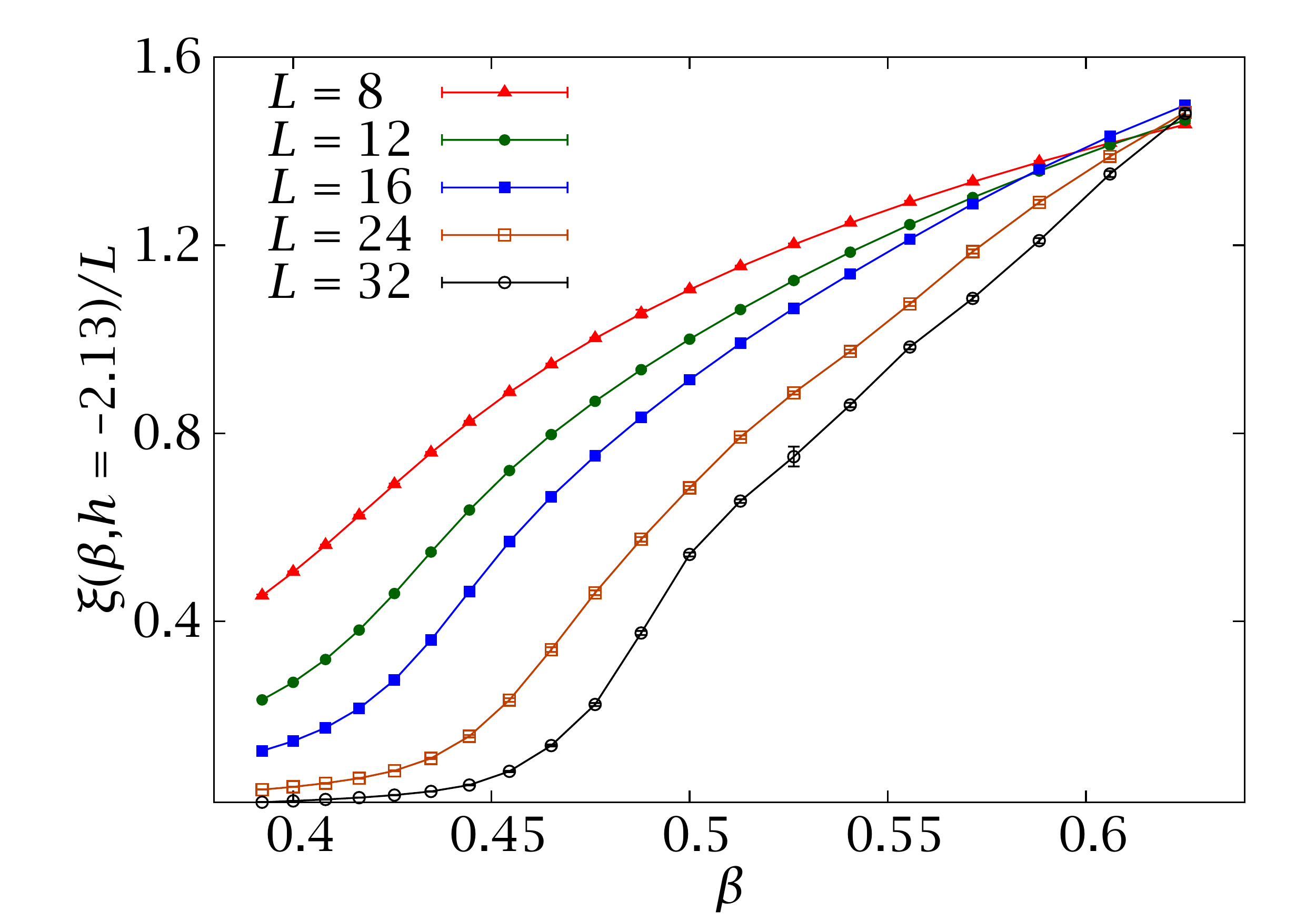}
\caption[$\xi/L$ as a function of $\beta$ for $h=-2.13$.]{Plot of
$\xi/L$ as a function of the inverse temperature
$\beta$ for $h=-2.13$. This is the approximate value 
of the critical field at $\beta=0.625$ (cf. Figure~\ref{fig:DAFF-xi-h}),
which explains the convergence at the end of our temperature
range.}
\label{fig:DAFF-xi-beta}
\end{figure}

Recalling definition~\eqref{eq:DAFF-phi-k} of the staggered Fourier transform
we define the following propagator\index{correlation function (equilibrium)!spatial}
\begin{equation}\label{eq:DAFF-F-k}
F_{\hat m}(\bk) = N \overline{\langle \phi(\bk) \phi(-\bk)\rangle}_{\hat m},
\end{equation}
which we use to define the second-moment correlation length
(explained in Section~\ref{sec:ISING-model})
\index{correlation length!DAFF}
\begin{equation}\label{eq:DAFF-xi}
\xi_2(h) =  \frac{1}{2\sin(\uppi/L)}\left[ \frac{F_{\hat m(h)}(0)}{F_{\hat m(h)}(\bk_\mathrm{min})}-1\right]^{1/2}.
\end{equation}
Notice that we write $\xi_2(h)$ instead of $\xi_2(\hat m)$, following
our discussion of ensemble equivalence in Section~\ref{sec:DAFF-disorder-average}.

The result of this computation is plotted in Figure~\ref{fig:DAFF-xi-h}.
In it we can see a clear occurrence of scale invariance, signalling
the onset of a second-order phase transition at $h\approx -2.13$.
As a welcome bonus, the 
dependence of $\xi$ on $h$ is linear, which will simplify the analysis.

We can obtain a complementary picture by  fixing the magnetic field and 
studying the temperature dependence. In Figure~\ref{fig:DAFF-xi-beta}
we show this for $h=-2.13$. For each temperature and  lattice, $\xi(h=-2.13)$
is obtained by linear interpolation in a plot analogous to Figure~\ref{fig:DAFF-xi-h}.
This value of $h$ causes the critical temperature
to appear at the very end of our simulated range. The reason for this choice
is that the $\hat m_\text{s}$ grid for each $\hat m$ was optimised
for $\beta=0.625$. As we move away from this temperature
the peaks are only irregularly sampled, which causes larger statistical errors 
and potentially discretisation biases. 
\index{tethered formalism!sampling}

\section{Computation of critical exponents}\label{sec:DAFF-quotients}
From the previous plots we can perform a
Finite-Size Scaling~\cite{amit:05,zinn-justin:05} analysis
in order to determine the critical exponents.
As we discussed in Section~\ref{sec:DAFF-phase-transition},
we need three independent
critical exponents in order fully to characterise the phase transition
of the DAFF, which we can choose as $\nu$, $\bar\eta$, $\theta$. In 
the following sections we shall compute these three exponents and then 
use additional observables to check the scaling and hyperscaling relations.

\index{quotients method}
\index{finite-size scaling}
We shall apply the quotients method~\cite{ballesteros:96},
which is  based on Nightingale's phenomenological
renormalisation~\cite{nightingale:75}.
Let us consider a system that experiences a second-order phase
transition at a critical point $X_\text{c}$ (where the parameter $X$
is, in our case, either the applied magnetic field $h$ or the 
temperature). If some physical observable $O$
behaves as $(X-X_\text{c})^{-y_o}=x^{-y_o}$ near the transition point, then
we expect the following dependence on a finite lattice (cf. Section~\ref{sec:INTRO-FSS}),
neglecting corrections to scaling,
\begin{equation}\label{eq:DAFF-FSS1}
O(L,x) = L^{y_o/\nu_x} f_O( xL^{1/\nu_x}).
\end{equation}
In this formula $\nu_x$ is the exponent of the correlation length,
$\xi\propto x^{-\nu_x}$ and $f_O$ is a smooth universal scaling 
function.

In order to define the quotients method, we write~(\ref{eq:DAFF-FSS1}) 
in the following form
\begin{equation}
O(L,x) = L^{y_o/\nu_x} F_O( \xi(L,x)/L).
\end{equation}
Now, for a pair of lattice sizes $L_1$ and $L_2=sL_1$ we 
identify the single value of $x$ such that 
$\xi(L_1,x^*)/L_1 =\xi(L_2,x^*)/L_2$. At this point, we have
\begin{equation}\label{eq:DAFF-quotients}
\frac{O(sL,x^*)}{O(L,x^*)} = s^{y_o/\nu_x}.
\end{equation}

In Table~\ref{tab:DAFF-quotients} we have applied this method to 
several observables, using $s=2$,
\begin{eqnarray}
\partial_x \xi &\quad\longrightarrow\quad&  y_o = \nu_x+1,\label{eq:DAFF-nu}\\
M_\mathrm{s}^2  &\quad\longrightarrow\quad&  y_o = \bar\gamma = 2\beta-3\nu_x,
\label{eq:DAFF-beta}\\
\partial_x M &\quad\longrightarrow\quad&  y_o = \alpha.\label{eq:DAFF-alpha}
\end{eqnarray}
In the following sections we shall analyse each critical exponent separately
and compare the results with previous numerical and experimental work
and with analytical conjectures.
\begin{table}
\small
\centering
\begin{tabular*}{\columnwidth}{@{\extracolsep{\fill}}clllll}
\toprule
 $L$ & \multicolumn{1}{c}{$h^*(L)$} & 
\multicolumn{1}{c}{$\beta/\nu_h$}&\multicolumn{1}{c}{$\nu_h$} & 
\multicolumn{1}{c}{$\alpha/\nu_h$}& 
\multicolumn{1}{c}{$\nu_\beta$}   \\
\toprule
8 & $-2.178(4)$ & 0.0125(7) & 0.887(5)& 0.0765(25)  & 1.07[7](5) \\
12& $-2.140(5)$ & 0.0104(5) & 0.790(9)& 0.0781(27)  & 1.01[3](3) \\
16& $-2.123(3)$ & 0.0119(4) & 0.742(7)& 0.224(4)    & 1.10[13](7) \\ 
\bottomrule
\end{tabular*}
\caption[Computation of the critical exponents with the quotients method]{Computation of the critical exponents with the 
quotients method, applied to pairs of lattices $(L,2L)$.
The first four columns give results from the quotients 
at $h^*(L)$ for fixed $\beta=0.625$. The last column
gives results for $\beta^*(L)$ at fixed $h=-2.13$.
For this last value we give two error bars, distinguishing
systematic and statistical errors.}
\index{quotients method|indemph}
\index{critical exponent!nu@$\nu$|indemph}
\index{critical exponent!beta@$\beta$|indemph}
\index{critical exponent!alpha@$\alpha$|indemph}
\label{tab:DAFF-quotients}
\end{table}

\subsection[Exponents $\nu$ and $\beta$]{Exponents \boldmath$\nu$ and $\beta$}\label{sec:DAFF-nu}
As a first application of the quotients method, we can compute
\index{critical exponent!nu@$\nu$}
the correlation length critical exponent, using
equations~(\ref{eq:DAFF-quotients}) and~(\ref{eq:DAFF-nu}).
This exponent can be computed either from $\partial_h \xi$
(Figure~\ref{fig:DAFF-xi-h}) or from $\partial_\beta \xi$ (Figure~\ref{fig:DAFF-xi-beta}).

In the first case, since the $h$ dependence is linear in all our simulated
range, we simply fit $\xi(h)$ to a straight line and approximate
$\partial_h \xi$ with its slope (as usual, we perform a different 
fit for each jackknife block) \index{jackknife method}.

As to the temperature dependence, the analysis is more delicate for 
two reasons. The first is that $\xi(\beta)$ is no longer a straight line, 
the smaller sizes showing a clear curvature even very close to the 
intersection point. The second is that, as discussed above, the value 
of the correlation length for temperatures other than $\beta=0.625$ has some
chaotic fluctuations due to discretisation error. Therefore, even if
the data are correlated (as is the case for parallel tempering simulations), \index{parallel tempering}
computing the temperature derivative is difficult. 

Our solution has been to fit the high-$\beta$ region to a quadratic
polynomial for $L\leq16$ and to a straight line for $L\geq24$. The 
aforementioned systematic effects cause the derivative to depend heavily
on the fitting range. We have accounted for this effect in
Table~\ref{tab:DAFF-quotients} by giving a second error bar, in square brackets.
\begin{figure}[h]
\centering
\begin{minipage}{.49\linewidth}
\includegraphics[height=\columnwidth,angle=270]{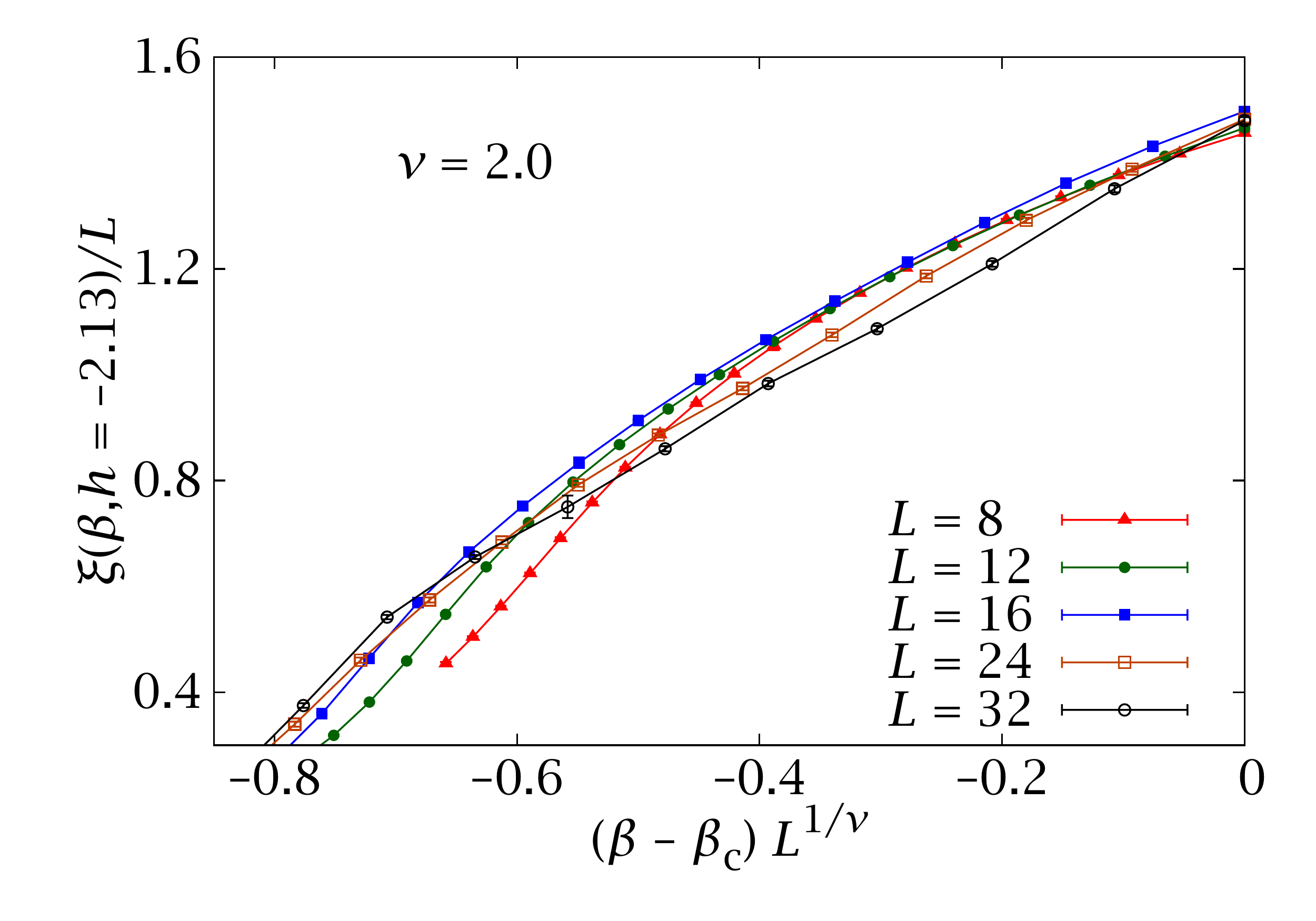}
\end{minipage}
\begin{minipage}{.49\linewidth}
\includegraphics[height=\columnwidth,angle=270]{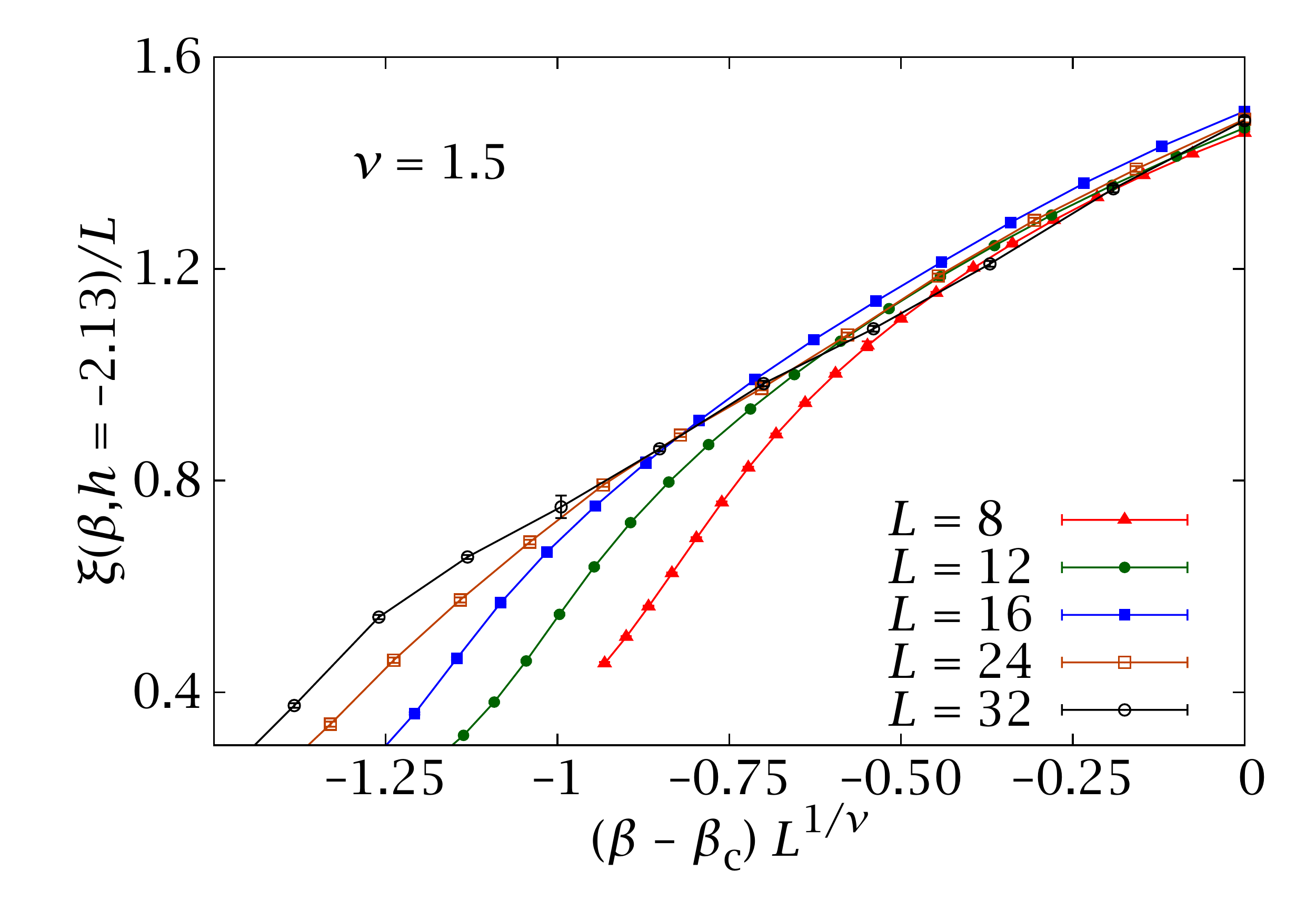}
\end{minipage}
\begin{minipage}{.49\linewidth}
\includegraphics[height=\columnwidth,angle=270]{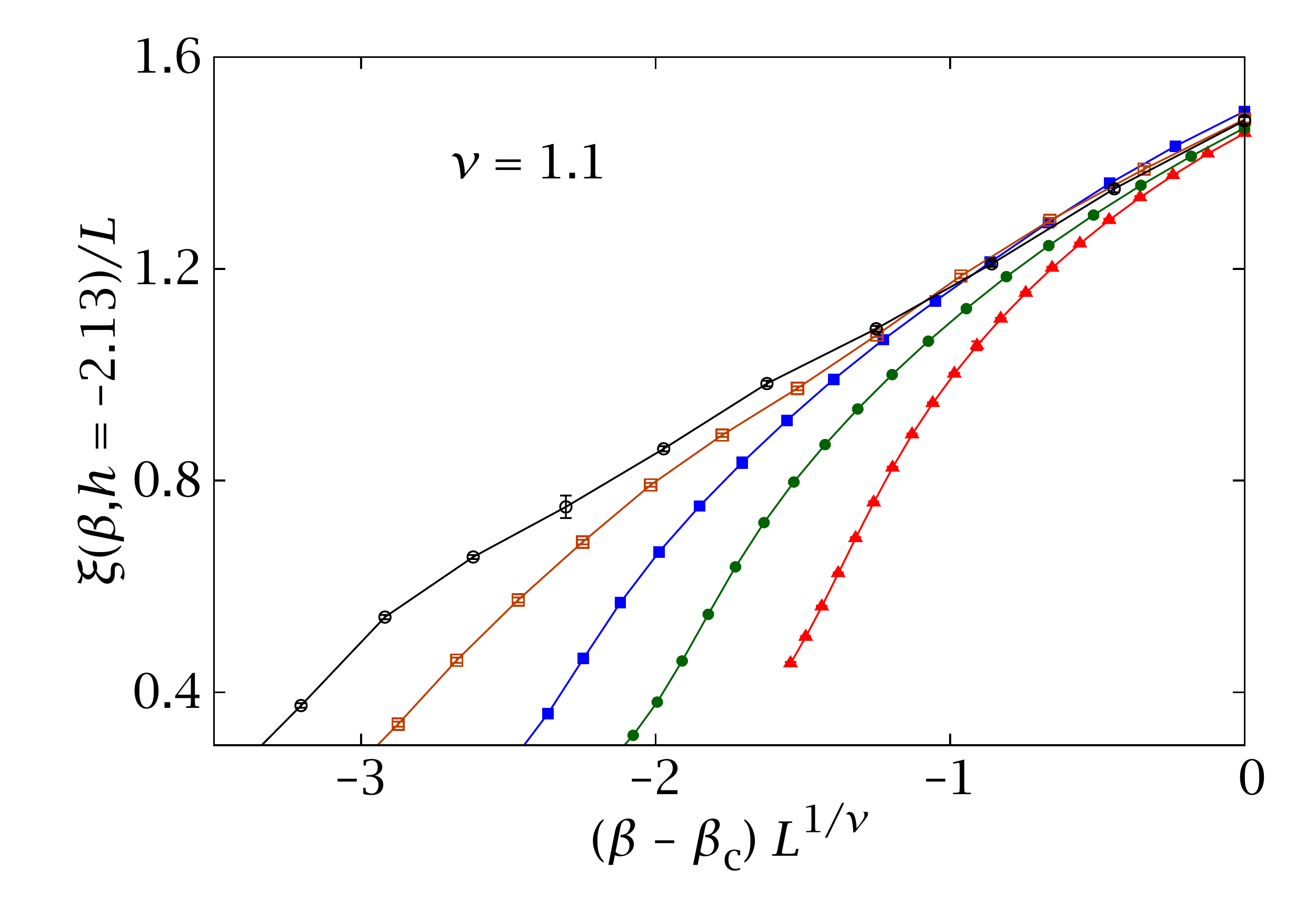}
\end{minipage}
\begin{minipage}{.49\linewidth}
\includegraphics[height=\columnwidth,angle=270]{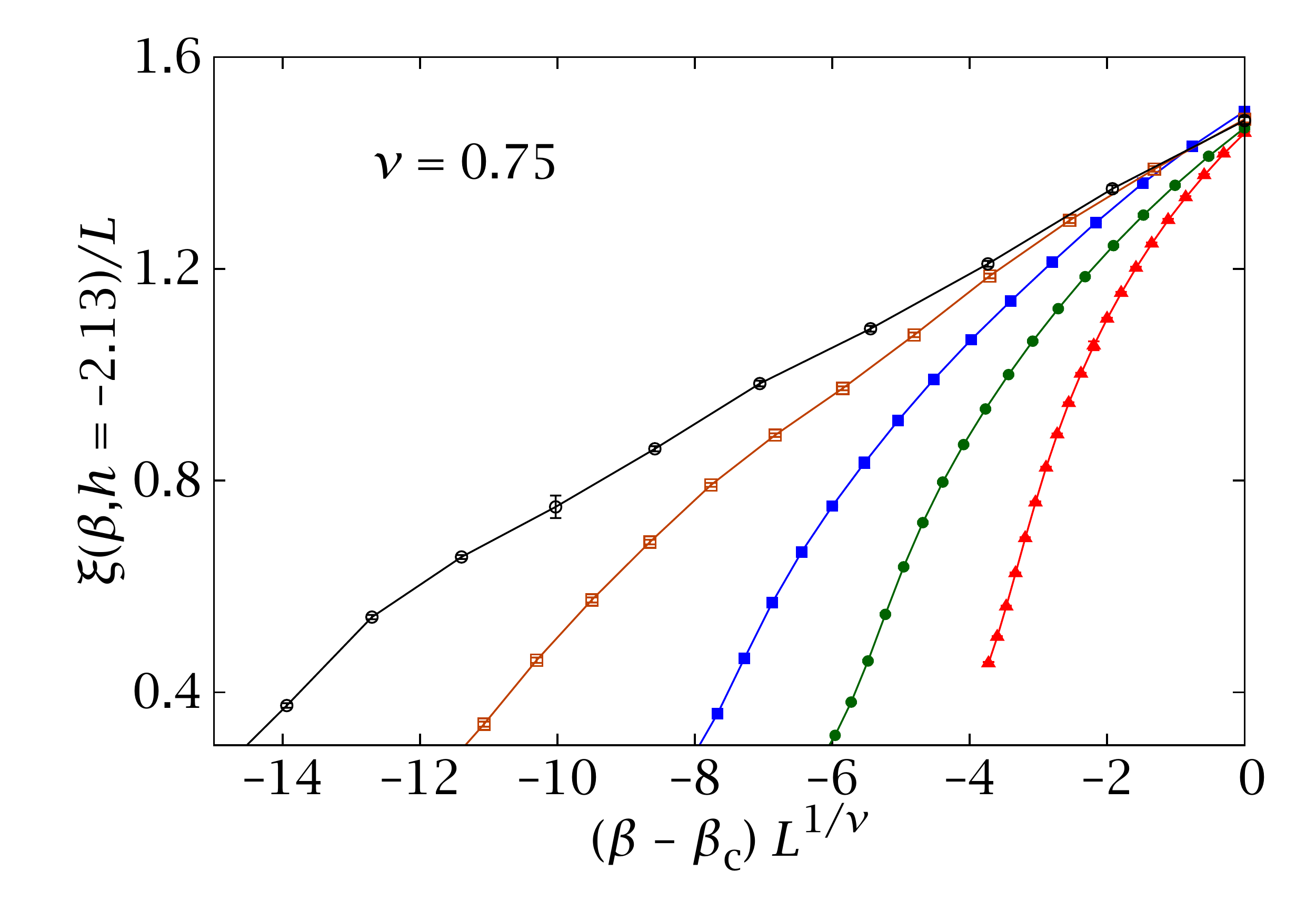}
\end{minipage}
\caption[Scaling plots of $\xi/L$]{Scaling plots
of $\xi/L$ for several values of the thermal 
critical exponent $\nu$. A very high value of $\nu$
causes a misleading collapse for low $L$, while
our preferred values (bottom panels) are better at
collapsing the larger sizes close to the intersection
points.
\index{finite-size scaling|indemph}
\index{correlation length!DAFF|indemph}
\label{fig:DAFF-xi-FSS}}
\end{figure}

Notice that exponents $\nu_h$ and $\nu_\beta$ should coincide, but we obtain $\nu_h\approx 0.75$
and $\nu_\beta\approx 1.05$. This discrepancy is mainly due to 
large corrections to scaling.\index{finite-size scaling!corrections}
In general, as discussed in the introduction 
to this chapter, attempts to determine $\nu$ have yielded values 
in a very broad range. We can see the reason in Figure~\ref{fig:DAFF-xi-FSS},
where we show scaling plots of $\xi(\beta)$ for values of $\nu$ in a very broad
range. For each $\nu$ there is a temperature range that shows a seemingly
good scaling. Therefore, attempting to estimate $\nu$ as the exponent 
that produces the best collapse, a method frequently employed for this
and other models, is not only imprecise but also dangerous. 
By precisely locating the critical point and using the quotients method \index{quotients method}
we minimise the effect of scaling corrections (but we do not eliminate them completely,
as evinced by the incompatible values of $\nu_h$ and $\nu_\beta$).

Notwithstanding these difficulties, notice that our range 
of values for $\nu$ is in good agreement to that defined 
by the best experimental estimates: 
from $\nu=0.87(7)$~\cite{slanic:99} 
to $\nu=1.20(5)$~\cite{ye:04}.

\index{critical exponent!beta@$\beta$}
We can determine the critical exponent $\bar\gamma/\nu$ by applying 
the quotients method to $\overline{\langle M_\text{s}^2\rangle}(h)$.
The results are quoted in Table~\ref{tab:DAFF-quotients}. According to the scaling
and hyperscaling relations (Section~\ref{sec:DAFF-phase-transition}),
$\bar\gamma/\nu = 4-\bar\eta = 3-2\beta/\nu$.
Our resulting value of $\beta/\nu$ is very low, but different from zero. 
Recall that this exponent controls the evolution of the peaks in $p(\hat m_\text{s};h)$
(cf. Sections~\ref{sec:ISING-betanu} and~\ref{sec:CLUSTER-results}).
In Figure~\ref{fig:DAFF-theta} we can see that, indeed, the zero in $\overline{\langle\hat b_\text{s}\rangle}_{\hat m,\hat m_\text{s}}$
shifts slowly but surely with increasing~$L$.
The figure is for fixed $\hat m=0.12\approx\hat m(h_\text{c})$. 

The extremely small value of $\beta/\nu$ is one of the reasons for 
past claims of a first-order phase transition and is consistent with
previous numerical studies. 

\subsection[The hyperscaling violations exponent $\theta$]{The hyperscaling violations exponent \boldmath $\theta$}\label{sec:DAFF-theta}
\begin{figure}
\centering
\includegraphics[height=.7\columnwidth,angle=270]{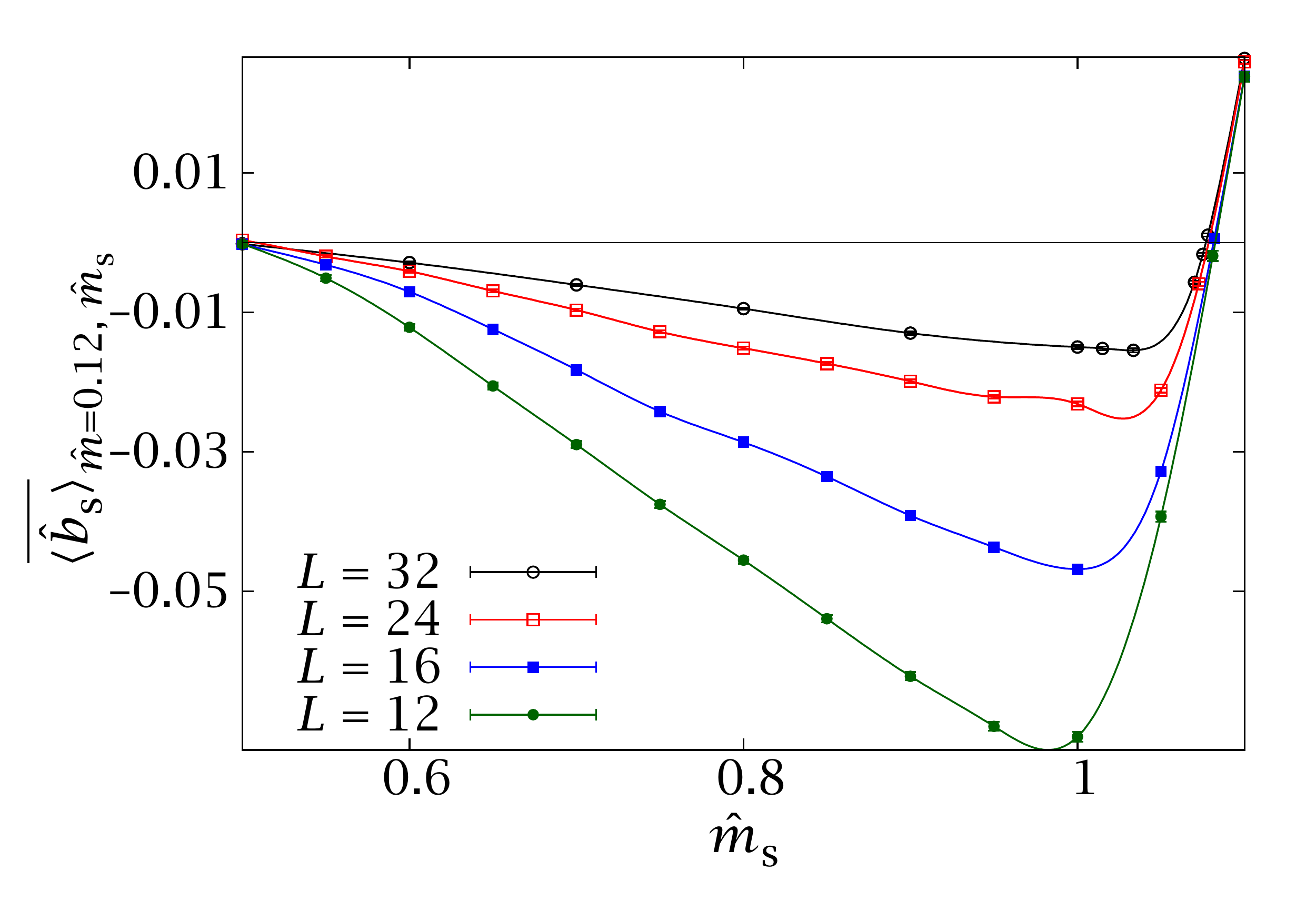}
\caption[Staggered component of the tethered field]{Plot
of the staggered component of the tethered magnetic field,
$\overline{\langle \hat b_\mathrm{s}\rangle}_{\hat m,\hat m_\mathrm{s}}$
for $\hat m=0.12\approx \hat m(h_\mathrm{c})$ at $\beta=0.625$.
This is a close-up of Figure~\ref{fig:DAFF-hatbs-samples}---bottom,
restricting the $\hat m$ range and eliminating the curve for $L=8$
in order better to see the $L$-evolution.
The area under the curve between the two zeros
gives the quantity $\Delta \overline F_1$, Eq.~(\ref{eq:DAFF-delta-F1}), which 
we can use to compute $\theta$. Notice the slow shift inwards of 
the antiferromagnetic zero, which marks the position
of the peak in $p(\hat m_\text{s}|\hat m)$.}
\label{fig:DAFF-theta}
\end{figure}
The third and last independent critical exponent is $\theta$, which gives
a measure of the violations to hyperscaling, Eq.~(\ref{eq:DAFF-modified-hyperscaling}).
This exponent cannot be obtained from the canonical averages of 
physical observables at $h_\text{c}$, as we did for $\beta$ and $\nu$.
Rather, its computation is an intrinsically tethered operation.
Following Vink et al.~\cite{vink:10,fischer:11}, we can relate $\theta$ to
the free-energy barrier between the two coexisting states at the critical
point,
\begin{equation}\label{eq:DAFF-deltaF-theta}
\Delta \overline F_N\propto L^{\theta-D},
\end{equation}
in this equation $\overline F_N$ is the free-energy density,
 as in equation~\eqref{eq:INTRO-legendre}.
\index{free energy}
This is a specially interesting quantity to measure, since in a 
first-order transition $\theta\geq D-1$. 
Indeed, in a first order 
scenario the system would tunnel between the two pure phases (paramagnetic and
antiferromagnetic) by building an interface. The free-energy barrier would then
be associated to a surface tension. This in turn would be proportional
to the dimension of the interface's surface, which must be at least $D-1$.
\index{interface}

We first notice that $\Delta \overline F_N$ is defined 
as the free-energy 
barrier between the two phases (disordered and antiferromagnetic) 
that form at the critical point $(\beta,h_\mathrm{c})$.
Recalling our study  of section~\ref{sec:DAFF-effective-potential}
\index{effective potential} we see that this is equivalent 
to computing the $\Delta \overline\varOmega_N^{(h)}$  between the paramagnetic saddle
point and one of the antiferromagnetic minima.  We denote these
two points by $(\hat m^{(1)}, \hat m^{(1)}_\mathrm{s})$ and 
$(\hat m^{(2)}, \hat m^{(2)}_\mathrm{s})$, respectively,
\begin{equation}
\Delta \overline \varOmega_N^{(h)} = \overline \varOmega_N^{(h)}(\hat m^{(1)}, \hat m^{(1)}_\mathrm{s})-\overline \varOmega_N^{(h)}(\hat m^{(2)}, \hat m^{(2)}_\mathrm{s}) .
\end{equation}

As always, this potential difference 
is simply the line integral of 
\begin{equation}
\hat{\boldsymbol B} = \bigl( \overline{\langle \hat b\rangle}_{\hat m,\hat m_\text{s}} -\beta h,\ \overline{\langle \hat b_\text{s}\rangle}_{\hat m,\hat m_\text{s}}\bigr)\end{equation}
along any path connecting $(\hat m^{(1)}, \hat m^{(1)}_\mathrm{s})$
and  $(\hat m^{(2)}, \hat m^{(2)}_\mathrm{s})$.

\begin{figure}[t]\centering
\includegraphics[height=0.7\linewidth,angle=270]{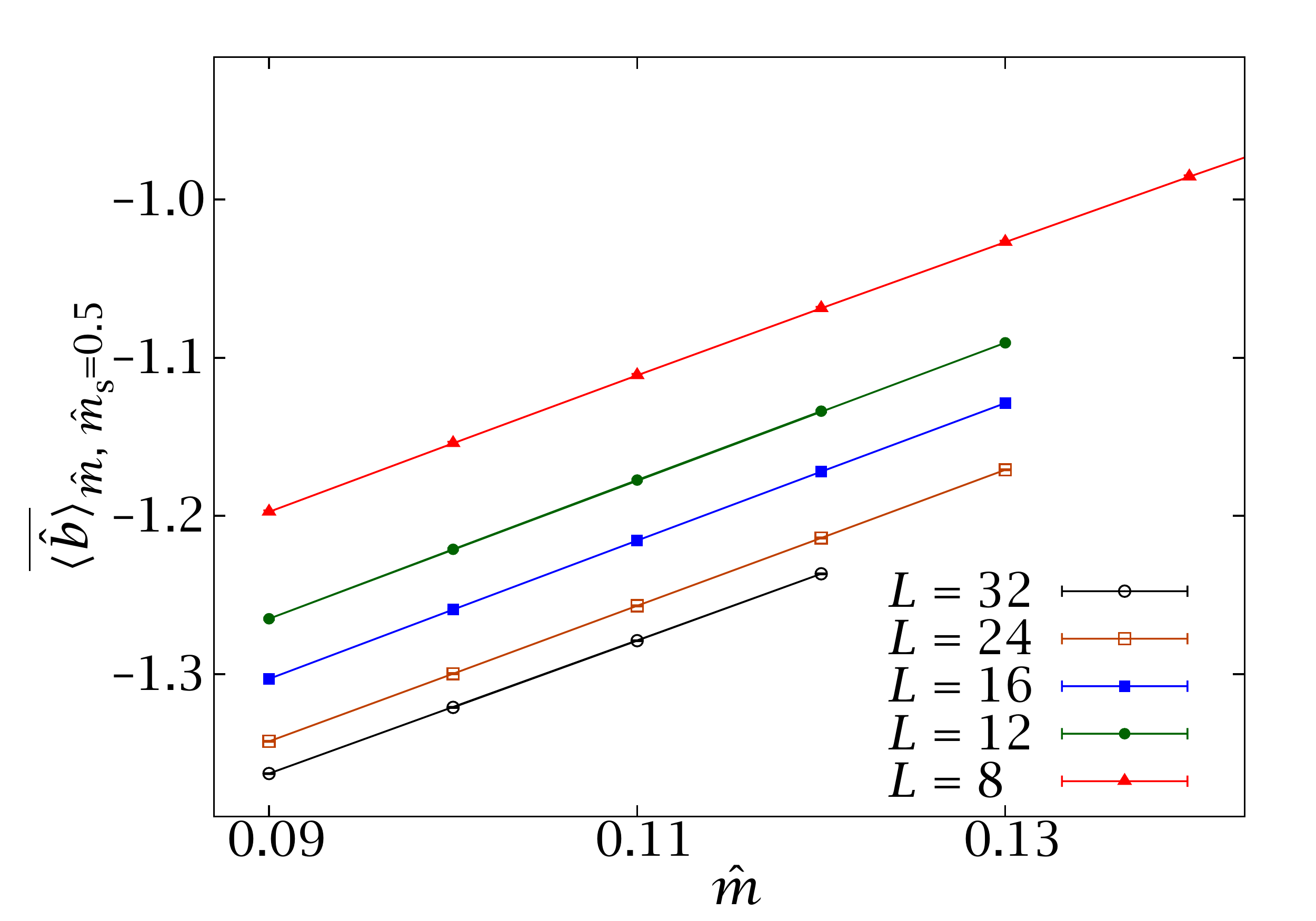}
\caption{Tethered average $\overline{\langle \hat b\rangle}_{\hat m,\hat m_\text{s}=0.5}$ 
for all our lattice sizes at $\beta=0.625$. }
\label{fig:DAFF-hatb-L}
\index{tethered field!DAFF|indemph}
\end{figure}

In our simulations we have not covered the $(\hat m,\hat m_\mathrm{s})$ plane
uniformly. Rather, we have simulated several slices with fixed $\hat m$. 
Therefore, the most convenient way to evaluate $\Delta \overline \varOmega_N^{(h)}$ 
is
\begin{enumerate}
\item At $\hat m=0.12 \approx \hat m_\text{c} = \hat m(h_\mathrm{c})$,
compute the potential difference between $(\hat m_\text{c},\hat m_\text{s}^*)$
and $(\hat m_\text{c},0.5)$, where 
$\hat m_\mathrm{s}^*$ is the position of the peak of $p(\hat m_\mathrm{s}| \hat m_\mathrm{c})$ 
Notice that, since we are working at constant $\hat m$,
\begin{equation}
\overline{\varOmega}_N^{(h)}(\hat m_\text{c},\hat m_\text{s}=0.5) 
- \overline{\varOmega}_N^{(h)}(\hat m_\text{c},\hat m_\text{s}^*)
=  
\overline{\varOmega}_N(0.5|\hat m_\text{c}) 
- \overline{\varOmega}_N(\hat m_\text{s}^*|\hat m_\text{c}).
\end{equation}
This is just the integral 
\begin{equation}\label{eq:DAFF-delta-F1}
\Delta F_1 = \overline\varOmega_N(0.5 | \hat m_\mathrm{c})-
\overline\varOmega_N(\hat m_\mathrm{s}^* | \hat m_\mathrm{c})
= \int \hat{\boldsymbol B} \cdot \dd \boldsymbol \ell_1=\int^{0.5}_{\hat m_\mathrm{s}^*} \dd\hat m_{\mathrm{s}}\ \overline{\langle\hat b_\mathrm{s}\rangle}_{\hat m_\mathrm{c},\hat m_\mathrm{s}}.
\end{equation}
In other words, $\Delta F_1$ is just the area under the curve in Figure~\ref{fig:DAFF-theta}.
We can average $\Delta F_1$ for the symmetric path in the region
with negative staggered magnetisation ($\hat m_\text{s}<1/2$).
\item The tethered average at $(\hat m_\text{c},\hat m_\mathrm{s}^*)$
defines a value of $\beta h_\mathrm{c}$ through the saddle-point equation,
\begin{equation}
\overline{\langle \hat b\rangle}_{\hat m_\mathrm{c}, \hat m_\mathrm{s}^*} = 
\beta h_\mathrm{c}\, .
\end{equation}
\item Consider the tethered values $\overline{\langle\hat b\rangle}_{\hat m,\hat m_\mathrm{s}=0.5}$
as a function of $\hat m$ and interpolate the value $\hat m^*$ that satisfies
\begin{eqnarray}
\overline{\langle \hat b\rangle}_{\hat m^*,\hat m_\mathrm{s}=0.5} = \beta h_\mathrm{c}\,.
\end{eqnarray}
This turns out to be very easy to do because the
$\overline{\langle\hat b\rangle}_{\hat m,\hat m_\mathrm{s}=0.5}$
fall on a straight line (Figure~\ref{fig:DAFF-hatb-L}).
\item Compute the potential difference between
$(\hat m_\mathrm{c},\hat m_\text{s}\!=\!0.5)$ and  $(\hat m^*,\hat m_\mathrm{s}\!=\!0.5)$.
\begin{equation}
\Delta F_2 = \int\hat{\boldsymbol B}\cdot\dd\boldsymbol \ell_2 =  \int^{\hat m^*}_{\hat m_\mathrm{c}}\dd \hat m\ 
\bigl(\overline{\langle \hat b\rangle}_{\hat m, \hat m_{\mathrm{s}}=0.5} -\beta h_\text{c}\bigr).
\end{equation}
\item The sum of $\Delta F_1$ and $\Delta F_2$ is equal to the integral 
of $\hat {\boldsymbol B}$ along a path joining the two saddle points
$(\hat m^{(1)}, \hat m_\mathrm{s}^{(1)})=(\hat m^*, 0.5)$ and 
$(\hat m^{(2)}, \hat m_\mathrm{s}^{(2)})=(\hat m_\mathrm{c}, \hat m_\mathrm{s}^*)$. 
i.e., the free-energy barrier we were looking for,
\begin{equation}\label{eq:DAFF-delta-F}
\Delta \overline F_N = \Delta F_1 + \Delta F_2.
\end{equation}
This integration path is depicted with a dashed line in Figure~\vref{fig:DAFF-flujos}.
Notice that, in accordance with the study of Section~\ref{sec:DAFF-effective-potential}, $\Delta F_1$ will be positive and $\Delta F_2$ negative.
\end{enumerate}

We have implemented this procedure for all our lattices.
Table~\ref{tab:DAFF-theta} shows the resulting $\Delta \overline F_N$ 
and fits to
\begin{equation}\label{eq:DAFF-theta-fit}
\Delta \overline F_N = A L^{3-\theta}.
\end{equation}
\begin{table}
\small
\centering
\begin{tabular*}{\columnwidth}{@{\extracolsep{\fill}}clllc}
\toprule
 $L$ &\multicolumn{1}{c}{$\Delta\overline F_N$} & Fit range & 
\multicolumn{1}{c}{$\theta$} & $\chi^2/\mathrm{d.o.f.}$\\
\toprule
 8  & 0.033\,82(29) & $L\geq 8\ $ & $1.448(9)$   & 5.56/3 \\
 12 & 0.017\,56(15) & $L\geq 12$ &  $\mathbf{1.469}(13)$  & $\mathbf{0.44\boldsymbol/2}$\\
 16 & 0.011\,38(9)  & $L\geq 16$ &  $1.461\mathbf{\boldsymbol(20\boldsymbol)} $     & 0.16/1\\
 24 & 0.006\,08(5)  &  \\
 32 & 0.003\,92(5)  & \\
\bottomrule
\end{tabular*}
\caption[Computation of $\theta$]{
Computation of the hyperscaling violations exponent $\theta$ from
the free-energy barriers $\Delta\overline F_N$. We report fits 
to~\eqref{eq:DAFF-theta-fit}, for different ranges.
Our preferred final estimate is $\theta=1.469(20)$, 
taking the central value of the fit for $L\geq12$ and the more conservative
error of the fit for $L\geq16$.  \label{tab:DAFF-theta}
\index{critical exponent!theta@$\theta$|indemph}
\index{free energy|indemph}
\index{effective potential!DAFF|indemph}}
\end{table}

As our preferred final result we can give the central value 
of the fit for $L\geq12$ with the error of the fit
for $L\geq16$. Therefore
\begin{equation}\label{eq:DAFF-valor-theta}
\theta = 1.469(20).
\end{equation}
Notice
that $\theta$ is close to the value $\theta= D/2$  that one would expect
from a naive 
Imry-Ma argument~\cite{imry:75}. In particular, $\theta < D-1$, 
and therefore the metastable states do not define stable 
phases, which is yet another argument against a first-order 
transition.

Notice that the value of $\theta>0$ is also the reason for the exponential
slowing down of this model. \index{critical slowing down}
Indeed, we saw in Section~\ref{sec:DAFF-effective-potential} that
the system gets trapped in local minima with escape times
$\tau \sim \exp[N \Delta \overline F_N] \sim \exp[L^\theta]$
causing a thermally activated critical slowing down~\cite{fisher:86b,nattermann:97}.

\subsection[Scaling relations: the specific heat and $\alpha$]{Scaling relations: the specific heat and \boldmath $\alpha$}\label{sec:DAFF-scaling} \index{specific heat!DAFF} \index{critical exponent!alpha@$\alpha$}
As discussed in the introduction to this chapter, there has long been disagreement
as to whether the specific heat in the DAFF is divergent,
as observed in experiments~\cite{belanger:83,belanger:98}. 
The specific heat critical exponent is generally difficult to obtain in numerical
simulations and the self-averaging violations in the DAFF have not made 
things any easier~\cite{hartmann:01,wu:06,malakis:06},

Since we already have a complete set of independent critical exponents, we 
can obtain $\alpha$ from the hyperscaling relation~\eqref{eq:DAFF-modified-hyperscaling}.\index{hyperscaling} \index{scaling relations} 
Using our range of values $0.75\leq \nu\leq 1.1$ and $\theta=1.469(20)$, 
we have the following bounds for $\alpha$,
\begin{equation}
0.32 \leq \alpha \leq 0.85.
\end{equation}
Clearly, the uncertainty in $\nu$ does not allow a good determination, yet 
we can safely exclude a non-divergent specific heat from our data
($\alpha< 0$  would imply $\nu> 1.3$). Furthermore, even the logarithmic
divergence ($\alpha=0$) suggested by experimental work seems excluded.
Notice, however, that the experimental values of $\nu$
together with hyperscaling relations, also imply $\alpha>0$. 
For instance, the result $\nu=0.87(7)$ of~\cite{slanic:99}, taken
with our $\theta=1.469(20)$, would give $\alpha\approx0.67(10)$.
Even the higher value $\nu=1.20(5)$ of~\cite{ye:04} would give $\alpha\approx0.16(8)$.

We can attempt a direct determination of $\alpha$ from our simulations.
We start by defining the specific heat as
\begin{align}\label{eq:DAFF-specific-heat}
C = \frac{\partial \overline{\langle m\rangle}}{\partial h}
\end{align}
In principle, at the critical field, $C\propto L^{\alpha/\nu}$, so we could \index{quotients method} 
compute $\alpha/\nu$ from $C$ using the quotients method. Unfortunately, the quotients
method is ill-suited to this quantity, which features a large analytical
background, so its scaling is more aptly described by $C\simeq A + B^{\alpha/\nu}$~\cite{ballesteros:98b}.
Therefore, one needs extremely large values of $L$ to reach the asymptotic 
regime where $C\sim L^{\alpha/\nu}$. 

The behaviour of the quotients shown in Table~\ref{tab:DAFF-quotients}
is consistent with the above expectation. Our results point 
to a divergent specific heat, but our estimates for the exponent $\alpha$ are clearly still very far 
from the asymptotic regime where~(\ref{eq:DAFF-modified-hyperscaling}) would be
satisfied. An additional source of error is that our determination of the specific heat
suffers from systematic effects due to the discretisation of $\hat m$.
Indeed, while our $\hat m$ grid was more than adequate for a determination of
$\partial_{\hat m}\xi$, which could be approximated by a straight line,
the specific heat shows a clear curvature as a function of $\hat m$. 
We have interpolated it with a quadratic polynomial in order to estimate
the derivative,  but we would have needed more resolution in $\hat m$
to make a safe computation.

\subsection[The two-exponent scenario and experimental work]{The
 two-exponent scenario and the experimental scattering line shape}
We can close our FSS analysis of the phase transition by considering 
the two-exponent scenario discussed in Section~\ref{sec:DAFF-phase-transition}.
Recall that according to this proposal,  $\theta = D/2-\beta/\nu$.
From~\eqref{eq:DAFF-valor-theta} and $\beta/\nu= 0.011\,1(8)$ (combining
the last two rows of Table~\ref{tab:DAFF-quotients}) we have
\begin{align}
\theta &= 1.469(20),\\
D/2-\beta/\nu &=1.488\,9(8).
\end{align}
\index{scattering line shape}
The two numbers are compatible. 

We can use our results for the critical exponents to shed some \index{DAFF!experiments}
light on the experimental situation. In an experimental study, 
the critical exponents are computed from fits to 
the scattering line shape or structure factor $S(k)$.
We can write this quantity as 
\begin{equation}
S(k) = S_\text{d}(k) + S_\text{c}(k),
\end{equation}
where $S_\text{d}$ and $S_\text{c}$ are the connected and disconnected
propagators defined in~\eqref{eq:DAFF-Sk}. Recall
that these diverge as 
\begin{align}
S_\text{d}(k) &\sim k^{-4+\bar \eta}+\ldots,\\
S_\text{c}(k) &\sim k^{-2+ \eta}+\ldots,
\end{align}
where the dots represent subdominant terms.

An experimental study of $S(k)$ faces, then, the considerable
challenge of distinguishing two different divergences at $k=0$.
One possibility is simply to disregard the difference and 
parameterise $S(k)$ with an effective anomalous dimension,
\begin{equation}
S(k) \sim k^{-2+\eta_\text{eff}}.
\end{equation}
This procedure was employed in~\cite{ye:04}, obtaining
$\nu=1.20(5)$ and $\eta_\text{eff}=-0.51(5)$.

An alternative way, supported by our results, is to 
adopt the two-exponent ansatz~\eqref{eq:DAFF-theta-equality}, which can be rewritten
as $2\eta = \bar \eta$. Then, the following approximation
can be made:
\begin{equation}\label{eq:DAFF-S-k-two-exponent}
S_\text{d}(k) = [S_\text{c}(k)]^2.
\end{equation}
This avenue was followed in~\cite{slanic:99}, yielding
$\nu=0.87(7)$ and $\eta=0.16(6)$. This value of $\nu$ is 
right in the middle of our numerical bounds. However, 
taking into account~\eqref{eq:DAFF-theta-equality},
we would have
\begin{equation}
\theta = 2-\eta = 1.84(6),
\end{equation}
which is clearly incompatible with our results.

The answer to this problem is that 
the approximation~\eqref{eq:DAFF-S-k-two-exponent}
is probably too naive. Even accepting that $\bar \eta = 2\eta$, 
approximating all of $S_\text{d}$, and not just its 
most divergent term, as the square of $S_\text{c}$ 
is an excessive simplification. Clearly, a better theoretical
parameterisation of $S(k)$ is needed. Our methods are 
well suited to a direct numerical approach to this question,
left for future work.

\section{Geometrical study of the critical configurations}\label{sec:DAFF-geometry}

The picture painted in the previous sections is that of a second-order
transition; but one with quite extreme behaviour. Among its peculiarities
we can cite an extremely small value of $\beta$ and large free-energy
barriers, features both that are reminiscent of first-order behaviour.
The free-energy barriers in the DAFF, however, diverge too slowly to be associated 
to the surface tension of well-defined, stable coexisting phases. But
this only raises the question of what kind of configurations can give rise
to such a behaviour.  
\begin{figure}[t]
\centering
\includegraphics[width=.7\columnwidth]{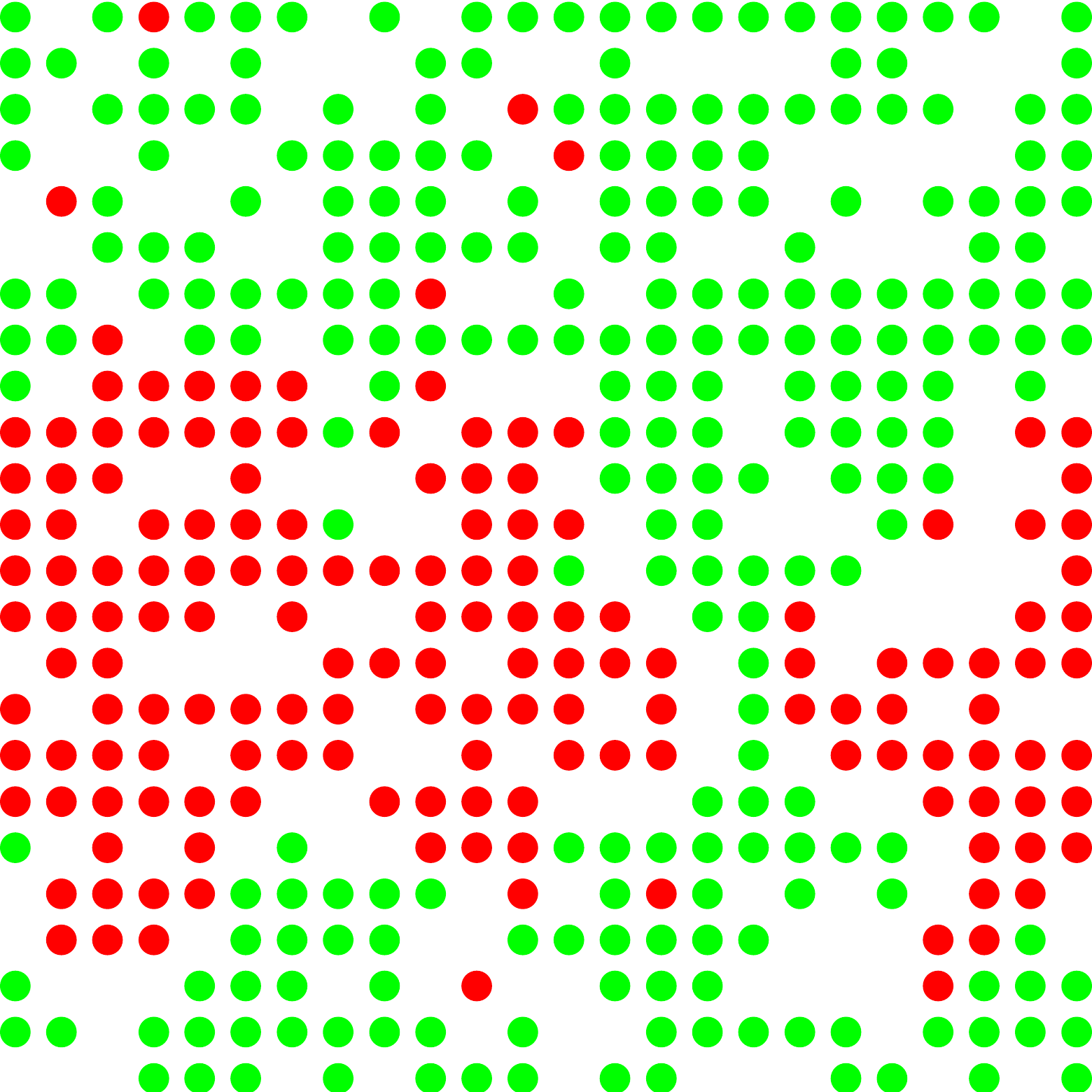}
\caption[Equilibrium configuration for $\hat m_\text{s}=0.5$]{Equilibrium configuration for an $L=24$ system
at $\beta=0.625$, $\hat m=0.12$, $\hat m_\mathrm{s} = 0.5$.}
\label{fig:conf}
\end{figure}

In this section we study the geometrical properties 
of the minimal-cost spin configurations joining the two 
\index{instanton}
ordered phases at the critical point. To this end, we 
consider simulations at $\beta=0.625$,
 $\hat m=0.12 \approx \hat m(h_\mathrm{c})$ and $\hat m_\mathrm{s} = 0.5$.
Recalling that $\hat m_\mathrm{s} \simeq m_\mathrm{s} +1/2$, this 
last condition expresses the fact that we are studying configurations
with no global staggered magnetisation. This is a good example of an 
`inherently tethered' study, that examines information hidden 
from a canonical treatment
(cf. Figure~\ref{fig:DAFF-p-hatms}, where
we show that the region considered here has a canonical probability density 
of $\sim10^{-70}$ for an $L=32$ system).

Figure~\ref{fig:conf} shows an example of such a configuration
for an $L=24$ system.
In order to make the different phases clearer, we are not representing
the spin field $s_\bx$, but the staggered field $s_\bx\pi_\bx$. 
As is readily seen, even if the global magnetisation is 
$m_\mathrm{s} \approx 0$, the system is divided into two phases
with opposite (staggered) spin. In geometrical terms, most of the 
occupied nodes of the system belong to one of two large clusters with 
opposite sign. All the other clusters are orders of magnitude
smaller, with most of the remaining spins forming
single-site clusters.

\index{interface|(}
At a first glance, this picture may seem consistent with a first-order
scenario, where the system is divided into two strips whose
surface tension gives rise to the free-energy barriers
(see~\cite{martin-mayor:07} for an example of these
geometrical transitions in a first-order setting). In order to test this
possibility, we can study the evolution of the interface mass with the
system size and compare it with the explicit computation of free-energy
barriers done in Section~\ref{sec:DAFF-theta}.
\begin{table}
\small
\centering
\begin{tabular*}{\columnwidth}{@{\extracolsep{\fill}}rrl}
\toprule
 $L$ & $\mN_\mathrm{samples}$ & $\mN_\mathrm{interface}$\\
\toprule
12 & 3000 & \phantom{1}160.29(25)\\
16 & 3000 & \phantom{1}304.62(41)\\
24 & 3000 & \phantom{1}755.74(94)\\
32 &  700 & 1446.1(39)\phantom{1}\\
\bottomrule
\end{tabular*}
\caption[Interface masses]{Masses of the interfaces for our equilibrium
configurations at $\beta=0.625$, $\hat m=0.12\approx\hat m_\mathrm{c}$,
$\hat m_\mathrm{s} =0 .5$. This mass grows as 
$\mN_\mathrm{interface} \propto L^c$. From a fit, $c=2.240\,2(24)$, which 
is incompatible with our estimate of $\theta=1.469(20)$, confirming
that the free-energy barriers are not associated to surface tensions.
\index{interface|indemph}}
\label{tab:interfase}
\end{table}

Given a configuration, we first trace all the geometric 
antiferromagnetic clusters. We then identify the largest and second
largest ones. Finally, we say that an occupied node belongs
to the `interface' if it belongs to the largest cluster and has at least
one first neighbour belonging to the second largest one. We have 
computed in this way the interface $\mN_\mathrm{interface}$ for 
our $700$ $L=32$ samples and for $3000$ samples for all our
smaller systems (we have run additional simulations just at this point).
Table~\ref{tab:interfase} shows the result of this computation.
A fit to
\begin{equation}\label{eq:interfase}
\mN_\mathrm{interface} = A L^c,
\end{equation}
for $L\geq12$ gives $c=2.240\,2(24)$ with $\chi^2/$d.o.f. $=3.62/2$.   \index{surface tension}
If the metastability of the system were to be associated 
to a surface tension, then the free-energy barriers would have \index{free energy!barriers}
to scale with $c$. However, we saw in Section~\ref{sec:DAFF-theta}
that the actual scaling is $N\Delta \overline F_N \propto L^{\theta}$, 
with $\theta=1.469(20)$.

To understand the reason why this large interfaces are not
responsible for metastable signatures,  we have 
to look more closely at the geometry of the configurations.
Indeed, we see in Figure~\ref{fig:conf} that there is a path
connecting the spins in the green strip \emph{across} the red one 
(there is only one green strip, since we are considering periodic
boundary conditions).
\begin{figure}
\centering
\includegraphics[height=.7\columnwidth,angle=270]{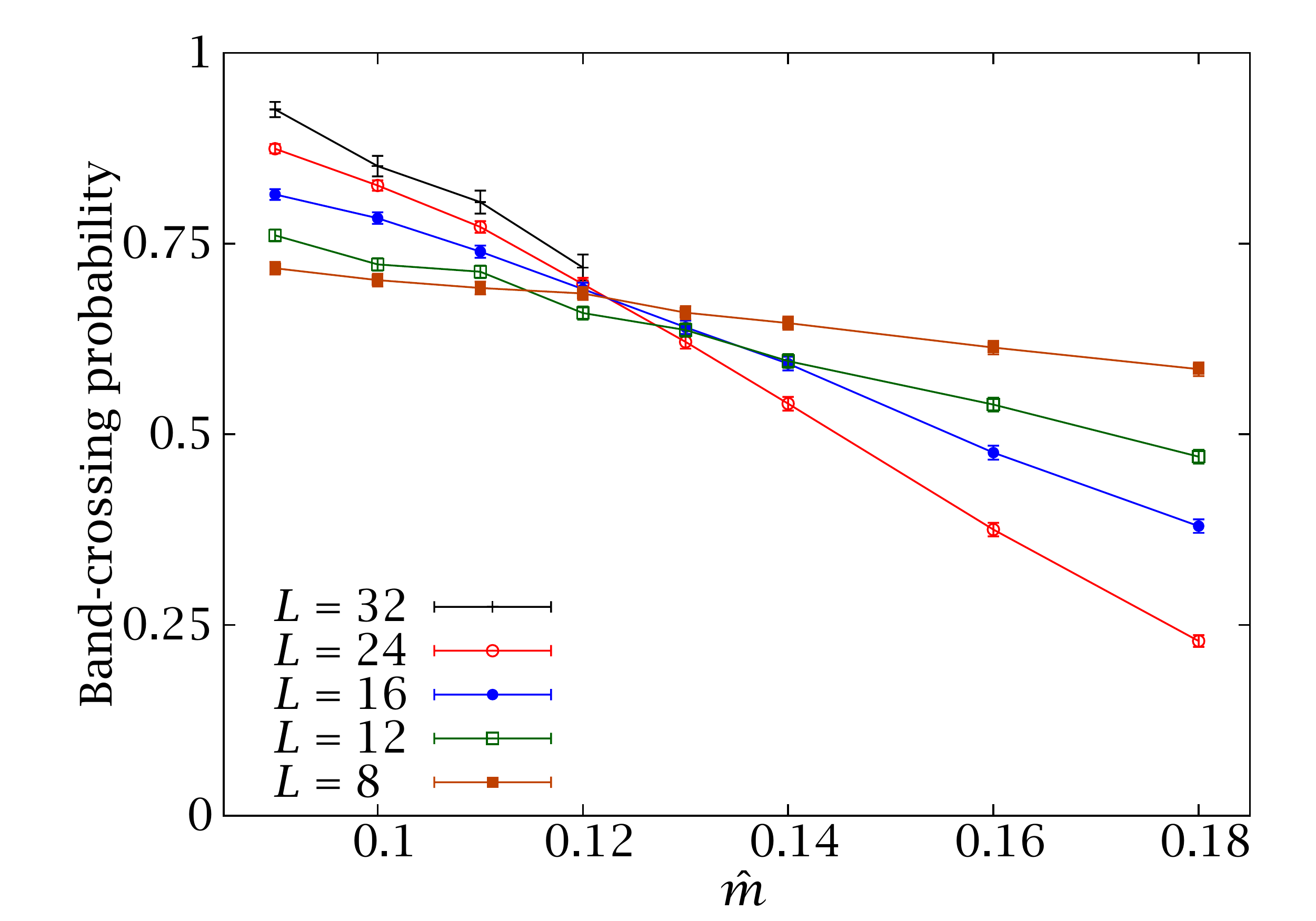}
\caption[Strip-crossing probability]{Probability of completing
a path with constant staggered spin across
  the strip with opposite staggered magnetisation, see Fig.~\ref{fig:conf},
  for our spin configurations at $\beta=0.625$, $\hat m_\mathrm{s}=0.5$.}
\label{fig:bandas}
\end{figure}
If we were to consider a complete tomography of this configuration, 
we would find several  of these paths (which, of course, need not be
contained in a plane). In other words, the phases are porous. This is
in clear contrast to a first-order scenario in which the phases have
essentially impenetrable walls. Now, this could be a peculiarity of the 
particular selected configuration. In order to make the analysis
quantitative, we shall examine all of our samples and determine the 
strip-crossing probability $P$. This is defined as the probability of
finding a complete path with constant staggered spin across the strip with 
opposite staggered magnetisation and we can compute it with the following
algorithm:
\begin{enumerate}
\item For each configuration, compute the staggered Fourier 
transform~\eqref{eq:DAFF-phi-k} of the spin field at 
$\bk_\text{min}^{(i)}$, for each of the three axes ($\phi_x$, $\phi_y$, $\phi_z$).
\item In a strip configuration, one of
the $\phi$ will be much larger
than the other two. 
Assume this is $\phi_x$, so the strips are perpendicular to the $OX$ axis. 
\item Measure the staggered magnetisation $M_\mathrm{s}^x$
 on each of the planes
with constant $x$ and identify the plane $x=x_\mathrm{max}$ 
with largest $|M_\mathrm{s}^x|$. This plane will be at the core
of one the strips.
\item Trace all the clusters that contain at least one spin on
the $x=x_\mathrm{max}$ plane, but severing the links between planes
$x=x_\mathrm{max}$ and $x=x_\mathrm{max}-1$. 
\item If any of the clusters reaches the plane $x=x_\mathrm{max}-1$
there is at least a path through the strip
with opposite magnetisation (the previous step has forced us to go
the long way around, so we know we have crossed the strip).
\end{enumerate}

We have plotted the strip-crossing probability $P(\hat m)$
in Figure~\ref{fig:bandas}. Notice that $1-P$ behaves as an order parameter.
If we keep increasing $\hat m$, so that we enter the ordered phase, the phases
eventually become proper impenetrable strips, hence $P=0$ for large enough
systems. On the other hand, for low $\hat m$, in the disordered phase, the
strips become increasingly porous, so that $P=1$ in the limit of large
systems. The crossover between these two regimes manifests in the crossing
of the $P(\hat m;L)$ at the phase transition point $\hat m \approx 0.12$.
\index{interface|)}
\section[Optimising tethered simulations]{Beyond \boldmath $L=32$: optimising TMC simulations}\label{sec:DAFF-optimization}
The simulations listed in Table~\ref{tab:DAFF-parametros-PT} required more
than $700$ years of total CPU time and were carried out using a combination
of large supercomputing facilities (\emph{Mare Nostrum} of the Red Española \index{Mare Nostrum@\emph{Mare Nostrum}}
de Supercomputación), large computing clusters (\emph{Terminus} at BIFI) \index{Piregrid@\emph{Piregrid}}
and grid resources (\emph{Piregrid}).\index{grid computing} 
Even with the parallel scheme explained in Section~\ref{sec:DAFF-numerical-implementation},
the wall-clock for some of the toughest points has been of \index{wall-clock} 
several months.\index{Terminus@\emph{Terminus}}

Essentially, by tethering the magnetisations we managed to remove
the largest free-energy \index{free energy!barriers}
barriers of the systems. However, for large sizes,
new barriers, associated  \index{reaction coordinate}
to different reaction coordinates, began to appear. In order
to eliminate these, one would have to tether additional quantities.

It may seem, then, that the methods described in this Chapter cannot be extended 
to larger lattices with current hardware. Actually, 
these simulations were intended to explore 
the physics of the DAFF thoroughly, perhaps sacrificing focus
for breadth. In a more targeted study, however,
we can take advantage of the model's characteristics 
to optimise the simulations. In particular, 
we can highlight two interesting facts:
\begin{itemize}
\item Only a very narrow region around the saddle points has any 
significant weight for reconstructing canonical averages (Figure~\ref{fig:DAFF-p-hatms}).
\item It is much harder to equilibrate the region far from these peaks
(Figure~\ref{fig:DAFF-histograma-hatm-012}).
\end{itemize}
Combining these two observations, it turns out that, if our only interest is
reconstructing canonical averages, we can achieve a qualitative reduction
in simulation time by simulating only a narrow range around the peak 
in $p(\hat m_\text{s}| \hat m)$ for each value of the smooth magnetisation
$\hat m$. In fact, all the results discussed  in this chapter, except
for Sections~\ref{sec:DAFF-theta} and~\ref{sec:DAFF-geometry},
could have been computed in this simplified fashion.

We have demonstrated this optimisation by simulating $400$ samples 
of an $L=48$ system
for $\hat m=0.12$. We use only $\mN_{\hat m_\text{s}}=6$ (but we have
to increase the number of temperatures in the parallel-tempering to $60$, in the 
same range,
in order to keep the exchange acceptance high). Table~\ref{tab:DAFF-picos}
shows the value of $\hat m_\text{s}^\text{peak}$ as a function of $L$.

\begin{figure}[p]
\centering
\includegraphics[height=.7\columnwidth,angle=270]{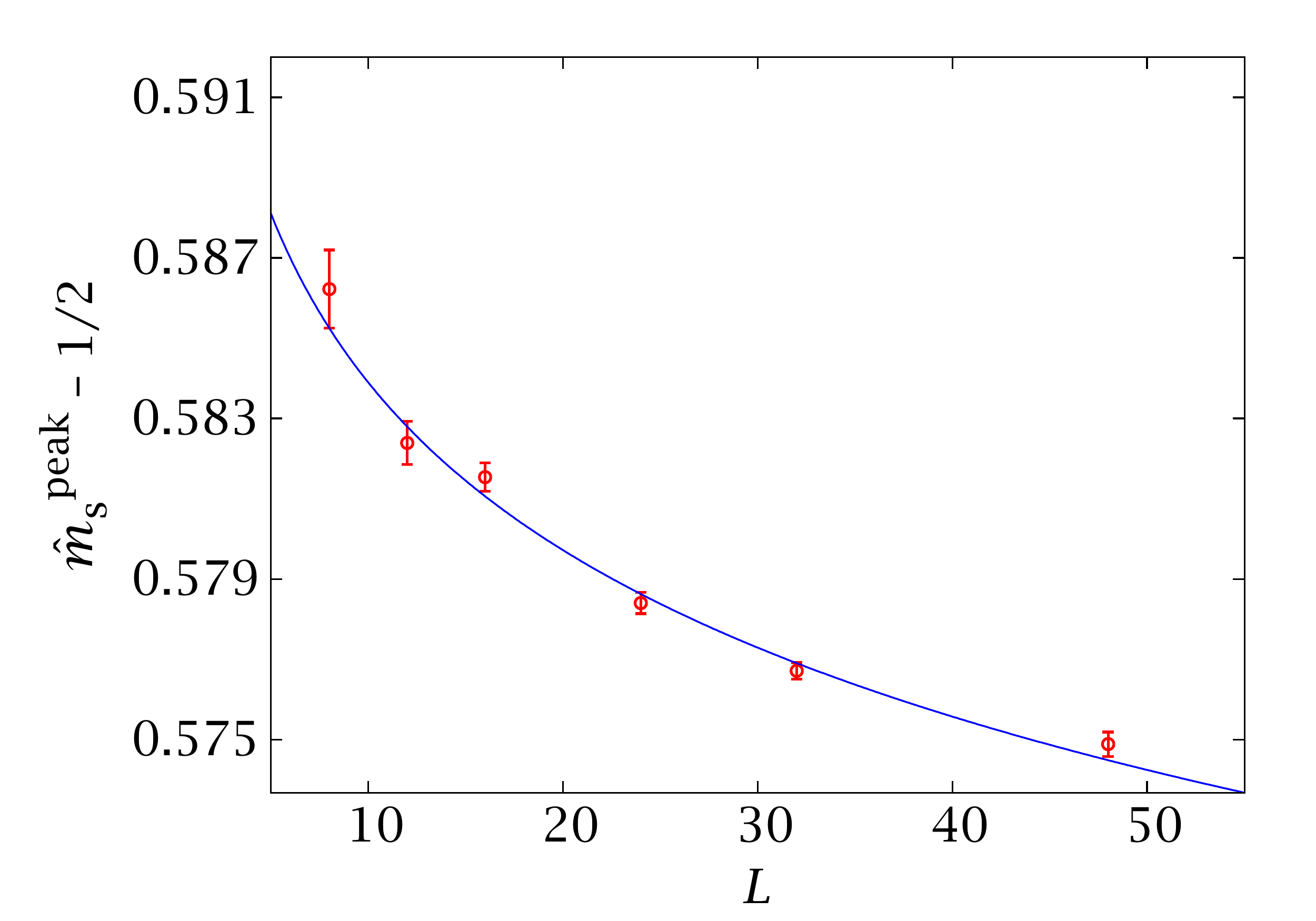}
\caption[Shift in the peak position with $L$]{Position of the peak against $L$ from Table~\ref{tab:DAFF-picos} 
with a fit to Eq.~(\ref{eq:DAFF-fit-picos}), giving $\beta/\nu=0.0111(7)$.}
\label{fig:DAFF-maximo}
\end{figure}

We can use the values of $\hat m_\text{s}^\text{peak}$ reported
in Table~\ref{tab:DAFF-picos} to attempt an alternative
determination of $\beta/\nu$, following the method described
in Section~\ref{sec:ISING-betanu}. In particular, 
we perform a fit to  \index{finite-size scaling} \index{critical exponent!beta@$\beta$}
\begin{equation}\label{eq:DAFF-fit-picos}
|\hat m_{\mathrm{s}}^\mathrm{peak} - 1/2| =  A L^{-\beta/\nu}.
\end{equation}
This is plotted in Figure~\ref{fig:DAFF-maximo}. The result of the fit,
with $\chi^2/$d.o.f. $= 2.98/3$, is
\begin{equation}
\beta/\nu = 0.011\,1(7).
\end{equation}
This result is compatible with 
our computation with the quotients method in Table~\ref{tab:DAFF-quotients}.
Notice, however, that it does not take into account the uncertainty in
$\hat m_\mathrm{c}$ (in Section~\ref{sec:ISING-betanu} we worked 
at the exact $\beta_\text{c}$ and even in Section~\ref{sec:CLUSTER-results}
we knew $\beta_\text{c}$ with many significant figures).

\begin{table}[p]
\centering
\small
\begin{tabular*}{\columnwidth}{@{\extracolsep{\fill}}rrrc}
\toprule
$L$ & $\mN_\text{samples}$ & $N_{\hat m_\text{s}}$ & $\hat m_\text{s}^\text{peak}-1/2$ \\
\toprule
8  & $1000\times2$ & 31 & 0.585\,85(87) \\
12 & $1000\times2$ & 35 & 0.582\,39(54) \\
16 & $1000\times2$ & 35 & 0.581\,54(36) \\
24 & $1000\times2$ & 33 & 0.578\,39(24) \\
32 &  $700\times2$ & 25 & 0.576\,72(20) \\
48 &  400 &  6 & 0.574\,91(33) \\
\bottomrule
\end{tabular*}
\caption[Peak position for $\hat m=0.12$]{Value of the peak position for our different system
sizes. The result for $L=48$ is of comparable accuracy, despite
being computed from only $6$ tethered simulations around 
the saddle point. Recall that for $L\leq 32$ we can average over 
the positive and negative peaks, so the number of samples 
for these systems is effectively double the value shown 
in Table~\ref{tab:DAFF-parametros-PT}.}
\label{tab:DAFF-picos}
\index{magnetisation!staggered|indemph}
\end{table}

\index{DAFF|)}

\part{The Edwards-Anderson spin glass}\label{part:sg}
\chapter{Spin glasses: an experimental challenge for theoretical physics}\label{chap:sg}\index{spin glass|(}

During the 20th century, the experimental study of magnetic alloys gradually
led to the identification of a class of materials that suffered a `freezing transition' 
at some critical temperature $T_\text{c}$. In particular, if we denote
by $\boldsymbol S_\bx$ the magnetic moment (spin) at site $\bx$ in a lattice\footnote{%
In this section we consider the general case where the spins are three-dimensional vectors.}
\begin{itemize}
\item Below $T_\text{c}$, each of the spins freezes 
in some orientation, $\braket{\boldsymbol S_\bx}_t \neq0$.\footnote{%
Here the $\braket{\cdots}_t$ denote an average over a very long measuring time.}
\item Though the freezing is collective, the spin orientations
are random. In particular, the  average orientation of the spins
is zero: $\frac1N \sum_\bx \braket{\boldsymbol S_\bx}_t = 0$. More generally, 
there is no long-range magnetic order of any kind
\begin{equation}\label{eq:SG-long-range}
\boldsymbol M_\bk = \frac1N \sum_\bx \ee^{-\ii \bk\cdot \boldsymbol r} \braket{\boldsymbol S_\bx}_t = 0,\qquad \forall \bk.
\index{long-range order}
\end{equation}
Notice that this expression includes both the ferromagnetic order parameter, $\bk=0$,
and the antiferromagnetic one, $\bk=(\uppi,\uppi,\uppi)$.
\item The dynamical evolution takes place at macroscopic times below $T_\text{c}$.
\end{itemize}
The position of the freezing temperature is typically signalled \index{susceptibility!spin glass}
by the appearance of a cusp in the imaginary part $\chi''(\omega)$ of the frequency-dependent
magnetic susceptibility, although it has been proved that this freezing corresponds 
to an actual phase transition (cf. Section~\ref{sec:SG-glass}).

These systems were soon seen to have many interesting physical 
properties, which in turn
generated a great deal of theoretical and experimental interest. 
In the early 1970s, the concept of `spin glass' was introduced,
and applied to random, mixed-interacting magnetic systems that experience
a random, yet cooperative, freezing of spins below some critical temperature.

The random and mixed nature of the magnetic interactions gives rise to frustration \index{frustration}
(recall Figure~\ref{fig:INTRO-frustrado}) 
and this, in turn, generates a rugged free-energy landscape.  \index{free energy!landscape}
Furthermore, the fact that spin glasses were particularly amenable to 
experimental investigation and theoretical modelling (see below) quickly
established them as quintessential examples of complex systems.

In this chapter we shall begin by giving some brief notes on 
spin glasses from an experimental point of view in Section~\ref{sec:SG-experiments}.
This will also serve as an introduction  to non-equilibrium physics and the concept 
of aging. \index{aging}
We then discuss the generally accepted Edwards-Anderson model in Section~\ref{sec:SG-EA},
stating the main conclusions of the different (and incompatible) theoretical pictures
that have been proposed for it. Finally, in Section~\ref{sec:SG-numerical} we motivate 
the numerical investigation of the problem that will be undertaken in Chapters~\ref{chap:sg-2}
and~\ref{chap:sg-3}.

\section{The experimental spin glass}\label{sec:SG-experiments}\index{spin glass!experiments|(}
\subsection{The canonical experimental spin glass: RKKY}
As we said above, in order to have a spin glass system we need at least disorder and a mixture
of ferromagnetic and antiferromagnetic interactions between the spins. In principle, 
we would be looking for a system whose spin Hamiltonian consisted of exchange  \index{exchange interaction}
interactions: 
\begin{equation}
\mathcal H_{\bx\by} = -J_{\bx\by} \boldsymbol S_\bx\cdot \boldsymbol S_\by,
\end{equation}
where we need to have both positive and negative $J_{\bx\by}$. 

The classical solution to these requirements is the Ruderman-Kittel-Kasuya-Yosida (RKKY)
interaction~\cite{ruderman:54,kasuya:56,yosida:57}. \index{RKKY interaction}
We consider a noble metal with magnetic impurities (for instance, \emph{Cu}Mn,
for copper with manganese impurities).
Then, the magnetic moments scatter the conducting electrons, which gives rise to
an indirect exchange interaction: $\mathcal H_{\bx,\bx+\boldsymbol r} =
 J(r) \boldsymbol S_\bx\cdot \boldsymbol S_{\bx+\boldsymbol r}$. For a large 
separation between impurities, we can write the coupling strength as
\begin{equation}\label{eq:SG-RKKY}
J(r) \simeq J_0 \frac{\cos (2k_\text{F} r + \phi)}{(k_\text{F} r)^3},
\end{equation}
($k_\text{F}$ is the Fermi momentum of the metal). Since the positions
of the impurities or, in other words, the distances between the spins,
are random, the oscillating nature of $J(r)$ will generate
precisely the random mixture of ferromagnetic and antiferromagnetic
couplings that we need.
\subsection{Aging and other non-equilibrium phenomena}
\begin{figure}
\centering
\includegraphics[width=0.7\linewidth]{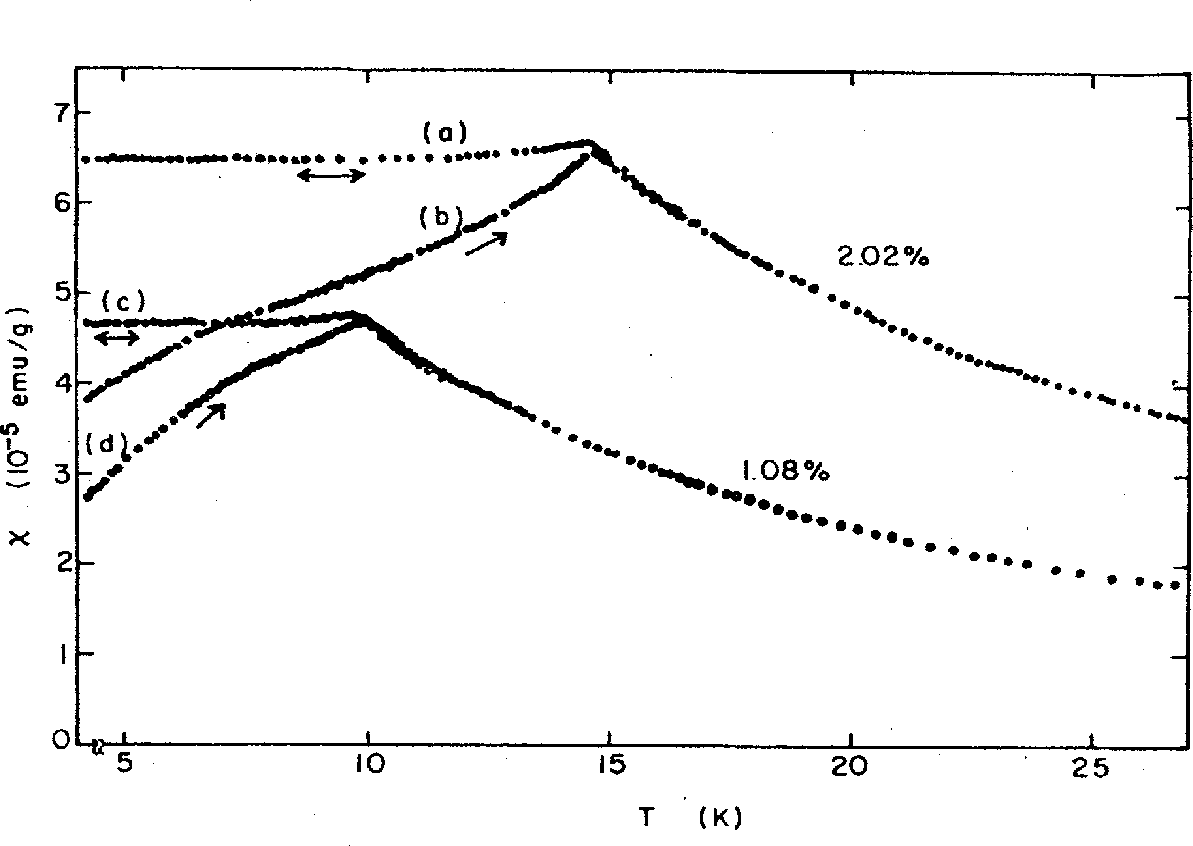}
\caption[Zero-field-cooled and field-cooled susceptibilities]{%
Comparison of the zero-field-cooled ($b$ and $d$ curves) and field-cooled
susceptibilities ($a$ and $c$) for two samples of \emph{Cu}Mn with different
concentration of impurities. Graph from~\cite{nagata:79}, as
quoted in~\cite{binder:86}.
\index{susceptibility!spin glass|indemph}
\index{susceptibility!FC and ZFC|indemph}
\label{fig:SG-FC-ZFC}}
\end{figure}

The physical properties observed in spin-glass simulations
are rich and surprising (see~\cite{binder:86,mydosh:93} for reviews).
The main feature is that, in general, the relaxation times below $T_\text{c}$ are not 
only macroscopic, but large enough for the system to be out of 
equilibrium for the whole of the experiment.  
When the spin glass evolves at fixed $T_1<T_\text{c}$, 
some frozen or spin-glass phase slowly forms. In this
situation, the system is said to age~\cite{vincent:97}: the response to a field variation depends \index{aging} 
on the time spent at $T_1$.

Let us see some examples of this. First we consider the experiment quoted 
in Figure~\ref{fig:SG-FC-ZFC}. In it, the authors measured the magnetisation 
$M$ induced in a spin glass by a small applied field $h$ and used it to
define a dc susceptibility\footnote{Notice that this is different
from the frequency-dependent, or ac, susceptibility we mentioned
earlier.} $\chi_\text{dc}= M/H$. They then consider \index{field cooling}
two experimental protocols. In the first one (field-cooling) \index{zero-field cooling}
they apply the field above $T_\text{c}$ and cool the system 
down to $T_1 < T_\text{c}$, without switching off the field. The resulting
curve is reversible if the temperature variation is cycled and it has the peculiarity
that the $\chi_\text{dc}^\text{FC}$ becomes constant below $T_\text{c}$.
In the second (zero-field
cooling) the temperature is lowered to $T_1<T_\text{c}$ without applying any field.
Then, the field is activated and the sample is reheated across the critical temperature.
Above $T_\text{c}$, $\chi_\text{dc}^\text{ZFC}$ coincides with the field-cooled
version, but this is not the case in the frozen phase. There, the curve is no longer
constant, but increasing with $T$. Furthermore, while the FC curve was reversible,
this one is not and it  also depends on the temperature-variation rate.
 We are seeing a system out of equilibrium. In fact, if, after zero-field-cooling
the system, we were to switch on the field and let the system evolve at fixed temperature,
the susceptibility would grow slowly, approaching the FC value but never reaching it, even for 
waiting times of several hours. 

A different, but related, example is the thermoremanent magnetisation.
 \index{magnetisation!thermoremanent}
In this case, one cools the system in a field down to $T_1<T_\text{c}$, then waits
for a time $t_\text{w}$, switches off the field and measures the 
decay of the remanent magnetisation when a time $t$ has passed after switching off the field.
It turns out that this magnetisation depends both on $t$ and on the waiting time $\tw$
that the system evolved in a field  (the decay is slower for longer $\tw$).
This phenomenon is called aging. \index{aging}
For a rapid cooling, the remanent
decays as a function of $[(t+\tw)/\tw^\mu]$~\cite{rodriguez:03}, 
so $\tw$ is the only recognisable time scale. If
$\mu=1$ we say that we have full aging.\index{full aging}

\begin{figure}
\centering
\includegraphics[width=.6\linewidth]{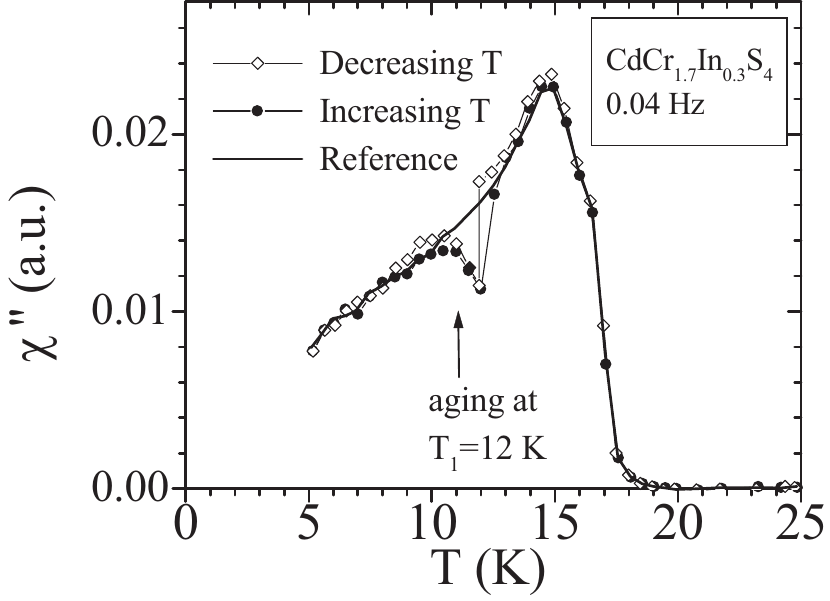}
\caption[Memory and rejuvenation in spin glasses]{%
Memory and rejuvenation effects in an experimental
spin-glass (see text for explanation). Figure taken from~\cite{jonason:98}.
\index{memory|indemph}
\index{rejuvenation|indemph}
\index{aging|indemph}
\index{susceptibility!ac|indemph}
\label{fig:SG-memory}
}
\end{figure}

Notice that the concept of aging 
also applies to the coarsening dynamics  \index{coarsening}
of ferromagnetic systems, although there the process is much faster~\cite{sicilia:08}.
However, let us stress that the differences between the dynamics 
of spin glasses and ferromagnets are much more profound than a simple
change of speed.
As an example of this, let us consider the experiment
in Figure~\ref{fig:SG-memory}, taken from~\cite{jonason:98}.
In this study, the authors took an experimental sample and measured the 
imaginary part $\chi''$ of the ac susceptibility as the temperature was varied.
The applied field had a low frequency of $\omega/2\uppi = 0.04$~Hz. 
In particular, they took the following three steps:
\begin{enumerate}
\item Starting at some temperature above $T_\text{c}$, they cool the system
at a constant slow rate of $0.1$ K/min and then reheat it at the same rate.
This cooling rate (in terms of the logarithmic derivative of the temperature), 
is very small compared with the frequency, $\omega$, which is equivalent
to considering a regime where $t\ll \tw$.
The resulting $\chi''(T)$ is seen to be essentially reversible and represented 
by a thick black line. Notice the cusp at the transition temperature $T_\text{c}\approx 15$~K.
\item Back at the maximum temperature, the system is cooled again at the same rate (curve 
with empty diamonds). Now, however, 
they stop the cooling at $T_1<T_\text{c}$ and let the system age for a few hours. This produces
a `dip' in the $\chi''$ curve.
\item They resume the cooling and the susceptibility quickly returns 
to the reference curve, as if it had not aged at all. This effect is often
named rejuvenation. \index{rejuvenation} 
\item Now, the system is reheated at the same rate, without stopping at $T_1$ (curve with black diamonds).
Surprisingly, the aging dip in the susceptibility  is reproduced. Therefore, the rejuvenation of the 
previous step was not complete: the system has retained some memory of its previous aging.
\end{enumerate}

No longer can we talk of a simple, temperature-independent spin-glass phase as 
in a ferromagnet. A picture of thermally activated barriers is also discarded, since
the time spent at $T_1$ does not contribute to aging at $T_2<T_1$. Instead, we have 
some sort of temperature chaos: \index{temperature chaos}
the appropriate spin-glass phase at $T_1$ looks completely random from the point of view of 
the system at $T_2$.

Still, even if the dynamical behaviour of spin glasses is very complex, 
there is some regularity. We shall see in Section~\ref{sec:SG-coarsening}
how it is possible to make experimentally testable predictions for 
several non-equilibrium quantities.

\subsection{Spin glasses as models for general glassy behaviour}\label{sec:SG-glass}\index{glasses|(}
\index{spin glass!vs. structural glasses|(} \index{viscosity}
Structural glasses are formed when the viscosity of a liquid
grows so much that the latter loses its ability to flow.
The system drops out of equilibrium and its microscopic structure
seems frozen. However, there is no long-range order: a still 
photo of the molecular arrangement would show no great difference 
from that of a dense liquid.

The above picture is, in qualitative terms, much the same as the 
one we painted at the beginning of this chapter, when introducing
the spin-glass transition. This is so despite the fact that the physical 
causes for the two kinds of freezing processes are very different.

This observation is, of course, the motivation for the name 
of `spin glasses' but, more than that, it is also the reason
for much of their physical relevance. Indeed, taking spin glasses
as model systems for investigating general glassy behaviour 
has many advantages, both from an experimental and from 
a theoretical point of view
\begin{itemize}
\item  The nature of glass formation is still poorly
understood, it has not yet been possible to
relate it to a reproducible thermodynamical
phase transition. On the other hand, the freezing
of magnetic moments in spin glasses is known to 
be a phase transition, both in experiments~\cite{gunnarsson:91}, 
where one observes a diverging non-linear susceptibility, \index{susceptibility!spin glass}
and in theoretical models~\cite{palassini:99,ballesteros:00}.
Knowing that one is really below a proper
critical temperature has many advantages from the point of view of analysis
and interpretation of results.
\item  In both classes of systems, the experimental
determination of characteristic length scales is very difficult
(the microscopic correlations are not accessible and one has to 
use an indirect determination via some susceptibility). 
However, spin-glass systems have two advantages:
(\textsc{i}) The size $\xi$ of the glassy domains is much larger in  \index{coherence length}
spin glasses, and therefore easier to access experimentally
(compare, e.g.,~\cite{joh:99, bert:04} with~\cite{berthier:05}). 
(\textsc{ii}) In experiments on structural glasses, the 
measurements are specially delicate and  taken with 
a variety of methods (in~\cite{berthier:05} they
estimate the enthalpy fluctuations from measurements 
of the dielectric susceptibility).
Spin-glass experiments, on the other hand,   \index{SQUID}
use SQUIDs, allowing for a far greater precision
and convenience. 
\item The free-energy barriers in fragile glasses grow
with a power law, while in spin glasses this growth 
is $\sim \log\xi$, as we shall see.
\item Finally, spin glasses are easier to model theoretically
(lattice models with simple interactions), 
making their subsequent analytical and numerical 
investigation more straightforward (although by no means easy,
as we shall see in Section~\ref{sec:SG-EA}).
\end{itemize}

Still, even if we accept that spin glasses are more
convenient model systems and that they have 
a qualitatively similar behaviour to structural glasses, 
we need some concrete, quantitative bridge between
both systems, if only to see how far we can take
the analogy.

In this sense, the study of dynamical heterogeneities \index{dynamical heterogeneities}
can provide a valuable common ground. This topic
has received much experimental attention in the last two decades~\cite{ediger:00,weeks:00,
kegel:00,vidal-russell:00,oukris:10, berthier:11} and, has, in fact, been 
identified as the key to understanding the statistical mechanics of amorphous 
solids. Indeed, even though in structural glasses there is no long-range order, \index{long-range order}
there are non-trivial spatiotemporal fluctuations, caused 
by the wide distribution of relaxation rates across the system.
A thorough understanding of these dynamical heterogeneities can therefore
give many clues on the nature of the glassy phase. 

This heterogenous dynamic behaviour has recently been studied
numerically in spin-glass systems~\cite{castillo:02,castillo:03,jaubert:07,aron:08}.
In this context, the dynamical heterogeneities can be characterised by 
a two-time  correlation length $\zeta(t,\tw)$,
although the time scales that  \index{correlation length!two-time}
have been traditionally reachable in simulations have not permitted
the authors to see correlation lengths greater than a couple of lattice spacings.
At the same time, some recent experimental works suggest~\cite{oukris:10} that $\zeta$
will soon be accessible experimentally. Therefore, there is every indication
that work on dynamical heterogeneities in structural and spin glass systems
is soon going to converge in a unified approach. In the following
chapters we shall dedicate much attention to $\zeta$,
which will be the basis of a finite-time scaling \index{finite-time scaling}
paradigm for the study of non-equilibrium dynamics.

\index{spin glass!vs. structural glasses|)}
\index{glasses|)}
\index{spin glass!experiments|)}
\newpage

\section{The theoretical spin glass: the Edwards-Anderson model}\label{sec:SG-EA}\index{Edwards-Anderson model|(}
The complexity of the experimental spin-glass phase is both an incentive
and a formidable challenge for theoretical physics. We need a system simple enough to 
allow for, at least, some level of analytical treatment yet complex enough
to explain the rich physics observed in experiments.

The first obstacle is the very definition of an order parameter \index{order parameter}
that could signal the onset of the spin-glass transition. By definition of 
spin glass ---Eq.~\ref{eq:SG-long-range}--- we cannot use anything based  on long-range order.
Moreover, the temperature chaos property \index{temperature chaos} \index{long-range order}
implies that the frozen state of the magnetic moments is going to depend on temperature.
Under these conditions, Edwards and Anderson~\cite{edwards:75}  proposed replacing
the spatial correlations of spins at different sites with a time correlation
at the same site. Considering a very large sample with $N$ nodes, we 
have\footnote{We go back to considering Ising spins, $s_\bx=\pm1$.
This does not constitute an unphysical simplification
(consider, for instance, an orthorhombic lattice
where one direction is favoured).} \index{order parameter!Edwards-Anderson}
\index{overlap!spin}
\begin{equation}
q = \lim_{t\to\infty} \frac{1}{N} \sum_\bx \braket{s_\bx(0) s_\bx(t)}_t.
\nomenclature[q]{$q$}{Spin overlap}
\end{equation}
In other words, they considered the overlap between the configuration at two different times
(in equilibrium). In the spin-glass phase, where the spins are frozen, this is non-zero.
In particular, we would expect $q=1$ at $T=0$ and $q\to0$ when $T\to T_\text{c}$ (as in a second-order
transition).

Now we can take the usual step of replacing the time average $\lim_{t\to\infty}\braket{\cdots}_t$ with 
an ensemble average and write
\begin{equation}\label{eq:SG-q1}
q = \frac{1}{N} \sum_\bx \braket{s_\bx}^2.
\end{equation}

Now that we know how to define a workable order parameter, we need a 
more tractable spin interaction than that defined by the RKKY theory. \index{RKKY interaction}
The Edwards-Anderson model is then defined in the simplest possible 
way: as a random mixture of positive and negative nearest-neighbour couplings,
\begin{equation}\label{eq:SG-Edwards-Anderson}
\mathcal H = -\sum_{\langle \bx,\by\rangle} J_{\bx\by} s_\bx s_\by.
\end{equation}
The quenched $J_{\bx\by}$ are extracted from some probability distribution
such that $\overline{J_{\bx\by}}=0$.  The actual shape of the coupling
distribution is not very important (universality), the most popular examples \index{universality}
are Gaussian and bimodal ($\pm J$) couplings. 

\subsection{The mean-field spin glass and the RSB picture}\label{sec:SG-RSB} \index{RSB|(} \index{mean field}
The Edwards-Anderson (EA) model is simple enough to formulate, but its 
solution is quite a different matter. In the DAFF,  \index{DAFF}
one considers the mean-field approximation, which turns
out to be trivial, and attempts to extend the solution 
to finite dimensions using the perturbative renormalisation group
(a difficult task, to be sure). In the case of the EA spin-glass, 
as we shall see, not even the mean-field case is easy. 
In the following we give an outline of the mean-field solution
for the EA spin glass. Our aim is not to present the
full derivation, which is well explained in several places (see, 
e.g., \cite{dotsenko:01,dedominicis:06}).  Rather, we shall simply
give the essential information to understand its testable (and exotic) predictions
for the nature of the spin-glass phase.

The mean-field version of the EA model is
the Sherrington-Kirkpatrick~\cite{sherrington:75}
model \index{Sherrington-Kirkpatrick model|(}
\begin{equation}\label{eq:SG-SK}
\mathcal H = -\sum_{i<j} J_{ij} s_i s_j.
\end{equation}
For the sake of simplicity, we have numbered the lattice sites, 
rather than denoting them by their position vector, as we have 
usually done. The interactions
are  now not only restricted to nearest-neighbours, but exist
between every pair of sites (this can be seen as an infinite-dimensional version of the EA model).

The couplings are extracted from the following conveniently
normalised Gaussian distribution
\begin{equation}\label{eq:SG-SK-J}
p(J_{ij}) = \prod_{i<j} \sqrt{\frac{N}{2\uppi}}\exp\left(-\frac N2 J_{ij}^2\right),
\end{equation}
so we have
\begin{align}
\overline{J_{ij}} &= 0, &
\overline{J^2_{ij}} &= \frac1N.
\end{align}
The factor $1/N$ has been chosen so that the total energy 
at fixed $\beta$ is proportional to $N$, to make the energy density independent
of $N$. 

Now, let us compute the Edwards-Anderson order parameter with the replica trick, 
which we introduced in Section~\ref{sec:INTRO-quenched}.\index{replica trick}
First, let us write this quantity in replica notation. We start
with the basic definition~\eqref{eq:SG-q1}, averaged over disorder
\begin{align}\label{eq:SG-q2}
\overline{q} &= \frac1N \sum_i \overline{\braket{s_i}^2}
= \frac1N \sum_i  \overline{\frac{ \left(\sum_{\{s\}} \ee^{-\beta \mathcal H_J(\{s\})}
 s_i\right)\left(\sum_{\{s\}} \ee^{-\beta \mathcal H_J(\{s\})}
 s_i\right)}{Z^2_J}}.
\end{align}
Now we multiply numerator and denominator by $Z^{n-2}_J$ and follow 
the same steps as in Section~\ref{sec:INTRO-quenched}. The numerator
again depends on $n$ replicas, two of which, ($\mu$ and $\nu$, say) 
are for the two spins that appear explicitly in~\eqref{eq:SG-q2}. 
Therefore,
\begin{align}
\overline{q} = \frac1N \sum_i\lim_{n\to0} \overline{ \sum_{\{s^a\}}
\ee^{-\beta \sum_a \mathcal H_J^{(a)}({s^a})} s_i^\mu s_i^\nu}.
\end{align}
After performing the average over the $J$, we have, 
in the notation of Section~\ref{sec:INTRO-quenched}
\begin{equation} \label{eq:SG-q3}
\overline{q} = \frac1N \sum_i \lim_{n\to0} \braket{s^\mu_i s_i^\nu}_n,
\end{equation}
Our main goal, then, is to compute the effective Hamiltonian that
defines the $\braket{\cdot}_n$.

The replica
partition function is (omitting irrelevant pre-exponential 
factors)
\begin{equation}
Z_n = \overline{Z_J^n} = \sum_{\{s^a\}} \int\mathrm{D}J_{ij}\
\exp\biggl(\beta \sum_{a=1}^n \sum_{i<j} J_{ij} s_i^a s_j^a
- \frac12 N \sum_{i<j} J_{ij}^2\biggr)
\end{equation}
The integration over the $J_{ij}$ can be performed explicitly
with~\eqref{eq:SG-SK-J} and gives
\begin{equation}
Z_n = \sum_{\{s^a\}} \exp\left[ \frac14 \beta^2 Nn 
+ \frac12 \beta^2 N \sum_{a<b}^n \biggl(\frac1N \sum_i s_i^a s_i^b\biggr)^2\right].
\end{equation}
Finally, one linearises the sum over the sites by introducing
the so-called replica matrix $Q_{ab}$,
\begin{equation}\label{eq:SG-Zn}
Z_n = \left(\prod_{a<b}^n\int \dd Q_{ab}\right) 
\sum_{\{s^a\}} \exp\left[ \frac14 \beta^2 Nn - \frac12 \beta^2 N 
\sum_{a<b}^n Q_{ab}^2 + \beta^2 \sum_{a<b}^n \sum_i Q_{ab}s_i^as_i^b\right].
\end{equation} 
By definition,
\begin{align}
Q_{ab}&=Q_{ba}, &
Q_{aa} &= 0.
\end{align}

After some algebra, one can represent~\eqref{eq:SG-Zn}
as 
\begin{equation}\label{eq:SG-Zn1}
Z_n = \int \mathrm{D}Q_{ab} \ee^{\mathcal L_n\{Q_{ab}\}},
\end{equation}
where the effective Lagrangian is
\begin{equation}
\mathcal L_n\{Q_{ab}\} = 
\frac{Nn}4 \beta^2 - \frac{N}{2} \beta^2 \sum_{a<b} Q_{ab}^2
N\log\left[ \sum_{s^a} \exp\left(\beta^2 
\sum_{a<b} Q_{ab} s^a s^b\right)\right]\, .
\end{equation}
Notice that we have eliminated the summation over $i$ and 
constructed a simpler single-site Lagrangian.
Finally, the equilibrium values of $Q_{ab}$
are defined by the equation
$\delta \mathcal L_n/\delta Q_{ab}= 0$ and turn out to be
\begin{equation}\label{eq:SG-Qab}
Q_{ab} = \frac1 N \sum_i \braket{s_i^a s_i^b}_n,\qquad a\neq b,
\end{equation}
where the average $\braket{\cdot}_n$ is defined by the
partition function in~\eqref{eq:SG-Zn1}.

It is here that the computation must stop, unless 
we introduce some ansatz for the replica matrix $Q_{ab}$.
Since the replicas are all equivalent, one could be
tempted to use a symmetric form
\begin{equation}\label{eq:SG-RS}
Q_{ab} = (1-\delta_{ab}) q.
\end{equation}
Using this hypothesis, the computation can be taken 
to the end. In particular, after performing the $n\to0$ 
limit, one finds that the 
value of $q$ is given by the saddle-point equation
\begin{equation}
q = \int_{-\infty}^\infty \frac{\dd z}{\sqrt{2\uppi}}\exp(-z^2/2)
\tanh^2(\beta z \sqrt q).
\index{saddle point}
\end{equation}
This equation has no non-trivial solution above $T=T_\text{c}=1$, \index{critical temperature!Sherrington-Kirkpatrick model}
while for $T<T_\text{c}$ it has a single solution $q_\text{EA}(T)$,
such that $\lim_{T\to 0} q_\text{EA}(T)= 1$. Even more, according to~\eqref{eq:SG-Qab},~\eqref{eq:SG-RS} and~\eqref{eq:SG-q3} we have
\begin{equation}
q_\text{EA}(T) = \frac1N \sum_i \lim_{n\to0} \braket{s^a_i s^b_i}_n
= \frac1N \sum_i \overline{\braket{s_i}^2}. 
\end{equation}
which is just the Edwards-Anderson order parameter that  we
introduced earlier. Therefore, it would seem that the 
replica-symmetric solution reproduces just the behaviour that
we want.

There are severe problems, though. The most serious being
that the entropy becomes negative at low temperatures
and that the spin-glass phase becomes unstable in 
the entire $T<T_\text{c}$ region.

Clearly, the replica symmetry must be broken somehow. 
The way to do this was explained by Parisi~\cite{parisi:79b,parisi:80}.
We consider the $n\times n$ replica matrix $Q_{ab}$. Before
breaking the replica symmetry, $Q_{ab}=(1-\delta_{ab})q_0$. We now 
divide the matrix into constant blocks of size $m_1\times m_1$
and set each diagonal block as a submatrix whose 
off-diagonal elements are all $q_1$. All the elements in the
off-diagonal blocks remain in their original value $q_0$.
We then take a second replica symmetry breaking (RSB) step
and subdivide the diagonal blocks, introducing a new overlap $q_2$.
See Figure~\ref{fig:SG-RSB} for a sketch of these two first steps.
This process is continued indefinitely. 
\begin{figure}
\small 
\begin{eqnarray}
&\left(\begin{array}{cccccccc}
0 & & & & \multicolumn{4}{c}{\multirow{4}{*}{\Huge $q_0$}}\\
 & 0 & &  & \\
 &  & 0 &  & \\
 & &  & 0 & \\
\multicolumn{4}{c}{\multirow{4}{*}{\Huge $q_0$}} & 0 &  &  &  \\
 & & & &   & 0 &  &   \\
& & &  &  &  & 0 &   \\
 & & & &  &  &  & 0
\end{array}\right) \quad \longrightarrow \quad 
\left(\begin{array}{cccc|cccc}
0 &  &\multicolumn{2}{r|}{\multirow{2}{*}{\Large $\ \ q_1$}} & \multicolumn{4}{c}{\multirow{4}{*}{\Huge $q_0$}}\\
 & 0 & &  & \\
\multicolumn{2}{c}{\multirow{2}{*}{\Large $q_1$}}  & 0 & & \\
 &  &  & 0 & \\
\hline
\multicolumn{4}{c|}{\multirow{4}{*}{\Huge $q_0$}} & 0 & & \multicolumn{2}{c}{\multirow{2}{*}{\Large $q_1$}}\\
 & & & &  & 0 &   &  \\
& & & & \multicolumn{2}{c}{\multirow{2}{*}{\Large $q_1$}}  & 0 &   \\
 & & & &  & &  & 0
\end{array}\right) \quad \longrightarrow \quad& \\
&\left(\begin{array}{cc|cc|cc|cc}
0 & q_2 &\multicolumn{2}{c|}{\multirow{2}{*}{\Large $\ \ q_1$}} & \multicolumn{4}{c}{\multirow{4}{*}{\Huge $q_0$}}\\
q_2 & 0 & &  & \\
\cline{1-4}
\multicolumn{2}{c|}{\multirow{2}{*}{\Large $q_1$}}  & 0 &q_2 & \\
 &  & q_2 & 0 & \\
\hline
\multicolumn{4}{c|}{\multirow{4}{*}{\Huge $q_0$}} & 0 &q_2 & \multicolumn{2}{c}{\multirow{2}{*}{\Large $q_1$}}\\
\multicolumn{4}{c|}{} &q_2  & 0 &   &  \\
\cline{5-8}
\multicolumn{4}{c|}{} & \multicolumn{2}{c|}{\multirow{2}{*}{\Large $q_1$}}  & 0 & q_2  \\
\multicolumn{4}{c|}{} &  & &q_2  & 0
\end{array}\right)  \quad \longrightarrow \quad {\Huge \cdots}&
\end{eqnarray}
\caption[Replica symmetry breaking]{%
Sketch of the first two steps of replica symmetry breaking.
\index{RSB|indemph}
\label{fig:SG-RSB}}
\end{figure}

Notice that the replica equivalence property is conserved, 
because all rows have the same elements (in different order). \index{replica equivalence}
Now the values of the matrix elements can be 
described by the pdf
\begin{align}
p(q) &= \frac{1}{n(n-1)} \sum_{a\neq b} \delta(Q_{ab} - q)\\
   &= \frac{n}{n(n-1)} \bigl[ (n-m_1) \delta(q-q_0)
+ (m_1-m_2) \delta (q-q_1)  \\
& \qquad  \qquad\qquad+(m_2-m_3) \delta(q-q_2) + \ldots
\bigr]
\end{align}
Finally, we have to take the limit $n\to0$,
\begin{equation}
p(q)  = m_1 \delta(q-q_0) + (m_2-m_1) \delta(q-q_1) 
+ (m_3-m_2) \delta(q-q_2) + \ldots
\end{equation}
This limit has inverted the order of the $m_i$, so
now we have $0<m_1<m_2<\ldots<1$.
In the limit of infinite RSB steps we obtain a continuous
variation, so $q_k\to q(x)$, with $x\in [0,1]$.

In other words, the spin-glass order parameter is no
longer a number, but a function. In particular,
\begin{equation}
\frac{\dd x}{\dd q} = p(q).
\end{equation}
There are infinitely many states, whose
overlap $q'$ is $0\leq q'\leq q_\text{EA} = q(1)$.
It can be seen that the RSB method cures the problems
of the symmetric ansatz and produces a stable spin-glass
phase. In fact, it has been proved rigorously
that the RSB scheme produces the correct
free energy for the Sherrington-Kirkpatrick
model~\cite{talagrand:06}.

Some of the main physical predictions of this mean-field 
solution are  (we shall see more in the following 
sections)
\begin{itemize}
\item Below the critical temperature, there 
are infinitely many states. This 
is represented by a non-trivial probability
distribution of the spin-glass order
parameter $q$, which can take any value
in $[-q_\text{EA},q_\text{EA}]$ with 
non-zero probability density.
\item The states are organised in a hierarchical structure,
giving rise to an ultrametric overlap space. This is best \index{ultrametricity}
seen by representing the RSB process as a branching tree.
At lowest order, $q_0$, all the elements are equal and 
contained in one branch. As we break up the ensemble of
states into more and more clusters we keep increasing the 
value of $q$. Then, to find the overlap $q_{ab}$ of two 
replicas belonging to different branches we simply
have to follow up the tree up to the encounter point.
Clearly enough, we have an ultrametric space 
where $q_{ac} \geq \min\{q_{ab},q_{bc}\}$.
This hierarchical organisation might explain 
the temperature chaos effect \index{temperature chaos}
observed in the experiments.
\item In the presence of an external magnetic
field, the spin-glass phase is not destroyed.
 \index{Sherrington-Kirkpatrick model|)}
\end{itemize}

Such is the mean-field solution to the EA model.
In principle, one could try to extend this 
description to finite dimensions by means of a perturbative
renormalisation group. One would expect
the $D=3$ spin-glass phase to share the essential
properties of the mean-field solution (infinity
of states, etc.) but differ in some details. 
In this way, one obtains the so-called RSB picture 
of the spin-glass phase
 (see~\cite{dedominicis:98,dedominicis:06,mezard:87,marinari:00}
 for some analytical results and more detailed physical predictions
of this theory).
\index{RSB|)}

\subsection{The droplet picture}\label{sec:SG-droplet}\index{droplet picture|(}
The above described RSB picture is not generally 
accepted as a faithful description of the spin-glass
phase in $D=3$.  The most popular alternative is 
the droplet theory~\cite{bray:84,
bray:87,mcmillan:84,fisher:86,fisher:88}, 
which stems from a Migdal-Kadanoff approximation
\index{Migdal-Kadanoff approximation}
to the EA model.

The droplet approach paints a completely different 
picture of the spin-glass phase. Instead of 
having a non-trivial order parameter distribution, 
it considers only two equilibrium states, with $q=\pm q_\text{EA}$.

From a phenomenological point of view, the model assumes
that the lowest-energy excitations of the system 
are compact domains, or droplets, of coherently flipped
spins. In particular, the typical 
free energy of an excitation of size $L$ 
scales with $L^y$, where $y$
is the so-called stiffness exponent. \index{stiffness exponent}
\nomenclature[y]{$y$}{Stiffness exponent of the droplet picture}
Alternatively, zero-energy droplets of size $L$ occur
with probability $L^{-y}$.  It can be proved that 
\begin{equation}\label{eq:SG-y-bound}
y < (D-1)/2.
\end{equation}

Notice that the stiffness exponent must be positive
in order for there to be a stable spin-glass phase.
The issue has been examined numerically in $D=3$, resulting
in values of $y\approx 0.27$~\cite{carter:02}
or $y\approx0.24$~\cite{boettcher:04,boettcher:05}.
In $D=2$ the exponent is negative and the spin-glass
phase disappears.

One of the main consequences of this picture is that the spatial
correlations decay with $y$,
\begin{equation}
C(r_{ij}) = \overline{\braket{s_is_j}^2}-
\overline{\braket{s_i}^2} \overline{\braket{s_j}^2}
\sim \frac{1}{r_{ij}^y}\ .
\end{equation}
For this reason, $\overline{q^2} - \overline{q}^2\to0$, that is, 
the $p(q)$ that we introduced in the previous section 
has zero variance. This is translated into 
the previously mentioned property of a trivial overlap, 
where $p(q) = \delta(q^2-q_\text{EA}^2)$.

Another point of departure of the droplet and RSB pictures
is the behaviour in an external magnetic field. 
The RSB predicts that the spin-glass phase can 
survive a small magnetic field. However, in the droplet 
picture we can generalise the Imry-Ma argument \index{Imry-Ma argument}
of Chapter~\ref{chap:daff-canonical}. The energy barrier
for flipping the droplet is $L^y - hL^{D/2}$. 
Since $y<(D-1)/2$, this makes the spin-glass phase
unstable in the presence of even an infinitesimal $h$.

Finally, let us mention that even if the droplet picture
is simple compared to the RSB one, it still can 
explain the complex physics of the experimental spin glass.
In particular, the temperature chaos \index{temperature chaos}
property is expected.  Indeed,  the droplet 
boundary scales as $L^{D_\text{s}}$, with $D-1\leq D_\text{s}<D$.
Therefore in order to produce the $L^y$ scale dependence, with $y<D_\text{s}$, different parts
of the droplet's boundary must give alternating contributions.
This gives rise to a delicate balance of energy and entropy
that is easily upset with a small change in temperature.

\index{droplet picture|)}
\subsection{The geometry of the excitations and the TNT picture}\label{sec:SG-TNT}
\index{TNT picture|(}
A final interesting issue concerns the geometry of the activated
domains. In the droplet picture we have assumed these are compact,
so their surface scales as $L^{D_\text{s}}$, 
where $D-1\leq D_\text{s}<D$, and their surface-to-volume
ratio goes to zero as $L\to\infty$. Furthermore, there are no
zero-energy excitations.

In the RSB picture, however, we do have to consider excitations 
with zero energy cost. Let us denote by $\{ s_\bx^{(0)}\}$
the original spin configuration and by $\{s_\bx^{(1)}\}$
the excited one. Then, we consider the
link overlap (as opposed to the spin overlap $q$)
\begin{equation}\label{eq:SG-link-overlap}
Q_\text{link} = \frac{1}{N_\text{l}} \sum_{\braket{\bx,\by}} 
s_\bx^{(0)} s_\by^{(0)} s_\bx^{(1)} s_\by^{(1)},
\nomenclature[Qlink]{$Q_\text{link}$}{Link overlap}
\index{overlap!link}
\end{equation}
where $N_\text{l}$ is the number of links. This may seem 
a pointless definition from the mean-field
point of view, since in the Sherrington-Kirkpatrick model
$Q_\text{link} = q^2$ trivially, but we shall see how 
we can get valuable information from this observable.
We note that in $D=3$ we no longer have this simple
relation, but the RSB picture still expects $Q_\text{link}$ to 
have a non-trivial distribution (i.e., be some 
function of $q^2$).

Let us now consider an excitation where we pass
from a ground state to another by
flipping $\mathcal O(N)$ spins. 
The link overlap, unlike $q$,
only changes for those links across the surface 
of the flipped domain. Its relative change is, therefore,
of order $L^{D_\text{s}-D}$.
In order to make this compatible with
a non-trivial distribution for $Q_\text{link}$
it follows that for RSB systems $D_\text{s}=D$. In other 
words, the excitations are space-filling.

The above considerations led some authors to consider 
the issue of zero-temperature excitations and their
geometry, as a means of deciding between the droplet
and RSB pictures. In the year 2000, Palassini
and Young~\cite{palassini:00} and Krzakala and Martin~\cite{krzakala:00}
reached the following conclusions from their numerical
simulations
\begin{itemize}
\item There exist excitations involving $\mathcal O(N)$ 
spins but having finite energy.
That is, the spin overlap distribution is non-trivial (as in RSB).
\item The surface of these excitations is $D_\text{s}<D$, 
so the link overlap distribution is trivial (as in the  droplet picture).
In particular~\cite{palassini:00,palassini:03}, $D-D_\text{s}\approx 0.44$
in $D=3$.
\end{itemize}
\index{TNT picture}
Therefore, this trivial-non-trivial (TNT) picture  emerges
as an intermediate alternative to droplet and RSB.
\index{TNT picture|)}

\subsection{Coarsening vs. non-coarsening dynamics}\label{sec:SG-coarsening}
\index{coarsening}
Throughout this chapter, there has been a disconnection between experiments 
and theory in that the former were considered in a non-equilibrium
regime and the latter was explained in terms of the equilibrium spin-glass phase.
The usual justification for this approach is that the equilibrium structures, though
unreachable in finite time, do condition the dynamical evolution of the system.

Still, the explicit theoretical study of the non-equilibrium evolution
can be very rewarding. To this end, the first (and most important) step
in the understanding of complex phenomena such as those of Figure~\ref{fig:SG-memory}
is surely understanding what happens during the isothermal aging.

In this sense, one considers the direct quench  experimental protocol~\cite{berthier:02,jimenez:05,
marinari:00b,kisker:96,rieger:93,jimenez:03,perez-gaviro:06,castillo:02,castillo:03,jaubert:07}.
We start by considering a starting configuration at a very high temperature
(i.e., completely random orientation of the spins) and instantaneously cool
it down to the working temperature $T<T_\text{c}$. We then let the system
age for some waiting time $\tw$ and probe its properties \index{aging}
at a later measuring time $t+\tw$. In these conditions, the main feature of the 
\nomenclature[tw]{$\tw$}{Waiting time}
\nomenclature[t]{$t$}{Measurement time}
dynamics is the formation and slow growth of coherent domains, whose
size is characterised by a coherence length $\xi(\tw)$. We remark \index{coherence length}
that this coherence length is accessible to experiments, through estimates
of Zeeman energies~\cite{joh:99}, where its extremely slow grow
results in $\xi\sim100$ lattice spacings, even close to $T_\text{c}$ 
and for several hours of evolution (the evolution grows ever slower as
we decrease the temperature).

This much is agreed by all, but the controversy soon begins when we consider
the nature of these coherent domains. On the one hand, in the droplet picture
the dynamics consists in the growth, or coarsening, of compact domains where the spin
overlap takes one of its two possible equilibrium values, $q=\pm q_\text{EA}$.
In this sense, the droplet spin glass behaves as a disguised ferromagnet.
The aging dynamics of all coarsening systems is qualitatively the same.

In particular, if one measures times and lengths in units of $\xi(\tw)$, no
non-equilibrium observable should depend on the quenched randomness and 
the resulting scaling functions should be identical for all 
coarsening systems~\cite{fisher:88}. This so-called superuniversality \index{superuniversality}
property then treats in exactly the same way systems as diverse  \index{coarsening}
as the ferromagnetic or site-diluted Ising model, \index{Ising model} \index{DAFF}
or even the DAFF that we studied in Part~\ref{part:daff}~\cite{aron:08}.
According to the droplet picture then, the isothermal aging of the spin 
glass would be no different.\footnote{%
We stress that we are talking of isothermal aging. When the temperature is varied,
the behaviour of a droplet spin glass is much more complicated than that
of a ferromagnet, due to temperature chaos. \index{temperature chaos}}

We are not aware of any investigation of the dynamical consequences of the TNT \index{TNT picture}
picture, but an antiferromagnetic analogy suggests that TNT systems will also
show coarsening behaviour.

The study of the dynamics in the mean-field limit was thoroughly 
undertaken in the 1990s~\cite{cugliandolo:93,cugliandolo:94}.  The emerging
picture for an RSB spin glass resulted much more complex than that 
of coarsening systems. The main difference is that in RSB systems
there exist equilibrium states with $q=0$, so the dynamics remains forever \index{Heisenberg ferromagnet} 
in this sector. Furthermore, the replicon, \index{replicon}
a Goldstone model analogous to magnons in Heisenberg ferromagnets, \index{magnons}
is present for all $T<T_\text{c}$~\cite{dedominicis:98,dedominicis:06}. 
As a consequence, the spin overlap is expected to vanish even inside
each domain, in the limit of large $\xi(\tw)$. Finally, the spin
overlap $q$ is not a privileged observable: the link overlap
exhibits the same aging behaviour (overlap equivalence). \index{overlap equivalence}

\begin{figure}
\centering
\includegraphics[width=0.5\linewidth]{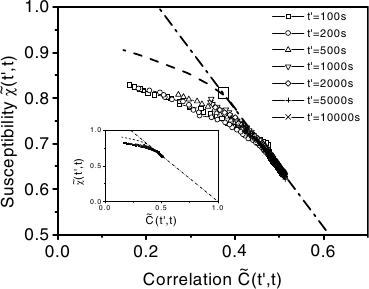}
\caption[Violation of fluctuation-dissipation
in an experimental spin glass]{Violation 
of the fluctuation-dissipation theorem (dotted-dashed line)
in an experimental spin glass (from~\cite{herisson:02}, 
see text for discussion).
In the figure, $t'$ corresponds to our $\tw$ and $t$ 
corresponds to our $t+\tw$.
\index{fluctuation-dissipation|indemph}
\label{fig:SG-FDT}}
\end{figure}

A good example of the relevance of these predictions 
for experimental work is the case of the mild 
violations of the \index{fluctuation-dissipation} 
classic fluctuation-dissipation theorem (FDT). This result, proven in~\cite{callen:51,kubo:57},
makes it possible to predict  the system's response 
to an external field from its equilibrium dynamics, but 
only if the system was already in 
equilibrium when the field was applied.
Still, the pattern of violations of the FDT for spin glasses
admits a very simple parameterisation.  Cugliandolo and Kurchan~\cite{cugliandolo:94b,cugliandolo:97}
proposed characterising the deviations from the FDT behaviour by studying a fluctuation-dissipation
ratio, representing the response as a function of the system correlations.

This procedure was followed experimentally in~\cite{herisson:02} and the 
result is represented in Figure~\ref{fig:SG-FDT}. In it, the authors 
consider the response function at time $t+\tw$ of a field that was 
switched off at time $\tw$, denoted by $\tilde\chi(t,\tw)$. This is 
represented against the temporal autocorrelation of the system $\tilde C(t,\tw)$.\footnote{%
The experimental definition 
and measurement of these quantities is somewhat involved,
 see~\cite{herisson:02} for the details.}
Should the systems obey FDT, the resulting plot $\tilde\chi(\tilde C)$
should fall on a straight line (dotted-dashed line in the figure).
In the spin glass, the linear behaviour is indeed reproduced for short
measuring times $t$, but the curves eventually deviate from this prediction 
at a $\tw$-dependent branching point.

In the  $\tw\to\infty$ limit, the RSB and droplet pictures expect a different
pattern of violations. In both theories, the measured curved should branch off
from the FDT prediction at $\tilde C = q_\text{EA}$. However, in the RSB
framework one expects that $\lim_{\tw\to \infty} \dd \tilde \chi/\dd \tilde C$
would correspond to the Parisi function $x(q)$ that
we introduced in Section~\ref{sec:SG-RSB}~\cite{franz:98,franz:99}.
In the droplet picture, on the other hand, the curve of violations should
be horizontal. In Figure~\ref{fig:SG-FDT} the authors have attempted
this extrapolation (dashed line).  

\section{The Edwards-Anderson spin-glass in numerical simulations}\label{sec:SG-numerical}
In order to be made quantitative, the different theories for the spin-glass
phase need input from numerical simulations. In this sense one can take 
two complementary approaches
\begin{itemize}
\item Simulate a very large system with a Monte Carlo update 
that reproduces the physical evolution (such as heat bath or Metropolis
dynamics). \index{heat bath algorithm} \index{Metropolis algorithm}
As long as $\xi(\tw)\ll L$, one can consider to be observing 
the same thermodynamical-limit behaviour as in a physical sample. \index{thermodynamical limit}
Then, take a long-$\tw$ limit to predict the physical evolution.
\item Simulate smaller lattices with some kind of optimised dynamics 
(typically parallel tempering) \index{parallel tempering}
to equilibrate the systems. Then perform a finite-size scaling
study to perform a large-$L$ extrapolation. Use the information
thus obtained on the equilibrium spin-glass phase to predict
the experimental non-equilibrium behaviour.
\end{itemize}

We stress that the goal of numerical work on the 
Edwards-Anderson model is ultimately to understand the nature of the spin-glass
phase. In this sense, the situation is complementary to that of the DAFF. \index{DAFF}
In the latter, the low-temperature phase is trivial (antiferromagnetic order), but 
the phase transition itself is poorly understood. In the spin glass, the phase
transition has been thoroughly examined numerically~\cite{palassini:99,ballesteros:00,
hasenbusch:08,hasenbusch:08b},\footnote{Actually, this is true
for the Ising spin glass that we consider here. For the version
with vector spins (Heisenberg spin glass), the nature \index{Heisenberg model}
of the phase transition is still controversial (see, e.g.,  \index{spin glass!Heisenberg}
\cite{fernandez:09b,viet:10}).} and it is the low-temperature phase, as far
from the critical point as possible, that has the most interest.

Of course, the extreme sluggishness of the dynamics poses a
problem for numerical investigations. For instance, the typical
time scales of experiments on spin glasses range from a few seconds 
to a few hours, but each MCS is equivalent to only 
$10^{-12}$ s~\cite{mydosh:93}. Therefore, one has to perform
an enormous number of lattice updates even to begin to approach
the physically relevant regime. In this sense, 
non-equilibrium simulations have reached only about $10^{-4}$ s.
In the equilibrium case, the lattice sizes that one can 
thermalise are very small, so the data are always suspect of
being dominated by finite-size effects and not
considered representative of the real physics.

On the other hand, the understanding of the spin-glass phase
is not yet ripe for a tethered study, because \index{tethered formalism!spin glass}
the appropriate reaction coordinate (which must be sample-dependent)
is not known. Still, some of the techniques developed
in the previous chapters will prove to be useful.

\subsection{Introduction to our work with \textsc{Janus}}
In the following chapters  we shall take a brute-force approach
to the numerical study of the spin-glass phase. In particular, we 
have carried out our simulations using  \textsc{Janus}, \index{Janus@\textsc{Janus}}
a special-purpose machine optimised for Monte Carlo
simulations of spin systems. For this restricted class of problems, 
\textsc{Janus} can outperform conventional computing architectures
by several orders of magnitude, thus providing a qualitative
jump in the time and length scales that can be accessed. 
In particular, we have been able to follow
the non-equilibrium dynamics up to times of $0.1$~s,
at the threshold of the experimental scale, and we have thermalised
lattices up to $L=32$ down to $T=0.64T_\text{c}$.

The characteristics of \textsc{Janus} are very briefly explained
in Appendix~\ref{chap:janus}. The Ising spin-glass simulations
carried out with \textsc{Janus}, whose parameters we report
in Appendix~\ref{chap:runs-sg}, have produced four
physics publications so far: \cite{janus:08b,janus:09b,janus:10,janus:10b}.
This work is the fruit of the collaboration of a large number
of people, including physicists and engineers. To reflect this, 
the following chapters include only those physical
studies where the author  carried out a major fraction
of the work.
In particular, we leave out the results corresponding to
Sections~5.2 and~6 in~\cite{janus:09b}, as well as those
corresponding to Sections~7.2 and~7.3 in~\cite{janus:10}.
Fortunately, the quoted sections are relatively independent
from the rest of the works, so our presentation will be
self-contained. 

We note that the order of presentation of our results in this and the next
chapter is markedly different from their original
development. In our original work  we were 
of course constrained by the running of the simulations
(we carried out the non-equilibrium ones first, while
our equilibrium simulation campaign took over a year).
In this dissertation, we shall try to present the results
in a logical fashion, rather than in a chronological one.

\index{Edwards-Anderson model|)}

\chapter[The statics-dynamics correspondence and finite-time scaling]{The statics-dynamics correspondence and the finite-time scaling paradigm}\label{chap:sg-2} \index{spin glass!dynamics|(}
In this chapter we tackle the second of our main topics: the relation 
between the (experimentally unreachable) equilibrium
phase and the off-equilibrium dynamics in the context of 
spin glasses.

After introducing the relevant physical observables and correlation
functions, we try to characterise the isothermal aging dynamics 
of the $D=3$ Edwards-Anderson spin-glass.  We begin by using
only non-equilibrium simulations and study issues
like the growing characteristic lengths or the thermoremanent
magnetisations. 

Eventually, however, the non-equilibrium study
faces an imposing stumbling block: in order
fully to understand the reasons for the dynamical behaviour
at finite time one must consider extrapolations to infinite time.
In particular, we observe a very interesting crossover
behaviour in the so-called dynamical heterogeneities 
(Section~\ref{sec:SG-dynamical-heterogeneities}) that we will 
be unable to quantify.

It is then that we introduce a complementary approach: that
of finite-size equilibrium simulations. First, we promote
the relation between the equilibrium phase and the dynamics
to a quantitative level, through a time-length dictionary.
This will allow us to introduce one of the most important
concepts (and results) of this study: the notion 
that experimental non-equilibrium physics is equivalent to 
the equilibrium behaviour, not of the thermodynamical limit,  \index{experimental scale}
but of a more modest lattice of $L\sim100$.

Finally, we resume the analysis of dynamical heterogeneities
from an equilibrium point of view. This allows us to 
perform a fully quantitative study and, through it, to
take the statics-dynamics one step further:
the finite-time scaling paradigm.

Throughout this chapter, we shall focus only on those
aspects of the spin-glass dynamics and equilibrium
about which the different theoretical pictures 
are in full agreement. For instance, we do not try
to decide whether the dynamics is coarsening-like
or not. We shall see that this neutral approach can 
actually take us very far.

In Chapter~\ref{chap:sg-3} we shall, on the other hand,
address the RSB-droplet-TNT controversy directly.
We shall then try to decide which of these is a better
description of the spin-glass phase.

\section{Physical observables for spin-glass dynamics}\label{sec:SG-observables-dynamics}
We work on the Edwards-Anderson model introduced in the previous
chapter, 
\begin{equation}
\mathcal H = -\sum_{\langle \bx,\by\rangle} J_{\bx\by} s_\bx s_\by.\tag{\ref{eq:SG-Edwards-Anderson}}
\end{equation}
We use bimodal couplings, $J_{\bx\by}=\pm1$ with 
$50\%$ probability. For this Hamiltonian, the phase transition \index{critical temperature!Edwards-Anderson model}
occurs at $T_\mathrm{c} = 1.109(10)$~\cite{hasenbusch:08,hasenbusch:08b}.

We study the dynamics in the direct quench protocol explained
in the previous chapter. We use heat-bath dynamics,
\index{heat bath algorithm}
which is in the universality class 
of physical evolution. We take $1$ MCS to correspond 
to $10^{-12}$ seconds~\cite{mydosh:93}.
When studying the non-equilibrium dynamics we shall consider that
we are in the thermodynamical limit, even if we simulate
in a finite lattice ($L=80$ in our case). In practice, this 
approximation is good as long as the length scale of the coherent
domains that grow during the evolution remains much smaller
than the system size. We shall check this issue
in Section~\ref{sec:SG-finite-size}

We represent by $O(\tw)$ the time evolution of some observable
$O$ for a single sample, and by $\overline{O(\tw)}$ its disorder
average. In principle one would have to average over many 
thermal histories. However, the main source of error is sample-to-sample
fluctuation, so it is better to run more samples
than more histories for each one. Still, we run two independent
thermal histories for each sample (two real replicas), for reasons
that will become apparent below. Whenever one observable
can be averaged over the two replicas, we shall do so, even if we
do not indicate it explicitly.

The first issue we have to consider when defining physical
observables is the gauge symmetry induced by the quenched
average~\cite{toulousse:77}. \index{gauge symmetry}
Indeed, the Hamiltonian is unchanged under the transformation
\begin{align}
s_\bx & \longrightarrow \alpha_\bx s_\bx,\\
J_{\bx\by} &\longrightarrow \alpha_\bx J_{\bx\by} \alpha_\by,
\end{align}
where $\alpha_\bx=\pm1$. In principle, this would mean that
one should have to consider for each observable its 
gauge average $2^{-L^D}\sum_{\{\alpha_\bx\}} O(\{\alpha_\bx s_\bx\})$.
Notice, however, that the gauge-transformed couplings
are as probable as the original ones, so the disorder average of
$O$ is the same as its average over the gauge orbit.

We need, therefore, gauge invariant observables. The usual way out of this problem is introducing real replicas. \index{replicas!real}
These are defined as statistically independent systems $\{s^{(i)}_\bx\}$
\nomenclature[Real replicas]{Real replicas}{Copies of the system that evolve with the same 
set of $J_{\bx\by}$}
that evolve with the same set of $J_{\bx\by}$. \index{overlap!spin}
Their overlap field is 
\begin{equation}
q_\bx = s_\bx^{(1)}(\tw)s_\bx^{(2)}(\tw),
\nomenclature[qx]{$q_\bx$}{Overlap field}
\end{equation}
obviously gauge-invariant.

The averaged density of the overlap field defines the 
spin-glass order parameter, the spin overlap $q$
\begin{equation}
q(\tw) = \frac1N \sum_\bx q_\bx(\tw).
\end{equation}
We can also consider the spin-glass susceptibility
\begin{equation}\label{eq:SG-chi-SG}
\chi_\text{SG}(\tw) = N \overline{q^2(\tw)},
\nomenclature[chiSG]{$\chi_\text{SG}$}{Spin-glass susceptibility}
\end{equation}
which, as can be seen using fluctuation-dissipation
relations,\index{fluctuation-dissipation} is
essentially the same as the non-linear 
susceptibility $\chi_4$ ---see~\eqref{eq:ISING-chi-2n}
and~\eqref{eq:ISING-m-h-Taylor}. In fact
(in equilibrium)
\begin{equation}
6\chi_4 = \chi_\text{SG} - 2/3.
\end{equation}

\subsection{Temporal correlations}\label{sec:SG-temporal-correlations}
Since we are working on an aging system, \index{aging}
we  need to consider two time scales, as we commented
in Chapter~\ref{chap:sg}. We are particularly
interested in the overlap between the configurations
at the waiting time $\tw$ and at the measuring time
$t+\tw$,
\begin{equation}\label{eq:SG-c}
c_\bx(t,\tw) = s_\bx^{(i)}(\tw)s_\bx^{(i)}(t+\tw).
\end{equation}
Notice that this quantity needs only one replica. 

We can use the two-time  overlap to define the 
temporal correlation function \index{correlation function (dynamics)!temporal}
\begin{equation}\label{eq:SG-Cttw}
C(t,\tw) = \overline{\frac1N \sum_\bx c_\bx(t,\tw)}.
\nomenclature[Cttw]{$C(t,\tw)$}{Two-time  correlation function}
\end{equation}
For fixed $\tw$, $C(t,\tw)$ is strictly decreasing in $t$, 
indicating the gradual but very slow decorrelation 
in the system. In fact, for fixed $\tw$,
the $C(t,\tw)$ can be interpreted as the thermoremanent  \index{magnetisation!thermoremanent}
magnetisation (Section~\ref{sec:SG-thermoremanent}).
This is because we can make a gauge transformation 
from the uniform configuration resulting from a strong
magnetic field to the random initial configuration
in our simulations. Therefore, the overlap with this 
random configuration plays the part of a magnetisation.

In the pseudoequilibrium regime, $t\ll\tw$, the 
(real part of the) magnetic susceptibility at 
frequency $\omega = 2\uppi/T$ is given by the 
fluctuation-dissipation formula \index{fluctuation-dissipation}
$\bigl(1-C(t,\tw)\bigr)/T$.

For any finite $\tw$, $\lim_{t\to\infty} C(t,\tw)=0$. However, 
in equilibrium $\tw\to\infty$, the overlap has a finite value
below $T_\mathrm{c}$. In particular, if we define
the stationary, or translationally invariant, part 
of the correlation,
\begin{equation}
C_\infty(t) = \lim_{\tw\to\infty} C(t,\tw),
\end{equation}
we get
\begin{equation}\label{eq:SG-limit-Cinf}
q_\text{EA} = \lim_{t\to\infty} C_\infty(t).
\end{equation}
This equation defines the value of 
the Edwards-Anderson order parameter.\footnote{Recall that in a droplet
setting this is the only equilibrium value of $q$, while  in 
an RSB setting it is the maximum of the Parisi function $q(x)$.}
 The dynamical computation of $q_\text{EA}$
from this limit is notoriously
difficult~\cite{perez-gaviro:06,iniguez:97}, even with the extremely
long times simulated on \textsc{Janus} (Section~\ref{sec:SG-qEA-dynamics}).
In this chapter, we shall consider an alternative, equilibrium, approach
(Section~\ref{sec:SG-phase-transition}). \index{order parameter!spin glass}
\index{order parameter!Edwards-Anderson}

The introduction of the stationary part of $C(t,\tw)$
can be generalised to a division of this correlation function in several
time sectors. We can define the $\mu$ sector by
\begin{equation}
C_\infty^{(\mu)}(s) = \lim_{\tw\to\infty}(s t^\mu_{\tw}, \tw).
\end{equation}
The translationally invariant sector would thus correspond
to $\mu=0$. Notice that $q_\text{EA}<C^{(0)}_\infty(s)<1$.
In order to cover the whole range of values $C\in[0,1]$, at least
another sector must be considered. The simplest (and experimentally
supported) one is the full-aging regime~\cite{rodriguez:03}, 
obtained by setting $\mu=1$. If full-aging holds, one would
\index{full aging}
expect $C^{(1)}_\infty(s)$ to cover the whole
range from $q_\text{EA}$ to $0$. We shall examine these claims
in Section~\ref{sec:SG-aging}. 

We finally note that some authors~\cite{castillo:02,castillo:03}
use different definitions
for these functions, so care must be exercised when 
consulting the literature. In particular, often $t$
is taken to be our $t+\tw$ and the $C_\infty(t)$ 
is defined so that it tends to zero (i.e., subtracting
$q_\text{EA}$ from it).

\subsection{Spatial correlations}
\index{correlation function (dynamics)!spatial}
The spin-glass dynamics is characterised by the growth 
of coherent domains both in the coarsening 
and the non-coarsening pictures (even if their shape and nature differ).
These can be studied through the spatial autocorrelation of the 
overlap field
\begin{equation}
C_4(\boldsymbol r,\tw) = \overline{\frac1N \sum_\bx q_\bx(\tw) q_{\bx+\boldsymbol r}(\tw)}\ .
\nomenclature[C4]{$C_4(\boldsymbol r,\tw)$}{Spatial autocorrelation of the overlap field (dynamics)}
\end{equation}
In particular, the long-distance decay of $C_4(\boldsymbol r, \tw)$
defines the coherence length $\xi(\tw)$ \index{coherence length}
\nomenclature[xitw]{$\xi(\tw)$}{Coherence length (dynamical)}
\begin{equation}\label{eq:SG-C4-long-distance}
C_4(\boldsymbol r,\tw) \xrightarrow{\ \ r\to\infty\ \ } \frac{1}{r^a}f\bigl(r/\xi(\tw)\bigr).
\nomenclature[a]{$a$}{Replicon exponent, equivalent to $\theta(0)$}
\end{equation}
At the critical point, $a$ 
is non-zero and related to the anomalous dimension, 
through $a(T_\mathrm{c}) = D-2+\eta$.
In $D=3$, $a(T_\mathrm{c})=0.625(10)$~\cite{hasenbusch:08,hasenbusch:08b}.

Below $T_\mathrm{c}$, the value of 
$a$ characterises the dynamics. For fixed $r/\xi(\tw)$, 
the correlation function of a coarsening system does not tend
to zero when $r\to\infty$, since the domains are compact. In 
a system described by the RSB picture, however, there exist 
equilibrium states with $q=0$, so the dynamics remains
forever with a vanishing order parameter. This is related 
to the existence of the replicon mode, a Goldstone
\index{replicon}%
\index{replicon exponent}%
\index{Goldstone boson}%
\index{magnons}%
\index{Heisenberg ferromagnet}%
boson (like magnons in Heisenberg ferromagnets or pions
in high-energy physics).
Therefore, $a\neq0$ for RSB systems. It has been 
conjectured that $a(T<T_\mathrm{c}) = a(T_\mathrm{c}) /2$~\cite{dedominicis:98}.
The exponent $a$ is probably discontinuous at $T_\mathrm{c}$~\cite{parisi:97}.

In short, the value of $a$ below the critical temperature
is a good marker to distinguish coarsening dynamics
(droplet or TNT) from the RSB predictions.
We shall consider this issue in Chapter~\ref{chap:sg-3}.

Notice that the presence of the damping function
\index{damping function}
is required due to causality considerations.
In the case of $C_4(\boldsymbol r,\tw)$, 
the damping function $f$ seems
to decay a little faster than an exponential~\cite{jimenez:05,marinari:00b},
\begin{equation}\label{eq:SG-C4-long-distance-2}
f(x)\sim e^{-x^\beta},\qquad \beta \approx 1.5.
\end{equation}

Notice that the spin-glass susceptibility~\eqref{eq:SG-chi-SG}
 can be defined as
\begin{equation}\label{eq:SG-chi-SG-integral}
\chi_\text{SG}(\tw) = N \overline{q^2(\tw)} =
\int \dd^D\boldsymbol r\ C_4(\boldsymbol r,\tw).
\end{equation}

In order to characterise the aging dynamics, one must
take dynamical heterogeneities into account~\cite{castillo:02,castillo:03,jaubert:07}.\index{dynamical heterogeneities}
This is best accomplished through the two-time  spatial 
correlation
\begin{equation}\label{eq:SG-C22}
C_{2+2}(\boldsymbol r,t,\tw) = \overline{\frac1N \sum_\bx\bigl[
c_\bx(t,\tw) c_{\bx+\boldsymbol r}(t,\tw) - C^2(t,\tw)\bigr]}
\nomenclature[C22]{$C_{2+2}(\boldsymbol r, t,\tw)$}{Two-time  spatial correlation}
\end{equation}
Notice that this is a connected correlation function. In principle, 
it may seem more natural to subtract $\overline{C(t,\tw)}^2$, 
but due to the self-averaging \index{self-averaging}
nature of $C$ (see below), both definitions are equivalent 
in the thermodynamical limit. We note that this 
correlation is natural for the structural glasses problem, where
the order parameter is not known (see, e.g., \cite{berthier:05}
and references therein).

$C_{2+2}$ seems complicated, but it actually has a simple probabilistic
interpretation. Let us call a defect a site
where $c_\bx(t,\tw)=-1$ and let $n(t,\tw)$ be 
the density of such defects. Trivially, $C(t,\tw) = 1-2n(t,\tw)$. 
Then, the conditional probability of finding a defect at
site $\bx+\boldsymbol r$, knowing that there is one at $\bx$, 
is $n(t,\tw)g(\boldsymbol r)$, where $g(\boldsymbol r)$ represents
the pair-correlation function of the defects. Hence, 
$C_{2+2}(\boldsymbol r,t,\tw)$ is just $4n^2(t,\tw)\bigl(1-g(\boldsymbol r)\bigr)$.

The long-distance decay of $C_{2+2}(\boldsymbol r,t,\tw)$ defines 
the correlation length $\zeta(t,\tw)$, the characteristic
length scale for dynamic heterogeneities,\footnote{Notice
that $C_{2+2}$ is the difference of two correlated quantities,
so its error is significantly reduced from what
one would expect from error propagation. We take this
into account, as always, with the jackknife method (Appendix~\ref{chap:correlated}).
\index{jackknife method}}
\begin{equation}\label{eq:SG-C22-long-distance}
C_{2+2}(\boldsymbol r,t,\tw) \xrightarrow{\ \ r\to\infty\ \ }
\frac{1}{r^b} g\bigl( r(\zeta(t,\tw)\bigr)\ .
\nomenclature[zeta]{$\zeta(t,\tw)$}{Two-time  correlation length}
\index{correlation length!two-time}
\end{equation}
Little is known in the literature about either $b$ of $g$, since, 
before \textsc{Janus}, the time scales that could be explored
reached only very small values of $\zeta(t,\tw)$.

Notice that we have called $\zeta$ a correlation length, 
while we said that $\xi$ was a coherence length. The distinction
stems from the fact that the former is computed from a connected
correlation function and the latter from a non-connected 
one. In particular, $\xi(\tw)$ diverges in the large-$\tw$ 
limit, while this may or may not be the case for $\zeta$.
We must point out, though, that this nomenclature is not 
universal.

In the RSB framework, we can distinguish two sectors in the dynamics: 
(\textsc{i}) relaxation within a single state, where $q_\text{EA}<C<1$,
and  (\textsc{ii}) exploration of new states, where $C<q_\text{EA}$.
These can be identified by the relation of $\zeta$ and $\xi$. 
In particular, for sector (\textsc{i}) we expect $\zeta(t,\tw)\ll \xi(\tw)$. 
We shall find $\zeta(t,\tw)/\xi(\tw)$ a valuable dynamical
variable (Section~\ref{sec:SG-FTS}).

Throughout this chapter, and the next, we shall need to consider
the evolution of the physical observables along many 
orders of magnitude in time, which complicates
their interpretation and analysis. Yet, recall that, 
for fixed $\tw$, $C(t,\tw)$ defines a one-to-one relation
between $C$ and $t$. Hence, we shall often eliminate
$t$ as an independent variable in favour of $C$.

A final note concerns the issue of self-averaging. \index{self-averaging}
Notwithstanding the discussion in Section~\ref{sec:INTRO-self-averaging}, 
most of the observables considered here are actually 
self-averaging. This is because, in the non-equilibrium regime,
the coherence length is much smaller than $L$, so we 
essentially have about $L^D/\xi^D$ independent blocks. The
exception is $\chi_\text{SG}$, which is non-local.
In fact, the central limit theorem \index{central limit theorem}
suggests that the probability distribution of $q(\tw)$ should
tend to a Gaussian when $L\to\infty$. Therefore, 
the variance of $\chi_\text{SG}(\tw)$ 
is $\sim 2\chi_\text{SG}^2(\tw)$ in the limit of large systems.

\subsection{Our non-equilibrium simulations}
We comment the parameters of our non-equilibrium simulations, 
as well as some technical details, in Appendix~\ref{chap:runs-sg}.
Let us summarise by saying that we have run simulations for
three subcritical
temperatures $T=0.6,0.7,0.8$ as well as for $T=1.1\approx T_\mathrm{c}$.
For the first three temperatures, we reached $\tw=10^{11}$, 
simulating $96$ samples for $T=0.6,0.8$ and $63$ samples for $T=0.7$.
At the critical point we reached $\tw=4.2\times10^9$ and simulated
$32$ samples.

These simulations were first presented in~\cite{janus:08b}.
Later, in~\cite{janus:09b}, we simulated
a new set of $768$ samples for $T=0.7$, which shall constitute our main 
working temperature. For these new simulations we reached only $\tw=10^{10}$, 
since for longer times we had observed finite-size effects (see Section~\ref{sec:SG-finite-size}). 
When discussing our results at $T=0.7$, we shall find it sometimes convenient
to use the longer $63$-sample simulation, instead of the shorter
but more precise one with $768$ samples.

\section[Aging and full aging in $C(t,\tw)$]{Aging and full aging in \boldmath $C(t,\tw)$}\label{sec:SG-aging}\index{aging|(}
\begin{figure}[t]
\centering
\includegraphics[height=0.7\linewidth,angle=270]{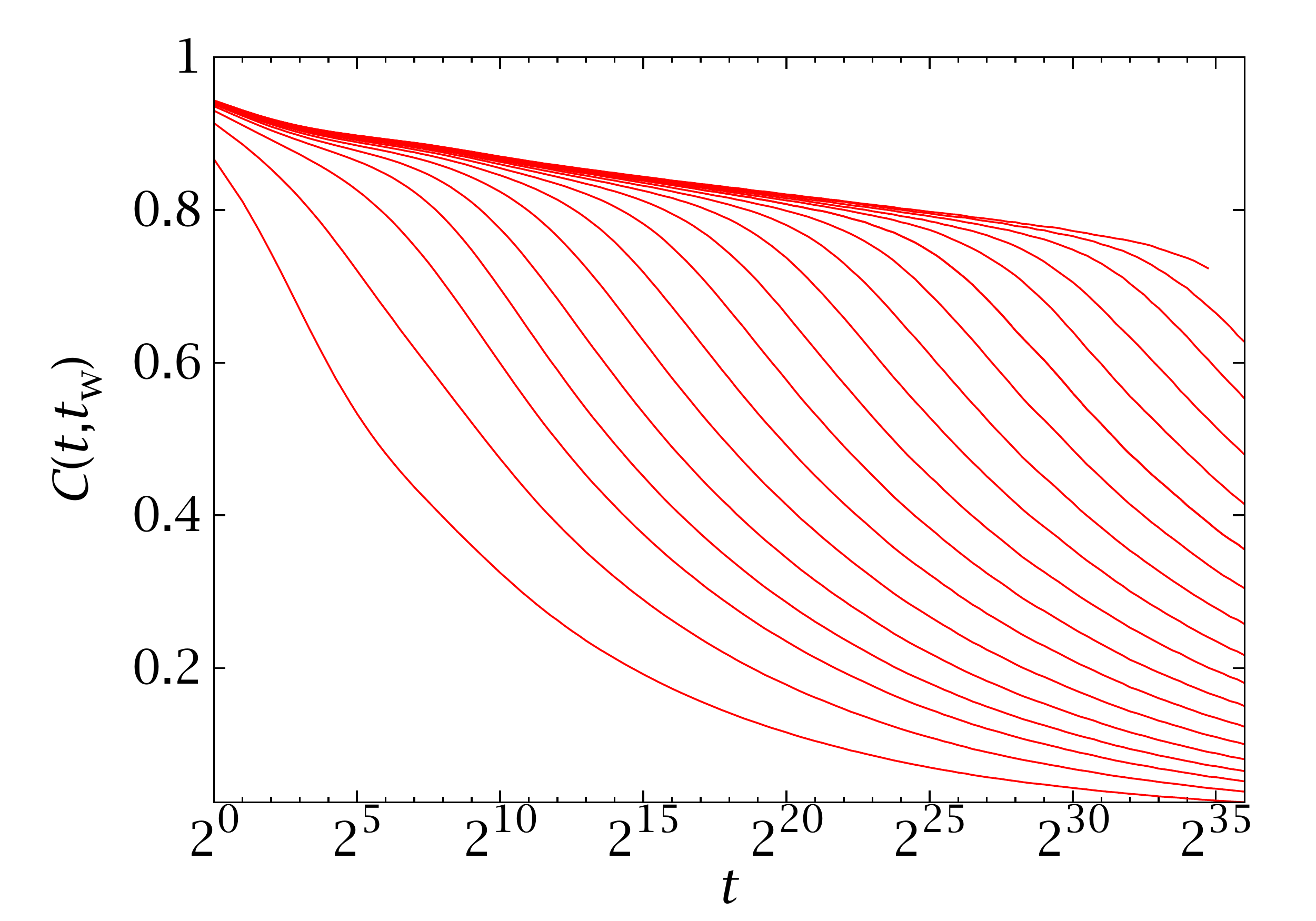}
\caption[Aging temporal correlation function]{Temporal correlation function $C(t,\tw)$, Eq.~\eqref{eq:SG-Cttw},
for $T=0.6$. We plot the curves as a function of $t$ for
$\tw=2^{2i}$, where $i=1,2,\ldots,18$ ($\tw$ grows from bottom 
to top). The errors are smaller than the width of the lines.
\index{correlation function (dynamics)!temporal|indemph}
\label{fig:SG-Cttw}}
\end{figure}
We start our study of the non-equilibrium dynamics by considering
the two-time correlation function. In Figure~\vref{fig:SG-Cttw}
we have plotted $C(t,\tw)$ for $T=0.6$ (our lowest temperature), 
for a range of $\tw$ across $11$ orders of magnitude 
(equivalent to physical times from picoseconds to one tenth of 
a second).  

We observe the qualitative claims made in the previous section, namely
\begin{itemize}
\item For each $\tw$, $C(t,\tw)$ is a strictly decreasing function 
of $t$.
\item $\lim_{t\to\infty} C(t,\tw)=0$.
\item As we increase $\tw$, we begin to see an enveloping 
curve. This is the $C_\infty(t)$ that we shall study in Section~\ref{sec:SG-qEA-dynamics}.
\end{itemize}
In the following we try to quantify these remarks. We begin by attempting 
to parameterise the aging behaviour. It has been claimed in experimental 
work~\cite{rodriguez:03} that $C(t,\tw)$ is well described by a 
full-aging behaviour, that is
\begin{equation}
C(t,\tw) = f\bigl( (t+\tw)/\tw\bigr), 
\end{equation}
at least in the range $10^{14}<\tw<10^{16}$. To check for this in a 
systematic way, we consider the functional form
\index{full aging}
\begin{equation}\label{eq:SG-full-aging}
C(t,\tw) = A(\tw) (1+t/\tw)^{-\alpha(\tw)}.
\end{equation}
If full aging is correct, the parameters $A(\tw)$ and $\alpha(\tw)$
should be constant for large $\tw$. This has been checked before~\cite{jimenez:03}, 
but with much smaller statistics.

\begin{figure}[h]
\includegraphics[height=\linewidth,angle=270]{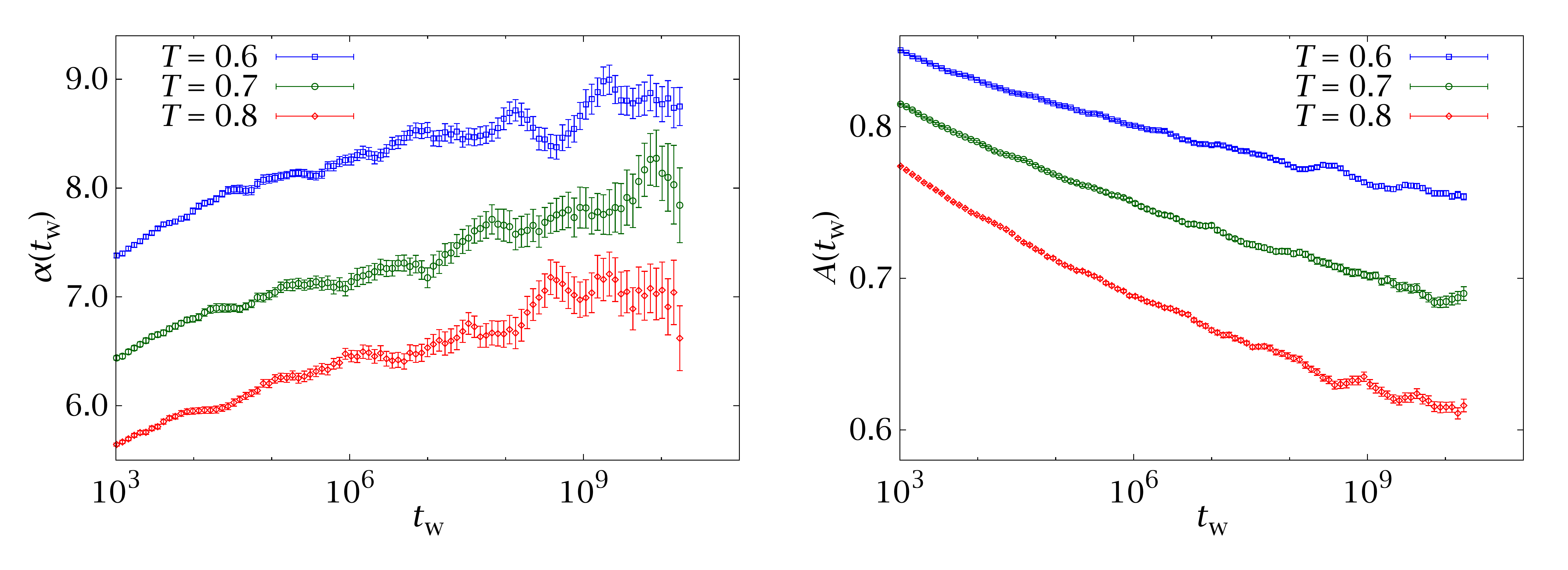}
\caption[Parameters in a full-aging fit]{Parameters in a fit of $C(t,\tw)$ to~\eqref{eq:SG-full-aging}
as functions of $\tw$. For $T=0.7$ we use our 
longer simulations with $63$ samples (cf.
Figure~\ref{fig:SG-full-aging-wiggles-2}). The evolution seems to slow down 
for large $\tw$, but we do not yet see the full aging behaviour,
in which these parameters are constant. 
\index{full aging|indemph}\label{fig:SG-full-aging}}
\end{figure}

We present the results of fits to~\eqref{eq:SG-full-aging} as
a function of $\tw$ for our three subcritical temperatures of $T=0.6,0.7,0.8$
in Figure~\ref{fig:SG-full-aging}.  For each $\tw$, we consider the range
$\tw\leq t\leq 10\tw$ (we want to avoid the short-times behaviour).
These fits, and their errors, 
have been computed with the jackknife method \index{jackknife method}
\index{fitting techniques}
explained in Appendix~\ref{chap:correlated}. In our simulated range, the exponent
$\alpha(\tw)$ is an increasing function of $\tw$, but its growth 
slows down noticeably. Still, the behaviour at $\tw= 10^{16}$ seems 
beyond reasonable extrapolation.

Notice that the $\alpha(\tw)$ exhibits some suspicious undulations
(especially for $T=0.6$). This is a visual effect caused 
by the data correlation, as we discuss in detail in Section~\ref{sec:CORR-wiggles}.

In Section~\ref{sec:SG-thermoremanent} we shall use
our fits to~\eqref{eq:SG-full-aging} to study
the thermoremanent magnetisation.
\index{magnetisation!thermoremanent}

\section{Spatial correlations and the coherence length}\label{sec:SG-xi}
In this section we consider the issue of growing length scales
in the spin-glass
dynamics. The computation of characteristic length scales 
is a recurring problem in statistical mechanics, as well as
in lattice gauge theory. In previous chapters we solved 
it by considering the 
second-moment definition~\eqref{eq:ISING-xi-second-moment}.
Notice that, in the thermodynamical limit, this second-moment estimator
is equivalent to
\begin{equation}\label{eq:SG-xi-second-moment}
\xi_2^{(\infty)}= \sqrt{\frac{\int \dd^D \boldsymbol r\ \boldsymbol r^2 C_4(\boldsymbol r, \tw)}{\int \dd^D\boldsymbol r\ C_4(\boldsymbol r,\tw)}}\ .
\index{correlation length!second moment}
\end{equation}
The denominator in this equation is $\chi_{\text{SG}}(\tw)$,
Eq.~\eqref{eq:SG-chi-SG-integral}, which
does not self-average. Therefore, this method is problematic in
non-equilibrium studies, where one considers very large systems
but not many samples (in our case $\mN_\text{samples}\approx100$ for
$T=0.6,0.8,T_\mathrm{c}$ and $\mN_\text{samples} = 768$ for $T=0.7$).

On the face of this problem, one could think of 
replacing the integral estimators by fits to
some functional form for $C_4(\boldsymbol r,\tw)$
featuring the $\xi(\tw)$. However, in the spin-glass
case we do not know this explicit functional form
---we have at best a reasonable ansatz as to its scaling, Eq.~\eqref{eq:SG-C4-long-distance}---
so this is not a viable alternative. In addition, we would face 
the ever present problem of statistical correlations, which complicate
fits and their interpretation.

Therefore, we must go back to the integral estimators. In this respect
notice that the $C_4(\boldsymbol r , \tw)$, for fixed $\boldsymbol r$,
is self-averaging. The lack of self-averaging in $\xi_2$ stems from the fact
that the integration volume grows with $L$. For a fixed size, 
it is the dismal signal-to-noise ratio in the long-distance tails which
spoils the precision. But this is precisely the problem
that one faces when trying to compute the integrated autocorrelation
time from a time series (Section~\ref{sec:THERM-recipes})
and we can borrow its solution: a self-consistent cutoff.


\index{aging|)}
\subsection{Integral estimators of characteristic length scales}\label{sec:SG-integral-estimators}
\index{coherence length|(}
We start our discussion by simplifying our definition of 
the spatial correlations to take into account only values of
$\boldsymbol r$ along the axes. We define
\begin{equation}
C_4(r,\tw) = \frac{1}{D}\sum_{i=1}^D C_4( r \hat{\boldsymbol u}_i, \tw),
\end{equation}
where the $\hat{\boldsymbol u}_i$ denote each of the lattice 
unit vectors. That is, we take the values of the correlation 
for a distance $r$ along the axes as representative of the
whole spherical shell of radius $r$.
We shall study the goodness of this approximation
in Section~\ref{sec:SG-isotropy},

Now, for positive integer $k$, we consider the integral,
\begin{equation}\label{eq:SG-I_k}
I_k(\tw) = \int_0^{L/2} \dd  r\ r^k C_4(r,\tw).
\end{equation}
Notice that our bound of integration is $L/2$, and not $L$, due
to the periodic boundary conditions. Since we are considering
the regime $\xi(\tw)\ll L$, this makes no difference.
In addition, the actual numerical computation of $I_k$ uses an
error-dependent cutoff: we only integrate up to the point $r_\text{cutoff}$
where the relative error in $C_4(r,\tw)$ rises above
some threshold. This is the same trick used in the computation
of the integrated autocorrelation time in Section~\ref{sec:THERM-recipes}.
We shall discuss this issue in more detail in Section~\ref{sec:SG-xi-estimators}.

With this notation, and assuming rotational invariance, the 
second-moment estimate is
\begin{equation}
\xi^{(\infty)}_2 \simeq \sqrt{\frac{I_{D+1}(\tw)}{I_{D-1}(\tw)}}\ .
\end{equation}
An alternative approach, considered in~\cite{kisker:96}, would be to
identify $\xi(\tw)$ with $I_0(\tw)$, which has the correct dimensions.
However, this is only a good definition if $a=0$ and we want 
a definition that would work for any scaling behaviour
of the type~\eqref{eq:SG-C4-long-distance}.
To this end, we note first that 
\begin{equation}
I_k(\tw) \propto \xi^{k+1-a}(\tw).
\end{equation}
Therefore, we can introduce the following definition
\begin{equation}\label{eq:SG-xi-k-k1}
\xi_{k,k+1}(\tw) = \frac{I_{k+1}(\tw)}{I_k(\tw)}\propto \xi(\tw) .
\end{equation}
Notice that the coherence length is only defined up to
a constant scale factor, so any of the $\xi_{k,k+1}(\tw)$
would work.

Still, there is a systematic error involved. Equation~\eqref{eq:SG-C4-long-distance} 
is valid only in the regime where $1\ll r\ll L$. The resulting bias
is minimised by considering large values of $k$, which suppress 
short distances in the integrals. However, too large a value
of $k$ would give too much weight to the tails of the correlation 
function, where the signal-to-noise ratio is worst. A  compromise
in the value of $k$ is clearly needed (cf. Figure~\ref{fig:SG-C4-tails}). 

In Section~\ref{sec:SG-xi-estimators} we shall compare different estimators
of the coherence length and demonstrate that, indeed, the integral 
ones produce a significant error reduction. Furthermore, the issue
of the choice of $k$ is not as critical as it may seem. 
For the moment, let us proceed with the physical discussion, using
as our preferred estimator $\xi_{1,2}(\tw)$.

\subsection{The dynamical critical exponent and finite-size effects}\label{sec:SG-finite-size}
\index{critical exponent!z@$z$}
\index{finite-size effects}
\begin{figure}
\centering
\includegraphics[height=.7\linewidth,angle=270]{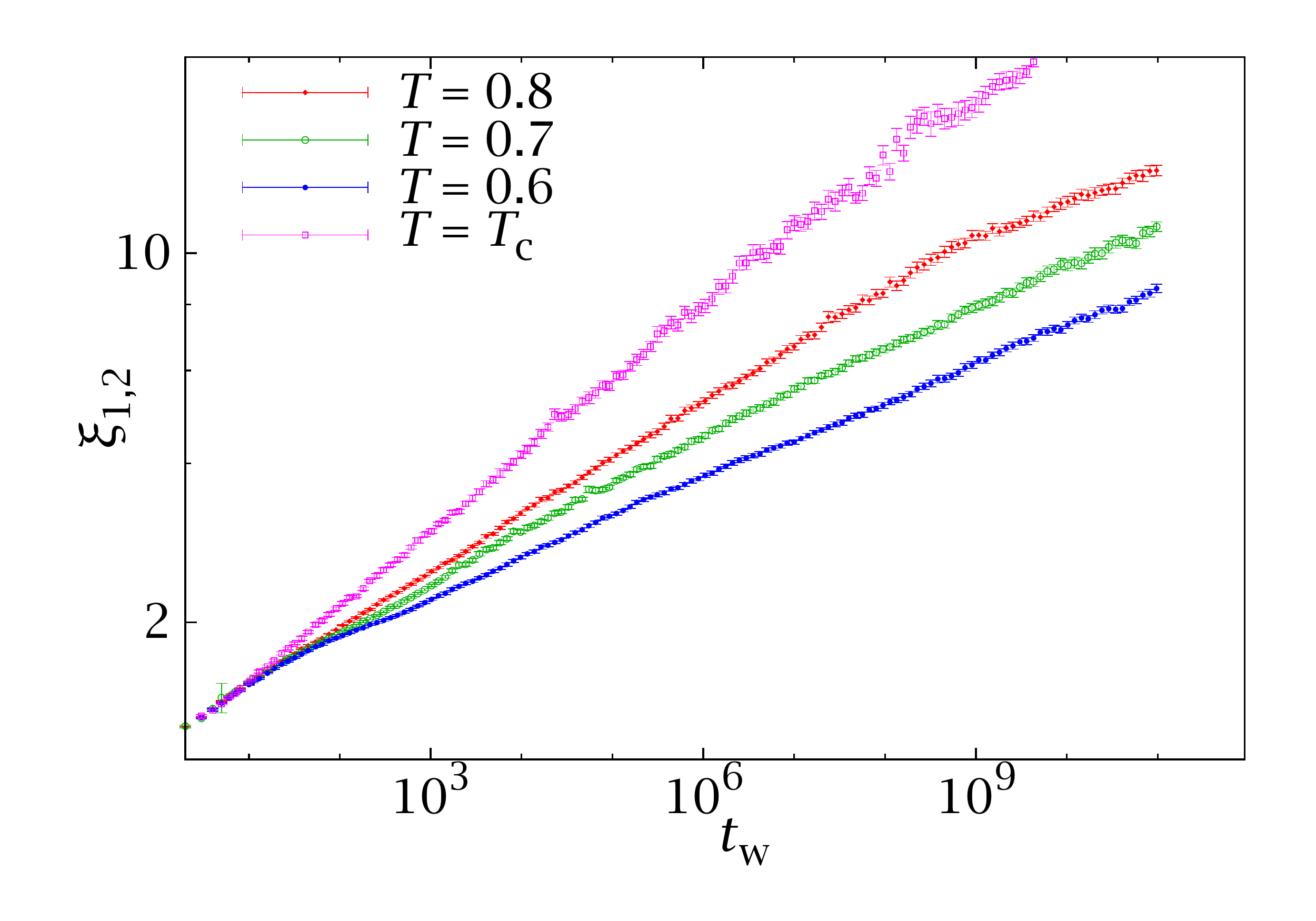}
\caption[Growth of the coherence length]{Growth of the 
coherence length for several subcritical temperatures
and for the critical point. For $T=0.7$ we use 
our longer simulations with $63$ samples.
\index{coherence length|indemph}
\index{critical exponent!z@$z$|indemph}
\label{fig:SG-xi-z}}
\end{figure}
We consider in Figure~\ref{fig:SG-xi-z} our coherence 
length $\xi_{1,2}(\tw)$ for $T=T_\mathrm{c}$ and for our 
three subcritical temperatures. For $T=0.7$ we have plotted 
our simulations with $63$ samples, since they reach longer times.

As expected, the growth of $\xi(\tw)$ is slower at lower temperatures.
Furthermore, our curves follow a power law for a wide
time range. There are some deviations at short times, due to lattice
discretisation effects (our definitions do not make much sense for 
$\xi\lesssim 3$). Some of the curves also present some deviations
for long times (this is most noticeable for $T=0.8$).

\begin{figure}
\centering
\includegraphics[height=0.7\linewidth,angle=270]{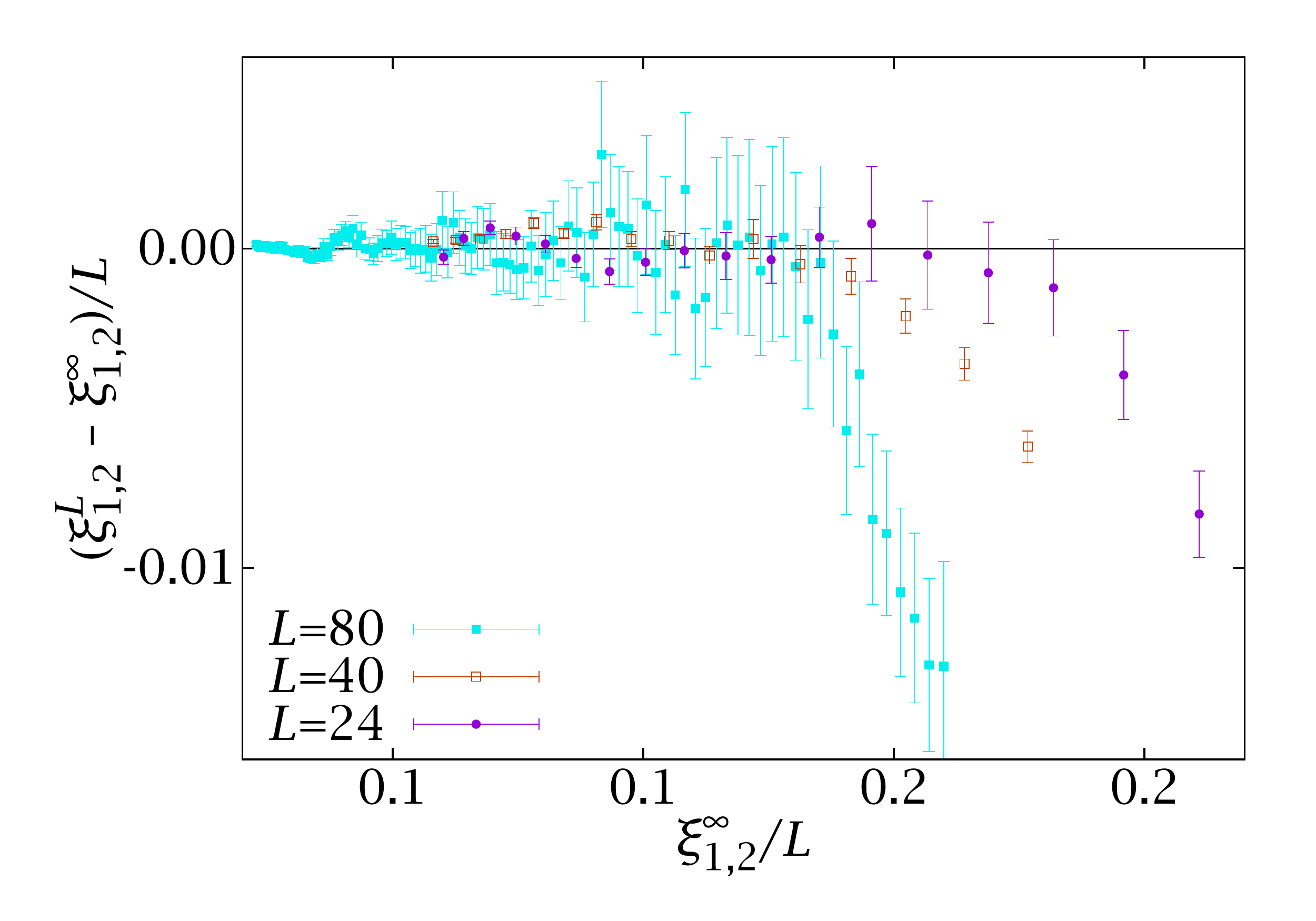}
\caption[Finite-size effects in $\xi(\tw)$]{Finite
size effects in $\xi(\tw)$,  obtained by computing
the difference between $\xi_{1,2}(\tw)$ for each of
our lattice sizes and a power-law extrapolation.
In particular, we define $\xi^\infty_{1,2}(\tw)$ as the result of 
applying the fit to~\eqref{eq:SG-xi-z}, computed for
$L=80$ in the range $\xi\in[3,10]$, to the whole temporal range.
Then, for each lattice size we consider the relative difference
 $[\xi_{1,2}^\infty(\tw)-\xi_{1,2}^{(L)}(\tw)]/\xi_{1,2}^\infty(\tw)$.
\index{coherence length|indemph}
\label{fig:SG-finite-size}}
\end{figure}

These can be interpreted as finite-size effects, even though we 
are simulating very large lattices ($L=80$). Indeed, when 
studying non-equilibrium dynamics we assume that our system is 
in the thermodynamical limit, which in principle
requires $\xi(\tw)\ll L$. In practice, the bound depends on 
the numerical accuracy. From finite-size scaling \index{finite-size scaling}
we would expect the largest $\tw$ for which $L=80$ still 
represents $L=\infty$ physics to scale as $L\geq k \xi_{1,2}(\tw)$.
We have estimated $k$ by computing $\xi_{1,2}(\tw)$ 
in simulations for $L=24,40$ and noting where they diverge
from their $L=80$ counterparts (Figure~\ref{fig:SG-finite-size}).
We conclude that at $T=0.8$
the safe range is $L\geq 7 \xi_{1,2}(\tw)$.
For $T_\mathrm{c}$, the equivalent bound is $L\geq6 \xi_{1,2}(\tw)$.

\begin{table}
\small
\centering
\begin{tabular*}{\columnwidth}{@{\extracolsep{\fill}}clcrlc}
\toprule
$T$ &
$\mN_\text{samples}$&
$[\xi_\text{min},\xi_\text{max}]$&
\multicolumn{1}{c}{ $z$} 
& $\chi^2_{\text{d}}$/d.o.f. & $\tw^\text{max}$\\
\toprule
\multirow{1}{*}{$0.6$}
&\multirow{1}{*}{96}
 & $[3,10]$ & 14.06(25) &  41.7/82 & $\approx9\times10^{11}$  \\
\midrule
\multirow{2}{*}{$0.7$}
&\multirow{2}{*}{63}
  &  $[3,10]$  & 11.84(22)& 82.7/81 & \multirow{2}{*}{$\approx2.2\times10^{10}$} \\
& &  $[3.5,10]$& 12.03(27) & 52.7/71 &\\
\midrule
\multirow{1}{*}{$0.8$}
&\multirow{1}{*}{96}
 & $[3,10]$   & 9.42(15)  & 17.1/63 & $\approx4.3\times10^8$ \\
\midrule
\multirow{1}{*}{$1.1$}
&\multirow{1}{*}{32}
 & $[3,10]$   & 6.86(16)  & 18.7/46 & $\approx3.5\times10^6$ \\
\midrule
\multirow{4}{*}{$0.7$}
&\multirow{4}{*}{768}
  & $[3,10]$   & 11.45(10) & 86.9/76 &\multirow{4}{*}{$\approx2.2\times10^{10}$} \\
&  & $[3.5,10]$ & 11.56(13) & 46.6/66&\\
&  & $[4,10]$   & 11.64(15) & 40.1/58&\\
&  & $[4.5,10]$ & 11.75(20) & 29.6/50&\\
\bottomrule
\end{tabular*}
\caption[The dynamical critical exponent]{Value of the dynamic critical exponent $z$ for several temperatures.
For our simulations with less than $100$ samples we have 
used the fitting range $\xi_{1,2}\in[3,10]$,
while for our $768$-sample simulation at $T=0.7$ 
we have needed to restrict the fitting range.
For $\xi_{1,2}(\tw)>10$, our simulations are affected by finite-size
effects. We also report the value of $\tw^\text{max}$ 
for which this bound is reached (for $T=0.6$ we never reach 
$\xi_{1,2}=10$, so this is an extrapolation).
\index{critical exponent!z@$z$|indemph}
\label{tab:SG-z}
}
\end{table}

In order to be safe, we have considered that $\xi_{1,2}(\tw)$
is physically meaningful for $3\leq \xi(\tw)\leq 10$. 
For $T=0.6$ the upper bound is not reached, while for $T=0.7$ 
it corresponds to $\tw\sim10^{10}$ and for $T=0.8$ 
it corresponds to $\tw\sim10^8$. In this
range, we have fitted the coherence length 
to a power law, in order to determine
the dynamical critical exponent $z$,
\begin{equation}\label{eq:SG-xi-z}
\xi(\tw) = A(T) \tw^{1/z(T)}\ .
\end{equation}
The results of these fits can be seen in Table~\ref{tab:SG-z}.
We also include the maximum waiting time safe from
 finite-size effects, defined 
as $\xi_{1,2}(\tw^\text{max}) = 10$.

The resulting values of $z$ are very large, going from $z(T_\mathrm{c}) = 6.86(16)$
at the critical point 
to $z(T=0.6)=14.06(25)$ at our lowest temperature. Also, we find that this exponent roughly follows the law
\begin{equation}
z(T) = z(T_\mathrm{c}) \frac{T_\mathrm{c}}{T}\, ,
\end{equation}
consistent with previous numerical and experimental studies~\cite{joh:99,marinari:00b}.

Furthermore, if we extrapolate the coherence length to a typical 
experimental scale of $100$ s ($\tw\sim 10^{14}$), we find
\begin{align}
\xi_{1,2}(\tw=10^{14}, T=0.6) &= 14.0(3), \\
\xi_{1,2}(\tw=10^{14}, T=0.7) &= 21.7(4), \\
\xi_{1,2}(\tw=10^{14}, T=0.8) &= 37.0(14), \\
\xi_{1,2}(\tw=10^{14}, T=T_\mathrm{c}\ ) &= 119(9),
\end{align}
which nicely compare with experiments~\cite{joh:99,bert:04}.

\subsection{Comparison of different estimators of the coherence length}\label{sec:SG-xi-estimators}
\begin{figure}
\centering
\includegraphics[height=\linewidth,angle=270]{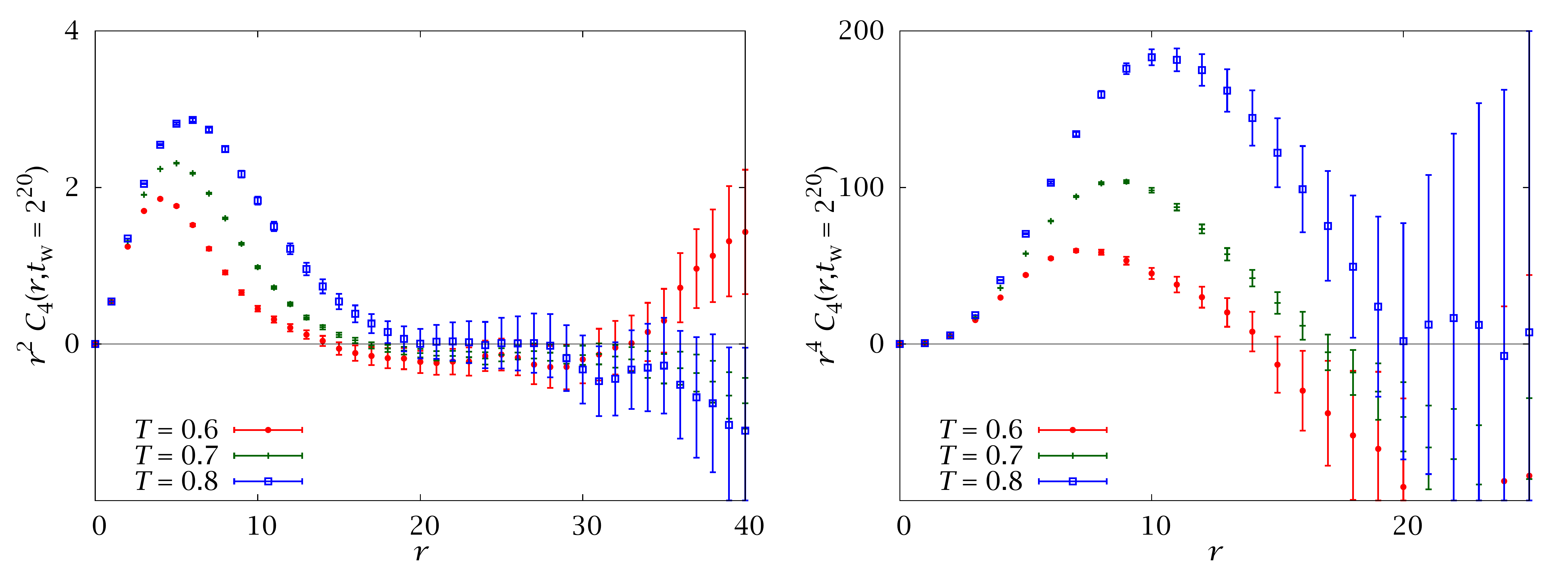}
\caption[The spatial correlation $C_4(r,\tw)$]{Plot
of the spatial correlation function $C_4(r,\tw)$ 
for several subcritical temperatures and $\tw=2^{20}$.
We show the correlation multiplied by $r^2$ (left) and 
$r^4$ (right), needed to compute $I_2(\tw)$ 
and $I_4(\tw)$, respectively.
The signal-to-noise ratio is the same 
in both cases, but for the latter the distances
with maximum weight are closer to the noise-dominated
region. The curve for $T=0.7$ is the average over $768$
samples, while those for $T=0.6,0.8$ were computed
with $96$ samples.
\index{correlation function (dynamics)!spatial|indemph}
\label{fig:SG-C4-tails}}
\end{figure}

In this section we consider some details on the implementation
of our integral estimators and present a comparison of the different
methods to compute $\xi(\tw)$.

A naive numerical implementation of~\eqref{eq:SG-I_k} to 
compute $\xi_{1,2}$ already
yields a considerable increase in precision with respect
to $\xi_2$. However, we can still improve the computation.
In Figure~\ref{fig:SG-C4-tails} we show how $r^2C_4(r,\tw)$ is very well
behaved up to the point where it becomes compatible with zero, where
the errors increase dramatically, spoiling the precision of the whole
integral. We take this into account with a self-consistent 
integration cutoff, much like what we describe 
for the evaluation of integral correlation times in Appendix~\ref{chap:thermalisation}.
We integrate our data (interpolated with a cubic spline, although this choice
is arbitrary) \index{cubic splines}
only up to the point where the relative error of $C_4(r,\tw)$ 
becomes greater than $1/3$. Thus, we reduce the statistical error
at the cost of introducing a, hopefully small, cutoff bias. We
can minimise this systematic effect by estimating the contribution 
of the tails with a fit to 
\begin{equation}
C_4(r,\tw) = \frac{A}{r^{0.4}}\exp\left[ -\bigl(r/\xi^\text{fit}(\tw)\bigr)^{1.5}\right].
\end{equation}
\begin{figure}
\includegraphics[height=\linewidth,angle=270]{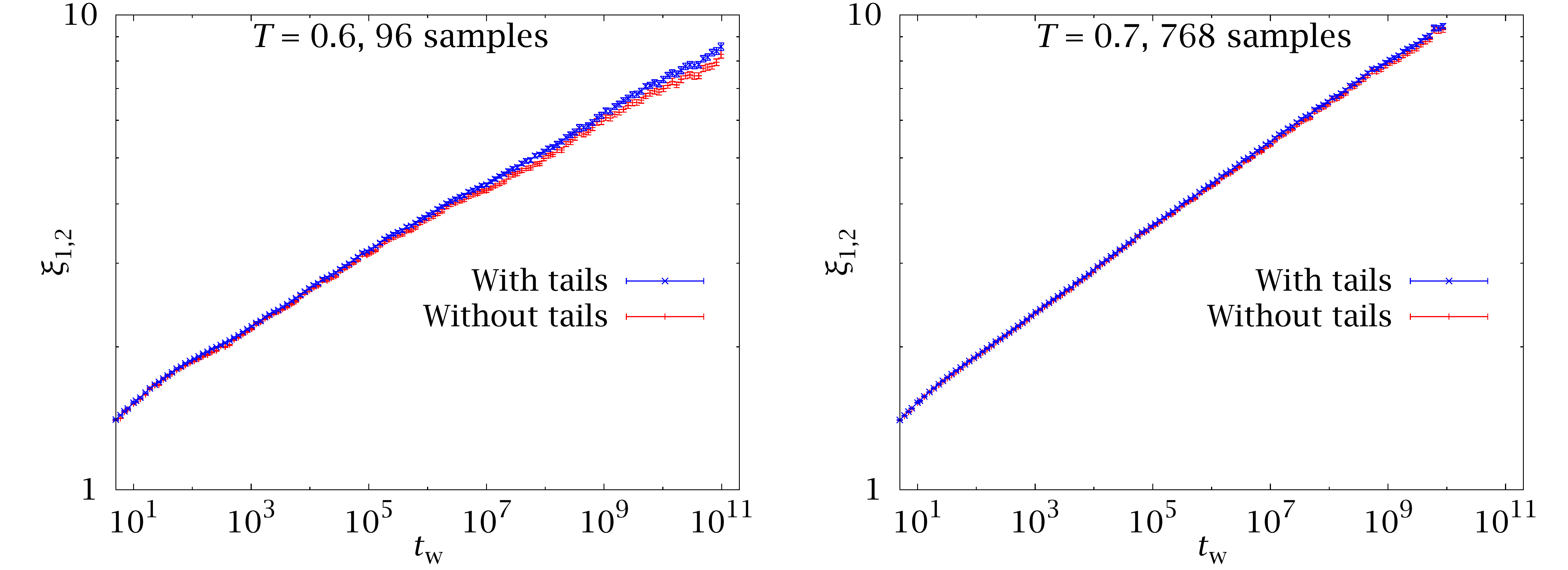}
\caption[Coherence length computed with and without the tail
contribution]{\emph{Left:} Result of computing
$\xi_{1,2}(\tw)$ in two different ways for our $96$ 
samples at $T=0.6$. In the red curve we stop the integration
at the cutoff point where the relative error in $C_4(r,\tw)$ is
greater than $1/3$. In the blue curve we add the contribution
of the tail from that point on, estimated with 
an extrapolation using a fit to~\eqref{eq:SG-C4-long-distance-2}.
The difference is small, but the second method maintains the power
law behaviour until longer times. \emph{Right:} Analogous
plot for our $768$ samples at $T=0.7$. Now the effect is much 
less noticeable, since the increase in statistics has shifted
the cutoff point.
\index{coherence length|indemph}
\label{fig:SG-xi-tails}}
\end{figure}

This is the scaling function~\eqref{eq:SG-C4-long-distance-2}, using 
$a=0.4$. The fit is computed for $3\leq r\leq \min\{15,r_\text{cutoff}\}$, 
where the signal is still good. This correction 
turns out to be only relevant for the largest times
and this only for our simulations with $\sim100$ samples.
For $T=0.7$, where we have $768$ samples, the cutoff 
distance is increased and the effect of the tail correction
is hardly noticeable (Figure~\ref{fig:SG-xi-tails}).
\begin{figure}
\centering
\includegraphics[height=.7\linewidth,angle=270]{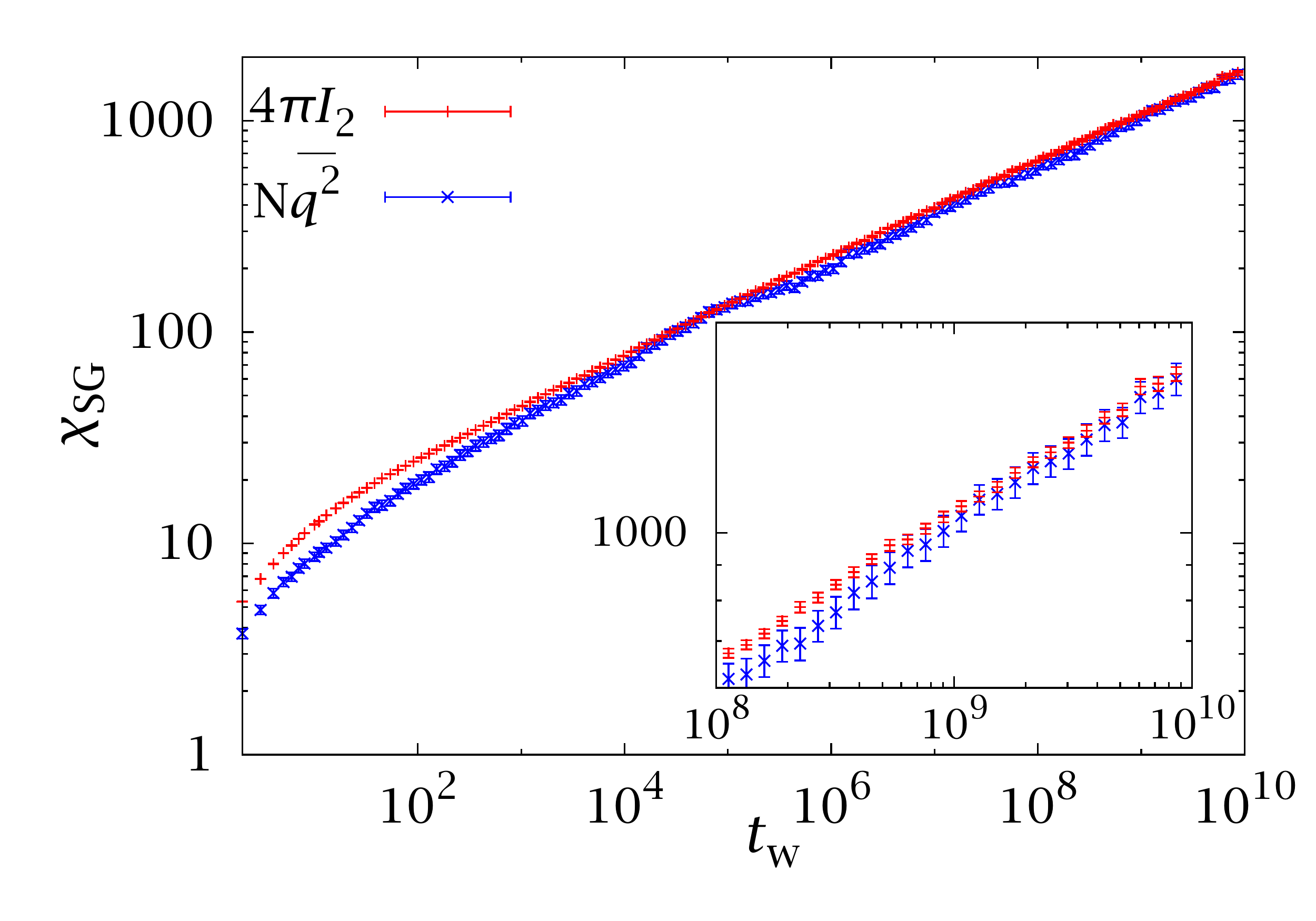}
\caption[Computation of the spin-glass susceptibility]{
Computation of the spin-glass susceptibility for
our $768$ samples at $T=0.7$ using both its 
straightforward definition~\eqref{eq:SG-chi-SG}
and the integral $I_2(\tw)$ of~\eqref{eq:SG-I_k}.
$\chi_\text{SG}(\tw) = N\overline{q^2(\tw)} = 4\uppi I_2(\tw)$.
The second determination has been carried out using 
a self-consistent cutoff, which noticeably increases
the precision. The inset details the upper-right corner of long times.
  \index{susceptibility!spin glass|indemph}
\label{fig:SG-I2-chi-SG}}
\end{figure}
\begin{figure}
\centering
\includegraphics[height=.7\linewidth,angle=270]{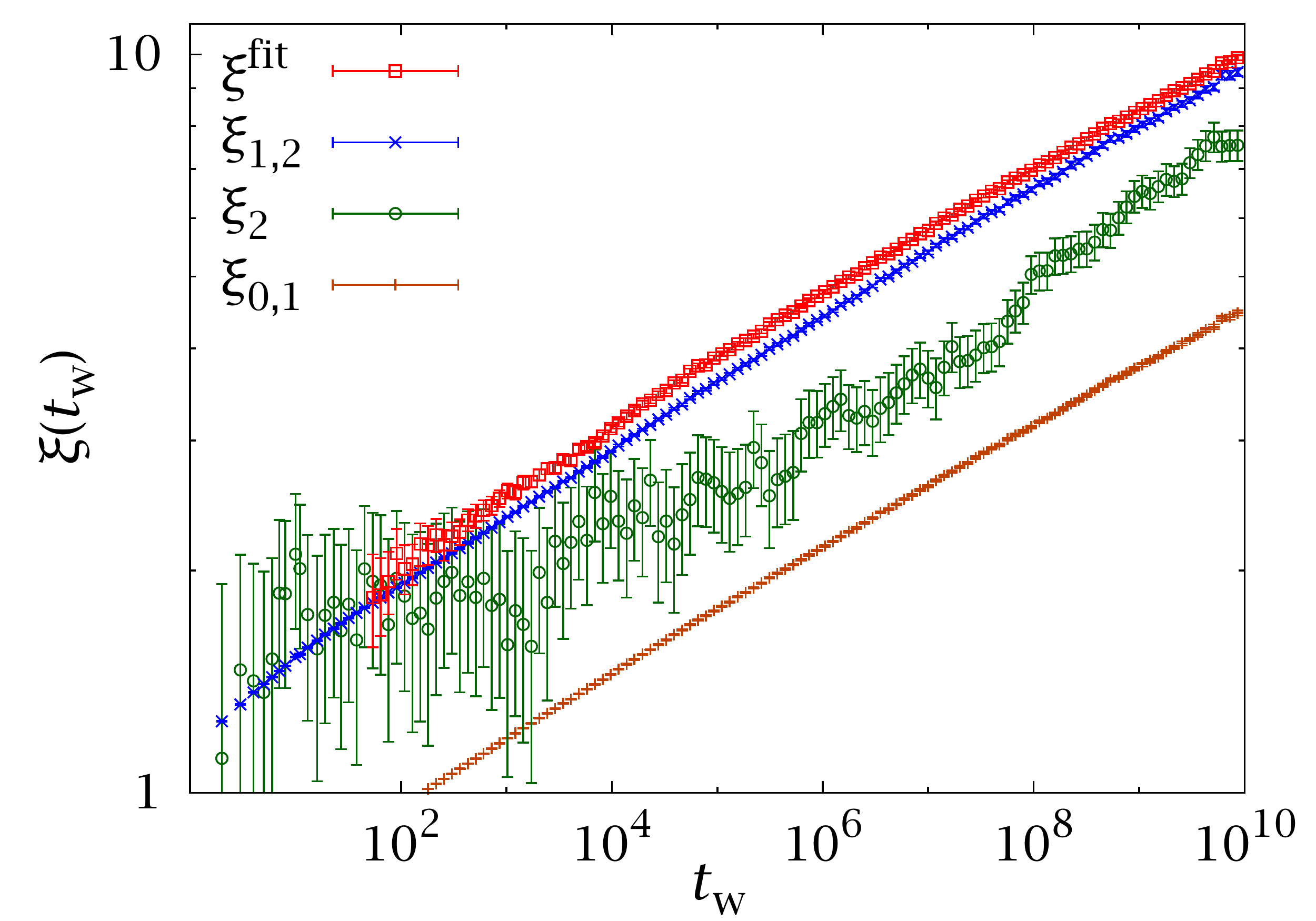}
\caption[Computation of $\xi(\tw)$ with several methods]{%
Comparison of several methods to estimate the coherence 
length $\xi(\tw)$ for our $768$ samples at $T=0.7$. We
include the integral estimators $\xi_{1,2}$ and $\xi_{0,1}$
of~\eqref{eq:SG-xi-k-k1}, the second-moment
estimate $\xi_2$ of~\eqref{eq:SG-xi-second-moment} 
and the result of a fit to~\eqref{eq:SG-C4-long-distance}
using~\eqref{eq:SG-C4-long-distance-2} and $a=0.4$.
All curves are parallel for large $\tw$, but the integral
estimators are much more precise.
\index{coherence length|indemph}
\label{fig:SG-xi-methods}}
\end{figure}
\begin{figure}
\centering
\includegraphics[height=.7\linewidth,angle=270]{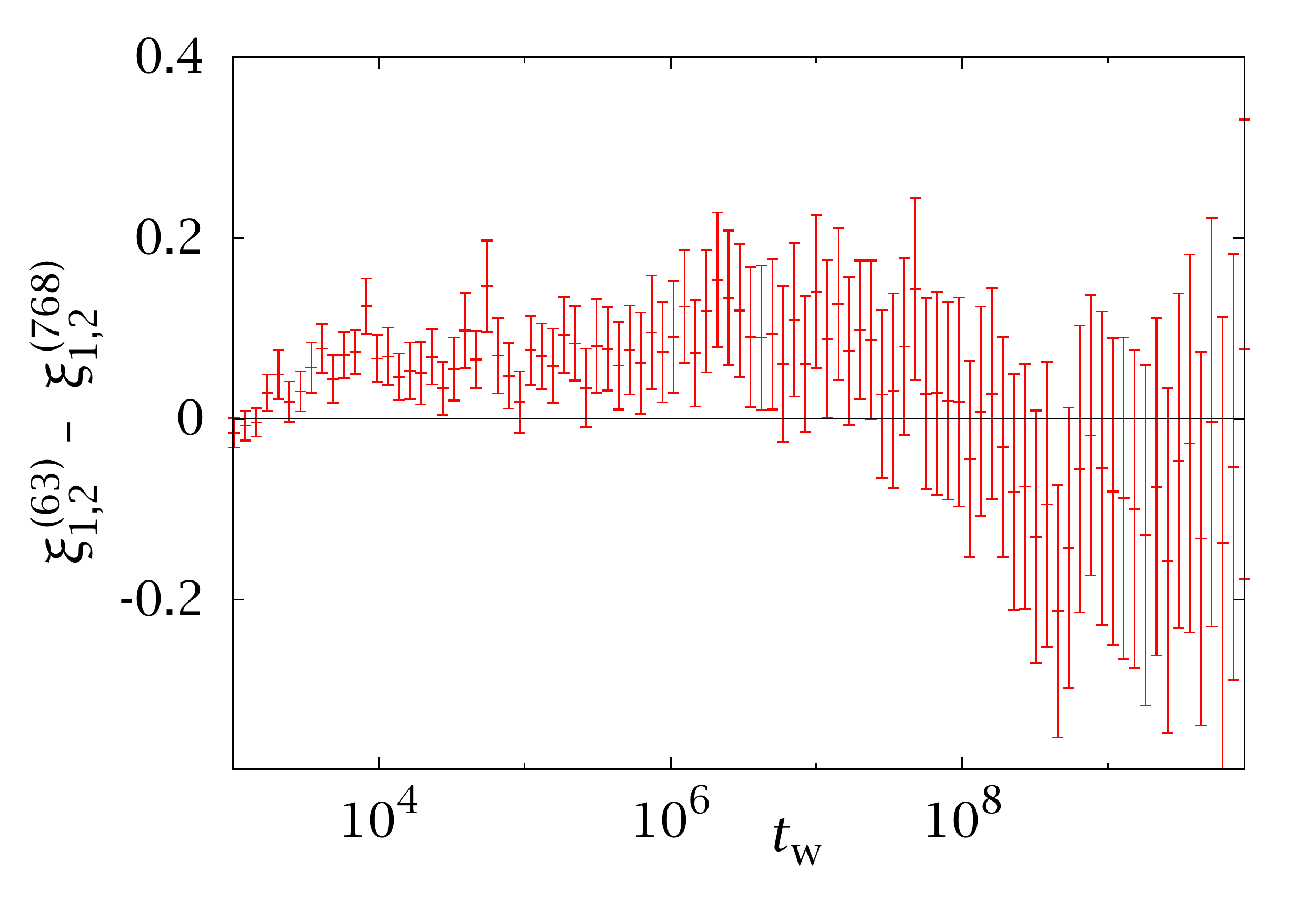}
\caption[$\xi(\tw)$ computed with $63$ and with $768$ samples]{%
Difference between the coherence length $\xi_{1,2}(\tw)$
computed with the 63 samples of~\cite{janus:08b}
and with the 768 of~\cite{janus:09b} (the errors
are the quadratic sum of those for each simulation).
The curves are
compatible in the whole time range. Mind
the dramatic statistical correlation in the sign
of this difference.}\label{fig:SG-xi-63-768}
\index{coherence length|indemph}
\end{figure}
\begin{figure}
\centering
\includegraphics[height=\linewidth,angle=270]{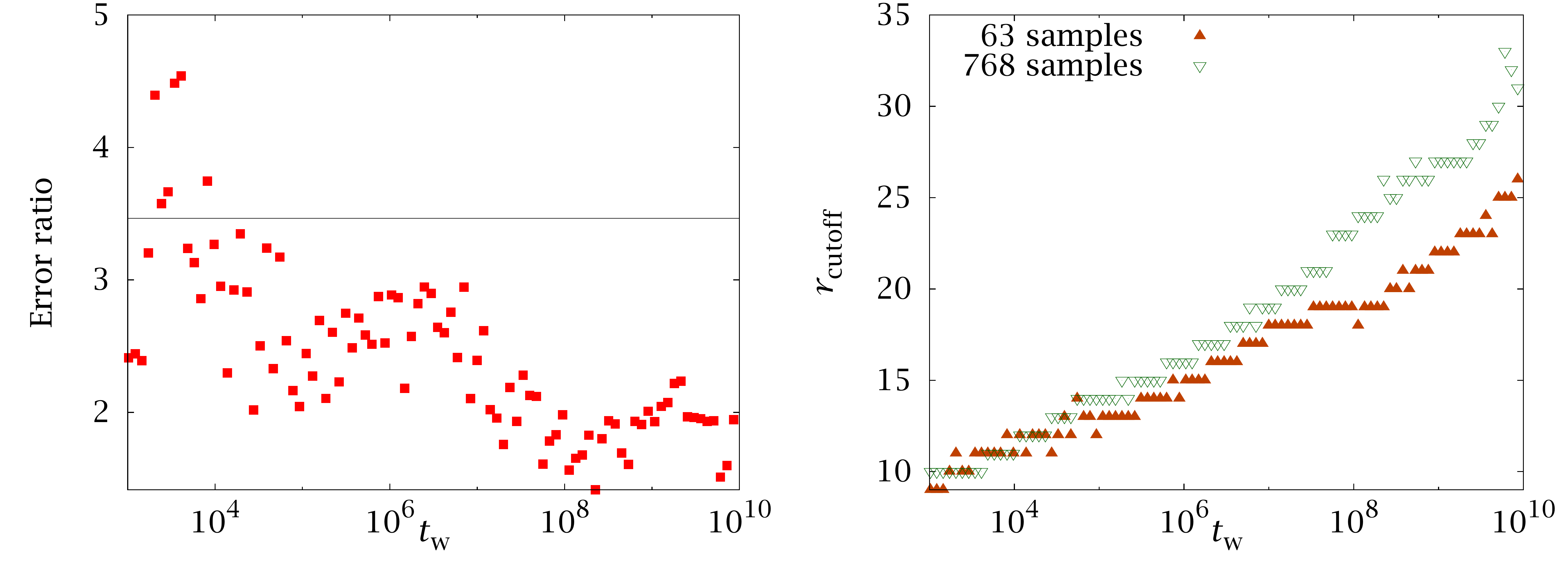}
\caption[Errors in $\xi(\tw)$ with $63$ and $768$ samples]{%
\emph{Left:} Ratio of the errors in $\xi_{1,2}(\tw)$ for the simulations
at $T=0.7$ with the $63$ samples of~\cite{janus:08b} and those of
the $768$ samples of~\cite{janus:09b} (see text for discussion).
The horizontal
line is $\sqrt{768/63}\approx3.46$. \emph{Right:}
Cutoff of the $I_k$ integrals as a function of time
for both simulations.}\label{fig:SG-xi-error-estimate}
\index{fitting techniques}
\end{figure}

We have claimed that this method yields a considerable increase
in precision with respect to the computation of $\xi_2(\tw)$,
or with the evaluation of $\xi$ from a fit. We offer two demonstrations
of this. First, in Figure~\ref{fig:SG-I2-chi-SG} we compare
$4\uppi I_2(\tw)$ with $N \overline{q^2(\tw)}$, two
quantities that should coincide (assuming rotational invariance).
As we can see, the former is much more precise for the whole
span of our simulation. 
As a second check, we have plotted in Figure~\ref{fig:SG-xi-methods}
the coherence length for $T=0.7$ (where we have the best statistics)
for several methods. All determinations are proportional, 
but the integral estimators are much more precise.

As a final test, we check not only our determination of $\xi$, but also 
its error estimate by comparing our $63$-sample computation of~\cite{janus:08b}
with the improved 768-sample one of~\cite{janus:09b} for $T=0.7$.
Figure~\ref{fig:SG-xi-63-768} shows the difference in both determinations
(the error in the difference is the quadratic sum of the individual errors, 
since both sets of samples are disjoint). We have excellent agreement 
for the whole time range. Notice that the points in this curve are
correlated in time, so the fluctuations of neighbouring points are not independent.

Finally, we consider the error estimates in Figure~\ref{fig:SG-xi-error-estimate}. The expected error reduction is $\sqrt{768/63}\approx 3.5$, but we 
do not reach this value for most of our times. Notice that 
the relative statistical error in our error 
estimate is $\sim 1/\sqrt{2\mN_\text{samples}}$, so the effect is not 
a fluctuation. The explanation is that the cutoff distance has increased
for the simulation with $768$ samples, which trades statistical error
for a reduction of systematic biases.

\subsection[The isotropy of $C_4(r,\tw)$]{The isotropy of \boldmath $C_4(r,\tw)$}
\label{sec:SG-isotropy}
\index{isotropy}
\begin{figure}
\centering
\includegraphics[height=0.45\linewidth]{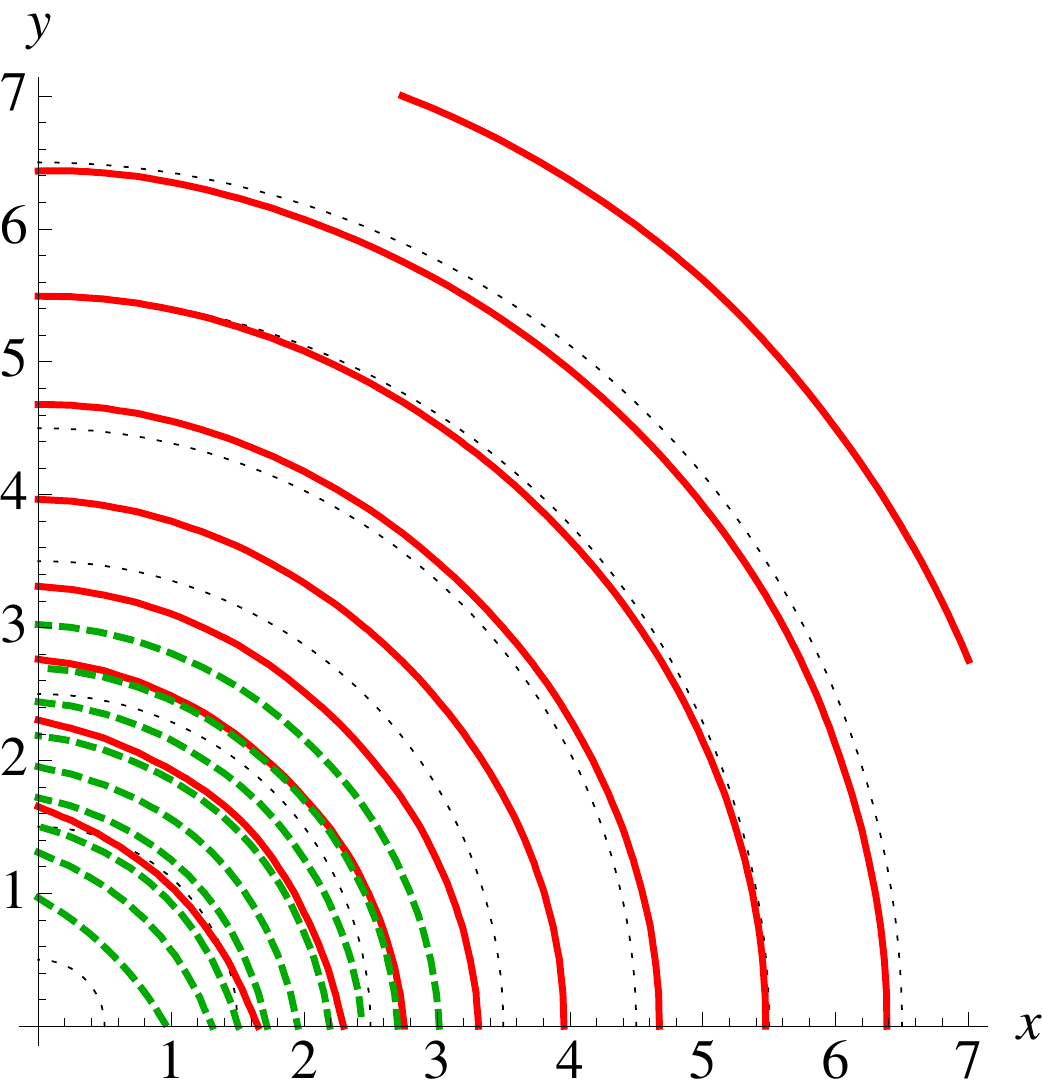}
\caption[Isotropy in the spatial correlations]{Level
curves $C_4(\boldsymbol r, \tw)=c$ 
for $c=0.3$ (dashed lines) and $c=0.1$ (solid lines)
at $T=0.6$ (dotted lines are circles, for visual reference). The plot
is restricted to the $z=0$ plane, for clarity. The innermost
curve corresponds in both cases to $\tw=4$ and the succeeding ones
correspond to geometrically growing times ($\tw=4\times 16^{i}$).
As we can see, the deviations from isotropy are mainly due
to lattice discretisation (i.e., functions of $r$).
The errors are smaller
than the thickness of the lines.
\index{correlation function (dynamics)!spatial|indemph}
\index{isotropy|indemph}
\label{fig:SG-C4-isotropy}}
\end{figure}
Throughout this section we have glossed over the issue 
of rotational invariance, assuming it is a good approximation. At 
all times we worked with correlations $C_4(r,\tw)$ restricted
to the directions along the axes, hence ignoring most of the $N$
points for a given $\tw$. The main motivation for this approach has traditionally
been to avoid the computation of the whole $C_4(\boldsymbol r,\tw)$, 
a task of $\mathcal O(N^2)$ in a naive implementation. However,
as detailed in Appendix~\ref{chap:thermalisation}, we can turn the
evaluation of an autocorrelation into the computation of two real 
Fourier transforms (the Wiener-Khinchin theorem).\index{Fourier transform}
\index{Wiener-Khinchin theorem}
This latter task is $\mathcal O(N\log N)$ with the FFT 
algorithm. \index{FFT}
Therefore, computing the whole $C_4(\boldsymbol r,\tw)$ implies
no increase in numerical effort.
\begin{figure}
\centering
\includegraphics[height=.7\linewidth,angle=270]{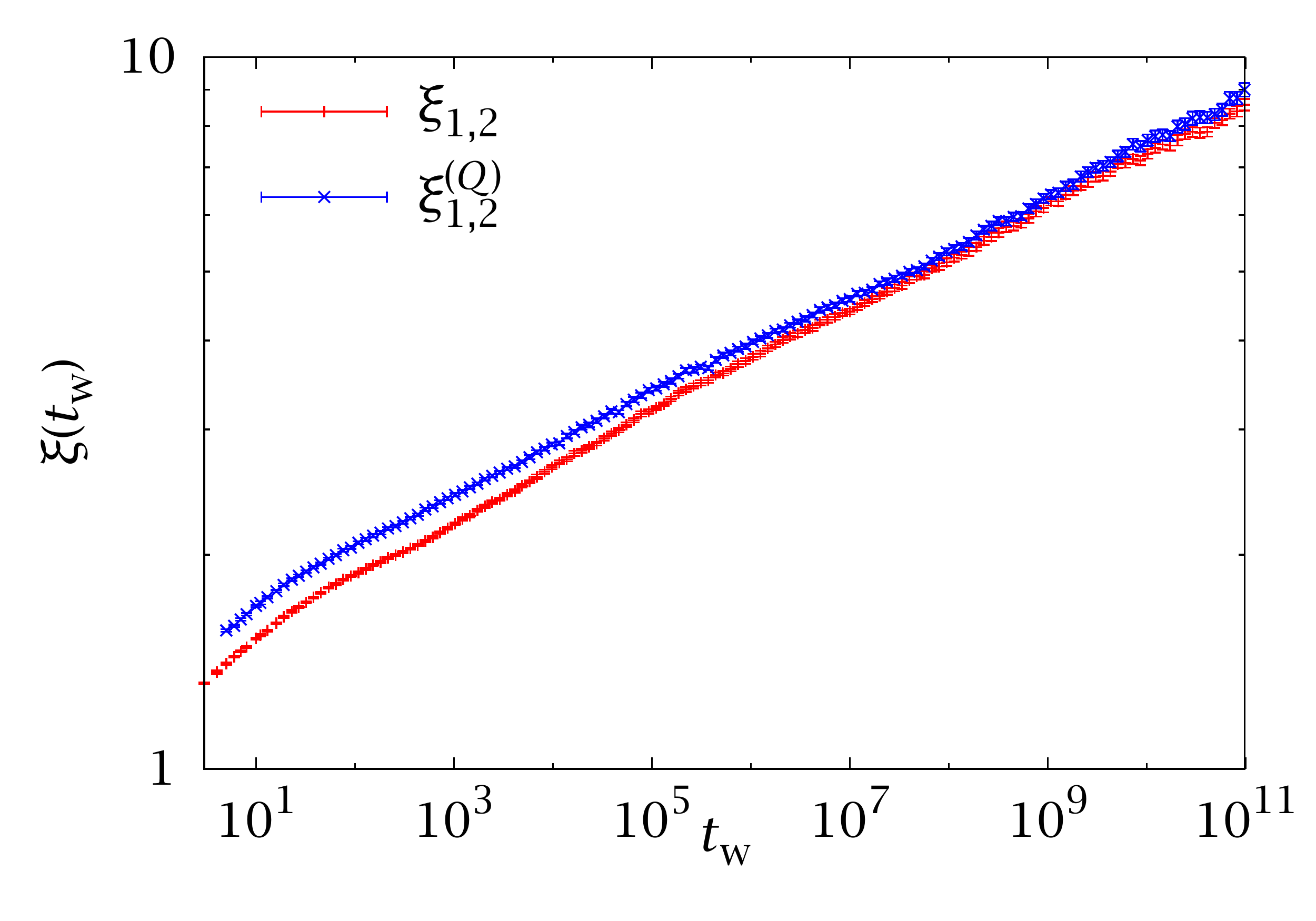}
\caption[Isotropy in the coherence length]{The coherence length
computed with the whole correlation functions, $\xi^{(Q)}_{1,2}$,
and with correlations
along the axes, $\xi_{1,2}$. Both estimates coincide for large
times and their errors are similar for
the whole range. Data for $T=0.6$.
\index{isotropy|indemph}
\index{coherence length|indemph}
\label{fig:SG-xi-isotropy}}
\end{figure}

We examine in this section whether this complete correlation function
is iso\-tro\-pic and whether we can take advantage of all the $N$
point to reduce statistical errors in the determination of 
$\xi(\tw)$. The first question is answered in Figure~\ref{fig:SG-C4-isotropy},
where we plot level curves $C_4(\boldsymbol r,\tw)=c$ for 
several values of $c$ and $\tw$. As we can see, isotropy
is recovered at quite small distances (remember we are
only interested in $\xi\gtrsim 3$).
\begin{figure}
\centering
\includegraphics[height=.7\linewidth,angle=270]{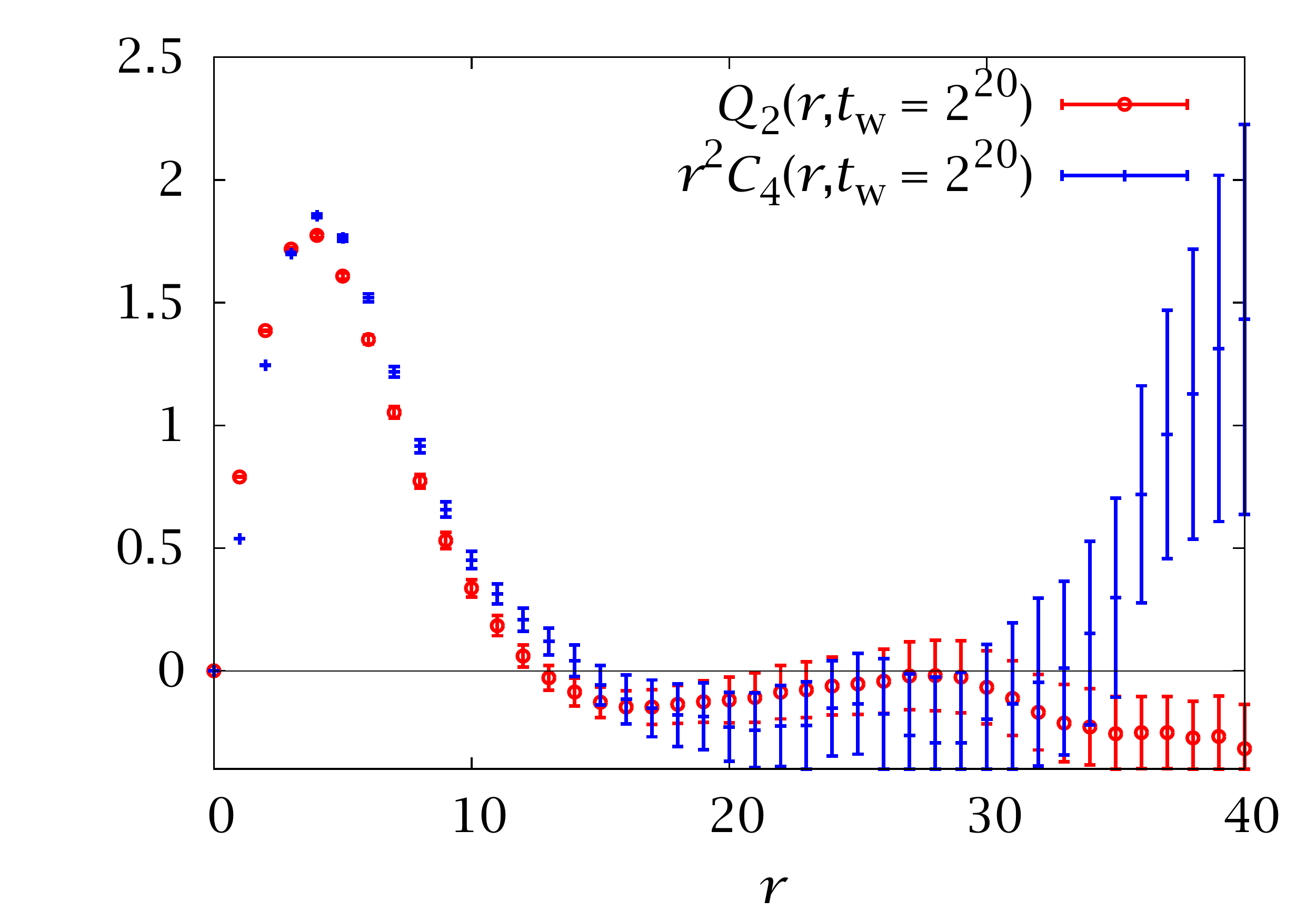}
\caption[The tails of the spatial correlations and isotropy]{We
compare the axes correlation $r^2C_4(r,\tw)$ with 
the shell-averaged correlation $Q_2(r,\tw)$. Both functions have
similar precision for small distances, but the second is better
behaved in the tails.
\index{isotropy|indemph}
\label{fig:SG-C4-isotropy-2}}
\end{figure}

We next consider the possible improvement of our determination
of $\xi(\tw)$ with the three-dimensional correlation 
function. In order to use 
our integral method we must first average over spherical
shells,
\begin{equation}
Q_k(n,\tw) = \frac{\sum_{|\boldsymbol r|\in[n,n+1)} |\boldsymbol r|^k C_4(\boldsymbol r,\tw)}{\sum_{|\boldsymbol r| \in [n,n+1)}1}\, .
\end{equation}
Notice that $Q_k(0,\tw) = r^k C_4(0,\tw) = \delta_{k0}$.
We can use $Q_k(m,\tw)$ as we did $r^kC_4(r,\tw)$ in our 
integral estimators. We would
expect the resulting coherence length $\xi^{(Q)}_{1,2}(\tw)$
to coincide with $\xi_{1,2}(\tw)$ for large $\tw$.
However, if the fluctuations of $C_4(\boldsymbol r,\tw)$ are
independent, $\xi^{(Q)}_{1,2}$ will have smaller errors, due
to the increase in statistics.
In fact, see Figure~\ref{fig:SG-xi-isotropy},
the statistical correlation between different parts of
 the lattice is so great that it renders
the gain in precision negligible.

In short, the usual approximation of considering only correlation along
the axes introduces neither a bias nor an increase in statistical errors.
Still, the computation of $C_4(\boldsymbol r,\tw)$ could be 
rewarding if one needed to consider $I_k$ with $k>2$ (Figure~\ref{fig:SG-C4-isotropy-2}).
\index{coherence length|)}
\section{The thermoremanent magnetisation}\label{sec:SG-thermoremanent}
\index{magnetisation!thermoremanent|(}
\begin{table}[h]
\small
\centering
\begin{tabular*}{\columnwidth}{@{\extracolsep{\fill}}ccllcl}
\toprule
$T$ &
$\tw$& 
\multicolumn{1}{c}{ $c(\tw)$} &
\multicolumn{1}{c}{ $A_0(\tw)\times10^3$}&
  $\chi^2_\text{d}$/d.o.f. &
\multicolumn{1}{c}{ $d(\tw)$}\\
\toprule
\multirow{4}{2.2cm}{$0.6\approx0.55T_\mathrm{c}$}
&2  &  $-0.1525(23)$ & 2.6(6)  & 14.7/64& $2.14(5)$ \\
&4  &  $-0.1495(22)$ & 2.8(8)  & 15.5/64& $2.10(5)$ \\
&8  &  $-0.1459(20)$ & 2.5(10) & 17.4/64& $2.05(4)$ \\
&16 &  $-0.1430(19)$ & 2.4(12) & 17.5/64& $2.01(4)$ \\
\midrule
\multirow{4}{2.2cm}{$0.7\approx0.64T_\mathrm{c}$}
&2 &  $-0.1787(14) $  & 1.47(25) & 23.3/50& $2.067(27)$\\
&4 &  $-0.1765(13) $  & 1.8(3)   & 18.4/50& $2.041(26)$\\
&8 &  $-0.1733(12) $  & 1.7(4)   & 18.9/50& $2.004(25)$\\
&16&  $-0.1704(12) $  & 1.6(5)   & 15.4/50& $1.971(25)$\\
\midrule
\multirow{4}{2.2cm}{$0.8\approx0.73T_\mathrm{c}$}
&2  &  $-0.210(8)$     & 1.7(10)  & 13.9/32& $1.98(9)$  \\
&4  &  $-0.212(7)$     & 2.8(12)  & 11.1/32& $2.00(8)$  \\
&8  &  $-0.208(7)$     & 3.0(14)  & 10.8/32& $1.96(8)$  \\
&16 &  $-0.205(6)$     & 3.0(18)  & 8.43/32& $1.93(7)$  \\
\bottomrule
\end{tabular*}
\caption[Thermoremanent magnetisation, power law decay]{Result of fitting the thermoremanent magnetisation
to~\eqref{eq:SG-C4-thermoremanent}, for our 
three subcritical temperatures. We give the temperatures 
in terms of $T_\mathrm{c}$, to facilitate comparison
with the experimental results of~\eqref{eq:thermoremanent-exp}.
For each fit we give the two parameters and the diagonal
chi-square per d.o.f. The last column 
shows the result of computing $d=-cz$, in order to
consider the scaling of equation~\eqref{eq:SG-C4-thermoremanent-2}.
The fits are computed for $10^6 < t<\tw^\mathrm{max}(T)$ (cf. Table~\ref{tab:SG-z}), 
which accounts for the different number of d.o.f.
at each temperature.
\label{tab:SG-thermoremanent}
\index{magnetisation!thermoremanent|indemph}
}
\end{table}
The thermoremanent magnetisation of a spin glass 
is one of the easiest quantities to study experimentally. Indeed, it has been 
known since the 1980s that it decays with a power law 
(see~\cite{granberg:87,prejean:88} and references therein),
even for temperatures very close to the critical point.\footnote{%
The only deviations from this simple behaviour were observed for 
$T>0.98T_\mathrm{c}$.} 

As we explained in Section~\ref{sec:SG-temporal-correlations}, the 
two-time correlation $C(t,\tw)$  \index{correlation function (dynamics)!temporal}
can be identified with the thermoremanent magnetisation 
(we have to choose $\tw\sim1$ as our initial `magnetised'
configuration and consider its overlap with the configuration
at time $t\gg\tw$). Following~\cite{parisi:97} we have fitted 
our data to the decay law
\begin{equation}\label{eq:SG-C4-thermoremanent}
C(t,\tw) = A_0(\tw) + A_1(\tw) t^{c(\tw)}.
\end{equation}
Notice that for a finite number of samples the correlation function
does not actually go to zero for long times, hence the need
for the finite asymptote $A_0(\tw)$. This is actually also a problem 
for experimental studies~\cite{granberg:87}.

Table~\ref{tab:SG-thermoremanent} shows the results of 
fits to~\eqref{eq:SG-C4-thermoremanent} for our 
subcritical temperatures. The fits 
where computed in the range $10^6\leq t \leq \tw^\mathrm{max}$, 
where $\tw^\text{max}$ was given in Table~\ref{tab:SG-z}.
The lower bound of the fitting region can be varied along 
several orders of magnitude with no effect. Notice that the values
of $\chi^2_\text{d}/\text{d.o.f.}$ are always much smaller 
than one, due to the correlations.

The asymptote $A_0(\tw)$ is in all cases of about $10^{-3}$,
small compared to the  the smallest value of $C(t,\tw)$ that we reach 
for our finite $t$, of about $10^{-2}$. The 
decay exponent is very small, $T$-dependent and exhibits
a slight, but systematic dependence on $\tw$
(a tendency that was already observed in~\cite{kisker:96}).

The experimental values for $c$ are, from~\cite{granberg:87},\footnote{%
The errors are small, about the size of the plotted data points
in Figure~3b of~\cite{granberg:87}.}

\begin{align}
c(0.55T_\mathrm{c}) &\approx -0.12, &
c(0.67T_\mathrm{c}) &\approx -0.14, &
c(0.75T_\mathrm{c}) &\approx -0.17,
\label{eq:thermoremanent-exp}
\end{align}
These values are slightly higher than our results of
Table~\ref{tab:SG-thermoremanent}. The reason may be that 
our values of $c$  have been computed in fits 
where $t$ and $\tw$ differ by as many as $10$ orders of magnitude,
while in experimental work $t/\tw\lesssim 10^4$.

We can recall here our study of full aging in Section~\ref{sec:SG-aging},  \index{full aging}
where we considered a fit in a narrower window of $\tw\leq t \leq 10\tw$.
We can extrapolate our $\alpha(\tw)$ to an experimental time
of $100$~s with a quadratic fit, such as we showed in Figure~\ref{fig:SG-full-aging-wiggles}.
We thus obtain
\begin{subequations}\label{eq:SG-thermoremanent-alpha}
\begin{align}
-1/\alpha(\tw=100\text{ s}) &\approx -0.11, & T=0.6\approx0.55T_\mathrm{c},\\
-1/\alpha(\tw=100\text{ s}) &\approx -0.12, & T=0.7\approx0.64T_\mathrm{c},\\
-1/\alpha(\tw=100\text{ s}) &\approx -0.14, & T=0.8\approx0.73T_\mathrm{c}.
\end{align}
\end{subequations}
These values still do not match the experimental values of~\eqref{eq:thermoremanent-exp}.
However, the difference between both sets seems to be roughly 
independent of temperature (this is best seen by plotting the parabolas
defined by each of the sets of exponents). We believe this is due
to an extrapolation error, with a similar bias for all temperatures.
\begin{table}[p]
\small
\centering
\begin{tabular*}{\columnwidth}{@{\extracolsep{\fill}}ccclll}
\toprule
$T$ &
$\tw$&
$t_\text{min}$ &
\multicolumn{1}{c}{$e(\tw)$} &
\multicolumn{1}{c}{$f(\tw)$} &
\multicolumn{1}{c}{$\chi^2$/d.o.f.}\\
\toprule
\multirow{8}{*}{$0.6$} 
&\multirow{2}{0.3cm}{$2$} &
      $10^3$ & $-0.236(7)$ & $0.873(9)$ & 52.2/104\\ 
& &   $10^6$ & $-0.30(6)$  & $0.82(5)$ & 13.9/64\\
\cmidrule{2-6}
&\multirow{2}{0.3cm}{$4$} &
      $10^3$ & $-0.203(6)$ & $0.909(8)$ & 47.9/104\\
 & &  $10^6$ & $-0.25(4)$  & $0.85(4)$ & 13.5/64\\
\cmidrule{2-6}
&\multirow{2}{0.3cm}{$8$} &
      $10^3$ & $-0.176(4)$ & $0.943(7)$ & 41.5/104\\
& &   $10^6$ & $-0.21(3)$  & $0.90(4)$ & 13.9/64\\
\cmidrule{2-6}
&\multirow{2}{*}{$16$} &
      $10^3$ & $-0.158(4)$ & $0.968(7)$ & 38.1/104\\
& &   $10^6$ & $-0.19(3)$ & $0.92(4)$ & 15.3/64\\
\midrule
\multirow{8}{*}{$0.7$} 
&\multirow{2}{0.3cm}{$2$} &
      $10^3$ & $-0.263(4)$ & $0.890(4)$ & 43.0/90\\
& &   $10^6$ & $-0.32(3)$ & $0.84(3)$ & 14.4/50 \\
\cmidrule{2-6}
&\multirow{2}{0.3cm}{$4$} &
      $10^3$   & $-0.230(3)$ & $0.921(4)$ & 71.9/90\\
& &   $10^6$   & $-0.29(3)$  & $0.862(25)$ & 12.8/50\\
\cmidrule{2-6}
&\multirow{2}{0.3cm}{$8$} &
     $10^3$    & $-0.2003(23)$ & $0.955(3)$ & 94.1/90\\
& &  $10^6$    & $-0.253(23)$  & $0.895(23)$ & 13.0/50\\
\cmidrule{2-6}
&\multirow{2}{*}{$16$} &
      $10^3$   & $-0.1768(20)$ & $0.985(3)$ & 138/90\\
& &   $10^6$   & $-0.226(19)$  & $0.921(22)$ & 10.6/50\\
\midrule
\multirow{8}{*}{$0.8$} 
&\multirow{2}{0.3cm}{$2$} &
      $10^3$   & $-0.302(16)$  & $0.891(16)$ & 45.1/72\\
& &   $10^6$   & $-0.5(3)$     & $0.77(15)$ & 14.3/32\\
\cmidrule{2-6}
&\multirow{2}{0.3cm}{$4$} &
      $10^3$   & $-0.257(12)$ & $0.934(14)$ & 63.6/72\\
& &   $10^6$   & $-0.6(4)$    & $0.71(17)$ & 11.4/32\\
\cmidrule{2-6}
&\multirow{2}{0.3cm}{$8$} &
      $10^3$   & $-0.223(10)$ & $0.970(13)$ & 69.8/72\\
& &   $10^6$   & $-0.49(24)$  & $0.76(12)$ & 11.1/32\\
\cmidrule{2-6}
&\multirow{2}{*}{$16$} &
      $10^3$   & $-0.192(8)$ & $1.008(12)$ & 65.9/72\\
& &   $10^6$   & $-0.40(19)$ & $0.81(12)$ & 8.49/32\\
\bottomrule
\end{tabular*}
\caption[Thermoremanent magnetisation, modified decay]{Parameters of a fit to Eq.~\eqref{eq:SG-C4-thermoremanent-3},
offering an alternative description of the 
thermoremanent magnetisation. For each temperature, 
we present fits in the range $t_\text{min}\leq t\leq \tw^\text{max}(T)$, 
where $t_\text{min}=10^3,10^6$ and $\tw^\text{max}(T)$ is given
in Table~\ref{tab:SG-z}.
\label{tab:SG-thermoremanent-2}
\index{magnetisation!thermoremanent|indemph}
}
\end{table}

Recall now that the coherence length $\xi(\tw)$ was also well 
represented by a power law. It follows that $C(t,\tw)$ should
be a power of $\xi(t+\tw)$, at least for the small $\tw$ 
of Table~\ref{tab:SG-thermoremanent}. In particular we
write
\begin{equation}\label{eq:SG-C4-thermoremanent-2}
C(t,\tw) \sim \bigl[\xi_{1,2}(t+\tw)\bigr]^{-d}.
\index{coherence length}
\end{equation}
We could perform a fit to~\eqref{eq:SG-C4-thermoremanent-2}
to find $d$, but we would have to contend with errors in both
the $x$ and $y$ coordinates (see Section~\ref{sec:CORR-xyerror}).
Rather, we combine equations~\eqref{eq:SG-xi-z} and~\eqref{eq:SG-C4-thermoremanent} to find
\begin{equation}
d = cz.
\index{critical exponent!z@$z$}
\index{coherence length}
\index{coherence length!experimental}
\end{equation}
Since both for $C(t,\tw)$ and for $\xi_{1,2}(\tw)$
 we are computing one fit for each jackknife block,
 \index{jackknife method}
we can apply this relation block by block to obtain $d$ and its
error estimate, without fitting for it directly.  
The result can be seen in Table~\ref{tab:SG-thermoremanent}.
We obtain $d\approx 2$, with a temperature dependence at the limit
of significance.

This is a potentially useful observation for experimental work
since, as we said before, the thermoremanent magnetisation 
is relatively easy to measure (compared with the coherence length, in any
case~\cite{joh:99}). Therefore, Eq.~\eqref{eq:SG-C4-thermoremanent-2}
could potentially be used as a definition of $\xi(\tw)$ for
experimental work.
\begin{figure}
\centering
\includegraphics[height=.7\linewidth,angle=270]{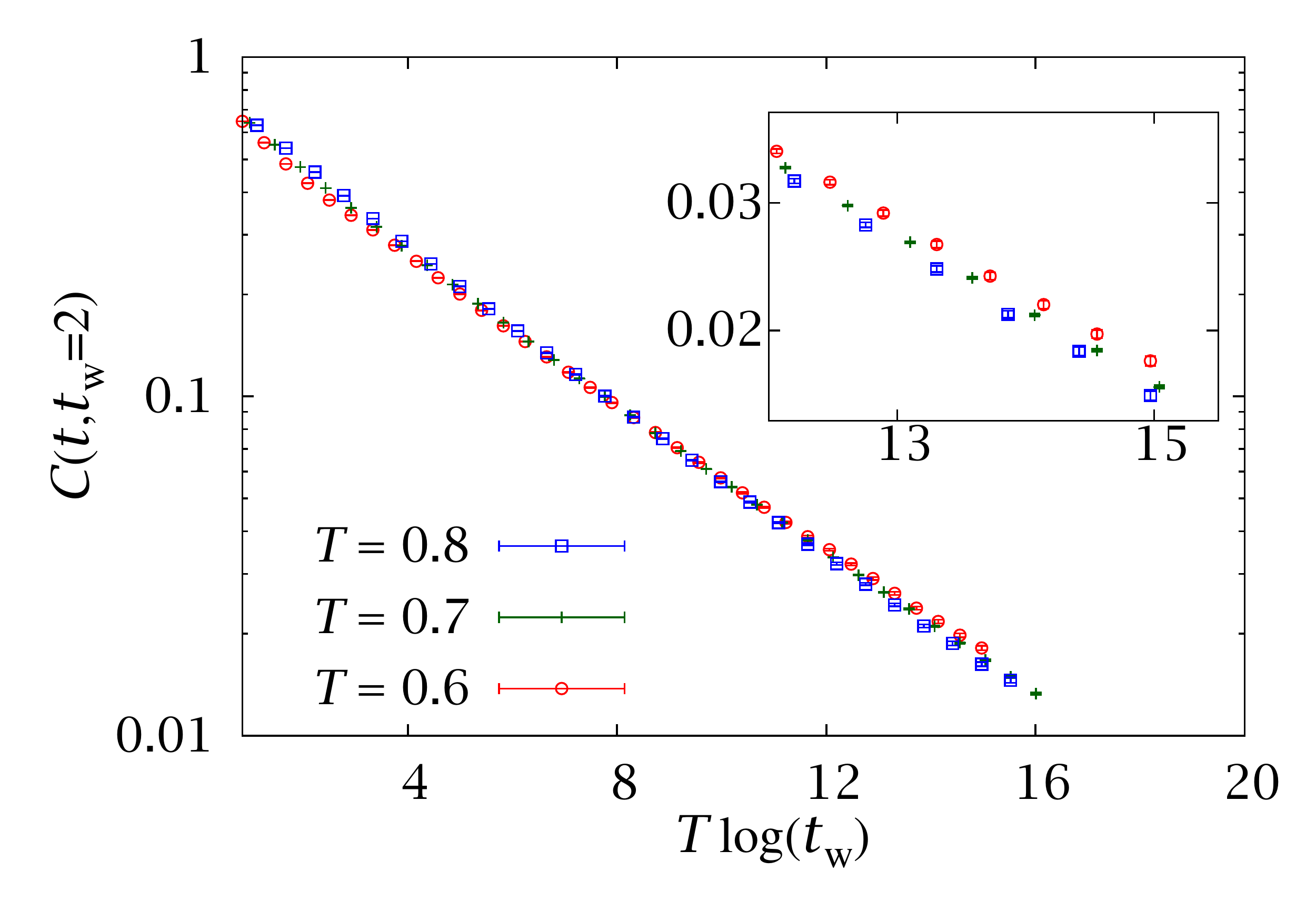}
\caption[Scaling of the thermoremanent magnetisation]{Coherence length $\xi_{1,2}(\tw)$ as a function
of $T\log \tw$, for three subcritical temperatures.
Even if the three curves are not equal within errors,
the overall scaling is suggestive.
\index{coherence length|indemph}
\label{fig:SG-thermoremanent}}
\end{figure}

As we have seen, there are several systematic effects in our results,
mainly the incompatibility of~\eqref{eq:thermoremanent-exp} 
and~\eqref{eq:SG-thermoremanent-alpha} and the need 
for a finite asymptote $A_0(\tw)$ in~\eqref{eq:SG-C4-thermoremanent}.
This suggests that~\eqref{eq:SG-C4-thermoremanent} is perhaps
not the best parameterisation. We can consider 
the alternative description
\begin{equation}\label{eq:SG-C4-thermoremanent-3}
C(t,\tw) = B(\tw) \exp\bigl[ e(\tw) (\log t)^{f(\tw)}\bigr].
\end{equation}
This would reproduce a power law if $f(\tw)=1$. Notice the
difference of this functional form with a 
 stretched exponential, discarded in~\cite{granberg:87} 
for our temperature range. The results of fits
to~\eqref{eq:SG-C4-thermoremanent-3} for several 
$\tw$ and fitting ranges can be seen in Table~\ref{tab:SG-thermoremanent-2}.
We observe that the fit parameters are very stable to variations 
in the fitting window. The value of $f(\tw)$ is incompatible 
with $1$, at least for $T=0.6,0.7$. However, from the point 
of view of the $\chi^2_\text{d}/\text{d.o.f.}$, both~\eqref{eq:SG-C4-thermoremanent} and~\eqref{eq:SG-C4-thermoremanent-3} are equally good.

Finally, let us go back to~\eqref{eq:SG-C4-thermoremanent} 
and Table~\ref{tab:SG-thermoremanent}. We see that the exponent $c(\tw)$ is 
roughly linear in $T$. This suggests that the thermoremanent 
magnetisation could be a temperature-independent function
of $T\log(t)$. We have tested this conjecture in Figure~\ref{fig:SG-thermoremanent}
and found it to be only approximate.
\index{magnetisation!thermoremanent|)}

\section{Dynamical heterogeneities}\label{sec:SG-dynamical-heterogeneities}
\index{dynamical heterogeneities|(}
Thus far, we have characterised the spin-glass dynamics globally
through the two-time correlation function. We have found 
this dynamics to be extremely slow and we have used
a growing coherence length to monitor the evolution.

However, the dynamics is actually heterogeneous 
across the system, with local regions behaving
differently from the bulk. As we said in Section~\ref{sec:SG-glass}, 
the study of these dynamical heterogeneities is a potential
point of contact between studies of spin glasses
and of general glassy behaviour. \index{glasses}

The traditional numerical approach to dynamic
heterogeneities has been the computation
of coarse-grained correlation functions. \index{correlation function (dynamics)!coarse-grained}
These are defined as an average over a cell of size $\ell^D$, with
$\ell\ll L$. The characteristic length scale for dynamical
heterogeneities can then be assessed through the probability
distribution of these coarse-grained correlations. For
$\ell$ much larger than the correlation length, one would 
expect the fluctuations to be averaged out, and the resulting
coarse-grained correlation would resemble the global 
one. For small $\ell$, however, strong deviations from a 
Gaussian behaviour are present~\cite{jaubert:07}. 
One defines, then, the correlation length as the crossover
length. 

Here we follow a different approach, 
consisting in the study of the two-time spatial correlation
 $C_{2+2}(\boldsymbol r,t,\tw)$ of Eq.~\eqref{eq:SG-C22}.
This observable was introduced in~\cite{jaubert:07}, but the 
time scales accessible to that work did not permit the 
measurement of correlation lengths greater than a couple
of lattice spacings.

\begin{figure}
\centering
\includegraphics[height=.7\linewidth,angle=270]{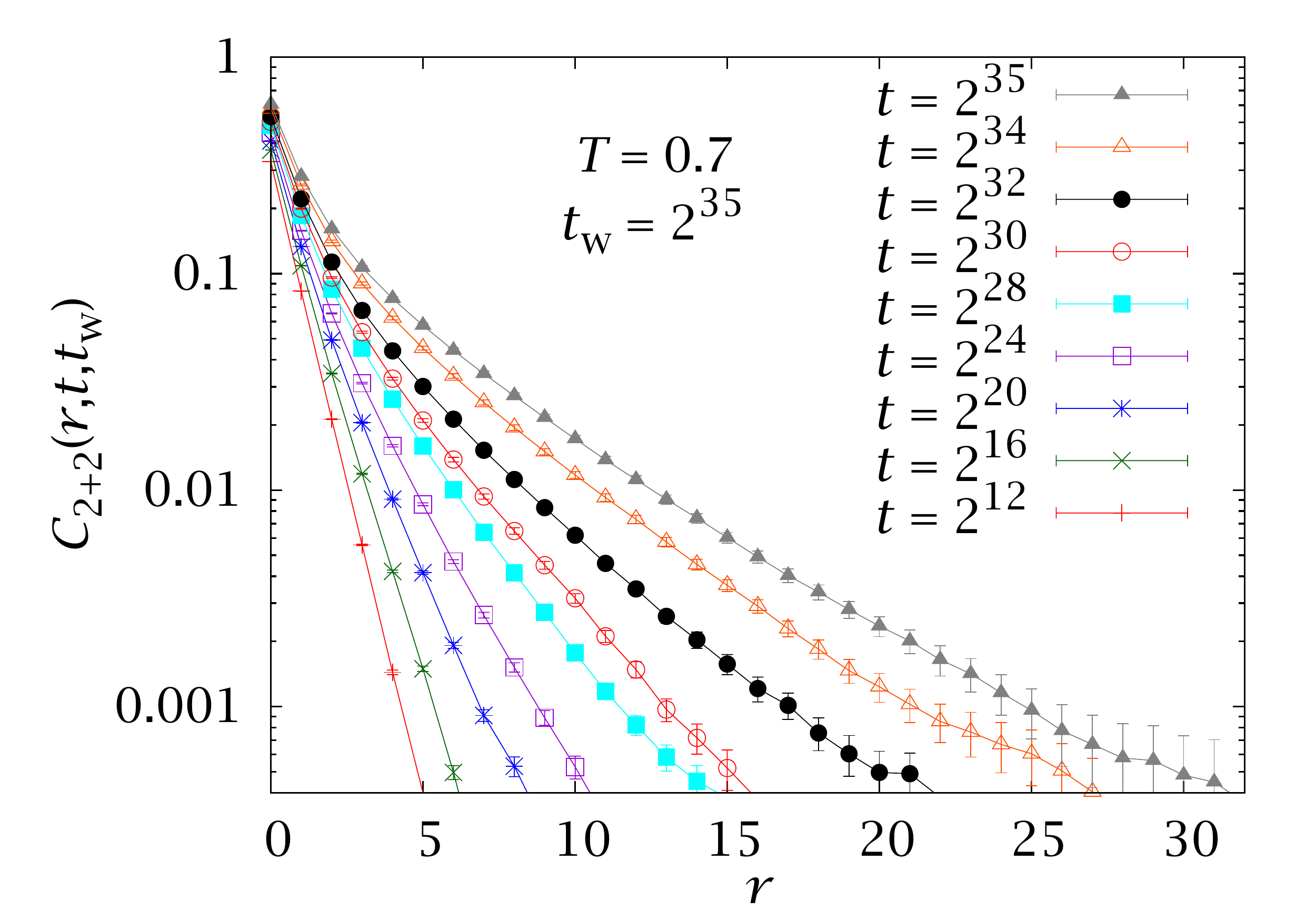}
\caption[$C_{2+2}(r,t,\tw)$]{The two-time
spatial correlation function $C_{2+2}(r,t,\tw)$ 
defined in~\eqref{eq:SG-C22} as a function of $r$, 
for several values of $t$. We consider a long waiting
time of $\tw=2^{35}$, for our $63$-sample simulations
at $T=0.7$.
\index{correlation function (dynamics)!spatial|indemph}
\label{fig:SG-C22}
}
\end{figure}

We have plotted the correlation length $C_{2+2}(r,t,\tw)$ (we use the 
same convention for axially oriented correlations as 
for $C_4$) in Figure~\ref{fig:SG-C22}, for a large waiting
time. From this plot we can see, qualitatively, that the correlation
reaches several lattice spacings for large times.

Recalling the one-to-one nature of $C(t,\tw)$ as a function
of $t$ for fixed $\tw$, we eliminate $t$ as independent
variable in favour of $C^2$, using a cubic spline
as our interpolating method. \index{cubic splines}
 We consider, then,
a correlation length $\zeta(C^2,\tw)$ from the long-distance
behaviour of $C_{2+2}(r,t(\tw,C^2),\tw)$, in complete
analogy with the case of $\xi(\tw)$ and $C_4(r,\tw)$,
 recall Eq.~\eqref{eq:SG-C22-long-distance}.
This correlation length is computed with the methods of
Section~\ref{sec:SG-integral-estimators}.
\begin{figure}
\centering
\includegraphics[height=\linewidth,angle=270]{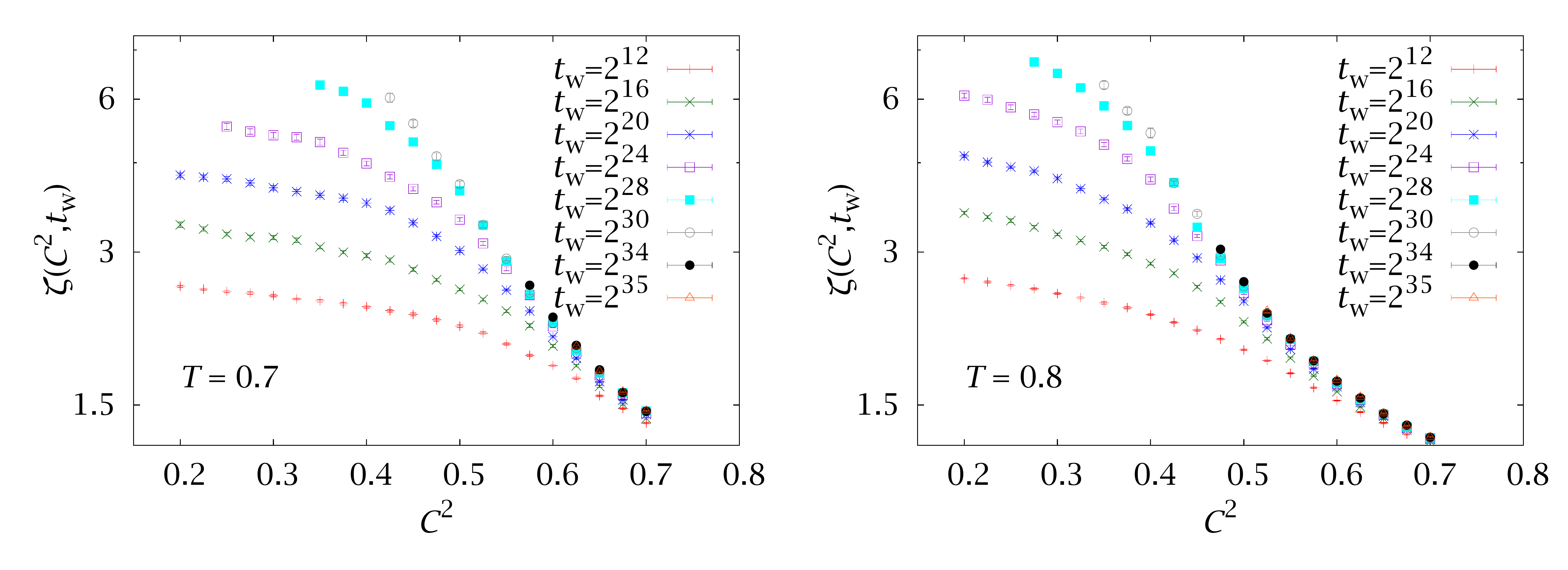}
\caption[Correlation length $\zeta(C^2,\tw)$]{%
Correlation length $\zeta(C^2,\tw)$ as a function of $C^2$
for several waiting times. We plot our results for
$T=0.7$ with $63$ samples and for $T=0.8$ ($96$ samples).
\label{fig:SG-zeta}
\index{correlation length!two-time|indemph}
}
\end{figure}

\begin{figure}[t]
\includegraphics[height=\linewidth,angle=270]{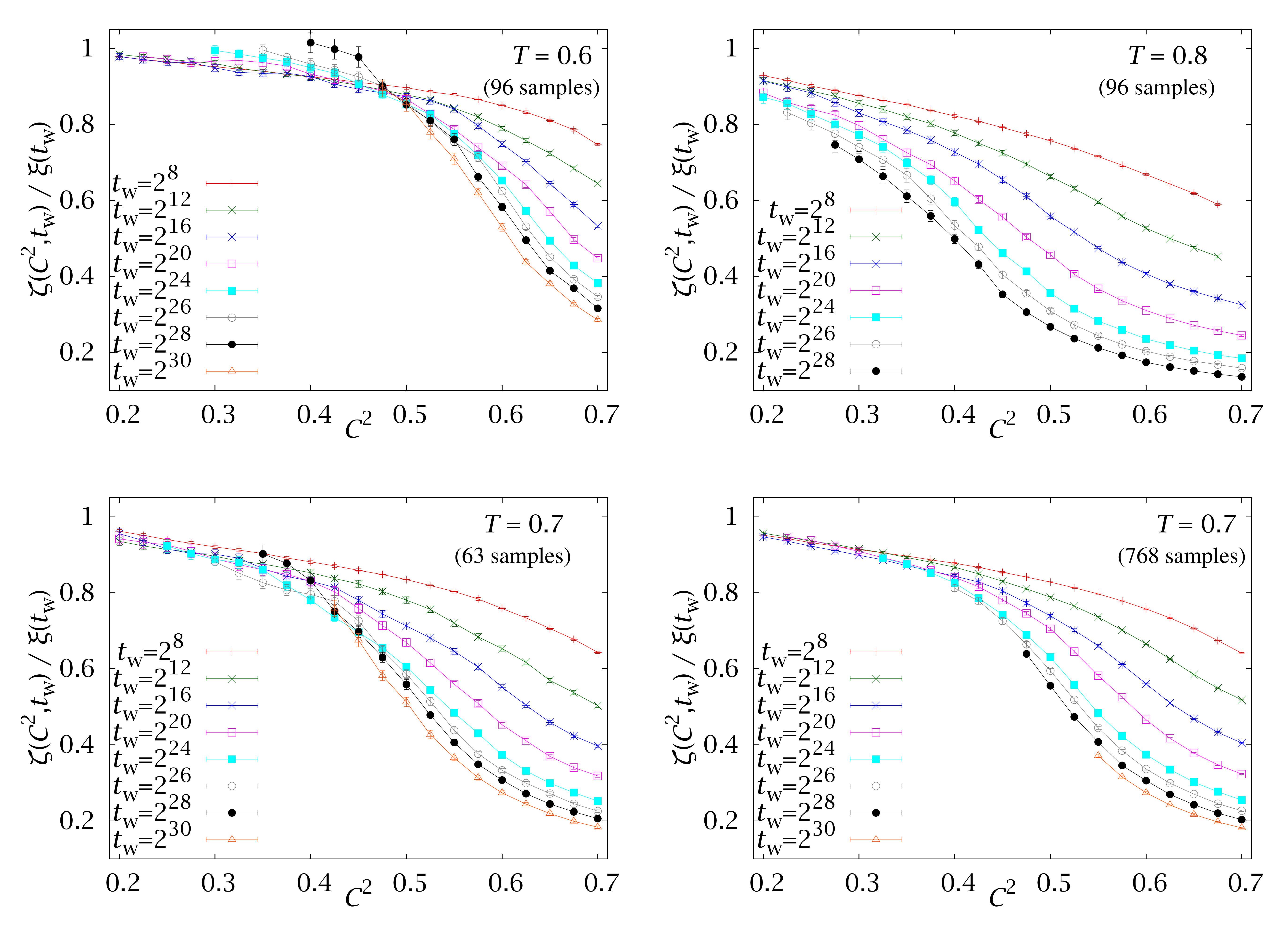}
\caption[Crossover behaviour in the dynamical heterogeneities]{%
Behaviour of the ratio $\zeta_{1,2}(C^2,\tw)/\xi_{1,2}(\tw)$
as a function of $C^2$ for several values of $\tw$. We observe
a crossover behaviour from a finite limit at small  $C^2$
to a vanishing limit for large $C^2$. Notice
that for $T=0.7$ we plot both our 
set with $63$ longer samples 
and our set with $768$ shorter ones.
\index{coherence length|indemph}
\index{correlation length!two-time|indemph}
\label{fig:SG-R}
}
\end{figure}

We have plotted $\zeta(C^2,\tw)$ as a function of $C^2$ 
for several $\tw$ at $T=0.7,0.8$ in Figure~\ref{fig:SG-zeta}.
We observe that the evolution in $\tw$ distinguishes 
two regimes. For large $C^2$  the correlation length 
approaches a $\tw$-independent enveloping curve. 
This corresponds with the low-$t$ regime, which
quickly approaches the translationally invariant
sector of the dynamics. When $C^2$ drops below $q_\text{EA}^2$,
however, we are in an aging regime, where the correlation
length diverges in the large-$\tw$ limit.

Since one would expect the divergence in the aging limit
to behave as $\xi(\tw)$, it is useful to consider 
the scaling variable
\begin{equation}
R(C^2,\tw) = \frac{\zeta(C^2,\tw)}{\xi(\tw)}\, .
\end{equation}
We expect
\begin{align}
\lim_{\tw\to\infty} R(C^2<q_\text{EA}^2) &> 0.\\
\lim_{\tw\to\infty} R(C^2>q_\text{EA}^2) &= 0.
\end{align}
We show this quantity for our three subcritical temperatures
in Figure~\ref{fig:SG-R}. For $T=0.7$ we show both our 
longer simulations with $63$ samples and our more precise
set with $768$.

Clearly, there is a crossover, which is best observed 
for $T=0.6$. However, we face a dilemma: for low
$T$ the growth of the coherence length is  very slow 
and we do not see the large-$C^2$ behaviour as clearly,
while for high $T$ the value of $q_\text{EA}$ is too 
low for us to appreciate the crossover at 
low $C^2$ (notice that the spin glass undergoes a second-order
transition, so $q_\text{EA}(T_\text{c}) = 0$).

\begin{table}
\small
\centering
\begin{tabular*}{\columnwidth}{@{\extracolsep{\fill}}ccrlcc}
\toprule
$T$ &
$C^2$ &
\multicolumn{1}{c}{ $z_\zeta$} &
 $\chi^2_{\zeta}$/d.o.f. & \multicolumn{1}{c}{ $z$} \\
\hline
$0.6$ & $0.200$ & $13.4(6)\ \ \ \ $ &  $0.01/2 $ &  $14.06(25)$\\
$0.7$ & $0.200$ & $11.14(20)$ & $0.69/3 $ & $11.56(13)$\\
$0.8$ & $0.100$ & $9.56(17)$ &  $3.73/5$  &   $\ \ 9.42(15)$\\
\bottomrule
\end{tabular*}
\caption[Dynamic exponent $z_\zeta$ of the two-time correlation length]{Value of the dynamic exponent $z_\zeta$ of Eq.~\eqref{eq:SG-zeta-z} for three subcritical temperatures. We also give our best
determination of the dynamic critical exponent $z$ 
from Table~\ref{tab:SG-z}.
\index{critical exponent!z@$z$|indemph}
\label{tab:SG-zeta-z}
}
\end{table}
In short, while our data include long enough times to appreciate
the behaviour qualitatively, they are not 
sufficiently clean for us to characterise the scaling behaviour 
at the crossover point. In Section~\ref{sec:SG-phase-transition}
we shall take on this problem again with equilibrium methods.
We shall show how the crossover in the behaviour of dynamical
heterogeneities can be interpreted as a proper phase 
transition and how its critical parameters, computed in equilibrium,
can be used to characterise the dynamics.

For now we address the simpler problem of the scaling 
of $\zeta(C^2,\tw)$ in the low-$C^2$ sector. Since in 
this situation $\zeta$ and $\xi$ differ only in a 
constant factor close to one, we consider the following ansatz
\begin{equation}\label{eq:SG-zeta-z}
\zeta(C^2<q_\text{EA}^2,\tw) = A_\zeta \tw^{1/z_\zeta},
\end{equation}
where $z_\zeta$ is expected to be similar to $z$.
For each $T$, we have performed fits to~\eqref{eq:SG-zeta-z}
for a value of $C^2$ expected to be below $q_\text{EA}$.
We note that the value of $q_\text{EA}$ has consistently
been overestimated
in the literature, as we shall show in the following sections.
In fact, in order to be safe we have simply fitted
the lowest value of $C^2$ for which we could obtain
an acceptable number of degrees of freedom in the fits.

The results are shown in Table~\ref{tab:SG-zeta-z}, where
we can see that $z_\zeta(T)$ is indeed similar to $z(T)$, 
although we can measure it with much worse precision.

\index{dynamical heterogeneities|)}
\newpage
\section[The translationally invariant sector and a first look at $q_\text{EA}$]{The translationally invariant sector}\label{sec:SG-qEA-dynamics}
\index{aging|(}\index{correlation function (dynamics)!temporal!stationary|(}
In Section~\ref{sec:SG-aging} we left pending the question of computing
the stationary part of the temporal correlation function. As we noted
there, a naive extrapolation is complicated,  because we have 
to consider the limit $\tw\to\infty$, which is also strongly 
$t$-dependent. 

Here we address this problem, taking advantage of the characteristic
lengths we have computed. In particular, we study the following
dynamical variable
\begin{equation}\label{eq:SG-x}
x(t,\tw) = \left(\frac{\zeta(t,\tw)}{\xi(\tw)}\right)^2\, .
\end{equation}
Following the discussion of the previous section, the 
translationally invariant sector is recovered
in the limit $x\to 0$ (where $x$ is essentially $\xi^{-2}(\tw)$
in its natural units for each $t$). However, this limit is
much easier to compute in a controlled way than $\tw\to\infty$.
\begin{figure}
\centering
\includegraphics[height=0.7\linewidth,angle=270]{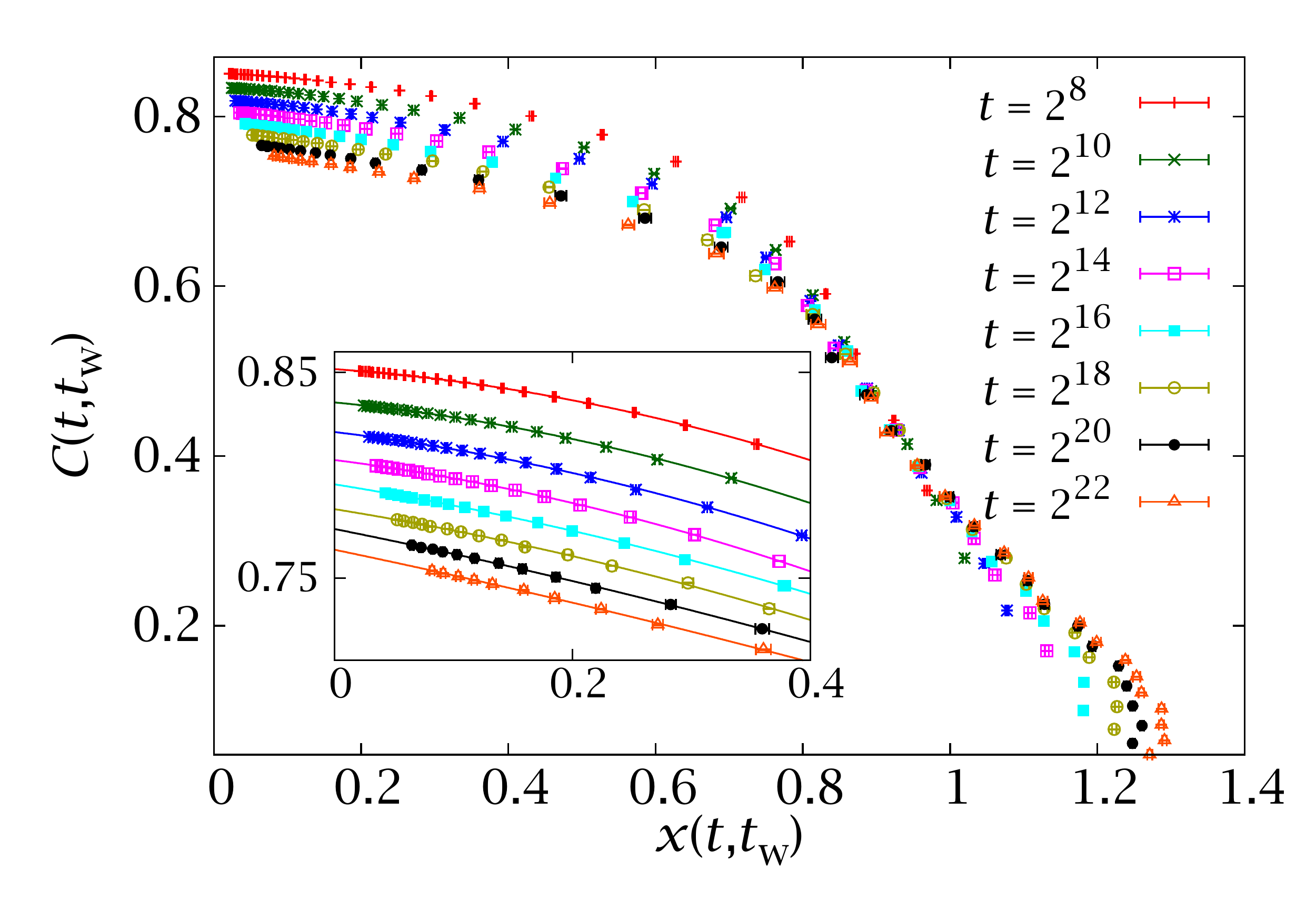}
\caption[Extrapolation of $C(t,\tw)$ to find $C_\infty(t)$]{%
The temporal correlation function as a function of $x(t,\tw)$, 
Eq.~\eqref{eq:SG-x}, with an inset showing a close-up of
the small-$x$ limit (equivalent to an extrapolation to $\tw\to\infty$).
Plot for our $768$-sample simulation at $T=0.7$.}
\label{fig:SG-C-x}
\end{figure}

We have plotted $C(t,\tw)$ against $x(t,\tw)$ for our simulations 
at $T=0.7$ in Figure~\ref{fig:SG-C-x}. We see that, indeed, the
curve is very smooth in the small-$x$ limit, and an extrapolation
seems feasible (see Inset). Furthermore, the curves
for different $t$ become parallel as $t$ grows, which 
suggests the existence of a smooth scaling function
$C(t,\tw)=C_\infty(t) + f(x)$. In particular, we consider
the following functional form
\begin{equation}\label{eq:SG-C-x}
C(t,x) = C_\infty(t) + a_1(t) x + a_2(t)x^2.
\end{equation}
However, a naive extrapolation using~\eqref{eq:SG-C-x} is still 
problematic. The reason is that we have errors both for $C$ and for 
$x$ (i.e., for both the $x$ and $y$ coordinates in the fit). Luckily, however, the errors
were similar in both directions and for all points. Therefore, we can 
use the method explained in Section~\ref{sec:CORR-xyerror}.
We compute the fits for $x\leq0.5$, using $\xi_{1,2}$ 
and $\zeta_{1,2}$ as our estimates of the characteristic lengths.
\begin{figure}
\centering
\includegraphics[height=0.7\linewidth,angle=270]{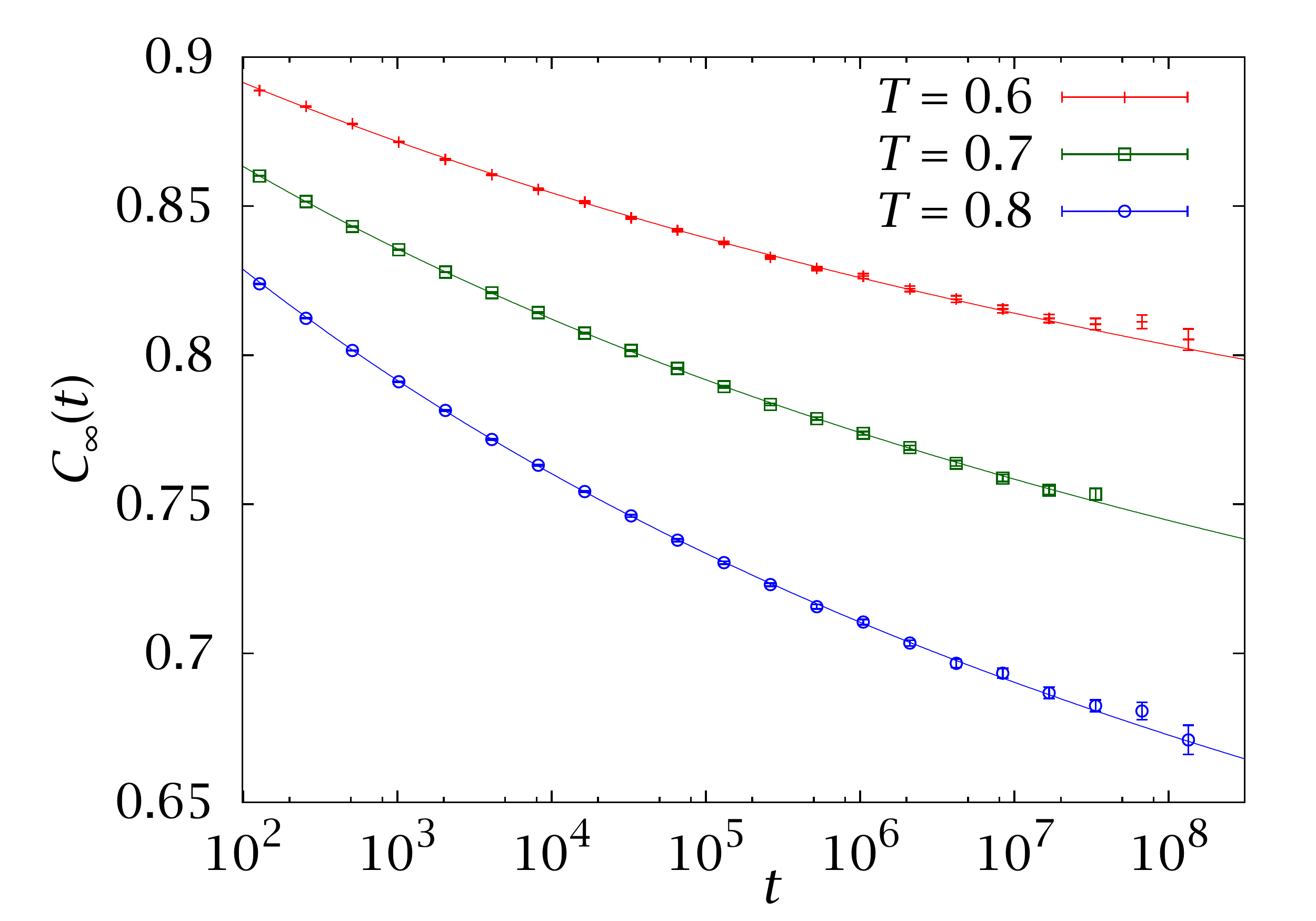}
\caption[Stationary part of the temporal correlation function]{%
Stationary part $C_\infty(t)$ of the two-time
correlation function $C(t,\tw)$ for $T=0.7,0.7,0.8$, obtained
from the extrapolations in Figure~\ref{fig:SG-C-x}.
\index{correlation function (dynamics)!temporal!stationary|indemph}
\label{fig:SG-Ct}}
\end{figure}

\index{spin glass!order parameter}
This method has allowed us to compute $C_\infty(t)$ with 
remarkable accuracy for $t\lesssim10^8$ (Figure~\ref{fig:SG-Ct}).
Now that we have this function, we could try a second extrapolation
to find $q_\text{EA}$, 
\begin{equation}
\lim_{t\to\infty} C_\infty(t) = q_\text{EA}.
\end{equation}
However, this second extrapolation turns out to be very 
difficult, because now we cannot avoid the issue of considering
the infinite limit explicitly. Furthermore, we lack a 
convincing theoretical expectation as to the functional
form of $C_\infty(t)$.

The simplest option is a power law decay, 
\begin{equation}\label{eq:SG-Ct-power}
C_\infty(t) = q_\text{EA} + At^{-B}.
\end{equation}
This equation yielded very good fits for $T=0.6,0.8$ (Table~\ref{tab:SG-qEA-dyn}, but 
the values of the exponents were very small, $B\sim0.05$.
The smallness of $B$ makes the extrapolation extremely risky,
since it means that $q_\text{EA}$ is going to be much smaller 
than our range of values for $C_\infty(t)$. We note that 
this slow evolution has led some authors to overestimate
$q_\text{EA}$ in the past (with larger errors, $C_\infty$
seems almost constant).
\begin{table}
\small
\centering
\begin{tabular*}{\columnwidth}{@{\extracolsep{\fill}}cclcllc}
\toprule
\multirow{2}{0.5cm}{$T$} &
\multirow{2}{1cm}{Fitting range}&
\multicolumn{2}{c}{ Logarithm} &
\multicolumn{3}{c}{ Power law} \\
\cmidrule{3-4} \cmidrule{5-7}
& &
\multicolumn{1}{c}{$q_\text{EA}$} & $\chi^2_\text{d}/\text{d.o.f}$ &
 \multicolumn{1}{c}{$q_\text{EA}$} &\multicolumn{1}{c}{$B\times10^2$} &  $\chi^2_\text{d}/\text{d.o.f}$ \\ 
\toprule
\multirow{3}{*}{$0.6$} &
  $[10^2,10^8]$   & 0.607(16) & 34.1/17 & 0.730(8)  & 5.7(4)  &  31.2/17 \\ 
& $[10^3,10^8]$   & 0.62(3)   & 7.23/14 & 0.733(14) & 5.8(7)  &  7.59/14 \\
& $[10^4,10^8]$   & 0.62(5)   & 6.25/10 & 0.726(24) & 5.4(12) &  6.32/10 \\
\midrule
\multirow{3}{*}{$0.7$} &                                     
  $[10^2,10^8]$   & 0.497(10) & 23.7/17 & 0.656(5)  & 6.16(18) &  32.6/17 \\
& $[10^3,10^8]$   & 0.474(21) & 18.9/14 & 0.637(11) & 5.5(3)   &  18.5/14 \\
& $[10^4,10^8]$   & 0.49(5)   & 15.0/10 & 0.63(3)   & 5.4(9)   &  15.3/10 \\
\midrule
\multirow{3}{*}{$0.8$} &                                     
  $[10^2,10^8]$   & 0.371(13) & 6.50/17 & 0.568(7)  & 6.56(20)  &  9.39/17 \\
& $[10^3,10^8]$   & 0.368(24) & 5.53/14 & 0.556(12) & 6.2(4)    &  4.27/14 \\
& $[10^4,10^8]$   & 0.40(6)   & 4.31/10 & 0.56(3)   & 6.4(11)   &  3.82/10 \\
\bottomrule
\end{tabular*}
\caption[Estimate of $q_\text{EA}$ from $C_\infty(t)$]{Estimate of
$q_\text{EA}$ for three subcritical temperatures,
using two different extrapolating functions. For $T=0.6,0.8$ both
are very good, but at $T=0.7$ (where we have better statistics)
they are somewhat forced.
 This suggests that the real $q_\text{EA}$ probably
lies in between our two estimates. For the power law
extrapolation, Eq.~\eqref{eq:SG-Ct-power},
we also quote the exponent $B$.
Notice that this exponent is not proportional to $T$.
See Section~\ref{sec:SG-phase-transition} for a considerably
more precise determination of this parameter.
\index{spin glass!order parameter|indemph}
\label{tab:SG-qEA-dyn}
}
\end{table}

Perhaps more disquieting, it may seem that a low
enough value of $B$ would allow for the 
possibility that $q_\text{EA}=0$. In order 
to dispel this notion, we have tried a second fit
to a logarithmic decay,
\begin{equation}\label{eq:SG-Ct-log}
C_\infty(t) = q_\text{EA} + \frac{A}{B+\log t}\ .
\end{equation}
We stress that this ansatz lacks theoretical basis, 
we merely take it as a lower bound on the value of $q_\text{EA}$.
Numerically, it turns out, the logarithmic fit is as 
good as the exponential one for $T=0.6,0.8$, even if 
it produces incompatible values of $q_\text{EA}$ (Table~\ref{tab:SG-qEA-dyn}).

Furthermore, if we try both fitting functions 
for our $T=0.7$ data, where we have much better statistics, 
we find that they were somewhat forced. This leads us to 
conclude that the real asymptotic behaviour of $C_\infty(t)$
is probably something in between~\eqref{eq:SG-Ct-power}
and~\eqref{eq:SG-Ct-log}. We can use 
the difference between both methods with a fitting 
window of $t\in[10^3,10^8]$ as our uncertainty
interval,
\begin{subequations}
\begin{align}
0.62 &\leq q_\text{EA}(T=0.6)\leq 0.733,\\
0.474 &\leq q_\text{EA}(T=0.7)\leq 0.637,\\
0.368 &\leq q_\text{EA}(T=0.8)\leq 0.556.
\end{align}
\end{subequations}
Even with our unprecedently long simulations, we 
are still at the threshold of being able to 
compute $q_\text{EA}$ with dynamical methods.
In Section~\ref{sec:SG-phase-transition} we shall 
perform an equilibrium computation of this quantity.
\index{correlation function (dynamics)!temporal!stationary|)}
\index{aging|)}

\section{Equilibrium analogues and the time-length dictionary}\label{sec:SG-statics-dynamics}
As we have seen, the non-equilibrium approach, while straightforward, 
is limited when one needs to perform some delicate analyses, such
as the estimation of $q_\text{EA}$. Therefore, one should complement
it with equilibrium computations.
Recall that we are working in a framework where the equilibrium phase
is unreachable for experimental systems, yet conditions
the  non-equilibrium dynamics. Remember, for instance, 
the relation between the fluctuation-dissipation ratio of  Figure~\ref{fig:SG-FDT} 
and the equilibrium $p(q)$~\cite{franz:98,franz:99}. \index{fluctuation-dissipation}

In this section we take the qualitative statics-dynamics relation one step farther.
Our proposal is that the equilibrium correlation functions
computed for systems of finite size should reproduce their
non-equilibrium counterparts in the thermodynamical limit 
for finite time. An infinite
system with a finite coherence length $\xi(\tw)$ can, very roughly, 
be considered as a collection of systems of size $L\sim \xi(\tw)$
in equilibrium, with some reservations. This relation
should establish a time-length dictionary $\tw\leftrightarrow L$, 
which is our objective in this section. 

In particular, we shall consider the equilibrium spatial autocorrelation,
\begin{equation}\label{eq:SG-C4-equilibrium}
C_4(\boldsymbol r) = \frac{1}{N} \sum_\bx \overline{\braket{q_{\boldsymbol x} q_{\boldsymbol x+\boldsymbol r}}} .
\nomenclature[C4]{$C_4(\boldsymbol r)$}{Spatial autocorrelation of the overlap field (equilibrium)}
\index{correlation function (equilibrium)!spatial}
\end{equation}

We want to relate $C_4(\boldsymbol r)$ to $C_{2+2}(\boldsymbol r,t,\tw)$.
As we have said, the waiting time $\tw$ can be related to $L$, but what of $t$?
The answer lies in the, already exploited, one-to-one relation between 
$t$ and the two-time overlap $C(t,\tw)$ at fixed $\tw$. In equilibrium, 
the dependence on $C$ is replaced by the computation of correlation 
functions conditioned to a fixed value of $q$. Finally, we note
that~\eqref{eq:SG-C4-equilibrium} is non-connected, unlike $C_{2+2}$, 
as we defined it in~\eqref{eq:SG-C22}. Therefore, we shall 
consider here a non-connected version of $C_{2+2}$,
\begin{equation}\label{eq:SG-C22-prime}
C_{2+2}'(\boldsymbol r,t,\tw) = 
\overline{\frac1N \sum_\bx
c_\bx(t,\tw) c_{\bx+\boldsymbol r}(t,\tw)}\ .
\end{equation}
Our objective, then, for this section, is finding a relation between
\begin{equation}
C_{2+2}'\bigl(\boldsymbol r, t(C(t,\tw)), \tw\bigr) \longleftrightarrow C_4(\boldsymbol r|q).
\end{equation}

We still have to provide a workable definition of the $q$-conditioned $C_4(\boldsymbol r |q)$.
In order to do this, let us step back for  a moment and consider the pdf
of the spin overlap,
\begin{equation}
p_1(q)  = \overline{\biggl\langle\delta\biggl(q-\frac1N \sum_\bx q_\bx\biggr)\biggr\rangle}.
\end{equation}
This definition is completely analogous to the $p_1(m)$ of~\eqref{eq:TMC-p1} we considered 
in Chapter~\ref{chap:tmc}. Like in that case, the $p_1(q)$ is not smooth 
for finite systems, but rather the sum of $N+1$ Dirac deltas. Even if we are not 
considering a tethered formalism here, \index{tethered formalism}
we can borrow the cure to this problem. In particular, we define our 
smooth $p(q)$ as the convolution of $p_1(q)$ with a Gaussian 
of width $1/\sqrt N$
\begin{align}\label{eq:SG-p-q}
p(q=c) &= \int_{-\infty}^\infty \dd q'\ p_1(q') \mathcal G_N(c-q')= \overline{\biggl\langle\mathcal G_N\biggl(c-\frac1N \sum_\bx q_\bx\biggr)\biggr\rangle}, \\
\mathcal G_N(x) &= \sqrt{\frac{N}{2\uppi}} \ee^{-N x^2/2}.
\nomenclature[p(q)]{$p(q)$}{Smoothed probability density function of the spin overlap}
\end{align}
The pdf $p(q)$ is an interesting observable in its own right, since, as we saw 
in Chapter~\ref{chap:sg}, its thermodynamical limit is very different 
in the droplet and RSB pictures. We shall study it in detail in Section~\ref{sec:SG-spin-overlap}.

Now, we can generalise our treatment of $p(q)$ to define conditional 
expectation values at fixed $q$
\begin{equation}\label{eq:SG-E}
\mathrm{E}(O|q=c) = \frac{\overline{\biggl\langle O \mathcal G_N\biggl(c-\frac1N \sum_\bx q_\bx\biggr)\biggr\rangle}}{\overline{\biggl\langle \mathcal G_N\biggl(c-\frac1N \sum_\bx q_\bx\biggr)\biggr\rangle}}.
\nomenclature[E(O)]{$\mathrm{E}(O\vert q)$}{Conditional expectation value of $O$ at fixed $q$}
\end{equation}
Of course, the standard expectation values can be recovered from the $\mathrm{E}(O|q)$,
\begin{equation}
\overline{\braket{O}} = \int_{-\infty}^\infty\dd q\ p(q)\mathrm{E}(O|q).
\end{equation}
Finally, we define the fixed-$q$ correlation function as
\begin{equation}\label{eq:SG-C4-q}
C_4(\boldsymbol r|q) = \mathrm{E}\left( \frac1N \sum_\bx q_\bx q_{\bx +\boldsymbol r}\middle| q\right).
\end{equation}

\begin{figure}
\centering
\includegraphics[height=.7\linewidth,angle=270]{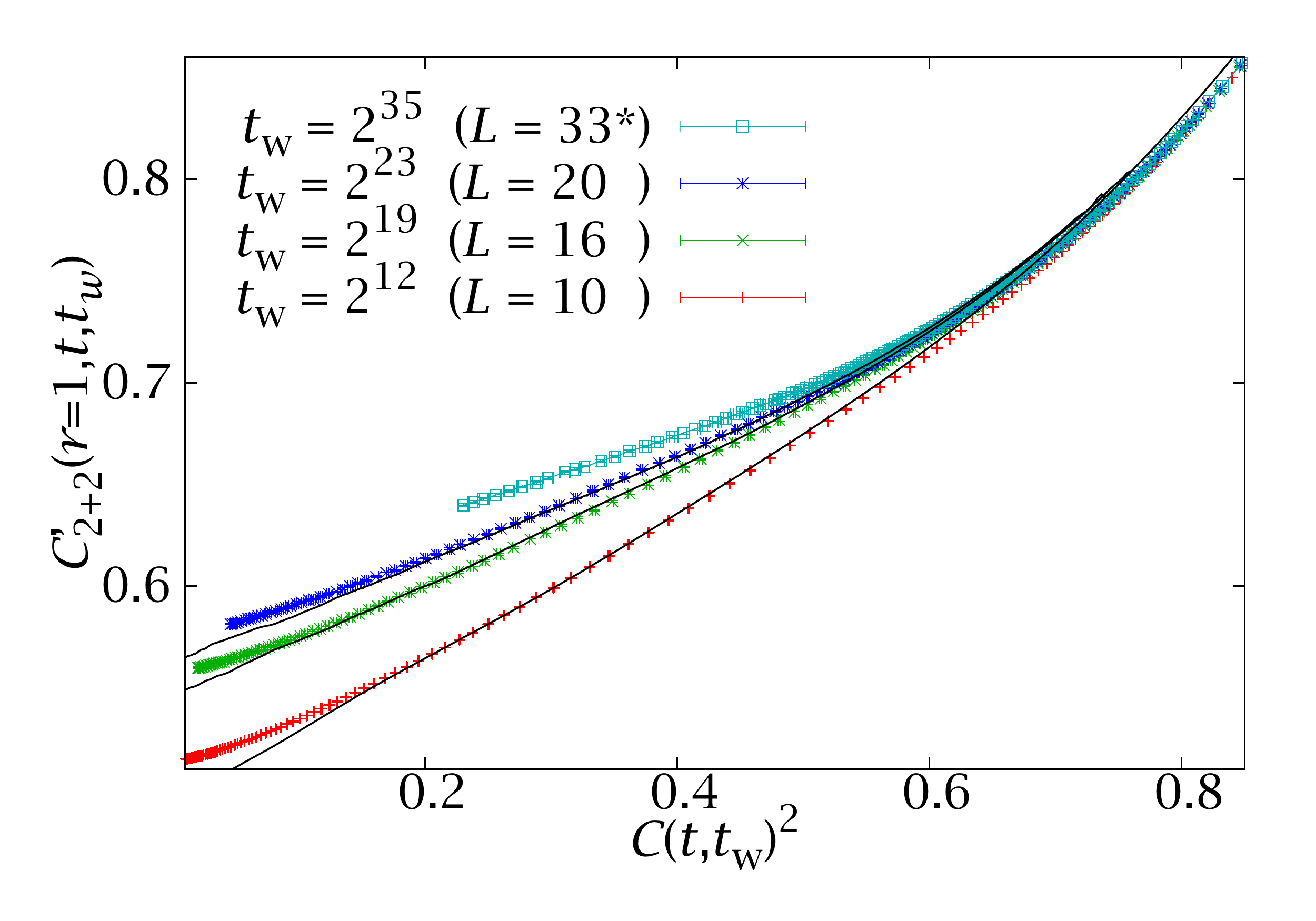}
\caption[Time-length dictionary for $r=1$]{Plot of the off-equilibrium 
non-connected correlation $C_{2+2}'(r=1,C^2,\tw)$ of Eq.~\eqref{eq:SG-C22-prime}
for several
waiting times at $T=0.7$. We compare it with the equilibrium
$C_4(r=1|q)$, Eq.~\eqref{eq:SG-C4-q}, using data from~\cite{contucci:07} (continuous lines), 
finding that they are accurately matched through the 
time-length dictionary $L\approx 3.7\xi(\tw)$. According
to this, the correlation for $\tw=2^{32}$ would correspond
to an $L=33$ system.
\index{correlation function (dynamics)!spatial|indemph}
\index{correlation function (equilibrium)!spatial|indemph}
\index{statics-dynamics equivalence|indemph}
\index{time-length dictionary|indemph}
\label{fig:SG-statics-dynamics-PRL}
}
\end{figure}

In~\cite{janus:08b} we compared our non-equilibrium $C_{2+2}'(r=1,t,\tw)$
with the equilibrium $C_4(r=1|q)$, using equilibrium data from~\cite{contucci:06,contucci:07},
at $T=0.7$. The result is plotted in Figure~\ref{fig:SG-statics-dynamics-PRL}.
We can see, how, indeed, the equilibrium correlation functions
reproduce the non-equilibrium ones through a consistent time-length 
correspondence. More precisely, the non-equilibrium 
correlations at time $\tw$ can be matched to the equilibrium 
ones with the time-length dictionary \index{time-length dictionary}
\begin{equation}
L(\tw)\approx3.7\xi_{1,2}(\tw).
\end{equation}
In~\cite{janus:08b} we used this equivalence to predict that $\tw=2^{32}$ 
would correspond to an $L=33$ system.  We note that, due to our 
discretisation of measuring times, we have only established
this dictionary up to the nearest power of $2$. 

In the previous analysis, we were limited by the equilibrium data, which 
only reached $L=20$ and did not consider correlations for $r>1$. 
We afterwards
ran our own equilibrium simulations with \textsc{Janus}, reaching $L=32$
and taking more thorough measurements (see Appendix~\ref{chap:runs-sg} for 
details on these simulations and thermalisation checks).
Our lowest temperature
for our largest lattices was precisely $T=0.7$. We simulated $L=8,12,16,24,32$.
We can now use this much more precise equilibrium simulation
to improve on the conclusions of Figure~\ref{fig:SG-statics-dynamics-PRL}.

\begin{figure}
\centering
\includegraphics[height=.7\linewidth,angle=270]{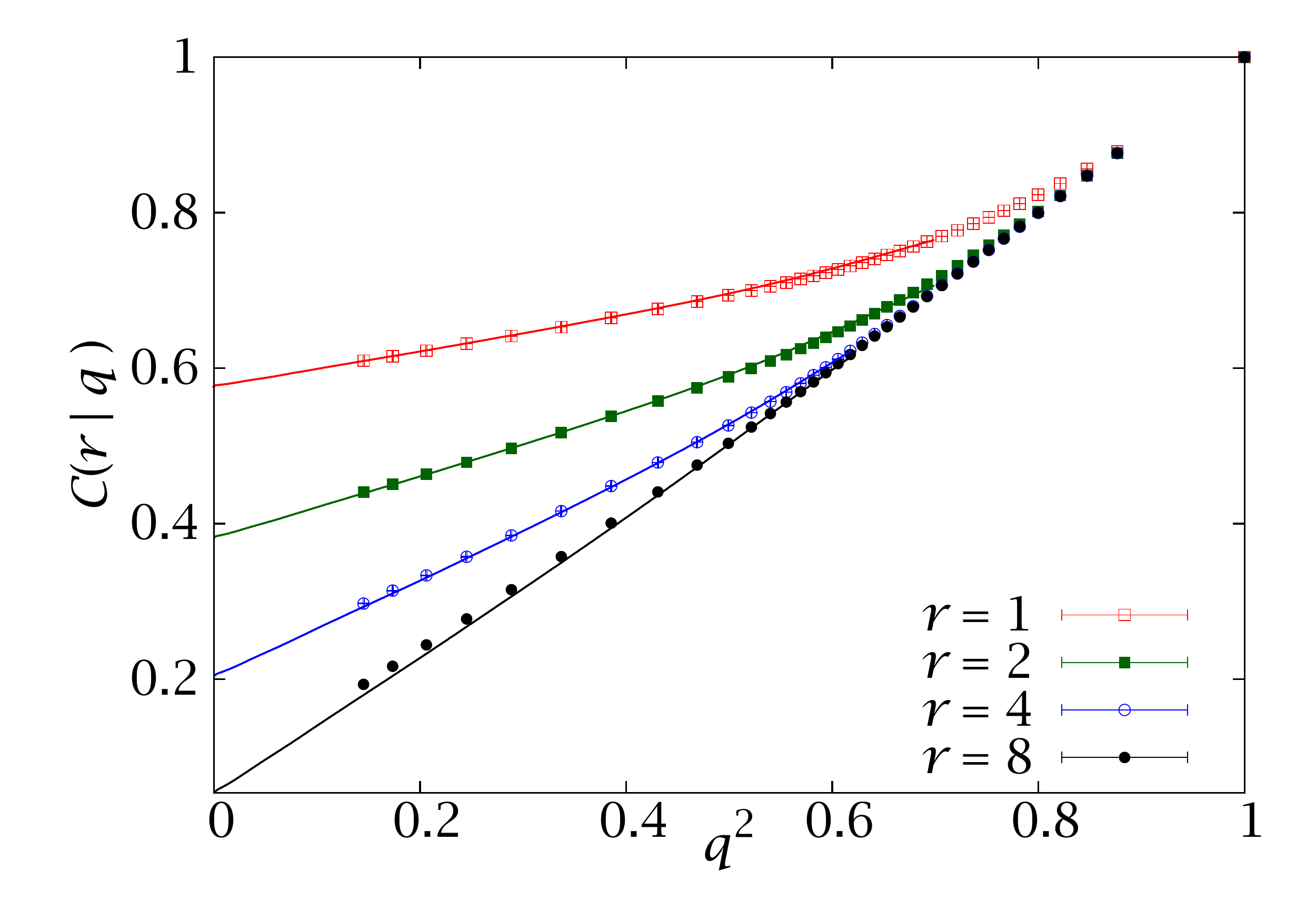}
\vspace*{1.5cm}

\includegraphics[height=.7\linewidth,angle=270]{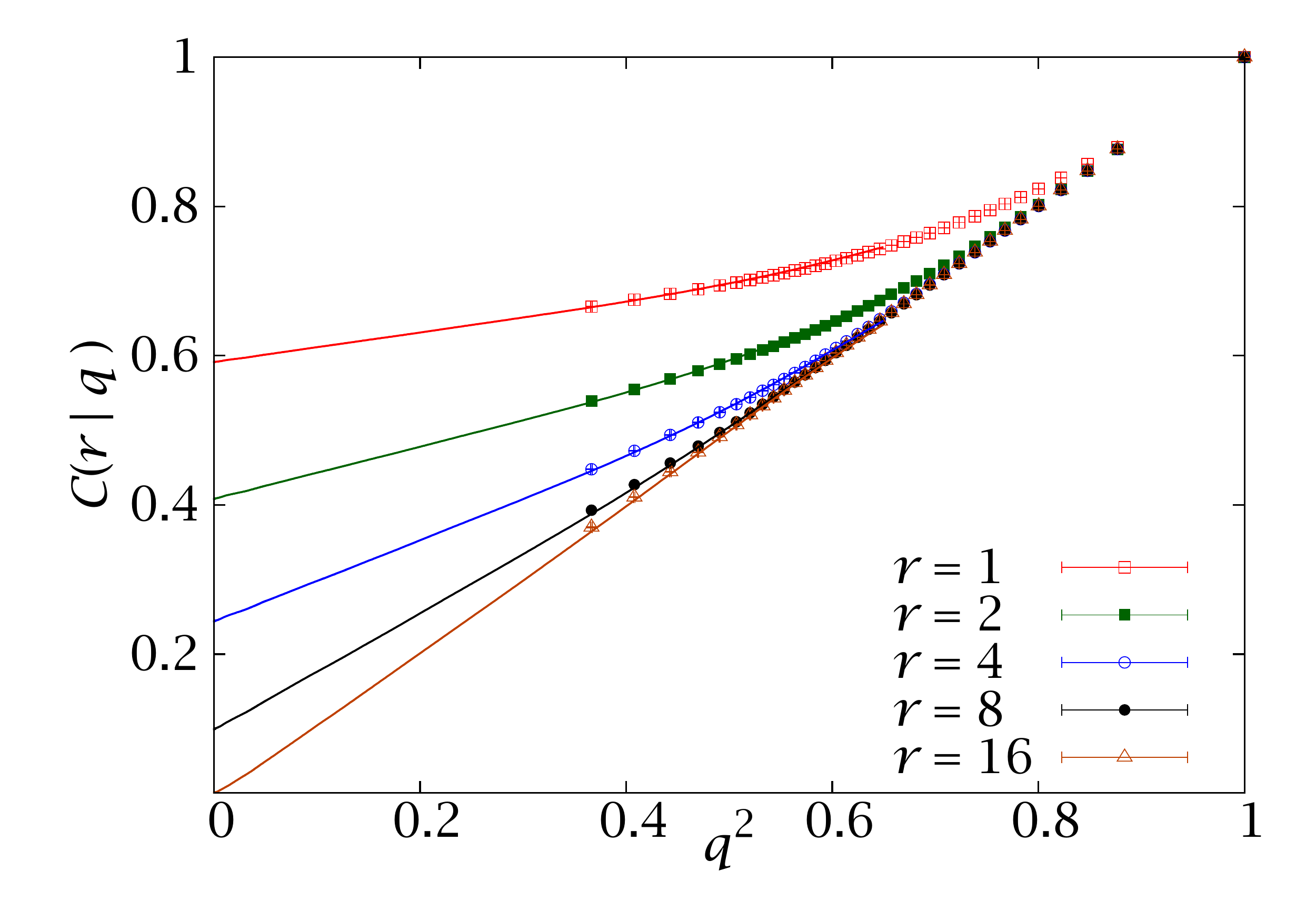}
\caption[The time-length dictionary for $r>1$]{%
Comparison of our non-equilibrium correlation functions
$C'_{2+2}(r,C^2,\tw)$ (points) and our equilibrium $C_4(r|q)$ (lines)
at $T=0.7$. In the top  panel we compare $L=24$ with $\tw=2^{26}$
and on the bottom one we consider $L=32$ and $\tw=2^{31}$. The errors
in both data sets are comparable and smaller than the points.
\index{statics-dynamics equivalence|indemph}
\label{fig:SG-statics-dynamics-JSTAT}
}
\end{figure}

According to the time-length dictionary, then, our $L=32$ 
simulations would correspond to $\tw\approx2^{31}$ and our 
$L=24$ ones would correspond to $\tw\approx2^{26}$.
We have plotted both cases in Figure~\ref{fig:SG-statics-dynamics-JSTAT}, 
this time for several values of $r$, corroborating our 
prediction of~\cite{janus:08b}. Notice that in a finite lattice
$C_4(r|q)$ can only be computed up to $r=L/2$, while
the non-equilibrium $C_{2+2}'(r,t,\tw)$ is defined 
for arbitrary distance. However, we find a very good matching 
even for relatively large values of $r$.

\subsection{The experimental length scale}\index{experimental scale|(}
As we can see, even relatively long times $2^{32} \text{ MCS}\sim10^{10}\text{ MCS}\sim 0.01\text{ seconds}$,
correspond to the equilibrium behaviour for quite small systems. This raises the question
of the relevance of the thermodynamical limit for experimental physics.
Indeed, for a typical experimental scale of $1$ hour, that is, 
for a $\tw\approx 3.6\times10^{15} \text{ MCS}$, we  can extrapolate
the expected coherence length with equation~\eqref{eq:SG-xi-z}
and the data in Table~\ref{tab:SG-z}.  We conclude
that 
\begin{equation}
L(\tw)\approx3.7\xi_{1,2}(\tw) \qquad \Longrightarrow\qquad 1 \text{ hour} \longleftrightarrow L =110
\end{equation}
That is, the equilibrium length scale relevant to  
the non-equilibrium experiments is not the thermodynamical limit, 
but rather $L\sim100$.

This concept of experimental length
scale will be extremely relevant in the next Chapter, where
we try to elucidate the nature of the spin-glass phase. 
We shall find there that making extrapolations to $L\to\infty$ 
is very difficult, but that extrapolating to $L=110$  is
safe.

As a final comment, we point out that some authors have proposed
a modified scaling for the coherence length, substituting
the power law with a more complex functional form~\cite{bouchaud:01}.
This was studied in Section~6 of~\cite{janus:09b} and found
to fit our non-equilibrium simulations (the underlying reason
being that $1/z$ is very small and it is very difficult
to distinguish a small power from a logarithm with 
numerical data). If this modified scaling were to hold
instead of the power-law behaviour, the length scale
equivalent to one hour would be smaller than $L=110$.
\index{experimental scale|)}
\section{The phase transition in the dynamical heterogeneities}\label{sec:SG-phase-transition}
The previous section presented one of our most important results:  the 
statics-dynamics equivalence can be made quantitative. We shall now take this observation
one step further and use the equilibrium correlation functions to 
understand the crossover in the behaviour of the dynamical heterogeneities.\index{dynamical heterogeneities}
Recall that, for large $C$, the two-time correlation length $\zeta(C,\tw)$ reached
a $\tw$-independent value, while for small $C$ it grew as the coherence length  $\xi(\tw)$. \index{correlation length!two-time} \index{coherence length}

Our dynamical study was precise enough for us to observe this phenomenon, but not
to study it quantitatively. One of the major limiting factors was that the crossover value, 
$q_\text{EA}$, was very low, so even with our long simulations we did not have enough statistics
in the low-$C$ sector. However, one of the advantages of an equilibrium study is that we can cover
the whole range of $q$. To this end we can consider the behaviour of the connected correlation 
function at fixed $q$. Below $q_\text{EA}$ (and for $T<T_\text{c}$), one expects
\begin{align}
\index{correlation function (equilibrium)!Fourier space}
C_4(\boldsymbol r|q) - q^2 &= \frac{A_q}{r^{\theta(q)}} + \ldots,& |q|&<q_\text{EA}.\label{eq:C4-theta}\\
\nomenclature[theta2]{$\theta(q)$}{Algebraic decay of the fixed-$q$ connected correlations}
\intertext{%
The droplet and RSB pictures have very different predictions for the structure exponent \index{theta@$\theta(q)$}
$\theta(q)$, which we shall study in detail in Section~\ref{sec:SG-structure-correlations}.
For the moment, we only need that $0\leq \theta(q)< D$, on which both theories agree.
In order to study these connected correlations in position space, one has to perform
a subtraction that complicates the analysis
(Section~\ref{sec:SG-structure-correlations} and cf.~\cite{contucci:09}).
Instead, 
in this Section we work in Fourier space. The scaling behaviour is now}
\hat C_4(\bk|q) &\propto k^{\theta(q)-D} + \ldots,& |q|&<q_\text{EA}.\label{eq:SG-C4-q-bajo}\\
\intertext{%
On the other hand~\cite{dedominicis:98,contucci:09},}
\hat C_4(\bk|q) &\propto \frac{1}{k^2 + \xi_q^{-2}},& |q|&>q_\text{EA}.\label{eq:SG-C4-q-alto}
\end{align}
This equation defines a correlation length $\xi_q$ ---cf. Eq.~\eqref{eq:ISING-free-field}---
which diverges when $|q|\to q_\text{EA}$ from above.

In what follows we shall use the notation:
\begin{align}\label{eq:SG-Fq}
F_q^{(n)} &= \hat C_4(n \bk_\text{min}| q), &
F_q &= F_q^{(1)}.
\end{align}

The two scalings \eqref{eq:SG-C4-q-bajo} and \eqref{eq:SG-C4-q-alto}
for $\hat C_4$ reproduce the crossover behaviour that we found when
studying $\zeta(C,\tw)$: $F_q\sim L^{D-\theta(q)}$ for $|q|<q_\text{EA}$
but $F_q\sim 1$ for $|q|>q_\text{EA}$. In the large-$L$ limit, 
the crossover becomes a phase transition where
\begin{equation}
\xi_q^{(\infty)} \propto \frac{1}{(q-q_\text{EA})^{\hat \nu}}\ .
\end{equation}
Here $\hat\nu$ is a critical exponent, in principle different 
from the thermal critical exponent at $T_\mathrm{c}$. 
This concept of a phase transition at $T<T_\mathrm{c}$ as we
vary $q$ may seem unorthodox for spin glasses. However, a 
very similar picture appears in the study of the equation
of state for  Heisenberg ferromagnets~\cite{brezin:73}.\index{Heisenberg ferromagnet}
\begin{figure}
\includegraphics[height=\linewidth,angle=270]{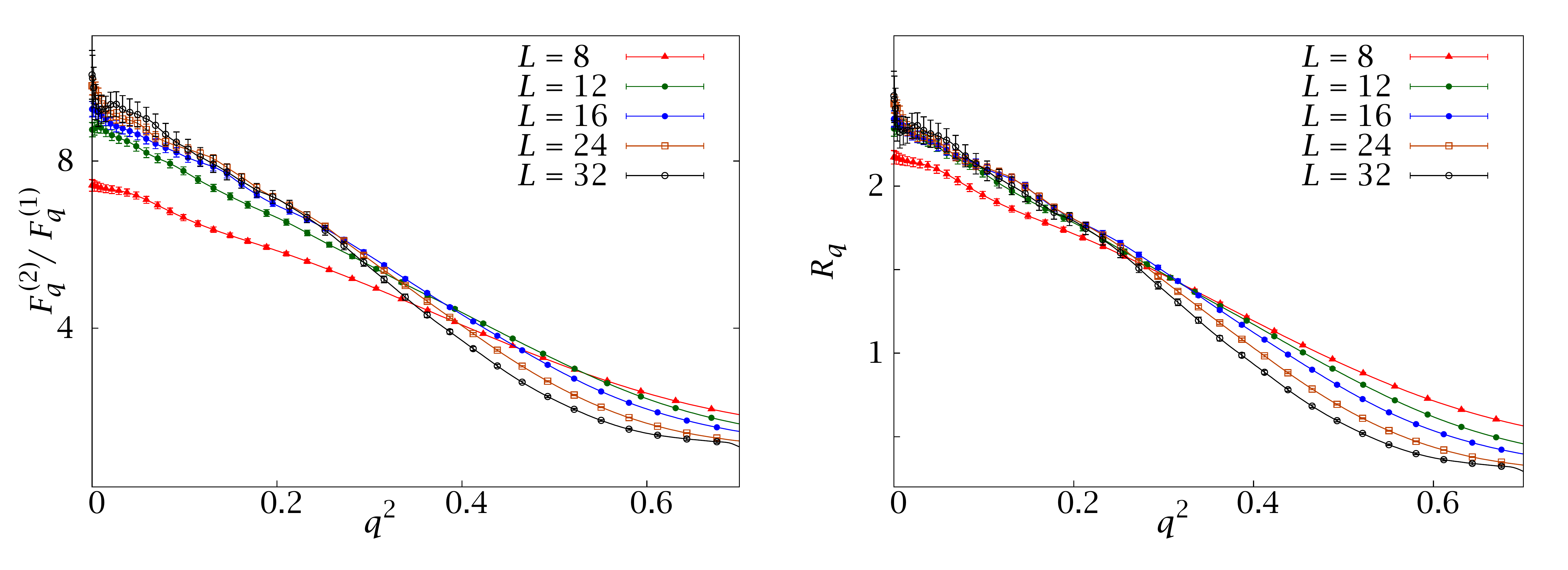}
\caption[$F_q^{(1)}/F_q^{(2)}$ against $q^2$]{\emph{Left:} Plot of the
 dimensionless quantity
$F_q^{(1)}/F_q^{(2)}$. \emph{Right:} Same plot, but
now we add factors of $\sin^2(n\uppi/L)$ to avoid trivial
free-field scaling corrections and plot $R_q= \bigl[F_q^{(1)}\sin(\uppi/L)\bigr]
/\bigl[F_q^{(2)}\sin(2\uppi/L)\bigr]$. In both 
cases the curves merge below $q_\text{EA}$ (as in a Kosterlitz-Thouless
transition), but we cannot distinguish the critical point with 
much precision.
\index{correlation function (equilibrium)!Fourier space|indemph}
\label{fig:SG-Rq}}
\end{figure}

We can study this phase transition using finite-size scaling,
 \index{finite-size scaling}
as we did for the Ising model and the DAFF. \index{Ising model} \index{DAFF}
Up to scaling corrections, we have
\begin{equation}\label{eq:SG-FSS-Fq}
F_q^{(n)} = L^{D-\theta(q_\text{EA})} G_n\bigl(L^{1/\hat \nu}(q-q_\text{EA})\bigr).
\end{equation}
Notice that this approach also has close parallels with the 
condensation transition, which can be studied\index{condensation}
with Ising ferromagnets.\footnote{%
In this case, the magnetisation density plays the role of $q$, 
the spontaneous magnetisation that of $q_\text{EA}$. Finally, 
$\theta(q_\text{EA}) = D$ and $\nu=(D+1)/D$~\cite{biskup:02}. The scaling
function $G_1(x)$ has, for this system, a discontinuity at finite $x$.}
In what follows, just as in the previous section, we shall work
at $T=0.703$.

As in previous chapters, we want to exploit  \eqref{eq:SG-FSS-Fq}
using phenomenological renormalisation. \index{quotients method}
The first step in those analyses was finding some suitable
dimensionless quantity and studying the intersections
of its curves for different $L$. The only 
obvious such quantity that we have here is $F^{(n)}_q/F^{(m)}_q$.
Unfortunately, these ratios do not show clear intersections.
Indeed, in the case where $\theta(|q|<q_\text{EA})>0$ the whole
phase $|q|<q_\text{EA}$ will be critical and we will be 
in a situation analogous to a Kosterlitz-Thouless transition~\cite{kosterlitz:73},
 where \index{Kosterlitz-Thouless transition}
the curves merge, rather than intersect, at the critical point.
Figure~\ref{fig:SG-Rq} suggests that this may be the case, but 
it is not possible to get a clear determination of $q_\text{EA}$ 
from these data.

\begin{figure}
\centering
\includegraphics[height=\linewidth,angle=270]{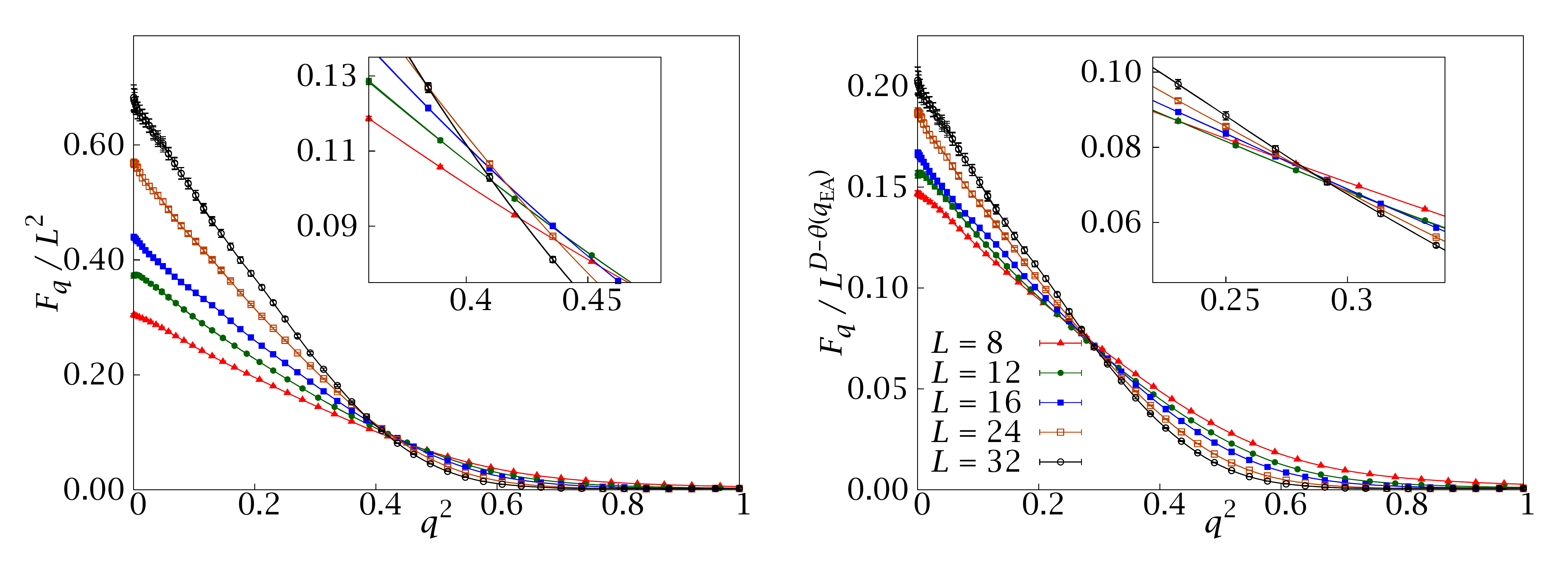}
\caption[$F_q/L^y$ against $q^2$, showing  clear crossings]{Plot of $F_q/L^y$ against $q^2$ for all our system
sizes at $T=0.703$ and two values of $y$. In the left panel
we consider the free-field scaling $F_q(\bk) \propto k^{-2}$, 
that is, $y=2$. In the right panel we consider $y=2.35\approx D-\theta(q_\text{EA})$
(see below). The insets are closeups of the crossing region and both are
shown at the same scale.
\index{finite-size scaling|indemph}
\index{correlation function (equilibrium)!Fourier space|indemph}
\label{fig:SG-Fq-y}
}
\end{figure}

We have therefore adopted an alternative approach. Rather than looking
for dimensionless quantities we consider the ratios $F_q/L^y$, 
where $y$ is a continuous parameter,
\begin{equation}\label{eq:SG-Fq-y}
F_q/L^y=L^{\epsilon-y} G_1\bigl(L^{1/\hat \nu} (q-q_\text{EA})\bigr), \qquad \epsilon = D-\theta(q_\text{EA}).
\end{equation}
Now, as long as $y< D-\theta(0)$, it follows that, in the large-$L$ limit,
$F_q/L^y$ vanishes for $|q|>q_\text{EA}$ and diverges for $|q|<q_\text{EA}$.
Therefore, the curves for these quantities as a function of $q$ for different $L$
will also intersect. We show two values of $y$ in Figure~\ref{fig:SG-Fq-y}.

Now we can once again analyse these intersections with the quotients method. \index{quotients method}
We consider pairs of lattices $(L,sL)$. Then, operating with \eqref{eq:SG-Fq-y}
we see that the crossing points $q_{L,y}^{(s)}$ scale as
\begin{align}\label{eq:SG-crossing-points}
q_{L,y}^{(s)} &= q_\text{EA} + A_y^{(s)}L^{1/\nu} + \ldots, &
A_y^{(s)} &= \frac{G_1(0)}{G'_1(0)} \frac{s^{\epsilon-y}-1}{s^{1/\nu}-s^{\epsilon-y}}\ ,
\end{align}
where the dots denote corrections to leading scaling.
A fit to this equation could produce the values of $q_\text{EA}$ 
and of $\hat \nu$. Recall that we tried, in vain, to estimate the former
parameter using non-equilibrium methods.
\index{spin glass!order parameter}
Notice that for an exponent $y=\epsilon=D-\theta(q_\text{EA})$ the intersection
point is constant in $L$ (neglecting scaling corrections, of course).

Since we have simulated $L=8,12,16,24,32$, our best option is
to choose $s=2$. In this case, for a fixed $y$, there will
be three intersection points: $(8,16)$, $(12,24)$, ($16,32$).
However, there are also three fit parameters ($q_\text{EA}$, $A_y^{(2)}$
and $\hat \nu$), so we would be left with no degrees of freedom.

We can get around this problem by considering $n$ values
of $y$ at the same time. This way, we can make a joint fit of all the 
resulting intersections, with fit parameters
$\{q_\text{EA},\hat \nu, A_{y_1}^{(2)},\ldots,A_{y_n}^{(2)}\}$
(that is, forcing the intersections for different $y$ to extrapolate
to the same $q_\text{EA}$ and with the same exponent).
Such a procedure may (should) raise an alarm: we are taking two
internally correlated curves and extracting many crossing points
between them by varying a free parameter. Fortunately, the 
potentially dangerous effects of correlations can in this
case be fully controlled by considering the complete covariance
matrix.

In particular, we have a set of measured data points $\{q_{L_a,y_i}^{(2)}\}$, 
labelled by their $L$ and their $y$. The sizes $L_a$ are $L_1=8$, $L_2=12$
and $L_3=16$, while the $y_j$ go from $y_1$ to $y_n$.
Then, the appropriate chi-square estimator is
\begin{equation}\label{eq:SG-chi-square}
\chi^2 = \sum_{i,j=1}^n \sum_{a,b=1}^3 \bigl( q_{L_a,y_i}^{(2)}
- q_\text{EA} - A_{y_i}^{(2)}L_a^{-1/\hat\nu}\bigr) \sigma_{(ia)(jb)}^{-1}
\bigl( q_{L_b,y_j}^{(2)}
- q_\text{EA} - A_{y_j}^{(2)}L_a^{-1/\hat\nu}\bigr) .
\index{chi-square test}
\end{equation}
In this equation, $\sigma_{(ia)(jb)}$
is the covariance matrix of the data (we need two coordinates to identify 
each point: its $L$ and its $y$). 
In this case, unlike in our
non-equilibrium analysis, the full covariance matrix 
is treatable. 
\begin{figure}[p]
\centering
\includegraphics[height=0.7\linewidth,angle=270]{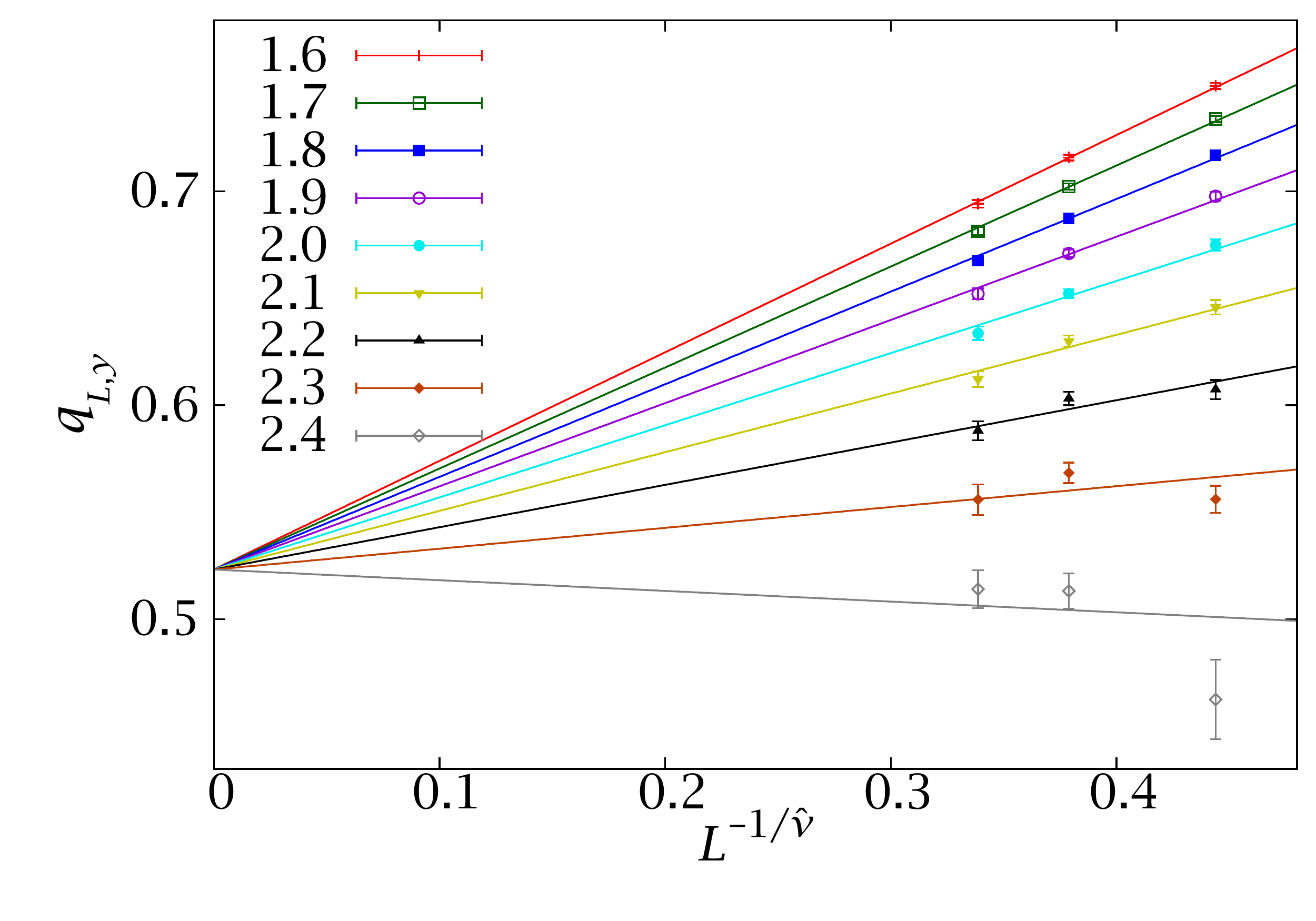}
\caption[Computation of $q_\text{EA}$ and $\hat\nu$]{Crossing points $q^{(2)}_{L,y}$ 
computed for pairs of lattices $(L,2L)$ and several values
of $y$, against $L^{-1/\hat\nu}$. The continuous lines
are fits to~\eqref{eq:SG-crossing-points} constrained
to yield common values of $\hat\nu$ and $q_\text{EA}$.
\index{spin glass!order parameter|indemph}
\index{critical exponent!nuhat@$\hat\nu$|indemph}
\label{fig:SG-qEA}}
\vspace*{1.5cm}

\centering
\includegraphics[height=0.7\linewidth,angle=270]{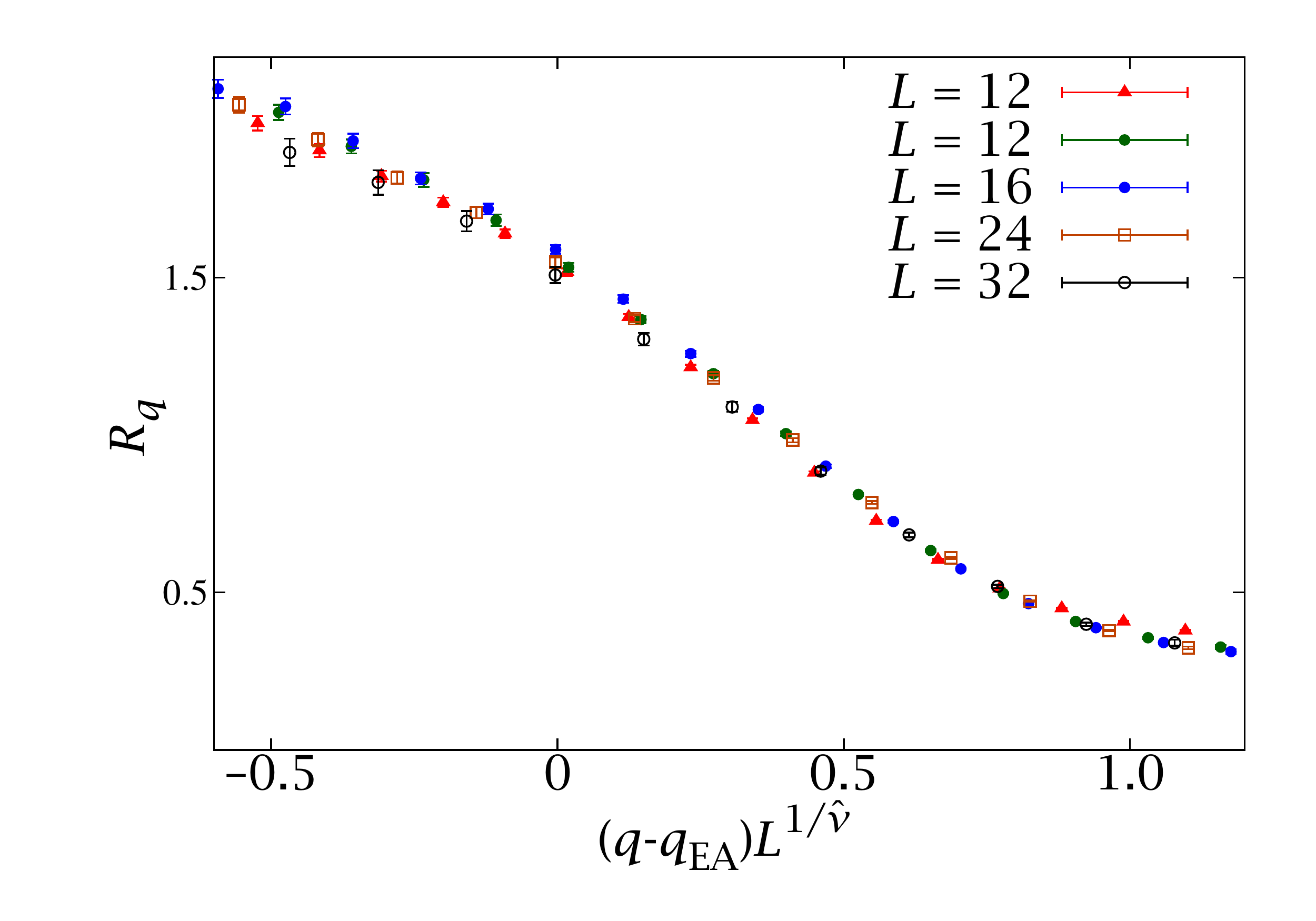}
\caption[Scaling plot of $F^{(1)}_q/F_q^{(2)}$]{Scaling plot of the quantity $R_q$ of Figure~\ref{fig:SG-Rq}, 
using the critical parameters computed in~\eqref{eq:SG-qEA}.
\index{scaling plot|indemph}
\label{fig:SG-Rq-scaling}}
\end{figure}

Figure~\ref{fig:SG-qEA} shows the result of this fitting procedure.
The selection of the number of $y$ is arbitrary:
the more $y$ we add, the more degrees of freedom (information)
 we have, but at the cost of increasing the effects of correlations and 
endangering the conditioning of the covariance matrix. Fortunately, 
what seems a delicate choice turns out to be arbitrary, both the fit
parameters and their errors being remarkably resilient to changes
in the $y_j$. For our choice of $y_j$ (shown in Figure~\ref{fig:SG-qEA}), 
the final results are
\begin{align}\label{eq:SG-qEA}
q_\text{EA} &= 0.52(3), & 1/\hat\nu &=0.39(5),
\end{align}
with $\chi^2/\text{d.o.f.} = 18.9/16$. Notice that 
our value of $q_\text{EA}$ is consistent with the (wide) bounds
that we obtained in our preliminary non-equilibrium analysis in 
Section~\ref{sec:SG-qEA-dynamics}.

As pointed out above, the result is very stable to changes in the 
values of $y$. For instance, removing the $L=8$ data for $y=2.3,2.4$
(the outliers) shifts our final values by one fifth of the error bar.
Notice that, from Figure~\ref{fig:SG-qEA}, $A_y^{(2)}$ 
changes sign at  $y\approx2.35$, so $\theta(q_\text{EA}) \approx 0.65$.
We shall return to this exponent in the next Chapter.

As a test of these critical parameters, we can attempt to produce 
a scaling plot of the ratio $R_q$ that we had plotted in Figure~\ref{fig:SG-Rq-scaling}.
This is represented in Figure~\ref{fig:SG-Rq-scaling},
which shows that, indeed, a good collapse is obtained.

Let us finally note that our value of $1/\hat \nu=0.39(5)$
is compatible with the best known result for the thermal
critical exponent $1/\nu=0.408(25)$~\cite{hasenbusch:08b}, so it may be that
the two are equal after all. We shall consider other 
conjectures for $1/\hat \nu$ and, in particular, its
relation with $\theta(q)$, in Section~\ref{sec:SG-structure-correlations}.
\index{critical exponent!nu@$\nu$}

\section{The finite-time scaling paradigm}\label{sec:SG-FTS}
\begin{figure}
\centering
\includegraphics[height=.7\linewidth,angle=270]{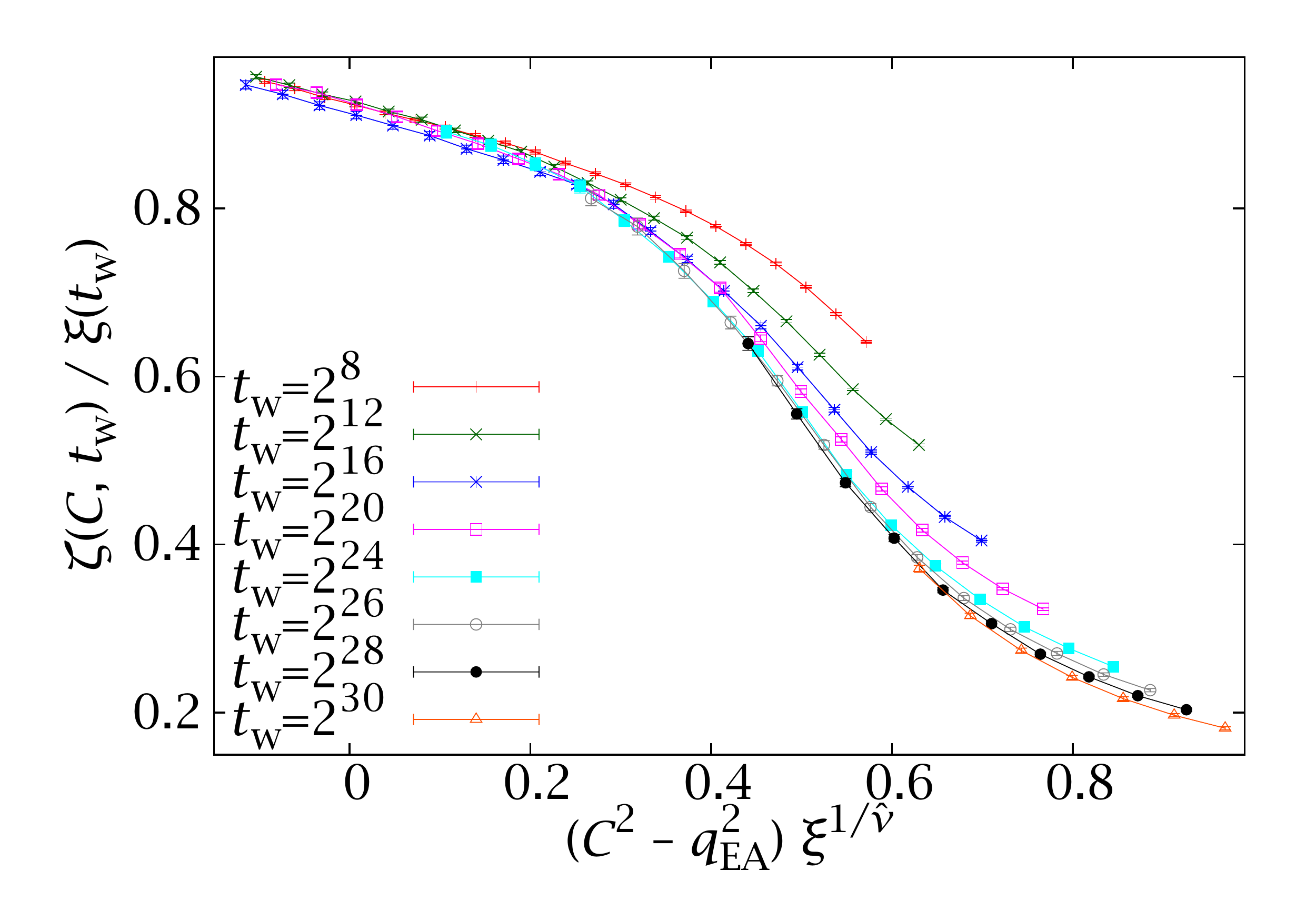}
\caption[Finite-time scaling]{Finite-time scaling plot: dimensionless
ratio $\zeta(C,\tw)/\xi(\tw)$ against the scaling variable
$(C^2-q_\text{EA}^2)\xi(\tw)^{1/\hat\nu}$.
\index{finite-time scaling|indemph}
\index{correlation length!two-time|indemph}
\index{coherence length|indemph}
\label{fig:SG-FTS}
}
\end{figure}

In the previous section we obtained an important result: the spin-glass
order parameter $q_\text{EA}$. This had eluded us in our previous non-equilibrium
study (Section~\ref{sec:SG-qEA-dynamics}) and had also never been 
done in a controlled way in equilibrium.\footnote{%
One naive way to estimate $q_\text{EA}$ is to consider
the evolution of the peaks in the $p(q)$ as $L$
grows. However, not knowing the exponent $\hat \nu$ 
that controls this finite-size evolution we would 
not have enough degrees of freedom for a reasonable fit.}

We also saw how the critical parameters thus 
obtained could produce scaling plots of dimensionless
observables in equilibrium.

We can now consider a complementary problem. Given the equivalence
between $L$ and $\xi(\tw)$ through the time-length dictionary,
should it not be possible to produce a dynamical scaling plot?  The answer
turns out to be yes. In Figure~\ref{fig:SG-FTS} we have plotted
the dimensionless ratio $\zeta(C,\tw)/\xi(\tw)$ as a function 
of the scaling variable $(C^2-q_\text{EA}^2)\xi(\tw)^{1/\hat\nu}$.
The values of $\hat\nu$ and $q_\text{EA}$ used were directly
those of~\eqref{eq:SG-qEA}. As we can see, we obtain a collapse 
of the curves for large $\tw$.

Figure~\ref{fig:SG-FTS} suggests that the correct manner to treat 
the statics-dynamics equivalence in a fully quantitative way
is to adopt a `finite-time scaling' (FTS) \index{finite-time scaling}
formalism.

The FTS approach provides a natural explanation for 
the extremely small exponents found in our 
power-law extrapolation of $C_\infty(t)$ (Section~\ref{sec:SG-qEA-dynamics}).
Indeed, FTS implies that
\begin{equation}
C_\infty(t) - q_\text{EA} \propto \zeta_\infty(t)^{-1/\hat \nu}, \qquad \zeta_\infty(t) = \lim_{\tw\to\infty} \zeta(t,\tw).
\end{equation}
Finally, we expect
\begin{equation}
\zeta_\infty(t) \propto t^{z_\zeta},
\end{equation}
and, in fact, (see Table~\ref{tab:SG-zeta-z}) the
evidence suggests that $z_\zeta$ 
is equal to the dynamic critical exponent $z$. \index{critical exponent!z@$z$}
Therefore, the exponent $B$ of Section~\ref{sec:SG-qEA-dynamics},
\begin{equation}
C_\infty(t) = q_\text{EA} + At^{-B},\tag{\ref{eq:SG-Ct-power}}
\end{equation}
is 
\begin{equation}
B = \frac{1}{z \hat \nu} \approx 0.0325.
\end{equation}
We have used the value of $z(T=0.7) =12.03(27)$
from Table~\ref{tab:SG-z}. This agrees at least 
in order of magnitude with the results of Table~\ref{tab:SG-qEA-dyn}
(which is about as much as we could have hoped for,
given the problems in estimating
$B$ from dynamical data).

We finally note that recent improvements in the experimental
measurement of response function with space-time resolution
suggest that $\zeta(t,\tw)$ may soon be experimentally
accessible~\cite{oukris:10}. Therefore, our finite-time
scaling framework has potential implications for 
experimental work.

\index{spin glass!dynamics|)}

\chapter[The structure of the $D=3$ spin-glass phase]{The structure of the \boldmath $D=3$ spin-glass phase}\label{chap:sg-3}

In this chapter we consider the finer details of the structure of the $D=3$ spin-glass phase. 
In particular, 
we try to decide which of the competing theories (RSB, droplet, TNT) offers a best
description. To this end, we begin by examining the would-be triviality in the spin and 
link overlaps which, as we explained in Chapter~\ref{chap:sg}, is the main differentiating
characteristic between the three pictures.

This analysis is carried out on the light of our previous results for the 
statics-dynamics equivalence. In particular, we stress the fact that the theory 
with most experimental relevance is the effective one at $L\sim100$, rather than
the one that best describes the thermodynamical limit. In this sense, we shall conclude
that the RSB framework provides the best picture of the experimental spin glass. 
Still, this large-$L$ 
limit is also interesting from a theoretical point of view. Therefore, we 
also attempt infinite-volume extrapolations, carefully checking for finite-size effects.

We conclude this chapter (and our study of spin glasses) with a detailed analysis of the 
structure of correlations in the spin-glass phase. 

\section{The spin overlap}\label{sec:SG-spin-overlap}\index{overlap!spin|(}
\begin{figure}
\centering
\includegraphics[height=.7\linewidth,angle=270]{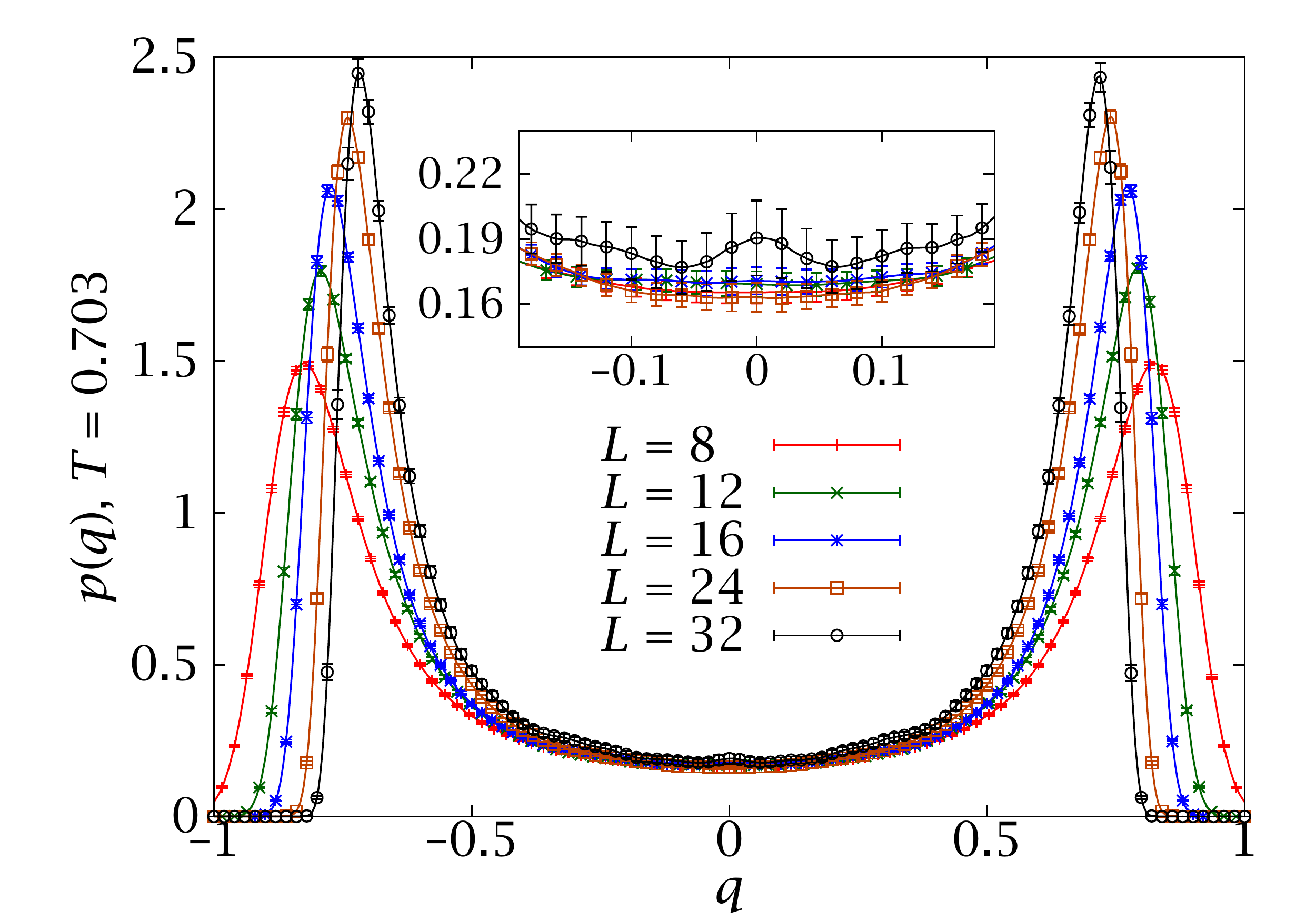}
\vspace*{1.3cm}

\includegraphics[height=.7\linewidth,angle=270]{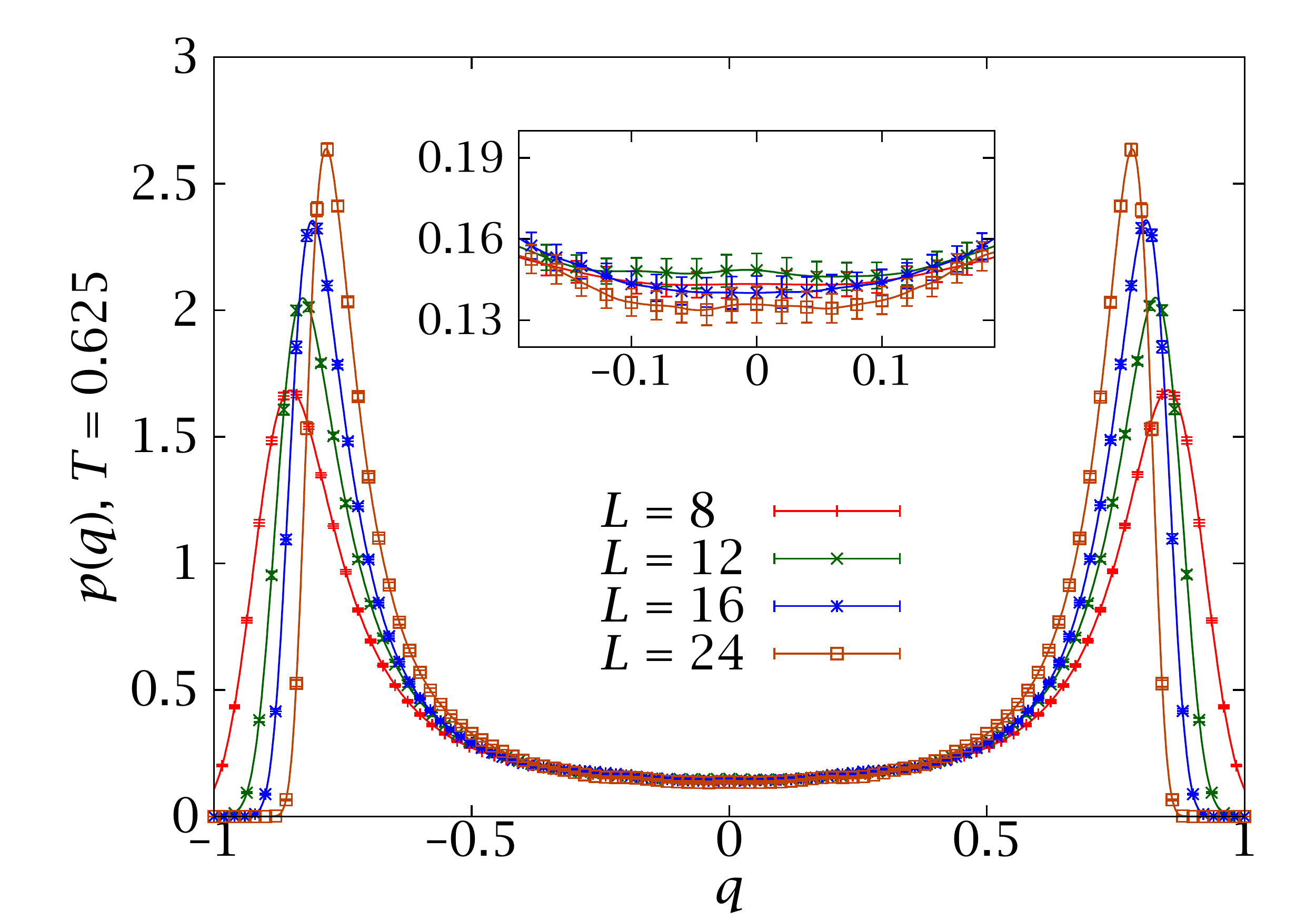}
\caption[Probability density function of the spin overlap]{%
Overlap  probability density function $p(q)$, Eq.~(\ref{eq:SG-p-q}), at $T=0.625$ and $T=0.703$.
Notice that for the central sector of $q\sim0$ the curves for the different system sizes quickly
reach a plateau with $p(q) > 0$.}
\label{fig:SG-p-q}
\index{overlap!spin|indemph}
\end{figure}

We begin our study by considering the most straightforward observable: the spin overlap $q$.
As we explained in Chapter~\ref{chap:sg}, the behaviour of this quantity is one of the 
clearest markers to distinguish the droplet picture from the RSB one.\footnote{The TNT picture
coincides with RSB in the predictions for this observable, but we shall not mention 
it explicitly in this Section.}
\index{droplet picture}
\index{RSB}
\index{TNT picture}
In particular, droplet predicts a trivial $p(q)$ in the thermodynamical limit (two deltas
at $\pm q_\text{EA}$), while RSB predicts a non-trivial behaviour (non-zero probability density
for all $|q|\leq q_\text{EA}$).

We have plotted the $p(q)$ of Eq.~\eqref{eq:SG-p-q},
in Figure~\ref{fig:SG-p-q} for $T=0.703$ (the lowest temperature for $L=32$)
and for $T=0.625$ (the lowest for $L=24$). The curves are very smooth, thanks
to the Gaussian convolution of~\eqref{eq:SG-p-q}. According to the previously mentioned theoretical 
predictions, one would expect the peaks to get narrower and closer together 
as $L$ grows. The shift in position is very clear (recall from the previous Chapter 
that $q_\text{EA}$ at $T=0.703$ is $\approx 0.52$) but a more careful
analysis is needed to assess the change in width (Section~\ref{sec:SG-peaks}).

\begin{figure}
\centering
\begin{minipage}{.48\linewidth}
\includegraphics[height=\linewidth,angle=270]{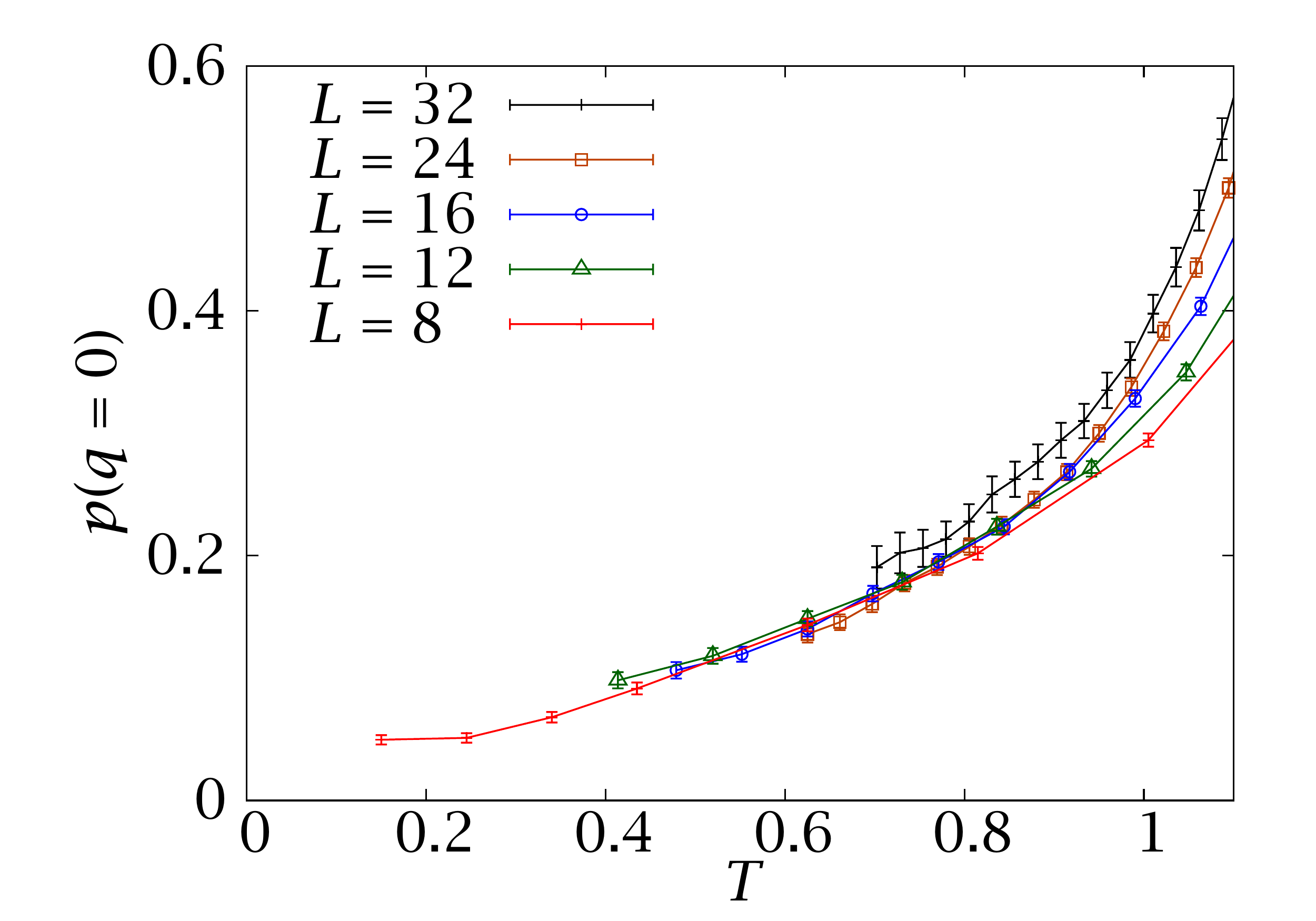}
\end{minipage}
\begin{minipage}{.48\linewidth}
\includegraphics[height=\linewidth,angle=270]{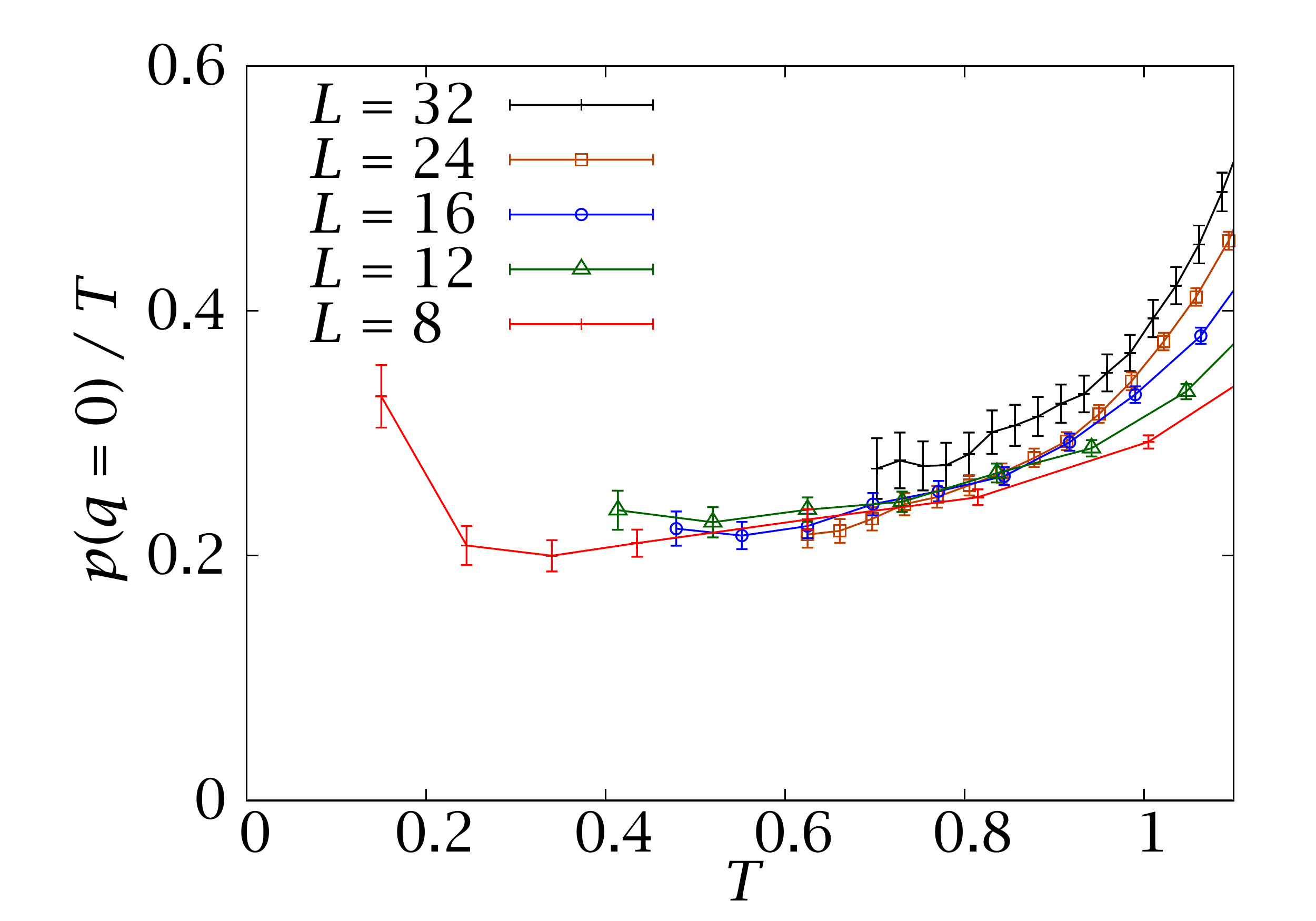}
\end{minipage}
\caption[Zero overlap probability density $p(q=0)$]{Overlap
 density distribution function at zero overlap as a function
of temperature. We observe an enveloping curve with a linear behaviour, 
as expected in an RSB setting.}
\label{fig:SG-p-q-zero}
\end{figure}
In addition, the probability density at $q=0$ shows no evolution, favouring the RSB 
scenario. In order to see it more clearly we have plotted $p(q=0)$ 
at $T=0.703$ in Figure~\ref{fig:SG-p-q-zero} for 
all our lattices. We can see a clear non-zero enveloping curve. More precisely, 
in a mean-field setting one expects this probability density 
to be linear in $T$ below $T_\mathrm{c}$~\cite{mezard:87}. This expectation
is checked in the right panel of Figure~\ref{fig:SG-p-q-zero}. The seemingly 
out of control value of $p(0)$ at the lowest temperature for $L=8$ is an artifact
of the binary nature of the couplings (a finite system always has a finite energy gap).
The fact that finite-size effects in $p(0)$ are stronger close to $T=0$ than at 
finite temperature has been studied in~\cite{palassini:01}.

From a droplet model point of view, Moore et al.~\cite{moore:98}
have argued that the apparent lack of a vanishing limit
for $p(0)$ in numerical work in the 1990s was
an artifact of critical fluctuations. In fact, at $T_\mathrm{c}$,
$p(0)$ diverges as $L^{\beta/\nu}$ while droplet theory
predicts that, for very large lattices, it vanishes with the stiffness 
exponent $y\approx0.2$ as $L^{-y}$, \index{stiffness exponent}
for all $T<T_\mathrm{c}$. These authors rationalise the numerical
findings as a crossover between these two limiting behaviours. 
However, a numerical study at very low temperatures
(so the critical regime is avoided) found for moderate system sizes a non-vanishing 
$p(0)$~\cite{katzgraber:01}.  Furthermore, we shall compute in Section~\ref{sec:SG-peaks}
a characteristic length for finite-size effects in the spin-glass phase, which turns 
out to be small at $T=0.703$.

In any case, the behaviour at the experimentally relevant scale of $L\sim100$ seems without
a doubt to be non-trivial (there is no room for a change of regime between our simulations
and $L\sim100$).

\index{overlap!spin|)}

\subsection{The Binder cumulant}\label{sec:SG-Binder}
\begin{figure}
\centering
\begin{minipage}{.48\linewidth}
\includegraphics[height=\linewidth,angle=270]{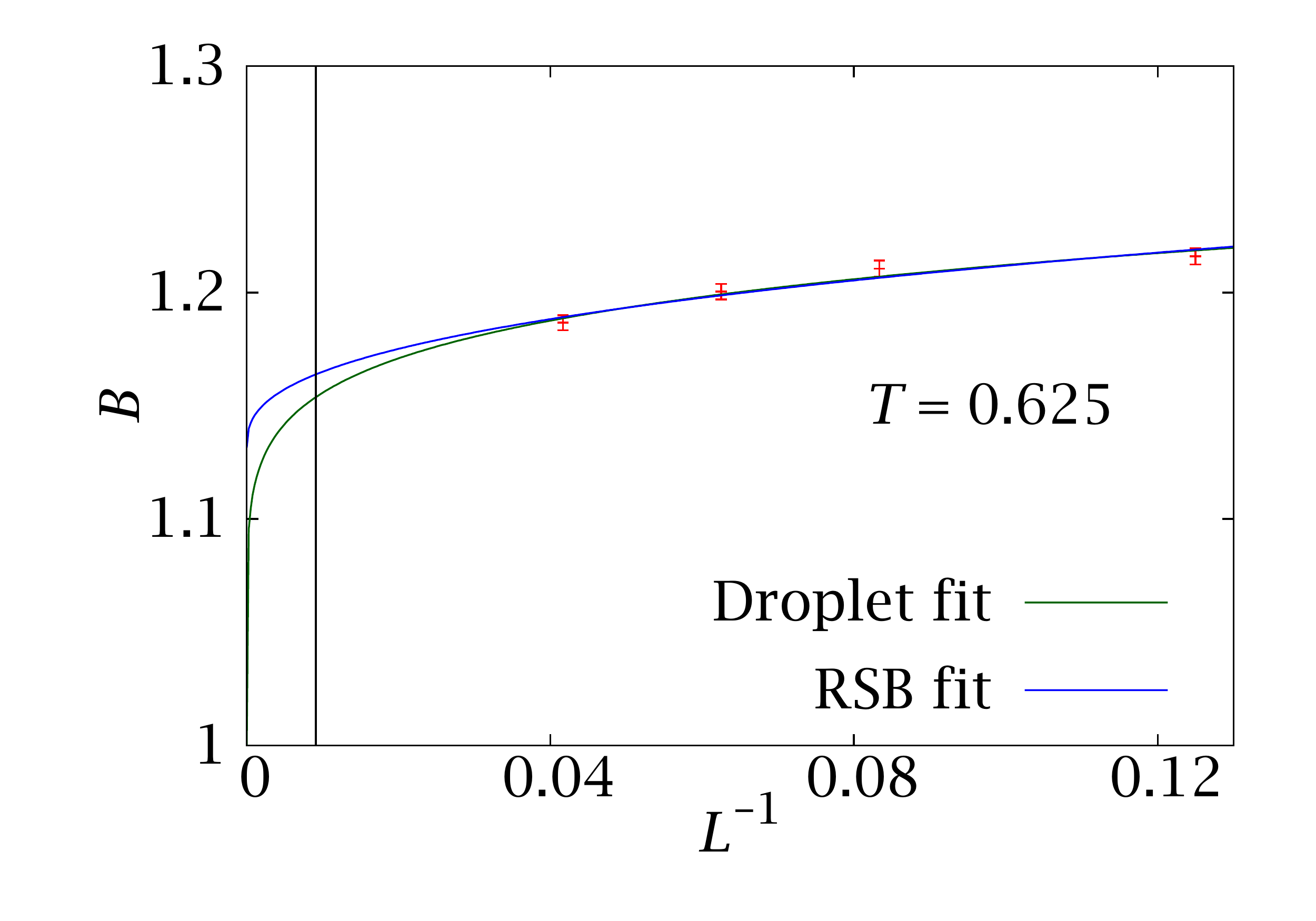}
\end{minipage}
\begin{minipage}{.48\linewidth}
\includegraphics[height=\linewidth,angle=270]{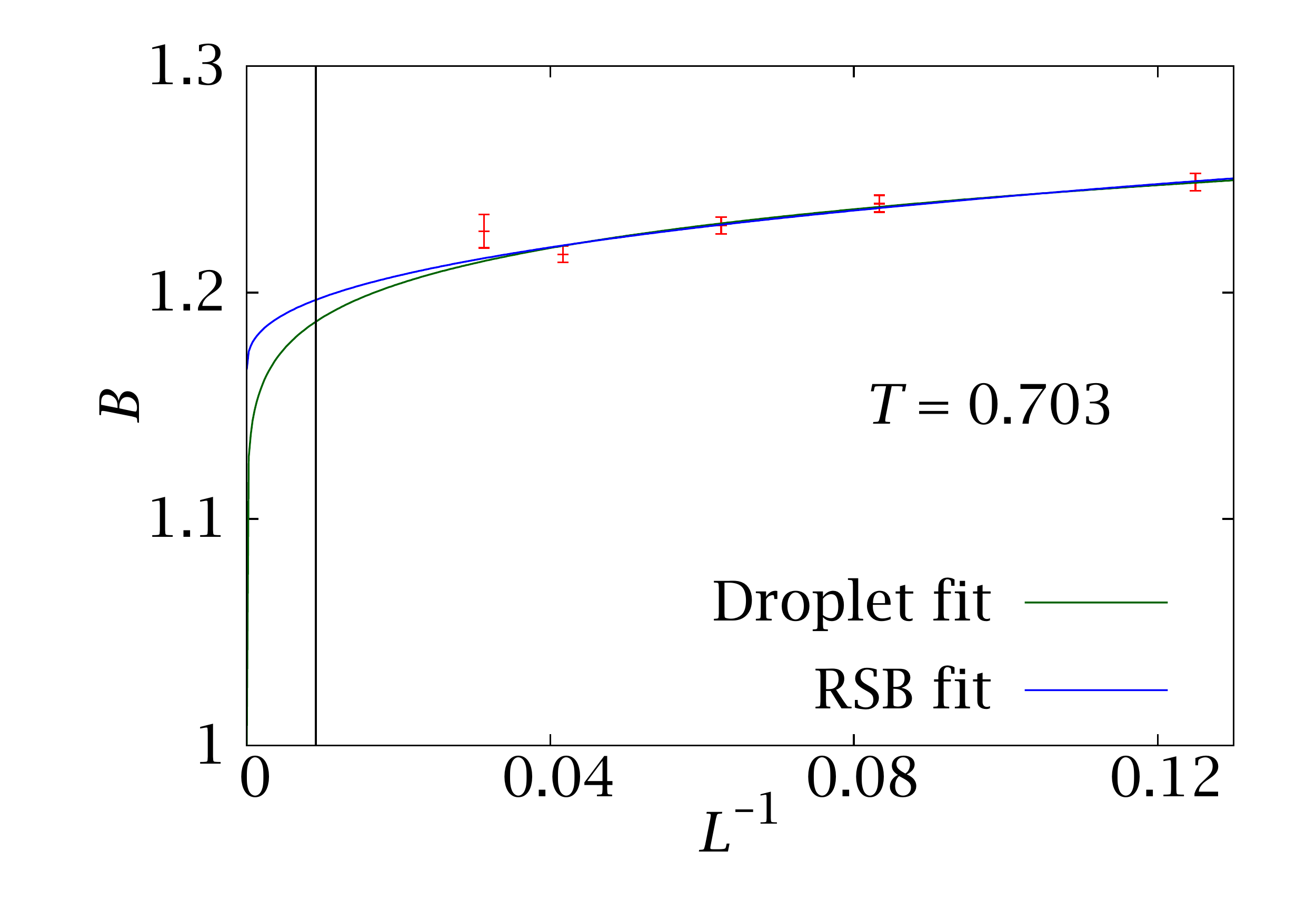}
\end{minipage}
\caption[Evolution of the Binder ratio with $L$]{%
Infinite volume extrapolation of the Binder parameter at $T=0.703$ and $T=0.625$ and 
fits to the behaviour expected in the RSB, Eq.~(\ref{eq:SG-Binder-RSB}),
 and droplet, Eq.~(\ref{eq:SG-Binder-droplet}), 
pictures (cf. Table~\ref{tab:SG-Binder}).
For the experimentally relevant scale of $L=110$ (dotted vertical line),
both fits are well above the $B=1$ value of a coarsening system.}
\label{fig:SG-Binder}
\index{Binder ratio!spin glass|indemph}
\end{figure}
\begin{table}
\small
\begin{tabular*}{\columnwidth}{@{\extracolsep{\fill}}cccccccc}
\toprule
\multirow{2}{1cm}{$T$} & \multicolumn{3}{c}{\bfseries Droplet fit} & &  \multicolumn{3}{c}{\bfseries RSB fit} \\
\cmidrule{2-4} \cmidrule{6-8}
& \multicolumn{1}{c}{ $\chi^2/\mathrm{d.o.f.}$} & \multicolumn{1}{c}{ $a$} & \multicolumn{1}{c}{ $y$} & &
 \multicolumn{1}{c}{ $\chi^2/\mathrm{d.o.f.}$} & \multicolumn{1}{c}{ $c$} & \multicolumn{1}{c}{ $d$}  \\
\toprule
0.703 & 3.78/3 & 0.312(17) & 0.110(17) & & 3.44/3 & 1.165(12)[34] & 0.186(34)[03]\\
0.625 & 2.00/2 & 0.289(16) & 0.134(21) & & 2.73/2 & 1.128(11)[33] & 0.193(28)[03]\\
\bottomrule
\end{tabular*}
\caption[Scaling of the Binder ratio]{Scaling of the Binder parameter and fit to the behaviour expected in the droplet, Eq.~(\ref{eq:SG-Binder-droplet}),  and RSB pictures, Eq.~(\ref{eq:SG-Binder-RSB}).}
\label{tab:SG-Binder}
\index{Binder ratio!spin glass|indemph}
\index{stiffness exponent|indemph}
\end{table}
We can define the Binder ratio for spin glasses just as we did for the Ising ferromagnet
in~\eqref{eq:ISING-Binder}
\index{Ising model}
\index{Binder ratio}
\begin{equation}\label{eq:SG-Binder}
B(T) = \frac{\overline{\braket{q^4}}}{\overline{\braket{q^2}}^2}\, .
\end{equation}
Notice that we compute these moments from the original $p_1(q)$, not 
from our smoothed version.
Above $T_\mathrm{c}$ the fluctuations of $q$ are expected to be Gaussian in the large-$L$ 
limit, hence 
\begin{equation}
\lim_{L\to\infty} B(T) = 3,\qquad T>T_\mathrm{c}.
\end{equation}
Below $T_\mathrm{c}$ the situation of course depends on whether the droplet or the 
RSB pictures are correct. In the former, $B(T<T_\mathrm{c})$ should 
approach $1$ in the large-$L$ limit (as in a ferromagnet).
 In the latter, one expects $1<B<3$.

Therefore, we have the following expectations
\begin{subequations}\label{eq:SG-Binder-theories}
\begin{align}
\mathrm{Droplet:}\qquad B(T;L) &= 1 + a L^{-y},\label{eq:SG-Binder-droplet} \\
\ \ \, \quad\mathrm{RSB:}\qquad B(T;L) &= c + d L^{-1/\hat\nu}.\label{eq:SG-Binder-RSB}
\end{align}
\end{subequations}
The droplet prediction~\eqref{eq:SG-Binder-droplet} depends on the stiffness exponent $y$, 
while in the RSB picture the finite-size evolution is controlled by the 
exponent $1/\hat \nu=0.39(5)$,  which we computed in the previous Chapter.
\index{critical exponent!nuhat@$\hat\nu$}

We have plotted the Binder ratio for $T=0.703$ and $T=0.625$ in Figure~\ref{fig:SG-Binder},
along with fits to~\eqref{eq:SG-Binder-theories}. The resulting parameters
are gathered in Table~\ref{tab:SG-Binder}. For the RSB fit we include two error
bars: the number enclosed in $(\cdot)$ comes from the statistical error
in a fit with fixed $\hat \nu$ and the one in $[\cdot]$ is the systematic
error due to our uncertainty in $\hat \nu$.

As it turns out, both fits have acceptable values of $\chi^2/\text{d.o.f.}$. However, 
the evolution of $B$ with $L$ is very slow, so in order to accommodate the droplet
behaviour we have needed a very small exponent $y\approx0.12$, smaller than 
the usual droplet prediction of $y\approx0.2$~\cite{bray:87}.

Let us finally note that, again, the extrapolation at $L\sim100$ is well above one even with the
droplet scaling.

\subsection[The peaks of  $p(q)$]{The peaks of \boldmath  $p(q)$}\label{sec:SG-peaks}
\begin{figure}
\centering
\begin{minipage}{.48\linewidth}
\includegraphics[height=\linewidth,angle=270]{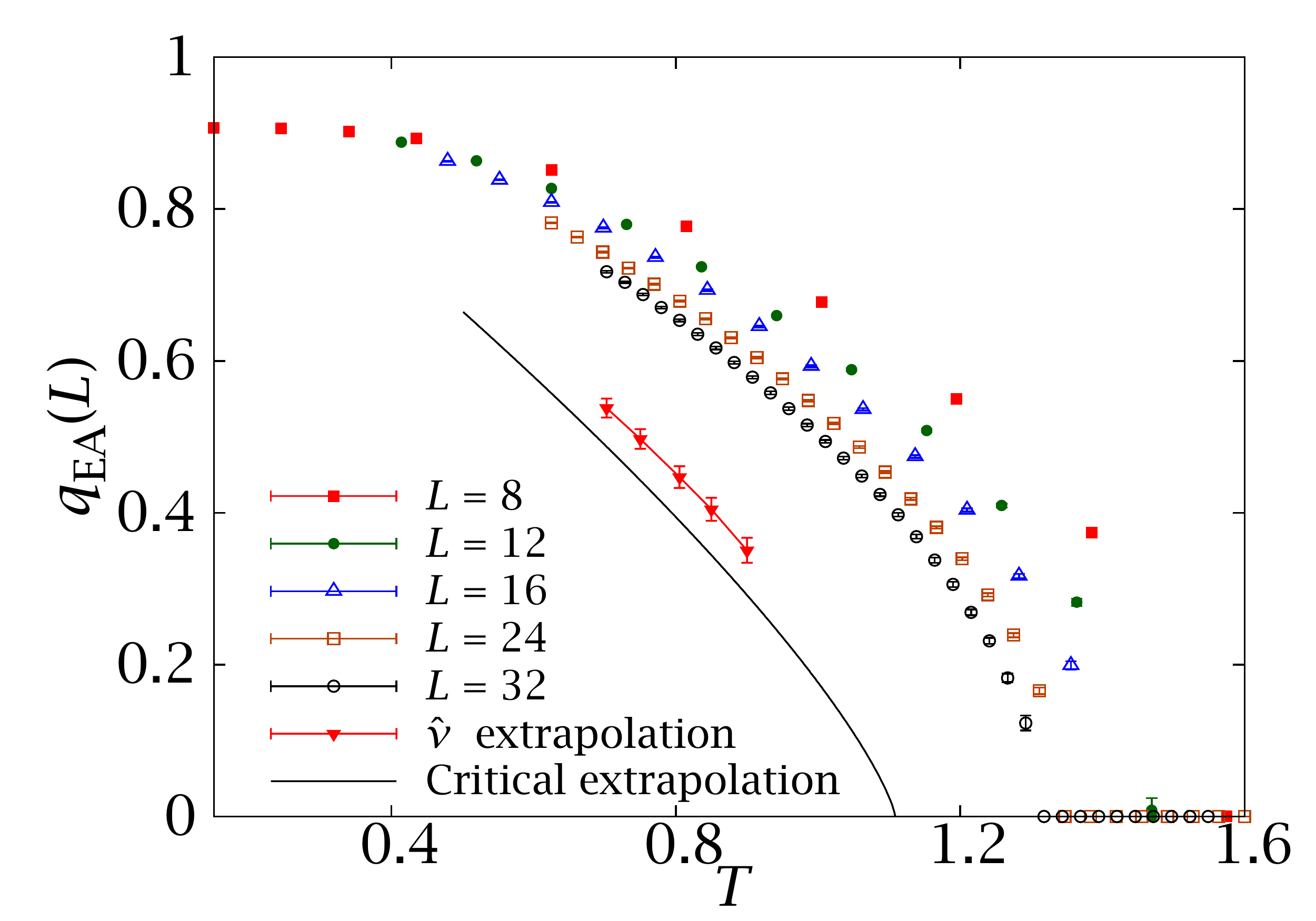}
\end{minipage}
\begin{minipage}{.48\linewidth}
\includegraphics[height=\linewidth,angle=270]{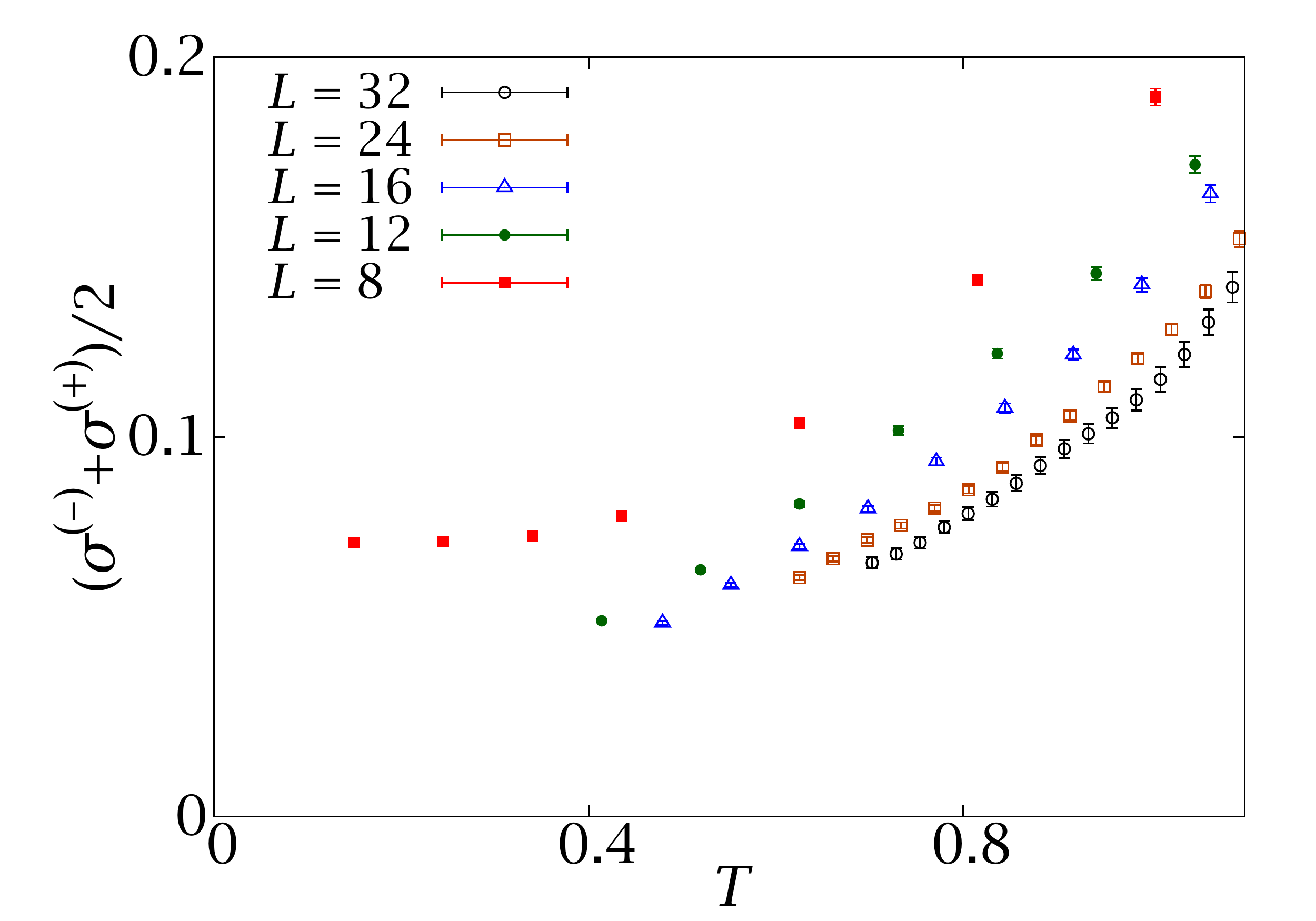}
\end{minipage}
\caption[Peak position and width in $p(q)$]{\emph{Left:} $q_\mathrm{EA}(L)$ as a function of the
  temperature. We include two different infinite-volume
  extrapolations. The first uses the dynamical heterogeneities exponent $1/\hat\nu$,
  cf.~\eqref{eq:SG-qEA-inf}  and~\eqref{tab:SG-qEA}. 
  The second is obtained from finite-size \index{dynamical heterogeneities}
  scaling arguments in the critical region, Eqs.~(\ref{eq:SG-qEA-FSS})
  and~(\ref{eq:SG-qEA-FSS2}).  \emph{Right:} Width of the peaks of $p(q)$,
  Eq.~(\ref{eq:SG-sigma}), as a function of $T$ for all our lattice
  sizes.}
\label{fig:SG-peaks}
\index{spin glass!order parameter|indemph}
\end{figure}
\begin{table}
\small
\begin{tabular*}{\columnwidth}{@{\extracolsep{\fill}}cccccc}
\toprule
\multirow{2}{1cm}{$L$}  
& \multicolumn{2}{c}{$T = 0.703$}&
& \multicolumn{2}{c}{$T = 0.805$}\\
\cmidrule{2-3}\cmidrule{5-6}
& $\sigma$ & $\sigma p\bigl(q_\mathrm{EA}(L)\bigr)$& &
$\sigma$ & $\sigma p\bigl(q_\mathrm{EA}(L)\bigr)$\\
\toprule
8  & 0.117\,7(20) &0.178\,4(10) & & 0.139\,1(25)& 0.183\,3(10) \\   
12 & 0.096\,3(21) &0.174\,0(12) & & 0.116\,5(25)& 0.180\,9(12) \\
16 & 0.081\,7(16) &0.169\,6(11) & & 0.100\,1(22)& 0.175\,6(11) \\
24 & 0.073\,5(16) &0.169\,0(12) & & 0.086\,0(19)& 0.172\,8(12) \\
32 & 0.066\,8(29) &0.163\,1(23) & & 0.079\,8(34)& 0.166\,9(22) \\
\midrule
$L_\mathrm{min}$         & 16         & & & 16     \\
$\chi^2/\mathrm{d.o.f.}$ & 0.43/1     & & &1.13/1  \\
$B$                      & $-0.278(28)$ & & & $-0.346(30)$ \\
\bottomrule
\end{tabular*}
\caption[Width of the peaks in $p(q)$]{Width $\sigma=\bigl(\sigma^{(+)} +\sigma^{(-)}\bigr)/2$ of the peaks in
$p(q)$ and fit to a power law $\sigma(L)=AL^B$ in the range $[L_\mathrm{min}, 32]$.
We also include the product $\sigma p\bigl(q_\mathrm{EA}(L)\bigr)$.}
\label{tab:SG-peak-width}
\end{table}
As we have seen, the droplet and RSB pictures have very different predictions
for $p(q)$ as a whole. However, they both agree in that the two symmetric peaks
observed at finite $L$ should eventually become two Dirac deltas at $\pm q_\text{EA}$,
a prediction that we can check.

Let us begin by defining $q_\text{EA}(L)$ as the position of the maximum of $p(q;L)$
(in the remainder of this section, we consider all overlaps positive and the pdf symmetrised).
Again, our Gaussian smoothing procedure makes this quantity easy to compute (we fit 
the neighbourhood of the peak to a third-order polynomial, since the  peak is
very asymmetric).

In order to test that the peak not only tends to $q_\text{EA}$, but also gets infinitely 
narrow, we can employ the half-widths at half height. Defining
$q^{(\pm)}$ through $p(q^{(\pm)}) = p\bigl(q_\text{EA}(L)\bigr)/2$, we have
\begin{equation}\label{eq:SG-sigma}
\sigma^{(\pm)} = \bigl| q^{(\pm)} -q_\text{EA}(L)\bigr|.
\end{equation}
Notice that, since we are considering the positive peak, $q^{(-)}<q_\text{EA}(L)<q^{(+)}$
(that is, $q^{(-)}$ is the inner width and $q^{(+)}$ the outer).

We have plotted $q_\text{EA}(L)$ and $\sigma^{(\pm)}$ in the left and right
panels of Figure~\ref{fig:SG-peaks}, respectively. We see that the width of 
the peaks does decrease slowly with $L$. However, the product $\sigma p\bigl(q_\text{EA}(L)\bigr)$
has a small dependence on $L$ (Table~\ref{tab:SG-peak-width}).

We can now consider the actual value of $q_\text{EA}$. Following Section~\ref{sec:SG-phase-transition},
we expect
\begin{equation}\label{eq:SG-qEA-inf}
q_\text{EA}(L,T) = q_\text{EA}(T) \left[1+\frac{A(T)}{L^{1/\hat \nu}}\right]\, .
\index{spin glass!order parameter}
\index{critical exponent!nuhat@$\hat\nu$}
\end{equation}
We cannot perform a three-parameter fit to~\eqref{eq:SG-qEA-inf}, due to the lack of
degrees of freedom, but we need to use our previously computed value of $1/\hat\nu=0.39(5)$.
Similar extrapolations were attempted in~\cite{iniguez:96}, but with less control
on $1/\hat\nu$ (and smaller sizes, $L\leq16$).

We present the values of $q_\mathrm{EA}(L)$ and the result of a fit to
Eq.~(\ref{eq:SG-qEA-inf}) in Table~\ref{tab:SG-qEA}.  As we can see,
the errors due to the uncertainty in the
exponent, denoted by $[\,\cdot\,]$, are greater than those caused by
the statistical error in the individual points, $(\,\cdot\,)$. In
fact, our data admit good fits for a very wide range of values in
$1/\hat\nu$. For instance, if we try to input the value of the
stiffness exponent obtained in the droplet-like extrapolation of the Binder
parameter, $y\sim0.12$ (see Eq.~(\ref{eq:SG-Binder-droplet}) and
Table~\ref{tab:SG-Binder}), we still obtain a good fit, even though the
extrapolated value for $q_\mathrm{EA}$ is almost zero at $T=0.703$ and
negative at $T=0.805$. Therefore, using the droplet exponent $y$
\index{stiffness exponent}
the spin-glass phase  would be non-existent. 
\begin{table}
\small
\begin{tabular*}{\columnwidth}{@{\extracolsep{\fill}}cll}
\toprule
\multicolumn{1}{c}{\bfseries $L$ } & \multicolumn{1}{c}{\bfseries $T = 0.703$}
& \multicolumn{1}{c}{\bfseries $T = 0.805$}\\
\toprule
8  & 0.824\,61(83)  & 0.781\,8(11) \\
12 & 0.793\,33(85)  & 0.741\,2(11) \\
16 & 0.773\,00(75)  & 0.716\,81(95)  \\
24 & 0.740\,27(71)  & 0.679\,05(83)  \\
32 & 0.717\,4(14) & 0.65\,35(16) \\
\midrule
$L_\mathrm{min}$         & 16           & 16     \\
$\chi^2/\mathrm{d.o.f.}$ & 1.83/1       & 0.98/1 \\
$q_\mathrm{EA}$          & 0.538[22](6) & 0.447[24](6) \\
Section~\ref{sec:SG-qEA-dynamics} & $0.474 \leq q_\mathrm{EA}\leq 0.637$ & $0.368\leq q_\mathrm{EA} \leq 0.556$ \\
\bottomrule
\end{tabular*}
\caption[Computation of $q_\text{EA}$ from the peaks 
in $p(q)$]{Extrapolation to infinite volume of $q_\mathrm{EA}(L,T)$
using the replicon exponent, Eq.~(\ref{eq:SG-qEA-inf}). We also include
the confidence interval previously obtained in the non-equilibrium study
of Section~\ref{sec:SG-qEA-dynamics}. The resulting estimate
for $T=0.703$ is compatible with our previous determination from the 
phase transition in the dynamical heterogeneities in Section~\ref{sec:SG-phase-transition}.
\label{tab:SG-qEA}
\index{spin glass!order parameter|indemph}}
\end{table}

Also included in Table~\ref{tab:SG-qEA} is the confidence interval for
this observable computed from non-equilibrium considerations
in Section~\ref{sec:SG-qEA-dynamics}. Notice that the equilibrium values are much more
precise, but consistent. The extrapolations included in this table
(and analogous ones for other values of $T$) are plotted in
Figure~\ref{fig:SG-peaks}. Finally, notice that the estimate of $q_\text{EA}(T=0.703)$
in Table~\ref{tab:SG-qEA} is compatible with our value of $q_\text{EA}(T=0.703)=0.52(3)$, 
obtained in Section~\ref{sec:SG-phase-transition}
with a completely different method.

\subsection{Critical and finite-size effects}
In order for our extrapolations to the thermodynamical limit to be meaningful, we 
have to check that we are not in a preasymptotic regime dominated 
by critical fluctuations. This is best done through a finite-size
scaling analysis of the $p(q)$. \index{finite-size scaling}
\begin{table}
\small
\begin{tabular*}{\columnwidth}{@{\extracolsep{\fill}}lclc}
\toprule
\multicolumn{1}{c}{$T$ } & \multicolumn{1}{c}{$L_\mathrm{c}^{1/\hat\nu}$} & \multicolumn{1}{c}{$L_\mathrm{c}$}
& \multicolumn{1}{c}{$L_\mathrm{c}(T_\mathrm{c}-T)^\nu$}\\
\toprule
0.703 & 1.253[20](32) & \ \ 1.78[8](11)  & 0.197[8](13)\ \ \  \\
0.75  & 1.448[24](34) & \ \ 2.58[12](16)  & 0.210[8](13)\ \ \ \\
0.805 & 1.731[28](44) & \ \ 4.08[18](27)  & 0.221[10](15)\\
0.85  & 2.023[32](54) & \ \ 6.09[26](42) & 0.222[10](15)\\
0.90  & 2.514[41](66) & 10.63[44](71) & 0.230[10](15)\\
\bottomrule
\end{tabular*}
\caption[Characteristic length for finite-size effects]{Determination of $L_\mathrm{c}$ from the fits to Eq.~\eqref{eq:SG-qEA-inf},
 using~\eqref{eq:SG-A-Lc}, for
  several temperatures below $T_\mathrm{c}$. Errors are given as in
  Table~\ref{tab:SG-qEA}.  The characteristic length $L_\mathrm{c}(T)$ scales
  as a correlation length when $T$ approaches $T_\mathrm{c}$
  ($\nu\approx2.45$ from~\cite{hasenbusch:08b}).  We warn the reader
  that the $\chi^2/\mathrm{d.o.f.}$ for the fits at $T=0.85$ and
  $0.90$ are, respectively, $2.6/1$ and $2.7/1$.}\label{tab:SG-Lc}
\index{correlation length|indemph}
\index{finite-size effects}
\end{table}

Let us first show that our estimate of $q_\text{EA}$ provides a determination
of the correlation length in the spin-glass phase. Notice that we are in a situation
were the correlations decay algebraically, so the concept of correlation 
length is delicate~\cite{josephson:66}. In particular, finite-size
effects are ruled by a crossover length $L_\text{c}(T)$, scaling as 
a correlation length: $L_\text{c} \propto (T_\text{c}-T)^{-\nu}$. In fact, one would
expect
\begin{equation}\label{eq:SG-qEA-Lc}
\frac{q_\text{EA}(T,L)}{q_\text{EA}(T)} = 1 + h[L/L_\text{c}(T)].
\end{equation}
We do not know the complete crossover function $h$, but we do know 
that for large $x$ it should behave as $h(x)\sim x^{-1/\hat\nu}$, so 
we make the simplest ansatz,
\begin{equation}
h(x) = x^{-1/\hat\nu}.
\end{equation}
Then, if we compare~\eqref{eq:SG-qEA-Lc} with~\eqref{eq:SG-qEA-inf}, we see
that the amplitude of the finite-size corrections in the latter can be written as
\begin{equation}\label{eq:SG-A-Lc}
A(T)=[L_\text{c}(T)]^{1/\hat\nu}.
\end{equation}
The resulting values of $L_\text{c}$ can be seen in Table~\ref{tab:SG-Lc}. Notice 
that this crossover length does scale with temperature as expected for the bulk behaviour,
which constitutes a weighty argument for the asymptotic nature of our results.

In order for our previous extrapolations for $q_\text{EA}$ to be valid, we need to stay in 
a temperature regime where $L\gg L_\text{c}(T)$, a condition that is amply
satisfied for our working temperatures of $T=0.703$ and $T=0.805$.

In the region close
to $T_\text{c}$, where $L$ becomes smaller than $L_\text{c}(T)$, we can use
finite-size scaling to extrapolate $q_\text{EA}(T)$. \index{finite-size scaling}
This somewhat unconventional use of finite-size scaling
was started in~\cite{luescher:91,kim:93,caracciolo:95,caracciolo:95b} and has
also been used in the spin-glass context~\cite{palassini:99,jorg:06}.
Most of the times, these ideas are used in the paramagnetic phase, but
we show below how to implement them in the low-temperature phase.

Close to $T_\mathrm{c}$, we know that
\begin{equation}\label{eq:SG-qEA-inf-FSS}
q_\mathrm{EA}^\infty(T) = \lambda (T_\mathrm{c}-T)^\beta [1+ \mu (T_\mathrm{c}-T)^{\omega\nu}+\ldots]\,.
\end{equation}
We have excellent determinations of $T_\mathrm{c}$ and $\beta$ from the work in~\cite{hasenbusch:08b}, 
so we need only estimate the amplitude $\lambda$. In fact, Wegner's confluent
corrections $(T_\mathrm{c}-T)^{\omega\nu}$ are small close to $T_\mathrm{c}$.
To proceed, we note that finite-size scaling tells us that
\begin{equation}\label{eq:SG-qEA-FSS}
q_\mathrm{EA}(L,T) = L^{-\beta/\nu} F(x)[1+ L^{-\omega} G(x)+\ldots],\qquad x= L^{1/\nu} (T_\mathrm{c}-T),
\end{equation}
where the critical exponents are (from~\cite{hasenbusch:08b}),
\begin{equation}
\nu = 2.45(15),\qquad \beta = 0.77(5),\qquad \omega = 1.0(1).
\end{equation}
In order to connect Eq.~\eqref{eq:SG-qEA-FSS} with the infinite-volume limit in
Eq.~\eqref{eq:SG-qEA-inf-FSS} the asymptotic 
behaviour of the scaling functions $F(x)$ and $G(x)$ must be for large $x$
\begin{equation}
F(x) \sim x^\beta,\qquad G(x)\sim x^{\omega\nu}.
\end{equation}

\begin{figure}
\centering
\includegraphics[height=0.7\linewidth,angle=270]{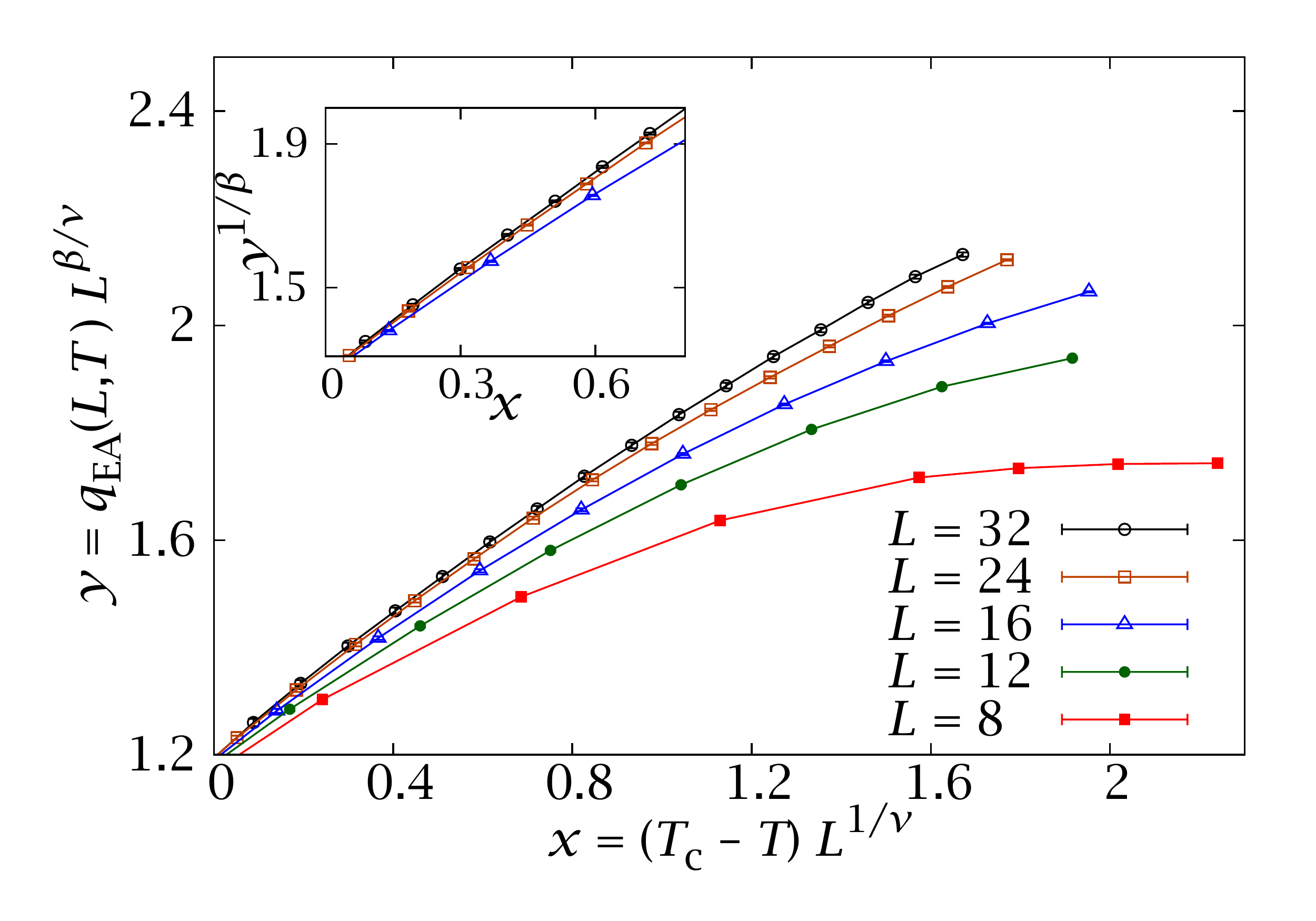}
\caption[Scaling plot of $q_\text{EA}(L,T)$]{Scaling plot of $y=q_\mathrm{EA}(L,T)L^{\beta/\nu}$ in the critical region below $T_\mathrm{c}$,
following Eq.~(\ref{eq:SG-qEA-FSS}) and using the values given in~\cite{hasenbusch:08b}
for the critical exponents and $T_\mathrm{c}$. \emph{Inset:} Close-up of the region near $T_\mathrm{c}$ 
in the representation of Eq.~(\ref{eq:SG-qEA-FSS2}), showing a linear behaviour for large $L$.}
\index{scaling plot|indemph}
\label{fig:SG-qEA-FSS}
\end{figure}

The resulting scaling plot is represented in Figure~\ref{fig:SG-qEA-FSS}. 
Varying the values of $T_\mathrm{c}$
and the critical exponents inside their error margins does not make significant  
changes in the plot. Notice 
how the curves collapse for small values of the scaling variable $x$ and large $L$, 
but how for our lowest temperatures scaling corrections become important. In fact, 
Eq.~\eqref{eq:SG-qEA-FSS} implies that when the temperature is lowered away from $T_\mathrm{c}$
the amplitude for scaling corrections grows 
as $x^{\omega\nu} \approx x^{2.45}$.

In order to estimate the amplitude $\lambda$ we 
shall concentrate on the small-$x$ region where finite-size scaling
corrections are smallest. Disregarding scaling corrections in~\eqref{eq:SG-qEA-FSS},
\begin{equation}\label{eq:SG-qEA-FSS2}
\bigl(q_\mathrm{EA}(L,T)  L^{\beta/\nu}\bigr)^{1/\beta}=F(x)^{1/\beta} \ \underset{x\to\infty}\longrightarrow\ x.
\end{equation}
The inset of Figure~\ref{fig:SG-qEA-FSS} shows that we reach this asymptotic behaviour
for $L\geq24$. Then, using the simplest parameterisation, $F(x) = (\lambda^{1/\beta} x+B)^\beta$,
\begin{equation}\label{eq:SG-qEA-FSS3}
q_\mathrm{EA}(L,T) = \lambda (T_\mathrm{c}-T)^{\beta} \left[ 1+ \frac{\beta B}{\lambda^{1/\beta} (T_\mathrm{c}-T) L^{1/\nu}}+\ldots\right]\ .
\end{equation}
We can fit our $L=32$ data for $x<0.4$ (where the curves for $L=24$ and $L=32$
are compatible) and use the resulting value of $\lambda$ to extrapolate in
Eq.~(\ref{eq:SG-qEA-FSS3}) to infinite volume. This extrapolation is represented
as a function of $T$ in Figure~\ref{fig:SG-peaks}. It is clear that this critical
extrapolation differs with the extrapolation from \eqref{eq:SG-qEA-inf} at most by
two standard deviations. The difference, if any, could be explained as
Wegner's confluent corrections. However, to make any strong claim on confluent
corrections, one would need to estimate the error in the critical
extrapolation. Unfortunately, we have found that this error estimate is quite
sensitive to the statistical correlation between $T_\mathrm{c}$, $\nu$, and
$\beta$ (as far as we know, the corresponding covariance matrix has not been
published).

One could be tempted to compare Eq.~\eqref{eq:SG-qEA-FSS3} with Eq.~\eqref{eq:SG-qEA-inf}
and conclude $\hat\nu=\nu$. In the previous Chapter we observed that, at the numerical level, 
$\nu=2.45(15)$~\cite{hasenbusch:08b} and $\hat\nu = 2.6(3)$. However, 
we do not regard this as fireproof. Indeed, it is a consequence of our somewhat arbitrary 
parameterisation $F(x) = (\lambda^{1/\beta} x+B)^\beta$. To investigate this issue 
further, the small-$x$ region is not enough. One is interested in the asymptotic behaviour
of $F(x)$ for large $x$ where unfortunately corrections to scaling are crucial. A careful
study of the crossover region can be done only by considering corrections to scaling
both at the critical temperature (at $q=0$) and below the critical temperature (at $q=q_\mathrm{EA}$).

Finally, the reader could worry about the applicability of
\eqref{eq:SG-qEA-inf-FSS} well below $T_\mathrm{c}$. The issue has been
considered recently within the framework of droplet
theory~\cite{moore:10}. It was found that \eqref{eq:SG-qEA-inf-FSS} is
adequate for all $T<T_\mathrm{c}$ (actually, no Wegner's scaling
corrections were discussed in~\cite{moore:10}). Thus, the fact that
our data are describable as scaling behaviour with leading Wegner's
correction does not imply that they are not representative of the
low-temperature phase.

\section{The link overlap and overlap equivalence}\label{sec:SG-overlap-equivalence}\index{overlap equivalence|(}
In the previous section we showed that our numerical simulations favour 
a non-trivial scenario for the spin overlap, at least for 
experimentally relevant scales. This is strong evidence against the droplet
picture of the spin-glass phase, but still leaves undecided the issue 
of TNT vs. RSB.  

The difference between these two theoretical descriptions is best 
examined through the link overlap, introduced in Section~\ref{sec:SG-TNT}. \index{replicas!real}
In the notation of real replicas $\{s_\bx^{(1)}\}$ and $\{s_\bx^{(2)}\}$, this is
\begin{equation}
Q_\text{link} = \frac{1}{N_\text{l}} \sum_{\braket{\bx,\by}} 
s_\bx^{(1)} s_\by^{(1)} s_\bx^{(2)} s_\by^{(2)}.
\index{overlap!link}
\end{equation}
Notice that in a system of $D$ spatial dimensions, the number of links
is $N_\text{l}=ND$. Also, 
\begin{equation}
C_4(r\!=\!1) = \overline{\braket{Q_\text{link}}}.
\end{equation}

The link overlap is arguably a better fundamental quantity
to describe the spin-glass phase in $D=3$ than
$q$~\cite{marinari:99,contucci:05b,contucci:06}. First, 
as was explained in Section~\ref{sec:SG-TNT}, it is more
sensitive than $q$ to the differences between RSB and the 
other theoretical scenarios for the spin-glass phase
(RSB expects a non-trivial behaviour for $Q_\text{link}$, 
which is trivial in both the  droplet and TNT pictures).
Second, while $Q_\text{link}$ is just $q^2$ 
in the mean-field model, the link overlap is actually more
amenable than $q$ to an analytical treatment in $D=3$ (within the 
RSB framework)~\cite{contucci:03,contucci:05,contucci:07}.
In particular, $Q_\text{link}$ is a more convenient quantity 
to study some properties of the spin-glass phase 
such as replica equivalence~\cite{parisi:98,parisi:00}
\index{replica equivalence}
\index{ultrametricity}
or ultrametricity~\cite{contucci:07b}.\footnote{We note that the replica
equivalence property is studied in Section~7.2
of~\cite{janus:10} using \textsc{Janus}' \index{Janus@\textsc{Janus}}
equilibrium simulations. We plan to study
the issue of ultrametricity in a future work~\cite{janus:xx}}

In short, for RSB system we expect to have overlap equivalence: 
fixing $q^2$ should also fix $Q_\text{link}$, even if the
relation between these two observables in $D=3$ is something
other than an identity.\footnote{In this respect we recall
a computation in a finite-connectivity mean-field model
that yielded $Q_\text{link} = a q^2+b$~\cite{fernandez:09f}.}

In this section we shall study the issue of overlap 
equivalence using both a non-equilibrium 
and an equilibrium approach.

\subsection{Non-equilibrium study}
From a non-equilibrium point of view, we can 
consider the two-time link overlap
\begin{equation}\label{eq:SG-Clink}
C_\text{link}(t,\tw) = \frac{1}{DN} \overline{\sum_{\braket{\bx,\by}} c_\bx(t,\tw)
c_\by (t,\tw)},
\nomenclature[Clink]{$C_\text{link}$}{Link correlation function}
\end{equation}
where $c_\bx(t,\tw)$ was defined in~\eqref{eq:SG-c}. 
Using the usual substitution of $C^2$ for $t$ as independent
variable, we can explore the would-be differences in the aging
of the spin and link overlap with this equations and, hence, 
distinguish between the coarsening dynamics of droplet and TNT
and the non-coarsening evolution of RSB. The analogue to
Eq.~\eqref{eq:SG-limit-Cinf} is now
\begin{equation}\label{eq:SG-Qlink-Clink}
\overline{\braket{Q_\text{link}}} = \lim_{t\to\infty} \lim_{\tw\to\infty} C_\text{link}(t,\tw)
= \lim_{C^2\to0} \lim_{\tw\to\infty} C_\text{link}\bigl( t(C^2,\tw), \tw\bigr).
\end{equation}

\begin{figure}
\centering
\includegraphics[height=.7\linewidth,angle=270]{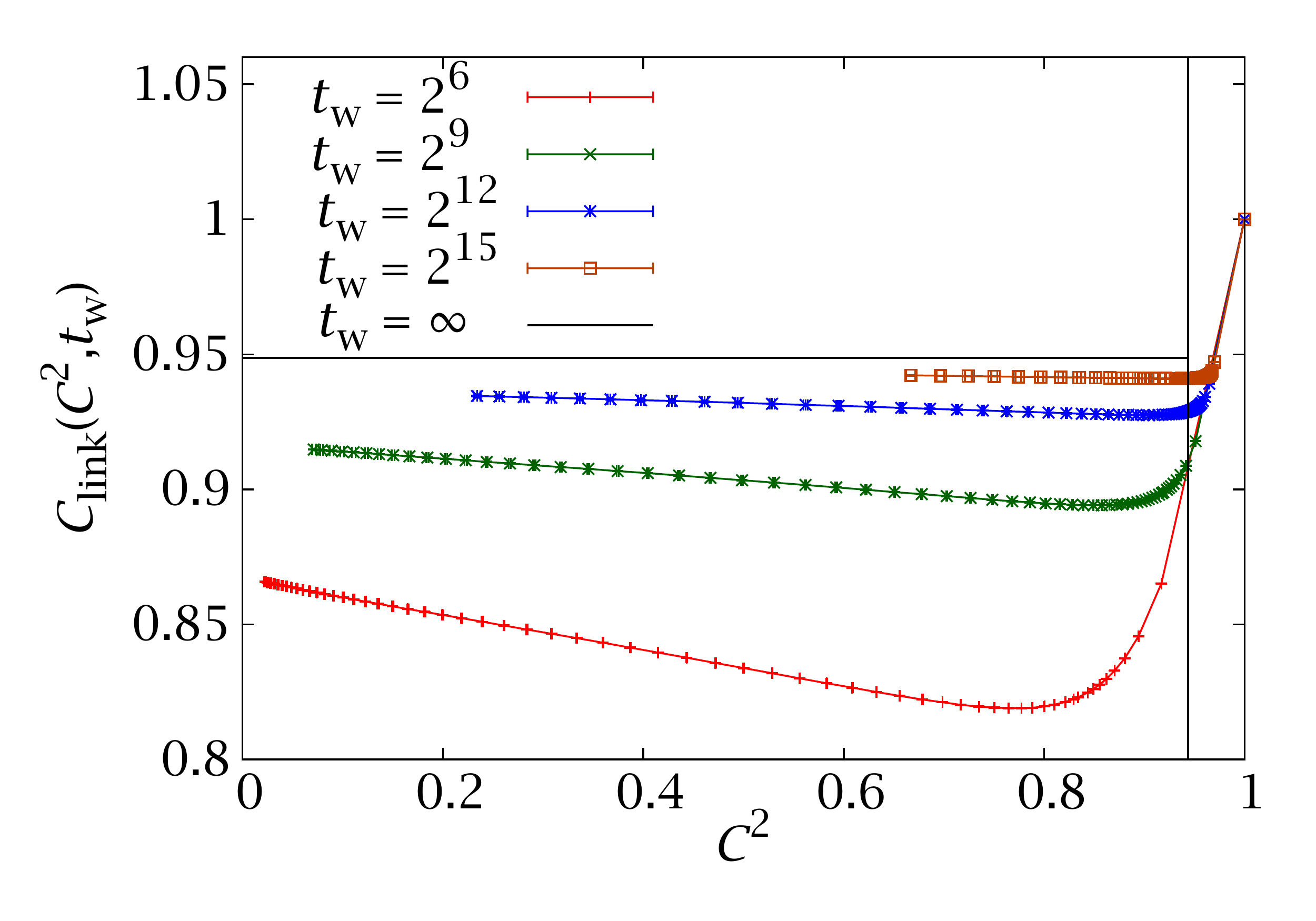}
\caption[$C_\text{link}$ of the $D=2$ Ising model]{%
Plot of $C_\text{link}(C^2,\tw)$ for the $D=2$ Ising model
at $T=0.66T_\text{c}$ (results from simulations of an $L=4096$ 
system, averaged over $20$ thermal histories). Below $C^2=q_\text{EA}^2$ 
the function approaches a $C$-independent value
for long times
---here $q_\text{EA}=m_\text{Y}^2$, where $m_\text{Y}$ is
Yang's magnetisation~\eqref{eq:ISING-m-yang}. \index{magnetisation!Yang}
\label{fig:SG-Clink-Ising}
\index{overlap!link!Ising model}
\index{correlation function (dynamics)!temporal!Ising model}
}
\end{figure}

In particular, for a coarsening system, $C_\text{link}(C^2,\tw)$
should be independent of $C^2$ for $C^2<q_\text{EA}^2$ and 
large $\tw$.  Indeed, in this regime, 
the relevant system excitations are the reversal of coherent domains
where the overlap is already $q_\text{EA}$ (as low as it can get, 
since there are no states for smaller $|q|$). Since these
domains have a vanishing surface-to-volume ratio
 (cf. Section~\ref{sec:SG-TNT}), these excitations do not 
induce a finite change in $C_\text{link}$, even if they
do modify $C^2$. Therefore, in the large-$\tw$ limit, 
$C_\text{link}$ should reach its equilibrium value $\overline{\braket{Q_\text{link}}}$
at $C^2=q_\text{EA}^2$ and not decrease any further.
This behaviour is illustrated in
Figure~\ref{fig:SG-Clink-Ising}, showing the
 $C_\text{link}(C^2,\tw)$ of the $D=2$ Ising model
\index{Ising model}
(the most straightforward coarsening system).

On the other hand, 
the overlap equivalence property of the RSB picture
translates into an equal aging for $C$ and $C_\text{link}$. 
Here, when $C^2$ decreases below $q_\text{EA}^2$  it is 
because new states are constantly being found, involving a 
change of a domain with a volume-filling surface and, hence, 
changing $C_\text{link}$.
Therefore, $C_\text{link}$ is a non-constant
function of $C^2$ for the whole range
(again, for the Sherrington-Kirkpatrick model,
$C_\text{link} = C^2$).
\begin{figure}
\centering
\includegraphics[height=0.7\linewidth,angle=270]{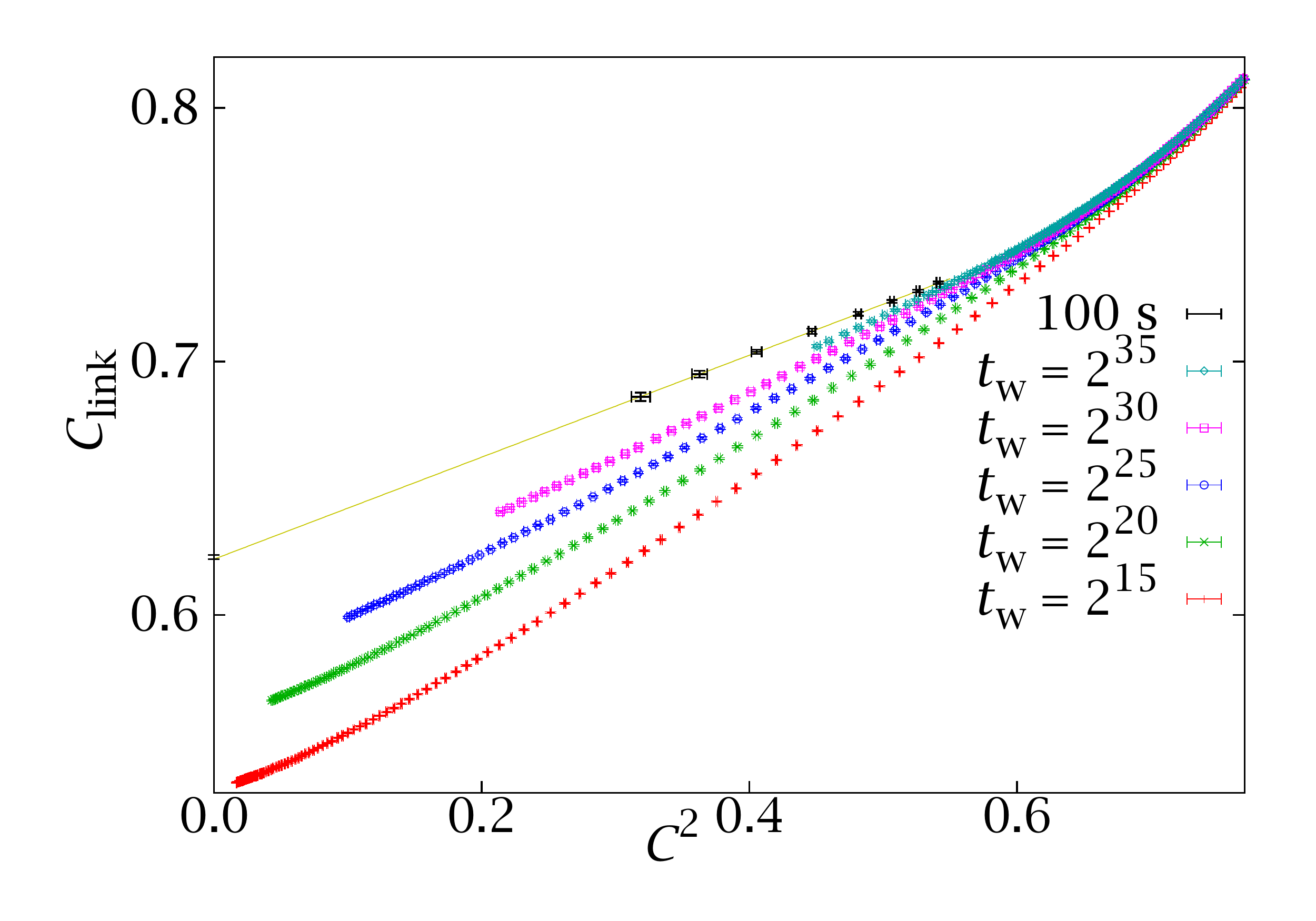}
\caption[$C_\text{link}(C^2,\tw)$ for the $D=3$
Edwards-Anderson model]{Plot of 
$C_\text{link}(C^2,\tw)$ for our spin-glass
simulations at $T=0.6$. The behaviour for our 
simulated times reproduces a non-coarsening
behaviour, even if the slope decreases with increasing
$\tw$. We also show an extrapolation for a typical
experimental time of $100$ s, which is still far 
from the coarsening behaviour of Figure~\ref{fig:SG-Clink-Ising}.
\label{fig:SG-Clink-EA}
\index{correlation function (dynamics)!temporal|indemph}
\index{overlap!link|indemph}
}
\end{figure}

Let us see where our simulations stand.
We have plotted the $C_\text{link}(C^2,\tw)$ 
of the Edwards-Anderson spin glass at $T=0.6$ in
Figure~\ref{fig:SG-Clink-EA}. The behaviour at our finite
times seems to reproduce the RSB prediction. Still, the slope
of the curve decreases (slowly) when $\tw$ increases. 
\begin{figure}
\centering
\includegraphics[height=.7\linewidth,angle=270]{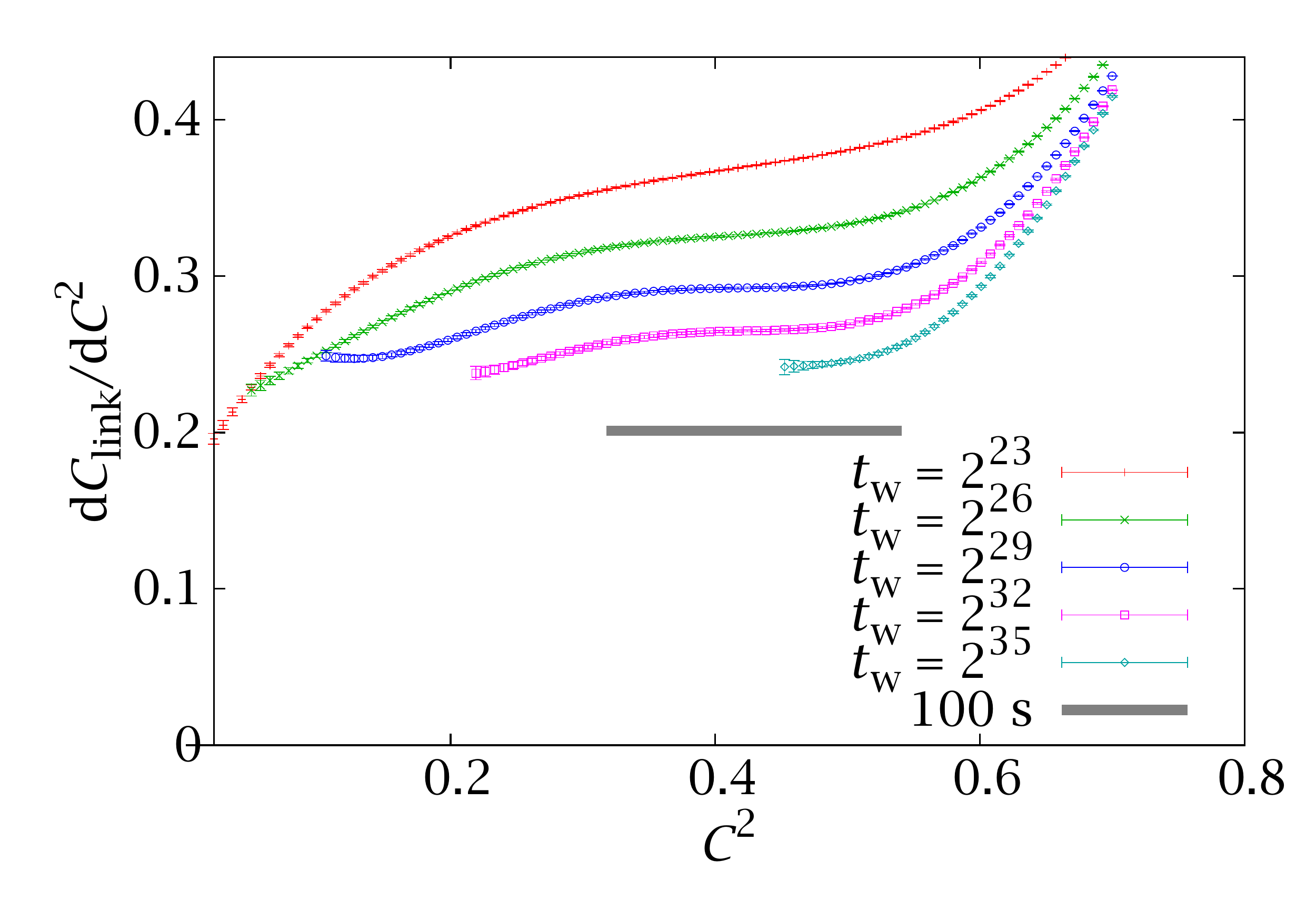}
\caption[Derivative of the link correlation function]{%
Derivative of the link correlation function, regarded 
as a function of $C^2$, for our 
spin-glass simulations at $T=0.6$. For fixed $C^2$, the
derivative decreases with time but an extrapolation to 
an experimental scale of $100$ s is well above zero
(grey line, the width represents our error bars).
\label{fig:SG-dClink}}
\end{figure}

In order to analyse this behaviour more quantitatively, we 
have interpolated the link correlation function at fixed $\tw$
with the lowest-order polynomial that allowed a fair 
fit (seventh order for $\tw\lesssim 2^{35}$, sixth
for longer times, where we cover a smaller $C^2$ range).
We then computed $\dd C_\text{link} / \dd(C^2)$ by differentiating
the interpolating polynomial (Figure~\ref{fig:SG-dClink}). 

Notice that for small $C^2$ the derivative seems to reach
a plateau for a widening $C^2$ range (suggesting that the 
relation between $C_\text{link}$ and $C^2$ becomes linear,
as in the mean-field model).\footnote{For $T=0.6$ we lack 
a good determination of $q_\text{EA}$ but, using the 
logarithmic extrapolation of Section~\ref{sec:SG-qEA-dynamics}
as a lower bound, 
we expect $q_\text{EA}^2\gtrsim 0.4$.}

Now, we would like to extrapolate these correlation 
functions and their derivatives to an experimentally
relevant time scale, to see whether or not they reach the 
coarsening behaviour. In order to do this, we consider 
$C(r\tw,\tw)$ and $C_\text{link}(r\tw,\tw)$ 
for $r=8,4,\ldots,\frac1{16}$ and fit 
each of these functions to $a_r + b_r \tw^{-c_r}$.
The parameters are stable to variations in the fitting
range, as long as $\tw>10^{5}$ and $c_r\approx 0.5$.
We then use the fits to extrapolate both the link and spin correlation
functions to $\tw=10^{14}\sim 100$ s. 
Additionally, recalling~\eqref{eq:SG-Qlink-Clink},
 the limiting behaviour
for $C^2=0$ is estimated by extrapolating $\overline{Q_\text{link}}(\tw)$
to the same experimental time.\footnote{We did not include
this point in the original study in~\cite{janus:08b}.
}
The resulting $C_\text{link}(C^2,\tw=10^{14})$ is plotted
with black crosses in Figure~\ref{fig:SG-Clink-EA}. 
The computed points fall on a straight line, whose slope 
we can use to estimate the derivative $\dd C_\text{link}/\dd(C^2)|_{\tw=10^{14}}$.
 This is plotted in Figure~\ref{fig:SG-dClink} with a thick grey
line (the width of the line represents our error interval).

In  short, the behaviour at a typical experimental time is still
far from that of a coarsening system. One could be tempted to use
the same procedure to extrapolate to $\tw\to\infty$,
but the resulting
errors are so large as to permit any kind of behaviour.

We need some more theoretical input 
on the scaling of $C_\text{link}$ in the coarsening
scenario in order 
to consider the infinite-time extrapolation.

We consider a large droplet of size $\xi(t+\tw)$ at time 
$t+\tw$ that, at time $\tw$, was made of $\mN_\text{C}$ smaller
droplets of size $\xi(\tw)$. The number of spins in the droplet's
boundary scales as $\xi(\tw)^{D_\text{s}}$. In a simple coarsening
system (such as a ferromagnet), $D_\text{s}=D-1$, but we
can actually have $D-1\leq D_\text{s}<D$~\cite{fisher:88,mcmillan:84,
bray:87, fisher:86,fisher:88b}. In fact, the numerical simulations
carried out in the TNT framework expect $D-D_\text{s}\approx0.44$~\cite{palassini:00,palassini:03}
in $D=3$.

The scaling of $\mN_\text{C}$ is straightforward:
$\mN_\text{C}\sim [\xi(t+\tw)/\xi(\tw)]^D$. The overlap of each of 
the droplets at time $\tw$ with the combined droplet
at $t+\tw$ is randomly $\pm q_\text{EA}$. Therefore, one expects
\begin{equation}\label{eq:SG-scaling-C}
C(t,\tw) \sim \sqrt{\mN_\text{C}} \left(\frac{\xi(\tw)}{\xi(t+\tw)}\right)^D \sim \left( \frac{\xi(\tw)}{\xi(t+\tw)}\right)^{D/2}\, .
\end{equation}
This equation is intuitively evident, but we note that for 
an Ising ferromagnet (the paragon of coarsening models)
it can be actually obtained with an explicit
computation. Indeed, \index{Ising model}
for an Ising model, it is shown in~\cite{cugliandolo:94} 
can be written as a series in a parameter $y$, where 
\begin{equation}
y = \frac{\xi(t+\tw)^{D/2}\xi(\tw)^{D/2}}{\bigl[ \xi(t+\tw)^2
+\xi(\tw)\bigr]^{D/2}}\, .
\end{equation}
Equation~\eqref{eq:SG-scaling-C} then
follows in the $\xi(t+\tw)\gg\xi(\tw)$ limit.

As to the link overlap, we expect
\begin{equation}\label{eq:SG-scaling-Clink-1}
C_\text{link}(t,\tw) = C_\text{link}^0 + \mN_\text{C} \frac{\xi(\tw)^{D_\text{s}}}{\xi(t+\tw)^D}\, .
\end{equation}
That is, $C_\text{link}$ cannot decrease below some finite
value $C_\text{link}^0$, which is nothing but the equilibrium
value $\overline{\braket{Q_\text{link}}}$. The excess over this value
at finite $t$ comes from the probability that a link belong 
to the surface of a droplet at time $\tw$.

We can now rewrite~\eqref{eq:SG-scaling-C} in the form
\begin{equation}\label{eq:SG-scaling-Nc}
\mN_\text{C} \sim \frac{g(C)}{C^2}\, ,
\end{equation}
where $g(x)$ is some continuous positive function, not necessarily
differentiable at $x=0$. Combining the previous expressions
we can rewrite~\eqref{eq:SG-scaling-Clink-1} as 
\begin{equation}\label{eq:SG-scaling-Clink-2}
C_\text{link}(t,\tw) = C_\text{link}^0 + C^1_\text{link}
g(C) \xi(\tw)^{D_\text{s}-D}.
\end{equation}

\begin{figure}
\centering
\includegraphics[height=0.7\linewidth,angle=270]{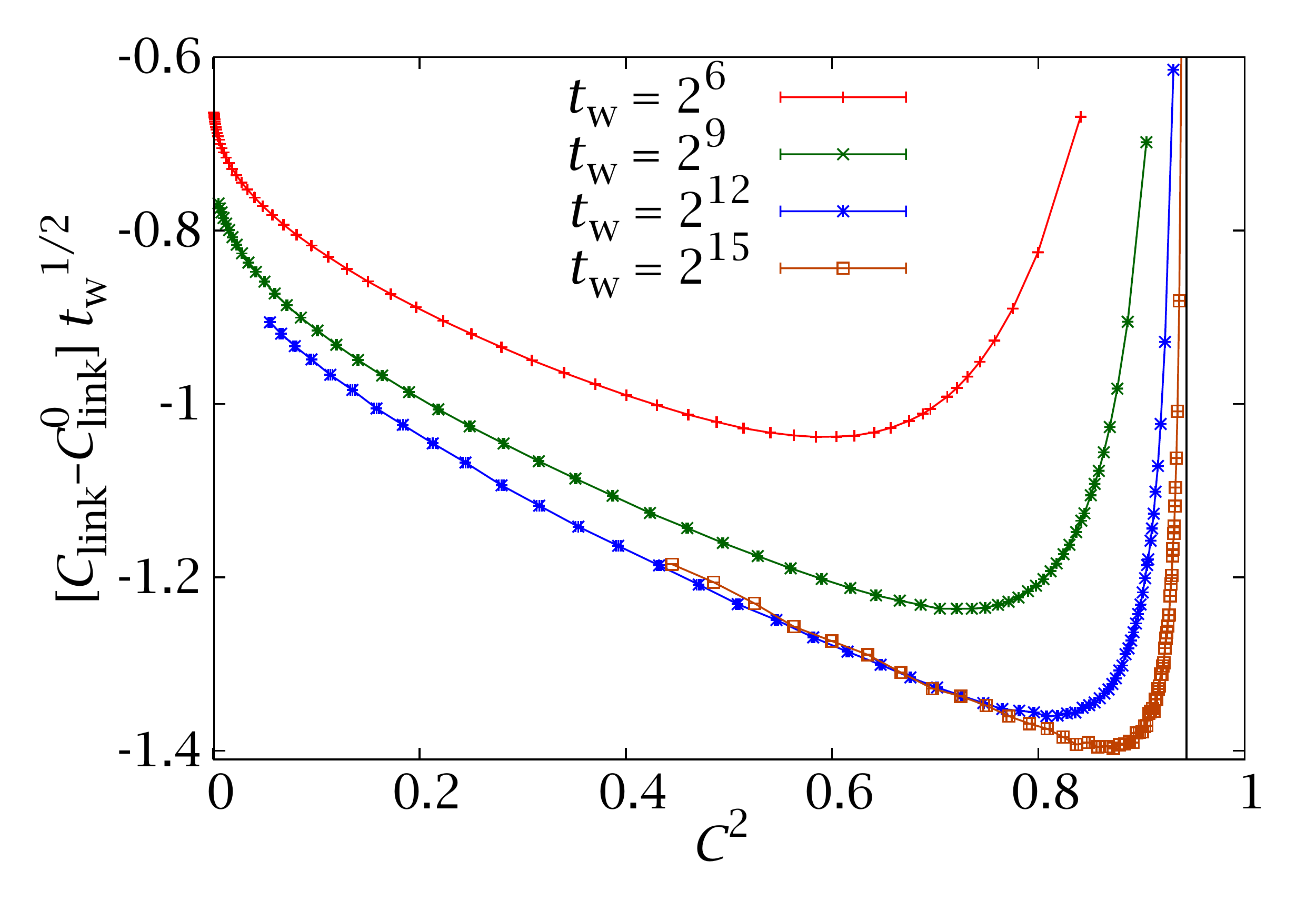}
\caption[Scaling of $C_\text{link}(C^2,\tw)-C_\text{link}^0$
for the Ising model]{Check of~\eqref{eq:SG-scaling-Clink-2}
for the Ising model (the vertical line is $C=q_\text{EA}$).
\label{fig:SG-Clink-Ising-xi}
}
\end{figure}

This equation explains the behaviour observed in Figure~\ref{fig:SG-Clink-Ising} 
for the Ising ferromagnet. \index{Ising model}
For this system, the  $D-D_\text{s}=1$, 
so we would expect $C_\text{link}(C^2,\tw)-C_\text{link}^0$ 
to scale with $\xi^{-1}$. In Figure~\ref{fig:SG-Clink-Ising-xi}
we can see that this is indeed the case ---notice that 
for the Ising model the coherence length scales as $\tw^{1/2}$
see, e.g.,~\cite{bray:94}.

\begin{figure}
\centering
\includegraphics[height=\linewidth,angle=270]{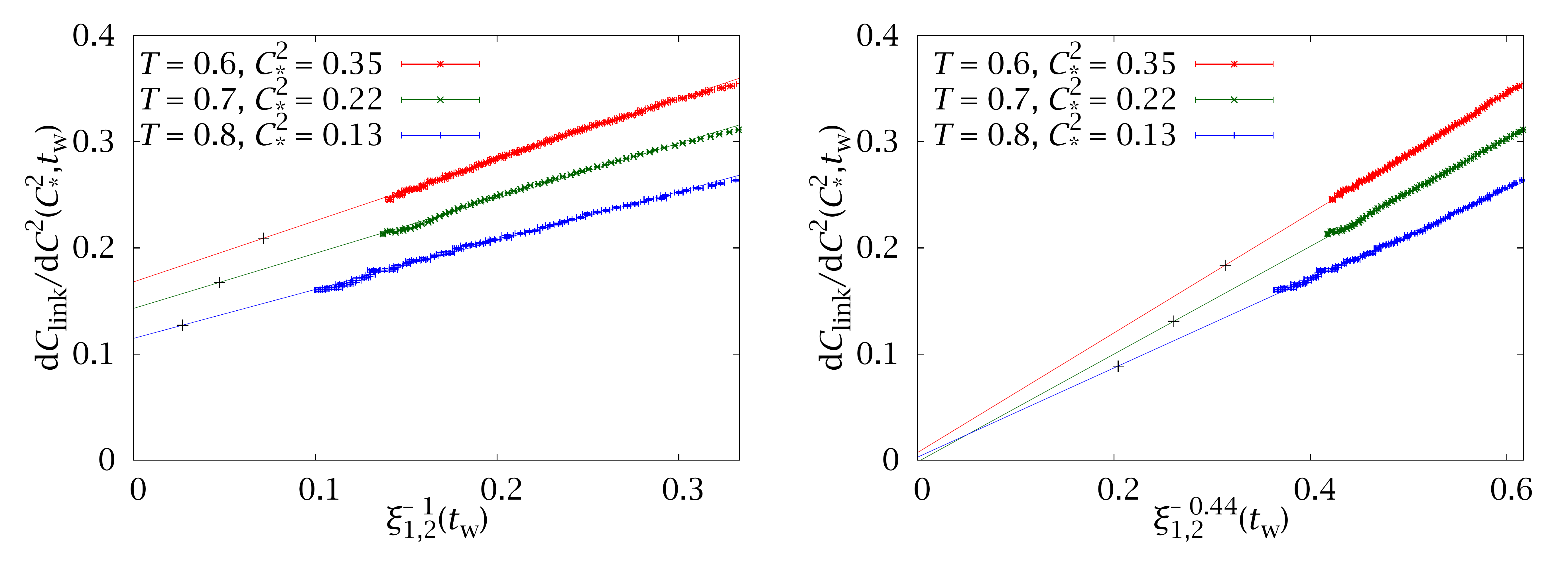}
\caption[Scaling of $\dd C_\text{link}/\dd(C^2)$]{%
Plot of $\dd C_\text{link}/\dd(C^2)$ against 
$\xi^{D-D_\text{s}}$ for $D_\text{s}=2$ (standard
coarsening, left) and $D_\text{s} = 2.55$
(TNT scaling, right). We compute the derivatives
at a value of $C_*$ below our lowest estimate for $q_\text{EA}$
at each temperature. We also mark by crosses our extrapolations
to an experimental scale of $100$ s.
\label{fig:SG-dClink-xi}}
\end{figure}

Equation~\eqref{eq:SG-scaling-Clink-2} suggests 
plotting $\dd C_\text{link}/\dd(C^2)|_{C=C_*}$ against 
$\xi^{-1}(\tw)$ and against $\xi^{-0.44}(\tw)$
(see Figure~\ref{fig:SG-dClink-xi}). It is important
to choose a value $C_*^2$ below $q_\text{EA}$ but not
too small, because otherwise the numerical derivative
would be unreliable for lack of data. Therefore, we have 
chosen values smaller than, but close to, our lower
bound for $q_\text{EA}$ in Section~\ref{sec:SG-qEA-dynamics}.
The two representations are linear within our errors. 
However, while a $\xi^{-1}(\tw)$ scaling compatible 
with standard coarsening seems falsified (i.e., its 
extrapolation is well above zero), the TNT scaling
of $\xi^{-0.44}(\tw)$ does extrapolate close 
to zero with our data.

\begin{figure}
\includegraphics[height=\linewidth,angle=270]{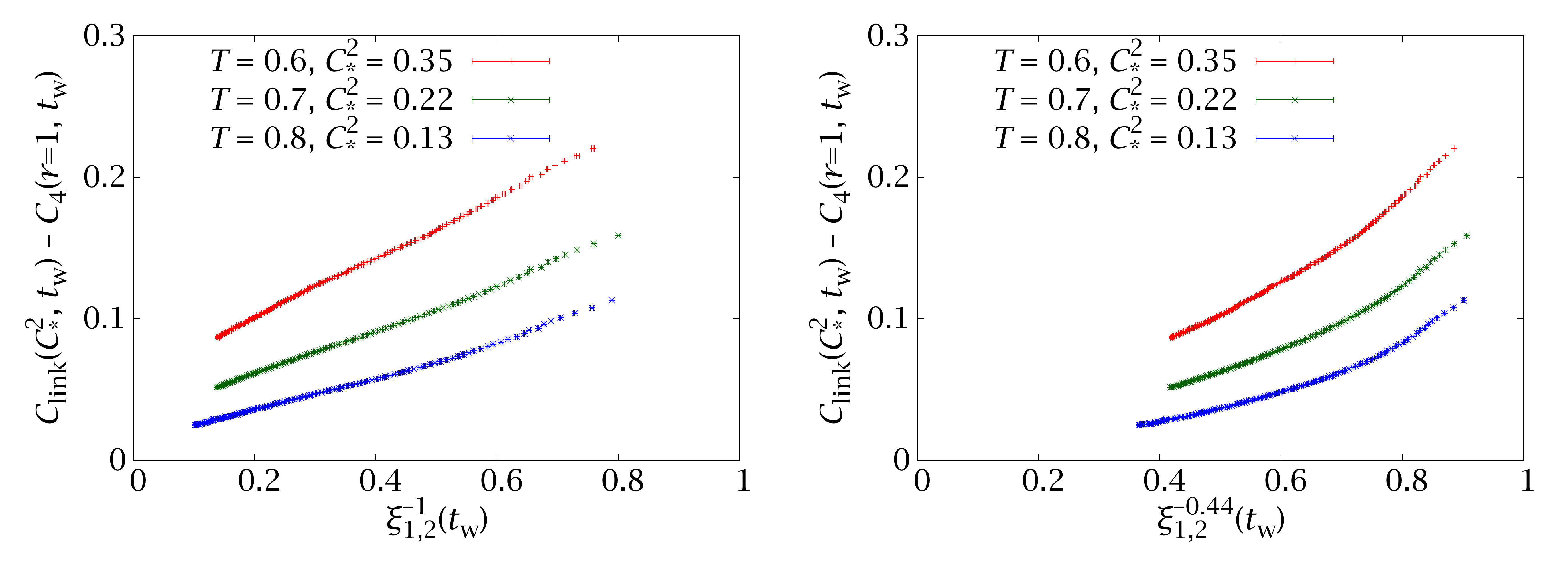}
\caption[Approach to $C_\text{link}^0$]{%
Difference $C_\text{link}(C^2_*,\tw) - C_4(r\!=\!1,\tw)$
against $\xi^{D_\text{s}-D}$ using the same values 
for $C_*$ and $D_\text{s}$ as in Figure~\ref{fig:SG-dClink-xi}.
An extrapolation to zero seems unlikely even in the $\xi^{-0.44}$
case.
\label{fig:SG-Clink-Clink0}
}
\end{figure}

It is also interesting to attempt a similar representation
to Figure~\ref{fig:SG-Clink-Ising-xi}
for the Edwards-Anderson model. Unlike the Ising case, 
now we do not know 
the exact value of $C_\text{link}^0$. However, by definition,
\begin{equation}
\lim_{\tw\to\infty} \overline{Q_\text{link}}(\tw) 
= \overline{\braket{Q_\text{link}}} = C_\text{link}^0.
\end{equation}
Of course, $\overline{Q_\text{link}}(\tw) = C_4(r\!=\!1,\tw)$.
Then, following~\eqref{eq:SG-scaling-Clink-2}, in 
a coarsening system
\begin{equation}
C_\text{link}(C^2_*,\tw) - C_4(r\!=\!1,\tw) \sim \xi(\tw)^{D_\text{s}-D},\qquad C_*^2 < q_\text{EA}^2.
\end{equation}
We have plotted this difference in Figure~\ref{fig:SG-Clink-Clink0}
for a value of $C_*$ below our lower bound for $q_\text{EA}$
in Section~\ref{sec:SG-qEA-dynamics}.  Again, we represent
both the standard $\xi^{-1}$ scaling and the TNT
scaling of $\xi^{-0.44}$. From the first one, it is easy 
to find a smooth extrapolation to a positive difference.
However, now the data are more precise than in Figure~\ref{fig:SG-dClink-xi}
and the TNT scaling of  $\xi^{-0.44}$ does not fit our data
and would only be possible if the whole of our simulation 
were in a preasymptotic regime.

There is another problem with the TNT scaling and it is
that~\eqref{eq:SG-scaling-Clink-2} relies on~\eqref{eq:SG-scaling-C},
an equation that is disproved by our computation of exponent $d$
in Section~\ref{sec:SG-thermoremanent} ---we obtained $d\approx2$, 
rather the $d=3/2$ compatible with~\eqref{eq:SG-scaling-C}.

In short, the infinite-time extrapolations remain
inconclusive, leaving room for a TNT-like coarsening behaviour.
However, in
all cases the extrapolations to the experimentally relevant time 
scale, much safer, show  that the dynamics is certainly non-coarsening 
there (thick line in Figure~\ref{fig:SG-dClink} and crosses
in Figure~\ref{fig:SG-dClink-xi}).

\subsection{Equilibrium study}
We can attempt an analogous study with our 
equilibrium simulations. The role of $C_\text{link}(C^2,\tw)$
is now played by $\mathrm{E}(Q_\text{link}|q)$, where
the conditional expectation value was defined in~\eqref{eq:SG-E}.

Just as in the non-equilibrium case, we expect $Q_\text{link}$ 
to be a strictly increasing function of $q^2$ below $q_\text{EA}^2$
for an RSB system, and to be constant for a TNT system.
Now the large-$\tw$ limit is of course replaced by a large-$L$
limit and the experimental scale is, as always, $L\sim100$. 

 
In Table~\ref{tab:SG-overlap-equivalence}
we give $C_4(r\!=\!1|q\!=\!0)$ and $C_4(r\!=\!1|q\!=\!0.523)$ 
for our simulations at $T=0.7$ (where $q_\text{EA}=0.523(32)$,
from Section~\ref{sec:SG-phase-transition}).\footnote{%
Remember that $C_4(r=1|q) = \mathrm{E}(Q_\text{link}|q)$.}
According to the TNT picture, these two quantities should
scale as $L^{-0.44}$ and have the same extrapolation to infinite
$L$.
As the reader can check, however, the data do not fit
the $L^{-0.44}$ behaviour, let alone extrapolate to the same
large-$L$ limit. The same conclusion holds if we replace $L$ by
\begin{equation}\label{eq:SG-ell}
\ell = \uppi/\sin(\uppi/L),
\end{equation}
more natural for lattice systems. Yet, it could be argued
that our data are preasymptotic, so we may try a TNT extrapolation
with corrections to scaling
\begin{equation}
C_4(r\!=\!1|q) = C_\infty + A_q L^{-0.44}(1+B_q L^{-x}).
\end{equation}
We have performed a joint fit of the data from Table~\ref{tab:SG-overlap-equivalence} to this equation,
 using a chi-square estimator
analogous to that of~\eqref{eq:SG-chi-square}. 
The fit parameters are the four amplitudes $A_0,B_0,A_{0.523},B_{0.523}$, 
the common scaling corrections exponent $x$ and the common
extrapolation $C_\infty$. In this case the correlation 
of the data for $q=0$ and $q=0.523$ at the same $L$ 
is very small ---see Table~\ref{tab:SG-overlap-equivalence}---
and, of course, there is no correlation between data 
at different $L$. The result is
\begin{align}
C_\infty & = 0.677_{-0.005}^{+0.012}, &
x &= 0.57^{+0.26}_{-0.08}, & \chi^2/\text{d.o.f.} &= 9.1/4.
\end{align}
Notice the highly asymmetric errors. The fit is not 
a very good one, its $\mathcal P$-value (cf. Appendix~\ref{chap:correlated})
is of only $6~\%$. 
\begin{table}
\small
\begin{tabular*}{\columnwidth}{@{\extracolsep{\fill}}cccr}
\toprule
\multicolumn{1}{c}{$L$ } & \multicolumn{1}{c}{ $C(1 |0)$}
& \multicolumn{1}{c}{ $C(1 |q_\mathrm{EA})$} & \multicolumn{1}{c}{$\mathcal R$}\\
\toprule
8  & 0.46138(82) & 0.57253(33) & 0.134\\
12 & 0.51649(71) & 0.60390(28) & 0.051\\
16 & 0.54552(60) & 0.62089(22) & 0.060\\
24 & 0.57573(77) & 0.63742(17) & $-0.119$ \\
32 & 0.59131(94) & 0.64579(24) & 0.063\\
\bottomrule
\end{tabular*}
\caption[$C(r=1|q)$ for $q=0$ and $q=q_\text{EA}$]{$C(r=1 | q)$ for $q=0$ and $q=q_\mathrm{EA}$ for all our system sizes 
at $T=0.703$. For each $L$, we include the correlation coefficient
between both values of $q$.}\label{tab:SG-overlap-equivalence}
\end{table}

Therefore, the TNT behaviour is seen to be, at best, very forced
with our numerical data. In the RSB setting, however, we 
expect $C_4(r\!=\!1|q)$ to scale as $\sim L^{-1}$, 
with a $q$-dependent infinite volume value $C_\infty(q)$.
Indeed, if we fit the data in Table~\ref{tab:SG-overlap-equivalence}
to $C(1|q) = C_\infty(q)+ A /\ell$ we obtain
\begin{align}
C_\infty(0) &= 0.634\,9(8), & \chi^2/\text{d.o.f.} &= 3.63/3,\\
C_\infty(q_\text{EA}) &= 0.671\,1(2), & \chi^2/\text{d.o.f.} &= 2.86/3.
\end{align}
Using $L$ instead of $\ell$, we would have had to restrict the fit
to the largest lattices, but the results would be similar (if with 
larger errors), as the reader can easily check.

\subsubsection{The variance of the link overlap}
We have seen that $Q_\text{link}$ is a strictly increasing
function of $q^2$. However, in order for these two variables
to be interchangeable, we need a second condition. That is, 
the conditional variance of $Q_\text{link}$ at fixed 
$q$ must vanish in the large-$L$ limit. In this way, fixing $q$ 
would unambiguously fix $Q_\text{link}$ also.

In general, we define the conditional variance of $O$ at
fixed $q=c$ as
\begin{equation}\label{eq:SG-Var}
\mathrm{Var}(O|q=c) = \mathrm{E}(O^2|c) - \mathrm{E}(O|c)^2.
\nomenclature[Var]{$\mathrm{Var}(O\vert q)$}{Variance of $O$, conditioned to fixed $q$}
\end{equation}
Notice that 
\begin{equation}\label{eq:SG-Var-width}
\overline{\braket{O^2}} - \overline{\braket{O}}^2
= \int_{-\infty}^\infty \dd q\ p(q) \bigl[
\mathrm{Var}(O|q) + \bigl( \mathrm{E}(O|q) - \overline{\braket{O}}\bigr)^2\bigr].
\end{equation}
\begin{figure}
\centering
\includegraphics[height=.7\linewidth,angle=270]{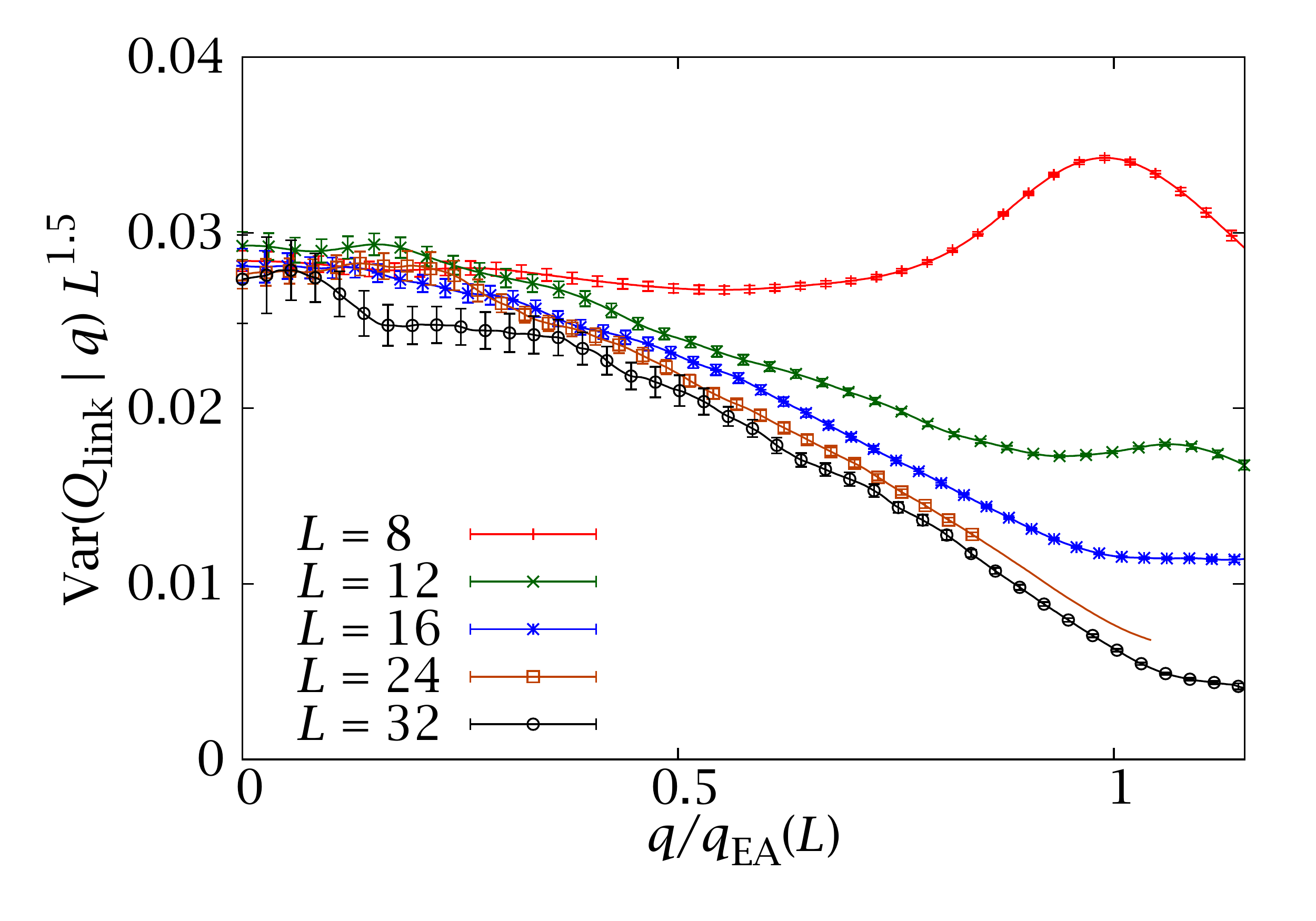}
\caption[Conditional variance $\mathrm{Var}(Q_\text{link} | q)$]{%
Conditional variance $\mathrm{Var}(Q_\text{link}|q)$ for all our lattice
sizes at $T=0.703$. We rescale the horizontal axis by the
 $q_\text{EA}(L)$, in order to see a better collapse.
\label{fig:SG-Var-Qlink}
\index{overlap equivalence|indemph}}
\end{figure}

In Figure~\ref{fig:SG-Var-Qlink} we plot $\mathrm{Var}(Q_\text{link}|q)$
for all our lattice sizes at $T=0.703$. We find 
a very clear decay of the variance. The
decay exponent  $\mathrm{Var}(Q_\text{link}|q)\sim L^{-c}$ 
is compatible with $c=D/2$, although we lack a theoretical
argument for this particular value.

\index{overlap equivalence|)}
\section{The structure of correlations in the spin-glass phase}\label{sec:SG-structure-correlations}
In the previous sections we examined two of the most straightforward\index{theta@$\theta(q)$|(}
characteristics of the spin-glass phase: the question of (non)-triviality
in the spin and link overlaps. We interpreted our study as a way of distinguishing
between the three competing pictures of the spin-glass phase: droplet, TNT and RSB.

Of the three, RSB fit our finite-size data better and also produced more consistent
extrapolations to the thermodynamical limit (assuming our data are representative 
of the asymptotic behaviour, a possibility supported by Section~\ref{sec:SG-finite-size}).
In any case, even accepting the possibility of a crossover at very large
sizes that would completely change the behaviour in the thermodynamical limit, the
conclusion that the RSB picture describes the spin-glass phase at
the experimentally relevant scale of $L\sim100$ seems safe.

Still, stating that the RSB framework seems more faithful to the physics of the 
Edwards-Anderson model would not end the discussion. Recall that this theory
was introduced as the solution of the mean-field version of the Edwards-Anderson 
model. In $D=3$ its more general points (such as the non-triviality of the spin
and link overlaps studied thus far) are expected to remain valid, but many 
fine details are certain to change.

Among these we have the issue of the structure of the correlations in the 
spin-glass phase. In Section~\ref{sec:SG-phase-transition} 
we said that the connected correlation function of an RSB system
was expected to decay as $C_4(\boldsymbol r|q^2\leq q_\text{EA}^2 ) -q^2\sim r^{-\theta(q)}$, where 
the structure exponent $\theta(q)$ is positive for $0\leq |q| \leq q_\text{EA}$.
This clustering property is a consequence of the existence of the replicon, \index{replicon}
\index{clustering property}
a critical Goldstone mode.
\index{Goldstone boson}
In particular, the RSB result above the upper critical dimension $(D>6)$ is~\cite{dedominicis:98,dedominicis:99}
\begin{subequations}\label{eq:SG-theta-MF}
\begin{align}
\theta(q=0) & = D-4,\\
\theta(0<|q|<q_\text{EA}) &= D-3.\\
\theta(|q|=q_\text{EA}) &=D-2.
\end{align}
\index{replicon exponent}
\end{subequations}%
These results
are expected to renormalise in lower dimensions~\cite{dedominicis:06}. In particular, 
if we want the spin-glass phase to have the clustering property,
these values are inconsistent
in $D<4$ (because then the correlation function would grow with distance).
 On the other hand, some authors~\cite{contucci:09} expect the mean-field
prediction for $q=q_\text{EA}$ not to change in $D=3$ (so the singularity 
in Fourier space remains $k^{-2}$, as for Goldstone bosons). Notice that the mean-field
study predicts a discontinuity at $q=0$,
which may or may not be present for $D=3$.
A theoretical conjecture~\cite{dedominicis:06},
suggests that $\theta(0)=(D-2+\eta)/2$, which, using
the anomalous dimension from~\cite{hasenbusch:08b}, 
would give $\theta(0)=0.313(5)$ in $D=3$. \index{critical exponent!eta@$\eta$}

Using our numerical data, we can try to estimate $\theta(q)$ directly for $D=3$
and, therefore, increase our knowledge of the RSB spin-glass phase. At the same 
time, a study of the connected correlations provides an additional way of 
distinguishing between the RSB and droplet pictures. In particular, a 
droplet system is not expected to have the clustering property. Rather, it
expects 
\begin{equation}\label{eq:SG-C4-droplet}
C_4(\boldsymbol r | q) = q_\text{EA}^2 f_{\boldsymbol r/r}(r/L),\qquad |q|<q_\text{EA}, 1\ll r\ll L,
\end{equation}
where $f_{\boldsymbol r/r}$ is a direction-dependent scaling function with $f_{\boldsymbol r/r}(0)=1$.
On the other hand, for $|q|=q_\text{EA}$, the connected correlation is expected
to vanish in the large-$r$ limit, decaying with the stiffness exponent $y$~\cite{bray:87}. \index{stiffness exponent}
We could, then, summarise the droplet expectation as
\begin{align}
\theta(|q|<q_\text{EA}) & = 0,\\
\theta(|q|=q_\text{EA}) &= y.
\end{align}

The exponent $\theta(q)$ is actually closely 
related with the critical exponent $1/\hat \nu$ that
we introduced in Section~\ref{sec:SG-phase-transition}.
Indeed, we can prove the following hyperscaling law \index{scaling relations}
\begin{equation}\label{eq:SG-qEA-hatnu}
\theta(q_\text{EA}) = 2/\hat \nu.
\end{equation}
In order to do this, let us consider a similar scaling
argument as we used for the DAFF in Section~\ref{sec:DAFF-phase-transition}. \index{DAFF}
We start by adding an interaction $h q L^D$ to the Hamiltonian. Then,
\begin{equation}\label{eq:SG-1}
\xi(h) \sim h^{-\nu_h} = h^{- \hat \beta \hat \delta/\hat \nu}.
\end{equation}
On the other hand, from the decay of the correlation function at the 
transition point
\begin{equation}
C_4(\boldsymbol r|q_\text{EA}) - q_\text{EA}^2 \sim  r^{-\theta(q_\text{EA})},
\end{equation}
combined with the definition~\eqref{eq:INTRO-eta}, we immediately 
read off
\begin{equation}\label{eq:SG-2}
\theta(q_\text{EA}) = D - 2+\hat \eta.
\end{equation}
Now, from~\eqref{eq:SG-1} and~\eqref{eq:SG-2} and the  \index{hyperscaling}
hyperscaling relation~\eqref{eq:INTRO-hyperscaling} we have
\begin{equation}\label{eq:SG-3}
\theta(q_\text{EA}) = 2 (D-\nu_h^{-1}).
\end{equation}
The field $h$ has the effect of changing $q_\text{EA}$,
\begin{equation}
\frac{\dd q_\text{EA}}{\dd h} \sim \xi^{D-\theta(q_\text{EA})},
\end{equation}
so 
\begin{equation}
[q_\text{EA}(h) - q_\text{EA}(0)] \sim h^{1-\nu_h \bigl(D-\theta(q_\text{EA})\bigr)} \sim
\xi^{(1-\nu_h D)/\nu_h},
\end{equation}
where we have used~\eqref{eq:SG-1} and~\eqref{eq:SG-3}. Finally, by definition of $\hat \nu$,
\begin{equation}
[q_\text{EA}(h) - q_\text{EA}(0)] \sim \xi^{-1/\hat\nu}.
\end{equation}
Comparing these last two equations we get
\begin{equation}
\frac{1}{\nu_h} + \frac{1}{\hat \nu} =  D,
\end{equation}
which, when plugged into~\eqref{eq:SG-3}, produces our sought
relation~\eqref{eq:SG-qEA-hatnu}. 
Notice that the combination 
of this hyperscaling law and our result $1/\hat\nu=0.39(5)$
gives $\theta(q_\text{EA})=0.78(10)$. Recall
that in Section~\ref{sec:SG-phase-transition}
we obtained a very 
rough estimate of $\theta(q_\text{EA}) \approx 0.65$, which
is therefore compatible with this scaling law.

However, both of the above values for $\theta(q_\text{EA})$
are incompatible with the droplet
prediction of $\theta(q_\text{EA}) = y$. Remember
that the values of $y$  in the literature
are typically close to $y\approx0.25$~\cite{carter:02,boettcher:04,
boettcher:05}. Furthermore, by forcing our data to follow
the droplet scaling in Section~\ref{sec:SG-Binder}
we obtained an even lower $y\approx0.12$.

Aside from this exact scaling law, we can formulate some additional
conjectures. Indeed, assume that $\theta(0^+)=\theta(0<|q|<q_\text{EA})<\theta(q_\text{EA})$. 
Then, we can write
\begin{align}
C_4(r|q) - q^2 &= \frac{A(q)}{L^{\theta(0^+)}}+\ldots,& 0&<|q| < q_\text{EA}.\label{eq:SG-landau1}\\
\intertext{On the other hand, }
C_4(r|q) - q^2 &= \frac{B(q)}{L^{\theta(q_\text{EA})}} + \ldots, & q&\simeq q_\text{EA}.\label{eq:SG-landau2}
\end{align}
Notice that the decay in~\eqref{eq:SG-landau1} is slower than that 
of~\eqref{eq:SG-landau2}. Therefore, $A(q_\text{EA})$ must 
be zero. Close to $q_\text{EA}$ we can consider, then the expansion
\begin{equation}\label{eq:SG-landau3}
C_4(r|q) - q^2 = \frac{(q-q_\text{EA})}{L^{\theta(0)}} + \ldots
\end{equation}
If $(q-q_\text{EA})$ is of order $L^{-1/\hat \nu}$, one expects both~\eqref{eq:SG-landau2}
and~\eqref{eq:SG-landau3} to hold and, therefore
\begin{equation}\label{eq:SG-landau}
\theta(0<|q|<q_\text{EA}) = \theta(0^+)< \theta(q_\text{EA}) \quad \Longrightarrow\quad
\theta(0^+) + \frac{1}{\hat \nu} = \theta(q_\text{EA}).
\end{equation}
Notice that the combination of this conjecture and the 
scaling law~\eqref{eq:SG-qEA-hatnu} gives the additional
conjecture
\begin{equation}
\theta(0^+) = 1/\hat\nu.
\end{equation}
In the previous equations, we have written $\theta(0^+)$ to 
accommodate the possibility of a discontinuity at $q=0$.

Let us finally note that, at the critical temperature, 
\begin{equation}\label{eq:theta-hasenbusch}
\theta(0) = 1+\eta=0.625(10),
\end{equation}
where we have used the value of $\eta$ from~\cite{hasenbusch:08b}.

In this section we shall use our spin-glass simulations to compute the 
 exponent $\theta(q)$ and decide between the RSB and droplet expectations.

\subsection{Non-equilibrium study}\label{sec:SG-replicon-dynamics}
As we have mentioned before, in the RSB scenario,
the non-equilibrium spatial correlation
gives us access to the $q=0$ physics (we start with $q=0$ at $\tw=0$
and remain there, since there are equilibrium states with vanishing order parameter).
Therefore, one expects the following long-distance behaviour of $C_4(r,\tw)$,
\begin{equation}\label{eq:SG-C4-theta}
C_4(\boldsymbol r,\tw) \xrightarrow{\ \ r\to\infty\ \ } \frac{1}{r^{\theta(0)}}f\bigl(r/\xi(\tw)\bigr).
\end{equation}
Notice that this is just~\eqref{eq:SG-C4-long-distance}, but now we have 
reinterpreted the decay exponent $a$ as $a=\theta(0)$.

The statement that $\theta(0)=0$ for droplet systems then translates in that the spatial 
correlation at fixed $r/\xi(\tw)$ should not vanish in the large-$\tw$ limit. This
is, of course, just the behaviour that one expects in a coarsening system.

\begin{figure}
\centering
\includegraphics[height=.7\linewidth,angle=270]{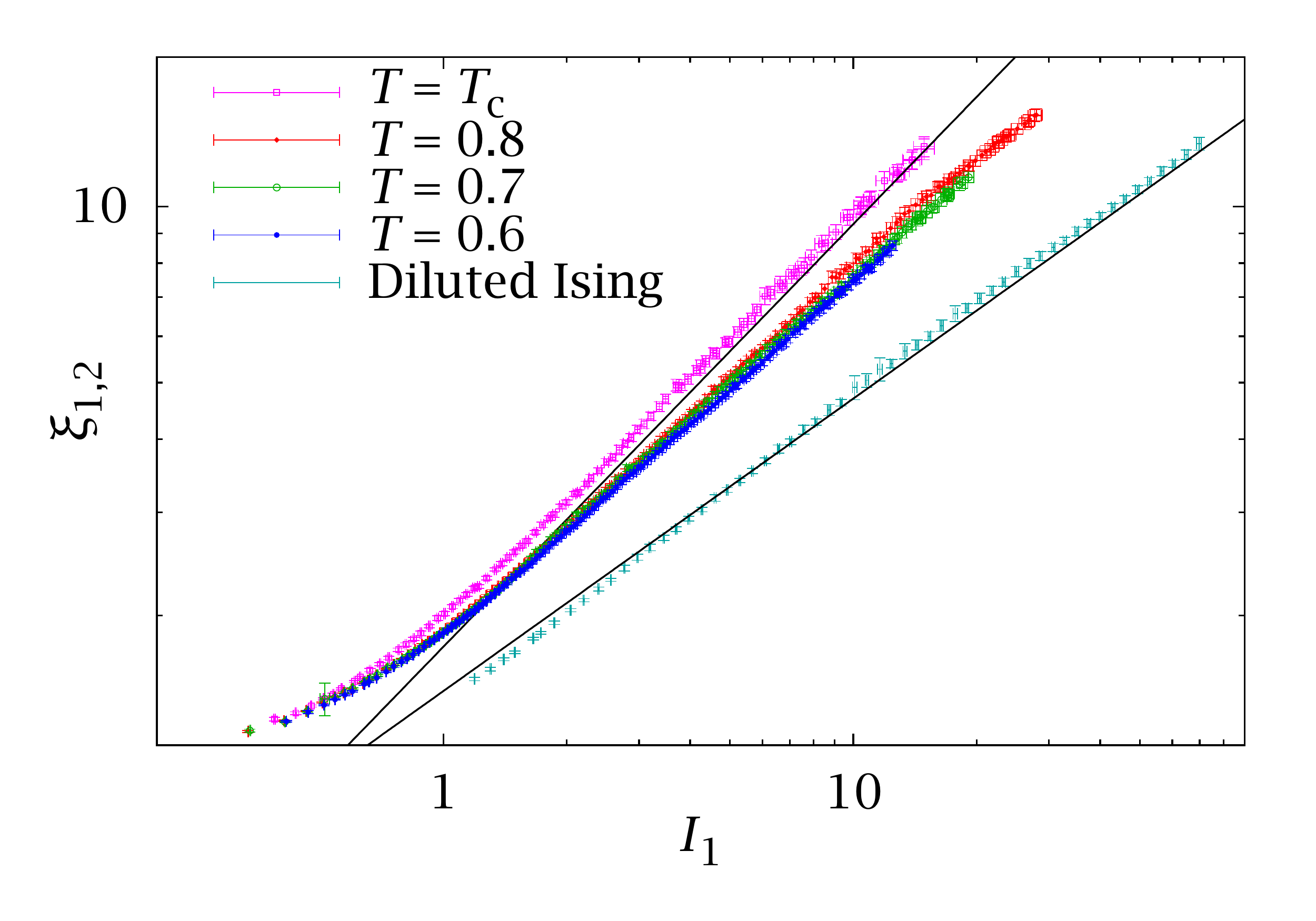}
\caption[Non-equilibrium computation of the replicon exponent]{%
Plot of $\xi_{1,2}(\tw)$ as a function of $I_1(\tw)$ for our 
three subcritical spin-glass runs and for our run at $T=1.1\approx T_\text{c}$.
We also include, as a comparison, the analogous plot 
for a $D=2$ site-diluted Ising model. We
expect a behaviour $\xi_{1,2}^{2-\theta(0)} \propto I_1$, 
with a different $\theta(0)$ in each case.
The diluted Ising is a coarsening system, so $\theta(0)=0$ and $\xi_{1,2}\propto I_1^{1/2}$.
The critical simulation is well represented by $\theta(0)=1+\eta$
(full line, taking the anomalous dimension from~\cite{hasenbusch:08b}).
On the other hand, the subcritical spin-glass simulations show an intermediate
value of $\theta(0)$, different both from the critical and the coarsening 
predictions.
\label{fig:SG-I1-xi12}
\index{replicon exponent}
}
\end{figure}
\begin{table}
\small
\centering
\begin{tabular*}{\columnwidth}{@{\extracolsep{\fill}}clcrlcc}
\toprule
$T$ &
$N_\text{samples}$&
$[\xi_\text{min},\xi_\text{max}]$&
\multicolumn{1}{c}{ $z$} &
\multicolumn{1}{c}{ $\theta(0)$} & $\chi^2_{\xi}$/d.o.f. & $\chi^2_{I_1}$/d.o.f.\\
\toprule
\multirow{1}{*}{$0.6$}
&\multirow{1}{*}{96}
 & $[3,10]$ & 14.06(25) & 0.359(13) & 41.7/82 & 49.0/82\\
$0.7$ & 768 & $[4,10]$   & 11.64(15) & 0.397(12) & 40.1/58 & 60.4/58\\
\multirow{1}{*}{$0.8$}
&\multirow{1}{*}{96}
 & $[3,10]$   & 9.42(15)  & 0.442(11) & 17.1/63 & 12.2/63\\
\multirow{1}{*}{$1.1$}
&\multirow{1}{*}{32}
 & $[3,10]$   & 6.86(16)  & 0.585(12) & 18.7/46 & 26.1/46\\
\bottomrule
\end{tabular*}
\caption[The replicon exponent with non-equilibrium methods]{%
Computation of the replicon exponent $\theta(0)$ with non-equilibrium
methods. For each temperature, we include the range of $\xi_{1,2}$ 
in which we computed the separate fits for $I_1$ and $\xi_{1,2}$ and 
the $\chi^2_\text{d}/\text{d.o.f.}$ for each fit. Notice that 
the fits for $\xi_{1,2}$ are the same ones reported in Table~\ref{tab:SG-z}.
\label{tab:SG-replicon}
\index{replicon exponent|indemph}
\index{critical exponent!z@$z$}}
\end{table}

Recalling the integrals $I_k(\tw)$, defined in~\eqref{eq:SG-I_k}, 
and our choice of  $\xi_{1,2}(\tw)$ for estimating the coherence length,  we have
\begin{equation}\label{eq:SG-I1-xi12}
I_1(\tw) \propto [\xi_{1,2}(\tw)]^{2-\theta(0)}.
\end{equation}
We have plotted $I_1$ against $\xi(\tw)$ in Figure~\ref{fig:SG-I1-xi12}
for $T=0.6,0.7,0.8, T_\text{c}$. We also include, as a comparison, the same 
plot for the $D=2$ site-diluted Ising model, a system that we know follows the $\theta(0)=0$ 
coarsening behaviour.\footnote{We could simply have used the ferromagnetic Ising model,
but this slightly less trivial model serves to illustrate the issue of superuniversality. \index{superuniversality}
The simulations are for an $L=4096$ lattice with a $25\%$ dilution, averaged over
$20$ samples at $T=0.64 T_\text{c}^{\text{Ising}}$.} 
 From this plot we see that the value of $\theta(0)$ for 
our subcritical temperatures is clearly different both from the coarsening behaviour
and from the critical value. On the other hand, the curve at $T=T_\text{c}$ seems to 
follow the expected $\theta(0)=1+\eta$ behaviour. Notice also that 
$\theta(0)$ is at least very similar for all the subcritical temperatures, in accordance
with our theoretical expectation (but in clear contrast with the behaviour of the dynamic
critical exponent $z$, which was inversely proportional to $T$).

In order to compute the actual value of $\theta(0)$ we could in principle fit $I_1$
to~\eqref{eq:SG-I1-xi12} as a function of 
$\xi_{1,2}$. Notice that these variables are highly correlated, which should 
reduce the statistical errors in $\theta(0)$. However, we would face the complicated
problem of fitting strongly correlated data to a functional form $y=f(x)$ 
with errors in both the $x$ and $y$ coordinates.
We faced quite the same difficulty when estimating the exponent $d$ measuring the decay
of the thermoremanent magnetisation in Section~\ref{sec:SG-thermoremanent} \index{magnetisation!thermoremanent}
and we can adopt the same solution. In particular, we fit $I_1(\tw)$ 
to a power law,
\begin{equation}
I_1(\tw) = B \tw^c
\end{equation}
and recall that $\xi_{1,2}(\tw)= A \tw^{1/z}$, Eq.~\eqref{eq:SG-xi-z}.  \index{critical exponent!z@$z$}
Therefore
\begin{equation}
\theta(0) = 2-cz,
\end{equation}
a relation that we can apply for each jackknife block. \index{jackknife method}

The results of following this procedure are quoted in Table~\ref{tab:SG-replicon}.\footnote{%
For $T=0.7$ we use our simulations with $768$ samples. See Section~\ref{sec:SG-finite-size}
for our choice for the fitting range.} As we can see, the values of $\theta(0)$ below $T_\text{c}$ 
are not actually compatible. However, the value at $T=0.8$ is probably affected by critical effects, 
while the values at $T=0.6,0.7$ are actually very close (see also Section~\ref{sec:CORR-bootstrap} for
some technical issues with these fits). Therefore, we use the interval between the computations at $T=0.6$
and $T=0.7$ as our confidence interval for the replicon exponent,
\begin{align}
\theta(0) &= 0.38(2), & T<T_\text{c}. \label{eq:SG-replicon-dynamics}\\
\intertext{%
This value is in good agreement with a previous (but much less precise) ground-state computation
giving $\theta(0)\approx0.4$ at $T=0$~\cite{marinari:01}. 
On the other hand, at $T=T_\text{c}$ we have}
\theta(0) &= 0.585(12), & T=T_\text{c}.
\end{align}
This value is a couple of standard deviations away from the estimate we quoted in~\eqref{eq:theta-hasenbusch}.
This difference is at the limit of statistical significance and could be due either
to corrections to scaling or to a small cutoff error in our computations of the coherence length
and $I_1$ (we have fewer samples at $T_\text{c}$, see Section~\ref{sec:CORR-bootstrap} for a discussion of this issue).

\subsection{Equilibrium study}
Let us now examine the scaling of the equilibrium connected 
correlations. We shall first carry out a study in real 
space, using the previously computed $\theta(0)$ as a starting
point. Then, as in Section~\ref{sec:SG-phase-transition}, 
we change into Fourier space, more convenient for a 
fully quantitative
analysis.

\subsubsection{Real space}
\begin{figure}
\includegraphics[height=\linewidth,angle=270]{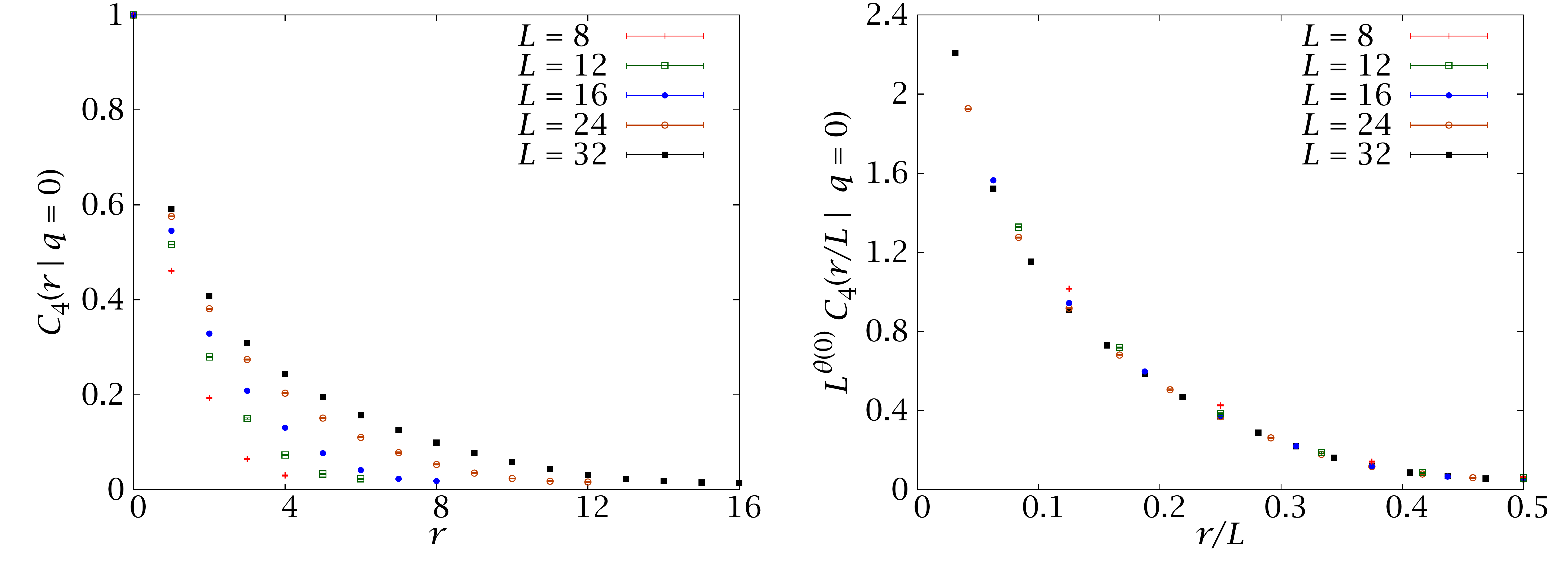}
\caption[Scaling of the spatial correlations at $q=0$]{%
Spatial correlation function $C_4(r|q=0)$ at $T=0.703$. 
We show on the right panel a rescaled version using the replicon
exponent $\theta(0)=0.38$ and the scaling variable $r/L$.
\label{fig:SG-C4-q0}
\index{correlation function (equilibrium)!spatial|indemph}
\index{replicon exponent|indemph}
}
\end{figure}

We first concentrate on $q=0$, the region where
the droplet and RSB prediction are most different. Here, see
Figure~\ref{fig:SG-C4-q0}, the correlation function
is seen to go to zero for large distances. Furthermore, in order
for the droplet scaling as $r/L$ of~\eqref{eq:SG-C4-droplet}
to work, we need to rescale our data by a factor $L^{\theta(0)}$,
using our previously computed value of $\theta(0)=0.38(2)$.

\begin{figure}
\centering
\includegraphics[height=0.7\linewidth,angle=270]{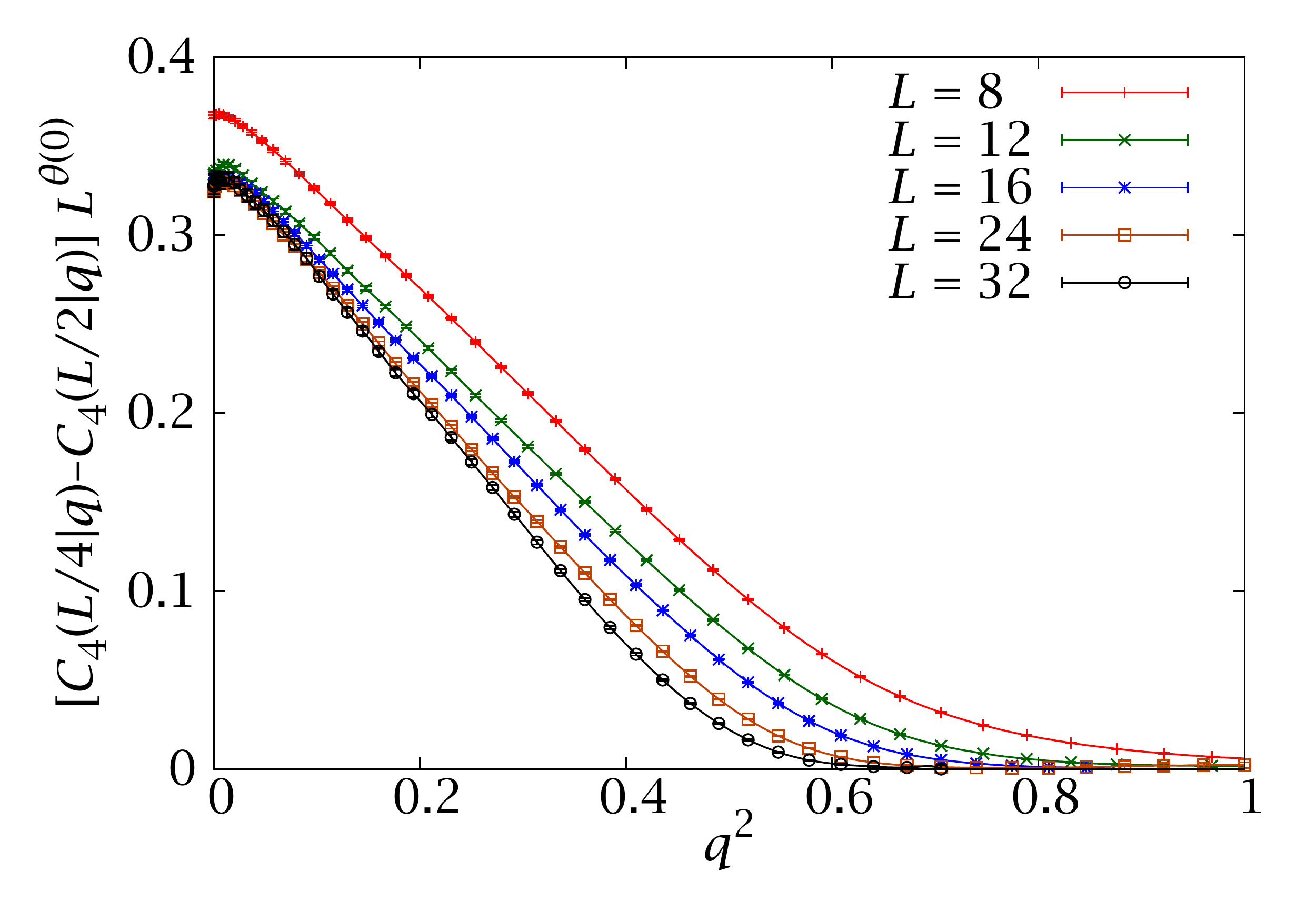}
\caption[Subtracted correlation functions and scaling
with $\theta(0)$]{Subtracted correlation functions in 
units of $L^{-\theta(0)}$ as a function of $q$.
We used $\theta(0)=0.38(2)$ from~\eqref{eq:SG-replicon-dynamics}.
\label{fig:SG-C4-L4-L2}
}
\end{figure}

For $|q|>0$, in principle we have to compute explicitly
the connected correlation 
$C_4(r|q)-q^2$. This subtraction is problematic for 
finite lattices~\cite{contucci:09}, so instead we take care
of the large-$r$ background by considering the differences
\begin{equation}
C_4(r=L/4|q) - C_4(r=L/2|q) \sim L^{-\theta(q)},
\end{equation}
We see in Figure~\ref{fig:SG-C4-L4-L2} that the above
differences scale with $L^{-\theta(0)}$ in a finite
$q$ range (approximately for $q^2<0.2$). This is a new piece
of evidence in favour of the clustering property, that is, 
of the algebraic decay of connected correlations.

On the other hand, for $q^2=q_\text{EA}^2\approx 0.3$
the exponent is clearly larger than $\theta(0)$. 
The fact that the scaling with $\theta(0)$ holds
for a finite range  and that, in particular, there
seems to be no discontinuity at $q=0$ suggests
the following scenario
\begin{equation}
\theta(0\leq |q|<q_\text{EA}) = \theta(0) < \theta(q_\text{EA}).
\end{equation}
Therefore, we are in the conditions for the 
scaling law~\eqref{eq:SG-landau} to hold and we 
should have $\theta(0) = 1/\hat\nu$. In fact, our results
are
\begin{align}
\theta(0) &= 0.38(2),\\
1/\hat\nu&= 0.39(5),
\end{align}
compatible with this expectation.
We still lack, however, one final piece
in the puzzle: the actual value of $\theta(q_\text{EA})$.
We shall attempt to compute it in Fourier space in 
the next section.

\begin{figure}[t]
\includegraphics[height=\linewidth,angle=270]{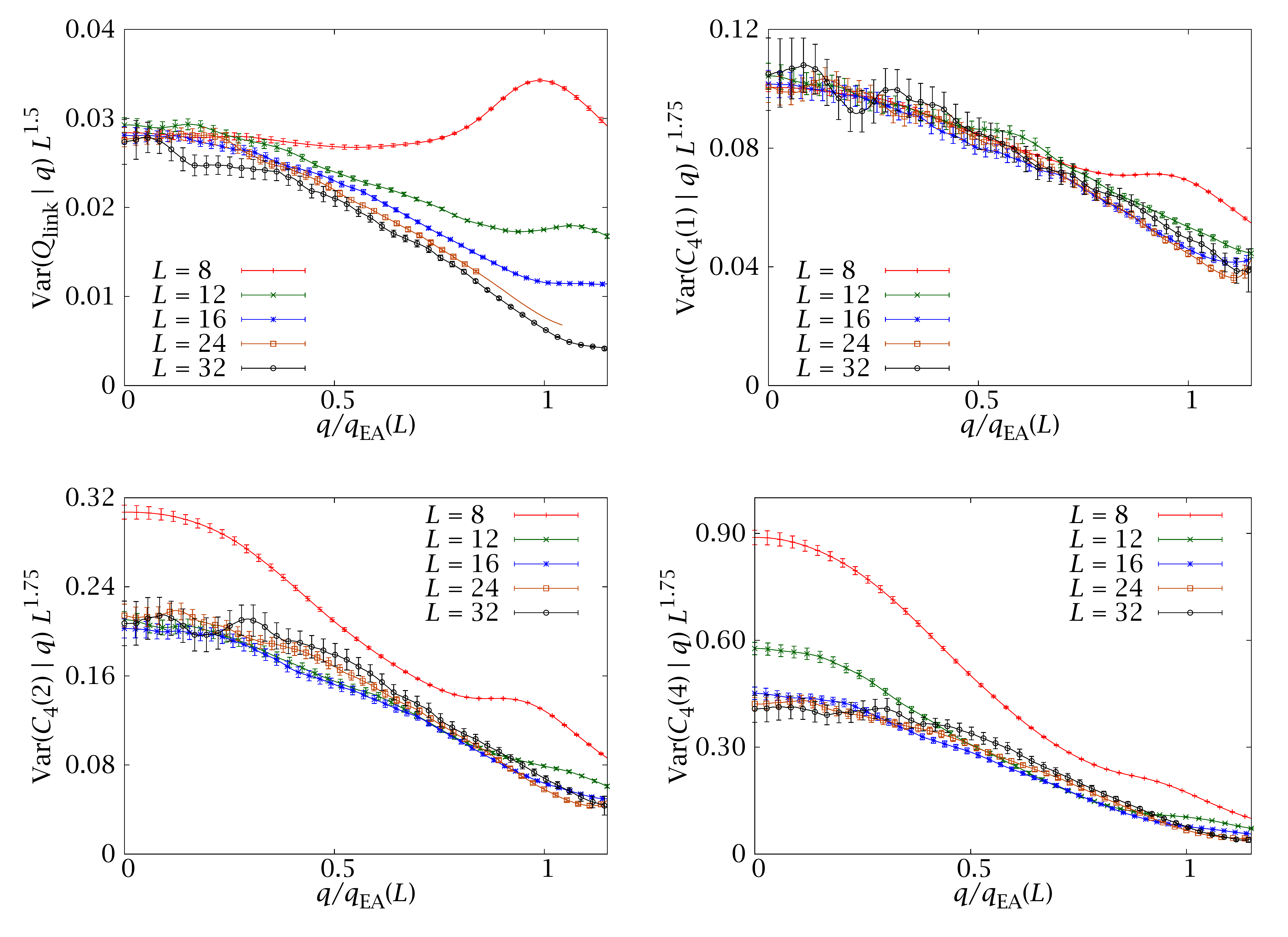}
\caption[Conditional variance of  the correlation functions]{%
Conditional variance of the spatial correlation functions 
at $T=0.703$, which is seen to scale as $\sim L^{-1.75}$,
faster than the $L^{-1.5}$ scaling of $\mathrm{Var}(Q_\text{link}|q)$.
\label{fig:SG-Var-C4}}
\end{figure}

Let us conclude the real-space study by examining the conditional
variance of the correlation function. We shall write 
$\mathrm{Var}\bigl(C_4(r)|q\bigr)$ in an
abuse of language ($C_4$ is already
defined as an averaged quantity, not a random variable) that
we hope will create no confusion. Notice that,
while $\mathrm{E}(Q_\text{link}|q) = C_4(r=1|q)$
the variance $\mathrm{Var}(Q_\text{link}|q)$
is different than $\mathrm{Var}\bigl(C_4(1) | q\bigr)$.
These variances are plotted in Figure~\ref{fig:SG-Var-C4}, 
where they are seen to decrease with $L$
even faster than $\mathrm{Var}(Q_\text{link} |q)$.

\subsubsection{Fourier space}
\begin{table}
\small
\begin{tabular*}{\columnwidth}{@{\extracolsep{\fill}}clcll}
\toprule
\multirow{2}{1cm}{$L$} &
\multicolumn{1}{c}{ $T = 1.109$} & &
\multicolumn{2}{c}{ $T = 0.703$} \\
\cmidrule{2-2} \cmidrule{4-5}
& \multicolumn{1}{c}{ $F_0$} & &  
                           \multicolumn{1}{c}{ $F_0$} & \multicolumn{1}{c}{ $F_{q_\mathrm{EA}}$}\\
\toprule
8  &16.126(64) & & 19.46(18) & \ \ 10.373(48)    \\
12 &40.59(18)  & & 53.65(62) & \ \ 25.79(15)     \\
16 &79.07(31)  & & 112.5(13) & \ \ 51.60(29)     \\
24 &204.05(83) & & 327.3(43) & \ \ 134.54(80)    \\
32 &404.4(29)  & & 699(23)   & \ \ 267.9(30)    \\
\midrule
$L_\mathrm{min}$         &  \multicolumn{1}{c}{8}          &  &
                             \multicolumn{1}{c}{8}         & \multicolumn{1}{c}{16}   \\
$\chi^2/\mathrm{dof}$  & \multicolumn{1}{c}{$0.90/3$}   & & 
                            \multicolumn{1}{c}{1.38/3}     & \multicolumn{1}{c}{0.13/1} \\
$\theta$             &  \multicolumn{1}{c}{$0.638\,2(44)$} &   &
                        \multicolumn{1}{c}{0.377(14)}  & \multicolumn{1}{c}{0.611(16)[60]}\\
\bottomrule
\end{tabular*}
\caption[Exponent $\theta(q)$ at $T=0.703$ and $T=T_\text{c}$]{$F_{q=0}$ 
for our different system sizes and temperatures $T\!=\!1.109\!\approx\! T_\mathrm{c}$
and $T\!=\!0.703$.
We report at the bottom power-law fits $F_q\! =\! A \ell^{D-\theta(q)}$ 
in the range $L\!\geq\!L_\mathrm{min}$.  At $T\!=\!0.703$ we 
also consider $F_{q}$ at $q\!=\!0.523\!\approx\! q_\mathrm{EA}$. The
second error bar, in square brackets, accounts  for the uncertainty induced by the 
determination of $q_\mathrm{EA}$ (a larger $q_\text{EA}$ produces a larger $\theta$).
\index{replicon exponent|indemph}
\index{critical exponent!eta@$\eta$|indemph}
\label{tab:SG-theta}}
\end{table}
In the previous section we saw that 
the equilibrium spatial correlations
scaled well at $q=0$ with the $\theta(0)$ 
computed out of equilibrium and that, indeed, 
this same exponent seemed to 
rule the scaling for a finite $q$ range.

Here we want to make a more quantitative study,
using the correlations in Fourier space (more convenient
at $q\neq0$, since they do not require a subtraction).
In particular, we shall obtain independent estimates 
of $\theta(0)$ and $\theta(q_\text{EA})$ and test
the hypothesis of constant $\theta(q)$ in the range 
$0\leq q<q_\text{EA}$.

To this end, let us recall definition~\eqref{eq:SG-Fq} and 
perform fits to 
\begin{equation}
F_q = A_q \ell^{D-\theta(q)},
\end{equation}
where $\ell$ was defined in~\eqref{eq:SG-ell}.
We have done this for $q=0$ and $q=q_\text{EA}$ 
at $T=0.703$ in Table~\ref{tab:SG-theta}. 
The scaling at $q=0$ is very good and we obtain
\begin{equation}
\theta(0) = 0.377(14),
\end{equation}
in excellent agreement with our non-equilibrium 
estimate~\eqref{eq:SG-replicon-dynamics}.
For $q=q_\text{EA}$, however, we need to
restrict our fitting range to $L\geq16$.
Furthermore, the uncertainty in $q_\text{EA}=0.52(3)$
induces a systematic error, which we denote with square 
brackets. Our final estimate is
\begin{equation}
\theta(q_\text{EA}) = 0.611(16)[60].
\end{equation}
This estimate is compatible with the scaling law
$\theta(q_\text{EA})=2/\hat\nu$ ---recall that $1/\hat\nu=0.39(5)$.

For the sake of completeness, and as a check of our procedure, 
we have also computed $\theta(0)$ at $T_\text{c}$.
Our result, $\theta(0)=0.638(4)$, is very different 
from the value at $T=0.703$ and compatible
with the best result in the literature~\cite{hasenbusch:08b},
$\theta(0)=0.625(10)$
(our error bar is deceptively small, because we have not 
taken scaling corrections into account, unlike~\cite{hasenbusch:08b}).

\begin{figure}
\centering
\includegraphics[height=0.7\linewidth,angle=270]{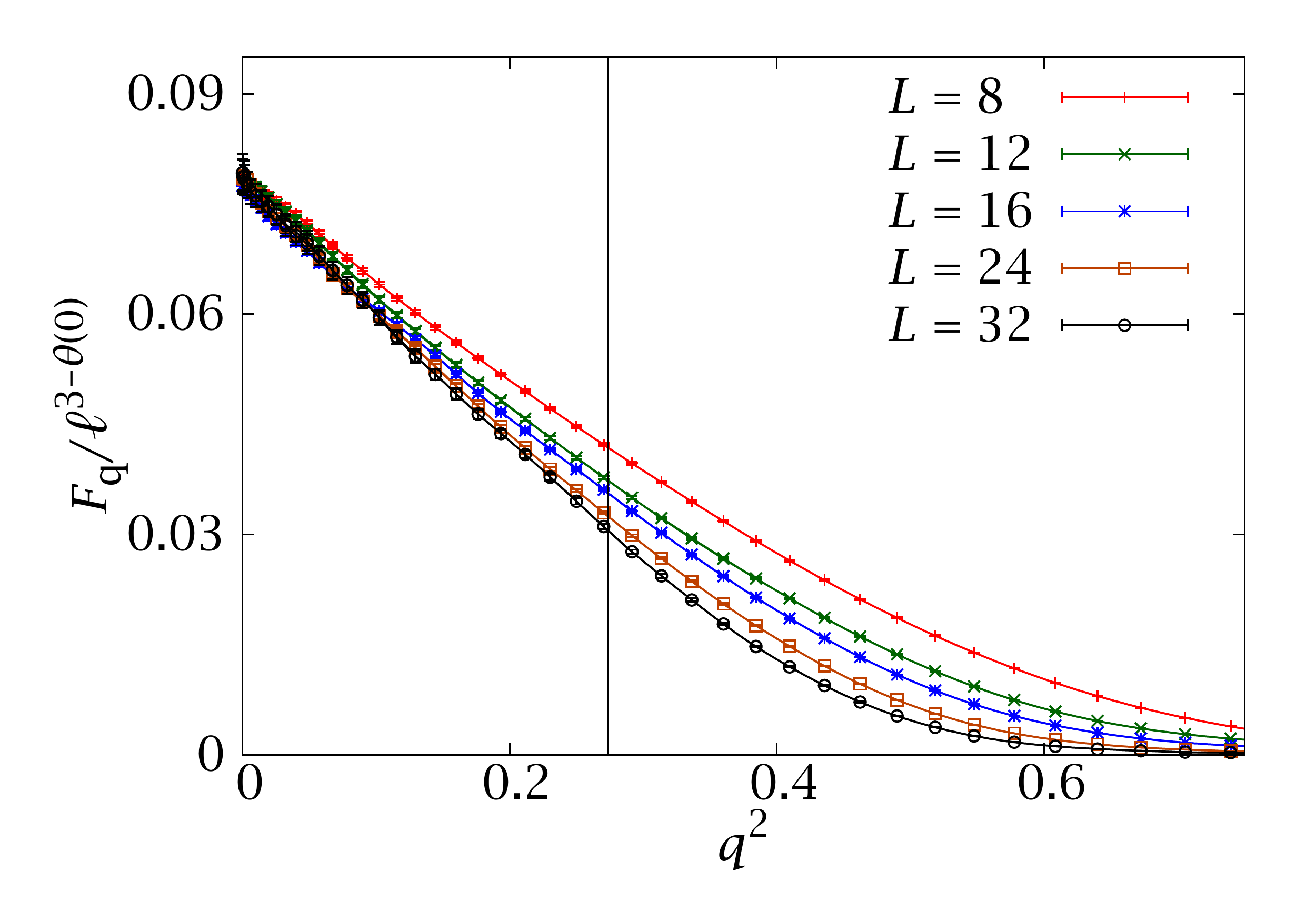}
\caption[Scaling of $F_q$ with $\theta(0)$]{Plot of $F_q$ 
at $T=0.703$ in units of $\ell^{D-\theta(0)}$. As in the real-space
representation of Figure~\ref{fig:SG-C4-L4-L2}, we can see a good
scaling for a finite range in $q$. The vertical line marks our 
estimate of $q_\text{EA}^2$.
\index{replicon exponent|indemph}
\index{correlation function (equilibrium)!spatial|indemph}
\label{fig:SG-Fq-theta}
}
\end{figure}

Finally, let us test the hypothesis of constant $\theta(q)$ for $|q|<q_\text{EA}$.
To this end, we show in Figure~\ref{fig:SG-Fq-theta} the analogous plot
to Figure~\ref{fig:SG-C4-L4-L2} in Fourier space. Not only do we see a good
scaling with $\theta(0)$ for a finite range, as in Figure~\ref{fig:SG-C4-L4-L2}, 
but now in our Fourier-space representation each of the curves 
is linear in $q^2$ below $q_\text{EA}^2$. Therefore, we have performed
the following fit
\begin{equation}\label{eq:SG-Fq-fit}
F_q-F_0 = -a_2 q^2 + a_4 q^4.
\end{equation}
Notice that, since the data at different $q$ are correlated, subtracting
the $q=0$ value yields a significant error reduction. The quadratic term 
is only included to control systematic effects in the computation of $a_2$
(and it probably causes us to overestimate errors). We report the resulting
vales of $a_2$ in Table~\ref{tab:SG-a2}.

\begin{table}
\small
\begin{tabular*}{\columnwidth}{@{\extracolsep{\fill}}cc}
\toprule
\multicolumn{1}{c}{$L$ } & \multicolumn{1}{c}{$a_2$}\\
\toprule
8  & 35.6(18)    \\
12 & 110.4(58)   \\
16 & 254(12)    \\
24 & 816(40)    \\
32 & 1823(216)  \\
\bottomrule
\end{tabular*}
\caption{Coefficient $a_2$ in a fit to~\eqref{eq:SG-Fq-fit}
for for $q^2<0.5$ at $T=0.703$.
\label{tab:SG-a2}
}
\end{table}

Now, in the case of discontinuous $\theta(q)$, one would expect a behaviour
\begin{equation}
a_2(L) = b \ell^c,
\end{equation}
with $c>D-\theta(0)$. If we perform this fit from the data in Table~\ref{tab:SG-a2}, 
we obtain $c=2.91(6)$, with $\chi^2/\text{d.o.f.} = 0.08/3$.
Yet, the extremely small $\chi^2$ is an 
indication that we have probably overestimated the errors in $a_2$. Therefore, let us take
the value of $c$ seriously and consider the scaling of $F_{q_*}$ with $q_*=0.1<q_\text{EA}$.
Then, in order for the would-be different scaling at $q_*$ to be noticeable, we need
a lattice size large enough so that
\begin{equation}
F_0-F_q \approx a_2 q_*^2 \sim F_0.
\end{equation}
The reader can check that this results in an enormous $L\sim 3\times 10^6$.
Therefore, we do not find any numerical evidence in favour of the perturbative 
prediction of a discontinuity at $q=0$.

Let us go back our favoured scenario of constant $\theta(q)$ in $|q|<q_\text{EA}$.
Then, the linear behaviour of $F_q$ with $q^2$ suggests a scaling
\begin{equation}
a_2(L) = e\ell^{D-\theta(0)} ( 1 - d \ell^{\theta(0)-\theta(q_\text{EA})}).
\end{equation}
In addition, in this hypothesis of constant $\theta(q)$
 we have $\theta(0)=1/\hat\nu=\theta(q_\text{EA})/2$
from~\eqref{eq:SG-qEA-hatnu} and~\eqref{eq:SG-landau}. Therefore, we can write
\begin{equation}
a_2(L) = e \ell^{D-\theta(0)} ( 1- d\ell^{-\theta(0)}).
\end{equation}
Fitting the data in Table~\ref{tab:SG-a2} to this expression, 
fixing $\theta(0)=0.38(2)$ and varying only the amplitudes, 
we obtain an excellent $\chi^2/\text{d.o.f.} = 0.35/3$.
Therefore, the simplest scenario consistent with the data is
\begin{align}
\theta(|q|<q_\text{EA}) &= \theta(0) = 1/\hat\nu = 0.38(2),\\
\theta(q_\text{EA}) &= 2\theta(0).
\index{theta@$\theta(q)$|)}
\index{spin glass|)}
\end{align}

\chapter{Conclusions and outlook}\label{chap:conclusions}
In this dissertation we have explored two main themes
in the context of the statistical mechanics of disordered systems,
\begin{enumerate}
\item It is worthwhile (and feasible) to consider a statistical ensemble 
tailored to the problem at hand, both from the point  \index{disorder}
of view of achieving thermalisation and in order to obtain  \index{complex systems}
the maximum amount of physical information about the system. \index{Monte Carlo method}
\item When working with disordered systems, one cannot help \index{thermalisation}
dealing with non-equilibrium phenomena. In this sense, a quantitative \index{statics-dynamics equivalence}
relation between the equilibrium phase and the non-equilibrium
evolution can, and should, be established.
\end{enumerate}
We took a Monte Carlo simulation approach to both issues, carrying out
simulations in a variety of computing systems (conventional computing clusters, \index{grid computing}
supercomputing facilities and grid resources) for a combined total
of several millions of CPU hours. In addition, we analysed the data from  \index{Janus@\textsc{Janus}}
about one year of non-stop production of \textsc{Janus}, a special-purpose
machine equivalent to many thousands of conventional computers. \index{special-purpose computers}

In regards to the first point we introduced the Tethered Monte Carlo \index{tethered formalism}
method, a general formalism to reconstruct the Helmholtz effective potential \index{effective potential}
associated to an arbitrary reaction coordinate. The method consists in performing individual 
Monte Carlo simulations at fixed values of the reaction coordinate, which
are then combined to construct the Helmholtz potential from  a \index{reaction coordinate}
fluctuation-dissipation formalism. \index{fluctuation-dissipation}
By choosing an appropriate reaction coordinate (i.e., one that labels the different relevant metastable
states) it is possible to use this method to obtain a comprehensive picture \index{metastability}
of the system's physics. 

We demonstrated that this approach is indeed workable in
physically relevant situations
by tackling a classic `hard' problem in  \index{DAFF}
the field of disordered systems: the phase transition of the 
diluted antiferromagnet in a field. We found that the tethered
approach unlocked much information that
remains hidden to a traditional method and we were 
able to obtain a comprehensive and consistent picture
of the critical behaviour in this model.

We explored our second point, the statics-dynamics relation, in 
the field of spin glasses. There we analysed large-scale \index{statics-dynamics equivalence}
simulations performed on \textsc{Janus}. \index{Janus@\textsc{Janus}}
These were carried out both in and out of equilibrium
and in both cases constituted a jump of several orders of magnitude
with respect to the state of the art. We were able to use
these high-quality data, combined with some novel analysis
techniques, to establish a quantitative statics-dynamics 
equivalence and to  \index{spin glass}
determine the nature of the spin-glass phase up
to the experimental scale. \index{experimental scale}

In the following subsections we detail our conclusions and outlook
for each of our three lines: Tethered Monte Carlo, the DAFF and 
the Edwards-Anderson spin glass.

\section{The Tethered Monte Carlo method}
We started by demonstrating our Tethered Monte Carlo method in the context
of Ising ferromagnets. In this comparatively simple problem \index{tethered formalism}
we found that our formalism, coupled to a local update
algorithm, was able to reconstruct the effective potential
without critical slowing down. We then showed that the method \index{critical slowing down}
can be combined with sophisticated update schemes (cluster algorithms),
in which case the critical slowing down disappeared even for  \index{cluster methods}
non-magnetic observables.

We then applied Tethered Monte Carlo to a system with a rugged \index{free energy!landscape}
free-energy landscape: the DAFF (see below for the physical results).
We found that the method was able \index{DAFF}
to eliminate the exponential slowing down caused by the free-energy \index{free energy!barriers}
barriers associated to our reaction coordinates (the magnetisation and staggered magnetisation).
\index{magnetisation}
\index{magnetisation!staggered}

Thus, we were able to thermalise safely much larger systems
that are possible with conventional methods, such as 
parallel tempering in the canonical ensemble. However, for our largest
systems we started to see metastable behaviour caused by new free-energy 
barriers. In this sense, the onset of exponential slowing down
was delayed, rather than outright eliminated.  \index{critical slowing down!exponential}

Dealing with these new free-energy barriers will probably 
require the introduction of additional reaction coordinates, 
capable of a further classification of metastable states.
Our method is in principle equipped to carry out this 
plan, but a practical implementation requires some 
work.

A different avenue for further work with Tethered Monte Carlo
is extending its application to new systems.  One interesting \index{condensation}
possibility is the study of the condensation transition~\cite{biskup:02,binder:03,macdowell:04,
nussbaumer:06,nussbaumer:08}, which
can be modelled with Ising systems. \index{Ising model}
We finally note that the tethered formalism has already been applied \index{crystallisation}
to the study of hard-spheres crystallisation~\cite{fernandez:11}.

\section{The diluted antiferromagnet in a field}
The use of the tethered formalism allowed us to obtain a consistent 
picture of the critical behaviour in the DAFF.  We obtained clear \index{DAFF}
evidence in favour of the second-order nature of the transition 
and computed the three independent critical exponents. In particular, we
obtained a precise determination, $\theta=1.469(20)$, for  \index{hyperscaling}
the elusive hyperscaling violations exponent. We also observed a clear \index{critical exponent!theta@$\theta$}
divergence of the specific heat, in accordance with experimental work. \index{specific heat}

Still, there is much work still to do. In particular, our data  \index{critical exponent!nu@$\nu$}
for the $\nu$ critical exponent was affected by finite-size effects.
Also, even if we determined that the specific-heat exponent $\alpha$ \index{critical exponent!alpha@$\alpha$}
is positive, we could not estimate its precise value very precisely.
To this end, the simulation of even larger systems would be required.

Finally, it would be very interesting to achieve a closer connection
to experimental work. The obvious contact point would be obtaining
a precise numerical determination of the scattering line shape, which
is the basic quantity explored in experiments to characterise the 
critical behaviour. Our methods are well suited to carry out this 
programme in the future. \index{DAFF!experiments}

\section{The Edwards-Anderson spin glass} \index{spin glass}
We studied the $D=3$ Edwards-Anderson spin glass in and out of \index{Edwards-Anderson model}
equilibrium and were able to establish a time-length dictionary,  \index{time-length dictionary}
relating the equilibrium phase of a system of size $L$ and 
the non-equilibrium state at time $\tw$ of a system
in the thermodynamical limit. In particular, we reached 
the conclusion that the equilibrium phase of a system of  \index{experimental scale}
size $L\sim100$, and not the thermodynamical limit, is the  \index{thermodynamical limit}
relevant one for understanding experimental work on spin glasses.

We later took this quantitative statics-dynamics \index{statics-dynamics equivalence}
connection one step further by establishing a finite-time scaling
framework. This last result, in which dynamical \index{finite-time scaling}
heterogeneity played a major role, has implications  \index{dynamical heterogeneities}
for experimental work.

The above studies were carried out in a neutral formalism, without
assuming the validity of any particular theory for 
the spin-glass phase. A more detailed analysis of the 
spin-glass phase showed clear evidence in favour 
of the RSB picture, at least for experimentally relevant  \index{RSB}
scales. We based this conclusion on three main observations: \index{overlap equivalence}
(\textsc{i}) non-triviality of the spin overlap,  \index{overlap!spin}
(\textsc{ii}) overlap equivalence and
(\textsc{iii}) the existence \index{replicon}
of a non-zero replicon exponent.

When trying to decide which is the relevant theory
in the thermodynamical limit
(equivalent to the limit of infinite experimental time), 
we found that our data still supported the RSB picture, 
but here our results are not as conclusive (in principle, 
there can always appear a change of regime for much larger
sizes or much longer times, even if we see no trace of it
with our data).

There are several promising avenues for future work in 
this field. First, deciding in favour of RSB for  \index{RSB}
the nature of the spin-glass phase does not automatically
grant one full knowledge of its characteristics. Much work
is needed in order to work out some very important details, 
such as the nature of temperature chaos. \index{temperature chaos}

Perhaps the most interesting prospect for future work
is trying to reproduce, as realistically as possible,
an actual spin-glass experiment. Our non-equilibrium
simulations only considered the simplest protocol 
of isothermal aging, but many interesting effects (memory, rejuvenation, etc.)
can only be observed by varying the temperature. In this
sense, our current simulations were limited by the onset of
finite-size effects. Indeed, for temperatures close to the 
critical point, our simulated size $L=80$ is not large  \index{finite-size effects}
enough to be representative of the non-equilibrium  \index{coherence length}
physics for experimentally relevant times, since the coherence length grows too quickly.
Let us stress that although this effect is restricted to the vicinity of the
critical temperature, any experimental cooling protocol should expend
a time close to $T_\text{c}$ which is very long in the microscopic scale.\footnote{Recall
that $1$ MCS is roughly equivalent to one picosecond and that our 
$L=80$ simulations at $T=T_\text{c}$ showed finite-size effects from about $10^6$ MCS.}

Simulating even larger lattices is, therefore, a must if one wants to 
reproduce sophisticated experimental protocols. We are very optimistic 
in this sense: we are already testing a code that allows us to simulate
lattices of size $L=256$ and we also have funding for \textsc{Janus II}, 
which will not only allow us to simulate even larger systems, but also
to probe longer times, taking us fully into the experimental regime.

\addtocontents{toc}{\protect\setcounter{tocdepth}{1}}
\appendix
\titleformat{\chapter}[display]
{\bfseries\LARGE} {\filleft\MakeUppercase{\chaptertitlename}
\Huge\Alph{chapter}} {2ex} {\titlerule
\vspace{1.5ex}%
\filright}
[\vspace{1.5ex}%
]
\renewcommand{\thesection}{\texorpdfstring{\textsc{\Alph{chapter}}.\oldstylenums{\arabic{section}}}{\Alph{chapter}.\arabic{section}}}
\renewcommand{\thesubsection}{\texorpdfstring{\textsc{\Alph{chapter}}.\oldstylenums{\arabic{section}.\arabic{subsection}}}{\Alph{chapter}.\arabic{section}.\arabic{subsection}}}
\renewcommand{\thesubsubsection}{\texorpdfstring{\textsc{\Alph{chapter}}.\oldstylenums{\arabic{section}.\arabic{subsection}.\arabic{subsubsection}}}{\Alph{chapter}.\arabic{section}.\arabic{subsection}.\arabic{subsubsection}}}
\renewcommand{\theequation}{\textsc{\alph{chapter}}.\oldstylenums{\arabic{equation}}}
\renewcommand{\thefigure}{\textsc{\alph{chapter}}.\oldstylenums{\arabic{figure}}}
\renewcommand{\thetable}{\textsc{\alph{chapter}}.\oldstylenums{\arabic{table}}}


\renewcommand{\sectionmark}[1]{\markright{\textit{\Alph{chapter}.\oldstylenums{\arabic{section}}}\ --- #1}}
\chapter{Thermalisation in Monte Carlo simulations}\label{chap:thermalisation}\index{thermalisation|(}%

This appendix is intended as a reference on thermalisation 
in Monte Carlo simulations, providing some definitions that 
are used throughout this thesis. Most of the definitions are standard, 
but subsection~\ref{sec:THERM-thermalisation-PT} 
contains some methods developed in part during this
thesis.

\section{Markov chain Monte Carlo}\label{sec:THERM-Markov-chain}\index{Markov chain}\index{Monte Carlo method!dynamic}
Let us begin by recalling some definitions and results relevant to 
the theory of dynamic Monte Carlo methods (this is not 
meant to be a rigorous or self-contained account, just a quick reference,
cf.~\cite{sokal:97,rubinstein:07}).
In general, we are interested in extracting configurations from some 
probability distribution $p(\{s_\bx\})$, often of the kind
$p(\{s_\bx\}) \propto \ee^{-\beta E(\{s_\bx\})}$.
In order to simplify the notation, we denote the system
configuration by a single random variable $X$ and our target 
probability distribution by $P_x=\mathcal P(X=x)$.

Then we define our Monte Carlo method as a random walk in configuration
space, that is, a sequence $X_t$ of random variables. The first one ---the starting 
configuration--- will follow some simple  probability distribution $P^{(0)}_x$
(completely random spins, for instance). We need the random walk 
to reach a stationary state such that  $\lim_{t\to\infty} P^{(t)}_x=P_x$. 

We consider a Markov chain: the probability distribution of $X_{t+1}$ depends
only on $X_{t}$ and not on the whole previous history. Then, we can 
specify the whole process by giving the initial distribution $P^{(0)}_x$
and a transition matrix $U_{xy}$. The latter is defined as 
\begin{equation}
U_{xy}=\mathcal P(X_{t+1}=y| X_{t}=x),
\end{equation}
that is, the probability that the system will be in the configuration
$y$ at time $t+1$ if it was at $x$ at time $t$.  Of course, we can iterate this 
definition
\begin{equation}
[U^n]_{xy} = \mathcal P(X_{t+n}=y| X_{t}=x).
\end{equation}
The objective is that, no
matter the initial distribution $P^{(0)}_x$, after a sufficiently long
number of iterations we obtain configurations distributed 
according to $P_x$. In other words
\begin{equation}\label{eq:THERM-stationary}
\lim_{n\to \infty} [U^n]_{xy} = P_x.
\end{equation} 

We need two ingredients for this
\begin{itemize}
\item First, we need our method to be capable of exploring the whole \index{irreducibility}
configuration space. To this end, we require the transition matrix to be \emph{irreducible}:
for all $x,y$ there exists an $n$ such that $[U^n]_{xy}>0$.
\item Naively, we could think of requiring that,
once we have reached the stationary distribution $P_x$,  the Markov process
should lose all sense of direction and be reversible. We should not be able to 
tell if it is running backwards or forwards. That is
\begin{equation}
P_x U_{xy}  = P_y U_{yx}.
\end{equation}
This is called the \emph{detailed balance} condition. \index{balance condition!detailed}
Actually, it turns out that this is too restrictive. We only need that the stationary
distribution of the Markov chain correspond to $P_x$, that is
\begin{equation}
P_y = \sum_x P_x U_{xy}.
\end{equation}
This is the \emph{balance condition}.
\end{itemize}
If $U_{xy}$ satisfies these two conditions, then~\eqref{eq:THERM-stationary} can be proved
and we have a legitimate Monte Carlo method.  Notice that the balance
condition is easier to check than it seems, since it does not 
depend on the normalisation of the $P_x$ (typically given by the unknown partition
function $Z$). \index{partition function}

Perhaps the easiest example of Monte Carlo dynamics is the heat bath. \index{heat bath algorithm}
Let us consider the update of a single  Ising spin at site $\bx$,  $s_\bx \to s_\bx'$, where
all the remaining spins in the lattice are kept fixed. Then, we consider the following
update probability
\begin{equation}
\mathcal P(X_{t+1}=\{s'\} | X_t=\{s\}) = \frac{\ee^{-\beta E(\{s'\})}}{\sum_{s''_\bx=\pm1} \ee^{-\beta E(\{s''\})}},
\end{equation}
Notice that all the terms in the energy that do not involve site $\bx$ factor
out. Therefore, it is obvious that the update of a single spin satisfies 
the detailed balance condition. In order to define a full Monte 
Carlo method we need to update all the spins. We can do this
by randomly choosing one spin each time and defining each Monte Carlo
step as $N$ spin updates (so that each spin in the lattice is updated
once, on average). In this case the whole transition matrix would also
satisfy detailed balance. We can also consider a scheme where 
we run through the lattice sequentially, visiting all the sites
in order. In this case, the process would clearly not be reversible, 
and the transition matrix would only satisfy balance.

\section{The autocorrelation times}\label{sec:THERM-tau}
As a general rule, the thermalisation of a Monte Carlo simulation
(i.e., how long it takes
to reach the stationary distribution)
should be 
\index{autocorrelation time}
assessed through the temporal autocorrelation functions~\cite{sokal:97}.
Let $O$ be an observable, defined as always as a real-valued function on
the configuration space, and assume we have equilibrated the system
for a very long time, so we have reached the stationary regime.
Then, we consider the equilibrium (stationary) evolution $O(t)$ and define
the equilibrium autocorrelation functions
\index{correlation function (equilibrium)!temporal}
\begin{equation}\label{eq:THERM-C-O-t}
C_O(t) =\bigl\langle[O(0)-\langle O\rangle][O(t)-\langle O\rangle]\bigr\rangle   ,\qquad
\rho_O(t) = \frac{C_O(t)}{C_O(0)}\ .
\end{equation}
The angle brackets here denote any kind of thermal average (e.g., canonical
or tethered).
The autocorrelation function decays as an exponential 
for long times, 
\begin{equation}
\rho_O(t) \xrightarrow{\ \ t\to\infty\ \ } A \exp[-t/\tau].
\end{equation}
This asymptotic behaviour suggests the definition of 
the exponential autocorrelation time,
\begin{align}
\tau_{\text{exp},O} &= \lim_{t\to\infty} \sup \frac{t}{-\log|\rho_O(t)|},\\
\tau_\text{exp} &= \sup_{O} \tau_{\text{exp},O} .
\end{align}
In general, we express the autocorrelation as a sum  of exponentials\footnote{%
Actually, this is only strictly true if the dynamics fulfils detailed balance \index{balance condition!detailed}
and is not only irreducible but aperiodic. \index{irreducibility} \index{aperiodicity}
Otherwise, there can be some modes in the form of damped oscillations.
However, these are never observed, within errors, in the applications
we consider in this dissertation.}
\begin{equation}\label{eq:THERM-rho-exponentials}
\rho_O(t) = \sum_i A_i \ee^{-t/\tau_i}.
\end{equation}
The exponential time, then, is the largest of the $\tau_i$ and characterises
the relaxation time of a certain observable. Actually, barring symmetry
considerations, the exponential time is the same for all observables.

Another useful concept is that of integrated autocorrelation time
\begin{equation}\label{eq:THERM-tau-int}
\tau_{\text{int},O}= \frac12 + \sum_{t=1}^\infty \rho_O(t),
\end{equation}
The integrated time indicates the minimum time difference so that
two measurements of some observable $O$ can be considered
independent (i.e., uncorrelated). In particular, if $\Var(O)$ is 
the variance of a single measurement of $O$ (after we have reached
equilibrium) and we average over $\mathcal N$ successive such measurements, 
the resulting  variance is
\begin{equation}
\Var\bigl(O^{(\mathcal N)}\bigr) = \frac{\Var(O)}{\mathcal N /(2\tau_{\text{int},O})},
\end{equation}
as opposed to the value $\Var(O)/\mathcal N$ that we would 
obtain for $\mathcal N$ statistically independent measurements.

The integrated time is less useful as a measure of thermalisation than
the exponential. In fact, notwithstanding the previous discussion, 
the determination of $\tau_{\text{int},O}$ 
is not even always needed to estimate statistical errors 
(see Appendix~\ref{chap:correlated}). It has, however, the considerable
advantage of being much easier to measure safely.
Notice  as well that, if the decomposition \eqref{eq:THERM-rho-exponentials}
contains a single exponential, $\tau_{\text{exp},O} = \tau_{\text{int},O}$,
This approximation can often be employed. Alternatively, if we
do not take measurements for every single MCS or, better, if we make
bins of consecutive measurements, we can neutralise the faster modes
in~\eqref{eq:THERM-rho-exponentials} and approach \index{binning}
the single-exponential limit. By `binning' we understand
\nomenclature[Binning]{Binning}{Averaging consecutive measurements 
of an observable in \\ blocks 
of either constant or geometrically growing length}
averaging groups of $n$ consecutive measurements. In this 
case, all the components with $\tau_i \ll n$ dissappear.
The same approach is taken in a different context, that 
of random variables strongly correlated in space, close
to a critical point (Kadanoff-Wilson blocks).
 
\subsection{Some computational recipes}\label{sec:THERM-recipes}
For a large-scale MC simulation, where we have to handle very long
time series (millions of time steps even after the binning procedure
mentioned above), the computation of $\rho_O(t)$ can be time-consuming.
Indeed, a naive implementation of definition~\eqref{eq:THERM-C-O-t}
requires $\mathcal O(\mathcal N^2)$ operations.

We can do better. For simplicity, and without loss of generality, 
let us consider an observable such that $\langle O\rangle=0$,
for which we take $\mathcal N$ measurements $O(t)$, $t=0,\ldots,\mathcal N-1$.
We also assume all this measurements have already been taken in the
equilibrium regime (i.e., we have discarded a sufficient number of 
measurements for thermalisation). Then we could estimate $C_O$ as\footnote{%
We use the notation $[O]$ for the numerical estimator
of some thermal average $\langle O\rangle$.} 
\begin{equation}\label{eq:THERM-C-O-est}
[C_O(t)] =\frac{1}{\mathcal N-t} \sum_{s=0}^{\mathcal N-1-t} O(s)O(s+t).
\nomenclature{$[O]$}{Numerical estimator for $\langle O\rangle$}
\end{equation}
Let us define
\begin{equation}\label{eq:THERM-S-O}
S_O(t) = (\mathcal N-t) C_O(t)
\end{equation}
and
\begin{equation}\label{eq:THERM-O-prime}
O'(t) =\begin{cases}
 O(t), & 0\leq t< \mathcal N,\\
0,     & \mathcal N \leq t < 2 \mathcal N,
\end{cases}
\end{equation}
and we consider $O'(t)$ to be extended periodically.

Then we can write our numerical estimate for the
unnormalised correlation of $O'$ as 
\newcommand{\Corr}{\mathrm{Corr}}
\begin{equation}\label{eq:THERM-S-O-prime}
[S_{O'}(t)] = \sum_{s=0}^{2\mathcal N-1} O'(s) O'(s+t)
\end{equation}
Notice that
\begin{equation}
[S_{O'}(t)]= [S_{O}(t)].
\end{equation}
Now consider the Fourier transform of $S_{O'}(t)$,
\index{Fourier transform}
\begin{align}
[\hat S_{O'}(q)] &= \sum_{t=0}^{2\mathcal N-1}\sum_{s=0}^{2\mathcal N-1} O'(s) O'(s+t) \ee^{-\ii qt}\\
\intertext{%
Notice that, 
since we are summing $t$ over a whole period of the function,
we can change $s+t\to t'$ in the second sum
}
[\hat S_{O'}(q)]&=\sum_{s=0}^{2\mathcal N-1} O'(s) \sum_{t'=0}^{2\mathcal N-1} O'(t') \ee^{\ii qs}\ee^{-\ii qt'}\\
               &= \biggl(\sum_{s=0}^{2\mathcal N-1} O'(s) \ee^{\ii qs} \biggr) \biggl(\sum_{t'=0}^{2\mathcal N-1} O'(t') \ee^{-\ii q t'}\biggr)
               =| \hat O(q) |^2.
\end{align}
That is, the Fourier transform of the autocorrelation is equal 
to the modulus of the Fourier transform of $O'$ (we have needed a periodic $O'$
to get this result). This is the Wiener-Khinchin theorem.

Therefore, we can obtain $S_{O'}(t) =S_{O}(t) = (\mathcal N-t) C_O(t)$
by computing the Fourier transform of $O(t)$, evaluating
its complex modulus and then computing the inverse Fourier 
transform of $|\hat O(q)|^2$. It may seem that we have gained 
nothing. However, the computation of discrete Fourier transforms, 
seemingly an $\mathcal O(\mathcal N^2)$ task,  can actually
be done in $\mathcal O(\mathcal N\log\mathcal N)$ operations
using the Fast Fourier Transform algorithm (see, e.g., \cite{press:92}). \index{FFT}
In the work reported herein we have always
followed this method to evaluate temporal 
autocorrelation functions.  We use the FFTW implementation
reported in~\cite{frigo:05}.

Notice that an analogous method can be followed to evaluate 
spatial correlations, as we will have ample occasion to 
do (especially in Part~\ref{part:sg}). In this case, even 
if the functions are three-dimensional, the procedure is even 
simpler, as the periodic boundary conditions are already built in.
Therefore, we do not have to pad our observables with zeros,
as in~\eqref{eq:THERM-O-prime}.

Once we have computed the $\rho_O(t)$ we need to estimate 
the autocorrelation times. For the exponential ones 
our only recourse is a fit to one or more
exponential modes (see below). The integrated time
must, in principle, be computed from the sum
of our numerical estimate $[\rho_O(t)]$ extended to all our values of $t$.
However, as is clear from our definition~\eqref{eq:THERM-C-O-est},
the final times have much smaller statistics, 
so their contribution is nothing but a white noise.
Therefore, $\tau_{\text{int},O}$ 
is best evaluated using a self-consistent window~\cite{sokal:97}
\begin{equation}\label{eq:THERM-self-consistent-window}
\tau_{\text{int},O} = \frac12 + \sum_{t=1}^{\varLambda} [\rho_O(t)],
\end{equation}
where  $\varLambda$ is chosen as $\varLambda=W\tau_{\text{int},O}$,
with $W$ a tunable parameter ($W=6$ works well).

\section{Thermalisation in disordered systems}\label{sec:THERM-disorder}
The safe computation of the autocorrelation times described in the previous
section requires a simulation time orders of magnitude larger than $\tau_\text{exp}$.
This is not a problem when simulating ordered systems, because extending the simulation
reduces the statistical error of the final results. With disordered systems, however,
the main source of statistical error is the sample-to-sample fluctuation. Reducing the 
thermal error for each sample beyond a certain, easily reachable, threshold is 
useless.  

Therefore, given a fixed computational budget, we would like to simulate 
each sample for the shortest possible time that ensures thermalisation.\footnote{%
Notice that this not only would allow us to maximise the number of samples
we can average for a fixed total CPU time, but, assuming we can run many samples
at the same time, would also minimise the wall-clock time (the physical
time we have to wait for the results).\index{wall-clock}}
Since so short a run time does not permit a computation of the $\tau_O$ for 
physical observables, researchers have traditionally employed less rigorous 
methods to check thermalisation.

In general, the typical method is to study the time evolution of the disorder-averaged 
physical observables. In particular, one of the most widespread practical recipes is
the so-called $\log_2$-binning. \index{log2 binning@$\log_2$-binning}
One divides the simulation time $\mathcal N$ in logarithmic intervals $\mathcal I_n$:
\begin{equation}
\mathcal I_n = (2^{-n+1} \mathcal N,\ 2^{-n}\mathcal N],
\end{equation}
Notice that with this definition time decreases with increasing $n$, so $\mathcal I_0$ corresponds 
to the second half of the run, $\mathcal I_1$ to the second quarter, etc.
Then, one evaluates the thermal average for each sample and the disorder average, for 
each of the time intervals. If the last few $\mathcal I_n$ show no evolution,
the simulations are considered to be thermalised.

This procedure is not optimal, because in many situations the thermalisation time is wildly
dependent on the sample (see Figures~\ref{fig:DAFF-histograma-hatm-012} and~\vref{fig:DAFF-histograma-hatms-08}
for the DAFF or Figures~\ref{fig:SG-tau-32} and~\vref{fig:SG-tau-24} 
\index{DAFF}
\index{spin glass}
for the Edwards-Anderson spin glass). Thus, if we choose a simulation time that would thermalise
even the slowest samples, we will spend most of our time extending already well thermalised 
runs for no benefit. Worse, a stationary average over many samples may hide the fact that a few samples
are still quite far from equilibrium (the thermalisation time can vary along several orders
of magnitude from one sample to another).

One could toughen the $\log_2$-binning test by taking the correlations into account~\cite{fernandez:07}. 
Indeed, since the samples considered for each $\mathcal I_n$ are always the same, the difference
between the resulting averages (even assuming complete temporal decorrelation) will typically
be much smaller than their statistical errors. This is because, as we said earlier, the thermal fluctuations
account for an almost negligible part of the total statistical error. 
We can take this into account by computing for each sample the differences between the 
average for $\mathcal I_0$ and each of the $\mathcal I_n$ and considering the evolution
of these correlated differences, which have a much smaller error 
(see Figure~\ref{fig:DAFF-log2}). To be sure, this is a more
stringent test, but ultimately an unphysical one (because it considers much smaller errors
than those of the physical results). Furthermore, it does not address the important 
problem of efficiency.

In the next section we shall see how we can provide a more rigorous method, taking advantage 
of the parallel tempering algorithm used to accelerate the thermalisation.

\subsection{Parallel tempering}
\label{sec:THERM-thermalisation-PT}
\index{parallel tempering|(}
Parallel tempering~\cite{hukushima:96,marinari:98b}
is one of the most common methods to accelerate thermalisation. Its motivation is the 
rugged free-energy landscape picture described in the General Introduction. \index{free energy!landscape}
Let us consider a system described in
the canonical ensemble by the following partition function
\index{partition function}
\begin{equation}
Z(\beta) = \sum_{\{s_\bx\}} \ee^{-\beta E}.
\end{equation}
At a physically interesting low temperature the system is trapped in deep free-energy valleys,
with extremely long escape times. At a higher temperature, however, the energy fluctuations 
are enough to overcome these barriers quickly. \index{disorder!quenched}

We can take advantage of this by simulating $\mN_T$ copies of the system
(in our case, $\mN_T$ starting configurations with the same disorder realisation).
Each copy starts the simulation at a different temperature, with $T_1<T_2<\ldots<T_{\mN_T}$.
We assume that $T_1$ is such that our chosen Monte Carlo dynamics would need
an impossibly long time to thermalise the system, while
$T_{\mN_T}$ is high enough for the same dynamics to achieve thermalisation
very quickly.  After updating each copy independently 
for one or more MCS we perform the parallel tempering update.

Then, we consider a configuration space with $\mN_T(V+1)$ degrees of freedom,
$\{T^{(i)}, \{s_\bx^{(i)}\}\}$. The temperature associated to copy 
$i$ of the system can change, but the ensemble of temperatures
$\{\beta^{(i)}\}$ must always be a permutation of
\begin{equation}
\{\beta_1,\ldots,\beta_{\mN_T}\} = \{1/T_1,\ldots,1/T_{\mN_T}\}.
\end{equation}
Now,  \index{partition function}
the partition function of the whole ensemble of systems is 
\begin{equation}\label{eq:THERM-Z-PT}
Z_{\mN_T} = \frac{1}{\mN_T!} \prod_{i=1}^{\mN_T} Z(\beta^{(i)}).
\end{equation}
We can compute the marginal distributions for the $\beta^{(i)}$. Summing
over the spins, we find that the marginal probability
density of the temperature configuration $\{\beta^{(i)}\}$ is
uniform,
\begin{equation}\label{eq:THERM-beta-ensemble}
p(\{\beta^{(i)}\}) = 1/(\mN_T!).
\end{equation}
Moreover, the marginal
probability for a single temperature is also uniform,
\begin{equation}\label{eq:THERM-beta}
p(\beta^{(i)})=1/\mN_T.
\end{equation}

The partition function~\eqref{eq:THERM-Z-PT} allows us to consider
a Monte Carlo dynamics where we simply try to exchange 
copies $i$ and $j$ of the system, initially at neighbouring temperatures
$T_{k}$ and $T_{k+1}$,
\begin{equation}
\begin{split}
\{\beta^{(i)} = \beta_k, \{s_\bx^{(i)}\}\}& \times\{\beta^{(j)} = \beta_{k+1}, \{s_\bx^{(j)}\}\}\ \longrightarrow\\
&\longrightarrow\ \{\beta^{(i)} = \beta_{k+1}, \{s_\bx^{(i)}\}\} \times\{\beta^{(j)} = \beta_{k}, \{s_\bx^{(j)}\}\}.
\end{split}
\end{equation}
The Metropolis acceptance probability (see Section~\ref{sec:tmc}) \index{Metropolis algorithm}
for this dynamics is
\begin{equation}\label{eq:THERM-acceptance-PT}
\mathcal P =  \min\{1, \omega_\text{new}/\omega_\text{old}\} =
\min\bigl\{ 1, \exp\bigl[ - (\beta_{k+1}-\beta_k)(E^{(j)}-E^{(i)})\bigr]\bigr\}.
\end{equation}
In order for this dynamics to be irreducible, \index{irreducibility}
a full parallel temperature update consists in attempting 
one exchange per temperature, going sequentially from the lowest to the highest
(so one configuration can actually jump several temperatures in a single update).

This way, the temperature of each copy performs a random walk 
in temperature space. Once the system reaches a high enough 
temperature, a few MCS are enough to make it cross the free-energy 
barriers so when it eventually returns to its original temperature 
it has been decorrelated from the starting configuration.

Notice that for this algorithm to work, we need to choose the 
participating temperatures close enough for their energy
histograms to overlap, so the acceptance probability~\eqref{eq:THERM-acceptance-PT}
is high enough.

\subsection{The temperature random walk}
In equilibrium, each copy should spend the same time at each temperature (recall~\eqref{eq:THERM-beta-ensemble}), 
so studying the temperature random walk can give information
about the thermalisation. In particular, several
round trips from lowest to highest temperature are needed 
for a system to be considered thermalised.
\begin{figure}[t]
\includegraphics[height=\linewidth,angle=270]{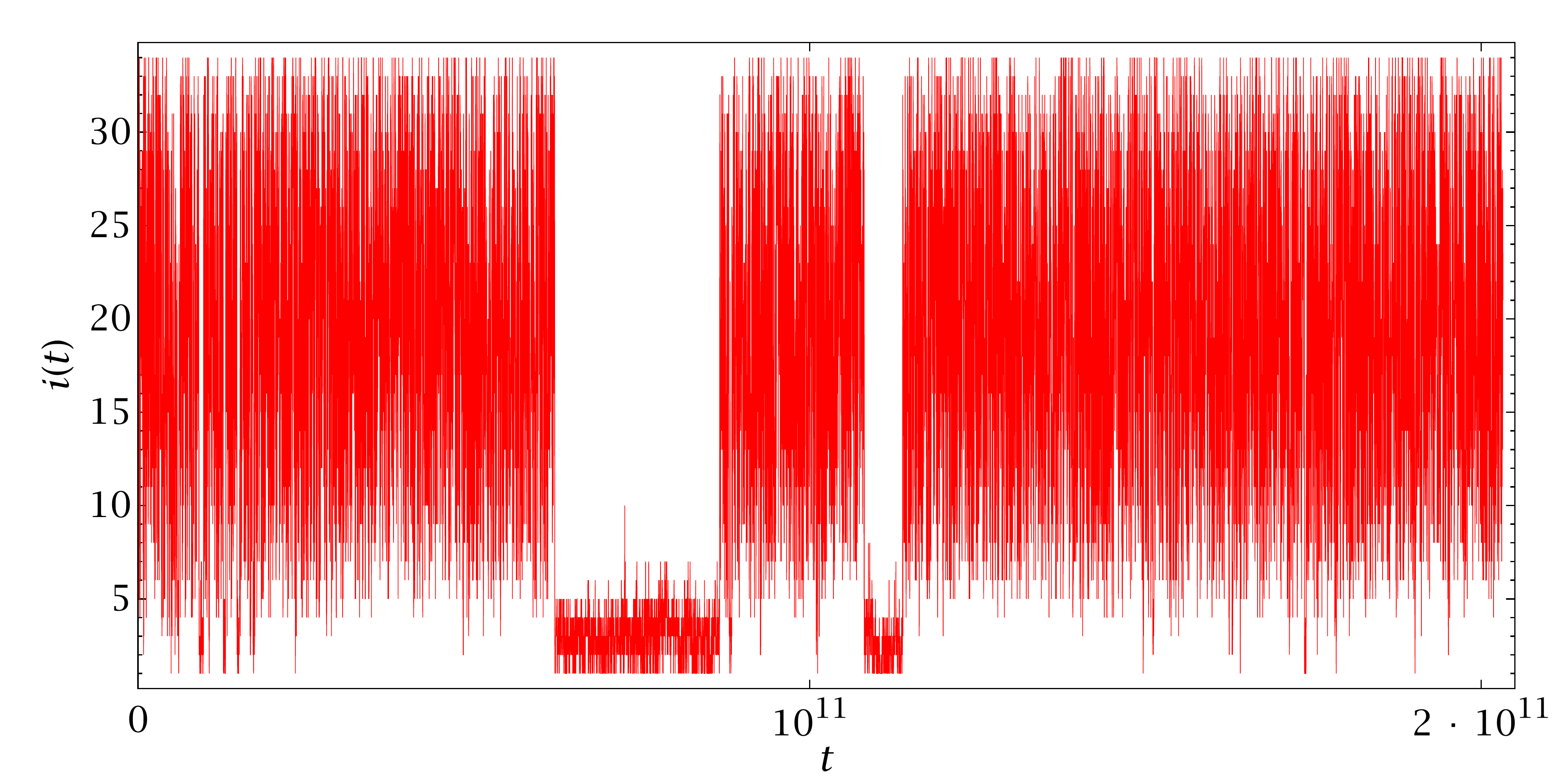}
\caption[Temperature random walk in a parallel tempering simulation]{Temperature random walk of a single configuration in one
of our spin-glass simulations of  Appendix~\ref{chap:runs-sg}.
We plot the temperature index $i(t)$ against the time measured
in heat bath updates. \index{heat bath algorithm}
The critical temperature corresponds to $i_\text{c}=17$ (in the middle
of the range).}
\label{fig:THERM-random-walk}
\index{temperature random walk|indemph}
\end{figure}

We can promote this idea to a fully quantitative and physically meaningful
level.\footnote{The method described
herein was first introduced in~\cite{fernandez:09b}
and we later refined it in~\cite{janus:10}.} Let us assume that 
our range of temperatures crosses the critical point and let 
us define $i_\text{c}$ such that
\begin{equation}
T_{i_\text{c}-1} < T_\text{c} < T_{i_\text{c}}.
\end{equation}
Now, for one of the $\mN_T$ copies of the system, we consider its
temperature random walk mapped to the range of the temperature 
index $i(t)\in \{1,\ldots,\mN_T\}$. Figure~\ref{fig:THERM-random-walk}
shows one example, taken from our spin-glass simulations.
Notice that the random walk is clearly not Markovian, as evinced
by the long plateaux, since the system
retains for a long time a memory that it started in the high (or low)
temperature phase.

Our goal is to construct a quantity that is both representative 
of the random walk and amenable to the computation of 
autocorrelation times described in the previous section.
To this end, we consider a mapping $f(i)$ such that
\begin{align}
f(i) &\geq 0, & \forall i \geq i_\text{c},\label{eq:THERM-cond1}\\
f(i) &< 0, & \forall i < i_\text{c},\label{eq:THERM-cond2}\\
\sum_{i=1}^{\mN_T} f(i) &= 0. \label{eq:THERM-sumf}
\end{align}
Recalling~\eqref{eq:THERM-beta}, we see that the last 
condition is just $\braket{f}=0$.
It is also convenient that $f$ be monotonic. Generally, one 
chooses the temperatures so that $i_\text{c}$ is in 
the middle of the range, so a simple linear $f$ 
suffices.

Now we can compute the autocorrelation $C_f(t)$
with Eq.~\eqref{eq:THERM-C-O-est}. Notice 
that $f$ has an advantage over the $C_O$ of the physical 
observables: we can average the correlations
over the $\mN_T$ copies in the parallel tempering. 
These are not, of course, completely independent 
(the most obvious constraint being that each must be at 
a different temperature), but still the averaged
correlation has a much smaller variance. Furthermore, in the 
case of spin glasses one simulates several statistically 
independent replicas, which provide an additional averaging
layer and allow an estimate of the statistical error.
\begin{figure}[p]
\includegraphics[height=\linewidth,angle=270]{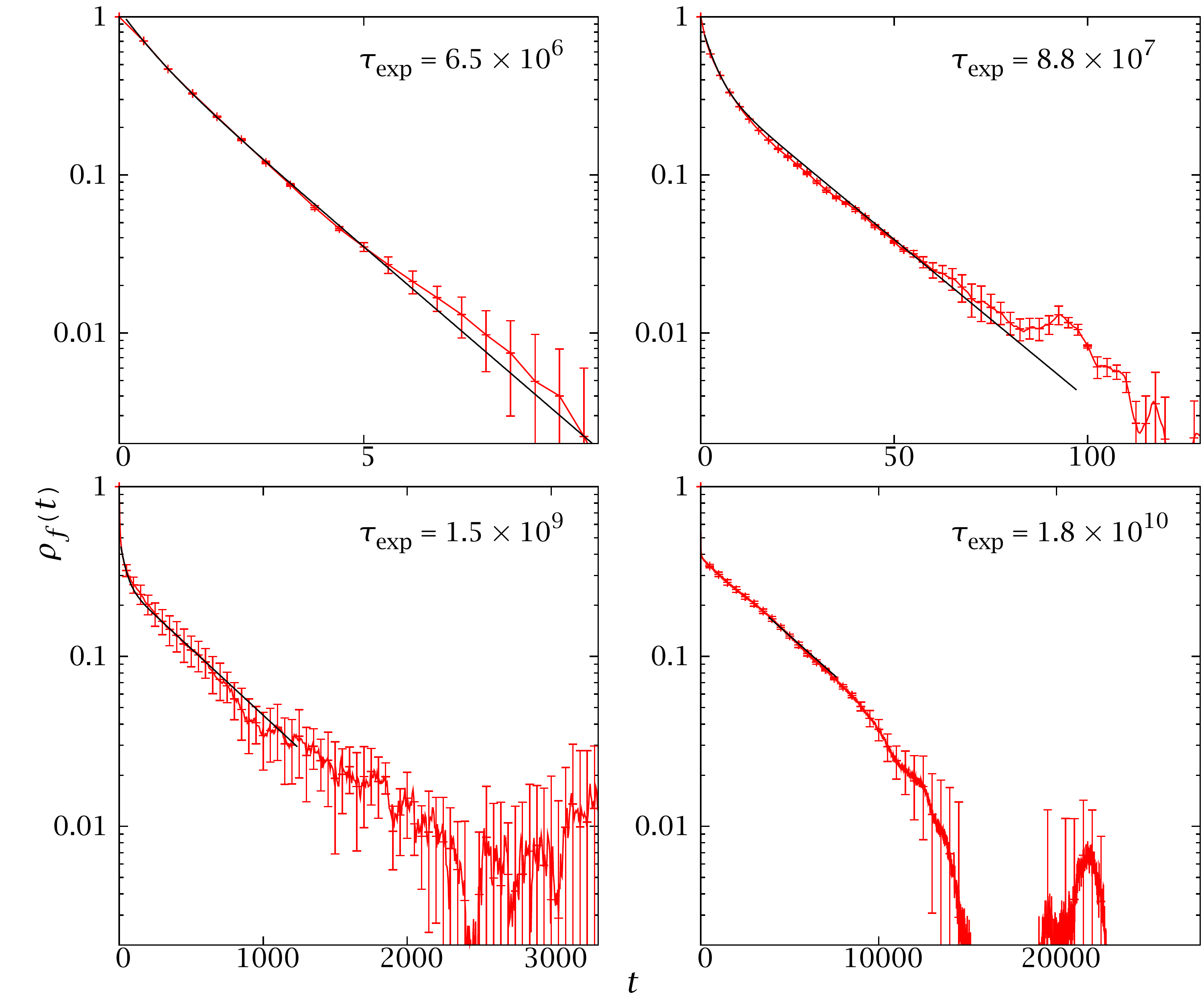}
\caption[Autocorrelation functions for spin-glass simulations]{Autocorrelation functions for some of our  equilibrium
spin-glass simulations of Appendix~\ref{chap:runs-sg}, 
showing samples with $\tau_\text{exp}$ across several
orders of magnitude, plotted against the simulation
time in units of $10^6$ MCS. We have plotted the range $[0,6\tau_\text{exp}]$
and also show the result of the automated fitting procedure
described in Section~\ref{sec:THERM-tau-fit}. In order 
to avoid cluttering the graphs we have only plotted a few times
(the actual correlation functions have many more points). In the last
panel the correlation function has a strong downwards fluctuation. Our fitting
code takes this into account by choosing a restricted range and fitting to 
a single exponential.}
\label{fig:THERM-corr-SG}
\index{correlation function (equilibrium)!temporal|indemph}
\end{figure}

As explained in Section~\ref{sec:THERM-tau}, the thermalisation 
of the system is characterised by the exponential autocorrelation 
time. Recall that, barring symmetry considerations, the exponential
time should be the same for all random variables in the simulation, 
including the physical observables. In general, the exponential 
autocorrelation time must be computed from a fit to an equation
such as~\eqref{eq:THERM-rho-exponentials}, with an appropriate
number of exponential modes. This is not a problem for ordered systems
simulations, where one can simply select the fitting range and functional
form manually. When simulating disordered systems, however, one would 
have to do thousands of such fits, each with a different fitting range
(in our study of the DAFF, for instance, we performed several 
hundred thousands independent parallel tempering simulations).

Clearly, an automated system must be found. The first
step is making the $C_f(t)$ as simple as possible. In this 
respect we notice that the choice of $f$ should not 
modify the exponential autocorrelation time, but will 
affect the relative sizes of the $A_i$. Our criteria~\eqref{eq:THERM-cond1}
and~\eqref{eq:THERM-cond2} hopefully select a family of 
functions that reduce the amplitude of the irrelevant fast modes.
A second simplification is afforded by the sheer length of the simulations
in practical cases. From a look at Tables~\ref{tab:DAFF-parametros-PT}
and~\ref{tab:SG-parameters-eq} we see that the runs can be as 
long as $10^{12}$ MCS. In this situation, averaging the $f\bigl(i(t)\bigr)$
over a large number of consecutive measurements implies no real
information loss (as long as the bins are much smaller than $\tau_\text{exp}$).
This binning procedure also has the welcome side effect of suppressing
the fastest modes.

With these considerations in mind, in practice it turns out that 
the $\rho_f(t)$ can be adequately described by considering just 
two exponential modes. As an example of this, we have plotted in 
Figure~\ref{fig:THERM-corr-SG} several autocorrelation 
functions from our spin-glass simulations. The depicted
correlation functions have been chosen randomly, with the only condition
that they represent samples with very different 
exponential autocorrelation times. 

\subsection{Thermalisation protocol for parallel tempering simulations}\label{sec:THERM-tau-fit}
We can summarise the previous discussion in the following three-step 
thermalisation procedure
\begin{enumerate}
\item Simulate for a minimum length of $\mN^{\text{min}}$ MCS, chosen 
to be enough to thermalise most samples. Notice that most published
parallel tempering studies stop here, assessing thermalisation only
through the evolution of disorder-averaged observables.
\item Discard the first sixth of the measurements and compute
the $\rho_f(t)$ from the remainder. Compute the integrated autocorrelation
time with the self-consistent window method of Eq.~\eqref{eq:THERM-self-consistent-window}, 
using $W=6$. Using this first estimate of the integrated time, we extend
the simulation until $\mN > A^\text{int} \tau_\text{int}$, always discarding
the first sixth. This criterion was introduced  in~\cite{fernandez:09b}.
\item Now that the simulation is reasonably dimensioned, estimate $\tau_\text{exp}$
(typically of the same order of magnitude as $\tau_\text{int}$, but always bigger).
Extend the simulations until $\mN > A^\text{exp} \tau_\text{exp}$.
\end{enumerate}
The actual thresholds $A^\text{int}$ and $A^\text{exp}$ are somewhat arbitrary and 
should be tuned to the system at hand (especially $A^\text{int}$, since the integrated
time, unlike the exponential one, depends on the choice of $f$ and the frequency 
of its measurements).

We introduced the third step in this protocol for our spin-glass 
simulations in~\cite{janus:10}, choosing $A^\text{int}=22$,
$A^\text{exp}=12$ (see \ref{sec:THERM-DAFF} for our criteria for the DAFF).
Notice that 
this implied performing non-linear fits with custom ranges for some $10^4$
autocorrelation functions. The first step was parameterising $\rho_f(t)$ 
by a double exponential decay (an assumption empirically justified by the shape
of these correlation functions, cf. Figure~\ref{fig:THERM-corr-SG}).
\begin{equation}\label{eq:THERM-two-exps}
\rho_f(t) \simeq A_1 \ee^{-t/\tau_1} + A_2\ee^{-t/\tau_2},\qquad \tau_1 = \tau_\text{exp} > \tau_2.
\end{equation}
The choice of fitting range of course depends on the values of the $\tau_i$, which
vary across several orders of magnitude from one sample to another. However, notice
that, if $A_2=0$, then $\tau_\text{exp} = \tau_\text{int}$. Therefore, $\tau_\text{int}$
is a good starting assumption for $\tau_\text{exp}$, so we took the following
steps:
\begin{enumerate}
\item[(\textsc{i})] Perform a first fit to a single 
exponential in the range $[2\tau_\text{int}, 3\tau_\text{int}]$, yielding
an amplitude $A$ and a time $\tau$.
\item[(\textsc{ii})] Using $\tau_1=\tau$, $\tau_2=\tau/10$, $A_1=A$ and 
$A_2=1-A$ as a starting point, perform a non-linear fit
to~\eqref{eq:THERM-two-exps}, with the fitting range $[\tau_\text{int}/10,10\tau_\text{int}]$.
\end{enumerate}
In most cases we can stop here. Sometimes, however, the second fit fails ---typically because 
one of the amplitudes is very small, so $\rho_f(t)$ is indistinguishable from a single exponential.
This can be detected, and solved, in a number of ways:
\begin{enumerate}
\item[(\textsc{iii}.a)] One of the $A_i$ is $A_i\gg1$, $\tau_1$ has an absurdly 
large value (indicating a fit to a short plateau in a noisy range) or the 
iterative fitting method breaks down. In these cases, a single exponential fits the data
better. We perform a fit to a single exponential in the range $[5\tau_\text{int},10\tau_\text{int}]$.
\item[(\textsc{iii}.b)] One of the $A_i$ is negative, indicating an unphysical downwards
fluctuation in the $\rho_f(t)$ for large times (cf. bottom-left panel of Figure~\ref{fig:THERM-corr-SG}).
We perform a fit to a single exponential in the reduced range $[2.5\tau_\text{int},3.5\tau_\text{int}]$.
\end{enumerate}
\subsubsection{Fail-safe mechanisms}\label{sec:THERM-fail-safe}
\begin{figure}[h]
\includegraphics[height=\linewidth,angle=270]{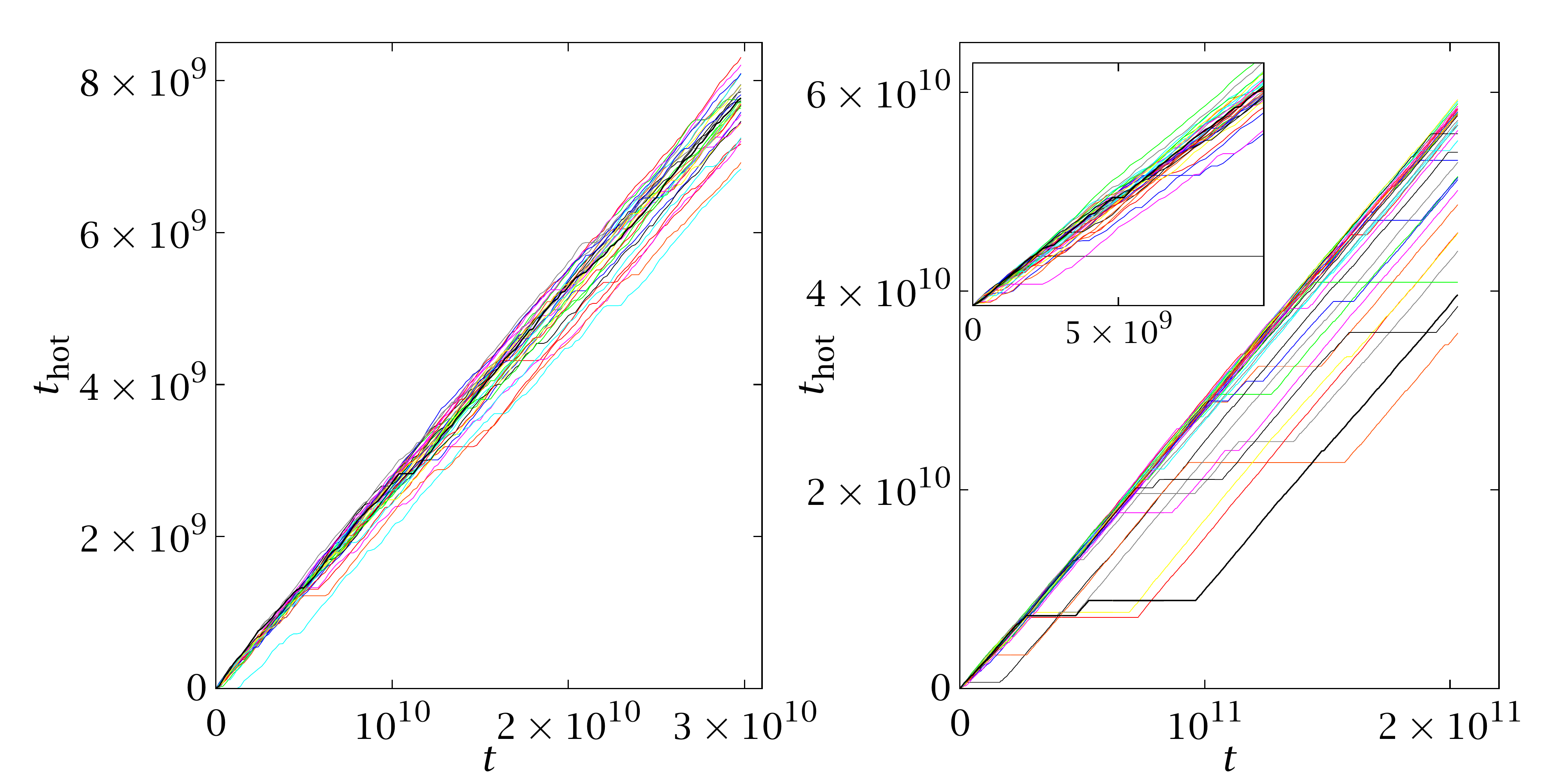}
\caption[Time spent above the critical temperature]{Time spent above the critical temperature against total simulation
time for two spin-glass samples. In each case we plot the $\mN_T$ copies
of the system. The left panel shows an `easy' sample, where the random walk 
is relatively fluid and all configurations progress gradually. The right
panel shows a `hard' sample, where the configurations get trapped in local
minima at low temperatures, producing long plateaux (notice the scales
in both plots). In a very short run time (inset), only one of the  copies
of the second system would betray its hard character.}
\label{fig:THERM-thot-SG}
\end{figure}

This automatic and fully quantitative procedure works for most samples, but there are some potential
pitfalls that can lead to our underestimating $\tau_\text{exp}$. For instance, 
the exponential time may sometimes be much larger than $\tau_\text{int}$. In these
situations, $A_1$ is very small and it is very difficult to fit the data properly.
Therefore, if our fitting method detects $\tau_\text{exp}>10\tau_\text{int}$, we increase
the measurement bins by a factor of $10$, making both $\tau_\text{int}$ and $A_1$ grow
and easing the fitting procedure.

More generally, we have assumed that our $\rho_f(t)$ has been measured in equilibrium, or 
close to it (i.e., that our simulation is at least a couple of $\tau_\text{exp}$ long).
In a very short run, however, some of the configurations may not even have had time to visit
the relevant minima, and get trapped in them, producing a misleadingly fluid random walk.
In particular, this happens when some of the copies have not crossed the critical temperature
in the parallel tempering dynamics. To prevent this from happening, we also measure $t_\text{hot}$, 
defined as the time that each copy has spent above the critical temperature. In case
one of the copies has a $t_\text{hot}$ smaller than a third of the median one, we consider
that the simulation time is woefully underestimated and we simply double it.  Notice that, 
unlike our other criteria, this is not completely quantitative. As a general rule, 
if many samples only raise the alarm because of the $t_\text{hot}$ check, we must 
conclude that our $\mN^\text{min}$ is too short and extend all the runs to a larger
minimum run time. We show two examples of $t_\text{hot}$ as a function of simulation 
time in Figure~\ref{fig:THERM-thot-SG}.

\subsubsection{The criteria for the DAFF simulations}\label{sec:THERM-DAFF}\index{DAFF!thermalisation}
The procedure described above for the computation of the $\tau_\text{exp}$ must be 
tuned for the physical system we want to study.
For example, consider the case of the DAFF simulations discussed in Chapter~\ref{chap:daff-tethered}. 
In contrast with our spin-glass simulations, which probed deep into the low-temperature region, 
these were restricted to the neighbourhood of the critical point. As a consequence, the resulting
$\rho_f(t)$ were generally simpler than those of the spin glass. Indeed, most 
of the time we could only see one exponential mode, so $\tau_\text{int}\simeq\tau_\text{exp}$.
 On the other hand, we only
simulated one replica, so we could not easily estimate the error in $\rho_f(t)$. This complicates
the evaluation of $\tau_\text{exp}$ (making it difficult to distinguish 
fluctuations from real effects), especially for the shorter runs, which have smaller
statistics.

To take this into account, we modified our criteria to give more weight
to the integrated time. We required a minimum run length of $100\tau_\text{int}$
(i.e., we chose $A^\text{int}=100$ in step \textsc{ii} above). In this way, even 
if $\tau_\text{exp}\simeq10\tau_\text{int}$, an upper bound in our
experience, we would still have a run length 
of about $10\tau_\text{exp}$.

\index{parallel tempering|)}
\index{thermalisation|)}

\chapter{Analysing strongly correlated data}\label{chap:correlated}
In a Monte Carlo computation, one has to deal both with systematic
and a statistical error sources. Furthermore, the strong correlation 
between the data makes the estimation of this latter error 
delicate. In this Appendix we present some recipes to handle
correlated data.

\section{Computing thermal averages from correlated data}\index{error analysis|(}
We start by resuming an issue introduced in Appendix~\ref{chap:thermalisation}:
how to estimate the error in the average of correlated measurements.
The methods contained in this section are well known, to the point
of being included in textbooks~\cite{amit:05}. It is, however, 
convenient to present a summarised account here, if only to
fix the notation for the more sophisticated methods of succeeding
sections.

Let us suppose we have $\mN$ independent measurements $O_i$ of 
some random variable (observable) $O$.
We want to use them to estimate the thermal 
average $\braket{O}$, for which our estimator $[O]$ is naturally
\begin{equation}
[O] =  \frac{1}{\mN}\sum_{i=1}^\mN O_i.
\end{equation}
This estimator is unbiased, meaning that $\braket{[O]}=\braket{O}$.
More than that, by virtue of the central limit theorem \index{central limit theorem}
we can expect it to behave as
\begin{equation}
[O] = \braket{O} + \eta \frac{\sigma}{\sqrt\mN} = \braket{O} + \eta \varDelta_O\, , \quad \sigma = \sqrt{\text{Var}(O)},
\end{equation}
where $\eta$ is a standard Gaussian random variable. We can likewise
estimate the error $\varDelta_O$ as
\begin{equation}
[\varDelta_O^2] = \frac{1}{\mN-1} \left[ \frac1\mN\sum_i O_i^2 - 
\left(\frac1\mN\sum_i O_i\right)^2\right]\, .
\end{equation}
Again, $[\varDelta_O^2]$ is an unbiased estimator for $\varDelta_O^2$.
It can be further shown that
\begin{equation}
[\varDelta_O] - \varDelta_O \sim \frac{\varDelta_O}{\sqrt{2\mN}}\ .
\end{equation}
So, in order to get a reliable error estimate, we need to average 
at least $50$ independent measurements.
Now consider that our $O_i$ are the values $O(t)$ that the observable
$O$ takes along a Monte Carlo simulation. The former formulas are no 
longer valid, because the $O(t)$ are no longer independent. In particular, 
it is straightforward to show that
\begin{equation}\label{eq:THERM-O2-bias}
\bigl\langle ([O]- \braket{O})^2\bigr\rangle = \frac{2\tau_{\text{int},O}}{\mN}\bigl(\braket{O^2}-\braket{O}^2\bigr)+\mathcal O(\mN^{-2}).
\end{equation}
Therefore, the effective number of independent measurements when estimating
errors is not $\mN$, but $\mN/2\tau_{\text{int},O}$.  
The computation of the $\tau_{\text{int},O}$ is costly and usually
not very precise, so we need a more robust method 
to estimate errors. One obvious possibility
is to consider $\mathcal M$ independent runs $O^{(i)}(t)$. Then we can compute
the individual $[O^{(i)}]$ and do
\begin{align}
[O] &= \sum_{i=1}^{\mathcal M} [O^{(i)}],\\
[\varDelta_O]^2 &= \frac{1}{\mN-1} \left[ \frac1\mN\sum_i [O^{(i)}]^2 - 
\left(\frac1\mN\sum_i [O^{(i)}]\right)^2\right]\, .
\end{align}
This strategy guarantees us independent data, but it is very cumbersome,
since we need many runs in order to get an acceptable error estimate.
It contains, however, the germ of a workable system: data binning. 
Indeed, let us combine our $\mN$ measurements of $O$ along a simulation
into $\mathcal B$ blocks of size $\mN/\mathcal B = n$:
\begin{equation}\label{eq:CORR-block}
\tilde O_i = \frac{1}{n} \sum_{t=(i-1)n+1}^{ni} O(t).
\end{equation}
Notice that
\begin{equation}
[\tilde O_i] = \frac{1}{\mathcal B}\sum_i \tilde O_i = 
\frac{1}{n\mathcal B} \sum_i \sum_{t=(i-1)n+1}^{ni} O(t) = [O].
\end{equation}
Now let us assume that the block length $n$ is $n\gg \tau_{\text{int},O}$,
so the $\tilde O_i$ can be considered independent from 
one another.\footnote{There are boundary effects, but
they are only $\mathcal O(n^{-1})$.} We have, then
\begin{equation}\label{eq:CORR-delta-block}
[\varDelta_O^2] = \frac{1}{\mathcal B-1} \left[ \frac1{\mathcal B}\sum_i \tilde O_i^2 - 
\left(\frac1{\mathcal B}\sum_i \tilde O_i\right)^2\right]\, .
\end{equation}
Of course, if we do not know $\tau_{\text{int},O}$ we cannot know
whether $n\gg\tau_{\text{int},O}$. However, we can just evaluate
$[\varDelta_O]$ for increasing block lengths $n$. For small
$n$ the error  will be underestimated, but we will eventually
reach a stationary regime. Once this is achieved, we should 
not let $n$ grow much more, in order to maximise the number 
of blocks $\mathcal B$ and, with it, the precision of our 
error estimate.

\section{Non-linear functions and the jackknife method}
\index{jackknife method|(}
Often, the interesting physical quantities are not restricted to 
thermal averages of observables, but include functions of these
thermal averages (susceptibilities, correlation lengths, etc.).
In general, we estimate $f(\braket{O})$ by $f([O])$. 
However, now  $f([O])$ is a biased estimator
for $f(\braket{O})$ (unless $f$ is linear). 
Fortunately, this bias is $\mathcal O(\mN^{-1})$, much
smaller than the statistical error which, as we have seen, goes
as $\mathcal O(\mN^{-1/2})$ (yet, see Section~\ref{sec:CORR-disorder}).
However, we still have the problem
of providing an error estimate for our $f([O])$. Naively, we would
do this with a linear error propagation but here the correlations
actually play in our favour, reducing the errors.

Throughout 
this thesis we have always employed the jackknife method. 
This is simple and robust but not, we must say, the only possible choice
(the other main alternative is the bootstrap, cf. Section~\ref{sec:CORR-bootstrap}). We consider a function $f$ of one or several thermal averages
and we define the following procedure
\begin{enumerate}
\item Estimate the central value $f(\braket{A},\braket{B}, \ldots, \braket{Z})$
by $f([A],[B],\ldots, [Z])$.
\item Bin each of the observables, as in~\eqref{eq:CORR-block}, with large 
enough $n$ and $\mathcal B$. Then form the jackknife blocks,
\begin{equation}
O_{\text{JK},i} = \frac{1}{\mathcal B-1} \sum_{j\neq i} \tilde O_i.
\end{equation}
\item Evaluate the function for each jackknife block,
\begin{equation}
f_{\text{JK},i} = f\bigl( A_{\text{JK},i},B_{\text{JK},i},\ldots,Z_{\text{JK},i}\bigr).
\end{equation}
\item Compute the jackknife error estimate
\begin{equation}
[\varDelta_f^2] = (\mathcal B -1) \left[\frac{1}{\mathcal B}
\sum_i f^2_{\text{JK},i} - \left(\frac{1}{\mathcal B} \sum_i f_{\text{JK},i}\right)^2\right]\, .
\end{equation}
\end{enumerate}
This error estimate coincides exactly with~\eqref{eq:CORR-delta-block}
for linear functions. For non-linear ones, it takes the correlations
into account and provides
a reliable estimate, since the blocks are allowed 
to fluctuate jointly.

\subsection{Disordered systems}\label{sec:CORR-disorder}
In this thesis, we are mainly concerned with disordered systems, where
one has to compute double averages of the type $\overline{\braket{O}}$.
In these cases, the application of the data binning procedure is even easier.
Since each thermal average $\braket{O}$ comes from a different, independent
sample we do not have to concern ourselves with the block length.
We simply consider one block for each sample, so $\tilde O_i$
is replaced by our estimate $[O_i]$ for the thermal average of 
the $i$-th sample. We can then form jackknife blocks with these $[O_i]$
as usual.

However, not all is simple. 
Sometimes we wish to compute disorder averages of non-linear functions
of thermal averages, that is, quantities of the type $\overline{f(\braket{O})}$.
In these cases, the bias produced by mocking $f(\braket{O})$ by $f([O])$
can have dangerous effects. There are general cures for this problem
(see, for instance, \cite{ballesteros:97}).
A  particularly elegant solution is afforded by the simulation of several
real replicas for each sample (for systems such as spin glasses, this is
something we have to do anyway).

For instance, let us consider a simple quadratic function, $\braket{O}^2$.
We already saw in~\eqref{eq:THERM-O2-bias} that $[O]^2$ is a biased
estimator of $\braket{O}^2$, with a bias of $\mathcal O(\mN^{-1}$).
Instead, if we run two independent simulations, resulting in time series
$O^\alpha$ and $O^\beta$ we have
\begin{equation}
\braket{O^\alpha O^\beta} = \braket{O^\alpha} \braket{O^\beta} = \braket{O}^2.
\end{equation}
However, the numerical estimate $[O^\alpha O^\beta]$ can be computed
without considering the square of an averaged quantity (just as if $O^\alpha O^\beta$
 where a single observable). Therefore,
\begin{equation}
\braket{[O^\alpha O^\beta]} = \braket{O^\alpha O^\beta} = \braket{O}^2
\end{equation}
\index{jackknife method|)}
\index{error analysis|)}
\section{Computing fits of correlated data}
\index{chi-square test}
\index{fitting techniques|(}
Let us suppose we have a series of $\mN$ points $(x_i,y_i)$, which we want
to fit to a functional form $y=y(x;a_1,\ldots,a_n)$, depending on some
unknown parameters $a_i$. For each point we have an estimate $\varDelta_i$ 
of the statistical error (standard deviation). Then, the maximum likelihood
estimator for the $a_i$ can be obtained by minimising the `chi-square'
\begin{equation}\label{eq:CORR-chi-square}
\chi^2 = \sum_{i=1}^{\mN} \left(\frac{y_i-f(x_i;a_1,\ldots,a_n)}{\varDelta_i}\right)^2.
\nomenclature[chi2com]{$\chi^2$}{Complete chi-square fit-goodness estimator}
\end{equation}
If our model function is linear in the $a_i$, and each of the $y_i$ are Gaussianly
distributed, it can be shown that the values of the minimum $\chi^2$ follow
the chi-square distribution for $\mN-n$ degrees of freedom.  
This distribution
has mean $\mN-n$ and standard deviation $\sqrt{2(\mN-n)}$.  Because of
the central limit theorem 
\index{central limit theorem}
it tends to a Gaussian distribution in the limit of large $\mN-n$.
Notice that this also guarantees that it will not be too wrong to 
treat a non-linear function of the $a_i$ in the same way, 
as is usually done.

The above results  mean that not only can we use the $\chi^2$ to obtain
the best fit parameters for a given functional form, but also 
to test whether this function is a good model for the data. Indeed, 
for a `good' fit, the value of $\chi^2$ per degree of freedom 
should be $\chi^2/\text{d.o.f.}\approx 1$. Too large a value of 
$\chi^2/\text{d.o.f.}$  would mean that the function is not a good
fit for the data (too low a value usually means that the $\varDelta_i$ are
overestimated, yet see below). To be more precise, the probability that 
$\chi^2$ should exceed a particular value $\mu$ by chance in a fit 
with $\nu=\mN-n$ degrees of freedom is given by the incomplete gamma
function, 
\begin{equation}
\mathcal P(\chi^2>\mu, \nu) = \frac{1}{\varGamma(\nu/2)}\int_{\mu/2}^\infty \ee^{-t} t^{\nu/2-1}\ \dd t.
\end{equation}
Notice that giving the value of $\chi^2/\text{d.o.f.}$ is not enough, we need 
to know $\chi^2$ and $\nu$ separately. For example, a fit 
with $2$ degrees of freedom and $\chi^2=4$ would have
$\chi^2/\text{d.o.f.}=2$ and $\mathcal P\approx0.135$ (not a very good 
\nomenclature[d.o.f.]{d.o.f.}{Degrees of freedom in a fit}
fit, but still acceptable). However, a fit with $10$ degrees of freedom
and $\chi^2=20$ would also have $\chi^2/\text{d.o.f.}=2$, yet 
now the $\mathcal P$-value would be $\mathcal P\approx0.029$, indicating a bad fit.

We can also use the $\chi^2$ test to estimate the errors in the fit 
parameters. Let us denote by $\chi^2_\text{min}$ and by $a_i^\text{min}$
the best fit parameters and the resulting $\chi^2$ value. Then, finding
the error in $a_i$ is a matter of perturbing this parameter 
and noting the change in $\chi^2$. The traditional recipe, followed here,
is to modify $a_i^\text{min}$ to $a_i^\text{min} + \varDelta a_i$ and perform
a new fit where the other parameters are varied, but $a_i$ is kept 
fixed. This new fit will have a $\chi^2=\chi^2_\text{min}+\varDelta \chi^2$. 
Since we are considering the fluctuations in $\chi^2$ caused by a single
(assumed to be) Gaussian variable $a_i$, a variation of one standard
deviation in $a_i$ is equivalent to $\varDelta \chi^2 =1$. 

We have hitherto considered the simplifying assumption that the $y_i$ are 
uncorrelated. This is sometimes the case, a typical example being fits
where the $x$ coordinate is the lattice size, so each $y_i$ comes 
from a completely independent simulation. Very often, however, the data
are strongly correlated. The rigorous way to take this into account
is to consider the full covariance matrix of the results. We define
\begin{align}
\sigma_{ij} &= \text{Cov}(y_i,y_j), & \delta_i = y_i - f(x_i;a_1,\ldots,a_n).
\end{align}
Let us mention in passing that the covariance $\text{Cov}(A,B)$ can be
estimated from our jackknife blocks as follows:
\begin{equation}\label{eq:CORR-cov}	
[\text{Cov}(A,B)] = \left(\mathcal B -1\right)\left[\frac{1}{\mathcal B}
\sum_i (A_{\text{JK},i} - [A]) (B_{\text{JK},i} - [B]) \right]\, .
\end{equation}
Then, the complete $\chi^2$ estimator is
\begin{equation}\label{eq:CORR-chi-square-completo}
\chi^2 = \sum_{i,j} \delta_i \sigma_{ij}^{-1} \delta_j,
\end{equation}
which, of course, if the data are not correlated is the same as~\eqref{eq:CORR-chi-square},
since $\sigma_{ii}= \varDelta_i^2$. The errors in the parameters are then 
estimated just as in the uncorrelated case.

The actual minimisation of $\chi^2$ can be performed in a number of ways. Let us first
note that, if $f$ is linear in the $a_i$, then~\eqref{eq:CORR-chi-square-completo}
is just a quadratic form in the $a_i$, whose minimisation yields a set 
of linear equations. If $f$ is non-linear, however, we must usually employ
an iterative method. In this thesis we have  used the Levenberg-Marquardt
algorithm~\cite{press:92}.

\subsection{Singularities in the covariance matrix}
Sometimes the data correlation is so extreme that the covariance matrix turns
out to be singular. This is the case, for instance, where we include 
$\mN_y$ random variables $y_i$ in a fit where we only have $\mN_\text{samples}<\mN_y$
samples (or jackknife blocks). In order to prove this, let  us 
denote each random variable with a Latin index $i=1,\ldots,\mN_y$
and each sample with a Greek index $\mu=1,\ldots,\mN_\text{samples}$.
We now define the matrix
\begin{equation}
T_{i\nu} = y_i^{\nu} - \frac1{\mN_\text{samples}}\sum_{\mu=1}^{\mN_\text{samples}} y_i^\mu.
\end{equation}
Now, since $\mN_\text{samples}<\mN_y$, it is clear that $T_{i\mu}$ has at most rank $\mN_\text{samples}$.
On the other hand, the covariance matrix is just
\begin{equation}
\sigma_{ij} = \frac{1}{\mN_\text{samples}(\mN_\text{samples}-1)} T T^\dagger.
\end{equation}
Therefore, 
\begin{equation}
\mathrm{rk}(\sigma) = \mathrm{rk}(TT^\dagger)= \mathrm{rk}(T) \leq \mN_\text{samples} < \mN_y
\end{equation}	
and the matrix $\sigma$ is singular.  The most obvious example of this are the fits of temporal 
correlations in Part~\ref{part:sg}, where we must consider some $10^4$ 
random variables extracted from a set of a few hundred samples.  

Throughout this thesis, we have followed an empirical procedure to address this issue. We find the best
fit parameters using only the diagonal part of the covariance matrix, i.e., minimising
the so-called diagonal chi-square
\begin{equation}\label{eq:CORR-chi-square-diagonal}
\chi^2_\text{d} = \sum_i \frac{\delta_i^2}{\sigma_{ii}}\ .
\nomenclature[chi2]{$\chi^2_\text{d}$}{Diagonal chi-square fit-goodness estimator}
\end{equation}
In order to take the correlations into account, we perform an independent fit for
\index{jackknife method}
each jackknife block and use their fluctuations to compute the errors
in the $a_i$. We have found this method to be reliable (see Section~\ref{sec:CORR-bootstrap}, below), even if it complicates
the interpretation of the $\chi^2_\text{d}$ values. Indeed, since the fitting 
function fluctuates coherently with the data, the values of $\chi^2_\text{d}/\text{d.o.f.}$
are typically much smaller than $1$. In effect, $\chi^2$ behaves as if there were
fewer degrees of freedom, so we cannot readily compute the $P$ values.

Throughout this thesis, we shall use $\chi^2$ to denote either the full $\chi^2$ 
estimator of~\eqref{eq:CORR-chi-square-completo} or the $\chi^2$ of a fit
of uncorrelated data and we shall denote by $\chi^2_\text{d}$ the diagonal
chi-square of~\eqref{eq:CORR-chi-square-diagonal}.

\subsection[Errors in the $x$ and $y$ coordinates]{Errors in the \boldmath $x$ and $y$ coordinates}\label{sec:CORR-xyerror}
Thus far we have assumed that there are only errors in the $y$ coordinate
of the data, which is often the case. If this is not true, we would have to estimate
the chi-square as (assuming no data correlation, for simplicity) 
\begin{equation}\label{eq:CORR-xyerror}
\chi^2 = \sum_{i=1}^\mN \frac{\bigl( y_i - f(x_i; a_1,\ldots,a_n)\bigr)^2}{\mathrm{Var}\bigl( y_i - f(x_i; a_1,\ldots,a_n)\bigr)}\, .
\end{equation}
Notice that now the $a_i$, which we must change during our 
iterative procedure to perform the fit, appear also in the denominator.
For non-linear $f$ the corresponding expressions rapidly get intractable with 
our usual numerical methods.

Sometimes one can skirt the issue by performing separate fits for
the $x_i$ and $y_i$ and relating the resulting fit parameters 
with those of the real fit to $y=f(x)$ (see, for instance, Section~\ref{sec:SG-thermoremanent}).
If this is not possible, one can often consider a simplified procedure
for minimising~\eqref{eq:CORR-xyerror}. Suppose that the errors
in $(x_i,y_i)$ are similar for both coordinates and for all values of $i$. 
Then the denominator in~\eqref{eq:CORR-xyerror} is almost constant in $i$
and  the values of the $a_i$ that minimise
\begin{equation}
\tilde \chi^2= \sum_{i=1}^\mN \bigl( y_i - f(x_i; a_1,\ldots,a_n)\bigr)^2
\end{equation}
should be very similar to those that minimise the real $\chi^2$
estimator. Now, this would give us an approximation to the values of the best-fit 
parameters, but not an idea of the fit's goodness. But the latter
can be estimated by an \emph{a posteriori} $\chi^2$ test
with~\eqref{eq:CORR-xyerror}, where we no longer vary any parameter. 

\subsection{Case studies}
In the remainder of this section we present two examples
of the previous methods, taken from our work.

\subsubsection[The replicon exponent]{Estimating the error in the replicon exponent}\label{sec:CORR-bootstrap}\index{jackknife method|(}
In this section we perform a detailed test of our technique for coping with singular
covariance matrices. We have chosen the computation of the replicon exponent $a$, detailed
in Section~\ref{sec:SG-replicon-dynamics}. There are two reasons for this: first, it is a particularly
tricky case, with both statistical and systematic error sources for the $y_i$. Second,
we actually have two independent determinations, one with ten times more data, so we
can check the reliability of the method.

Let us recall the steps of the computation. We want to estimate the exponent $a$ in an 
equation
\begin{equation}
I_1(\tw) = A \xi^{2-a}(\tw).
\end{equation}
The quantities $I_1$ and $\xi$ are estimated from integrals of the spatial correlation
function for each time, using a dynamical cutoff procedure:
\begin{equation}
[I_k(\tw)] = \int_0^{\varLambda} \dd r\ r^k [C_4(r,\tw)],
\end{equation}
where the cutoff $\varLambda$ is chosen as the point where the relative 
error in $[C_4(r,\tw)]$ becomes larger than $50\%$. The coherence length
is  estimated as
\begin{equation}
[\xi(\tw)] = [I_2(\tw)]/[I_1(\tw)].
\end{equation}
We then perform two independent fits, using the diagonal chi-square method,
\begin{align}
I_1(\tw) &=  B \tw^c, & \xi(\tw)&=C\tw^{1/z},
\end{align}
and we finally estimate $a$ as
\begin{equation}
a = 2-cz,
\end{equation}
a relation that is used, of course, for each jackknife block in order to estimate errors.

We first performed this procedure in~\cite{janus:08b}, for simulations 
at $T=0.7$ with $63$ samples, obtaining the result
\begin{equation}
a = 0.355(15).
\end{equation}
The computation was then repeated in~\cite{janus:09b}, this time with 
$768$ samples,  producing
\begin{equation}
a = 0.397(12).
\end{equation}
The discrepancy in the results could be due to a simple statistical fluctuation, but they 
are different enough to make this unlikely. Another hint that something may be amiss
is that the error has not decreased nearly the factor $\sqrt{768/63}$ that 
one would expect from the increase in statistics.
\begin{figure}[t]
\includegraphics[height=\linewidth,angle=270]{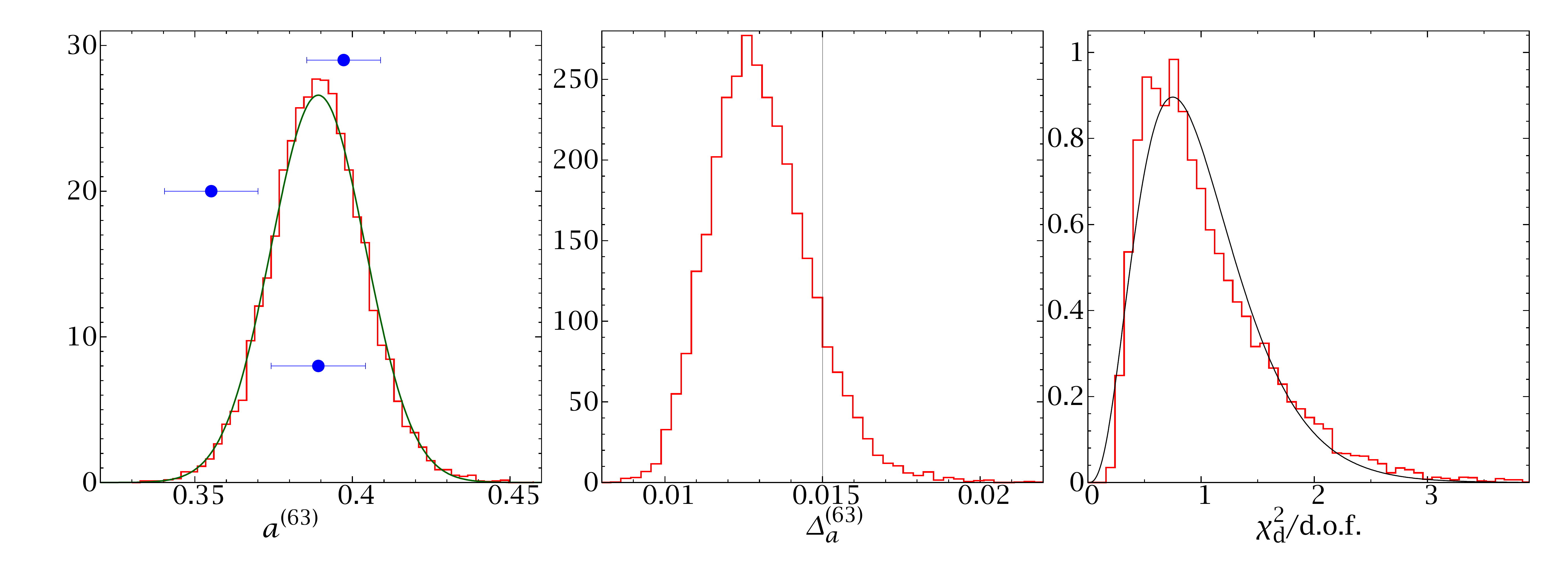}
\caption[Histograms of our replicon exponent estimates]{\textit{Left:} Probability density function of $a^{(63)}$,
the estimate of exponent $a$ 
as obtained from a set of 63 samples. The dots with
\emph{horizontal} error bars are, from top to bottom:
our best estimate with 768 samples, the value
with the 63 samples of~\cite{janus:08b}
and the mean and standard deviation of $a^{(63)}$.
The continuous line is a Gaussian distribution
with the same mean and variance.
\textit{Centre:} As in left panel, for the jackknife
errors $\varDelta_a^{(63)}$. The vertical line marks
the standard deviation of $a^{(63)}$.
\textit{Right:} Histogram of the $\chi^2_\text{d}$/d.o.f.
for the $\xi$ fit, which had $80$ degrees of freedom. 
We see that there is a very large variance, indicating
that the effective number of degrees of freedom is much smaller.
As a comparison, we have plotted with a continuous line the 
chi-square distribution (per d.o.f.) for only $8$ degrees of freedom.} 
\label{fig:CORR-histograms}
\end{figure}

Therefore, we may have underestimated our errors, or there may be some bias
in our methods (or in our statistics). The first source of systematic error
is the dynamic cutoff method. Naturally, for the simulations with $768$ samples
the cutoff point $\varLambda$ is larger. This has the effect of trading statistical
uncertainty (including more noisy data at the tails) for a reduction 
of systematic errors, and is probably at the core of the small decrease 
in our error estimate.

\index{chi-square test|(}
However, the effect is large enough to merit a more detailed study, which we shall
\index{bootstrap method}
carry out with a bootstrap procedure (see, e.g.,~\cite{efron:94}).
Our objective is to estimate the probability 
distribution of $a^{(63)}$, defined as the result of computing
$a$ from a statistic of $63$ samples. 
This can be easily done, since
from our second set of simulations we have $768$ samples. Therefore, we can pick
random sets of $63$ samples and determine for each set the estimate of $a^{(63)}$
and of its error $\varDelta_a^{(63)}$.
There are $\binom{768}{63}\sim10^{93}$
possible combinations, so our distribution of $a^{(63)}$ computed
from a set of $768$ samples can be considered the `real' one. 

We have used $10\,000$ random sets, with which we have computed normalised histograms
of $a^{(63)}$ and its errors, plotted in Figure~\ref{fig:CORR-histograms}. From the
first histogram we can take several conclusions:
\begin{enumerate}
\item The distribution of $a^{(63)}$ is Gaussian, with a very good
approximation. We denote its mean by $\mathrm{E}(a^{(63)})$.
\item We were unfortunate in~\cite{janus:08b}, in that
the $63$ samples used therein are `atypical'. More precisely, the samples
 of~\cite{janus:08b} produce an estimate of $[a^{(63)}]$ that is $2.2$ standard
deviations away from its true expectation value $\mathrm{E}(a^{(63)})$ (there is only a $3\%$ probability
to obtain a larger fluctuation in Gaussianly distributed results).
\item The random variable $a^{(63)}$ seems to be a slightly biased
estimator of $a$, $\mathrm{E}(a^{(63)})$ showing a small deviation from our estimate
of $a$ with $768$ samples. This is probably due from the cutoff effects discussed
earlier, although the fact that $a$ is obtained from a non-linear operation
may also produce a small bias in the fitting procedure.
\end{enumerate}
From the second histogram, representing the pdf of $\varDelta^{(63)}_a$, 
we see that the jackknife method tends to underestimate the errors
somewhat (compare the mean of this distribution with the vertical line, marking
the true standard deviation of $a^{(63)}$).

As a final result, we can look at the histogram of the $\chi^2_\text{d}/\text{d.o.f.}$
obtained in the $10\,000$ fits during the bootstrapping procedure. Each of these 
had $80$ degrees of freedom, yet the standard deviation of the distribution 
is much larger than that of a chi-square distribution with $80$ degrees
of freedom. This supports our initial claim that fits with the diagonal 
chi-square procedure have a small effective number of degrees of freedom. 

\index{chi-square test|)}
\index{jackknife method|)}
\index{fitting techniques|)}

\subsubsection{The full-aging exponent}\label{sec:CORR-wiggles} \index{full aging}
Let us now consider our analysis of full aging in Section~\ref{sec:SG-aging}.
The $\alpha(\tw)$ that we plotted in Figure~\vref{fig:SG-full-aging} showed 
suspicious fluctuations (this is especially noticeable for $T=0.6$).
This may lead the reader to think we have underestimated our errors
when computing fits to~\eqref{eq:SG-full-aging}.
\begin{figure}
\includegraphics[height=\linewidth,angle=270]{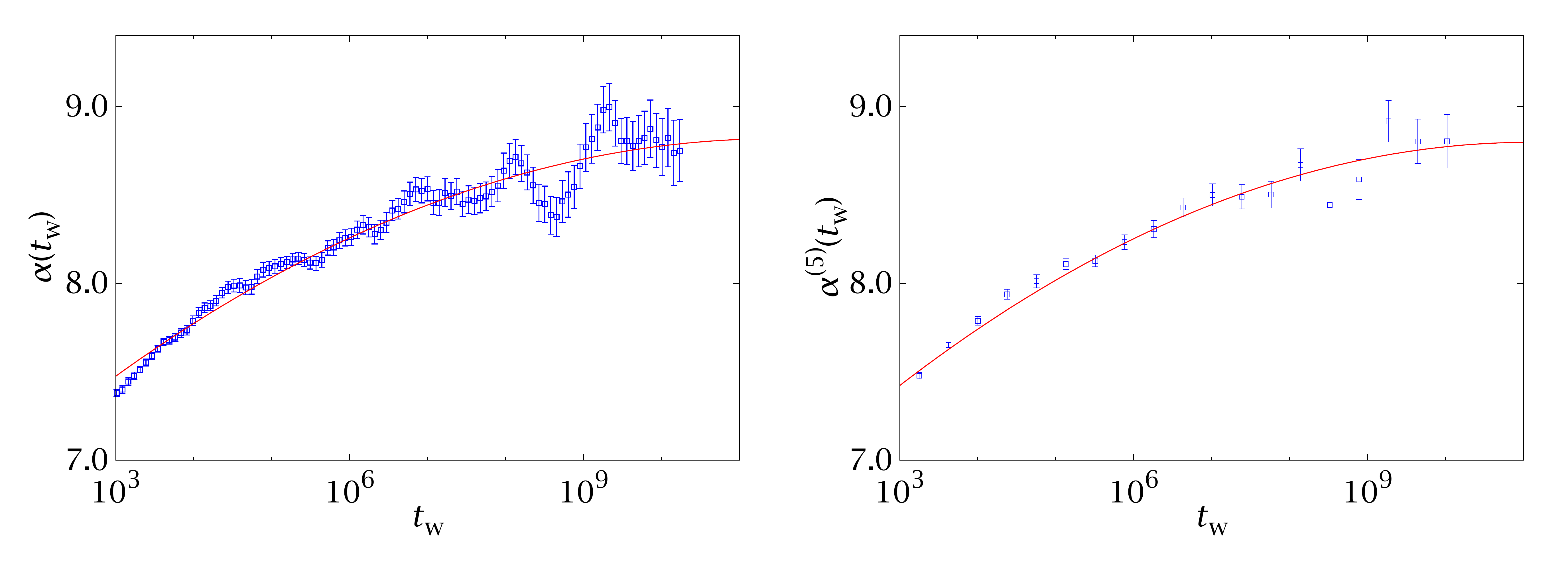}
\caption[Parameters in a full-aging fit]{Exponent $\alpha(\tw)$ 
for our $T=0.6$ simulations, with a fit to a quadratic polynomial
in $\log\tw$. We show on the left panel the  same data as in Figure~\ref{fig:SG-full-aging}
and on the right one the result of binning the data in blocks of $5$ consecutive
times.
\index{full aging|indemph}
\index{fitting techniques}
\label{fig:SG-full-aging-wiggles}
}
\end{figure}

We can check that the errors are actually correctly estimated by performing 
a fit of $\alpha(\tw)$ to a very smooth function. We have done this in the worst
case ($T=0.6$, Figure~\ref{fig:SG-full-aging-wiggles}---left) with a quadratic fit
\begin{equation}
\alpha(\tw) = \alpha_0 + \alpha_1 \log\tw + \alpha_2 \log^2 \tw.
\end{equation}
For $\tw>10^5$ we have 
\begin{align}
\alpha_0 &= 6.4(6), &
\alpha_1 &= 0.19(8), &
\alpha_2 &= -0.003\,5(20),
\end{align}
with a diagonal chi-square of $\chi^2_\text{d}/\text{d.o.f.}=66.26/63$.

As an additional check we show in Figure~\ref{fig:SG-full-aging-wiggles}---right
the same plot, but now binning the data in blocks of $5$ consecutive times.
The resulting parameters are
\begin{align}
\alpha_0^{(5)} &= 6.2(6), &
\alpha_1^{(5)} &= 0.20(8), &
\alpha_2^{(5)} &= -0.003\,9(24),
\end{align}
with $\chi^2_\text{d}/\text{d.o.f.} = 11.58/10$. 
Therefore, neither the parameters nor the value of $\chi^2_\text{d}/\text{d.o.f.}$ are affected by the binning procedure.
Notice that in Figure~\ref{fig:SG-full-aging-wiggles}---right
the wiggles in the curve seem just what they 
are: random fluctuations.

\begin{figure}
\centering
\includegraphics[height=\linewidth,angle=270]{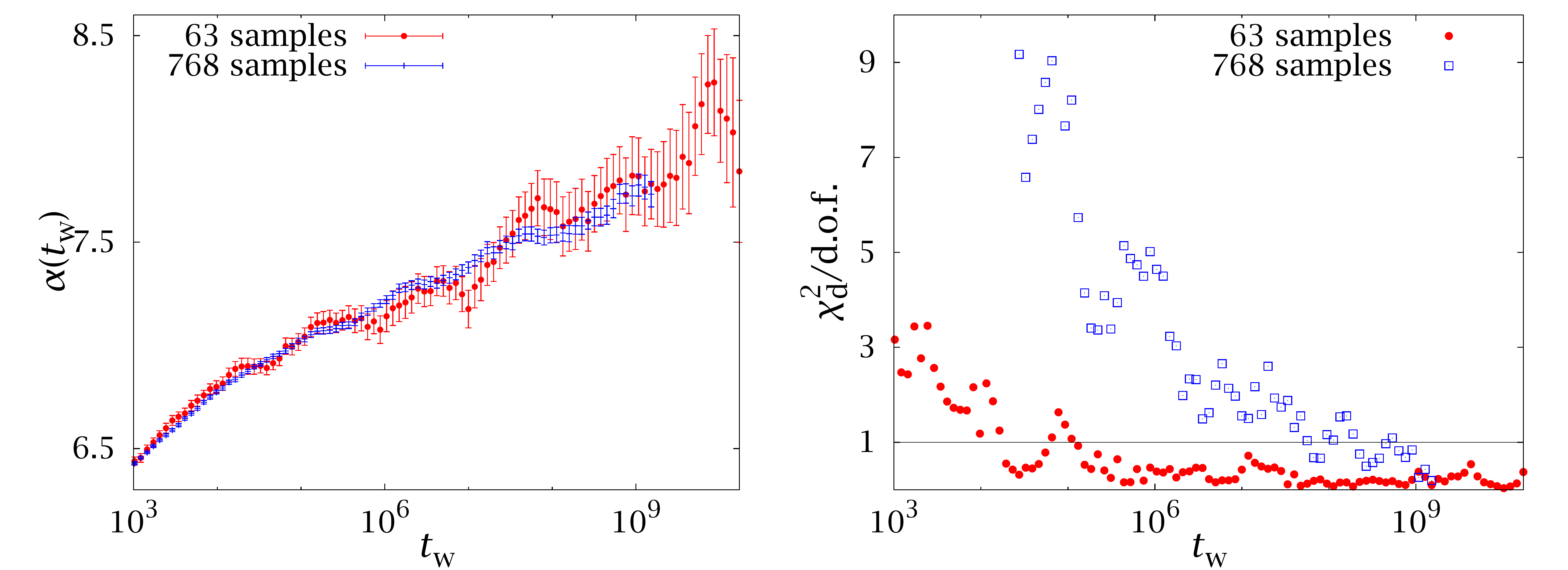}
\caption[Full-aging exponent $\alpha(\tw)$]{\emph{Left:} Estimate of $\alpha(\tw)$ in~\eqref{eq:SG-full-aging}
for our $T=0.7$ simulations with $63$ and $768$ samples. \emph{Right:} Values
of $\chi^2_\text{d}/\text{d.o.f.}$ for both series.
\index{full aging|indemph}
\label{fig:SG-full-aging-wiggles-2}}
\end{figure}

A second check is provided by our simulations for $T=0.7$. For this
temperature, we performed two series of runs (cf. Appendix~\ref{chap:runs-sg}).
The first, reaching $\tw=10^{11}$, included $63$ samples. The second, 
for $768$ samples, reached $\tw=10^{10}$. We have plotted $\alpha(\tw)$ 
for both sets in Figure~\ref{fig:SG-full-aging-wiggles-2}. We see 
that, increasing the number of samples, the period of the oscillations
does not change, but their amplitude decreases proportionally to the 
error. However, see right panel, the $\chi^2_\text{d}/\text{d.o.f.}$,
which is reasonable from $\tw\sim10^4$ for $63$ samples, is not 
good with $768$ until $\tw\sim10^8$. This indicates that there probably
is a systematic deviation from~\eqref{eq:SG-full-aging}. Nevertheless, 
the estimates for $\alpha(\tw)$ are compatible in a much wider range.

\section[Control variates]{Using correlations to improve the statistics: control variates}\index{control variates|(}
In the above section, we have discussed the difficulties posed by data correlation. 
However, sometimes correlations can be exploited to our advantage.
A good example is the use of control variates
 (see~\cite{rubinstein:07} and \cite{fernandez:09c} for
a review, applied to Monte Carlo simulations).

The idea is the following, assume we want to compute some parameter $A$ and 
that we have a random variable $\mathcal A$ such that $\mathrm{E}(\mathcal A)=A$ (that is, $\mathcal A$ is 
an unbiased estimator for $A$). Here
$\mathrm{E}$ denotes the mathematical expectation, which in a Monte Carlo context 
can be either a thermal or a disorder average. In addition, we have an
estimator $\mathcal B$ (the control variate)
whose expectation value $B = \mathrm{E}(\mathcal B)$ is known 
in advance. Then $\mathcal A'=\mathcal A + \alpha (\mathcal B -B)$ is another 
unbiased estimator for $A$, $\mathrm{E}(\mathcal A') = \mathrm{E}(\mathcal A) = A$.
Furthermore,
\begin{equation}
\Var(\mathcal A') = \Var(\mathcal A) + \alpha^2 \Var(\mathcal B) + 2\alpha \Cov(\mathcal A,\mathcal B),
\end{equation}
where $\Var$ denotes the variance and $\Cov(\mathcal A,\mathcal B)$ the covariance 
of $\mathcal A$ and $\mathcal B$. It is easy to show that choosing 
\begin{equation}
\alpha = -\frac{\Cov(\mathcal A,\mathcal B)}{\Var(\mathcal B)}
\end{equation}
minimises the variance of $\mathcal A'$, so that
\begin{align}\label{eq:DAFF-var-control-variate}
\Var(\mathcal A') &= (1-\mathcal R_{\mathcal A,\mathcal B}^2) \Var(\mathcal A), &
\mathcal R_{\mathcal A,\mathcal B} = \frac{\Cov({\mathcal A,\mathcal B})}{\sqrt{\Var({\mathcal A})\Var({\mathcal B})}},
\end{align}
where $\mathcal R_{\mathcal A,\mathcal B}$ is the correlation coefficient of $\mathcal A$ and $\mathcal B$.
Notice that this process not only reduces statistical errors in the 
sought parameter, but also corrects finite-statistics biases.

\begin{figure}[t]
\centering
\includegraphics[height=\columnwidth,angle=270]{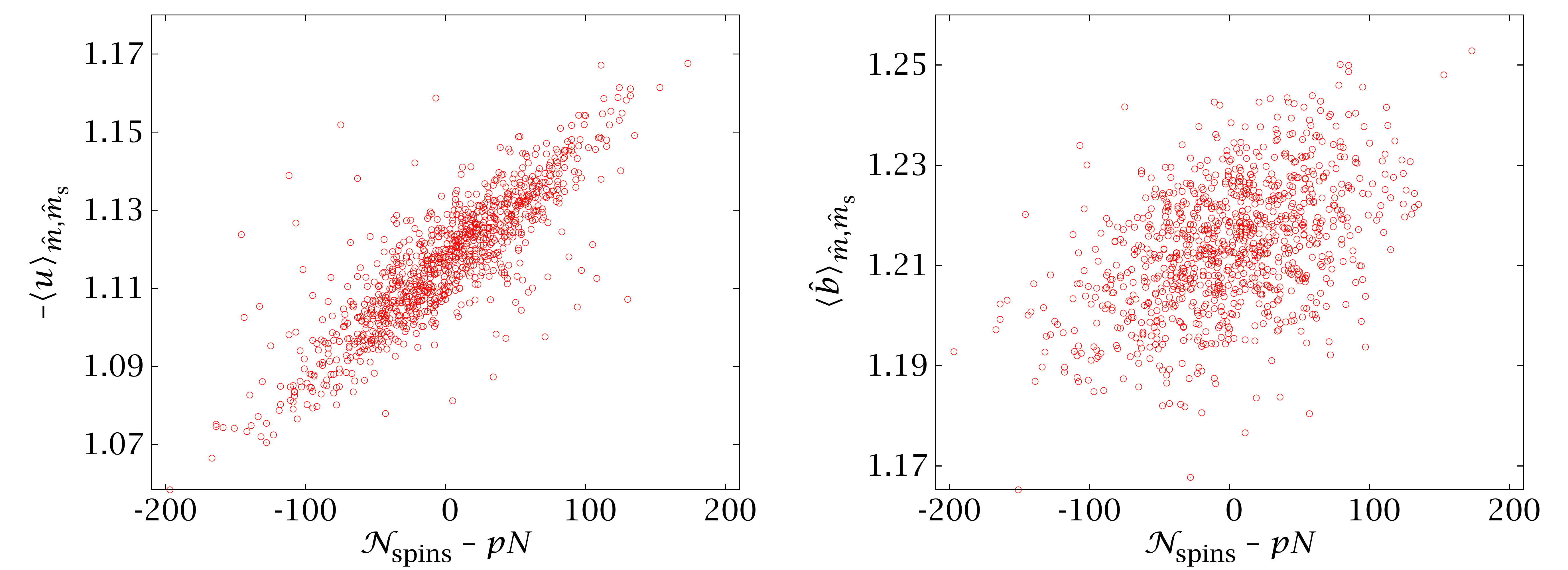}
\includegraphics[height=\columnwidth,angle=270]{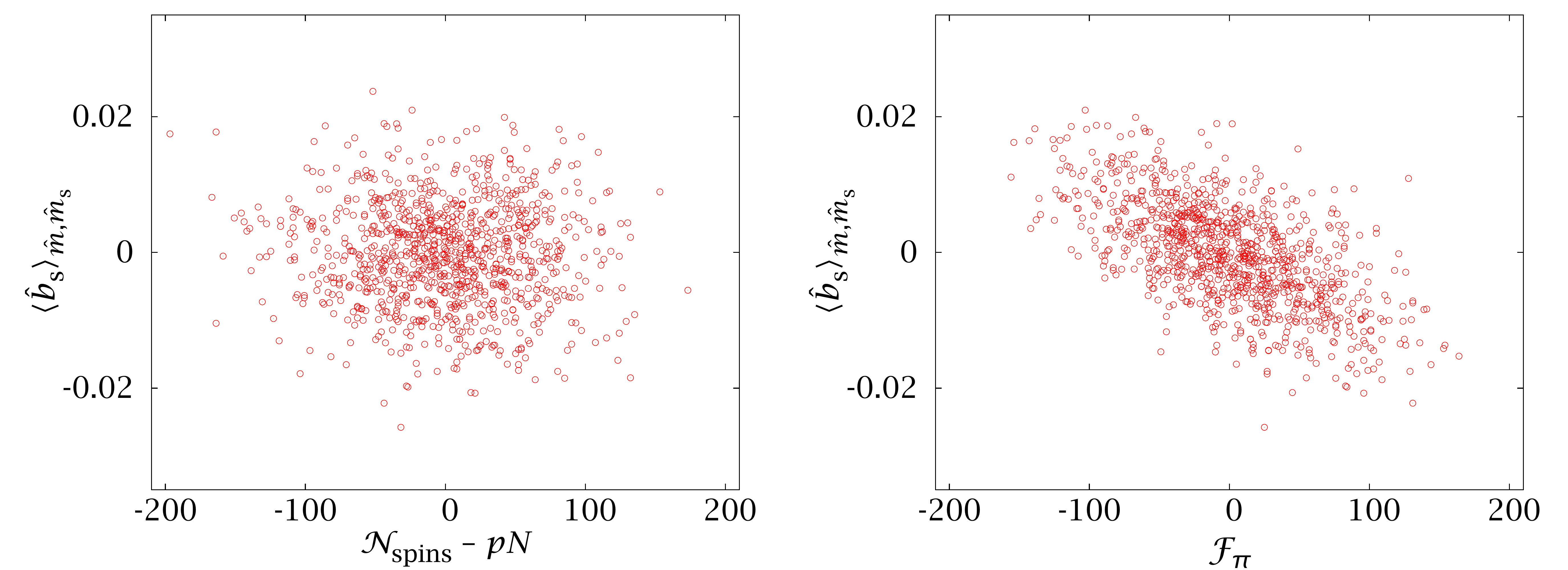}
\caption[Scatter plots of physical quantities against control variates]{Scatter
plots of the energy and tethered field 
of individual samples against the number of spins (top panels 
and bottom left panel). The bottom right panel plots the staggered
component of the tethered magnetic field against the $\mathcal F(\bk)$ of
Eq.~\eqref{eq:DAFF-F-huecos} for $\bk=(\uppi,\uppi,\uppi)$. All panels
show our $1000$ samples
for $L=24$ at $\beta=0.625$, $\hat m= 0.12$ and $\hat m_\text{s}=0.5$.}
\label{fig:DAFF-scatter}
\index{energy!DAFF|indemph}
\index{tethered field!DAFF|indemph}
\index{control variates|indemph}
\end{figure}
We have applied this method in order to minimise 
the variance in the disorder average of physical quantities \index{DAFF}
in our DAFF simulations of Chapter~\ref{chap:daff-tethered}.
 For potential control variates we can consider the 
distribution of the vacancies $\epsilon_\bx$ in the lattice. Indeed, we know that, for instance,
the average value of spins in the lattice is 
\begin{equation}
\overline{\mN_\text{spins}} = p N.
\end{equation} 
More generally, the distribution of empty nodes on the lattice should be completely
random, with no periodicity or correlation. We can, therefore, consider
the Fourier transform of the $\epsilon_\bx$,
\begin{equation}\label{eq:DAFF-F-huecos}
\mathcal F(\bk) = \sum_\bx \epsilon_\bx \ee^{-\ii \bx\cdot \bk}.
\end{equation}
We have, then,
\begin{align}
\overline{\mathcal F}(0)&=\overline{\mN_\text{spins}} = p N, &
\overline{\mathcal F}(\bk) &= 0, \quad \bk \neq 0.
\end{align}

Once we have found viable candidates for control variates, we have to check
whether they are correlated with any physical quantity.  Let us start with
$N_\text{spins}$, for which we show scatter plots in  the first three 
panels of Figure~\ref{fig:DAFF-scatter}. This quantity has a strong correlation
with the energy of the sample (top left) and a slight correlation with the
regular component of the tethered field (top right), but no correlation 
at all with $\hat b_\text{s}$ (bottom left). However, see bottom right panel,
this latter quantity does show a correlation with $\mathcal F_\uppi=\mathcal F\bigl( \bk=(\uppi,\uppi,\uppi)\bigr)$.

Therefore, we have used $\mN_\text{spins}$ as a control variate for the
energy and $\hat b$ and $\mathcal F_\uppi$
as a control variate for $\hat b_\text{s}$. The correlation coefficients
for an $L=24$ system at $\beta=0.625$, $\hat m=0.12$ and $\hat m_\text{s}=0.5$ 
are\footnote{The numbers are similar for other lattice sizes and magnetisations.}
\begin{align}
\mathcal R_{u,\mN_\text{spins}} &=  -0.85, &
\mathcal R_{\hat b,N_\text{spins}} &=  0.47, &
\mathcal R_{\hat b_\text{s},\mathcal F_\uppi} &=  -0.62.
\end{align}
For the energy, the resulting reduction, Eq.~\eqref{eq:DAFF-var-control-variate},
is of about $50\%$. Even for $\hat b_\text{s}$ the error reduction is of
about $20\%$.  Recalling that the statistical error
otherwise goes as $\sim1/\sqrt{\mN_\text{samples}}$, we see that using control 
variates for $\hat b_\text{s}$ is equivalent to an increase of $\approx65\%$ 
in the number of samples (and, hence, in the simulation time).

\index{control variates|)}

\chapter{Fine tuning Tethered Monte Carlo}\label{chap:recipes}
\index{tethered formalism!numerical implementation|(}
This chapter contains some technical details on our tethered simulations,
as well as some practical recipes for the correct usage of TMC.
We use the example of the $D=2$ Ising model, as in Chapter~\ref{chap:Ising}, 
except for Section~\ref{sec:DAFF-numerical-implementation}, which explains
our numerical implementation for the DAFF. 

\section[Numerical implementation of the Metropolis scheme]{Numerical implementation of the Metropolis scheme for the \boldmath $D=2$
Ising model}\label{sec:ISING-optimisation}
We have used the Metropolis algorithm explained in Section~\ref{sec:TMC-Metropolis}.
\index{Metropolis algorithm}
A naive implementation is completely straightforward and no different than that
of canonical Metropolis (see~\cite{amit:05}, for instance, for a complete working example).
We can, however, significantly enhance the performance in a simple way. Indeed, using
the explicit form~\eqref{eq:TMC-omega} of the tethered weight, the spin reversal probability is
\begin{equation}
P(s_\bx \to -s_\bx) = \begin{cases} 0, & M'>\hat M,\\
\min\{1, \ee^{\Delta S}\}, & M'\leq M,
\end{cases}
\end{equation}
where
\begin{equation}
\Delta S = -\beta(U'-U)+(M'-M)- \left(\frac N2-1\right)\log\left(1-\frac{M'-M}{\hat M-M}\right)\ .
\end{equation}
Now, in a Metropolis simulation, most of the CPU time is going to be dedicated
to evaluating $\exp[\Delta S]$, which includes computing a logarithm,
and therefore needs a large number of elementary operations. However, there are only a finite number
of possible values for $\Delta S$. 
\begin{itemize}
\item $(U'-U)=-Hs_\bx$, where $H$ is the local field around $\bx$ and can only take
$5$ different values for this model, $H=-4,-2,0,2,4$. 
\item $M'=M-2s_\bx$ and $M$ can only take $(N+1)$ values.
\item The spin $s_\bx$ itself can take only two values.
\end{itemize}
In all, we have $10(N+1)$ possible values of the Metropolis update probability (many
of which are going to be zero or one, because of the constraints).  This is 
a manageable number, so we can compute them all and store them in memory in 
a look-up table before starting the simulation. 
This way, we remove the need for computing
\index{look-up table}
logarithms on the fly and greatly accelerate the code.

A second note concerns the choice of the pseudorandom number generator. We originally
used a $64$-bit congruential generator reported in~\cite{ossola:04}. We ran through the lattice
sequentially, extracting one random number per site (although, in principle, we would only
need one when $\Delta S < 0$ and $\hat M \geq M'$). This reproduces the conditions studied
in~\cite{ossola:04,ossola:04b}, where significant deviations from the expected values
where found using the same generator in a $D=3$ Ising model for $N\geq 128^3$. Perhaps
unsurprisingly, we also obtained wrong results for our larger lattices $(N=1024^2)$.
We solved the problem by repeating all our simulations with a more sophisticated generator,
computed as the sum (modulo $2^{64}$) of the congruential generator and a $64$-bit
\index{random number!generation}
version of the shift-register method reported in~\cite{parisi:85}.

\section[Sampling the $\hat m$ space]{Sampling the \boldmath $\hat m$ space}\label{sec:ISING-hatm-grid}
The first practical decision in a TMC computation is the sampling 
\index{tethered formalism!sampling}
of the coordinate space, that is, how many values of $\hat m$ to simulate 
and where to put them. This is an important matter because the reconstruction 
of the canonical partition function (and of the canonical expectation values) 
involves an integral over $\exp[-N \varOmega_N]$, where $\varOmega_N$ is itself
an integral of the discretised $\braket{\hat b}_{\hat m}$. A good sampling
should make the discretisation errors negligible, as well as minimise the statistical ones.

Recalling the definition~\eqref{eq:TMC-hatM} of $\hat M$, we see that this 
parameter can, in principle take values in the interval $[-1,\infty)$. In practice, 
though, $\hat m\simeq m+1/2$, so the pdf $p(\hat m)$ is going to be completely
negligible outside a finite range (roughly $[-1/2,3/2]$, except for very small lattices,
where we need a broader scope). In addition, see Figure~\ref{fig:ISING-p-hatm},
$p(\hat m)$ is a two-peaked distribution, so values of $\hat m$ close to the probability
maxima contribute more towards the canonical average. The peaks get narrower and closer
together as $L$ grows (notice that $\varOmega_N$ has to be a convex function in the
thermodynamical limit). It may seem that the choice of $\hat m$ is quite delicate
(especially considering we do not know the pdf until we have run the simulation).
Finally, given a fixed computational budget, is it better to compute 
$\braket{\hat b}_{\hat m}$ at more points or more precisely  at each point 
with a coarser grid?

Let us assume we have measured $\braket{\hat b}_{\hat m}$ at $K$ points
in a grid. Our numerical estimator $[\hat b]_{\hat m}$
at point $k$ is 
\begin{equation}
[\hat b]_k = \braket{\hat b}_k + \eta_k,\qquad k=1,\ldots, K.
\end{equation}
Here $\eta_k$ are the statistical errors, expected to follow a Gaussian
distribution and to be of similar size and statistically uncorrelated.
Therefore, our numerical estimate for $\varOmega_N$ will be given
by some quadrature formula
\begin{equation}
\varOmega_N \simeq [\varOmega_N] = \sum_{k=1}^N g_k \braket{\hat b}_k 
+ \sum_{k=1}^K g_k \eta_k.
\end{equation}
The first term in this equation is subject to systematical error
while the second one represents the statistical fluctuations.
Naively, the quadrature coefficients $g_k$ scale as $1/N$, so the 
error will scale as $1/\sqrt{N}$. This suggests that doubling
the number of points is equivalent to doubling the simulation
time in each one. The analysis of canonical expectation values is,
however, more involved, since the errors in $\varOmega_N$ are 
going to be highly correlated for different $\hat m$. Therefore,
we perform a numerical experiment.

We report in Table~\ref{tab:ISING-grid} the values for the energy
density $\braket{u}$ at the critical temperature for an $L=128$
lattice, obtained in two different series of runs. In the first
column, we take $10^6$ Monte Carlo sweeps
(MCS) on each point, while in the second we perform $10^7$ such updates.
In both cases the points are evenly spaced.  In the third and fourth
columns we report analogous simulations, but now with a greater
density of points inside the peaks. We can reach 
several conclusions from this table:
\begin{table}
\small
\centering
\begin{tabular*}{\columnwidth}{@{\extracolsep{\fill}}lllll}
\toprule
\multirow{2}{*}{$\mN_\text{points}$} 
& \multicolumn{2}{c}{Uniform sampling} 
& \multicolumn{2}{c}{Improved sampling}\\
\cmidrule{2-3} \cmidrule{4-5}
& \multicolumn{1}{c}{$10^6$ MCS} &\multicolumn{1}{c}{$10^7$ MCS}&
 \multicolumn{1}{c}{$10^6$ MCS} &\multicolumn{1}{c}{$10^7$ MCS}\\
\toprule
12    & 1.437\,28(33)  &1.437\,38(11)   &                &\\               
23    & 1.419\,25(22)  &1.419\,117(6)   &                & \\               
46    & 1.419\,21(13)  &1.419\,117(43)  & 1.419\,08(11)  & 1.419\,107(38)   \\
91    & 1.419\,14(10)  &1.419\,093(36)  & 1.419\,13(8)   & {\bf 1.419\,128(31)}\\
181   & 1.419\,14(7)   &\textbf{1.419\,095(28)}  & 1.419\,03(5)  & \\ 
451   & 1.419\,06(5)   &                & 1.419\,073(39) & \\       
901   & 1.419\,065(33) &                & \textbf{1.419\,077(26)} &   \\
1801  & \textbf{1.419\,062(24)}                   \\   
Exact & \multicolumn{4}{c}{1.419\,076\,272\,086\dots}\\
\bottomrule
\end{tabular*}
\caption[Optimising the $\hat m$ sampling]{Final value for $-\langle \hat u\rangle$ as we
change the number of points for the reconstruction of $\varOmega_N$ 
and their precision. MCS = Monte Carlo Sweeps for each point. 
The runs labelled `uniform sampling' consist of $\mN_\text{points}$
values of $\hat m$ evenly distributed in the range $[-0.4,1.4]$.
The runs labelled `improved sampling' have $2/3\mN_\text{points}$ points evenly
distributed in the same range, plus and additional $\mN_\text{points}/3$ 
inside the peaks, effectively doubling the density in the dominant regions.
The last results of each column have a similar precision, but those
computed with uniform sampling required twice the simulation time.
The exact value has been computed from the results of~\cite{ferdinand:69}.
\index{tethered formalism!sampling|indemph}
\label{tab:ISING-grid}}
\end{table}
\begin{itemize}
\item If we use to low a number of points we will see a significant
systematic error, no matter how precise they are. 
\item Once the systematic error is under control, increasing the 
number of evenly distributed points has an effect of $1/\sqrt{\mN_\mathrm{points}}$
in the statistical error. This is best seen in the leftmost column.
The effect is roughly the same if we increase
the number of MCS in each point by the same factor (the errors
of the first column are $\sim\sqrt{10}$ times greater than the
corresponding ones of the second).
\item If we add more points inside the peaks, the error
may decrease faster than $1/\sqrt{\mN_\mathrm{points}}$. 
The errors with $\mN_\text{points}$ and uniform sampling are 
only slightly smaller than those with $\mN_\text{points}/2$ 
and improved sampling.
\end{itemize}
We can summarise this analysis with the following prescription
for the choice of $\hat m$:
\begin{enumerate}
\item Run a first simulation with enough uniformly sampled $\hat m$ 
so that the systematic error is small or unnoticeable (i.e., so
that the peaks of the distribution can be roughly reconstructed
and the tails are reliably sampled).
We have used $\sim 50$. This may seem a big number, but we must
take into account that we have only looked at the energy 
in Table~\ref{tab:ISING-grid}. Other quantities, such as high 
moments of the magnetisation or observables at a nonzero
magnetic field, require that the tails of the distribution
be reasonably well sampled.
\item Add a similar number of points inside the peaks of the pdf to 
eliminate the systematic error and reduce the statistical error.
\end{enumerate}
We have found that the second step is not always necessary. In fact, 
for lattices up to $L=256$ we have limited ourselves to 
computing $51$ evenly distributed $\hat m$. For bigger
lattices the peaks are steeper and we have added
another $26$ points inside them. 

This general prescription can be generalised to more complicated
situations (such as the DAFF studied in Chapter~\ref{chap:daff-tethered}). \index{DAFF}
The rule of thumb is that the final results should not change, within
the statistical errors, even if we remove several points.

A final comment concerns computation time. Obviously, it takes as much
total CPU time to run $10$ simulations of length $T$ as
one simulation of length $10T$. The wall-clock, defined as the physical
time we have to wait for the results, is, however, $10$ times
smaller in the former case, since the different tethered simulations 
\index{parallelisation}
\index{wall-clock}
are trivially parallelisable. Therefore, in a large-scale computation
and so long as thermalisation has been ensured, it is generally better
to add more points rather than extend the original ones.
\newpage
\section[Integrating over $\hat m$ space]{Integrating over \boldmath $\hat m$ space}
The numerical implementation of~\eqref{eq:TMC-tethered-to-canonical} can be 
done in several different ways of similar numerical accuracy. Here we briefly 
explain our choices. Once we have the $\braket{O}_{\hat m_k}$ we need
to take three steps
\begin{enumerate}
\item Interpolate the tethered averages as a smooth function of $\hat m$.
\item Integrate $\braket{\hat b}_{\hat m}$ to get $\varOmega_N$.
\item Use~\eqref{eq:TMC-canonical-average} to compute the canonical averages.
\end{enumerate}  

For the first step we have used a cubic spline interpolation (see, e.g.,~\cite{press:92} 
for an implementation). \index{interpolation!cubic splines}
We have not used the so called
natural spline, which imposes vanishing curvature for $\braket{\hat b}_{\hat m}$
at $\hat m_\text{max}$ and $\hat m_\text{min}$. Instead, we 
have estimated the derivative of this function at both ending points
with a parabolic interpolation.

In order to compute the potential $\varOmega_N$ we have to integrate 
$\braket{\hat b}_{\hat m}$ for our whole $\hat m$ range. In principle we could compute
$\varOmega_N$ from
\begin{equation}\label{eq:ISING-IN-1}
I_N(\hat m) = \int_{\hat m_\text{min}}^{\hat m}\dd\hat m'\ \braket{\hat b}_{\hat m'},
\end{equation}
$I_N$ defines $\varOmega_N$ up to an additive normalisation constant.
This expression has the disadvantage of not treating symmetrically the $\hat m$ range.
In particular, 
values of $\hat m$ close to $\hat m_{\text{min}}$ will have smaller errors 
than those close to $\hat m_\text{max}$.  In order to avoid this problem, 
we use  the average of the integral in both directions,
\begin{equation}\label{eq:ISING-IN-2}
I_N(\hat m) = \frac12\left(\int_{\hat m_\text{min}}^{\hat m}\dd \hat m'\ \braket{\hat b}_{\hat m'} - 
\int_{\hat m}^{\hat m_\text{max}}\dd \hat m'\ \braket{\hat b}_{\hat m'}\right).
\end{equation} 
Notice that, assuming $\hat m_\text{min}$ and $\hat m_\text{max}$ are far enough 
as to have negligible probability, both~\eqref{eq:ISING-IN-1} and~\eqref{eq:ISING-IN-2}
are equivalent, considering the potential nature of $\Omega_N$. It is only when we consider
numerical issues that the second is preferable.

We now normalise the pdf, 
\begin{equation}
\mathcal C = \int_{\hat m_\text{min}}^{\hat m_\text{max}}\dd \hat m\ \exp[-NI_N(\hat m)],
\end{equation}
so that
\begin{equation}
\varOmega_N(\hat m) = I_N(\hat m) - \frac1N \log \mathcal C.
\end{equation}

Some comments are in order. First of all, naively applying this interpolation scheme
for the exponential, $p(\hat m)=\exp[-N \varOmega_N(\hat m)]$,
could introduce strong integration errors. Fortunately, 
this can be easily solved by accurately representing $\varOmega_N$, which
is a smooth function (recall the curve in logarithmic scale of Figure~\ref{fig:TMC-tethered}).

In particular, since we have represented the tethered field with a cubic spline, 
$\varOmega_N$ is a fourth order polynomial between each pair of simulated points,
which can be exactly computed. To avoid losing precision, 
we evaluate $\varOmega_N(\hat m)$ in an extended grid
that includes $3$ equally spaced intermediate points between each 
pair of simulated values of $\hat m$.  This way the Lagrange 
interpolating polynomial for each segment of 
the extended lattice (two original points plus three
intermediate ones) represents the exact integral of our spline.

Of course, the pdf,
$\exp[-N \varOmega_N(\hat m)]$, is not a polynomial anymore. It is, 
however, a smooth function, so a self consistent 
Romberg method~\cite{press:92} provides an estimate of the
integral~\eqref{eq:TMC-canonical-average}
with any required numerical accuracy. Notice that this yields
the basically  exact results for a given interpolation 
of $\braket{\hat b}_{\hat m}$, but it does not cure
any discretisation errors introduced by the spline,
which should be minimised with the prescription of the previous section.

Throughout this process, we estimate errors with the jackknife method (see Appendix~\ref{chap:correlated}).
\index{jackknife method} For instance, we compute and integrate a different cubic 
spline for each jackknife block. This way, data correlation is safely taken
into account when estimating the statistical errors.

Typically, even with a very moderate effort, the Romberg integration error
has been much smaller than the statistical one. There is one exception:
the fluctuation-dissipation formulas, such as~\eqref{eq:ISING-C}, because of 
the large cancellations between the two terms.
\index{fluctuation-dissipation}
\index{specific heat}
To solve this problem, we have computed the fluctuation-dissipation formula 
as a sum of two squares:
\begin{align}
N^{-1} C= \braket{\hat u^2} - \braket{\hat u}^2 &= 
      \int\dd \hat m\ p(\hat m) \left[ \braket{\hat u^2}_{\hat m} - \braket{\hat u}^2_{\hat m}
                                 + \braket{\hat u}^2_{\hat m} - \braket{\hat u}^2\right]\nonumber\\
         &= \int\dd\hat m\ p(\hat m) \left[ \braket{( \hat u -\braket{\hat u}_{\hat m})^2}_{\hat m}
                                + \left(\braket{\hat u}_{\hat m} - \braket{\hat u}\right)^2                           \right].
\end{align}
In spite of this, as a consistency check, we have also employed the original equation 
and forced the Romberg integral to yield the same value, by 
reducing its tolerance.

It is sometimes interesting to compute
high moments of the magnetisation. One obvious possibility
is to measure the instantaneous values
for $m^\ell(\hat m; \{\sigma_{\boldsymbol x}\})$ during the simulation
and then compute the 
tethered and canonical averages as usual. But TMC
provides an alternative way of calculating
$\langle m^\ell\rangle$. Indeed, we have 
the whole pdf $p(\hat m)$ and we know
that $\hat M = M + R$. Now, the moments
for $R$  can be easily obtained analytically, so
it suffices to compute $\langle \hat m^\ell\rangle$ 
to reconstruct $\langle m^\ell\rangle$ without 
any need for individual measurements of $m^\ell$.
For example,
\begin{align}
\langle m^2\rangle &= \langle (\hat m-1/2)^2\rangle - \frac{1}{2N}\ .\\
\langle m^4\rangle &= \langle (\hat m-1/2)^4\rangle
                              -\frac{3}{N} \langle m^2\rangle 
                              - \frac{3}{4N^2}+\frac{3}{N^3}\ .\label{eq:ISING-moments}
\end{align}
These formulas are valid for the symmetric phase, where $\langle m\rangle = 0$. We have computed the moments of the magnetisation up to
$\langle m^8\rangle$, both from individual measurements and with this
procedure, and the results are identical. This will not 
be at all surprising once we see Section~\ref{sec:ISING-numerical-performance}, where
it is shown that the correlation time for $\langle m\rangle_{\hat m}$
is $<1$ (which means that the uncertainty in $p(\hat m)$ is 
going to determine the total error). 

\section[Numerical implementation for the DAFF]{Numerical implementation of TMC for the DAFF}\label{sec:DAFF-numerical-implementation}
The tethered simulations of the DAFF were also carried 
out with a Metropolis algorithm. \index{Metropolis algorithm}
In principle we could have used the, potentially more efficient, 
cluster method studied in detail in Appendix~\ref{chap:cluster}.
Indeed,  the formulation
\index{cluster methods}
for the DAFF follows the same steps as that for the Ising model \index{Ising model},
with the exceptions that the clusters are now of constant staggered spin 
and that we now have two tethered quantities, so now the cluster 
flipping weight is
\begin{equation}\label{eq:DAFF-cluster-prob}
\omega(\{S_i\}) \propto \gamma(\hat m,m) \gamma(\hat m_\text{s},m_\text{s}),
\end{equation}
where $\gamma(\hat x,x)$ was defined in~\eqref{eq:TMC-gamma}.
Unfortunately, for the relevant temperature and magnetisation regimes
for the DAFF, the spin configurations are not well suited to cluster
methods. In particular (see Section~\ref{sec:DAFF-geometry}), the whole
configuration is dominated by two very large clusters of opposite staggered
spin, of similar size for $\hat m_\text{s}=0.5$ and with one dominant 
cluster in magnetised regions. What few spins remain outside them
are scattered in a few very small clusters and a large number of single-spin
ones. In these conditions, and given the stringent constraints on $(m,m_\text{s})$
imposed by~\eqref{eq:DAFF-cluster-prob}, the cluster flipping would only
invert loose spins and leave the large clusters untouched.

We must, therefore, stick with the Metropolis scheme. Still, there are a few 
things we can do to optimise the simulation. First of all, we avoid computing
logarithms when evaluating the Metropolis acceptance probability with 
the preparation of look-up tables\index{look-up table}, as discussed
in Section~\ref{sec:ISING-optimisation}. Now we have to consider $\mathcal O(N^2)$ values for
$(M,M_\text{s})$, 
so we cannot fit all the possibilities in a single array. However, since the contributions of $U$, $M$ and $M_\text{s}$ to the update probability are factorised, 
we can simply prepare one look-up table for each of these quantities and evaluate
the whole Metropolis probability with a simple multiplication.

The peculiarities of the DAFF in the tethered formalism afford 
us an additional possible optimisation. It turns out that the Metropolis acceptance 
is very low, of about $10\%$ in most cases (this is mainly due to the regular
magnetisation, which is in a very constrained regime due to the high applied 
fields that we consider). In addition, with the look-up table
method, the computation of the acceptance probability is faster than the generation
of a single pseudorandom number.\footnote{We use
the same combination of congruential and shift-register generator
as in Section~\ref{sec:ISING-optimisation}.} \index{random number}
\index{Metropolis algorithm!survival scheme}

With this in mind, let us suppose we are performing a sweep of the whole
lattice, updating nodes $0$ to $N-1$.\footnote{Empty nodes are simply skipped, 
we run through the lattice sequentially}
Let us call $p_i$ the probability
of flipping spin $i$ and $q_i=1-p_i$ the probability of rejecting
the change.
Then, the probability of leaving the spins from $i=k$ to
 $i=s$ unchanged 
\begin{equation}
Q_{k,s} = \prod_{j=k}^{s} q_{j} = Q_{k,s-1} q_{s}.
\end{equation}
Therefore, we can define a `survival' Metropolis scheme 
\index{Metropolis algorithm!survival scheme}
in the following way:
\begin{itemize}
\item For $k=0,\ldots,N-1$ do
\begin{enumerate}
\item Generate a single pseudorandom number $\mathcal R \in [0,1)$.
\item Compute $Q_{k,s}$ until $Q_{k,s-1}>\mathcal R \geq Q_{k,s}$.
Then flip spin $s$. 
\item  Set $k=s+1$.
\end{enumerate}
\end{itemize}
This algorithm obviously satisfies the balance condition for a 
Monte Carlo method and should be more efficient than the standard
Metropolis for low acceptance probabilities. In practice, modern 
compilers make this a strongly device-dependent optimisation. We have
found that for modern Intel processors, using the Intel C compiler \texttt{icc},
the gain is negligible 
for the acceptance of $10\%$ found in our DAFF simulations.
However, more than half of our simulations (over $3.5$ million CPU hours)
were carried out on the \emph{Mare Nostrum} supercomputer, of different 
\index{Mare Nostrum@\emph{Mare Nostrum}}
design, where the survival version of the Metropolis scheme nets us a 
$30$ to $40$\% acceleration. 

Finally, we take one parallel tempering (Section~\ref{sec:THERM-thermalisation-PT})
step for each lattice update, attempting
to exchange configurations for the same $(\hat m,\hat m_\text{s})$ 
but different $\beta$. In the tethered 
weight~\eqref{eq:TMC-omega}, the factors that depend on the magnetisation
are independent of $\beta$, so the parallel tempering acceptance probability
is exactly the same as that for a canonical simulation, given by~\eqref{eq:THERM-acceptance-PT}.
\index{parallel tempering}
For our longest simulations, we use a parallel version of our code that 
runs the Metropolis update of each of
 the participating configurations in a different processor. When the whole
lattice has been updated, the energies and temperatures of each configuration are
sent to the master process (with MPI, Message Passing Interface),
\index{MPI}
which performs the parallel tempering update
and sends back the new temperatures. In this way the total CPU time
\index{wall-clock}
is of course unchanged, but the wall-clock
is reduced by a factor $\mN_T$.

\index{tethered formalism!numerical implementation|)}

\chapter{The \textsc{Janus} computer}\label{chap:janus}\index{Janus@\textsc{Janus}|(}
The Monte Carlo simulation of complex disordered systems has an unquenchable
thirst for computing power. Any acceleration due to innovations in computer design 
is immediately exhausted.  In addition, the terrible scaling properties of the 
characteristic times (often exponential, as we pointed out in Chapter~\ref{chap:tmc})
mean that, in order to obtain a qualitative improvement in the data, one needs
an acceleration of orders of magnitude.

In these conditions, the investigation on new, optimised, Monte Carlo methods in not
only necessary but also rewarding, as we saw with our examination 
of the Tethered Monte Carlo method during Part~\ref{part:tmc} of \index{tethered formalism}
this dissertation. Sometimes, however, devising ingenious dynamical
algorithms is not an option.

A prime example of this issue is the non-equilibrium dynamics of
spin glasses. \index{spin glass!dynamics}
Here, we face the following constraints
\begin{itemize}
\item The interesting physical phenomena occur at
macroscopic times, with the typical experimental 
scale ranging from a few seconds to a few hours.
\item The physical dynamics can be faithfully simulated
with heat bath or Metropolis dynamics,   \index{heat bath algorithm} \index{Metropolis algorithm}
but each MCS is equivalent to only 
$\sim 10^{-12}$~s~\cite{mydosh:93}.
\item One has to simulate very large lattices, to keep
the system in a truly off-equilibrium regime, free
of finite-size effects. Therefore, each MCS takes  \index{finite-size effects}
some time.
\end{itemize}
Notice that it is not easy to see how one could improve 
this situation with a clever update algorithm, since we want to reproduce the physical
dynamics.

\begin{figure}
\centering
\includegraphics[width=0.5\linewidth, trim=0 40 0 40]{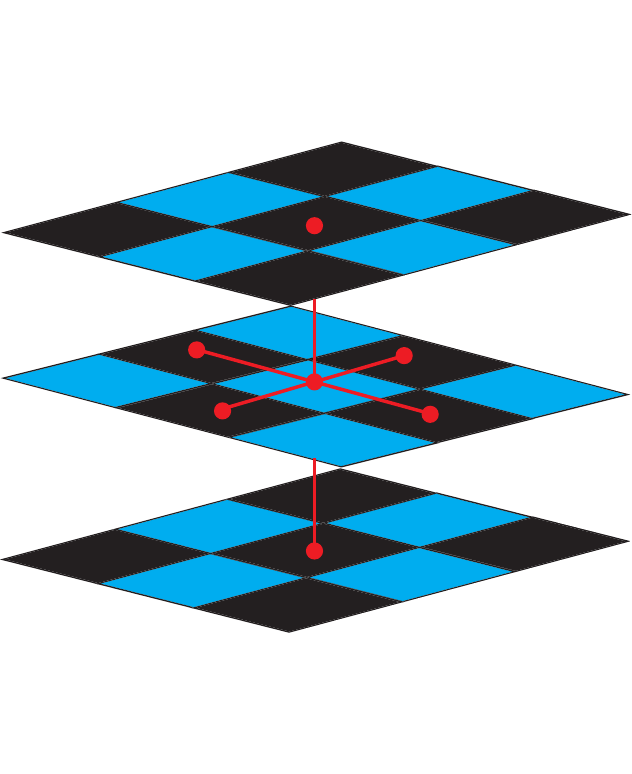}
\caption[Checkerboard scheme for the nearest-neighbour interactions]{%
Division of the spin-glass lattice in a checkerboard scheme. The first
neighbours are always the opposite colour of the site, so sites of the 
same colour can be updated at the same time.
\label{fig:JANUS-damas}
}
\end{figure}

Thankfully, the physics of the spin glass provide a different avenue
for acceleration: parallelisation.   \index{parallel algorithms}
Indeed, the nearest-neighbours nature of the interaction means that 
we can divide each plane of the lattice in a checkerboard scheme (Figure~\ref{fig:JANUS-damas}),
with sites of the same colour being statistically independent from one another. That is, 
one could in principle update the whole lattice in two steps: first all the black 
sites at once and then all the white sites at once.

The problem with this idealised scheme is that conventional computers (CPUs) 
are simply not equipped to handle it:
\begin{itemize}
\item CPUs are optimised for operations on long data words, but we need
(\textsc{i}) operations on single bits (the spins) (\textsc{ii}) variables that only 
appear in a small number of states (the local field needed to 
compute the update probability).
\item The memory architecture does not permit the processor
to gather all the necessary information quickly enough.
\item We need one random number for each node that we update, 
so random-number generation is another bottleneck. \index{random number!generation}
\end{itemize}
Sharing the computation across several CPUs (making each handle one 
portion of the lattice) does not work due to communication limitations:
even if we only need small chunks of data, they have to be accessed extremely
often.

One possible answer to this problem is designing a special-purpose computer, 
whose architecture is optimised for simulations of spin systems.
We note that this idea is not new, custom computers have been used in 
statistical mechanics for some time~\cite{ogielski:85,cruz:01}.
The \textsc{Janus} computer, fruit of a collaboration between physicists 
and engineers in the universities of Zaragoza, Complutense de Madrid
and Extremadura (in Spain) and Ferrara and Rome 1 `La Sapienza' (in Italy), is
one such custom machine.

Since the author was not involved in 
the hardware design or the low-level programming of the machine, just in the 
physical analysis  of the simulations, this is not 
the place to give a detailed account of the machine. We shall only consider
some brief notions to understand why \textsc{Janus} can outperform 
conventional computers. The interested reader is referred to~\cite{janus:06,janus:09}
for details on the architecture and to~\cite{janus:08} for an explanation
of the implementation of the Monte Carlo simulation. See also the doctoral
dissertation of F. Mantovani~\cite{mantovani:08}.

The solution adopted by \textsc{Janus} consists in replacing
the conventional CPUs by Field Programmable Gate Arrays (FPGAs) \index{FPGA}
as computing nodes.  These devices offer a large number of reconfigurable
logic resources, so they can be divided into many update engines, each 
taking care of the update of one spin.

\begin{figure}[p]
\centering
\includegraphics[width=0.6\linewidth]{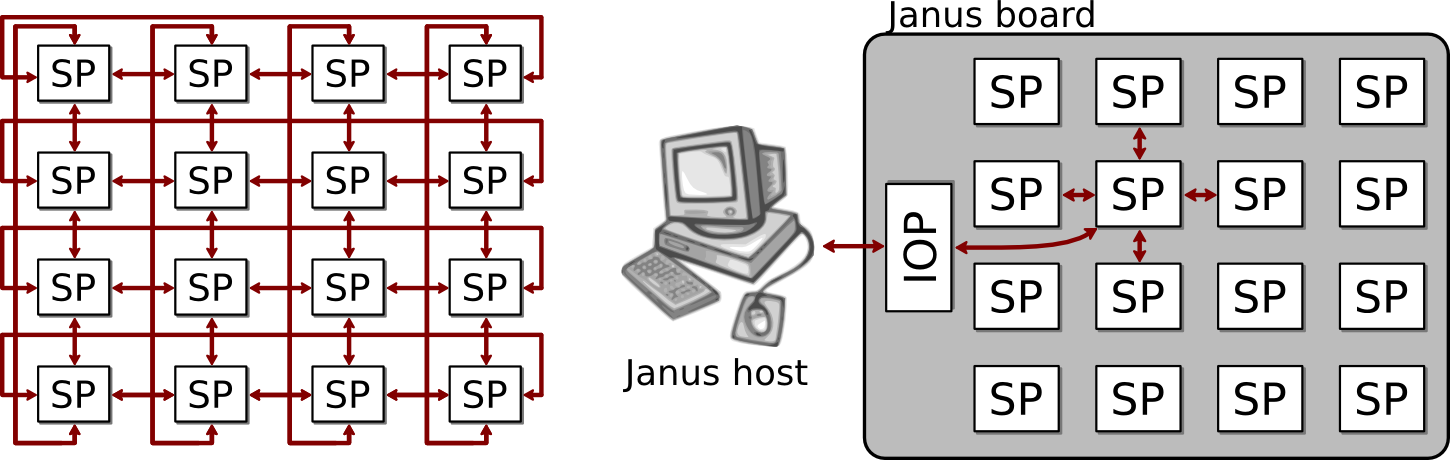}
\caption[\textsc{Janus} board]{Board of the \textsc{Janus} computer.
Each of the $16$ scientific processors (SPs) is an FPGA. These 
are connected along nearest-neighbour links, with periodic boundary 
conditions. The board includes an input/output processor.
\index{Janus@\textsc{Janus}|indemph}
\label{fig:JANUS-board}
}

\includegraphics[width=0.4\linewidth]{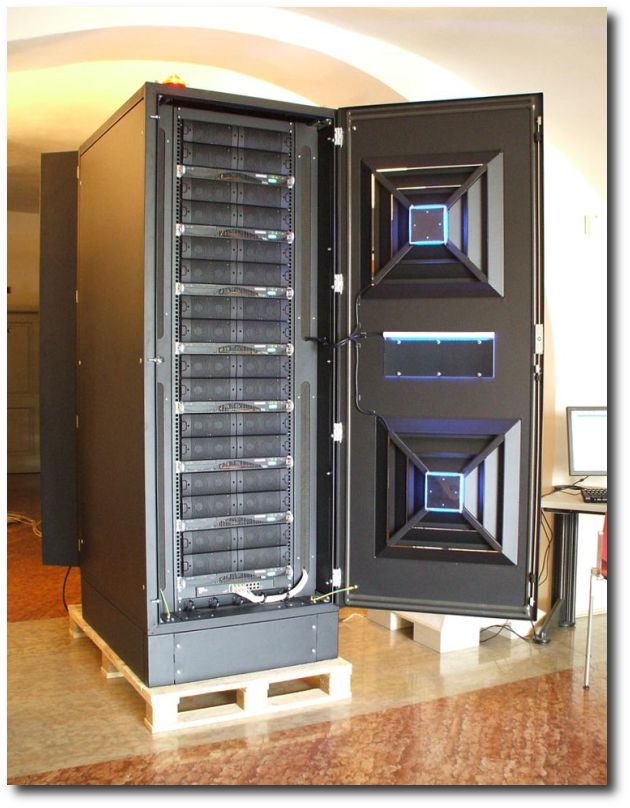}
\caption[The \textsc{Janus} computer]{The complete
\textsc{Janus} computer.
\label{fig:JANUS-rack}
}
\end{figure}

The FPGAs are grouped in $4\times4$ arrays,
forming boards (Figure~\ref{fig:JANUS-board}). Each board includes
data links among nearest neighbours and an input/output processor (IOP), whose
job is to distribute work among the computing nodes and gather the results.
There is a standard PC (called host) for each two boards, which acts as the 
interface and storage element. The whole \textsc{Janus} computer
consists  of $16$ boards (a total of $256$ processors) in one rack (Figure~\ref{fig:JANUS-rack}).

In the basic implementation for non-equilibrium dynamics, each FPGA simulates 
two replicas of one sample, divided in the checkerboard scheme.\footnote{Actually, 
for optimisation purposes,
two mixed lattices are formed, one containing the white nodes of replica
1 and the black ones of replica 2 and the other containing the complementary
sites. Then all the spins within the same mixed lattice can be updated
at the same time.} The implementation seeks to maximise the number of 
spins that can be updated in each clock cycle. In particular, each FPGA
can be divided in $1024$ update engines. Each of them takes 
the following steps
\begin{enumerate}
\item Reads as input $6$ nearest neighbours and the corresponding
$6$ couplings.
 \item Computes the local field and addresses a pre-computed
look-up table (cf. Section~\ref{sec:ISING-optimisation}). \index{look-up table}
\item Compares the result with a freshly generated random number.
\item Sets the new value for the spin.
\end{enumerate}
This whole process is pipelineable so that each of the $1024$ update engine updates
one spin per clock cycle (notice that this requires a clever memory architecture, 
so the needed information can be accessed with no latency).

Another potential bottleneck is the generation of random numbers. \index{random number!generation}
\index{Parisi-Rapuano generator}
As we indicated above, we need one random number for each spin of the $1024$ spins
that we update in each clock cycle. For these, we use the Parisi-Rapuano shift-register
generator~\cite{parisi:85}. We have a wheel $I(k)$ of $62$ numbers ($32$-bit integers). Then, in order
to generate a random number $R(K)$ we perform the following operation
\begin{align}\label{eq:JANUS-PR}
R(k) &= \bigl(I(k-24) + I(k-55)\bigr) \otimes I(k-61).
\end{align}
Then, the wheel is shifted (we increase $k$) and update the $k$ position with 
the previously computed sum
\begin{equation}\label{eq:JANUS-PR2}
I(k) = I(k-24) + I(k-55).
\end{equation}
The value of $k$ is, of course, always taken modulo $62$.

\begin{figure}
\centering
\begin{minipage}{.49\linewidth}
\includegraphics[width=\linewidth]{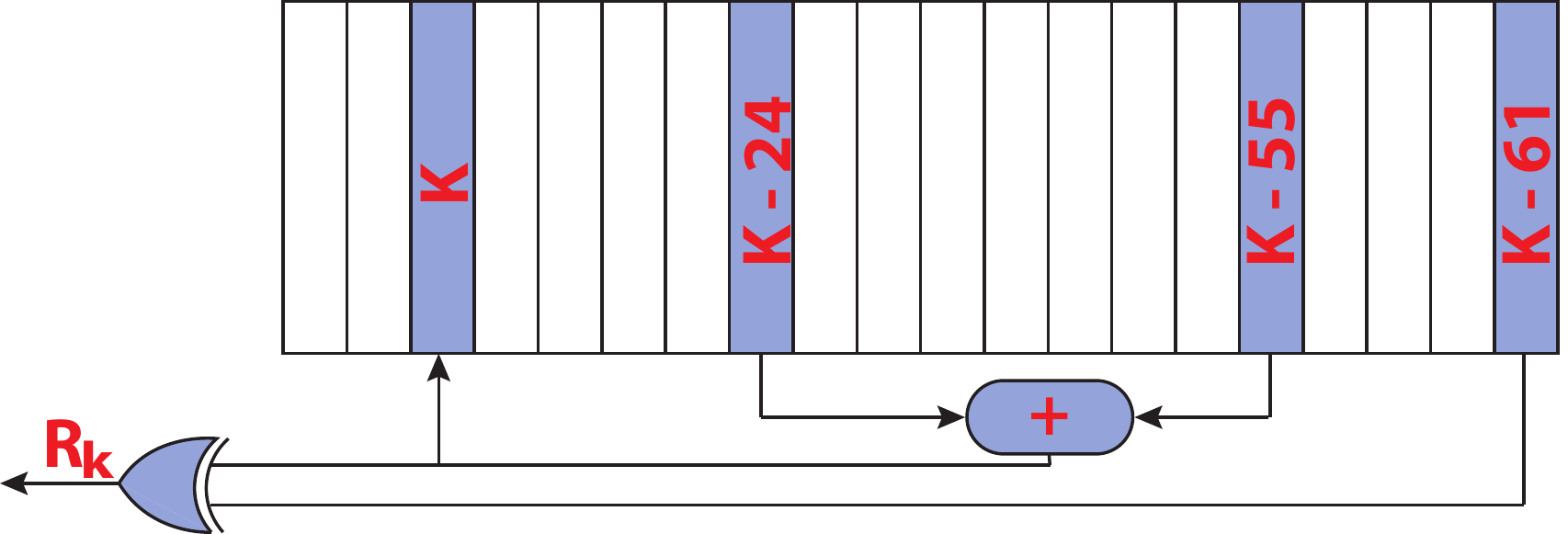}
\end{minipage}
\begin{minipage}{.49\linewidth}
\includegraphics[width=\linewidth]{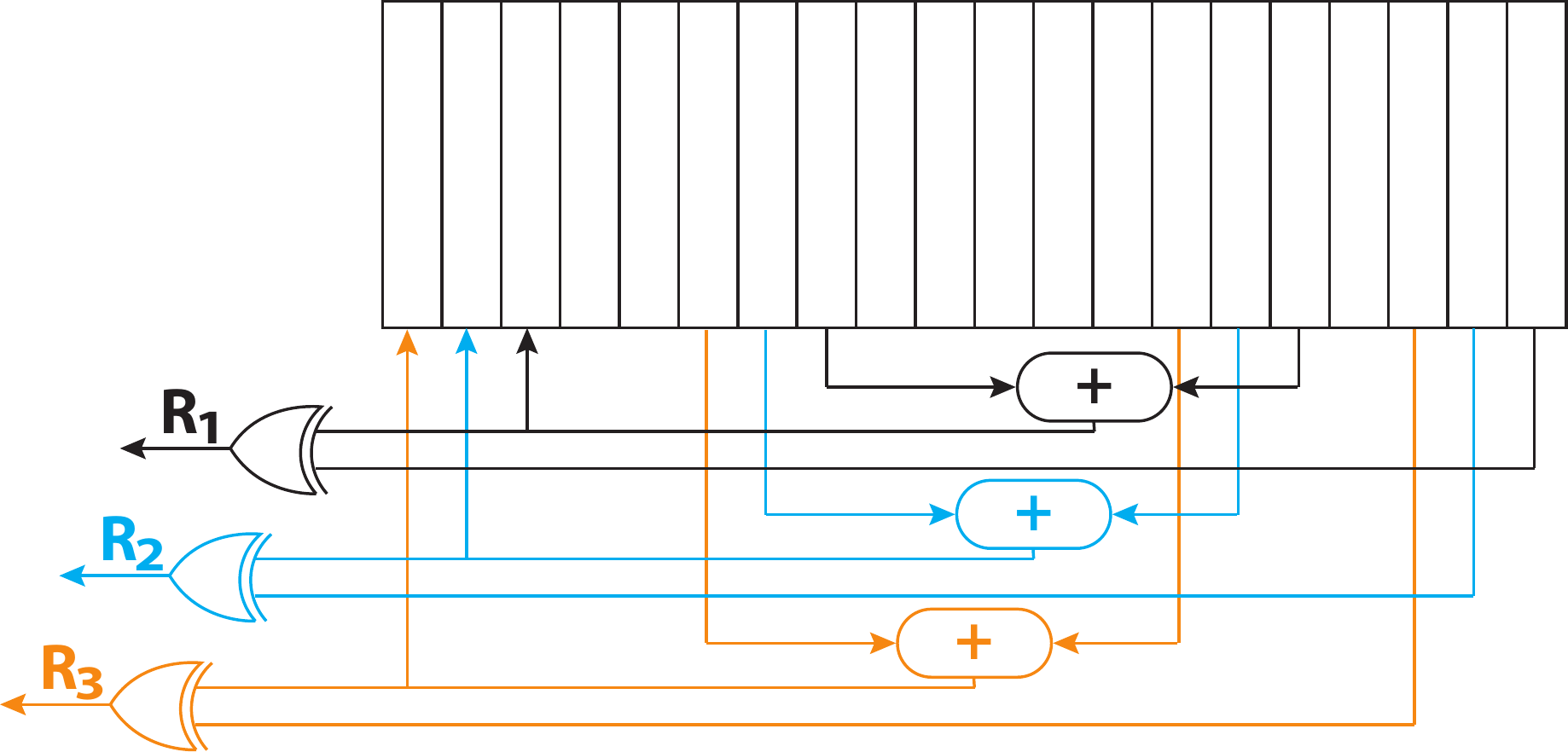}
\end{minipage}
\caption[Parisi-Rapuano generator]{\emph{Left:} One step of the 
Parisi-Rapuano generator, implemented through a logical 
circuit. \emph{Right:} Nested generation of three random 
numbers at once with the same Parisi-Rapuano wheel. At 
the cost of complicating this circuit, \textsc{Janus}
generates up to $96$ random numbers per wheel per
clock cycle.
\index{Parisi-Rapuano generator|indemph}
\index{random number!generation|indemph}
\label{fig:JANUS-PRNG}
}
\end{figure}
\begin{figure}[t]
\centering
\includegraphics[width=0.9\linewidth]{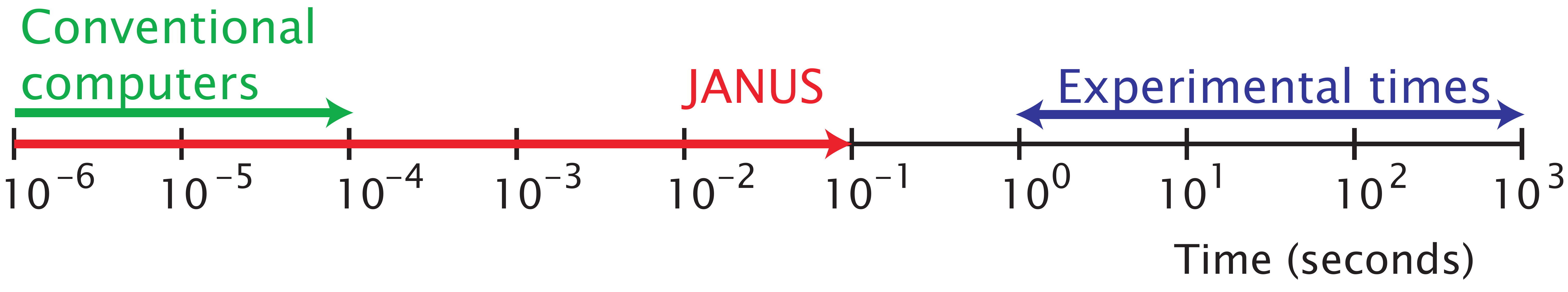}
\caption[Comparison between \textsc{Janus} and conventional
computers]{Comparison of the time scales accessible
to \textsc{Janus} and to conventional computers. 
\label{fig:JANUS-tiempos}
}
\end{figure}
The interesting feature of this generator is that it can be easily implemented through
a logic circuit (Figure~\ref{fig:JANUS-PRNG}---left). More than that, at the 
cost of complicating the circuit, we can consider the simultaneous generation 
of many random numbers from the same wheel ---several iterations of~\eqref{eq:JANUS-PR} and~\eqref{eq:JANUS-PR2}--- 
by setting up a cascade-structured combinatorial logic. This is represented
for the simultaneous generation of three random numbers in Figure~\ref{fig:JANUS-PRNG}---right.
In the actual implementation, \textsc{Janus} is capable of generating 
$96$ random numbers for each wheel in each clock cycle (we still need 
to keep several wheels turning at the same time to keep up with the 
needs of the update engines).

\section{Performance}
\textsc{Janus} is capable of simulating a system of size up to $L=96$ in each FPGA.\footnote{%
The FPGAs are Xilinx Virtex$4$-LX$200$.} The above-outlined implementation is capable
of exhausting $88\%$ of the available logic resources. The clock frequency of the machine
is of $62.5$~MHz, so the performance of each FPGA is
\begin{equation}
\frac{16 \text{ ns}}{1 \text{ cycle}}\times\frac{1 \text{ cycle}}{1024 \text{ spins}} \approx 16 \text{ ps/spin}
\end{equation}
This number is, however, too abstract. One can update many spins per second by simply running
many independent simulations at once, without accelerating any of them. The real 
measure of the performance of a computer (or an implementation) is the new physics 
to which it grants us access. In this sense, in the above implementation
\textsc{Janus} is capable of simulating $256$ samples of an $L=80$ Edwards-Anderson
spin glass for $10^{11}$ MCS in less than a month, a time long enough
to understand what happens at the experimental scale, as we 
see in Chapter~\ref{chap:sg-2} (see Figure~\ref{fig:JANUS-tiempos}).
We also note that one can simulate even larger systems by sharing the workload
among several FPGAs. For instance, a \textsc{Janus} board can simulate an
$L=256$ lattice for $10^{10}$ MCS in less than ten days. Notice that
this means that fewer samples can be simulated at once, but this is \index{self-averaging}
of no consequence due to the self-averaging nature of the physical observables.\footnote{%
At the time of running our non-equilibrium simulations, the code for the
distributed simulation of an $L=256$ system was still not ready, so we ran 
$L=80$ systems instead.}

Let us stress that \textsc{Janus} is optimised for accelerating the 
simulation of a single sample. If all one needs is to run many samples
for a comparatively short time each ---as is commonly the case for critical-point studies---
\textsc{Janus} is also going to be
very efficient, but there are simpler alternatives (the most popular
being asynchronous multi-spin coding in CPUs or even graphic cards).  \index{multi-spin coding}
In this dissertation, we consider also equilibrium simulation with 
\textsc{Janus}, but these are in the low-temperature regime where 
the thermalisation time for each sample is exceedingly long (cf. Appendix~\ref{chap:runs-sg}).

We finally note that, aside from the simulations for the Edwards-Anderson spin glass
with Ising spins reported in Appendix~\ref{chap:runs-sg}, \textsc{Janus}
has been used to simulate the Potts spin glass (where the spins can take \index{Potts model}
several values, not just $\pm1$)~\cite{janus:09c,janus:10c}. \index{spin glass!Potts} 
In this case, the gain factor with respect to conventional PCs is even greater.

\index{Janus@\textsc{Janus}|)}

\chapter{Our spin-glass simulations: parameters and thermalisation}\label{chap:runs-sg}
\index{spin glass!simulations|(}

We report in this Appendix the parameters of our spin-glass simulations
and give some technical details on our logistics and (in the case
of the ones in equilibrium) thermalisation tests.

\section{Non-equilibrium simulations}
\begin{table}
\small
\begin{tabular*}{\columnwidth}{@{\extracolsep{\fill}}clcc}
\hline
$L$ &\multicolumn{1}{c}{$T$} & $\mathcal N_\text{HB}$ & $\mathcal N_\text{samples}$\\
\hline
$80$ & $0.6$   & $10^{11}$          & $96$ \\
$80$ & $0.7$   & $10^{11}$          & $63$ \\
80 &  0.7   & $10^{10}$ & 768 \\
80 &  0.8   & ${10}^{11}$          & 96 \\
$80$ & $1.1$   & $4.2\times 10^{9}$  & $32$ \\
$40$ & $0.8$   & $2.2\times 10^8$    & $2218$ \\
\hline
\end{tabular*}
\caption[Parameters of our non-equilibrium simulations]{Parameters the
non-equilibrium simulations studied in this dissertation.
The overall wall-clock time needed was less than six weeks.
\label{tab:SG-runs-off}
}
\end{table}
We have simulated the direct quench \index{direct quench}
protocol explained in Chapter~\ref{chap:sg} for 
several temperatures. We have used 
heat bath dynamics, \index{heat bath algorithm}
in the highly parallelised 
implementation described in Appendix~\ref{chap:janus}.
The parameters of these runs are included in Table~\ref{tab:SG-runs-off}.

We have used lattices of linear size $L=80$, simulating
two real replicas for each sample. \index{replicas!real}
In a first simulation campaign, originally reported in~\cite{janus:08b},
we ran one sample in each of our $256$ FPGAs \index{FPGA}
for just under one month, resulting in $96$ samples 
for $T=0.6,0.8$, $63$ samples for $T=0.7$ and $32$
 for $T=1.1\approx T_\text{c}$.\footnote{One of the FPGAs was faulty (it has since been replaced),
so the total number of samples was actually $255$.}
Our wall-clock time of $24$ days was enough to reach $10^{11}$ MCS for our \index{wall-clock}
subcritical simulations. At the critical point we only ran for $4.2\times10^9$ 
MCS, using the freed-up time to run some shorter simulations for smaller 
lattices. These were used exclusively to check for finite-size effects (cf. 
Section~\ref{sec:SG-finite-size}). \index{finite-size effects}

We were interested in two kinds of dynamics: those at the critical temperature
and those representative of the low-temperature phase. In the latter, 
one is interested in a temperature low enough not to be dominated by
critical effects. However, the lower the temperature, the slower the 
dynamics  (where the speed is measured by the growth
rate of the coherence length). \index{coherence length}
Since one is interested in seeing as large a coherence length as possible, one
should try to find a middle ground.
After analysing the results of this 
first batch of runs in~\cite{janus:08b}, we decided that $T=0.7$ provided
the best compromise. Therefore, we carried out a new set of simulations at 
this temperature, increasing the number of simulated samples by 
an order of magnitude. In these new runs, first reported in~\cite{janus:09b},
we took only $10^{10}$ MCS,
because after that time we could begin to resolve some finite-size effects
in the coherence length growth at that temperature. 
When giving non-equilibrium results at $T=0.7$, we shall always refer 
to this larger set of runs, unless we say otherwise.

We note that $T=0.7$ is also the lowest temperature
at which we could safely thermalise our largest samples in our 
equilibrium simulations (cf. Section~\ref{sec:SG-runs-eq}, below).
Therefore, this should be considered as our main working temperature
and will, in particular, be the one we consider when establishing 
the quantitative statics-dynamics equivalence, one of our main results. \index{statics-dynamics equivalence}

Finally, let us give some logistical notes. Some of the physical quantities
we study are time-consuming to measure and some are even functions 
of the configurations at different times. Therefore, rather than taking 
complex measurements during the simulation (online) we saved to disk 
the spin configurations at logarithmically spaced times. In particular, 
we stored all the information at times $[2^{i/4}]+[2^{j/4}]$,
with integer $i$ and $j$ (the square brackets denote the integer part).
This choice was made so that our $t$ and $\tw$ (cf. Chapter~\ref{chap:sg})
could be both studied in a logarithmic scale. 

\section{Equilibrium simulations}\label{sec:SG-runs-eq}
\subsection{Set-up of the simulations}
For our equilibrium simulations we used the parallel tempering \index{parallel tempering}
algorithm described in Appendix~\ref{chap:thermalisation}. In this
case we simulated four real replicas \index{replicas!real}
for each sample (in order to have better statistics for some fine 
analyses, as well as for constructing unbiased estimators for some
sophisticated observables).  

The parallel tempering was coupled to heat bath dynamics. \index{heat bath algorithm}
The implementation in this case was a little different than that
explained for the non-equilibrium dynamics in Appendix~\ref{chap:janus}.
In particular, since we had to run many temperatures for each replica, 
each FPGA simulated just one replica, so each sample was spread across 
four computing cores. \index{FPGA} \index{Janus@\textsc{Janus}}
Due to the special architecture of \textsc{Janus}, the parallel
tempering step is not costless, as was the case for the DAFF. \index{DAFF}
This is because the FPGAs first have to compute the total energy
for each temperature. We thus equilibrate the computational cost of both
updates by performing $10$ heat bath steps for each parallel tempering 
update. This hardly affects the efficiency of the parallel tempering
scheme. 

As was the case in our off-equilibrium dynamics simulations,  most 
of the analyses were performed offline. In particular, for the shortest
simulations (the samples that do not need extending after applying our
thermalisation criteria) \index{thermalisation}
we stored on disk about $\sim 100$ evenly spaced configurations. This 
number grew proportionally to the length in the case of
extended runs. Notice that this is in stark contrast with the DAFF,  \index{DAFF}
where  we just saved the last configuration to act as a checkpoint
for possible extensions.\footnote{Actually, these configurations were
also used for the geometrical study of Section~\ref{sec:DAFF-geometry}.}

In particular, we needed many configurations on disk to perform
the analysis of fixed-$q$ correlation functions, as we explain
in Section~\ref{sec:SG-offline}.

Finally, for a few specific samples (one for $L=24$ and four for $L=32$)
the wall-clock time required to fulfil our thermalisation criteria
(cf. Appendix~\ref{chap:thermalisation}) was exceedingly  \index{wall-clock}
long, more than six months. For these cases we used a special low-level code that spread
the different temperatures for each replica across several FPGAs, \index{replicas!real} \index{FPGA}
thus speeding up the simulation (recall that \textsc{Janus} \index{Janus@\textsc{Janus}}
has very fast connections between the FPGAs on the same board). \index{FPGA}

For the smaller lattices ($L\leq12$) we ran the simulations on \textit{Terminus}, 
the computing cluster of the BIFI. \index{Terminus@\emph{Terminus}}
Even if these were very small systems, we thermalised them down to extremely 
low temperatures, so the simulation time was far from negligible.

\subsection{Parameters of our simulations}
\begin{table}
\centering
\small
\label{tab:parameters}
\begin{tabular*}{\columnwidth}{@{\extracolsep{\fill}}ccccrlllcc}
\toprule
 $L$ &  $T_{\mathrm{min}}$ &  $T_{\mathrm{max}}$ &
  $\mN_T$ &  $\mN_\mathrm{mes}$ &  $\mN_\mathrm{HB}^\mathrm{min}$ &
\multicolumn{1}{c}{ $\mN_\mathrm{HB}^\mathrm{max}$}& \multicolumn{1}{c}{ $\mN_\mathrm{HB}^\mathrm{med}$} &
 $\mN_\mathrm{s}$ & System\\
\toprule
 8 & 0.150 & 1.575 & 10  & $10^3$ & $5.0\!\times\! 10^6$ & $8.30\!\times\!10^8$    & $7.82\!\times\!10^6$ & 4000 & PC    \\ 
 8 & 0.245 & 1.575 &  8  & $10^3$ & $1.0\!\times\! 10^6$ & $6.48\!\times\!10^8$    & $2.30\!\times\!10^6$ & 4000 & PC    \\
12 & 0.414 & 1.575 & 12  & $5\!\times\! 10^3$ & $1.0\!\times\! 10^7$ & $1.53\!\times\!10^{10}$ & $3.13\!\times\!10^7$ & 4000 & PC \\
16 & 0.479 & 1.575 & 16  & $10^5$ & $4.0\!\times\! 10^8$ & $2.79\!\times\!10^{11}$ & $9.71\!\times\!10^8$ & 4000 & Janus \\
24 & 0.625 & 1.600 & 28  & $10^5$ & $1.0\!\times\! 10^9$ & $1.81\!\times\!10^{12}$ & $4.02\!\times\!10^9$ & 4000 & Janus \\
32 & 0.703 & 1.549 & 34  & $2\!\times\!10^5$ & $4.0\!\times\! 10^9$ & $7.68\!\times\!10^{11}$ & $1.90\!\times\!10^{10}$ & 1000 & Janus \\
32 & 0.985 & 1.574 & 24  & $2\!\times\!10^5$ & $1.0\!\times\! 10^8$ & $4.40\!\times\!10^9$ & $1.16\!\times\!10^8$ & 1000 & Janus \\
\bottomrule
\end{tabular*}
\caption[Parameters of our parallel tempering
simulations]{Parameters of our spin-glass parallel tempering simulations. \index{parallel tempering|indemph}
 In all cases we have simulated four independent real replicas per
  sample. The $\mathcal N_T$ temperatures are uniformly distributed between
  $T_\mathrm{min}$ and $T_\mathrm{max}$ (except for the runs of the
  first row, which have all the temperatures of the second one plus
  $T=0.150$ and $T=0.340$).  In this table $\mN_\mathrm{mes}$ is the
  number of Monte Carlo Steps between measurements (one MCS consists
  of $10$ heat-bath updates and $1$ parallel-tempering update). The
  simulation length was adapted to the thermalisation time of each
  sample, using the methods of Appendix~\ref{chap:thermalisation}.
 The table   shows the minimum, maximum and medium simulation times
  ($\mN_\mathrm{HB}$) for each lattice, in heat-bath steps.  Lattice
  sizes $L=8,12$ were simulated on conventional PCs, while sizes
  $L=16,24,32$ were simulated on \textsc{Janus}. Whenever we have two runs with
  different $T_\mathrm{min}$ for the same $L$ the sets of simulated samples
  are the same for both. The total spin updates for all lattice sizes sum
  $1.1\times 10^{20}$. It is interesting to compare the length 
  of these simulations with the ones for the DAFF in Table~\ref{tab:DAFF-parametros-PT}.
  \index{DAFF}
  \label{tab:SG-parameters-eq}}
\end{table}
In our DAFF simulations,  \index{DAFF}
we set-up our parallel tempering scheme \index{parallel tempering}
with a very high maximum temperature, where the copies of 
the system achieved decorrelation very fast.

Here, however, we are not interested in the critical point, \index{critical point}
but rather in the low-temperature physics, which makes thermalisation 
all the more difficult. For that reason, the simulations are several 
orders of magnitude longer than in the DAFF's case and, hence, for
the same ratio of autocorrelation time at $T_\text{max}$ over total simulation time
we do not need to reach such high temperatures.
Still, we can perform a quantitative analysis in order to justify our parameters (which
can be seen in Table~\ref{tab:SG-parameters-eq}).

Following Ogielski~\cite{ogielski:85}, the equilibrium autocorrelation time in the 
thermodynamic limit is taken from a power law to a critical divergence
\begin{equation}\label{eq:SG-tau-high}
\tau_\text{HB}(T) \sim (T-T_\text{c})^{-z \nu}\, .
\index{critical exponent!z@$z$}
\index{critical exponent!nu@$\nu$}
\end{equation}
For instance, for the maximum temperature used in our largest lattice, 
Ogielski found $\tau_\text{HB}\sim 10^5$. This is several orders
of magnitude shorter than our shortest simulations in Table~\ref{tab:SG-parameters-eq}.

We chose the minimum temperature so that the whole simulation campaign
took about $200$ days of the whole \textsc{Janus} machine \index{Janus@\textsc{Janus}}
and so that the lowest temperature scaled as $T_\text{c}-T_\text{min} \sim L^{-1/\nu}$.
With $4000$ samples for $L=16,24$ and $1000$ for $L=32$, this resulted 
in $T_\text{min}=0.479,0.625,0.703$, respectively. 

As we cautioned in Appendix~\ref{chap:thermalisation}, trying to optimise
the choice of the remaining parameters (number and distribution of the intermediate
temperatures) is an unrewarding task. We dedicated several weeks to
testing several combinations, mainly by modifying the number $\mathcal N_T$ of 
temperatures so that the acceptance of the parallel tempering exchanges \index{parallel tempering}
varied between $7\%$ and $30\%$. Perhaps against conventional wisdom,
taking into account that the computational effort
at fixed $\mathcal N_\text{HB}$ is proportional to $\mathcal N_T$, 
we found that the efficiency hardly changed. Eventually, we settled for a spacing 
of temperatures that produced acceptances of $\approx 20\%$. This both 
avoided  unconventionally low acceptances and saved disk space.

\subsection{Offline evaluation of observables}\label{sec:SG-offline}
As we indicated before, most of our analyses were performed offline, 
on previously stored spin-glass configurations. 
In particular,  the computation of conditional correlation
 functions $C_4(\vn{r}|q)$ ---Eq.~\ref{eq:SG-C4-q}--- was not
only time-consuming but also required a lot of unrefined data.
This was a problem, due to the scarcity of stored configurations.
 In fact, for the samples that were
simulated only for the minimum simulation time, we had only
$\mN_\mathrm{conf} \sim 100$ configurations stored on disk (ranging from
$\mN_\mathrm{conf}=10$ for $L=12$ to $\mN_\mathrm{conf} = 200$ in the case
$L=32$).  We regard the second half (in a Monte Carlo time sense) of
these configurations as well thermalised. Yet, when forming the
overlap field, one needs only that the two \index{overlap!spin}
spin configurations, $\{s_{\vn{x}}^{(1)}\}$ and
$\{s_{\vn{x}}^{(2)}\}$, be thermalised and independent. Clearly
enough, as long as the two configurations belong to different real \index{replicas!real}
replicas and belong to the second half of the Monte Carlo history they
will be suitable. There is no need that the two configurations were
obtained at the same Monte Carlo time (as it is done for the online
analyses). Furthermore, the four real replicas offer us 6 pair
combinations.  Hence, we had at least $6\times (\mN_\mathrm{conf}/2)^2
\sim 10000$ ($60000$ for $L=32$) measurements to estimate the overlaps
and the correlation functions.  We used the Fast Fourier Transform to \index{FFT}
speed up the computation of the spatial correlations.  For those
samples that had more configurations (because their total simulation
time exceeded $\mN_\mathrm{min}^\mathrm{HB}$), we considered
nevertheless $\mN_\mathrm{conf}/2$ configurations evenly spaced along
the full second half of the simulation.

For some quantities, such as the spin overlaps, we did have a large number of online
measurements. Therefore,  When some quantity, for
instance the $p(q)$, could be computed in either way, online or
offline, we have compared them. The two ways turn out to be not only
compatible, but also equivalent from the point of view of the
statistical errors. As an example of this let us compute the following
quantity:
\begin{equation}
\sigma_\mathrm{link}^2 = \overline{\langle Q_\mathrm{link}^2\rangle} - \overline{\langle Q_\mathrm{link}\rangle}^2.
\end{equation}
For $L=32$, $T = 0.703$, the value of $\sigma_\mathrm{link}^2$ computed from 
online measurements of $Q_\mathrm{link}$ and $Q_\mathrm{link}^2$ is
\begin{equation}
N \sigma_\mathrm{link, online}^2 = 50.88(90).
\end{equation}
We could now recompute this value from offline measurements of $Q_\mathrm{link}$
and $Q_\mathrm{link}^2$. Instead, we are going to use~\eqref{eq:SG-Var-width},
which involves the intermediate step of computing
conditional expectation values and variances at fixed $q$ 
and then integrating with $p(q)$. This will serve as a test both 
of the offline measurements' precision  and of our Gaussian convolution method
for the definition of fixed-$q$  quantities. The result is
\begin{equation}\label{eq:SG-sigma-link}
N \sigma_\mathrm{link, conf}^2 = 50.81(90),
\end{equation}
The precision of $\sigma_\mathrm{link, online}^2$ and $\sigma_\mathrm{link, conf}^2$
is the same and the difference less than $10\%$ 
of the error bar, even though we only analysed $100$ configurations per 
sample for the second one. Of course, both determinations are very highly 
correlated, so the uncertainty in their difference is actually much smaller than their individual
errors. Computing the difference for each jackknife block we see that
\begin{equation}
N [\sigma_\mathrm{link,conf}^2 - \sigma_\mathrm{link, online}^2] = -0.065(79),
\end{equation}
which is indeed compatible with zero.

A final issue is the comparison of data computed in different system
sizes at the same temperatures.  Unfortunately the grids of
temperatures that we used for the different $L$ differ. Hence we have
interpolated our data by means of a cubic spline. \index{cubic splines} \index{interpolation}

\subsection{Thermalisation}\index{thermalisation|(}
\begin{figure}
\centering
\includegraphics[height=0.7\linewidth,angle=270]{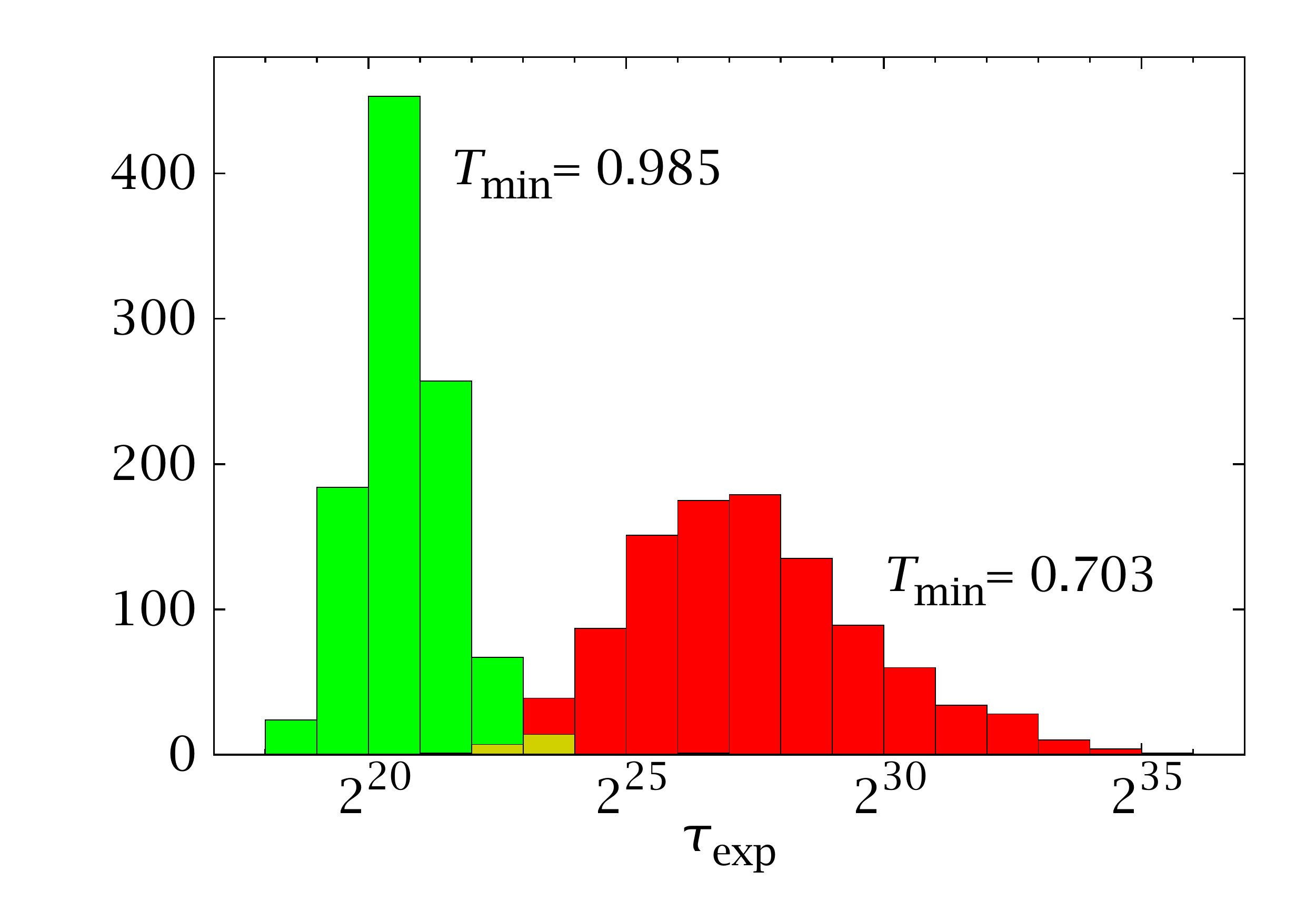}
\caption[Autocorrelation times for $L=32$]{Histogram of exponential autocorrelation times for our simulations of
the $L=32$ lattice (1000 samples).
\index{autocorrelation time!spin glass|indemph}
\label{fig:SG-tau-32}
}
\end{figure}
\begin{figure}
\centering
\includegraphics[height=0.7\linewidth,angle=270]{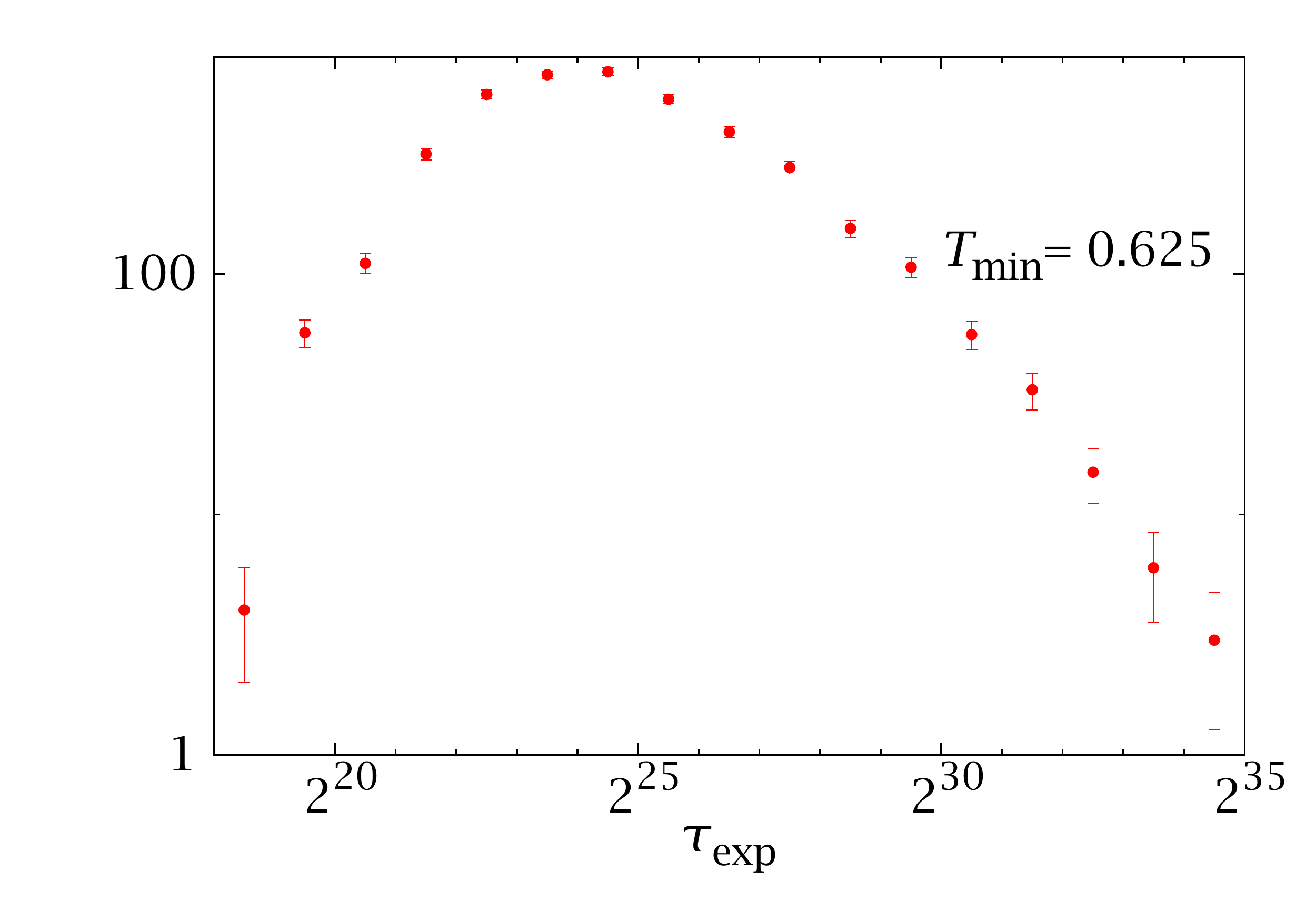}
\caption[Autocorrelation times for $L=24$]{Logarithm of the histogram of exponential autocorrelation
  times for our simulations of the $L=24$ lattice (4000 samples). Mind
  the  behaviour of the long-times tail.
\label{fig:SG-tau-24}
\index{autocorrelation time!spin glass|indemph}}
\end{figure}

We have followed the thermalisation protocol detailed in Section~\ref{sec:THERM-tau-fit}.
Figure~\ref{fig:SG-tau-32} shows the histogram of exponential  autocorrelation times
for our $L=32$ simulations. As with the DAFF, \index{DAFF}
we need to use logarithmic bins. Notice the dramatic increase of the $\tau_\text{exp}$
when decreasing the minimum temperature of the simulation. In Figure~\ref{fig:SG-tau-24}
we plot the logarithm of the histogram to show the exponential behaviour of the long-times
tail. This result gives confidence that rare events with extremely large correlation
times are at least exponentially suppressed. 

\subsubsection{Thermalisation tests}
\begin{figure}
\centering
\includegraphics[height=0.7\linewidth,angle=270]{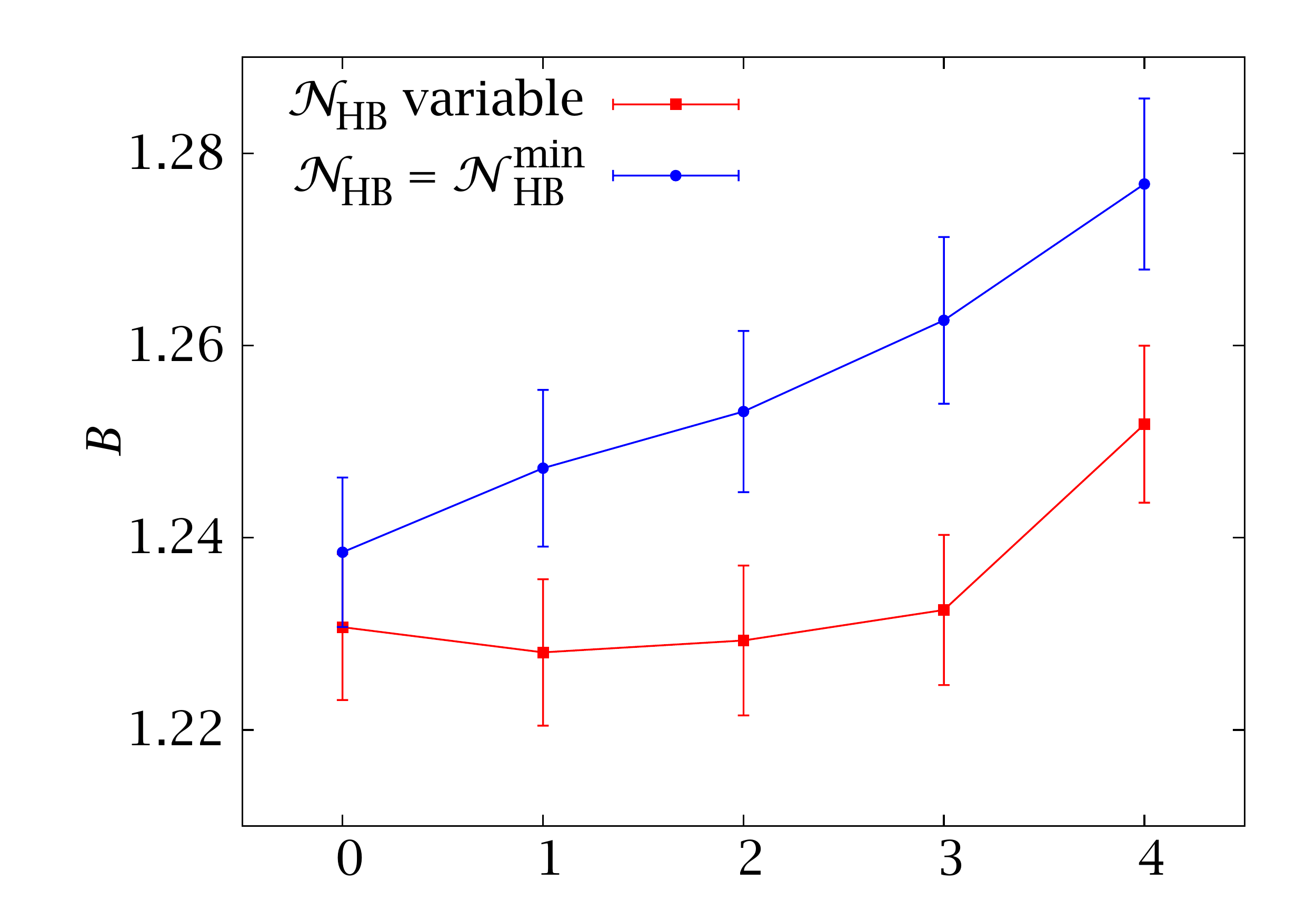}
\caption[$\log_2$-binning for the Binder ratio]{Evolution
   of the Binder parameter for $L=32$, $T=0.703$ using
  $\log_2$ binning. The blue curve
  (circles) is the result of stopping at step 1 of our thermalisation
  protocol (i.e., all samples simulated for a fixed time of $4\!\times\! 10^9$
  heat-bath updates). The red curve (squares) is the result of completing all
  the steps, which implies an increase of roughly 150\% in simulation time.
\index{Binder ratio!spin glass|indemph}
\index{log2 binning@$\log_2$-binning|indemph}}
\label{fig:SG-log2}
\end{figure}
We consider in this section thermalisation tests directly based on physically
meaningful quantities. 

We start by the traditional $\log_2$-binning procedure of Section~\ref{sec:THERM-disorder}, \index{log2 binning@$\log_2$-binning}
illustrated in Figure~\ref{fig:SG-log2} for the Binder ratio of Eq.~\eqref{eq:SG-Binder}, \index{Binder ratio}
which is especially sensitive to rare events and, therefore, to possible
thermalisation biases. In the figure we show two curves, both for $L=32$
and $T=T_\text{min}=0.703$. The blue one, with circles,
is the result of stopping at step one in our thermalisation protocol
of Section~\ref{sec:THERM-tau-fit} and shows a poorly thermalised
ensemble. The situation improves dramatically if we follow
the simulation protocol to the end, simulating
each sample for a time proportional to its autocorrelation time. \index{autocorrelation time}
Notice that, thanks to our choice of $\mN^\text{min}_\text{HB}$, 
the simulation time for most samples has not increased.  If we first rescale 
the data according to total simulation length and average for equal
rescaled time, the $\log_2$-binning gives four steps of stability 
within errors. That is to say: we obtain the Binder parameter 
without thermalisation bias just discarding $1/16$ of the history.
Regarding $B$, then, our requirement of $12\tau_\text{exp}$ has 
been excessive.

In retrospect, shorter simulations would
have produced indistinguishable physical results for most observables.
We do not regret our choices, however, as we plan to use these
thermalised configurations in the future~\cite{janus:xx} for very
delicate analyses (such as temperature chaos), which are
probably much more \index{temperature chaos}
sensitive to thermalisation effects.

\begin{figure}
\centering
\includegraphics[height=0.7\linewidth,angle=270]{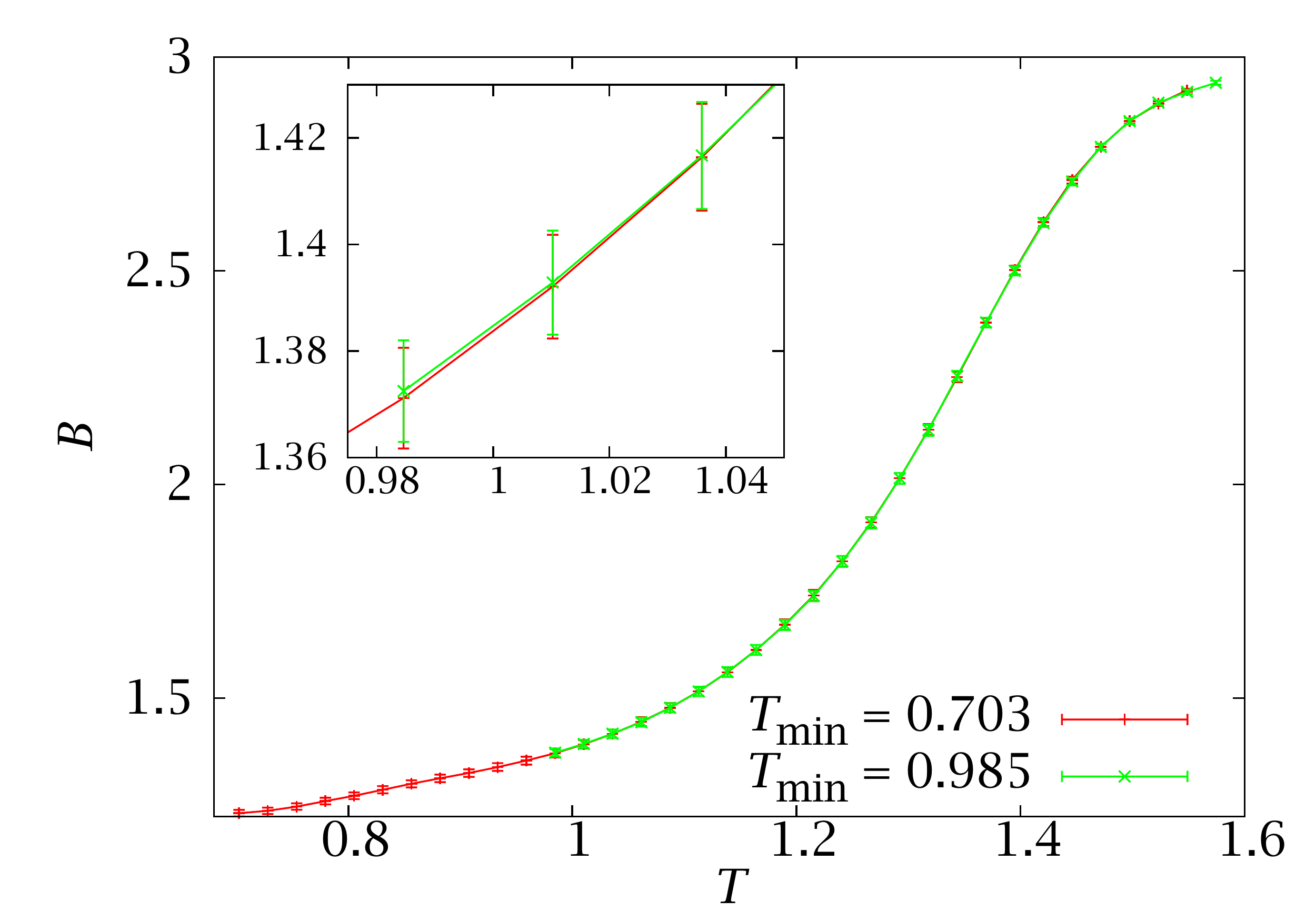}
\caption[Binder ratio for two sets of runs with different $T_\text{min}$]{Binder
   ratio as a function of the temperature for $L=32$. The good
  overlap between two different simulations (one of them in the much easier
  critical region) is a further thermalisation check. We use the same set of
  1000 samples.
\index{Binder ratio!spin glass|indemph}
\label{fig:SG-Binder-T}
}
\end{figure}

A different test can be performed by comparing the difficult
low-temperature simulations of our largest lattice with simulations
of the same samples in the much easier critical region. A faulty thermalisation
(for instance, a configuration remains trapped at low temperatures)
could be observable as inconsistencies in the values of quantities in
common temperatures. In Figure~\ref{fig:SG-Binder-T} we show the Binder
parameter as a function of temperature for the two simulations with
$L=32$ (see Table~\ref{tab:SG-parameters-eq}). The agreement between both
simulations is excellent.

\begin{figure}
\centering
\includegraphics[height=\linewidth,angle=270]{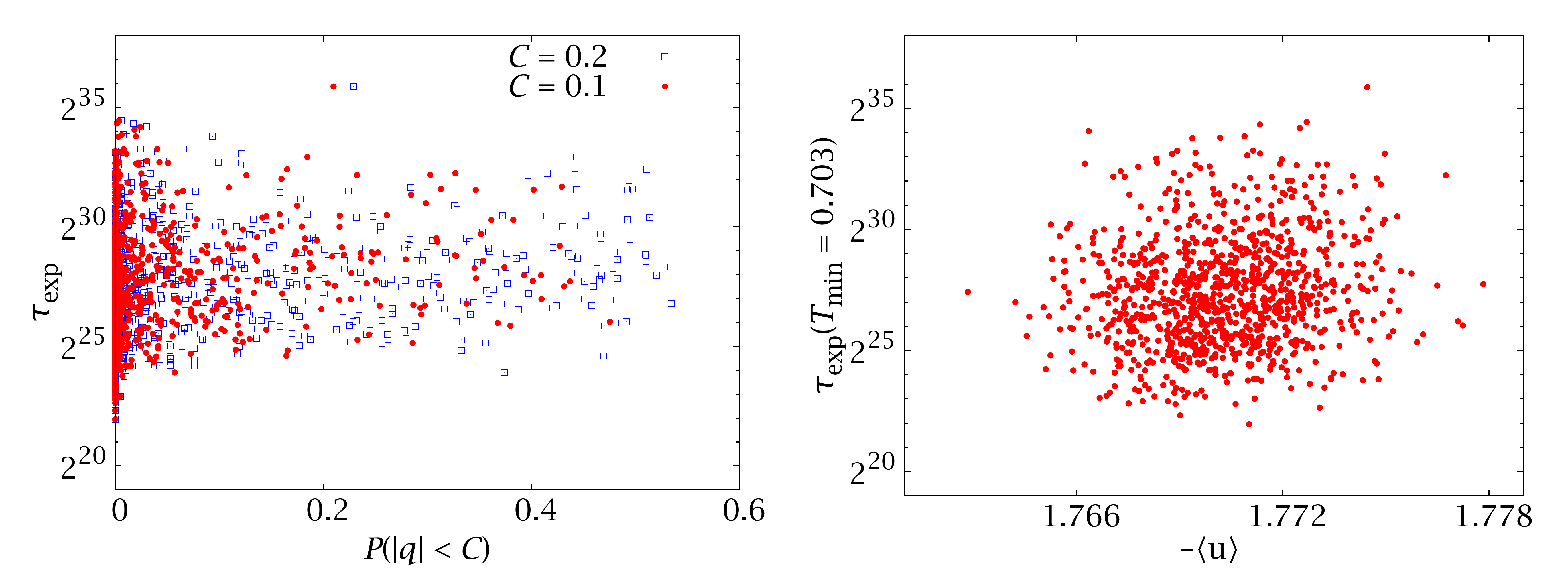}
\caption[Scatter plot of $\tau_\text{exp}$ vs. physical 
observables]{\emph{Left:} Scatter plot of the exponential
   autocorrelation time against the probability that the overlap
   be less than a small quantity at $T=0.703$, for all our $L=32$ samples.
   \emph{Right:} Scatter plot of the autocorrelation time 
   and the energy at $T=0.703$ for the same samples.
   The thermalisation time does not seem to be correlated
   with these physically relevant quantities.
\label{fig:SG-tau-vs-prob}
\index{energy!spin glass|indemph}
\index{autocorrelation time!spin glass|indemph}}
\end{figure}

We carefully avoided making decisions during thermalisation based on
the values of physical quantities. However, one could worry about the
possibility of important statistical correlations between the
temperature random walk and interesting quantities. Such correlation
could originate some small biases that would be difficult to
eliminate. Fortunately, we have not found any correlation of this
type. In Figure~\ref{fig:SG-tau-vs-prob} we show the correlation between
$\tau_\mathrm{exp}$ and two important quantities: probability of the
overlap being small and the energy.

\index{thermalisation|)}
\index{spin glass!simulations|)}



\cleardoublepage
\phantomsection
\nolinenumbers
\addcontentsline{toc}{chapter}{\protect\numberline{}Bibliography}
{\small\bibliography{biblio}}
\bibliographystyle{tesisalphnum}


\index{Tethered Monte Carlo |see {tethered formalism}}
\index{tethered @tethered|see{tethered formalism}}
\index{ensemble equivalence @ensemble equivalence|seealso{saddle point}}
\index{Helmholtz potential|see{effective potential}}
\index{cluster methods @cluster methods|seealso{Swendsen-Wang algorithm}}
\index{thermalisation @thermalisation|seealso{autocorrelation time}}
\index{Gibbs free energy|see{free energy}}
\index{replicon exponent @replicon exponent|seealso{$\theta(q)$}}
\index{y@$y$|see{stiffness exponent}}
\index{Edwards-Anderson model @Edwards-Anderson model|seealso{spin glass}}
\index{RFIM @RFIM|seealso{DAFF}}
\index{ferromagnets @ferromagnets|seealso{Ising model}}
\index{Gaussian bath|see{demons}}

\scriptsize
\cleardoublepage
\phantomsection
\addcontentsline{toc}{chapter}{\protect\numberline{}Alphabetic index}
\printindex[default][{\small Boldface page numbers refer to
mentions in a figure or table.}]

\end{document}